\pdfoutput=1

\newif\ifcameraready
\camerareadytrue

\newcounter{version}
\ifcameraready
\else
\fi

\newcommand{\thesisTitleFrontmatter}{ENABLING EFFICIENT AND SCALABLE\\DRAM READ DISTURBANCE MITIGATION\\
VIA NEW EXPERIMENTAL INSIGHTS\\INTO MODERN DRAM CHIPS}

\newcommand{\thesisTitlePlain}{Enabling Efficient and Scalable DRAM Read Disturbance Mitigation via New Experimental Insights into Modern DRAM Chips}

\newcommand{\thesisAuthor}{Abdullah Giray Ya{\u{g}}l{\i}k{\c{c}}{\i}}
\newcommand{\thesisUni}{\protect{ETH Z\"urich}}

\newcommand{\thesisYear}{2024}
\newcommand{\thesisVersion}[0]{\theversion.0}

\documentclass[12pt,oneside,a4paper]{ethzthesis}

\usepackage{siunitx}
\DeclareSIUnit\byte{B}
\usepackage{algorithm2e}
\usepackage{xcolor}

\usepackage{multirow}
\usepackage{float}
\usepackage{bm}
\usepackage{amsmath}
\usepackage{glossaries}
\usepackage{booktabs} 
\usepackage{tcolorbox}
\usepackage{adjustbox}
 
\usepackage[textwidth=20mm, prependcaption,textsize=tiny]{todonotes}
\usepackage{fancyhdr}

\usepackage{cleveref}
\crefformat{section}{Section #2#1#3}
\crefformat{subsection}{Section #2#1#3}
\crefformat{subsubsection}{Section #2#1#3}
\crefformat{appendix}{Appendix #2#1#3}

\definecolor{amber}{rgb}{1.0, 0.49, 0.0}
\definecolor{awesome}{rgb}{1.0, 0.13, 0.32}
\definecolor{dollarbill}{rgb}{0.52,0.73,0.4}
\definecolor{moegi}{rgb}{0.357, 0.537, 0.188}
\definecolor{burgundy}{rgb}{0.5, 0.0, 0.13}
\definecolor{ballblue}{rgb}{0.13, 0.67, 0.8}
\definecolor{ups-truck}{rgb}{0.53, 0.28, 0.21}
\definecolor{airforceblue}{rgb}{0.36, 0.54, 0.66}
\definecolor{cadmiumgreen}{rgb}{0.0, 0.42, 0.24}
\definecolor{darkcyan}{rgb}{0.0, 0.55, 0.55}
\definecolor{caribbeangreen}{rgb}{0.0, 0.8, 0.6}
\definecolor{flamingopink}{rgb}{0.99, 0.56, 0.67}
\definecolor{jazzberryjam}{rgb}{0.65, 0.04, 0.37}
\definecolor{mediumpersianblue}{rgb}{0.0, 0.4, 0.65}
\definecolor{coolblack}{rgb}{0.0, 0.18, 0.39}
\definecolor{bleudefrance}{rgb}{0.19, 0.55, 0.91}
\definecolor{ao}{rgb}{0.0, 0.0, 1.0}
\definecolor{babyblueeyes}{rgb}{0.63, 0.79, 0.95}
\definecolor{antiquefuchsia}{rgb}{0.57, 0.36, 0.51}

\newcommand{\agy}[2]{\ifnum#1=\value{version}\textcolor{blue}{#2}\else{#2}\fi}
\newcommand{\agycomment}[2]{\ifnum#1=\value{version}\todo[size=\scriptsize, linecolor=orange, bordercolor=orange, backgroundcolor=white]{\textcolor{blue}{\textbf{@gy:} #2}}\else{}\fi}

\newcommand{\copiedlabel}[2]{\ifnum#1=\value{version}\todo{\tiny{\textcolor{burgundy}{\textbf{Copied from} #2}}}\else{}\fi}
\newcommand{\om}[2]{\ifnum#1=\value{version}\textcolor{blue}{#2}\else{#2}\fi}

\newcommand*\circled[1]{\tikz[baseline=(char.base)]{
    \node[shape=circle,fill,inner sep=1pt] (char) {\textcolor{white}{\textbf{#1}}};}}

\newcommand{\param}[1]{#1}
\newcommand{\numchips}[1]{\textcolor{red}{<NUMBER OF CHIPS>}}
\newcommand{\nummodules}{\textcolor{red}{<NUMBER OF MODULES>}}
\newcommand{\numworkloadmixes}{\textcolor{red}{<NUMBER OF WORKLOAD MIXES>}}
\newcommand{\secref}[1]{§\ref{#1}}

\newcommand{\tabref}[1]{Table~\ref{#1}}
\newcommand{\algref}[1]{Alg.~\ref{#1}}
\newcommand{\figref}[1]{Fig.~\ref{#1}}

\newcommand{\expref}[1]{Exp.~\ref{#1}}
\newcommand{\expsref}[1]{Exps.~\ref{#1}}
\newcommand{\obsref}[1]{Obsv.~\ref{#1}}
\newcommand{\obssref}[1]{Obsvs.~\ref{#1}}

\newcommand{\tabhead}[2][10em]{
  \rotatebox{90}{\parbox{#1}{\raggedright \textbf{#2}}}}
\newcommand{\thead}[1]{\tabhead[30mm]{#1}}

\providecommand{\footref}[1]{\textsuperscript{\ref{#1}}}

\newcommand{\vshort}[0]{\vspace{0em}}

\newcounter{obs}
\newcommand\observation[1]{\refstepcounter{obs}
   \noindent
   \colorbox{gray!20}{\textbf{Observation \theobs.}} \emph{#1}}
   
\newcounter{take}
\newcommand\takebox[1]{\refstepcounter{take}
    \begin{tcolorbox}[colback=white!25!white,colframe=black!65!white, arc=1pt, boxrule=0.5pt, left=2pt,right=2pt,top=0pt,bottom=0pt,  title=\textbf{{Takeaway \thetake.}}]
        \emph{{#1}}
   \end{tcolorbox}
}
\newcommand\take[1]{\refstepcounter{take}
    \begin{tcolorbox}[colback=white!25!white,colframe=black!65!white, arc=1pt, boxrule=0.5pt, left=2pt,right=2pt,top=0pt,bottom=0pt,  title=\textbf{{Takeaway \thetake.}}]
        \emph{{#1}}
   \end{tcolorbox}
}

\newcommand{\rowhammer}[0]{RowHammer}
\newcommand{\blockhammer}[0]{BlockHammer}
\newcommand{\nind}[0]{\noindent}
\newcommand{\head}[1]{\noindent\textbf{#1.}}

\newcommand{\hcfirst}[0]{HC_{first}}
\newcommand{\wlcnt}{\numworkloadmixes{}}
\newcommand{\tras}{t_{RAS}}
\newcommand{\trp}{t_{RP}}

\newacronym{nrh}{$N_{RH}$}{}
\newacronym{wcdp}{$WCDP$}{worst-case data pattern}
\newacronym{trc}{$t_{RC}$}{row activation cycle}
\newacronym{trcd}{$t_{RCD}$}{row activation latency}
\newacronym{tcl}{$t_{CL}$}{column access latency}
\newacronym{tcwl}{$t_{CWL}$}{column write latency}
\newacronym{tfaw}{$t_{FAW}$}{four row activation window}
\newacronym{trefw}{$t_{REFW}$}{refresh window}
\newacronym{trefi}{$t_{REFI}$}{refresh interval}
\newacronym{trrslack}{$t_{RefSlack}$}{{the maximum delay between the time a {periodic}/{preventive} refresh is generated and the time the refresh is performed}}
\newacronym{tapa}{$t_{APA}$}{the latency of issuing $ACT-PRE-ACT$ command sequence}
\newacronym{ref}{$REF$}{refresh}
\newacronym{rfm}{$RFM$}{refresh management}
\newacronym{act}{$ACT$}{activate}
\newacronym{pre}{$PRE$}{precharge}
\newacronym{rd}{$RD$}{read}
\newacronym{wr}{$WR$}{write}
\newacronym{trfc}{$t_{RFC}$}{refresh latency}
\newacronym{iqr}{$IQR$}{interquartile range}
\newacronym{cv}{$CV$}{the coefficient of variation}
\newacronym{hc}{$HC$}{hammer count}
\newacronym{pth}{$p_{th}$}{{PARA's probability threshold}}
\newacronym{pf}{$p_{failure}$}{failure probability over a sufficiently long time}
\newacronym{prh}{$p_{RH}$}{reliability target for a \gls{trefw}}
\newacronym{cchip}{$D_{chip}$}{chip density}
\newacronym{rbcpki}{$RBCPKI$}{row buffer conflicts per kilo instruction}
\newacronym{mpki}{$MPKI$}{misses per kilo instruction}
\newacronym{vdd}{$V_{DD}$}{supply voltage}
\newacronym{vpp}{$V_{PP}$}{wordline voltage}
\newacronym{vppmin}{$V_{PPmin}$}{the lowest \gls{vpp} at which the DRAM module can successfully communicate with the FPGA}
\newacronym{vwl}{$V_{PP}$}{wordline voltage}
\newacronym{gnd}{$GND$}{ground}
\newacronym{vgs}{$V_{GS}$}{gate-to-source voltage}
\newacronym{vthresh}{$V_{TH}$}{the voltage threshold that the bitline voltage should exceed for the activation to be reliably completed}

\newacronym{hcfirst}{\textit{HC\textsubscript{first}}}{the minimum hammer count {value} at which the first bit error is observed}
\newacronym{ber}{$BER$}{the fraction of DRAM cells in a DRAM row that experience a bitflip, referred to as bit error rate}
\newacronym{taggon}{$t_{AggOn}$}{the time that an aggressor row stays active, {i.e., aggressor row's on-time}}
\newacronym{taggoff}{$t_{AggOff}$}{the time that {the bank} stays precharged, {i.e., aggressor row's off-time}}
\newacronym{tras}{$\tras{}$}{\agy{6}{the minimum time that a row should stay open after being activated}}
\newacronym{trp}{$\trp{}$}{precharge latency}
\newacronym{trcdmin}{$t_{RCDmin}$}{{the minimum time delay {required}}}
\newacronym{trasmin}{$t_{RASmin}$}{the minimum latency required}
\newacronym{kde}{KDE}{kernel density estimate}

\newacronym{svard}{Svärd}{spatial variation aware read disturbance solutions}
\newacronym{hira}{HiRA}{hidden row activation}
\newacronym{hiramc}{HiRA-MC}{HiRA memory controller}
\newcommand{\rhmemisolationrefs}[0]{\cite{aga2017when, agarwal2018rowhammer_for, barenghi2018softwareonly, bhattacharya2016curious, bhattacharya2018advanced, bosman2016dedup, brasser2017cant, burleson2016invited, carre2018openssl, cohen2022hammerscope, cojocar2019exploiting, cojocar2020arewe, deridder2021smash, fahrjr2022when, fournaris2017exploiting, frigo2018grand, frigo2020trrespass, genssler2022onthe, gruss2015rowhammerjs_a, gruss2018another, hassan2021uncovering, hong2019terminal, jang2017sgxbomb, jattke2022blacksmith, ji2019pinpoint, khan2018analysis, kim2014flipping, kim2020revisiting, kogler2022halfdouble, kwong2020rambleed, li2014write, lipp2018nethammer, liu2022generating, mutlu2017therowhammer, mutlu2019rowhammer_a, mutlu2023fundamentally, ni2018write, orosa2021adeeper, orosa2022spyhammer, park2016experiments,lim2017active, park2016statistical, pessl2016drama_exploiting, poddebniak2018attacking, qiao2016anew, rakin2022deepsteal, razavi2016flip, ryu2017overcoming, safari2014rowhammer, seaborn2015exploiting, tatar2018defeating, tatar2018throwhammer, tobah2022spechammer, tol2022toward, tol2023dont, vanderveen2016drammer_deterministic, vanderveen2018guardion, walker2021ondram, weissman2020jackhammer, xiao2016onebit, yaglikci2022understanding, yang2019trapassisted, yao2020deephammer, yun2018study, zhang2018triggering, zhang2020pthammer, zhang2022implicit, zheng2023trojvit}}

\newcommand{\exploitingRowHammerAllCitations}[0]{\cite{fournaris2017exploiting, poddebniak2018attacking, tatar2018throwhammer, carre2018openssl, barenghi2018softwareonly, zhang2018triggering, bhattacharya2018advanced, seaborn2015exploiting, kim2014flipping, safari2014rowhammer, vanderveen2016drammer_deterministic, gruss2015rowhammerjs_a, razavi2016flip, pessl2016drama_exploiting, xiao2016onebit, bosman2016dedup, bhattacharya2016curious, burleson2016invited, qiao2016anew, brasser2017cant, jang2017sgxbomb, aga2017when, mutlu2017therowhammer, tatar2018defeating, gruss2018another, lipp2018nethammer, vanderveen2018guardion, frigo2018grand, cojocar2019exploiting,  ji2019pinpoint, mutlu2019rowhammer_a, hong2019terminal, kwong2020rambleed, frigo2020trrespass, cojocar2020arewe, weissman2020jackhammer, zhang2020pthammer, yao2020deephammer, deridder2021smash, hassan2021uncovering, jattke2022blacksmith, tol2022toward, tol2023dont, kogler2022halfdouble, orosa2022spyhammer, zhang2022implicit, liu2022generating, cohen2022hammerscope, zheng2023trojvit, fahrjr2022when, tobah2022spechammer, rakin2022deepsteal, aydin2022cyber, mus2022jolt, wang2022research, lefforge2023reverse,fahr2022theeffects, kaur2022workinprogress, cai2022onthe, li2022cyberradar, li2023fphammer, roohi2022efficient, staudigl2022neurohammer, yang2022sociallyaware, islam2022signature, tomita2022extracting, france2022modeling, kurmus2017from}}

\newcommand{\trrCitations}[0]{\cite{jedec2020jesd794c, jedec2020jesd2095a,lee2014green,micron2016ddr4, yaglikci2021security, bennett2021panopticon, devaux2021method}}

\newcommand{\understandingRowHammerAllCitations}[0]{\cite{redeker2002aninvestigation, kim2014flipping, park2014activeprecharge, park2016statistical, yang2016suppression, park2016experiments, lim2017active, ryu2017overcoming, yang2017scanning, lim2018study, yun2018study, yang2019trapassisted, gautam2019rowhammering, walker2021ondram, kim2020revisiting, jiang2021quantifying, orosa2022spyhammer, cohen2022hammerscope, khan2018analysis, agarwal2018rowhammer_for, li2014write, ni2018write, genssler2022onthe, mutlu2023fundamentally, he2023whistleblower, baeg2022estimation, frigo2020trrespass, mutlu2017therowhammer, mutlu2018rowhammer, mutlu2019rowhammer_a, olgun2023anexperimental, olgun2023dram_bender, zhou2023doublesided, zhou2024unveiling, zhou2024understanding, li2024understanding, luo2023rowpress, lang2023blaster}}

\newcommand{\understandingRowHammerRowPressCitations}[0]{\agy{6}{\cite{kim2014flipping, park2016statistical, kim2020revisiting, orosa2021adeeper, yaglikci2022understanding, luo2023rowpress, olgun2024read, luo2024experimental, olgun2023anexperimental, yang2017scanning, lim2018study, yun2018study, yang2019trapassisted, walker2021ondram, mutlu2017therowhammer, mutlu2018rowhammer, mutlu2019rowhammer_a, olgun2023dram_bender, zhou2023doublesided, zhou2024unveiling, zhou2024understanding, li2024understanding, mutlu2023fundamentally, lang2023blaster}}}

\newcommand{\understandingRowHammerSimulationCitataions}[0]{\agy{6}{\cite{redeker2002aninvestigation, ryu2017overcoming, yang2016suppression, yang2017scanning, yang2019trapassisted, gautam2019rowhammering, jiang2021quantifying, walker2021ondram, li2024understanding, zhou2023doublesided, zhou2024understanding, zhou2024unveiling}}}

\newcommand{\mitigatingRowHammerAllCitations}[0]{\cite{apple2015about, enterprise2015hpmoonshot,lenovo2015rowhammer,greenfield2012throttling, kim2014flipping, kim2014architectural,  bains2015method, bains2016rowhammer, bains2016distributed, jedec2020jesd794c, jedec2020jesd795, aichinger2015ddrmemory, aweke2016anvil, gomez2016dram_rowhammer, yang2016suppression, son2017making, seyedzadeh2017counterbased, seyedzadeh2017mitigating, seyedzadeh2018mitigating, irazoqui2016mascat, ryu2017overcoming, yang2017scanning, you2019mrloc, lee2019twice, park2020graphene, yaglikci2021security, yaglikci2021blockhammer, frigo2020trrespass, kang2020cattwo, hassan2021uncovering, qureshi2022hydra, saileshwar2022randomized, brasser2017cant, konoth2018zebram, vanderveen2018guardion, vig2018rapid, hassan2019crow, gautam2019rowhammering, kim2022mithril, lee2021cryoguard, marazzi2023protrr, zhang2022softtrr, joardar2022learning, juffinger2023csirowhammercryptographic, yaglikci2022hira, saxena2022aqua, manzhosov2022revisiting, ajorpaz2022evax, naseredini2022alarm, joardar2022machine, hassan2024acase, zhang2020leveraging,loughlin2021stop, devaux2021method, han2021surround, fakhrzadehgan2022safeguard, saroiu2022theprice, saroiu2022howto, loughlin2022moesiprime, zhou2022ltpim, hong2023dsac, mutlu2023fundamentally, marazzi2022rega, didio2023copyonflip, sharma2022areview, woo2023scalable, park2022rowhammer_reduction, wi2023shadow, kim2023a11v, guderamarao2023defending, guha2022criticality, france2022modeling, france2022reducing, bennett2021panopticon, enomoto2022efficient, arikan2022processor, tomita2022extracting, saxena2023ptguard, zhou2023dnndefender, bostanci2024comet, olgun2024abacus, vanderveen2024dynamic, vanderveen2024dram, verma2023defense}}

\newcommand{\hwBasedRowHammerMitigations}[0]{\cite{apple2015about, enterprise2015hpmoonshot,lenovo2015rowhammer,greenfield2012throttling, kim2014flipping, kim2014architectural, bains2015rowhammer,  bains2015method, bains2016rowhammer, bains2016distributed, aichinger2015ddrmemory, aweke2016anvil, son2017making, irazoqui2016mascat, ryu2017overcoming, yang2017scanning, seyedzadeh2017counterbased, seyedzadeh2017mitigating, seyedzadeh2018mitigating, you2019mrloc, lee2019twice, park2020graphene, yaglikci2021blockhammer, frigo2020trrespass, kang2020cattwo, hassan2021uncovering, qureshi2022hydra, saileshwar2022randomized, brasser2017cant, konoth2018zebram, vanderveen2018guardion, vig2018rapid, kim2022mithril, lee2021cryoguard, zhang2022softtrr, joardar2022learning, juffinger2023csirowhammercryptographic, yaglikci2022hira, saxena2022aqua, enomoto2022efficient, manzhosov2022revisiting, ajorpaz2022evax, joardar2022machine, hassan2024acase, zhang2020leveraging, loughlin2021stop, devaux2021method, han2021surround, fakhrzadehgan2022safeguard, saroiu2022theprice, saroiu2022howto, loughlin2022moesiprime, zhou2022ltpim, mutlu2023fundamentally, didio2023copyonflip, sharma2022areview, woo2023scalable, park2022rowhammer_reduction, wi2023shadow, kim2023a11v, guderamarao2023defending, guha2022criticality, france2022modeling, france2022reducing, arikan2022processor, tomita2022extracting, saxena2023ptguard, zhou2023dnndefender, yang2016suppression, bostanci2024comet, bennett2021panopticon, gomez2016dram_rowhammer, gautam2019rowhammering, wang2021discreetpara, hassan2019crow, yaglikci2021security, woo2023scalable, olgun2024abacus, marazzi2023protrr}}

\newcommand{\swBasedRowHammerMitigations}[0]{\cite{enomoto2022efficient, konoth2018zebram, vanderveen2018guardion, brasser2017cant, bock2019riprh, aweke2016anvil, zhang2022softtrr}}

\newcommand{\mcBasedRowHammerMitigations}[0]{\cite{apple2015about, enterprise2015hpmoonshot,lenovo2015rowhammer,greenfield2012throttling, kim2014flipping, kim2014architectural, bains2015method, aichinger2015ddrmemory, aweke2016anvil,  bains2016distributed, bains2016rowhammer, son2017making, irazoqui2016mascat, ryu2017overcoming, yang2017scanning, seyedzadeh2017counterbased, seyedzadeh2017mitigating, seyedzadeh2018mitigating, you2019mrloc, lee2019twice, park2020graphene, yaglikci2021blockhammer, frigo2020trrespass, kang2020cattwo, hassan2021uncovering, qureshi2022hydra, saileshwar2022randomized, brasser2017cant, konoth2018zebram, vanderveen2018guardion, vig2018rapid, gautam2019rowhammering, kim2022mithril, lee2021cryoguard, zhang2022softtrr, joardar2022learning, juffinger2023csirowhammercryptographic, yaglikci2022hira, saxena2022aqua, enomoto2022efficient, manzhosov2022revisiting, ajorpaz2022evax, joardar2022machine, hassan2024acase, zhang2020leveraging, loughlin2021stop, devaux2021method, han2021surround, fakhrzadehgan2022safeguard, saroiu2022theprice, saroiu2022howto, loughlin2022moesiprime, zhou2022ltpim, mutlu2023fundamentally, didio2023copyonflip, sharma2022areview, woo2023scalable, park2022rowhammer_reduction, wi2023shadow, kim2023a11v, guderamarao2023defending, guha2022criticality, france2022modeling, france2022reducing, arikan2022processor, saxena2023ptguard, zhou2023dnndefender}}

\newcommand{\inDRAMRowHammerMitigations}[0]{\cite{jedec2017jesd794b, jedec2020jesd794c, gomez2016dram_rowhammer, seyedzadeh2017counterbased, seyedzadeh2017mitigating, seyedzadeh2018mitigating, yang2016suppression, yaglikci2021security, marazzi2023protrr, hassan2019crow, hong2023dsac, marazzi2022rega, bennett2021panopticon}}

\newcommand{\refreshBasedRowHammerDefenseCitations}[0]{\cite{lenovo2015rowhammer, enterprise2015hpmoonshot, lee2019twice, seyedzadeh2017counterbased, seyedzadeh2017mitigating, seyedzadeh2018mitigating, vig2018rapid, irazoqui2016mascat, kang2020cattwo, park2020graphene, kim2022mithril, kim2014architectural,  bains2015method, bains2016distributed, bains2016rowhammer, aweke2016anvil, apple2015about, kim2014flipping, son2017making, you2019mrloc, yaglikci2021security, frigo2020trrespass, hassan2021uncovering, loughlin2021stop, qureshi2022hydra, devaux2021method, wang2021discreetpara, marazzi2023protrr, zhang2022softtrr, joardar2022learning, yaglikci2022hira, saroiu2022howto, bostanci2024comet, olgun2024abacus, joardar2022machine, vanderveen2024dynamic, vanderveen2024dram}}

\newcommand{\throttlingBasedRowHammerDefenseCitations}[0]{\cite{greenfield2012throttling, yaglikci2021blockhammer}}

\newcommand{\integrityBasedRowHammerDefenseCitations}[0]{\cite{saxena2023ptguard, juffinger2023csirowhammercryptographic, fakhrzadehgan2022safeguard, ajorpaz2022evax, didio2023copyonflip, manzhosov2022revisiting, guderamarao2023defending, woo2023rampart}}

\newcommand{\isolationBasedRowHammerDefenseCitations}[0]{\cite{saileshwar2022randomized, wi2023shadow, loughlin2021stop, zhou2023dnndefender, woo2023scalable, bock2019riprh,konoth2018zebram, vanderveen2018guardion, brasser2017cant, saxena2022aqua, woo2023scalable, hassan2019crow}}

\newcommand{\migrationBasedRowHammerDefenseCitations}[0]{\cite{saileshwar2022randomized, wi2023shadow, saxena2022aqua, woo2023scalable, hassan2019crow}}

\newcommand{\circuitBasedRowHammerDefenseCitations}[0]{\cite{gautam2018improvement, park2022rowhammer_reduction, kim2023a11v, gautam2019rowhammering, yang2016suppression, hassan2019crow, gomez2016dram_rowhammer, han2021surround, ryu2017overcoming, yang2017scanning, zhou2022ltpim, lee2021cryoguard, hassan2024acase}}

\newcommand{\deviceLevelDefenses}[0]{\cite{park2022rowhammer_reduction, gautam2018improvement, gautam2019rowhammering, yang2016suppression, han2021surround, ryu2017overcoming, yang2017scanning}}

\newcommand{\dramStandardCitations}[0]{\cite{jedec2008ddr3, jedec2008jesd79f, jedec2012ddr4, jedec2010jesd218, jedec2010jesd219, jedec2013standard, jedec2014lowpower, jedec2020jesd795, jedec2021jesd250c, jedec2016jesd232a,jedec2022jesd238, jedec2021jesd235d, micron2014tn4003, micron2014sdram}}

\newcommand{\inDRAMRowAddressMappingCitations}[0]{\agy{6}{\cite{kim2014flipping, smith1981laser, horiguchi1997redundancy, keeth2001dram_circuit, keeth2007dram_circuit, itoh2001vlsi, liu2013anexperimental,seshadri2015gatherscatter, khan2016parbor, khan2017detecting, lee2017designinduced, cojocar2020arewe, tatar2018defeating, barenghi2018softwareonly, patel2020bitexact, patel2024rethinking, patel2024rethinkingarxiv}}}

\newcommand{\rowHammerGetsWorseCitations}[0]{\cite{kim2014flipping, kim2020revisiting, frigo2020trrespass, mutlu2017therowhammer, mutlu2018rowhammer, mutlu2019rowhammer_a, cojocar2020arewe, mutlu2023fundamentally, luo2023rowpress}}

\newcommand{\rowHammerDefenseScalingProblemsCitations}[0]{\cite{kim2020revisiting, park2020graphene, mutlu2017therowhammer, mutlu2018rowhammer, mutlu2019rowhammer_a, mutlu2023fundamentally, hassan2021uncovering}}

\newcommand{\rhdefrefresh}[0]{\cite{apple2015about, kim2014flipping, kim2014architectural,  bains2015method, aweke2016anvil, bains2016rowhammer, bains2016distributed, son2017making, seyedzadeh2017counterbased, seyedzadeh2017mitigating, seyedzadeh2018mitigating, you2019mrloc, lee2019twice, park2020graphene, yaglikci2021security, frigo2020trrespass, kang2020cattwo, hassan2021uncovering, qureshi2022hydra, kim2022mithril, devaux2021method, lee2021cryoguard, marazzi2023protrr, zhang2022softtrr, joardar2022learning}}

\newcommand{\rhdef}[0]{\cite{apple2015about, kim2014flipping, kim2014architectural, aweke2016anvil,  bains2015method, bains2016rowhammer, bains2016distributed, son2017making, seyedzadeh2017counterbased, seyedzadeh2017mitigating, seyedzadeh2018mitigating, you2019mrloc, lee2019twice, park2020graphene, yaglikci2021security, yaglikci2021blockhammer, frigo2020trrespass, kang2020cattwo, hassan2021uncovering, qureshi2022hydra, saileshwar2022randomized, brasser2017cant, konoth2018zebram, vanderveen2018guardion, greenfield2012throttling, kim2022mithril, lee2021cryoguard, marazzi2023protrr, zhang2022softtrr, joardar2022learning, juffinger2023csirowhammercryptographic}}

\newcommand{\rhsafe}[0]{\cite{frigo2020trrespass, lee2014green, micron2014sdram}}

\newcommand{\rhworse}[0]{\cite{kim2014flipping,mutlu2017therowhammer,mutlu2019rowhammer_a,kim2020revisiting, frigo2020trrespass, hassan2021uncovering, orosa2021adeeper, jattke2022blacksmith, yaglikci2022understanding, deridder2021smash}}

\newcommand{\rhdefworse}[0]{\cite{kim2020revisiting,hassan2021uncovering, orosa2021adeeper, yaglikci2021security, yaglikci2021blockhammer, park2020graphene, qureshi2022hydra}}

\newcommand{\salprefs}[0]{\agy{6}{\cite{kim2012acase, chang2014improving, wang2020figaro, zhang2014cream, seshadri2013rowclone, seshadri2018rowclone}}}

\makeatletter
\newcommand\requiredelimiter[2][########]{%
  \ifdefined#2%
    \def\@temp{\def#2#1}%
    \expandafter\@temp\expandafter{#2}%
  \else
    \@latex@error{\noexpand#2undefined}\@ehc
  \fi
}
\@onlypreamble\requiredelimiter
\makeatother

\newcommand{\cmark}{\ding{51}}%
\newcommand{\xmark}{\ding{55}}%

\begin{document}
\frenchspacing
\raggedbottom
\selectlanguage{english}
\pagenumbering{roman}
\pagestyle{plain}

\bstctlcite{IEEEexample:BSTcontrol}
\setbiblabelwidth{1000} 

\bstctlcite{IEEEexample:BSTcontrol}

\begin{titlepage}
    \large
    \begin{center}
        \begingroup
        \MakeUppercase{Diss. ETH No. 30214}
        \endgroup
    
        \hfill

        \vfill

        \begingroup
            \textbf{\thesisTitleFrontmatter}
        \endgroup

        \vfill

        \begingroup
            A thesis submitted to attain the degree of\\
            \vspace{0.5em}
            \MakeUppercase{Doctor of Sciences}\\
            \vspace{0.5em}
            (Dr. sc. \thesisUni) \\
            
        \endgroup

        \vfill

        \begingroup
            presented by\\
            \vspace{0.5em}
            ABDULLAH G\.{I}RAY YA\u{G}LIK\c{C}I\\
            \vspace{0.5em}
            M.Sc., The University of Notre Dame du Lac\\
            \vspace{0.5em}
            born on 9 August 1988\\
        \endgroup

        \vfill

        \begingroup
            accepted on the recommendation of\\
            \vspace{0.5em}
            Prof.\ Dr.\ Onur Mutlu, examiner\\
            \vspace{0.5em}
            Prof. Dr. Daniel Gruss, co-examiner \\
            \vspace{0.5em}
            Prof. Dr. Norbert Wehn, co-examiner \\
            \vspace{0.5em}
            Dr. Stefan Saroiu, co-examiner \\
        \endgroup

        \vfill

        \thesisYear%

        \vfill
    \end{center}
\end{titlepage}

\thispagestyle{empty}

\hfill

\vfill

\noindent\thesisAuthor: \textit{\thesisTitlePlain,}
\textcopyright\ \thesisYear
\setstretch{1.3}
\cleardoublepage
\thispagestyle{empty}

\vspace*{3cm}

\begin{center}
    \agy{8}{\textit{To my loving and supportive parents -- Hatice and Ya\c{s}ar,\\
    my wife -- Bet{\"{u}}l,\\
    my sister -- Nefise Gizem,\\
    and our dearest Ey\"{u}p Alper}}
\end{center}

\medskip

\clearpage

\chapter*{Acknowledgments}
\addcontentsline{toc}{chapter}{Acknowledgments}


\agy{8}{I received tremendous help from many people throughout the six years I spent at ETH Zürich.}

\agy{8}{First and foremost, I thank my advisor, Onur Mutlu, for his continuous support and guidance throughout my PhD journey. Under his supervision, I have grown as a researcher, benefiting from the environment he cultivated in SAFARI and his trust in me. As my PhD advisor, he has always been patient with me, even when my early research projects failed. His constructive criticism and advice have been invaluable, helping me identify and prioritize the most important research problems and insightful solutions to them. Without his guidance, I might not have been able to conduct the research that made this dissertation possible.}

\agy{8}{I would like to express my sincere gratitude to my PhD committee members, Daniel Gruss, Norbert Wehn, and Stefan Saroiu, for their time and efforts to review my thesis and provide valuable feedback.}
\agy{8}{In addition, I thank Moinuddin Qureshi, Saugata Ghose, O\u{g}uz Ergin, Victor van der Veen,\agycomment{8}{does this negatively affect the medal application?} Salman Qazi, and Christian Weis for their feedback on my research, which greatly helped me improve this dissertation's quality.}

\agy{8}{I am grateful to all SAFARI members and alumni for providing a stimulating intellectual environment and for their friendship. In particular, I thank Kevin Chang, Jeremie S. Kim, Minesh H. Patel, Lois Orosa, Hasan Hassan, Rachata Ausavarungnirun, Lavanya Subramanian, Jisung Park, Jawad Haj-Yahya, and Donghyuk Lee for their support, mentorship, and being always reachable and helpful even after we scattered across the globe. I thank Geraldo Francisco de Oliveira Junior, Can F{\i}rt{\i}na, and Ataberk Olgun for their great friendship, being invaluable psychological anchors, and their professional support. Surviving PhD would not be possible without them.
In addition, I thank all my friends, co-authors, and colleagues in SAFARI, including Konstantinos Kanellopoulos, Mohammad Sadrosadati, {I}smail E. Y\"{u}ksel, F. Nisa Bostanc{\i}, Nika Mansouri Ghiasi, Rahul Bera, Haocong Luo, Yahya C. Tu\u{g}rul, O\u{g}uzhan Canpolat, Z\"{u}lal Bing\"{o}l, Nastaran Hajinazar, Jo\"{e}l Lindegger, Roknoddin Azizibarzoki, Haiyu Mao, Juan Gómez Luna, Lukas Breitwieser, Konstantina Koliogeorgi, Klea Zambaku, Rakesh Nadig, and many others for their support, company, fruitful discussions, and insightful comments.
I thank Tracy Ewen et al. for their great help throughout my PhD, including but not limited to improving my research's visibility in the community, navigating me in the mazes of Swiss bureaucracy, and kindly offering practical solutions when life as an immigrant becomes more confusing than the cutting-edge challenges in the computer architecture research.}

\agy{8}{I acknowledge the support of our industrial partners (especially Google, Huawei, Intel, and Microsoft), which have been instrumental in enabling my research on DRAM read disturbance and memory research. My PhD research was in part supported by the Google Security and Privacy Research Award and the Microsoft Swiss Joint Research Center.}

\agy{8}{I thank Robert Zemeckis, Bob Gale, Steven Spielberg, and Christopher Lloyd for creating the fictional character Dr. Emmett Brown, who inspired my love for science at a young age.}

\agy{8}{I am eternally grateful to my family and friends for their unconditional support that helped me to reach this far in my journey. 
I thank H. Harun Ser\c{c}e, \"{O}mer S. Ta\c{s}koparan, Tuna \c{C}. G\"{u}m\"{u}\c{s}, H. Do\u{g}aner S\"{u}merkan, M. Tahir K{\i}lavuz, Enes Calay{\i}r, Vehbi E. Bayraktar, Ahmet Kurtulu\c{s}, Aamir A. Khan, Muhammet \"{O}zg\"{u}r, and many others who helped me through difficult times and walked with me side-by-side spiritually.}

\agy{8}{Finally, my most important gratitude is to my parents, Hatice and Ya\c{s}ar Ya\u{g}l{\i}k\c{c}{\i}, whom I owe who I am today. I am grateful for the unconditional support, love, and patience at all times from my parents, my sister, Nefise Gizem Ser\c{c}e, and my wife, Bet\"{u}l Ta\c{s}koparan Ya\u{g}l{\i}k\c{c}{\i}, which enabled me to breathe as a human being and function as a researcher.}

\clearpage
\chapter*{\vspace{-2em}Abstract\vspace{-1em}}
\addcontentsline{toc}{chapter}{Abstract}

Improvements in main memory storage density are primarily driven by technology node scaling, which causes DRAM cell size and cell-to-cell distance to reduce significantly. Unfortunately, technology scaling negatively impacts \agy{1}{the reliability of DRAM chips} by exacerbating DRAM read disturbance, i.e., accessing a row of DRAM cells can \om{3}{cause} bitflips in data stored in other physically nearby DRAM rows. DRAM read disturbance 1)~can be reliably exploited to break memory isolation, a fundamental principle of security and privacy in memory subsystems, and 2)~existing defenses against DRAM read disturbance are either ineffective or prohibitively expensive. 
Therefore, it is critical to 
mitigate DRAM read disturbance efficiently
to ensure robust (reliable, secure, and safe) execution in future DRAM-based systems.
\agy{4}{We define two research problems to address this challenge.
First, protecting DRAM-based systems becomes increasingly more expensive over generations as technology node scaling exacerbates the vulnerability of DRAM chips to DRAM read disturbance. 
Second, many previously proposed DRAM read disturbance \agy{1}{solutions} are limited to systems that can obtain proprietary DRAM circuit design information \agy{1}{about the physical layout of DRAM rows}.}

This dissertation \agy{4}{tackles these two problems in three sets of works.}
First, we build a detailed understanding of DRAM read disturbance by rigorously characterizing the read disturbance vulnerability of off-the-shelf modern DRAM chips under varying \agy{1}{properties} of 1)~temperature, 2)~memory access patterns, 3)~spatial features of victim DRAM cells, and 4)~voltage. Our novel observations demystify the large impact of these \agy{1}{four properties} on DRAM read disturbance and explain their implications on future DRAM read disturbance-based attacks and solutions.
Second, we propose new memory controller-based mechanisms that mitigate read disturbance bitflips efficiently and scalably \agy{1}{by}
\agy{4}{leveraging insights into DRAM chip internals and memory controllers. Our mechanisms significantly reduce the performance overhead of maintenance operations that mitigate DRAM read disturbance by leveraging 1)~subarray-level parallelism and 2)~variation in read disturbance across DRAM rows in off-the-shelf DRAM chips.
Third, we propose a novel solution that does not require proprietary knowledge of DRAM chip internals to mitigate DRAM read disturbance efficiently and scalably. Our solution selectively throttles unsafe memory accesses that might cause read disturbance bitflips.}

\agy{3}{We} demonstrate that \agy{3}{it is possible to mitigate DRAM read disturbance efficiently and scalably \agy{1}{with worsening DRAM read disturbance vulnerability over generations} by 1)~}building a detailed understanding of DRAM read disturbance, \agy{4}{2)~}\agy{1}{leveraging insights into DRAM chips and} memory controllers, \agy{3}{and \agy{4}{3})~devising novel solutions that do not require proprietary knowledge of DRAM chip internals.} \agy{3}{We believe our experimental results and architecture-level solutions will enable and inspire future works targeting better reliability, performance, fairness, and energy efficiency in DRAM-based systems while DRAM-based memory systems continue to \agy{4}{scale to higher density and become more vulnerable to read disturbance.}
We hope and expect that future works will explore avenues on how to leverage the insights and solutions we provide in this dissertation to enable such advancements in DRAM-based systems.}

\clearpage
\vspace{-3em}\chapter*{Zusammenfassung}
\addcontentsline{toc}{chapter}{Zusammenfassung}

Verbesserungen in der Hauptspeicher-Speicherdichte werden hauptsächlich durch die Skalierung der Technologieknoten vorangetrieben, was dazu führt, dass die DRAM-Zellgröße und der Abstand zwischen den Zellen erheblich reduziert werden. Leider beeinträchtigt die Technologieskalierung die Zuverlässigkeit von DRAM-Chips negativ, indem sie die DRAM-Lesestörung verschärft, d.h. das Zugreifen auf eine Reihe von DRAM-Zellen kann Bitflips in Daten verursachen, die in anderen physisch nahegelegenen DRAM-Reihen gespeichert sind. DRAM-Lesestörungen können 1) zuverlässig ausgenutzt werden, um die Speicherisolation zu durchbrechen, ein fundamentales Prinzip der Sicherheit und Privatsphäre in Speichersubsystemen, und 2) bestehende Abwehrmaßnahmen gegen DRAM-Lesestörungen sind entweder ineffektiv oder prohibitively teuer. Daher ist es entscheidend, DRAM-Lesestörungen effizient zu mindern, um eine robuste (zuverlässige, sichere und sichere) Ausführung in zukünftigen DRAM-basierten Systemen zu gewährleisten. Wir definieren zwei Forschungsprobleme, um diese Herausforderung anzugehen. Erstens wird der Schutz von DRAM-basierten Systemen im Laufe der Generationen zunehmend teurer, da die Skalierung der Technologieknoten die Anfälligkeit von DRAM-Chips für DRAM-Lesestörungen verschärft. Zweitens sind viele der bisher vorgeschlagenen Lösungen für DRAM-Lesestörungen auf Systeme beschränkt, die proprietäre Informationen über das physische Layout der DRAM-Reihen erhalten können.


Diese Dissertation befasst sich mit diesen beiden Problemen in drei Arbeiten. Erstens entwickeln wir ein detailliertes Verständnis von DRAM-Lesestörungen, indem wir die Anfälligkeit handelsüblicher moderner DRAM-Chips für Lesestörungen unter verschiedenen Eigenschaften rigoros charakterisieren: 1) Temperatur, 2) Speicherzugriffsmuster, 3) räumliche Merkmale der Opfer-DRAM-Zellen und 4) Spannung. Unsere neuartigen Beobachtungen entmystifizieren den großen Einfluss dieser vier Eigenschaften auf DRAM-Lesestörungen und erklären ihre Auswirkungen auf zukünftige DRAM-Lesestörungs-basierte Angriffe und Lösungen. Zweitens schlagen wir neue speichercontroller-basierte Mechanismen vor, die Lesestörungs-Bitflips effizient und skalierbar mindern, indem sie Erkenntnisse über die internen Strukturen von DRAM-Chips und Speichercontrollern nutzen. Unsere Mechanismen reduzieren die Leistungsüberhänge von Wartungsoperationen, die DRAM-Lesestörungen mindern, erheblich, indem sie 1) Subarray-Level-Parallelismus und 2) Variation in der Lesestörung über DRAM-Reihen in handelsüblichen DRAM-Chips nutzen. Drittens schlagen wir eine neuartige Lösung vor, die kein proprietäres Wissen über die internen Strukturen von DRAM-Chips erfordert, um DRAM-Lesestörungen effizient und skalierbar zu mindern. Unsere Lösung drosselt selektiv unsichere Speicherzugriffe, die Lesestörungs-Bitflips verursachen könnten.


Wir demonstrieren, dass es möglich ist, DRAM-Lesestörungen effizient und skalierbar zu mindern, trotz der zunehmenden Anfälligkeit für DRAM-Lesestörungen über Generationen hinweg, indem wir 1) ein detailliertes Verständnis von DRAM-Lesestörungen entwickeln und Erkenntnisse über DRAM-Chips und Speichercontroller nutzen und 2) neuartige Lösungen entwickeln, die kein proprietäres Wissen über die internen Strukturen von DRAM-Chips erfordern. Wir glauben, dass unsere experimentellen Ergebnisse und Architektur-Level-Lösungen zukünftige Arbeiten ermöglichen und inspirieren werden, die auf bessere Zuverlässigkeit, Leistung, Fairness und Energieeffizienz in DRAM-basierten Systemen abzielen, während DRAM-basierte Speichersysteme weiterhin zu höherer Dichte skalieren und anfälliger für Lesestörungen werden. Wir hoffen und erwarten, dass zukünftige Arbeiten erkunden werden, wie die Erkenntnisse und Lösungen, die wir in dieser Dissertation bereitstellen, genutzt werden können, um Fortschritte in DRAM-basierten Systemen zu ermöglichen.

\pagestyle{headings}
\cleardoublepage
\tableofcontents
\newpage
\listoffigures
\newpage
\listoftables

\cleardoublepage
\pagenumbering{arabic}%

\chapter{Introduction}

Dynamic random access memory (DRAM), {first introduced in the late 1960s~\cite{dennard1968fieldeffect, dennard1974design, markoff2019ibms, electronics2018memory}, has served as the de-facto standard main memory technology across a broad range of computing systems for decades. This is primarily due to its unique design point in the trade-off space of capacity, \om{3}{access latency, and} cost-per-bit \agy{3}{(}e.g., DRAM's \om{3}{cost-per-bit} is \om{3}{greatly} lower than SRAM, \om{3}{which is used for on-chip caches,} and \om{3}{DRAM} has a \om{3}{much} lower access latency compared to \om{3}{NAND} Flash~\cite{hennessy2011computer, meena2014overview, mutlu2013memory, mutlu2015themain, chang2017thesis}\agy{3}{, which is used for solid-state drive storage)}}.

\agy{3}{With the advancements in manufacturing technology, DRAM chip manufacturers continue to reduce cost-per-bit across DRAM generations} by shrinking DRAM cells and cell-to-cell distances~\cite{kang2014coarchitecting, mandelman2002challenges, childers2015achieving, mutlu2014research, mutlu2017therowhammer, mutlu2019rowhammer_a, mutlu2019rowhammer_and, mutlu2023fundamentally}.
\agy{3}{As an artifact of increased density,} modern DRAM chips are susceptible to a widespread {phenomenon}, called \emph{DRAM read disturbance}: accessing (reading) a row of DRAM cells ({i.e., a} DRAM row) degrades the data integrity of other physically close but \emph{unaccessed} DRAM rows~\rhmemisolationrefs{}.
\agy{1}{Many prior works~\exploitingRowHammerAllCitations{} exploit} DRAM read disturbance \agy{1}{to} reliably break a fundamental building block \agy{1}{of system} robustness, \agy{1}{i.e.,} \emph{memory isolation} \om{3}{across different locations where data is shared}: accessing a memory address should \emph{not} \om{3}{have} unintended side-effects on data stored on other addresses~\cite{silberschatz2018operating}.
\emph{RowHammer}~\cite{kim2014flipping} and \emph{RowPress}~\cite{luo2023rowpress} are two prime examples of {the} {DRAM read {disturbance} phenomenon {where} a DRAM row (i.e., victim row) can experience bitflips when a nearby DRAM row (i.e., aggressor row) is} 1)~repeatedly opened (i.e., hammered) or 2)~kept open for a long period (i.e., pressed), respectively.
Many prior works demonstrate attacks on a wide range of systems {that} exploit DRAM read disturbance {to escalate {privilege}, leak private data, and manipulate critical applications~\exploitingRowHammerAllCitations{}. Therefore, it is critical to mitigate DRAM read disturbance to ensure robust \om{3}{operation of} DRAM-based systems in terms of reliability, security, and safety. 

\section{Problem \agy{3}{Definition}}
\label{sec:intro:problem}

In this dissertation, we tackle two \agy{6}{major problems that critically affect DRAM-based computing systems}. 

First, protecting DRAM-based systems becomes increasingly more expensive over generations as technology node scaling exacerbates the vulnerability of DRAM chips to DRAM read disturbance. Prior works~\rowHammerGetsWorseCitations{} experimentally demonstrate that DRAM chips from newer generations, including chips that are marketed as RowHammer-safe~\cite{frigo2020trrespass, hassan2021uncovering}, are \agy{1}{\emph{significantly}} more susceptible to read disturbance}.
For example, chips manufactured in {2018-2020} can experience \agy{1}{bitflips as a result of \om{3}{the} RowHammer effect (\om{3}{i.e.,} RowHammer bitflips)} at an order of magnitude fewer hammers \om{3}{than} chips manufactured in {2012-2013}~\cite{kim2020revisiting}. 
As {read disturbance {in} DRAM chips} worsens, ensuring {robust (i.e.,} {reliable, secure, and safe}{)} operation {becomes} more expensive in terms of performance overhead, energy consumption, and hardware complexity~\rowHammerDefenseScalingProblemsCitations{}. This is because exacerbated DRAM read disturbance leads to bitflips at fewer hammers, and preventing these bitflips requires \agy{1}{state-of-the-art DRAM read disturbance solutions to act} more aggressively. As \agy{1}{these solutions} become more aggressive, they consume a larger 1)~\agy{6}{hardware} area to detect \agy{1}{aggressor rows}, \om{3}{which increases hardware cost,} and/or 2)~memory bandwidth and energy to perform necessary maintenance operations to prevent bitflips (i.e., preventive actions) more often, \agy{1}{e.g., refreshing potential victim rows more frequently, \om{3}{which increases performance and energy overheads of DRAM read disturbance solutions}}.

Second, many previously proposed DRAM read disturbance \agy{1}{solutions}~\refreshBasedRowHammerDefenseCitations{} are limited to systems that can obtain proprietary DRAM circuit design information \agy{1}{about the physical layout of DRAM rows}. These \agy{1}{solutions} must identify \om{3}{\emph{all}} potential victim rows based on the aggressor row's address. However, DRAM communication protocols~\dramStandardCitations{} obfuscate the physical layout of DRAM rows by allowing DRAM chips to internally {translate} memory-controller-visible row addresses to physical row addresses, \agy{1}{and DRAM manufacturers classify this information highly proprietary}~\inDRAMRowAddressMappingCitations{}.
As a result, read disturbance solutions are limited to systems that can {1)}~obtain proprietary DRAM circuit design information on in-DRAM row address mapping or {2)}~modify \om{3}{the internals of and/or the interfaces to} DRAM chips.

\pagebreak
\section{Our Goal}
\label{sec:intro:goal}

{Our goal is twofold: 1)~build a detailed understanding of DRAM read disturbance, and 2)~mitigate DRAM read disturbance efficiently and scalably without requiring proprietary knowledge of DRAM chip internals by leveraging our detailed understanding.}

\section{Thesis Statement}
\label{sec:intro:thesis}

The following thesis statement encompasses our approach:

\begin{center}
\parbox{15cm}{\textit{{We can mitigate DRAM read disturbance efficiently and scalably by 1)~building a detailed understanding of DRAM read disturbance, 2)~leveraging insights into modern DRAM chips and memory controllers, and 3)~devising novel solutions that do not require proprietary knowledge of DRAM chip internals.}}}
\end{center}

\section{Our Approach}

We build a detailed understanding of DRAM read disturbance by testing real off-the-shelf DRAM chips under various environmental conditions and testing parameters. By leveraging the insights we obtained via these real chip experiments, we develop new architecture-level methods that advance the state-of-the-art in DRAM read disturbance mitigation.

\subsection{Building a Detailed Understanding of DRAM Read Disturbance}
We present an experimental characterization {using} 272 real off-the-shelf DRAM chips that implement various chip densities and die revisions from four major DRAM manufacturers. Our characterization study demonstrates how the RowHammer effects vary with four fundamental properties: 1)~{DRAM chip} temperature, 2)~\agy{1}{memory access pattern}, 3)~{victim} DRAM cell's physical location, and 4)~voltage.
We highlight that a RowHammer {bitflip} is more likely to occur 1)~in a bounded temperature range, {specific {to each DRAM cell}} {(e.g., 5.4\% of the vulnerable DRAM cells exhibit errors in \agy{1}{a range from} \SI{70}{\celsius} to \SI{90}{\celsius})}, 2)~if the aggressor row {is} active for longer time {(e.g., RowHammer vulnerability {increases} by 36\% {if an aggressor row stays} active for 15 column accesses {when it is activated})}, 3)~in {certain physical regions} of {the} DRAM module {under attack} {(e.g., 5\% of the rows are $2\times$ more vulnerable than the remaining 95\% of the rows)}, and 4)~when the aggressor row is activated by using a higher voltage (e.g., lowering voltage decreases the RowHammer bit error rate by up to 66.9\% with an average of 15.2\% across all tested chips without significantly affecting reliable DRAM operation).
{Our study has important practical implications on future RowHammer attacks and defenses, \agy{1}{which we} {describe} and analyze.}
Our observations and analyses on DRAM read disturbance's sensitivity to temperature, memory access patterns, and victim cell's physical location are published at MICRO 2021~\cite{orosa2021adeeper} and wordline voltage's effect on DRAM reliability and read disturbance at DSN 2022~\cite{yaglikci2022understanding}.\footnote{\agy{3}{The author of this thesis is one of the two co-first authors with equal contribution in the MICRO 2021 paper~\cite{orosa2021adeeper} and the sole first author of the DSN 2022 paper~\cite{yaglikci2022understanding}.}} 
\agy{6}{Our findings on temperature's and memory access patterns' effects on DRAM read disturbance already led to the first practical attack that can spy on DRAM temperature, SpyHammer~\cite{orosa2022spyhammer}, and the discovery of a new read disturbance phenomenon, RowPress~\cite{luo2023rowpress}, respectively. SpyHammer~\cite{orosa2022spyhammer} exploits RowHammer's temperature sensitivity and wields a DRAM cell's vulnerable temperature range as a measurement tool to measure a DRAM chip's temperature. RowPress~\cite{luo2023rowpress} is a new DRAM read disturbance phenomenon where keeping a DRAM row open for a long time induces bitflips in physically adjacent rows. RowPress~\cite{luo2023rowpress} can induce bitflips on real systems by forcing the memory controller to keep a DRAM row open by accessing different columns in the row at significantly lower hammer counts than RowHammer attacks (even with one row activation in extreme cases).}


\subsection{Leveraging Insights into Modern DRAM Chips and Memory Controllers}
\subsubsection{Spatial Variation-Aware Read Disturbance Solutions}
We tackle the \agy{1}{challenge of scalability with worsening DRAM read disturbance that existing DRAM read disturbance solutions face. To this end, our goal is to reduce their} performance overheads by leveraging the spatial variation in read disturbance across different memory locations in real DRAM chips.
To do so, we
1)~present the first rigorous real DRAM chip characterization study of spatial variation of read disturbance and
2)~propose Svärd, a new mechanism that dynamically adapts the existing solutions \agy{1}{to behave more or less aggressively} based on the read disturbance \agy{1}{vulnerability of a potential victim row}. 
Our experimental characterization on 144 real DDR4 DRAM chips representing 11 chip densities and die revisions demonstrates a large variation in DRAM read disturbance vulnerability across different memory locations. \agy{1}{For example, compared to the part of the memory with the least read disturbance vulnerability,} the most vulnerable part experiences 1)~up to $2\times$ the number of bitflips and 2)~bitflips at an order of magnitude fewer accesses.
We showcase that Svärd {leverages {this} variation to reduce} the overheads of {five} state-of-the-art read disturbance {solutions}, {and thus significantly increases system performance (by $1.23\times$, {$2.65\times$}, $1.03\times$, $1.57\times$, and $2.76\times$, for AQUA~\cite{saxena2022aqua},} {BlockHammer~\cite{yaglikci2021blockhammer},} Hydra~\cite{qureshi2022hydra}, PARA~\cite{kim2014flipping}, and RRS~\cite{saileshwar2022randomized}, respectively, on average across 120 multiprogrammed memory-intensive workloads).
\agy{3}{Our observations and mechanism design are published at HPCA 2024~\cite{yaglikci2024spatial}.\footnote{\agy{3}{The author of this thesis is the first author of the HPCA 2024 paper~\cite{yaglikci2024spatial}.}}}

\subsubsection{Leveraging Subarray-Level Parallelism in Off-the-Shelf DRAM Chips}
As DRAM chip density increases with technology node scaling, DRAM refresh operations are performed more frequently because:
1)~{the number of DRAM rows in a chip increases}; and 2)~DRAM cells {need additional refresh operations to mitigate bit failures caused by DRAM read disturbance.} 
{Thus,} it is critical to {enable} refresh {operations} at low performance overhead.
To this end, we \agy{6}{first} propose \agy{1}{a new operation, \agy{6}{\gls{hira}, that refreshes a DRAM row concurrently with refreshing or accessing another row in the same DRAM bank in off-the-shelf DRAM chips by violating two timing constraints.}
\gls{hira} \agy{6}{\gls{hira} does so by leveraging the subarray-level parallelism}
with \emph{no} modifications to off-the-shelf DRAM chips unlike prior works~\salprefs{}. To do so, it leverages the new observation that two rows in the same \agy{1}{DRAM} bank can be activated without data loss if the rows are connected to different charge restoration circuitry.
We experimentally demonstrate on 56 real off-the-shelf DRAM chips that \gls{hira} can reliably parallelize a DRAM row's refresh operation with refresh or access of any of the 32\% of the rows within the same bank. By doing so,
\gls{hira} reduces the time spent for refresh operations by \SI{51.4}{\percent}. 
\agy{6}{Second, we propose} a memory controller-based mechanism, \gls{hiramc}, that schedules and performs} \gls{hira} operations.
\gls{hiramc} modifies the memory request scheduler \agy{1}{(in the memory controller)} to perform \gls{hira} when a refresh operation can be {performed concurrently} with a memory access or another refresh.
Our system-level evaluations show that, \agy{1}{compared to the baseline that does \emph{not} leverage subarray-level parallelism,}
\gls{hiramc} increases system performance by \SI{12.6}{\percent} and $3.73\times$ as it reduces the performance degradation due to periodic refreshes and refreshes for DRAM read disturbance mitigation (i.e., preventive refreshes){, respectively,} for future DRAM chips with increased density and DRAM read disturbance vulnerability.
\agy{3}{Our observations and mechanism design are published at MICRO 2022~\cite{yaglikci2022hira}.\footnote{{The author of this thesis is the first author of the MICRO 2022 paper~\cite{yaglikci2022hira}.}}}

\subsection{Preventing RowHammer Bitflips without Proprietary Knowledge of DRAM Chip Internals}
We tackle two challenges that DRAM read disturbance \agy{1}{solutions} face: scalability with worsening DRAM read disturbance vulnerability and compatibility with off-the-shelf DRAM chips \emph{without} the \agy{1}{proprietary} knowledge of DRAM rows' physical layout. 
We show that it is possible to efficiently and scalably prevent RowHammer bitflips without knowledge of or {modification} to DRAM internals. {To this end, we} introduce \blockhammer{}, a low-cost, effective, and easy-to-adopt {RowHammer mitigation} mechanism that prevents all \rowhammer{} bitflips while overcoming the two key challenges.
\blockhammer{} selectively throttles memory accesses that could {otherwise potentially cause} \rowhammer{} bitflips.
{The key idea of BlockHammer is to 1)~{track row activation rates using} area-efficient Bloom filters, and 2)~use {the tracking data} to ensure that no row is ever activated rapidly enough to induce RowHammer bitflips.}
By guaranteeing that no DRAM row ever {experiences} \agy{1}{an activation rate that can cause RowHammer bitflips} (i.e., \rowhammer{}-{un}safe activation rate), \blockhammer{} {1)}~makes it impossible for a \rowhammer{} bitflip to {occur} {and 2)~{greatly} reduces a \rowhammer{} attack's impact on the performance of \agy{1}{other concurrently running threads that would \emph{not} cause RowHammer bitflips (i.e.,} benign applications}). \agy{1}{BlockHammer introduces a new metric called RowHammer likelihood index (RHLI), which enables the memory controller (and optionally the system software) to distinguish a thread performing a RowHammer attack from a benign thread.}
Our evaluations across a comprehensive range of {280~workloads} show that, compared to {the best of six} state-of-the-art \rowhammer{} mitigation mechanisms (all of which require knowledge of or {modification} to DRAM internals), \blockhammer{} provides 1)~competitive performance and energy when the system is not under {a} RowHammer attack and 2)~significantly better performance and energy when the system is under a RowHammer attack. We \agy{1}{open-source} BlockHammer~\cite{safari2021blockhammer} and also implement as part of Ramulator 2.0, a cycle-level memory simulator~\cite{luo2023ramulator, safari2023ramulator2}.
\agy{3}{BlockHammer is published at HPCA 2021~\cite{yaglikci2021blockhammer}.\footnote{{The author of this thesis is the first author of the HPCA 2021 paper~\cite{yaglikci2021blockhammer}.}}}

\section{Contributions}

This dissertation makes the following contributions:

\begin{enumerate}
\item We build a detailed understanding of how RowHammer vulnerability changes with four fundamental properties:
1)~DRAM chip temperature,
2)~memory access pattern,
3)~victim DRAM cell's physical location, and 4)~voltage.
To our knowledge, our study, encompassing \agy{6}{detailed} experimental characterization and statistical analyses~\cite{orosa2021adeeper, yaglikci2022understanding, yaglikci2024spatial}, is the first to rigorously study DRAM read disturbance under these four properties.
In doing so, we 
1)~test over hundreds of real DRAM chips,
2)~introduce new statistical analyses to explain how DRAM read disturbance changes with these four properties,
and 3)~the implications of our findings on future DRAM read disturbance attacks and defenses. 
    \begin{enumerate}
        \item We present the first rigorous experimental study that examines temperature's effects on RowHammer bitflips in modern DRAM chips. Our results demonstrate that {a DRAM cell experiences bitflips in a specific and bounded range of temperature, and the RowHammer vulnerability tends to worsen as temperature {increases}.}
        \item We experimentally demonstrate how DRAM read disturbance changes with the active time of the aggressor rows. {Our results show that as the} aggressor row remains activated longer {(e.g., by 5$\times$), 1)~more DRAM cells (6.9$\times$ on average) experience RowHammer bitflips at a given hammer count and 2)~a DRAM row experiences RowHammer bitflips at a smaller hammer count (by 36\% on average), \agy{1}{compared to the memory access pattern where the aggressor row is opened and closed as frequently as possible}.}
        This finding has ignited a thorough examination by researchers to investigate its underlying causes, ultimately leading to the revelation of a new DRAM read disturbance phenomenon, denoted as \emph{RowPress}~\cite{luo2023rowpress}. 
        \item We investigate the variation of DRAM read disturbance vulnerability across rows in a DRAM module and show a {significant and irregular} variation {in} DRAM read {disturbance} vulnerability across rows: in the part of memory with the worst read disturbance vulnerability, 1)~up to $2\times$ the number of bitflips can occur and 2)~bitflips can occur at an order of magnitude fewer accesses, compared to the memory locations with the least vulnerability to read disturbance.
        \item We {present the first {experimental} study of} voltage's {effect on} DRAM read disturbance, access latency, charge restoration, and {data} retention time{.} {Our experiments} {on real DRAM chips} show that {when a DRAM module is operated at a {reduced} \gls{vpp},}
        {1)~an attacker $i$)~needs to hammer a row {in the module more {times} (by 7.4\%/85.8\%)} to induce a bitflip, and $ii$)~can cause fewer (15.2\%/66.9\%) {read disturbance} bitflips {in the module} {(on average /} {at maximum} across {all} tested {modules)};} and 2)~DRAM access latency, charge restoration process, and data retention time are slightly worsened, but 
        $i$)~most (208 out of 272) DRAM chips still reliably {operate}, but the erroneous chips reliably operate using increased row activation latency, simple error correcting codes, {or doubling the refresh rate \emph{only} for \SI{16.4}{\percent} of the rows.}
        \item Based on our {new} observations {on DRAM read disturbance's sensitivities to temperature, aggressor row's active time, and a victim DRAM cell's physical location in the DRAM chip}, we {describe and} analyze {three} future RowHammer attack {and six future RowHammer} defense {improvements}.
    \end{enumerate}
\pagebreak
\item We leverage our insights \agy{4}{into modern DRAM chips and controllers in two aspects.} 
    \begin{enumerate}
        \item We leverage the spatial variation of DRAM read disturbance across DRAM rows to reduce the performance overheads of existing DRAM read disturbance solutions.
        \begin{enumerate}
            \item We propose \emph{Svärd}~\cite{yaglikci2024spatial}, a new mechanism that dynamically adapts the aggressiveness of an existing read disturbance solution to the vulnerability level of the potential victim row.
            \item We showcase Svärd's integration with \param{five} different state-of-the-art read disturbance solutions. Our results show that Svärd reduces the performance overhead of these \param{five} state-of-the-art solutions, leading to large system performance benefits. 
        \end{enumerate}

        \item We leverage \agy{4}{the subarray-level parallelism in off-the-shelf DRAM chips. We} experimentally demonstrate that 1)~real off-the-shelf DRAM chips can refresh a DRAM row concurrently with refreshing or activating another row within the same DRAM bank using subarray-level parallelism~\salprefs{}
        and 2)~doing so is beneficial to reduce the performance overhead of both {\emph{periodic}} refresh operations (required for reliable DRAM operation) and \emph{preventive} refresh operations (required for preventing DRAM read disturbance bitflips).
        \begin{enumerate}
            \item We show that \agy{1}{parallelizing a refresh operation with other} refresh or access \agy{1}{operations} within a bank is possible {in} off-the-shelf DRAM chips by issuing a carefully-engineered sequence of row activation and precharge commands, which we call \emph{\gls{hira}}~\cite{yaglikci2022hira}.  
            \item We experimentally demonstrate on 56 real {DDR4} DRAM chips that {\gls{hira}} 1)~reduces the latency of refreshing two rows back-to-back by \param{\SI{51.4}{\percent}}, and 2)~{reliably parallelizes a DRAM row's {refresh operation} {with refresh or {activation} of any of the} 32\%} of the rows in the same bank.
            \item We {design \agy{1}{a memory controller-based mechanism called}} {\gls{hiramc} to perform \gls{hira} operations}. We show that {\gls{hiramc} significantly {improves system performance by \SI{12.6}{\percent} and $3.73\times$ {as it reduces}} the performance degradation {due to} periodic refreshes and {preventive} refreshes}{,} respectively. 
        \end{enumerate}
    \end{enumerate}
\item We show that it is possible to prevent DRAM read disturbance bitflips efficiently and scalably with \emph{no} knowledge of or modifications to DRAM chips.
    \begin{enumerate}
        \item {We introduce the first mechanism that} efficiently and scalably {prevents} RowHammer bitflips \emph{without} the knowledge of or {modifications} to DRAM internals. {Our mechanism,} \blockhammer{}~\cite{yaglikci2021blockhammer}, provides {competitive} performance and energy  with {existing {\rowhammer{} mitigation} mechanisms} when the system is \emph{not} under {a} RowHammer attack, and \emph{significantly} better performance and energy {than} existing {mechanisms} when the system \emph{is} {under a RowHammer} attack.
        \item We show that we can greatly reduce the performance degradation and energy wastage a \rowhammer{} attack inflicts on \agy{1}{other concurrently running threads that would \emph{not} cause read disturbance bitflips} by accurately identifying the \rowhammer{} attack thread and reducing its memory bandwidth usage. We introduce a new metric called the \emph{\rowhammer{} likelihood index}, which enables the memory controller to distinguish a \rowhammer{} attack from a benign thread.
        \item We enable proactive throttling \agy{1}{of memory accesses} as a practical DRAM read disturbance solution. To do so, we employ a variant of counting Bloom filters that 1)~avoids the area and energy overheads of per-\agy{1}{DRAM}-row counters used by prior proactive throttling mechanisms, and 2)~never fails to detect a \rowhammer{} attack.
    \end{enumerate}
\end{enumerate}

\section{Outline}

This dissertation is organized into \agy{6}{nine} chapters. 

Chapter~\ref{chap:bg} gives relevant background information about DRAM organization, operation, timing constraints, DRAM read disturbance phenomenon, DRAM read disturbance mitigations, and in-DRAM row address mapping. 

Chapter~\ref{chap:related} provides an overview of the related prior work.

\agy{4}{Chapters~\ref{chap:deeperlook} and~\ref{chap:hammerdimmer} build a detailed understanding of RowHammer vulnerability.}
Chapter~\ref{chap:deeperlook} presents 1)~our experimental study of DRAM read disturbance in real DRAM chips under varying temperatures, memory access patterns, and victim cell locations and 2)~the implications of our findings on future RowHammer attacks and defenses~\cite{orosa2021adeeper}.
Chapter~\ref{chap:hammerdimmer} presents our experimental characterization study on the effect of voltage scaling on DRAM read disturbance, \agy{4}{access latency, charge restoration, and data retention time}~\cite{yaglikci2022understanding}.

\agy{4}{Chapters~\ref{chap:svard} and~\ref{chap:hira} shows that it is possible to enable efficient and scalable read disturbance mitigation by leveraging two insights into modern DRAM chips and memory controllers: the spatial variation of DRAM read disturbance across DRAM rows (Chapter~\ref{chap:svard}) and subarray-level parallelism in off-the-shelf DRAM chips (Chapter~\ref{chap:hira}).}
Chapter~\ref{chap:svard} presents a more detailed investigation of the spatial variation in DRAM read disturbance across DRAM rows and introduces \gls{svard}, a new mechanism that leverages the spatial variation in read disturbance vulnerability across DRAM rows to reduce the performance overhead of existing DRAM read disturbance solutions~\cite{yaglikci2024spatial}.
Chapter~\ref{chap:hira} introduces 1)~\gls{hira}, a new DRAM operation that enables leveraging subarray-level parallelism to reduce the performance overhead of periodic and read disturbance preventive refresh operations, and 2)~\gls{hiramc}, a memory controller-based mechanism that dynamically schedules \gls{hira} operations from within the memory controller~\cite{yaglikci2022hira}. 

Chapter~\ref{chap:blockhammer} \agy{4}{shows that it is possible to prevent DRAM read disturbance bitflips efficiently and scalably with \emph{no} proprietary knowledge of or modifications to DRAM chips. To this end, Chapter~\ref{chap:blockhammer}} introduces BlockHammer, a new DRAM read disturbance mitigation mechanism that selectively throttles unsafe memory accesses that might otherwise have led to read disturbance bitflips~\cite{yaglikci2021blockhammer}.

\agy{6}{Chapter}~\ref{chap:conc} provides a summary of this dissertation as well
as future research \agy{6}{directions} and concluding remarks.
\chapter{Background}
\label{chap:bg}

{This chapter provides an overview of the background
material necessary to understand our discussions, analyses, and contributions.
\secref{sec:background_dramorg} reviews DRAM organization.
\secref{sec:dram_operation_and_timings} explains DRAM operation and timing constraints.
\secref{sec:background_rowhammer} provides background material about DRAM read disturbance. 
\secref{sec:background_mitigation} briefly summarizes the existing DRAM read disturbance solutions and two outstanding key challenges for those solutions. 
\secref{sec:background:dramvoltage} provides a brief background on the DRAM voltage control and its relevance to both DRAM operation and read disturbance.

\section{DRAM Organization}
\label{sec:background_dramorg}

{\figref{fig:dram-organization} shows the organization of DRAM-based memory systems. A memory channel connects the processor (CPU) to a set of DRAM chips, called {\emph{DRAM rank}}.}
{Chips in a DRAM rank operate in lock-step.}
{Each chip has multiple DRAM banks, each consisting of multiple DRAM cell arrays {(called \emph{subarrays})} and their local I/O circuitry. {Within a subarray,} DRAM cells are organized as a two-dimensional array {of} DRAM rows and columns.}
{A} DRAM cell {stores one bit of data} {in the form of} electrical charge in {a} capacitor, which can be accessed through an access transistor.  
A wire called {wordline} drives the gate of all DRAM cells' access transistors in a DRAM row{. A} wire called {\emph{bitline}} connects all DRAM cells in a DRAM column to a common differential sense amplifier. Therefore, when a wordline is asserted, each DRAM cell in the DRAM row is connected to its corresponding sense amplifier. The set of sense amplifiers in a subarray is called {\emph{the row buffer}}, where the data of {an activated} DRAM row is buffered {to serve a column access.}

\begin{figure}[t] \centering
    \includegraphics[width=\linewidth]{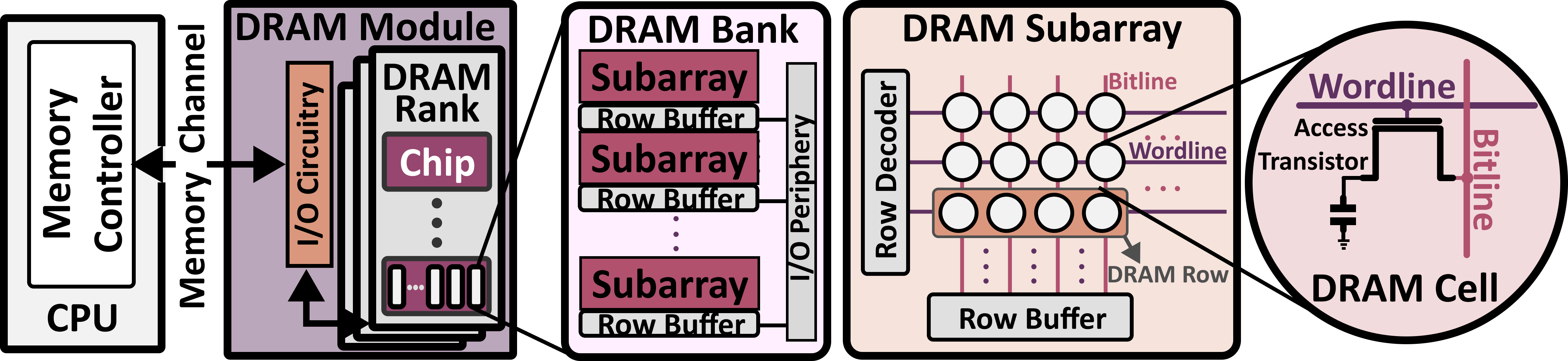}
    \caption{DRAM organization.}
    \label{fig:dram-organization}
    \vshort{}
\end{figure}

\section{DRAM Operation and Timing Constraints}
\label{sec:dram_operation_and_timings}
{The memory controller serves memory access requests by issuing DRAM commands, e.g., row activation ($ACT$), bank precharge ($PRE$), data read ($RD$), data write ($WR$), and refresh ($REF$) {while respecting certain timing parameters to guarantee correct operation~\dramStandardCitations{}.}
To read or write data, the memory controller first {needs to activate the corresponding row. To do so, it} issues an $ACT$ command alongside the bank address and row address corresponding to the memory request's address. 
When a row is activated, its data is copied {to and temporarily stored at} the row buffer.
{The latency from the start of a row activation until the data
is reliably readable{/writable} in the row buffer is called {the} \emph{\gls{trcd}}.}
{During the row activation process, a DRAM cell loses its charge, and thus, its initial charge needs to be restored ({via} a process called \emph{charge restoration}). 
The latency from the start of a row activation until the completion of the DRAM cell's charge restoration is called {\emph{\gls{tras}}}.
}
The memory controller can read/{write} data from/to the row buffer using $RD$/$WR$ commands.
The changes are propagated to the DRAM cells in the open row. 
Subsequent accesses to the same row can be served quickly {from the row buffer (i.e., called a \emph{row hit})} without issuing another $ACT$ to the same row. 
{The latency of performing a read/write operation is called \gls{tcl}/\gls{tcwl}}. 
To access another row in {an already} activated DRAM bank, the memory controller must issue a $PRE$ command {to} close the opened row {and prepare the bank for a new activation}.
{When the $PRE$ command is issued}, the DRAM chip de-asserts the active row's wordline and precharges the bitlines. The relevant timing parameter is the {\emph{\gls{trp}}}.}

{{A} DRAM cell {is} inherently leaky and thus {loses its} stored electrical charge over time. To maintain data integrity, a DRAM cell {is periodically refreshed} with a {time interval called {the \emph{\gls{trefw}}}, which is typically} \SI{64}{\milli\second} (\SI{32}{\milli\second}) {at normal operating temperature up to \SI{85}{\celsius} (above \SI{85}{\celsius} up to \SI{95}{\celsius})}~\dramStandardCitations{}.  
To {timely} refresh all cells, the memory controller {periodically} issues a refresh {($REF$)} command with {an interval called} {the \emph{\gls{trefi}}}, {typically} \SI{7.8}{\micro\second} (\SI{3.9}{\micro\second}) {at normal (extended) operating temperature range~\dramStandardCitations{}.} When a rank-/bank-level refresh command is issued, the DRAM chip internally refreshes several DRAM rows, during which the whole rank/bank is busy \agy{4}{for a time window, called the \emph{\gls{trfc}}}.} 

\section{DRAM Read Disturbance}
\label{sec:background_rowhammer}
Read disturbance is the phenomenon that reading data from a memory or storage device causes physical disturbance (e.g., voltage deviation, electron injection, electron trapping) on another piece of data that is \emph{not} accessed but physically located nearby the accessed data. Two prime examples of read disturbance in modern DRAM chips are RowHammer~\cite{kim2014flipping}, and RowPress~\cite{luo2023rowpress}, where repeatedly accessing (hammering) or keeping active (pressing) a DRAM row induces bitflips in physically nearby DRAM rows, respectively. In RowHammer and RowPress terminology, the row that is hammered or pressed is called the \emph{aggressor} row, and the row that experiences bitflips is called the \emph{victim} row.
{For read disturbance bitflips to occur, 1)~\agy{6}{an} aggressor row needs to be activated more than a certain threshold value, {defined as \gls{hcfirst}~\cite{kim2020revisiting} {and/}or 2)~\gls{taggon}~\cite{luo2023rowpress} need to be large-enough~\understandingRowHammerRowPressCitations{}. To avoid read disturbance bitflips, systems take preventive actions, e.g., {they refresh} victim rows~\refreshBasedRowHammerDefenseCitations{}, selectively {throttle} accesses to aggressor rows~\throttlingBasedRowHammerDefenseCitations{}, and physically {isolate} potential aggressor and victim rows~\isolationBasedRowHammerDefenseCitations{}. These {solutions} aim to perform preventive actions before the cumulative effect of an aggressor row's \emph{activation count} and \emph{on time} causes read disturbance bitflips.}}
\section{DRAM Read Disturbance Mitigation}
\label{sec:background_mitigation}
Given the severity of DRAM read disturbance, {various} mitigation methods have been proposed, which we classify into four {high-level} approaches:
$i$)~\emph{increased refresh rate}, which {refreshes \emph{all}} rows more frequently to reduce the probability of a successful {bitflip}~\cite{apple2015about, kim2014flipping};
$ii$)~\emph{physical isolation}, which {physically separates sensitive data from {any} potential attacker's memory space (e.g., by adding buffer rows between sensitive data regions and other data)~\isolationBasedRowHammerDefenseCitations{};}
$iii$)~\emph{reactive refresh}, which observes row activations
and refreshes the potential victim rows as a reaction to rapid row activations~\refreshBasedRowHammerDefenseCitations{}; and
{$iv$)~\emph{proactive throttling}, which {limits} {row activation} rates~\throttlingBasedRowHammerDefenseCitations{} to {\rowhammer{}-safe levels}.}
Unfortunately, each of these four approaches faces at least one of two major challenges towards effectively mitigating \rowhammer{}.

\head{Challenge 1: Efficient Scaling as RowHammer Worsens}
{As DRAM chips become more vulnerable to \rowhammer{} {(i.e., \rowhammer{} {bitflips} can occur at significantly lower row activation {counts} than before)}, {{mitigation mechanisms} need to act more aggressively.}} 
{A} \emph{scalable} mechanism {should} exhibit acceptable performance, energy, and area overheads as its design is reconfigured for more vulnerable DRAM chips. Unfortunately, as chips become more vulnerable to \rowhammer{}, {most} state-of-the-art mechanisms of all four approaches either cannot easily adapt because they are based on fixed design points, or their performance, energy, and/or area overheads become increasingly significant.
$i$)~{Increasing the refresh rate further {in order to prevent all \rowhammer{} {bitflips}} is 
{prohibitively expensive,}
even for {existing} DRAM chips~\cite{kim2020revisiting}, due to the large number of rows that {must} be refreshed within a refresh window.
{$ii$)}~}Physical isolation mechanisms {must} provide {greater} isolation (i.e., increase the physical distance) between sensitive data and a potential attacker's memory space {as DRAM chips become denser and more vulnerable to \rowhammer{}}. 
This is because 
denser chip designs bring circuit elements closer together,
which increases the {number of} rows across which the hammering {of} an aggressor row can induce \rowhammer{} {bitflips}~\cite{kim2014flipping, mutlu2017therowhammer, yang2019trapassisted, kim2020revisiting, qazi2021halfdouble, qazi2021introducing, kogler2022halfdouble}. Providing {greater} isolation (e.g., increasing the number of buffer rows between sensitive data and an attacker's memory space) both wastes increasing amounts of memory capacity and reduces the fraction of physical memory that can be protected from \rowhammer{} attacks.
{$iii$)}~Reactive refresh mechanisms need to increase the rate at which they refresh potential victim {rows.} Prior {work~\cite{kim2020revisiting}} shows that state-of-the-art {reactive refresh} \rowhammer{} mitigation mechanisms
{lead to prohibitively large performance overheads with increasing RowHammer vulnerability.}
{$iv$)}~{Previously proposed proactive} throttling approaches {must throttle {activations} at a more aggressive rate to counteract the increased \rowhammer{} vulnerability.}
{This requires either throttling} benign applications' row activations or {tracking} per-row activation rates for the entire refresh window, 
{incurring prohibitively-expensive performance or area overheads} {even {for} existing} DRAM chips~\cite{kim2014flipping, mutlu2018rowhammer}.

\head{Challenge 2: Compatibility with Commodity DRAM Chips} 
Both physical isolation~{($ii$)} and reactive refresh~{($iii$)} mechanisms require the ability to either 1)~identify \emph{all potential victim rows} that can be affected by hammering a given row or 2)~modify the DRAM chip such that {either} the potential victim rows are internally isolated within the {DRAM} chip or the \rowhammer{} mitigation mechanism can accurately issue reactive refreshes to {all} potential victim {rows}.
{Identifying all potential victim {rows} requires knowing the mapping {schemes} that {the DRAM chip} uses to internally {translate} memory-controller-visible row addresses to physical row addresses}.
{Unfortunately,} DRAM vendors consider their in-DRAM row address mapping schemes to be highly \emph{proprietary} and do not reveal any details in {publicly-available} documentation, as these details contain insights into the chip design and manufacturing quality~\inDRAMRowAddressMappingCitations{}
(discussed in Section~\ref{sec:background_physicallayout}). 
As a result, both physical isolation and reactive {refresh} are limited to systems that can {1)}~obtain such proprietary information on {in-}DRAM row address mapping or {2)}~modify DRAM chips internally.

\section{In-DRAM Row Address Mapping}
\label{sec:background_physicallayout}
DRAM vendors {often use {DRAM-internal} mapping} {schemes} to internally translate memory-controller-visible row addresses to physical row {addresses~\inDRAMRowAddressMappingCitations{} for two reasons:}
(1)~to optimize their chip design for density, performance, and power {constraints}; and (2)~{to improve factory yield by mapping the addresses of faulty rows to more reliable spare rows} {(i.e., post-manufacturing row repair).}
{Therefore, row {mapping} schemes can vary {with} (1) chip design {variation} across different vendors, DRAM models, and generations and (2) manufacturing process variation across different chips of the same design}. 
{State-of-the-art \rowhammer{} mitigation mechanisms must account for {both sources of variation in order to}
be able to {accurately} identify all potential victim rows that are {physically nearby}
an aggressor row.}
{{Unfortunately}, DRAM vendors {consider their in-DRAM row address mapping schemes to be highly proprietary and ensure not to reveal mapping details in any public documentation because exposing the row {address} mapping {scheme} can reveal insights into the chip design and factory yield}~\inDRAMRowAddressMappingCitations{}}.
\section{DRAM Voltage}
\label{sec:background:dramvoltage}
Modern DRAM chips {(e.g., DDR4~\cite{jedec2020jesd794c}, DDR5~\cite{jedec2020jesd795}, GDDR5~\cite{jedec2016jesd212c}, GDDR5X~\cite{jedec2016jesd232a}, GDDR6~\cite{jedec2021jesd250c}, HBM2~\cite{jedec2021jesd235d}, and HBM3~\cite{jedec2022jesd238} {standard compliant ones})} use two separate voltage rails: {1)~\gls{vdd}, which is used to operate the core DRAM array and peripheral circuitry (e.g., the sense amplifiers, row/column decoders, precharge and I/O logic), and 2)~\agy{6}{wordline voltage ($V_{PP}$)}, which is exclusively used to assert {a} wordline during a DRAM row activation.} 
\gls{vpp} is generally significantly higher (e.g., 2.5V~\cite{kim2012acase, micron2016ddr4,micron2016ddr4,micron2014sdram}) than \gls{vdd} (e.g., 1.25--1.5V~\cite{kim2012acase, micron2016ddr4,micron2016ddr4,micron2014sdram}) in order to ensure 1)~full activation of {all} access {transistors of a row} when the wordline is {asserted} and 2)~low leakage when the wordline is {de-asserted}. 
\gls{vpp} is internally generated from \gls{vdd} in older DRAM chips (e.g., DDR3~\cite{jedec2012jesd793d}). However, newer DRAM chips (e.g., DDR4 onwards~{\cite{jedec2020jesd794c, jedec2020jesd795, jedec2016jesd232a, jedec2021jesd250c, jedec2016jesd212c, jedec2022jesd238}}) expose \emph{both} \gls{vdd} and \gls{vpp} rails to external pins, allowing 
to drive them
with different voltage {sources}.\agycomment{6}{HBM has a separate VPP pin, but we cannot control it as far as I understand. So, I don't think that we can study it.}\agy{6}{\footnote{\agy{6}{\gls{vpp} is \emph{not} exposed to external pins in LPDDRX DRAM chips~\cite{jedec2012lowpower, jedec2017jesd2094b, jedec2020jesd2095a}.}}} 


\subsection{Wordline Voltage's Impact on DRAM Read Disturbance}
As explained in \secref{sec:background_rowhammer}{,}
a larger \gls{vpp} exacerbates both {electron injection / diffusion / drift}
and capacitive crosstalk mechanisms. Therefore, we hypothesize that the RowHammer vulnerability of a DRAM chip increases {as} \gls{vpp} {increases}. Unfortunately, \agy{6}{\emph{no} prior work tests} this hypothesis and quantifies the effect of \gls{vpp} on real DRAM chips' RowHammer vulnerability.

\subsection{Wordline Voltage's Impact on DRAM Operations}
{An access transistor turns on (off) {when its gate voltage is higher (lower) than a threshold.}} {An access transistor's gate is connected to a wordline (\figref{fig:dram-organization}) and driven by \gls{vpp} (ground) when the row is {activated} ({precharged}).\footnote{{To increase DRAM cell retention time, modern DRAM {chips} may apply a negative voltage to the wordline~\cite{frank2001device, lee2011simultaneous} when {the wordline} is not {asserted}{. Doing so} {reduces} the leakage current {and this improves data retention}.}}} 
{Between \gls{vpp} and ground, a larger access transistor gate voltage} forms a stronger channel between the bitline and the capacitor.
{A} strong channel allows {fast DRAM row activation and full charge restoration.}
Based on these properties, we hypothesize that a \emph{larger} \gls{vpp} provides \emph{smaller} row activation latency and increased {data} retention time, leading to {more reliable DRAM operation}.\footnote{{Increasing/decreasing \gls{vpp} does \emph{not} affect the reliability of $RD$/$WR$ and $PRE$ operations since the DRAM {circuit components} involved in these operations are powered using \emph{only} \gls{vdd}.}}
Unfortunately, {there is \emph{no} prior work that tests this hypothesis and quantifies \gls{vpp}'s effect on real DRAM chips' {reliable operation (i.e.,} row activation and charge restoration {characteristics)}.}

\chapter{Related Work}
\label{chap:related}

Many prior works study DRAM read disturbance, develop DRAM read disturbance exploits to mount system-level attacks, and propose solutions to DRAM read disturbance. This chapter provides an overview of closely related works \agy{6}{and provides a brief background of read disturbance in other memory technologies.}

\section{Detailed Understanding of DRAM Read Disturbance}

Our DRAM read disturbance characterization study (Chapters~\ref{chap:deeperlook} and~\ref{chap:hammerdimmer}) is the first {work} {that} rigorously analyzes how RowHammer vulnerability changes with four fundamental properties: 1)~DRAM chip temperature, 2)~aggressor row active time, 3)~victim DRAM cell's physical location, and 4)~voltage used for activating DRAM rows. We divide prior work on building a detailed understanding of DRAM read disturbance into two {categories}:  
{1})~characterization of real DRAM chips,
{and 2})~circuit-level simulation-based studies.
{Three works~\cite{mutlu2017therowhammer, mutlu2019rowhammer_a, mutlu2023fundamentally} provide an overview of {the} RowHammer {literature}, and project the effect of increased RowHammer vulnerability in future DRAM chips and memory systems.} 

\subsection{Major DRAM Read Disturbance Characterization Works} 

{\agy{5}{Six} major works~\cite{kim2014flipping, kim2020revisiting, luo2023rowpress, olgun2023anexperimental, olgun2024read, luo2024experimental}} extensively characterize DRAM read disturbance {using} real DRAM chips. 

The first work~\cite{kim2014flipping}, published in 2014, 1)~investigates the vulnerability of {129} commodity {DDR3} DRAM {modules} to various RowHammer attack models, 2)~{demonstrates for the first time} that RowHammer is a real problem for commodity DRAM chips, {3)~characterizes RowHammer's sensitivity to refresh rate and activation rate in terms of \gls{ber}, \gls{hcfirst}, and {the physical distance between aggressor and victim rows}}, and {4}) {examines various potential solutions and} proposes a {new} low-cost mitigation mechanism. 

{The second} work~\cite{kim2020revisiting}, published in 2020, conducts {comprehensive} {scaling} experiments on a {wide} range of {1580 DDR3, DDR4, and LPDDR4 commodity DRAM chips from different DRAM generations and technology nodes}, clearly {demonstrating} that DRAM read disturbance {has} {become {an even} more serious} problem {over {DRAM} generations.}
Even though these two works rigorously characterize {various aspects of DRAM read disturbance in} real DRAM chips, {they} do not analyze {the effects of temperature, aggressor row active time, and victim DRAM cell's physical location on the} RowHammer {vulnerability}.
Our work {complements and furthers the analyses} o{f} these two papers~\cite{kim2014flipping, kim2020revisiting} {by} {1)~rigorously {analyzing} how these three properties affect the RowHammer vulnerability, and 2)~{providing} new {insights} {in}to crafting more effective {and efficient} RowHammer attacks and defenses.}

Third, building on our insights into RowHammer's sensitivity to memory access patterns~\cite{orosa2021adeeper}, Luo et al.~\cite{luo2023rowpress} are the first to experimentally demonstrate and analyze RowPress, a new read disturbance phenomenon, in 164 real DDR4 DRAM chips. They show that RowPress is different {from} RowHammer, since it 
1)~affects a different set of DRAM cells from RowHammer and 
2)~behaves differently from RowHammer as temperature and access pattern change.  
This work is a successor of our analysis.

\agy{5}{Fourth, Olgun et al.~\cite{olgun2023anexperimental} test {one} HBM2 DRAM chip's RowHammer vulnerability and data retention characteristics on a subset of DRAM rows in a DRAM bank with a focus on the spatial variation of RowHammer vulnerability {across rows and HBM channels} and the characteristics of on-die target row refresh mechanisms.}

\agy{5}{Fifth, Olgun et al.~\cite{olgun2024read} extend their analysis by testing six HBM2 DRAM chips for RowHammer and RowPress. This extended paper presents the first analysis on the hammer count to induce up to 10 bitflips in a DRAM row and more rigorously delves into the analyses of spatial variation in read disturbance and the characteristics of on-die target row refresh mechanisms.}

\agy{5}{Sixth, Luo et al.~\cite{luo2024experimental} present a comprehensive study of the read disturbance effect of a combined RowHammer and RowPress access pattern in 84 DDR4 DRAM chips. The analysis shows that time to bitflip can be reduced significantly by combining RowHammer and RowPress access patterns.}

\agy{5}{Our work, {in contrast,} 1)~\emph{rigorously} {analyzes} the effects of all four properties by {testing} a significantly large set of {272} DRAM chips, and 2)~{provides insights into resulting} RowHammer attack and defense improvements.}

\subsection{Simulation-based Studies} 

{Prior works}~\understandingRowHammerSimulationCitataions{} attempt {to} explain the error mechanisms {that cause} DRAM read disturbance {bitflips} {through circuit-level simulations of capacitive-coupling and charge-trapping mechanisms{, without testing real DRAM chips.} {These works, {some of which we discuss in \secref{deeperlook:sec:temperature_circuit} and~\secref{deeperlook:sec:temporal_circuit}}}{,} are complementary to our experimental study on real DRAM chips.}

\subsection{Other Experimental Studies of Memory Devices}
\label{sec:rel:experimental_studies}

Many prior works conduct experimental error-characterization studies using real
memory devices to understand the error mechanisms involved. This
section reviews these works.

\subsubsection{Other DRAM Read Disturbance Characterization Works}
{Three other} works~\cite{park2014activeprecharge, park2016experiments, park2016statistical} {present \emph{preliminary}} experimental {data from only three~\cite{park2014activeprecharge, park2016statistical} or five~\cite{park2016experiments} DDR3 DRAM chips to build models {that explain} how {the RowHammer vulnerability of DRAM cell{s}} varies with the three properties we analyze. Unfortunately, the {experimental data {provided by} these works} is limited due to {1)}~their {extremely small} sample set {of DRAM cells, rows, and chips} and {2)}~the lack of analysis {of} system-level implications.}

\subsubsection{Other DRAM Chip Characterization Works}

Many other works perform their own experimental studies of real DRAM chips that focus on various areas of interest, including
data-retention~{\cite{jung2014optimized, hamamoto1995well, hamamoto1998onthe, yaney1987ametastable, shirley2014copula, weis2015retention, jung2015omitting, weis2015thermal, baek2014refresh, khan2014theefficacy, liu2013anexperimental, venkatesan2006retentionaware, hassan2017softmc, weis2017dramspec_a, khan2016acase, khan2016parbor, qureshi2015avatar, sutar2016dpuf, kim2009anew, kong2008analysis, lieneweg1998assessment, patel2017thereach}},\agycomment{6}{ISCA'17 is already here~\cite{patel2017thereach}}
access latency~{\cite{chang2017thesis, lee2016reducing, chandrasekar2014exploiting, chang2016understanding, lee2015adaptivelatency, kim2018thedram, kim2018solardram, kim2019drange, talukder2019exploiting, talukder2018ldpuf, talukder2019prelatpuf, mukhanov2020dstress, hassan2017softmc}}, 
on-DRAM-die error mitigation and correction schemes~{\cite{frigo2020trrespass, jung2016reverse, hassan2021uncovering, patel2019understanding,  patel2020bitexact, patel2021enabling, patel2021harp}}\agycomment{6}{U-TRR, BEER, and TRRespass are already here~\cite{frigo2020trrespass, patel2020bitexact, hassan2021uncovering}}, 
power consumption~\cite{ghose2018what, david2011memory}, 
{voltage~\cite{david2011memory, deng2011memscale, chang2017understanding}}, 
and the effects of issuing non-standard command sequences~\cite{gao2019computedram,
olgun2021quactrng, olgun2022pidram, gao2022fracdram, yuksel2023pulsar, yuksel2024functionallycomplete}. 
Two works~\cite{nam2024dramscope, marazzi2024hifidram} propose methodologies to reverse-engineer the organization of DRAM cells via 1)~performing tests of read disturbance, data retention, and in-DRAM row copy (also known as RowClone~\cite{seshadri2013rowclone})~\cite{nam2024dramscope} and 2)~scanning electron microscopy with focused ion beam~\cite{marazzi2024hifidram}. 
Other studies~{\cite{schroeder2009dram_errors, hwang2012cosmic, sridharan2012astudy,
sridharan2015memory, sridharan2013feng, meza2015revisiting,
bautistagomez2016unprotected, siddiqua2013analysis, meza2018large,
zhang2021quantifying}} examine failures observed in large-scale systems. All these studies are complementary to our work.

\subsubsection{Studies of Other Memory Devices}

Significant work has {examined} other memory technologies, including
\agy{6}{SRAM (e.g.,~\cite{maiz2003characterization, autran2009altitude,
radaelli2005investigation}), NAND flash memories (e.g.,~{\cite{meza2015revisiting,
cai2011fpga, cai2015read, luo2015warm, cai2015data, cai2014neighborcell,
cai2013program, cai2013error, cai2013threshold, cai2012flash, cai2012error, 
meza2015alargescale, schroeder2016flash, luo2016enabling, narayanan2016ssdfailures,
fukami2017improving, cai2017error, cai2017vulnerabilities, luo2018heatwatch,
luo2018improving, kim2020evanesco, park2021reducing, luo2018thesis, cai2012nand}}), hard
disks (e.g.,~\cite{bairavasundaram2008ananalysis, bairavasundaram2007ananalysis,
pinheiro2007failure, schroeder2007understandingdisk, schroeder2007understanding}) and emerging memories such as
phase-change memory (e.g.~\cite{pirovano2004reliability, zhang2012memory}).} These works
are also complementary to the experimental studies that we perform.

\section{DRAM Read Disturbance Solutions}
{Many prior works propose hardware-based~\hwBasedRowHammerMitigations{} and software-based~\swBasedRowHammerMitigations{} {techniques} to mitigate DRAM read disturbance, including various approaches, e.g., 
{1)~refreshing potential victim rows \refreshBasedRowHammerDefenseCitations{};
2)~selectively throttling memory accesses {that might cause bitflips} \throttlingBasedRowHammerDefenseCitations{};
3)~physically isolating an aggressor row from sensitive data \isolationBasedRowHammerDefenseCitations{};}
\agy{6}{4)~employing integrity check schemes to detect and correct read disturbance bitflips~\integrityBasedRowHammerDefenseCitations{},}\agycomment{6}{RAMPART and other similar mechanisms are here}
and 5)~enhancing DRAM circuitry to mitigate RowHammer effects~\circuitBasedRowHammerDefenseCitations{}. Each approach has different trade-offs {in terms of} performance impact, {energy overhead, hardware complexity}, and security guarantees. Among these works, we evaluate the most relevant state-of-the-art solutions in Sections~\ref{svard:sec:evaluation}, \ref{hira:sec:doduopara}, and~\ref{blockhammer:sec:hardware_complexity}--\ref{blockhammer:sec:qualitative_analysis}. and~\ref{blockhammer:sec:evaluation}. 
The rest of this section explains 1)~in-DRAM reactive refresh-based solutions, 2)~solutions that focus on fundamentally improving the reliability of DRAM chips, 3)~other uses of memory access throttling, and 4)~reducing the performance overhead of DRAM refresh.

\subsection{Improving the Reliability of DRAM Chips}
\label{sec:relatedwork_betterchips}
Several prior works implement architecture- and device-level improvements that make DRAM chips stronger against read disturbance. 

\head{Architecture-level improvements~\cite{hassan2019crow, gomez2016dram_rowhammer, bennett2021panopticon, devaux2021method, yaglikci2021security, jedec2024jesd795c}}
CROW~\cite{hassan2019crow} maps potential victim rows into dedicated \emph{copy rows} and mitigates \rowhammer{} bitflips by serving requests from copy rows.
Gomez et al.~\cite{gomez2016dram_rowhammer} place \emph{dummy cells} in DRAM rows that are engineered to be more susceptible to \rowhammer{} than regular cells, and monitor dummy cell charge levels to detect a \rowhammer{} attack.
Panopticon~\cite{bennett2021panopticon} implements an activation counter for each DRAM row within a small DRAM array in the DRAM chip to detect RowHammer attacks and performs preventive refresh operations using the time slack that might be available within the latency of a periodic refresh ($REF$) command ($t_{RFC}$).
Silver Bullet~\cite{devaux2021method, yaglikci2021security} implements a set of activation counters, each of which is mapped to a set of consecutive DRAM rows, i.e., a subbank. When a counter's value reaches a threshold value, Silver Bullet refreshes \emph{all} potential victim rows in the subbank and the subbank's close proximity.
\agy{5}{DDR5 introduces Refresh Management (RFM)~\cite{jedec2020jesd795} to provide the DRAM chip with time to perform its countermeasures. Refresh Management advises the memory controller to issue a DRAM command called RFM periodically with the number of activations the bank receives. This feature supports the in-DRAM read disturbance mitigation mechanisms but can lead to large performance overheads~\cite{canpolat2024understanding}.}
\agy{5}{Self-Managing DRAM (SMD)~\cite{hassan2024selfmanaging} is a DRAM architecture that eases the adoption of new in-DRAM maintenance mechanisms, including read disturbance mitigation mechanisms with \emph{no} modification to the DRAM communication protocols except a one-time addition of a signal called ACT-NACK. ACT-NACK is a signal that the DRAM chip sends to the memory controller to deny, i.e., not acknowledge (NACK), the most recent row activation command. Self-Managing DRAM uses the ACT-NACK signal to partially block memory to perform in-DRAM maintenance mechanisms.}
\agy{5}{\agy{6}{Per-Row Activation Counting (PRAC)~\cite{jedec2024jesd795c, kim2023a11v}, introduced into the DDR5 DRAM standards in April 2024,} is an in-DRAM read disturbance mitigation framework that 1)~implements the activation count of each DRAM row within the row to accurately track row activation counts and 2)~introduces a new alert back-off signal to allow the DRAM chip to request time from the memory controller to preventively refresh potential victim rows. DDR5 documentation does \emph{not} include a concrete implementation for a PRAC-based read disturbance mitigation mechanism. Two concurrent \agy{6}{academic} works~\cite{qureshi2024moat, canpolat2024understanding} propose concrete implementations of PRAC-based mitigations. Among these works, Canpolat et al.~\cite{canpolat2024understanding} open-source their PRAC implementation~\cite{safari2024ramulator2} and identify PRAC's two outstanding research problems and potential solutions: 1)~non-negligible performance overheads and 2)~exploitability for memory performance attacks~\cite{canpolat2024understanding}.}
\agy{5}{PRiDE~\cite{jaleel2024pride} and MINT~\cite{qureshi2024mint} are on-DRAM-die probabilistic read disturbance mitigation mechanisms that implement variations of probabilistic row activation~\cite{kim2014flipping} with respect to the challenges and the limitations of DDR5 communication protocol~\cite{jedec2020jesd795, jedec2024jesd795c}. ImPress~\cite{qureshi2024impress} extends existing counter-based read disturbance mitigation mechanisms to address RowPress, by accounting for a DRAM row's open time as a new row activation.}
\agy{5}{These works can be combined with our works 1)~HiRA~\cite{yaglikci2022hira}\agy{6}{,} to reduce the overall time spent for periodic and RowHammer/RowPress-preventive DRAM refresh operations and 2)~Svärd~\cite{yaglikci2024spatial}\agy{6}{,} to reduce the number of preventive refreshes by leveraging the variation of read disturbance across DRAM rows.} 

\head{Device-level improvements~\deviceLevelDefenses{}} \cite{park2022rowhammer_reduction}
{Three other works propose manufacturing process enhancements {or {implantation of} additional dopants in transistors to reduce} wordline {crosstalk}.}
Ryu et al.~\cite{ryu2017overcoming} propose an annealing process that reduces the traps in a DRAM array, thereby reducing wordline cross-talk. 
Han et al.,~\cite{han2021surround} propose to implement the DRAM cell's access transistor using a vertical pillar transistor (VPT), which has VPT has a longer physical gate and higher density, compared to the current saddle FinFET. The authors conduct TCAD simulations and show that using VPT significantly reduces the DRAM cell's RowHammer vulnerability.
\agy{6}{Yang et al.,~\cite{yang2016suppression, yang2017scanning} propose improvements to the doping profile methodology that lead to a significantly higher resolution in the doping profile and implant phosphorus (P) between wordlines for better isolation {between rows}, reducing RowHammer vulnerability.
Gautam et al.,~\cite{gautam2018improvement, gautam2019rowhammering} introduce metal nanoparticles and metal nanowires at the interface of gate metal and gate oxide of the DRAM cell's access transistor.
Park et al.,~\cite{park2022rowhammer_reduction} analyze and propose using partial isolation type buried channel array transistor. The proposed transistor significantly reduces the DRAM cell's RowHammer vulnerability due to its shallow drain-body junction.}

{Although} these methods reduce the effects of DRAM read disturbance, they 1)~\emph{cannot} be applied to already-deployed commodity DRAM chips and 2)~can be high cost {due to {the} required {extensive modifications to the} DRAM chip design and manufacturing process.} In contrast, our works~\cite{yaglikci2022hira, yaglikci2024spatial, yaglikci2021blockhammer} propose solutions that do \emph{not} require changes to the DRAM circuitry. 

\subsection{Proprietary In-DRAM Solutions} 


A subset of DRAM standards (e.g.,~\cite{jedec2017jesd794b, jedec2020jesd794c}) support a mode called \emph{target row refresh (TRR)}, allowing DRAM manufacturers to implement DRAM read disturbance mitigation mechanisms without disclosing their proprietary design. 
Recent works~\cite{frigo2020trrespass, hassan2021uncovering, jattke2022blacksmith, deridder2021smash, luo2023rowpress, qazi2021halfdouble, qazi2021introducing, kogler2022halfdouble} demonstrate that existing proprietary implementations of TRR are not sufficient to prevent DRAM read disturbance bitflips: many-sided RowHammer attacks reliably induce and exploit bitflips in state-of-the-art DRAM chips that already implement {TRR}. \agy{5}{PRAC~\cite{jedec2024jesd795c, canpolat2024understanding, qureshi2024moat} already replaces  TRR starting from the DDR5 specification \agy{6}{in April 2024}~\cite{jedec2024jesd795c}. \secref{sec:relatedwork_betterchips} discusses PRAC.}


\subsection{Other Uses of Memory Access Throttling} 

Prior quality-of-service- and fairness-oriented works propose selectively throttling main memory accesses to provide latency guarantees and/or improve fairness across applications \agy{6}{(e.g., \cite{rixner2000memory, rixner2004memory, moscibroda2007memory, mutlu2007stalltime,  mutlu2008parallelismaware, lee2008prefetchaware, kim2010atlas, kim2010thread,  subramanian2014theblacklisting, subramanian2016bliss, ausavarungnirun2012staged, ebrahimi2011parallel, ebrahimi2010fairness, ebrahimi2011prefetchaware, nychis2012onchip, nychis2010next, chang2012hatheterogeneous, usui2016dash, lugo2022survey, kim2016bounding, hassan2015aframework, zhou2016mitts, farshchi2020bru, sun2015response}).
These mechanisms are \emph{not} designed to prevent DRAM read disturbance and thus do \emph{not} interfere with the memory accesses of a RowHammer attack when there is \emph{no}} contention between memory accesses. 
In contrast, our throttling-based mechanism, BlockHammer, {prevents \rowhammer{} bitflips based on DRAM row activation counts.} As such, \blockhammer{} is {complementary} to these mechanisms and can work together with them.

\subsection{Reducing the Performance Overhead of DRAM Refresh}

We cluster prior works that tackle the performance overhead of DRAM refresh into five categories. All of these works can be combined with our works Svärd (Chapter~\ref{chap:svard}) and HiRA (Chapter~\ref{chap:hira}) as they tackle the same problem from different aspects. 

{\head{Eliminating unnecessary refreshes~{(e.g., {\cite{liu2012raidr, lin2012secret, nair2013archshield,baek2013refresh, qureshi2015avatar,patel2017thereach,venkatesan2006retentionaware,khan2017detecting,khan2016parbor,khan2014theefficacy,isen2009eskimoenergy,liu2011flikker,khan2016acase, jung2015omitting,jafri2021refresh, katayama1999faulttolerant, wilkerson2010reducing, ghosh2007smart, song2000method, emma2008rethinking, mukhanov2019workloadaware, hong2018eareccaided, kraft2018improving, park2011power, wang2018content, kim2020chargeaware}})}}
Various prior works eliminate unnecessary refresh operations by leveraging the heterogeneity in the retention time of DRAM cells to reduce the rate at which {some or all} DRAM rows are refreshed.
Our {work} differs from these {works} in {two} key aspects.
{First,} {most of} these works rely on identifying cells or rows with worst-case retention times, which is a difficult problem~\cite{mrozek2010analysis,mrozek2019multirun, patel2017thereach, liu2012raidr, qureshi2015avatar, liu2013anexperimental}. {In contrast,} our work on reducing the time spent for refresh operations, HiRA,} uses a {relatively simple} {and well-understood} one-time {experiment} to {1)~identify \gls{hira}'s coverage and 2)~verify that \gls{hira} reliably works, presented in \secref{hira:sec:experiments_coverage} and \secref{hira:sec:experiments_chargerestoration}, respectively. Second, these works} focus on reducing the performance overhead of periodic refreshes, {but \emph{not} the increasingly worsening performance overhead of preventive refresh {operations} of RowHammer defenses~\rhdefworse{}. In contrast, Svärd eliminates unnecessary preventive refresh operations, and HiRA reduces the time spent on both periodic and preventive refresh operations. 

\head{Circuit-level modifications to reduce the {performance impact} of refresh operation{s} {(e.g., \cite{hassan2019crow, luo2020clrdram, kim2000dynamic, kim2003blockbased,yanagisawa1988semiconductor,ohsawa1998optimizing,nair2014refresh,orosa2021codic, choi2020reducing})}}
{{These} works develop DRAM-based techniques that 1)~reduce the latency of a refresh operation~\cite{hassan2019crow, luo2020clrdram}, {2)~3)~reduce the rate at which some or all DRAM rows are refreshed~\cite{kim2000dynamic,kim2003blockbased,yanagisawa1988semiconductor,ohsawa1998optimizing}, and 3)~implement a new refresh command that can be interrupted to quickly perform main memory accesses~\cite{nair2014refresh}}. As opposed to HiRA, these techniques} {1)}~are \emph{not} compatible with {existing} DRAM chips as they require modifications {to} DRAM circuitry,
{and} 2)~\emph{cannot} hide DRAM access latency in the presence of refresh operations. 

{\head{Memory {access} scheduling techniques to reduce the performance impact of refresh operations (e.g.,~\cite{mukundan2013understanding,stuecheli2010elastic, chang2014improving, pan2019thecolored, pan2019hiding, kotra2017hardwaresoftware})}}
{{Several} works {propose 
{issuing} $REF$ commands} during \emph{DRAM idle time} {(where no memory access requests {are} scheduled)} {to reduce the performance impact of refresh operations.} {{Most of these} works leverage} {the flexibility of delaying a $REF$ command for {multiple refresh intervals (e.g., for \SI{70.2}{\micro\second} in DDR4~\cite{jedec2020jesd794c})}}. In contrast, \gls{hira} overlaps the latency of a refresh operation with other refreshes or memory accesses and thus can reduce the performance impact of refresh operations \emph{without} relying on {DRAM} idle time and Svärd eliminates unnecessary preventive refresh operations based on the DRAM chip's read disturbance vulnerability profile.}

{\head{Modifications to the DRAM architecture to leverage subarray-level parallelism {(e.g.,~\salprefs{})}} 
Several works partially overlap the latency of refresh operations or memory accesses via modifications to the DRAM architecture. {The \gls{hira} operation builds on the basic idea{s} of subarray-level parallelism introduced in~\cite{kim2012acase} and {refresh-access parallelization introduced in}~\cite{chang2014improving,zhang2014cream}}. {However,} unlike our work on leveraging subarray-level parallelism, HiRA, these works require modifications to DRAM chip design, {and} thus they are \emph{not} compatible with {off-the-shelf} DRAM chips.} In contrast, {HiRA} uses {existing} {\gls{act} and \gls{pre}} commands{, {and we demonstrate that it works on real off-the-shelf DRAM chips}}.


{\head{{Reducing the latency of {major} DRAM operations} {(e.g.,~\cite{hassan2016chargecache, das2018vrldram, wang2018reducing, zhang2016restore, shin2014nuat, lee2013tieredlatency, lee2015adaptivelatency, kim2018solardram,chang2016understanding, chang2017understanding, lee2017designinduced,chandrasekar2014exploiting,shin2015dramlatency_optimization, mathew2017using, mathew2020using})}}} 
{{Many} {works}} {develop techniques that} {\emph{reduce}} the {latency of {major} DRAM operations (e.g., \gls{act}, \gls{pre}, \gls{rd}, \gls{wr}) by leveraging {1)}~temporal locality in workload access patterns~\cite{hassan2016chargecache,das2018vrldram,wang2018reducing,zhang2016restore,shin2014nuat}{,} 2)~the guardbands in manufacturer-recommended DRAM timing parameters~\cite{lee2013tieredlatency,lee2015adaptivelatency,kim2018solardram,chang2016understanding,lee2017designinduced, chang2017understanding, mathew2017using, mathew2020using}{, and 3)~variation in DRAM {latency} due to temperature dependence~\cite{chandrasekar2014exploiting,lee2015adaptivelatency,kim2018solardram}.}}
{{These latency reduction techniques} can improve system performance by alleviating the performance impact of refresh operations (e.g., a refresh can be performed {faster} with a reduced charge restoration latency).}
{These techniques can be combined with HiRA to further alleviate the performance impact of refresh operations, as HiRA \emph{overlaps} 
the latency of refreshing a DRAM row 
with the latency of refreshing or {activating} another row {in the same bank}.}

\section{{{RowHammer} Attacks}}
{Many previously proposed RowHammer attacks~\exploitingRowHammerAllCitations{} {identify} the rows at the desired vulnerability level (e.g., templating) to increase {the attack's} success probability. This step fundamentally differs from our characterization because it is enough for an attacker to identify a few rows at the desired vulnerability level. In contrast, as a defense-oriented work, we identify the vulnerability level of {\emph{all}} DRAM rows {in a DRAM bank} to configure read disturbance solutions correctly without compromising their security guarantees. Therefore, \agy{7}{our experimental observations present} a significantly more {detailed chip-level} characterization compared to {related attacks}.}

\section{Read Disturbance in Other Memory Technologies}

Although this dissertation focuses on DRAM read disturbance, \agy{7}{the disturbance phenomenon does \emph{not} exclusively happen on DRAM chips. On the contrary, other scaled memory and storage technologies, including SRAM~\cite{chen2005modeling, guo2009largescale, kim2011variationaware}, flash~\cite{lee2008drain, cai2012error, cooke2007theinconvenient, grupp2009characterizing, cai2012flash, ha2013areaddisturb, mielke2008biterror, sugahara2014memory, cai2015read, cai2017error, cai2017errors, watanabe2018system, cai2018errors, mutlu2018guest} and hard disk drives~\cite{jiang2003crosstrack, tang2008understanding, wood2009thefeasibility}, exhibit such disturbance problems. Similarly, phase-change memory~\cite{lee2009architecting, zhou2009adurable, qureshi2009scalable, qureshi2009enhancing, wong2010phase, raoux2008phasechange, lee2010phase, lee2010phasechange, yoon2012rowbuffer, yoon2014efficient}, STT-MRAM~\cite{chen2010advances, kultursay2013evaluating, khan2018analysis}, and resistive memory (RRAM/ReRAM/memristors)~\cite{wong2012metaloxide, staudigl2024itsgetting, staudigl2022neurohammer, kumar2023fault} are likely to exhibit read disturbance-related \agy{6}{robustness} issues~\cite{meza2013acase, mutlu2013memory, mutlu2017therowhammer, mutlu2018guest, mutlu2023fundamentally}.
Although the circuit-level error mechanisms are different in different technologies, the high level root cause of the problem, \emph{cell-to-cell interference}, i.e., that the memory cells are too close to each other, is a fundamental issue that appears and will appear in any technology that scales down to small enough technology nodes. Thus, we expect such problems to continue as we scale other memory technologies, including emerging ones, to higher densities~\cite{mutlu2017therowhammer, mutlu2023fundamentally}.}
\agy{7}{We hope and expect that our experimental insights into DRAM read disturbance and architecture-level solutions that mitigate read disturbance efficiently will inspire future research on understanding and solving disturbance problem in other memory technologies.}

\setcounter{version}{99}
\chapter[A Deeper Look into RowHammer's Sensitivities]{A Deeper Look into RowHammer}
\label{chap:deeperlook}




\hyphenation{cryp-to-graphy succeeding}

\DeclareRobustCommand{\bgyellow}[1]{{{\sethlcolor{soulyellow}\hl{#1}}}}
\DeclareRobustCommand{\bgblue}[1]{{{\sethlcolor{soulcyan}\hl{#1}}}}
\DeclareRobustCommand{\bgred}[1]{{{\sethlcolor{red}\hl{#1}}}}
\DeclareRobustCommand{\bggreen}[1]{{{\sethlcolor{flamingopink}{\hl{#1}}}}}


\newcommand{\squishlist}{
 \begin{list}{$\circ$}
  { \setlength{\itemsep}{0pt}
     \setlength{\parsep}{0pt}
     \setlength{\topsep}{0pt}
     \setlength{\partopsep}{0pt}
     \setlength{\leftmargin}{1em}
     \setlength{\labelwidth}{1em}
     \setlength{\labelsep}{0.5em} } }

\newcommand{\squishsublist}{
\begin{list}{$\rightarrow$}
 { \setlength{\itemsep}{0pt}
    \setlength{\parsep}{0pt}
    \setlength{\topsep}{-10em}
    \setlength{\partopsep}{-3pt}
    \setlength{\leftmargin}{1em}
    \setlength{\labelwidth}{1em}
    \setlength{\labelsep}{0.5em} } }

\newenvironment{squishenum}
{\begin{enumerate}[leftmargin=*]
  \setlength{\itemsep}{0em}
  \setlength{\parskip}{0pt}
  \setlength{\parsep}{0pt}
  \setlength{\topsep}{3pt}
  \setlength{\partopsep}{0pt}}
{\end{enumerate}}

\newcommand{\squishend}{
  \end{list}  }

\newcommand{\figlbls}[0]{Figs.}
\newcommand{\figrefs}[1]{\figlbls{}~\ref{#1}}

\newcommand{\seclbls}[0]{§§}
\newcommand{\secrefs}[1]{\seclbls{}~\ref{#1}}

\newcommand{\obslbls}[0]{Obsvs.}
\newcommand{\obsrefs}[1]{\obslbls{}~\ref{#1}}

\newcommand{\lois}[1]{{#1}} 
\newcommand{\jk}[1]{{#1}}
\newcommand{\js}[1]{{#1}}
\newcommand{\hluo}[1]{{#1}}
\newcommand{\mpo}[1]{{#1}} 
\renewcommand{\om}[1]{{#1}}
\newcommand{\atb}[1]{{#1}}

\newcommand{\atbc}[1]{}
\newcommand{\loiscomment}[1]{}
\newcommand{\onurcomment}[1]{}
\newcommand{\hluocomment}[1]{}
\newcommand{\jktodo}[1]{}
\newcommand{\mptodo}[1]{}

\newcommand{\agyl}[1]{}
\newcommand{\oml}[1]{}
\newcommand{\hluol}[1]{}
\newcommand{\ignore}[1]{}

\renewcommand{\numchips}{272}
\newcommand*{\myalign}[2]{\multicolumn{1}{#1}{#2}}

\section{Motivation and Goal}
\label{deeperlook:sec:motivation}

To enable RowHammer-safe operation in future DRAM-based computing systems in an effective {and} efficient way, it is {critical} to {rigorously} gain {detailed} insights into the RowHammer vulnerability {and its sensitivities to} varying attack properties. Unfortunately, despite the existing research efforts expended towards understanding RowHammer{~\understandingRowHammerAllCitations{}}, scientific literature lacks rigorous experimental observations on how {the} RowHammer vulnerability varies with three fundamental properties:
1)~DRAM chip {temperature}, 2)~{aggressor row active time}, 
and 3)~{victim DRAM cell's physical location}. This {lack of understanding} raises very practical and important concerns {{as to} how the {effects of these} three fundamental properties can be {exploited} to improve both RowHammer attacks and defense mechanisms.}

{\textbf{{O}ur goal}} in this \agy{0}{chapter} is to {rigorously} evaluate and understand how the RowHammer vulnerability of {a} real DRAM chip at the circuit level changes with {1)~}temperature, {2)~aggressor row active time, and 3)~victim} DRAM cell{'s physical} location {in the DRAM chip}. Doing so provides us {with} a deeper understanding of RowHammer to enable future research {on improving the effectiveness of existing RowHammer attacks and defense mechanisms. We hope that these analyses will pave the way for {building} RowHammer-safe systems {that use increasingly more vulnerable} DRAM chips.}
{To achieve this goal, we rigorously characterize how the RowHammer vulnerability of \param{248}~DDR4 {and \param{24}~DDR3 {modern} DRAM chips} from \param{four} major DRAM {manufacturers} vary with these three properties.}

\section{Methodology}
\label{deeperlook:sec:methodology}

We describe our methodology and infrastructure for characterizing the RowHammer vulnerability in real DRAM modules. 

\subsection{Testing Infrastructure}
\label{deeperlook:sec:testing_infrastructure}

We experimentally study {\param{248}} DDR4 and \param{24} DDR3 DRAM chips across a {wide} range of {testing} conditions.
{We use two different testing infrastructures: 1)~SoftMC~\cite{hassan2017softmc, safari2017softmc}, capable of precisely controlling temperature and command timings of DDR3 DRAM modules and 2)~\agy{3}{DRAM Bender~\cite{olgun2023dram_bender, safari2022dram_bender}} that supports DDR4 chips,}
also used in~\cite{kim2020revisiting, frigo2020trrespass, olgun2021quactrng, hassan2021uncovering}.

\figref{deeperlook:fig:infrastructure} shows one of our \agy{3}{DRAM Bender} setups for testing DDR4 modules (\figref{deeperlook:fig:infrastructure}a). 
We use {two} types of {Xilinx} FPGA boards: {1)}~Alveo U200~\cite{xilinxu200} (\figref{deeperlook:fig:infrastructure}b) to test DDR4 DIMMs~\cite{jedec2020jesd794c}, and {2)}~ML605~\cite{xilinx2012ml605} to test DDR3 SODIMMs.
{This infrastructure enables precise control over both DDR4 and DDR3 timings at the granularity of \SI{1.5}{\nano\second} and \SI{2.50}{\nano\second}, respectively.}
We use a host machine, connected to our FPGA boards through a PCIe port~\cite{xilinx2011virtex6} (\figref{deeperlook:fig:infrastructure}c) to
{1})~perform the RowHammer tests that we describe in \secref{deeperlook:sec:testing_methodology}
{and 2})~monitor and adjust the temperature of DRAM chips in cooperation with the temperature controller (\figref{deeperlook:fig:infrastructure}d).

\begin{figure}[htbp] \centering
    \includegraphics[width=\linewidth,trim={0 1.5cm 0 0},clip]{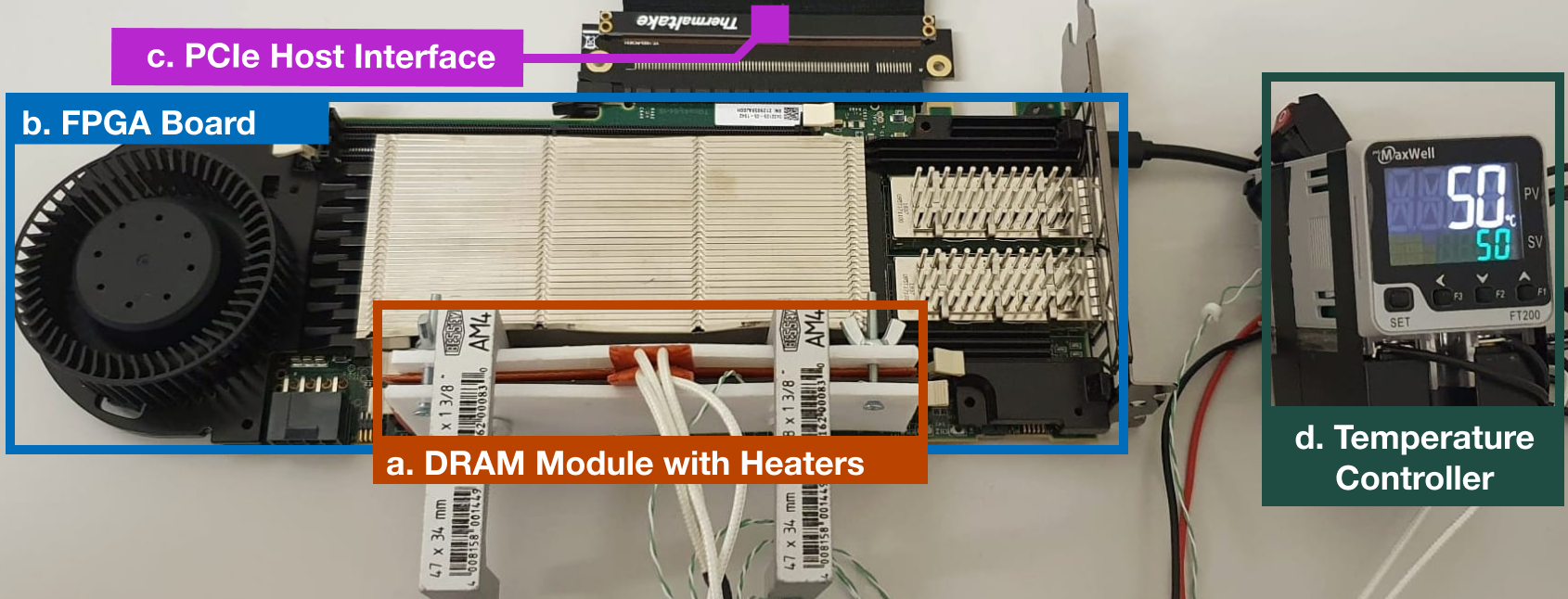}
    \caption{\agy{3}{DRAM Bender} Infrastructure: (a)~DRAM module under test clamped with heater pads, (b)~Xilinx Alveo U200 FPGA board~\cite{xilinxu200}, programmed with \agy{3}{DRAM Bender~\cite{hassan2017softmc, safari2017softmc, olgun2023dram_bender, safari2022dram_bender}}, (c)~PCIe connection to the host machine, and (d)~temperature controller.
    }
    \label{deeperlook:fig:infrastructure}
\end{figure}

\head{{Temperature Controller}}
To regulate the temperature in DRAM modules, we use silicone rubber heaters pressed to both sides of the {DRAM} module (\figref{deeperlook:fig:infrastructure}a). 
We use a thermocouple, placed on the DRAM  chip to measure the chip's temperature {(similar to JEDEC standards~\cite{jedec1995eiajesd511})}. A Maxwell FT200 temperature controller~\cite{maxwellft20x} (\figref{deeperlook:fig:infrastructure}d) 1)~monitors a DRAM chip's temperature using a thermocouple, and 2)~keeps the temperature stable by heating the chip with heater pads. The temperature controller 1)~communicates with our host machine 
via an RS485 channel~\cite{instruments2014rs485} to get a reference temperature and to report the instant temperature, and 2) controls the heater pads using a closed-loop PID controller.
In our tests using this infrastructure, we measure temperature with an error of {at} most $\pm$\SI{0.5}{\celsius}. {{We believe {that} our {temperature measurements from the DRAM package's surface accurately represent the DRAM die's real temperature} because the temperature of the DRAM package and the DRAM internal components are {strongly} correlated~\cite{micron2002tn0008}.}}

\subsection{Testing Methodology}
\label{deeperlook:sec:testing_methodology}

\head{Disabling Sources of Interference} 
Our goal is to directly observe the circuit-level bitflips such that we can make conclusions about DRAM's vulnerability to RowHammer at the circuit technology level rather than {at} the system level. {To this end, we minimize all possible sources of interference with the following steps. 
First,} we disable all DRAM self-regulation events {(e.g., DRAM Refresh}~\cite{jedec2020jesd794c,hassan2017softmc,xilinxultrascale}) except {calibration related events (e.g., ZQ calibration} for signal integrity~\cite{jedec2020jesd794c,hassan2017softmc}). {Second, we ensure that all RowHammer tests are conducted within a relatively short period of time such that we do \emph{not} observe retention errors~\cite{liu2013anexperimental,khan2014theefficacy,meza2015revisiting,patel2017thereach,qureshi2015avatar}}.  
{Third, we use the SoftMC/DRAM Bender memory controller~\cite{hassan2017softmc, safari2017softmc, olgun2023dram_bender, safari2022dram_bender} so that we can 1)~issue DRAM commands with precise control (i.e., our commands are \emph{not} impeded by system-issued accesses), and 2)~study the RowHammer vulnerability on DRAM chips without interference from existing system-level RowHammer protection mechanisms (e.g.,~\cite{apple2015about, bains2016distributed, bains2016rowhammer}).}
{Fourth, we test DRAM modules that do {\emph{not}} implement error correction codes (ECC){~\cite{hamming1950error,bose1960onaclass,hocquenghem1959codes,reed1960polynomial,cojocar2019exploiting,kim2016allinclusive}}. {Doing so ensures} that {neither} on-die~\cite{patel2020bitexact,patel2019understanding,micron2017eccbrings,nair2016xedexposing,patel2021harp} {nor} rank-level~\cite{cojocar2019exploiting,kim2016allinclusive} ECC  {can} alter {the RowHammer bi{t f}lips we observe and analyze}.}
{Fifth}, we prevent known on-DRAM-die RowHammer defenses (i.e., TRR~\trrCitations{}) from working by \emph{not} issuing refresh commands throughout our tests~\cite{frigo2020trrespass, kim2020revisiting}.

\head{{RowHammer}} 
All our tests use double-sided {RowHammer}~\cite{kim2014flipping, kim2020revisiting, seaborn2015exploiting}, which activates, in an alternating manner, each of the two rows (i.e., aggressor rows) that are physically-adjacent to {a} victim row. {We call this victim row a double-sided victim row.} {We define single-sided victim rows as the rows that are hammered in a single-sided manner by the two aggressor rows (i.e., rows with +2 or -2 distance from victim row).} We define one hammer as a pair of activations to the two aggressor rows.
We perform double-sided {hammering} with the maximum activation rate possible within DDR3/DDR4 command timing {specifications}~\cite{jedec2020jesd794c,jedec2012jesd793d}. Prior works report that this is the most effective access {pattern} for RowHammer attacks on DRAM chips when {RowHammer mitigation mechanisms} are disabled~\cite{kim2014flipping, kim2020revisiting, frigo2020trrespass, cojocar2020arewe, seaborn2015exploiting}{.}{\footnote{Our analysis of aggressor row active time uses a different access sequence that introduces additional delays between row activations. {{S}ee \secref{deeperlook:sec:aggressor_active_time} for details.}}}
{W}e use 150K hammers (i.e., 300K activations) 
in {our} \gls{ber} experiments.\footnote{We find that 150K hammers is low enough to be used in a system-level RowHammer attack in a real system~\cite{frigo2020trrespass}, and it is high enough to provide a large number of bitflips in all DRAM modules we tested{.}} {We} use up to 512K hammers (i.e., the maximum number of hammers so that our hammer tests run for less than 64ms) in {our} \gls{hcfirst} experiments. Due to time limitations, {we repeat each test five times,} and we study the effects of the RowHammer attack on the 1) first \param{8K} rows, 2) last \param{8K} rows, and 3) middle \param{8K} rows of {a bank} in each DRAM chip (similar to ~\cite{kim2014flipping}).

\head{{Logical-to-Physical} Row Mapping} 
DRAM manufacturers {use {DRAM-internal} mapping} {schemes} to internally translate memory-controller-visible row addresses to physical row {addresses~\inDRAMRowAddressMappingCitations{}{,} which can vary across different DRAM modules}.
We {reverse-engineer} this mapping,
so {that} we can identify and hammer aggressor rows that are physically adjacent to a victim row. We reconstruct the mapping by 1) performing single-sided RowHammer attack on each DRAM row, 2) inferring that the two victim rows with the most RowHammer bitflips are physically adjacent to the aggressor row, and 3) deducing the address mapping after analyzing the aggressor-victim row relationships across all studied DRAM rows. 

\head{Data Pattern} We conduct our experiments on a DRAM module by using the module's \gls{wcdp}.
We identify the \gls{wcdp} {for each module} as the pattern that results in the largest number of bitflips among \param{seven} different data patterns used in prior works on DRAM characterization~\cite{khan2014theefficacy,liu2013anexperimental,patel2017thereach, kim2020revisiting, chang2016understanding, chang2017understanding, lee2017designinduced,khan2016parbor,khan2016acase,khan2017detecting,lee2015adaptivelatency}{, presented in Table~\ref{deeperlook:tab:data_patterns}:} 
{colstripe, checkered, rowstripe{, and} random {(we also test the {complements} of the first three)}.} 
For each RowHammer test, we write the corresponding data pattern
to the victim row ($V$ in Table~\ref{deeperlook:tab:data_patterns}), and to the 8 previous ($V - [1...8]$) and next ($V + [1...8]$) physically-adjacent rows.

\begin{table}[h]
    \centering
    \caption{Data patterns used in our RowHammer {analyses.} }
    \vspace{-10pt}
    \begin{tabular}{l|cccc}
    \toprule
           \bf{Row {Address}}
           &  \bf{Colstripe{$^\dagger{}$}} & \bf{Checkered{$^\dagger{}$}} & \bf{Rowstripe{$^\dagger{}$}} & \bf{Random}\\
        \midrule
        $V^* \pm [0,2,4,6,8]$ &\verb$0x55$&\verb$0x55$ &\verb$0x00$&\verb$random$\\
        $V^* \pm [1,3,5,7]$ &\verb$0x55$&\verb$0xaa$&\verb$0xff$&\verb$random$\\
        \bottomrule
    \end{tabular}
    \begin{flushleft}
       $\quad^*V$ is the physical address of the victim row \\
       $\quad^\dagger{}${We also test the {complements} of these} patterns
    \end{flushleft}  
    \label{deeperlook:tab:data_patterns}
    \vshort{}
\end{table}

\glsresetall 

\head{Metrics} We measure two metrics in our tests: 1)~\gls{hcfirst} and 2)~\gls{ber}. {A {\emph{lower}} \gls{hcfirst} {or \emph{higher} \gls{ber}} value indicates {higher} RowHammer vulnerability{.}}
{To {quickly} identify \gls{hcfirst}, we perform a binary search {where} we use an initial hammer count of 256K. We repeatedly increase (decrease) the hammer count by $\Delta$ if we observe {(do not observe)} bitflips in the victim row. The initial value is $\Delta = 128K$, and we halve it for each test until it reaches $\Delta = 512$ (i.e., we identify \gls{hcfirst} with an accuracy of 512 row activations){.}}

\head{Temperature {Range}} To study the effects of temperature, we test DRAM chips across a wide range of temperatures, from \SI{50}{\celsius} to \SI{90}{\celsius}, with a step size of \SI{5}{\celsius}. 

\subsection{Characterized DRAM Chips}
\label{deeperlook:sec:characterized_dram_chips}
\label{deeperlook:sec:characterized_region}

Table~\ref{deeperlook:tab:dram_chips} summarizes \agy{3}{and Table~\ref{deeperlook:tab:detailed_info} provides a more detailed information about} the \param{\numchips{}} DDR4 and \param{24} DDR3 DRAM chips we test from four major manufacturers. We use a {diverse set of} modules {with} different {chip} densities, die revisions and chip {organizations}.

\begin{table}[h]
    \caption{{Summary of {DDR4 (DDR3)} {DRAM} chips tested.}}
    \label{deeperlook:tab:dram_chips}
    \vshort{}
    \centering
    \footnotesize{}
    \setlength\tabcolsep{3pt} 
    \begin{tabular}{cccrlcl}
        \toprule
            {{\bf Mfr.}} & \parbox{2cm}{\centering\bf{{DDR4}}\\\#DIMMs} & \parbox{2cm}{\centering\bf{{DDR3}}\\\#SODIMMs} & {{\bf  \#Chips}}  & {{\bf Density}} & {{\bf Die}}& {{\bf Org.}}  \\
        \midrule
        {Mfr. A}&9 &  1&144  (8)&{8Gb} (4Gb)&{B} (P)&{x4} (x8)\\
        {Mfr. B}&4 &  1& 32  (8)&{4Gb} (4Gb)&{F} (Q)&{x8} (x8)\\
        {Mfr. C}&5 &  1& 40  (8)&{4Gb} (4Gb)&{B} (B)&{x8} (x8)\\
        {Mfr. D}&4  & --& 32 (--)&{8Gb} (--)&{C} (--)&{x8} (--)      \\
        \bottomrule
    \end{tabular}
\end{table}

\begin{table}[ht]
\centering
\caption{Characteristics of the tested DDR4 and DDR3 DRAM modules.}
\label{deeperlook:tab:detailed_info}
\begin{adjustbox}{angle=0}
\resizebox{\linewidth}{!}{
\begin{tabular}{l|ll|ccccccc}
\textbf{Type}         & \textbf{\begin{tabular}[c]{@{}l@{}}Chip Mfr.\\ Module Vendor\end{tabular}}  & \textbf{\begin{tabular}[c]{@{}l@{}}Chip Identifier\\ Module Identifier\end{tabular}}                                                                                                    & \textbf{\begin{tabular}[c]{@{}c@{}}Freq. \\ (MT/s)\end{tabular}} & \textbf{\begin{tabular}[c]{@{}c@{}}Date \\ Code\end{tabular}} & \textbf{Density}     & \textbf{\begin{tabular}[c]{@{}c@{}}Die \\ Rev.\end{tabular}} & \textbf{Org.}       & \textbf{\#Modules} & \textbf{\#Chips} \\ \hline
\multirow{9}{*}{DDR4} & \multirow{3}{*}{\begin{tabular}[c]{@{}l@{}}A: Micron\\ Micron\end{tabular}} & \multirow{3}{*}{\begin{tabular}[c]{@{}l@{}}MT40A2G4WE-083E:B\\ MTA18ASF2G72PZ-\\ 2G3B1QG$\sim$\cite{micron2016ddr4mta18asf2g72pz}\end{tabular}}                        & \multirow{3}{*}{2400}                                            & 1911                                                          & \multirow{3}{*}{8Gb} & \multirow{3}{*}{B}                                           & \multirow{3}{*}{x4} & 6                  & 96               \\ 
                      &                                                                             &                                                                                                                                                                                         &                                                                  & 1843                                                          &                      &                                                              &                     & 2                  & 32               \\ 
                      &                                                                             &                                                                                                                                                                                         &                                                                  & 1844                                                          &                      &                                                              &                     & 1                  & 16               \\ 
                      & \begin{tabular}[c]{@{}l@{}}B: Samsung\\ G.SKILL\end{tabular}                & \begin{tabular}[c]{@{}l@{}}K4A4G085WF-BCTD$\sim$\cite{samsung2021k4a4g085wfbctd}\\ F4-2400C17S-8GNT$\sim$\cite{gskill2021f42400c17s8gnt}\end{tabular} & 2400                                                             & 2021 Jan $\star$                                              & 4Gb                  & F                                                            & x8                  & 4                  & 32               \\ \cline{2-10} 
                      & \begin{tabular}[c]{@{}l@{}}C: SK Hynix\\ G.SKILL\end{tabular}               & \begin{tabular}[c]{@{}l@{}}DWCW (Partial Marking) $\dag$\\ F4-2400C17S-8GNT$\sim$\cite{gskill2021f42400c17s8gnt}\end{tabular}                                          & 2400                                                             & 2042                                                          & 4Gb                  & B                                                            & x8                  & 5                  & 40               \\ \cline{2-10} 
                      & \begin{tabular}[c]{@{}l@{}}D: Nanya\\ Kingston\end{tabular}                 & \begin{tabular}[c]{@{}l@{}}D1028AN9CPGRK $\ddag$\\ KVR24N17S8/8$\sim$\cite{kingston2021kvr24n17s88}\end{tabular}                                                       & 2400                                                             & 2046                                                          & 8Gb                  & C                                                            & x8                  & 4                  & 32               \\ \hline
\multirow{6}{*}{DDR3} & \begin{tabular}[c]{@{}l@{}}A: Micron\\ Crucial\end{tabular}                 & \begin{tabular}[c]{@{}l@{}}MT41K512M8DA-107:P$\sim$\cite{crucial2017mt41k512m8da107p}\\ CT51264BF160BJ.M8FP\end{tabular}                                               & 1600                                                             & 1703                                                          & 4Gb                  & P                                                            & x8                  & 1                  & 8                \\ \cline{2-10} 
                      & \begin{tabular}[c]{@{}l@{}}B: Samsung\\ Samsung\end{tabular}                & \begin{tabular}[c]{@{}l@{}}K4B4G0846Q\\ M471B5173QH0-YK0$\sim$\cite{samsung2013m471b5173qh0}\end{tabular}                                                              & 1600                                                             & 1416                                                          & 4Gb                  & Q                                                            & x8                  & 1                  & 8                \\ \cline{2-10} 
                      & \begin{tabular}[c]{@{}l@{}}C: SK Hynix\\ SK Hynix\end{tabular}              & \begin{tabular}[c]{@{}l@{}}H5TC4G83BFR-PBA\\ HMT451S6BFR8A-PB$\sim$\cite{hynix2013ddr3lHMT451S6BFR8A}\end{tabular}                                                     & 1600                                                             & 1535                                                          & 4Gb                  & B                                                            & x8                  & 1                  & 8                \\ \hline
\end{tabular}

}
\end{adjustbox}
\begin{flushleft}
$\star$ We use the date marked on the modules due to the lack of date information on the chips.

$\dag$ A part of the chip identifier is removed on these modules. We infer the DRAM chip manufacturer and die revision information based on the remaining part of the chip identifier.

$\ddag$ We extract the DRAM chip manufacturer and die revision information from the serial presence detect (SPD) registers on the modules.
\end{flushleft}
\end{table}
\section{Temperature Analysis}
\label{deeperlook:sec:temperature}

{We} {1)}~provide the first rigorous experimental characterization {of the effects of temperature} on the RowHammer vulnerability using real DRAM chips and {2)}~present new observations and insights based on our results.

\subsection{{Impact} of Temperature {on} DRAM Cells}

We analyze the relation between {temperature and the} RowHammer vulnerability of a DRAM cell {using the methodology described in Section~\ref{deeperlook:sec:testing_methodology}}. To {do so, {we first}} cluster vulnerable {DRAM} cells by their {\emph{vulnerable temperature range} (i.e., the {minimum and maximum temperatures within}} which a cell {experiences at least one} {RowHammer} bitflip across all {experiments}). 
{{Second,} we analyze {\emph{how}} the RowHammer bitflips of DRAM cells manifest within their vulnerable temperature range.}
Table~\ref{deeperlook:tab:temperature_gaps} shows the percentage of vulnerable cells that flip in \emph{all} temperature {points} of {their} vulnerable temperature {ranges}.

\begin{table}[h]
    \centering
    \caption{{Percentage of vulnerable DRAM cells that flip in all temperature points within the vulnerable temperature range of the cell.}}
    \vshort{}
    \begin{tabular}{rrrr}
        \toprule
        {\bf{Mfr. A}} & {\bf{Mfr. B}} & {\bf{Mfr. C}} & {\bf{Mfr. D}}\\
        \midrule
        {99.1\%} & {98.9\%} & {98.0\%} & {99.2\%} \\ 
        \bottomrule
    \end{tabular}
    \label{deeperlook:tab:temperature_gaps}
\end{table}

\observation{A DRAM cell is, with a very high probability, vulnerable to RowHammer in a continuous temperature range specific to the cell.\label{deeperlook:temp:bounded}}

For example, only \param{0.9\%} of the vulnerable DRAM cells in Mfr. A  do {\emph{not}} exhibit bitflips in at least one temperature point within their vulnerable temperature range. Hence, our experiments demonstrate that a cell {exhibits} bitflips  {with {very} high probability} in a {continuous} temperature range {that is specific to the cell}.

{To analyze the diversity of vulnerable temperature ranges across DRAM cells, we cluster all vulnerable DRAM cells according to their vulnerable temperature {ranges}. \figref{deeperlook:fig:tempIntervals} shows each cluster's {size} as a percentage of the full population of vulnerable cells.}
{The x-axis (y-axis) indicates the lower (upper) bound of the vulnerable temperature range.}
{Because we do not test temperatures higher (lower) than \SI{90}{\celsius} (\SI{50}{\celsius}), the vulnerable temperature ranges with an upper (lower) limit of \SI{90}{\celsius} (\SI{50}{\celsius}) include cells that {also} flip  at higher (lower) temperatures. For example, 5.4\% of the vulnerable DRAM cells in Mfr. A fall into the range \SIrange{70}{90}{\celsius}, which includes cells with \emph{actual} vulnerable temperature ranges of} \SIrange{70}{95}{\celsius}, \SIrange{70}{100}{\celsius}, etc.

\begin{figure}[ht] 
    \centering
    \includegraphics[width=0.9\linewidth]{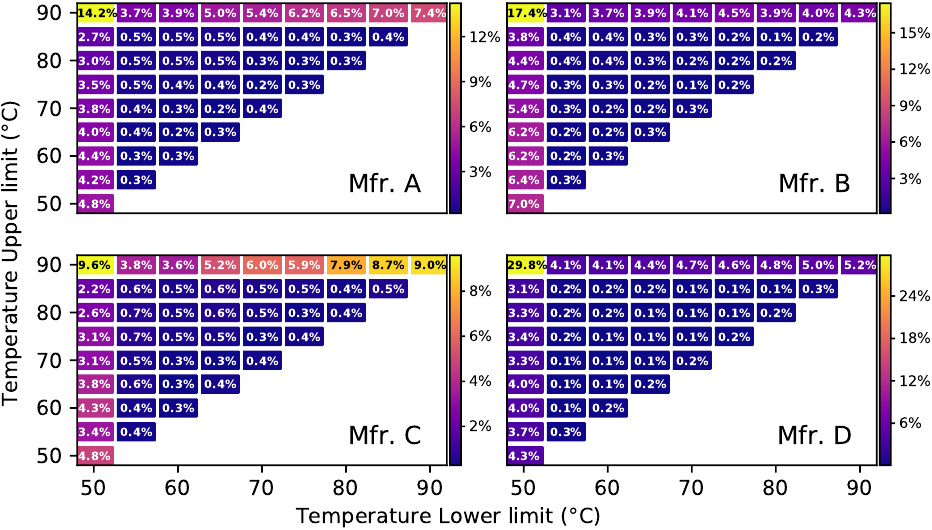}
    \vshort{}
    \caption{Population of {vulnerable DRAM cells}, clustered by vulnerable temperature range.}
    \label{deeperlook:fig:tempIntervals}
    \vshort{}
\end{figure}

\observation{A significant fraction of vulnerable DRAM cells exhibit bitflips at all tested  temperatures.\label{deeperlook:temp:wide}}

We observe that {between \param{9.6\%} and \param{29.8\%} of the cells (x-axis}=\SI{50}{\celsius}, {y-axis}=\SI{90}{\celsius} in  \figref{deeperlook:fig:tempIntervals}) are vulnerable to RowHammer {across {\emph{all}} tested} temperatures {(\SIrange{50}{90}{\celsius})} for {the four} DRAM manufacturers. We also verify {(not shown)} that {\obsref{deeperlook:temp:wide}} holds for the three SODIMM DDR3 modules described in Table~\ref{deeperlook:tab:dram_chips}.

\observation{A small fraction of all vulnerable DRAM cells are vulnerable to RowHammer only in a very narrow temperature range.\label{deeperlook:temp:narrow}}

For example, {\param{0.4\%} of all vulnerable DRAM cells {of {Mfr. A}},} are {only vulnerable to RowHammer at \param{\SI{70}{\celsius}} (i.e., a single tested temperature value).} {Note} that inducing even a single bitflip can be critical for system security, as shown by prior works~\cite{xiao2016onebit, frigo2018grand, gruss2018another, razavi2016flip}. Our experimental results show that 2.3\%, 1.8\%, 2.4\%, and 1.6\%  of all tested DRAM cells  for {Mfrs.} A, B, C, and D, respectively, experience a RowHammer bitflip within a temperature range as narrow as \SI{5}{\celsius}. {We conclude that {some} DRAM cells {experience} RowHammer bitflips at localized {and narrow} temperature {ranges}.}

{We exploit {\obsrefs{deeperlook:temp:bounded}--\ref{deeperlook:temp:narrow} in \secref{deeperlook:sec:implications}}.}

\take{{To ensure that a DRAM cell is not vulnerable to RowHammer, we must characterize the cell at all operating temperatures.}}

\subsection{Impact of Temperature on DRAM Rows}
\label{deeperlook:sec:temperature_rows}
We  analyze  the  relation  between  a DRAM row's RowHammer vulnerability and temperature in terms of both \gls{ber} and \gls{hcfirst}.

\head{\emph{BER Analysis}}
\figref{deeperlook:fig:ber_temp} shows {how the} \gls{ber} {changes} {as temperature increases}, {compared to the mean {\gls{ber}} value across all the samples at \SI{50}{\celsius}}, for four DRAM manufacturers. In each {plot}, we {use a point and error bar}\footnote{{Each point and error bar represent the mean and the 95\% confidence interval across the samples, respectively.}} to show the \gls{ber} {change} for the victim row (i.e., distance from the victim row = 0), and the \gls{ber} {change} for {the two single-sided victim rows (i.e., distance $\pm$2 from the victim row), across all rows we test}. 
\begin{figure}[h] \centering
    \includegraphics[width=0.9\linewidth]{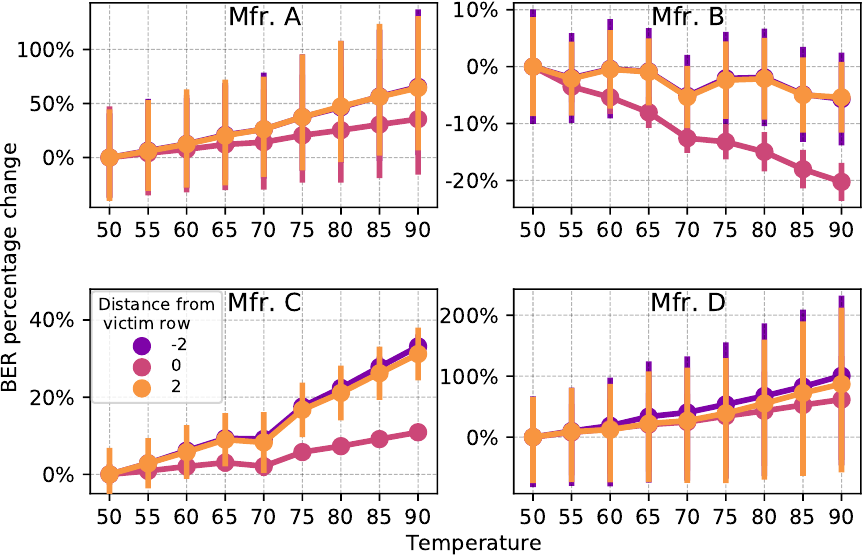}
    \vshort{}
    \caption{{{Percentage} change {in $BER$ (RowHammer bitflips)} with increasing temperature, compared to $BER$ at $50^\circ$C.}}
    \label{deeperlook:fig:ber_temp}
\end{figure}

\observation{{A DRAM row's} \gls{ber} can either increase or decrease with temperature depending on the DRAM manufacturer.\label{deeperlook:temp:ber_vs_temp}} 

{We observe that {the average \gls{ber} of} all three victim rows (one {double-sided} victim row and two single-sided victim rows),}
 from {Mfrs.} A, C, and D {increases with temperature}, whereas {the \gls{ber} of} rows from Mfr. B {decreases} as temperature increases. 
We hypothesize that the difference between these trends is caused by a combination of DRAM circuit design and manufacturing process technology differences {(see \secref{deeperlook:sec:temperature_circuit})}.

\head{\gls{hcfirst} Analysis}
\figref{deeperlook:fig:hcfirst_variation} shows the distribution of {the change in} \gls{hcfirst} {(in percentage)} when temperature {increases} from \param{\SI{50}{\celsius}} to \param{\SI{55}{\celsius}}, and from \param{\SI{50}{\celsius}} to \param{\SI{90}{\celsius}}, for the vulnerable rows of {the} \param{four} manufacturers. 
The x-axis {represents the percentage of all} vulnerable rows, sorted from {the} {most} positive \gls{hcfirst} {change} to {the} {most} negative \gls{hcfirst} {change}.
For each curve, we mark the {x-axis} {point at which the curve crosses the y=0\% line. This represents} the percentile of rows {whose} \gls{hcfirst} {increases with temperature;} 
e.g., for Mfr. A, when temperature {increases} from \SI{50}{\celsius} to {\SI{90}{\celsius}}, {only} 45\% (P45) of the tested rows have a higher \gls{hcfirst} {(indicating reduced vulnerability for that fraction of rows); i.e., most rows {from Mfr. A} are more vulnerable {at \SI{90}{\celsius} than at \SI{50}{\celsius}}}. For clarity, we only show two temperature changes (i.e., from \SI{50}{\celsius} to \SI{55}{\celsius} and from \SI{50}{\celsius} to \SI{90}{\celsius}), but our observations are consistent across all intermediate temperature changes we tested (i.e., from \SI{50}{\celsius} to 50+$\Delta ^\circ$C, for all $\Delta$'s that are {multiples} of \SI{5}{\celsius}). 

\begin{figure}[h] \centering
    \includegraphics[width=0.9\linewidth]{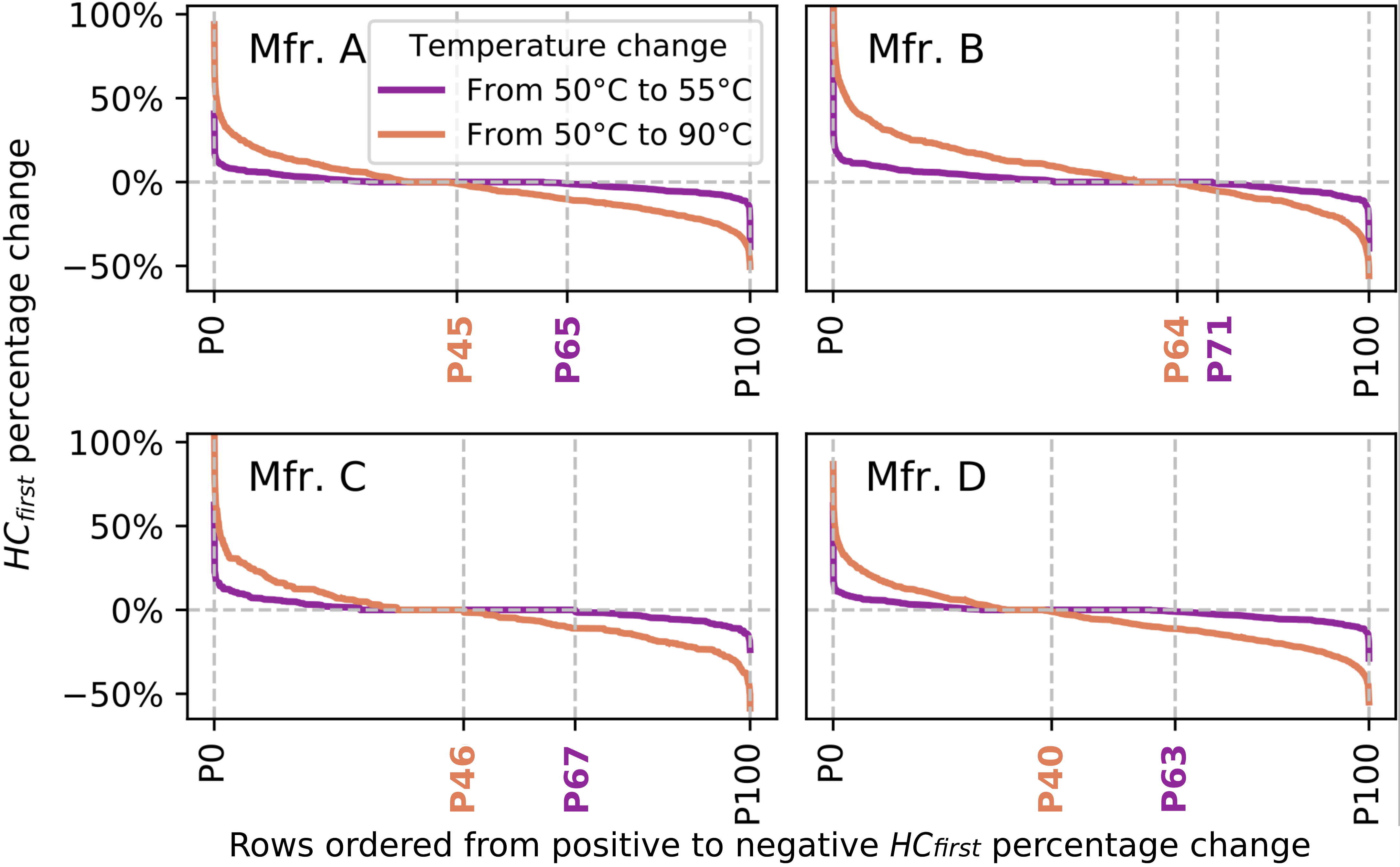}
        \vshort{}
    \caption{Distribution of {the change in} $HC_{first}$ across vulnerable DRAM rows {as} temperature increases.}
    \vshort{}
    \label{deeperlook:fig:hcfirst_variation}
\end{figure}

\observation{{DRAM rows can show either higher or lower \gls{hcfirst} when temperature increases.}\label{deeperlook:temp:hcfirst_vs_cells}}

{We observe that, for all four manufacturers, a significant {fraction} of rows can show either higher or lower \gls{hcfirst} when temperature increases. For example, when the temperature changes from  \SI{50}{\celsius} to \SI{55}{\celsius} in Mfr.~A, 65\% of the rows show higher \gls{hcfirst}, {while} 35\% of the rows show lower \gls{hcfirst}. We conclude that \gls{hcfirst} changes differently depending on the DRAM row.}

\observation{\gls{hcfirst} tends to generally decrease as temperature {change increases}.\label{deeperlook:temp:hcfirst_vs_temperature}}

We observe that, for all four manufacturers,
{fewer rows have a {higher} \gls{hcfirst} when the temperature delta is larger;} {i.e., the point at which each curve crosses the y=0\% {point} shifts left when the temperature change increases.}
For example, {for Mfr. D,} the fraction of {vulnerable} cells with {a higher} \gls{hcfirst} is much {larger} when temperature increases from \SI{50}{\celsius} to \SI{55}{\celsius} (63\% of cells) than when the temperature increases from \SI{50}{\celsius} to \SI{90}{\celsius} (40\% of  cells). {We conclude that the dominant trend is for a row's \gls{hcfirst} to decrease when {the temperature delta is larger}.} 

\observation{The change in \gls{hcfirst} tends to be larger as the temperature change is larger.\label{deeperlook:temp:hcfirst_vs_temperature_larger}}

{The \gls{hcfirst} distribution curve exhibits higher absolute magnitudes when temperature changes from \SI{50}{\celsius} to \SI{90}{\celsius}, compared to when temperature changes from \SI{50}{\celsius} to \SI{55}{\celsius} {(i.e., the curve {generally} rotates right {and has much higher peaks at its edges} when the temperature change increases, {i.e., going from orange to purple in the figure})}. {We quantify this observation by calculating the cumulative magnitude change (i.e., the sum of the absolute values of {the} \gls{hcfirst} change from all rows). Our results show that the cumulative magnitude change {(not shown in the figure)} is  4.2$\times$, 3.9$\times$, 3.8$\times$ and 4.3$\times$ larger in Mfrs. A, B, C, and D, respectively, when the temperature changes from \SI{50}{\celsius} to \SI{90}{\celsius}, compared to
\SI{50}{\celsius} to \SI{55}{\celsius}.}} {We conclude that a larger change in temperature causes a larger change in \gls{hcfirst}.} 

\take{{RowHammer vulnerability (i.e., both \gls{ber} and \gls{hcfirst})} {tend} to worsen {as DRAM} temperature increases. However, individual DRAM rows can exhibit behavior {different} {from} this dominant trend.}

\subsection{{Circuit-level Justification}}
\label{deeperlook:sec:temperature_circuit}
We hypothesize that {our observations on} {the relation between RowHammer vulnerability and temperature } {are} caused by the non-monotonic behavior of charge trapping characteristics of DRAM cells.
{Yang et al.~\cite{yang2019trapassisted} show a DRAM charge trap model simulated using a 3D~TCAD tool ({\emph{without}} real DRAM chip experiments).}
{The model} shows that  \gls{hcfirst} decreases as temperature increases, until a {temperature inflection} point where \gls{hcfirst} {starts} to increase as temperature increases.
{According to this model, a cell is more vulnerable to RowHammer at temperatures close to its temperature inflection point. 
We hypothesize that rows within a DRAM chip might have a {wide} variety of temperature {inflection} points, {and} thus the average temperature inflection point of a DRAM chip would determine whether the {average} RowHammer vulnerability increases or decreases with temperature
{(\obsrefs{deeperlook:temp:bounded}--\ref{deeperlook:temp:hcfirst_vs_temperature_larger})}.}
Park et al.~\cite{park2014activeprecharge, park2016experiments} also {show} {an analysis of the relation between }\gls{hcfirst} {and DRAM temperature. Their observations are similar to ours,} but they consider {only} a small number 
of DDR3 DRAM cells.

Unlike simulations and {limited} results {reported} by~\cite{yang2019trapassisted,park2014activeprecharge, park2016experiments}, our {comprehensive} experiments with {\param{{272}}} DRAM chips show that 
{the temperature inflection points {for RowHammer vulnerability}  are very diverse {across DRAM cells and chips}}. 
\section{Aggressor Row Active Time Analysis}
\label{deeperlook:sec:aggressor_active_time}
{We} provide the first rigorous characterization of RowHammer considering the time that the aggressor row stays in the row buffer (i.e., \emph{{aggressor row active time}}).
{Prior {works}~\cite{park2014activeprecharge, park2016experiments, walker2021ondram} propose} circuit models and {suggest that RowHammer vulnerability of a victim row can depend on the aggressor row active time based on} preliminary data on a {very} small number of DRAM cells (i.e., only one carefully-selected DRAM row from each manufacturer)~\cite{park2014activeprecharge, park2016experiments}.
However, none of these {works} conduct a rigorous analysis {of how} RowHammer vulnerability {varies} with aggressor row active time across a significant population of DRAM rows from real off-the-shelf DRAM modules. 

{\figref{deeperlook:fig:tagg_on_off_procedure} describes the three test{s} we {perform} in our experiments:}  
{1) {\emph{Baseline Test}}, where we use $\tras{}$ as \gls{taggon}, and we use $\trp{}$ as \gls{taggoff},} 2)~\emph{Aggressor On Tests}, where we increase \gls{taggon} before the row is precharged (compared to {$\tras{}$ in Baseline Test}), and {3)}~\emph{Aggressor Off Tests}, where we increase \gls{taggoff} {before the aggressor row {is activated} (compared {to $\trp{}$ in Baseline Test})}.
Therefore, {for a given hammer count $HC$,} the overall attack time is $(t_{AggOn} + t_{RP}) \times HC$ and $(t_{RAS} + t_{AggOff}) \times HC$ for Aggressor On and Off Tests, respectively, while it is $(t_{RAS} + t_{RP}) \times HC$ for the baseline tests.
{Our experiments in this section are conducted at \SI{50}{\celsius} on the first 1K rows, the last 1K rows, and the 1K rows in the middle of a bank in our DDR4 chips.}

\begin{figure}[ht] 
    \centering
    \includegraphics[width=0.85\linewidth]{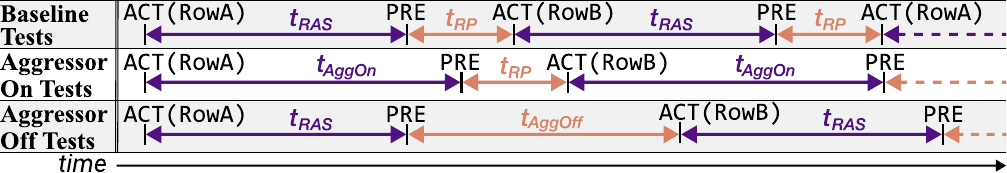}
    \vshort{}
    \vshort{}
    \caption{DRAM command timings for aggressor row active time ($t_{AggOn}$/$t_{AggOff}$) experiments. Purple/Orange color indicates that an aggressor row is active/precharged.}
    \label{deeperlook:fig:tagg_on_off_procedure}
    \vspace{-3pt}
    
\end{figure}

\label{deeperlook:sec:temporal}
\subsection{Impact of Aggressor Row's {On-}Time}
{\figref{deeperlook:fig:ber_vs_tagg_on} and \figref{deeperlook:fig:hcf_vs_tagg_on} show the RowHammer bitflips per row (\gls{ber}) and \gls{hcfirst} distributions using box plots\footnote{{In a {box plot~\cite{Tukey1977Exploratory}}, {the} box shows the lower and upper quartile of the data (i.e., the box spans the $25^{\mathrm{th}}$ to the $75^{\mathrm{th}}$ percentile of the data). The line in the box represents the median. The bottom and top whiskers each represent {an} additional $1.5\times$ the \emph{inter-quartile range} (IQR, the range between the bottom and the top of the box) beyond the lower and upper quartile, respectively.}} and letter-value plots,\footnote{{In a letter-value plot~\cite{hofmann2017lettervalue}, the widest box shows the lower and upper quartile of the data. The line in the box represents the median. The narrower box extended from the bottom of the widest box shows the lower octile ($12.5^{\mathrm{th}}$ percentile) and the lower quartile of the data, and the narrower box extended from the top of the widest box shows the upper octile and the upper quartile of the data, etc. Boxes are plotted until all remaining data are outliers. Outliers are defined as the 0.7\% extreme values in the dataset, and are plotted as fliers in the plot.}} respectively, across all DRAM chips, as we vary \gls{taggon} {from \SI{34.5}{\nano\second} {($\tras{}$)} to \SI{154.5}{\nano\second}}.}

\begin{figure}[t] 
    \centering
    \includegraphics[width=0.8\linewidth]{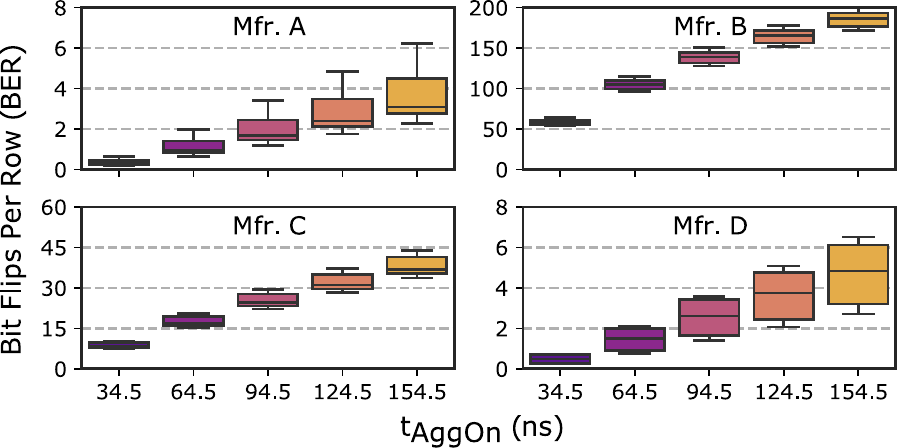}
    \caption{Distribution of the average number of bitflips per victim row {across chips as aggressor row} on-time ($t_{AggOn}$) {increases}.}
    \label{deeperlook:fig:ber_vs_tagg_on}
\end{figure}

\begin{figure}[ht] 
    \centering
    \includegraphics[width=0.8\linewidth]{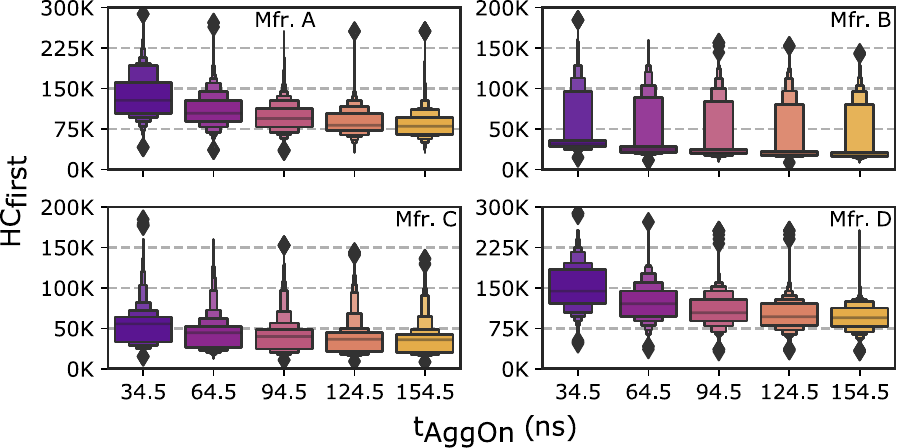}
    \caption{Distribution of {per-row} $HC_{first}$ {across chips as aggressor row} on-time ($t_{AggOn}$) {increases}.}
    \label{deeperlook:fig:hcf_vs_tagg_on}
    \vshort{}
\end{figure}

\observation{As the aggressor row stays active longer (i.e., \gls{taggon} increases), more DRAM cells experience RowHammer bitflips and they experience RowHammer bitflips at lower hammer counts.\label{deeperlook:taggon:vulnerability}}

{We observe that increasing \gls{taggon} from \SI{34.5}{\nano\second} to \SI{154.5}{\nano\second} \emph{{significantly}}  1)~}{increases \gls{ber} by $10.2\times,\ 3.1\times,\ 4.4\times,\ \text{and } 9.6\times$ on average}
{and 2)}~{decreases \gls{hcfirst}  by 40.0\%, 28.3\%, 32.7\%, and 37.3\% on average, in DRAM chips from Mfrs. A, B, C and D, respectively.}

\observation{RowHammer vulnerability consistently worsens as \gls{taggon} increases in DRAM chips from all four {manufacturers}.\label{deeperlook:taggon:bervariation}}

To see how  RowHammer vulnerability changes {as \gls{taggon} increases}, we examine \gls{cv}\footnote{$CV={standard~deviation/average}$~\cite{biran1998thecambridge}.} values {of} {the} \gls{ber} and \gls{hcfirst} distributions {(not shown in the figures)}. We find that \gls{cv} {decreases} by around 15\% and 10\% for \gls{ber} and \gls{hcfirst}, respectively, across all four {manufacturers\om{,} as \gls{taggon} increases from \SI{34.5}{\nano\second} to \SI{154.5}{\nano\second}}. {This indicates that increasing {the} aggressor {row} active time {consistently worsens} RowHammer vulnerability across {the} DRAM chips {we test}.}

{{We conclude from }\obsrefs{deeperlook:taggon:vulnerability} and~\ref{deeperlook:taggon:bervariation} that increasing \gls{taggon} makes victim DRAM cells {much} more vulnerable to {a} RowHammer attack. {We exploit these observations in \secref{deeperlook:sec:implications}.}}

\take{{As {an aggressor row stays} active longer}, {victim DRAM cells become more vulnerable to RowHammer.}
\label{deeperlook:taggon:takeaway}}

\subsection{Impact of Aggressor Row's Off-Time}
\figrefs{deeperlook:fig:ber_vs_tagg_off} and~\ref{deeperlook:fig:hcf_vs_tagg_off} show the \gls{ber} and \gls{hcfirst} distributions, {respectively, as we vary} \gls{taggoff} {from \SI{16.5}{\nano\second} {($\trp{}$)} to \SI{40.5}{\nano\second}}.{\footnote{Statistical configurations {of the box} and letter-value plots in \figrefs{deeperlook:fig:ber_vs_tagg_off} and~\ref{deeperlook:fig:hcf_vs_tagg_off} are identical {to those in} \figrefs{deeperlook:fig:ber_vs_tagg_on} and~\ref{deeperlook:fig:hcf_vs_tagg_on}, respectively.}}

\begin{figure}[h!] 
    \centering
    \includegraphics[width=0.8\linewidth]{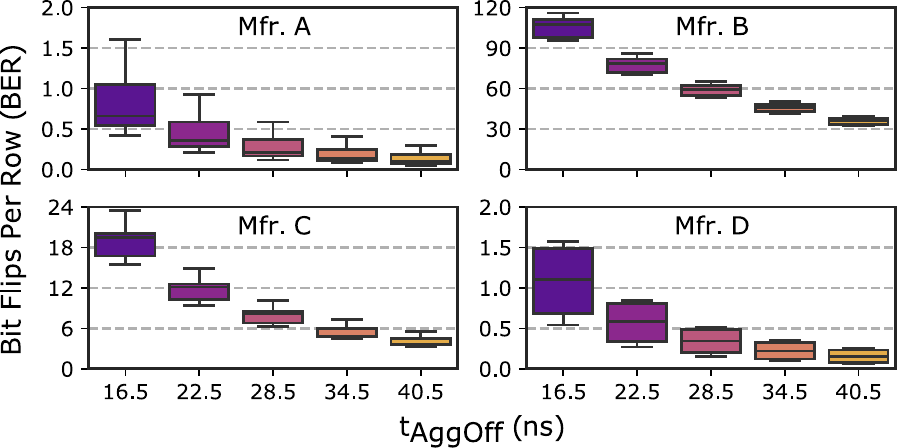}
    \caption{Distribution of the average number of {bitflips} per victim row {across chips as aggressor row} off-time ($t_{AggOff}$) {increases}.}
    \label{deeperlook:fig:ber_vs_tagg_off}
\end{figure}

\begin{figure}[ht] 
    \centering
    \includegraphics[width=0.8\linewidth]{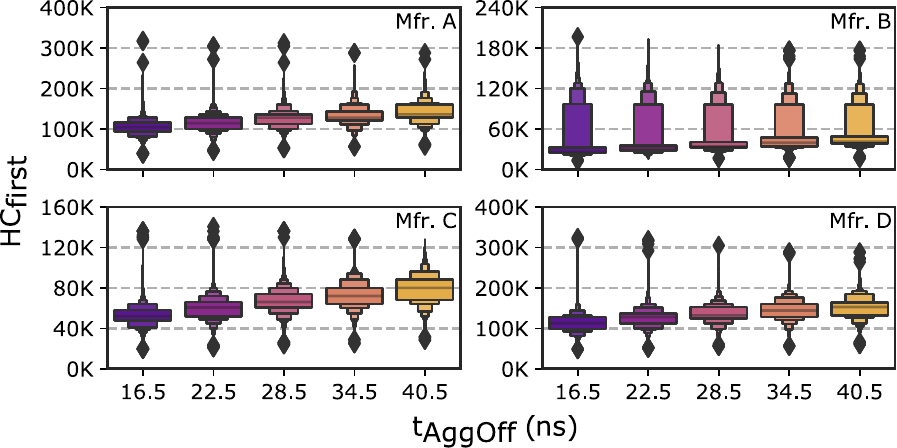}
    \caption{Distribution of {per-row} $HC_{first}$ {across chips as aggressor row} off-time ($t_{AggOff}$) {increases}.}
    \label{deeperlook:fig:hcf_vs_tagg_off}
    \vshort{}
\end{figure}

\observation{
As the bank stays precharged longer (i.e., \gls{taggoff} increases), fewer DRAM cells experience RowHammer bitflips and they experience RowHammer bitflips at higher hammer counts.
\label{deeperlook:taggoff:vulnerability}}

{We observe that {increasing \gls{taggoff} {from \SI{16.5}{\nano\second} to \SI{40.5}{\nano\second}}} {\emph{significantly}} 1)~decreases \gls{ber} {by} $6.3\times,\ 2.9\times,\ 4.9\times,\text{ and } 5.0\times$ {on average,} and 2)~increases \gls{hcfirst} by 33.8\%, 24.7\%, 50.1\%, and 33.7\% {on average,} in DRAM chips from Mfrs. A, B, C, and D, respectively{.}
}

\observation{
RowHammer vulnerability consistently reduces {as \gls{taggoff} increases} in DRAM chips from all four manufacturers.\label{deeperlook:taggoff:bervariation}}

We observe that {the} \gls{cv} of \gls{hcfirst} {(not shown in the figures)} does not increase for any manufacturer
{as we increase }\gls{taggoff}. 
{Hence}, the level of reduction in RowHammer vulnerability is {similar} across {different rows'} \emph{most vulnerable cells}.
{In contrast,} {the} \gls{cv} of \gls{ber} increases by 18\% {on average} {for all} four manufacturers{, indicating {that the level} of reduction in RowHammer vulnerability {is different} across different rows}. 

{We conclude from} \obsrefs{deeperlook:taggoff:vulnerability} and~\ref{deeperlook:taggoff:bervariation} that increasing \gls{taggoff} makes it harder for a RowHammer {attack} to be successful. 
{We exploit this to improve} {RowHammer} {defense mechanisms} in \secref{deeperlook:sec:implications_defense}.

\take{{RowHammer vulnerability of victim cells decreases} when the {bank is precharged for a longer time}.}

\subsection{{Circuit-level Justification}}
\label{deeperlook:sec:temporal_circuit}
{{Prior work explains two circuit- and device-level mechanisms, causing RowHammer bitflips:} 
1) electron injection into the victim cell
~\cite{walker2021ondram, yang2016suppression}, and 2) {wordline{-to}-wordline} cross-talk noise between {aggressor and victim rows that occurs} when the aggressor row is being activated~\cite{walker2021ondram, ryu2017overcoming}. 
{We hypothesize that increasing the aggressor row's active time (\gls{taggon}) has a larger impact on exacerbating electron injection {to} the victim cell, 
compared to the reduction {in} cross-talk noise due to
lower activation frequency. Thus, RowHammer vulnerability worsens when \gls{taggon} {increases,} as our \obsrefs{deeperlook:taggon:vulnerability} and~\ref{deeperlook:taggon:bervariation} show.}
}

{{On the other hand, }{increasing a bank's precharged time (\gls{taggoff})}
{decreases} RowHammer vulnerability (\obsrefs{deeperlook:taggoff:vulnerability} and~\ref{deeperlook:taggoff:bervariation})
because longer \gls{taggoff} reduces the effect of cross-talk noise {without affecting electron injection} (since \gls{taggon} is unchanged). We leave the detailed device-level analysis and explanation of our observations to future works.}
\section{{Spatial Variation Analysis}}
\label{deeperlook:sec:spatial}

{We} provide the first {rigorous} {spatial variation} analysis of RowHammer across DRAM rows, subarrays, and columns. Prior work~\cite{kim2014flipping, kim2020revisiting, park2014activeprecharge, park2016experiments, park2016statistical} analyzes RowHammer vulnerability {at the} DRAM bank granularity across many DRAM modules without {providing analysis of} the variation of this vulnerability across rows, subarrays, and columns. We provide this analysis and show that it is useful for improving both attacks and defense mechanisms.
{Our experiments in this section are conducted at \SI{75}{\celsius}.}

\subsection{Variation Across DRAM Rows} 
\label{deeperlook:sec:spatial_acrossrows}
\figref{deeperlook:fig:hcfirst_across_rows} shows the {distribution of} \gls{hcfirst} {values} across {all vulnerable} DRAM rows {among the rows we test (\secref{deeperlook:sec:testing_methodology}). {For each row, we plot the minimum \gls{hcfirst} value observed across 5 repetitions of the test.} Each subplot {shows} DRAM modules from a different manufacturer, and each {curve} {corresponds to} a different DRAM module.}
The x-axis {shows} all the tested rows, sorted by decreasing \gls{hcfirst} {and marked with percentiles ranging from P1 to P99}.

\begin{figure}[h!] 
    \centering
    \includegraphics[width=\linewidth]{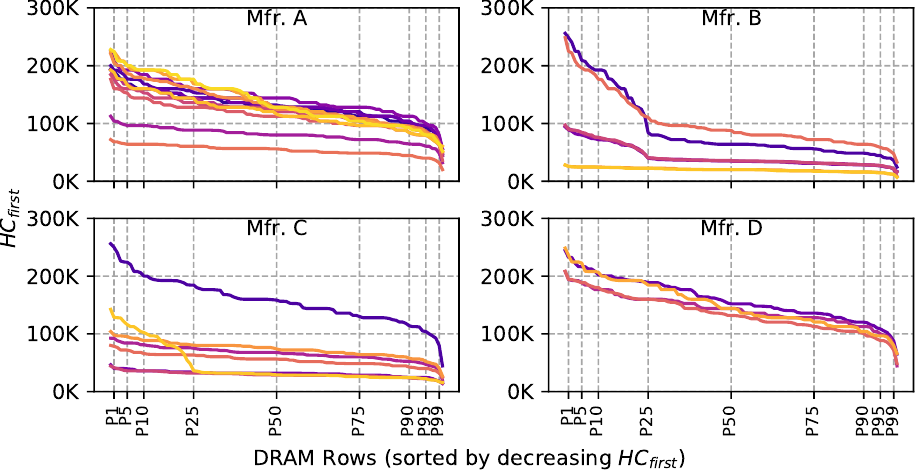}
    \vshort{}
    \vshort{}
    \caption{Distribution of $HC_{first}$ across {vulnerable} DRAM rows. {Each curve represents a different tested DRAM module}.}
    \label{deeperlook:fig:hcfirst_across_rows}
\end{figure}

\observation{A small fraction of DRAM rows are significantly more vulnerable to RowHammer than the vast majority of the rows.\label{deeperlook:spatial:hcfirst_across_rows}}

\gls{hcfirst} {varies significantly} across {rows}. We observe that 99\%, 95\%, and 90\% {of tested rows} {exhibit \gls{hcfirst} values {that are at least}}
\param{1.6$\times$}, \param{2.0$\times$}, and \param{2.2$\times$} {greater than} the {most vulnerable row's} \gls{hcfirst}{,} on average across all \param{four} {manufacturers}.
For example, 
{the lowest \gls{hcfirst} across all tested rows in a DRAM module from Mfr.~B is \param{33K},}
while 99\%, 95\%, and 90\% of the rows in the {same} {module} 
{exhibit \gls{hcfirst} values {equal to or greater than}}
\param{48.5K, 60.5K, and 64K}, respectively. {Therefore, we conclude that a small fraction of DRAM rows are significantly more vulnerable to RowHammer than the vast majority of the rows.}

The {large} variation {in} \gls{hcfirst} across DRAM rows can 
enable future improvements in
low-cost RowHammer {defenses} 
{(\secref{deeperlook:sec:implications_defense})}. 

\subsection{Variation Across Columns} 
\label{deeperlook:sec:spatial_acrosscols}
\figref{deeperlook:fig:ber_across_columns} shows
the distribution of {the number of} RowHammer bitflips across columns in {eight} representative DRAM {chip}s from {each of} \param{all four} manufacturers.
For each DRAM chip {(y-axis)}, we count the bitflips in each column {(x-axis)} across all {24K tested} rows.
The color-scale next to each subplot shows the bitflip count: a brighter color {indicates} more bitflips.

\begin{figure}[h!] 
    \centering
    \includegraphics[width=\linewidth]{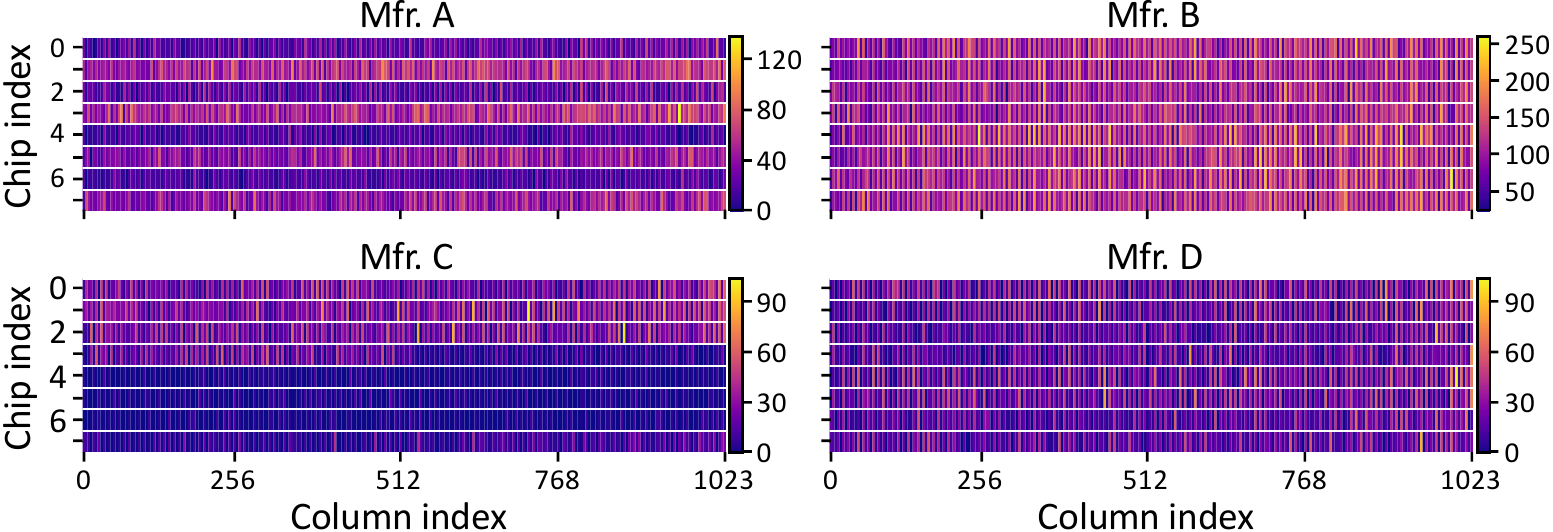}
    \caption{RowHammer bitflip distribution across columns in representative DRAM chips from four different manufacturers.}
    \label{deeperlook:fig:ber_across_columns}
    \vshort{}
\end{figure}

\observation{Certain columns are significantly more vulnerable to RowHammer than other columns.\label{deeperlook:spatial:columns}}

All chips show {significant variation} {in \gls{ber}} across columns. 
For example, the difference between the maximum and the minimum bitflip counts {per column} is larger than {\param{100} in modules from all four manufacturers.}
{Except for the module from Mfr.~B, where every column shows at least 6 bitflips, all the other tested modules have a considerable fraction of columns where \emph{no} bitflip occurs (27.80\%/31.10\%/9.96\% in Mfr.~A/C/D),
along with a very small faction of columns with more than 100 bitflips
(0.59\%/0.01\%/0.61\% in Mfr.~A/C/D).} {Therefore, we conclude that certain columns are significantly more vulnerable to RowHammer than other columns.}

{To better understand this column-to-column variation, we study how RowHammer vulnerability varies between columns \emph{within} a single DRAM chip and \emph{across} different DRAM chips.}
{{Understanding this variation can provide insights into the {impact of circuit design} on a column's RowHammer vulnerability, {which is {important} for understanding and overcoming RowHammer}. A smaller variation in a column's RowHammer vulnerability across chips indicates a stronger influence of design-induced variation~\cite{lee2017designinduced, kim2018solardram}, while a larger variation across {chips that implement} the same design indicates a stronger influence of manufacturing process variation~\cite{lee2015adaptivelatency, chang2016understanding, chang2017understanding, liu2012raidr, patel2017thereach, liu2013anexperimental,kim2019drange,kim2018thedram}.} {{To differentiate between these two sources of variation in our experiments}, we cluster {every column in a given DRAM module}
based on
two metrics.
The first metric is 
{the} column's \emph{relative RowHammer vulnerability}, defined as the column's \gls{ber}, normalized to the maximum \gls{ber} across all columns in the same module.
The second metric is \emph{the RowHammer vulnerability variation} at a column address.
We quantify the variation using 
the coefficient of variation (\gls{cv}) of the relative RowHammer vulnerability in columns with the same column address from different DRAM chips.}} \figref{deeperlook:fig:colvulnerability_across_chips} {shows a two-dimensional histogram with the \emph{relative RowHammer vulnerability} (y-axis) and \emph{Rowhammer vulnerability variation} (x-axis) uniformly quantized into 11 buckets each (i.e., 121 total buckets across each subplot).\footnote{We plot the x-axis as saturated at 1.0 because a \gls{cv} $>$ 1 means that the standard deviation is larger than the average, i.e., the variation is very large across chips.} Each bucket is illustrated as a rectangle containing a percentage value, which shows the percent of all columns that fall within the bucket. Empty buckets are omitted for clarity.}

\begin{figure}[ht!] 
    \centering
    \includegraphics[width=0.85\linewidth]{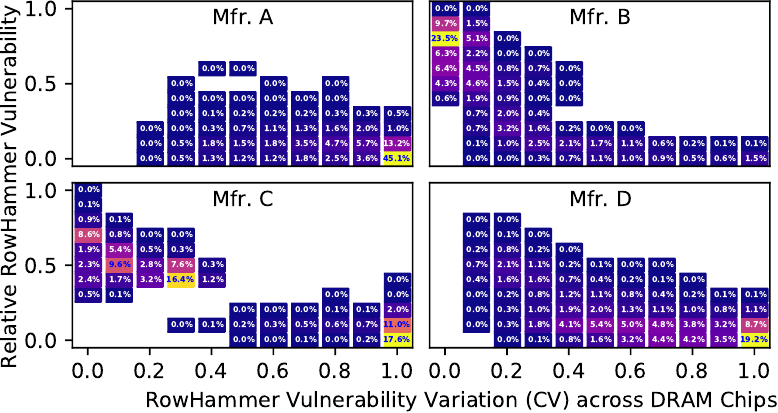}
    \vshort{}
    \vshort{}
    \caption{{Population of DRAM columns, clustered by relative RowHammer vulnerability. }
    }
    \label{deeperlook:fig:colvulnerability_across_chips}
    \vshort{}
\end{figure}

\observation{Both design and manufacturing {processes} may affect {a DRAM column's RowHammer vulnerability.}\label{deeperlook:spatial:design_and_process}}

{We find that} {\param{50.9\%}} and {\param{16.6\%}}\footnote{{These numbers represent the population of columns whose \gls{cv} across chips is zero, i.e., sum of all annotated percentage values where \gls{cv}=0.}} of all vulnerable columns in DRAM modules from {Mfrs. B and C have \gls{cv}=0.0, which indicates that {each of these columns} {exhibit} the same level of RowHammer vulnerability {consistently} across {\emph{all}} DRAM chips {in a module}.} This {consistency} across chips
implies {that \emph{systematic variation} is {present}}, induced by {a chip's design{~\cite{lee2017designinduced, kim2018solardram,lee2013tieredlatency,son2013reducing,vogelsang2010understanding,chang2016lowcost,lee2015adaptivelatency,chang2016understanding}}}.
{In contrast, \param{59.8\%, 30.6\%, and 29.1\%} of vulnerable columns in DRAM modules from {Mfrs.} A, C, and D show a {very} large variation across chips 
(\gls{cv}=1.0).}
This large variation 
across chips suggests {that
\emph{manufacturing process} variation is {\emph{also} a significant factor in} determining} a given DRAM column's RowHammer vulnerability.

{We conclude from} \obsrefs{deeperlook:spatial:hcfirst_across_rows}--\ref{deeperlook:spatial:design_and_process} that there is significant {variation} {in} RowHammer vulnerability across DRAM rows, columns, and chips. These observations {are} useful for 1)~crafting attacks that target vulnerable {locations} ({see} \secref{deeperlook:sec:implications_attack}) or 2)~improving defense mechanisms and error correction schemes that exploit the heterogeneity {of vulnerability} across DRAM rows and columns {({see} \secref{deeperlook:sec:implications_defense}).}

\take{RowHammer vulnerability significantly varies across DRAM rows and columns due to both design{-induced} and manufacturing{-}process{-induced} variation.}

\subsection{Variation Across Subarrays}
\label{deeperlook:sec:spatial_acrosssubarrays}
We analyze the RowHammer vulnerability of individual subarrays across DRAM chips. Since subarray boundaries are not publicly available, we {conservatively}
{assume a subarray size of 512~rows as reported in prior work~\cite{kim2012acase, lee2017designinduced, chang2014improving, kim2018solardram, vogelsang2010understanding}.{\footnote{{We verify this for some of our chips} {by performing 1)~single-sided RowHammer attack tests~\cite{kim2014flipping, kim2020revisiting} that induce bitflips in both rows adjacent to the aggressor row if the aggressor row is \emph{not} at the {edge} of a subarray and 2)~RowClone tests~\cite{seshadri2013rowclone, olgun2021quactrng, gao2019computedram} that can successfully copy data {only} between two rows within the same subarray.}}}}

\figref{deeperlook:fig:hcfirst_across_subarrays} shows the variation of \gls{hcfirst} characteristics {in a DRAM bank} across subarrays {both 1) in a} DRAM module {and 2) across modules from the same manufacturer}. {Each color-marker pair represents a} different DRAM module. We represent the \gls{hcfirst} of a subarray in terms of {1)}~the {average} (x-axis) and {2)~the minimum} (y-axis) of \gls{hcfirst} across {the subarray's rows}. {For each manufacturer, }{we annotate a dashed line {that fits to the data via linear regression {with the specified $R^2$-score~\cite{wright1921correlation}.}}}

\begin{figure}[t] 
    \centering
    \includegraphics[width=0.85\linewidth]{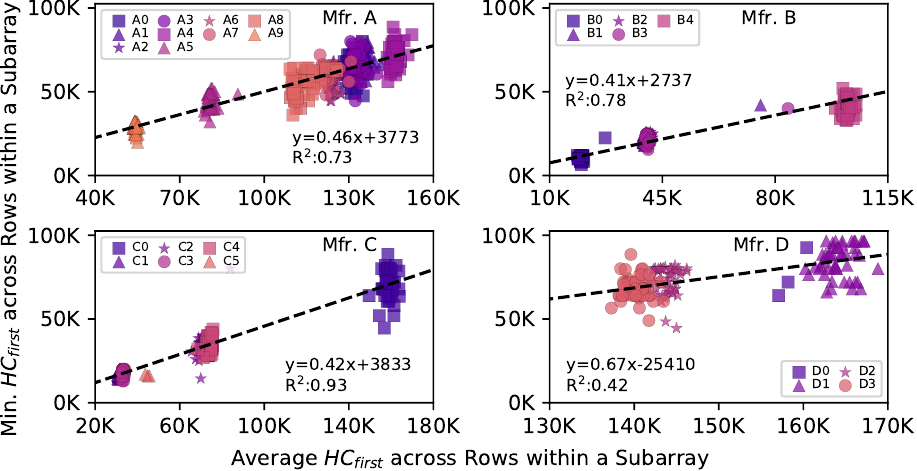}
    \vshort{}
    \vshort{}
    \caption{$HC_{first}$ variation across subarrays. Each subarray is represented by the {average} (x-axis) and the {minimum} (y-axis) $HC_{first}$ across the rows within the subarray.}
    \label{deeperlook:fig:hcfirst_across_subarrays}
    \vshort{}
    \vshort{}
\end{figure}

\observation{The most vulnerable DRAM row in a subarray is significantly more vulnerable than the other rows in the subarray.
\label{deeperlook:spatial:subarray_vs_bank}}

{We make two observations from \figref{deeperlook:fig:hcfirst_across_subarrays}.
First,
the average \gls{hcfirst} across all rows in a subarray is {on the order of 2$\times$} the most vulnerable row's \gls{hcfirst}, i.e., the minimum \gls{hcfirst}.
Therefore, the most vulnerable row in a subarray is \emph{significantly} more vulnerable than the other rows in the same subarray.
}
{{Second,} this relation between the minimum and average \gls{hcfirst} values is similar across subarrays from different modules from the same manufacturer, and thus can {be} {modeled using a linear regression}.
For example, {the minimum \gls{hcfirst} value in a subarray from Mfr.~C can be estimated using {a well-fitting} linear model with a $R^2$-score of 0.93.} 
{This observation is important because it indicates an underlying relationship between the average and minimum \gls{hcfirst} values across subarrays. For example, although subarrays in module~C0 have significantly larger \gls{hcfirst} values than subarrays from module~C3, a the linear model accurately expresses the relationship between both subarray's minimum and average \gls{hcfirst} values. Therefore, given {a} module from {Mfr.} C, the data shows that it may be possible to {predict} the minimum {(worst-case)} \gls{hcfirst} values of {another module's} subarrays, given {the} average {\gls{hcfirst}} values {of those subarrays.}}}

{We conclude from these two observations that 1)~the most vulnerable DRAM row in a subarray is significantly more vulnerable than the other rows in the subarray {and~2)~the}} worst-case \gls{hcfirst} in a subarray {can be predicted based on the average \gls{hcfirst} values and the linear models we provide.}

{To analyze and quantify the similarity between the RowHammer vulnerability of different subarrays, we statistically compare each subarray against all other subarrays from the same manufacturer. To compare two given subarrays, we first compare their \gls{hcfirst} distributions using Bhattacharyya distance ($BD$)~\cite{bhattacharyya1943onameasure}{,} which is used to measure the similarity of {two} statistical distributions. Second, for each pair of subarrays ($S_A$ and $S_B$), we normalize $BD$ to the $BD$ {between the first subarray $S_A$ and itself}: $BD_{norm} = BD(S_A, S_B) / BD(S_A, S_A)$. {Therefore}, $BD_{norm}$ is 1.0 if two distributions are identical, while $BD_{norm}$ value gets farther from 1.0 as the variation across two distributions increases.}
{\figref{deeperlook:fig:hcfirst_bd_across_subarrays} shows the cumulative distribution of $BD_{norm}$ values for subarray pairs from 1)~the same DRAM module and 2)~different DRAM modules. We annotate {P5, P95, and the central P90}
of the total population ({y-axis}) to show the range of $BD_{norm}$ values {in common-case.}}

\begin{figure}[t] 
    \centering
    \includegraphics[width=0.85\linewidth]{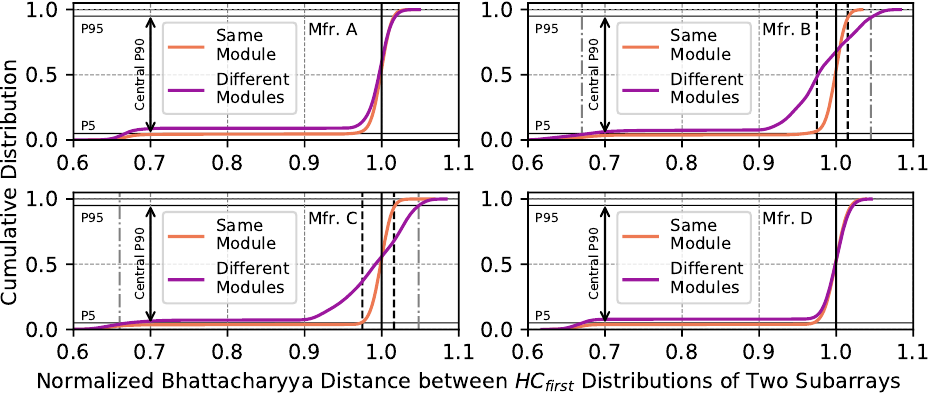}
    \vshort{}
    \vshort{}
    \caption{{Cumulative {distribution} of normalized Bhattacharyya distance {values between $HC_{first}$ distributions {of} different subarrays} from 1) the same DRAM module and 2) different DRAM modules}}
    \label{deeperlook:fig:hcfirst_bd_across_subarrays}
    \vshort{}
\end{figure}

\observation{\gls{hcfirst} distributions of subarrays within a DRAM module exhibit significantly more similarity to each other than \gls{hcfirst} distributions of subarrays from different modules.\label{deeperlook:spatial:bdist}}

{We observe that, when both $S_A$ and $S_B$ are from the same DRAM module (orange curves), the central 90th percentile {(i.e., between 5\% and 95\% of the population, as marked in~\figref{deeperlook:fig:hcfirst_bd_across_subarrays})}
of all subarray pairs exhibit $BD_{norm}$ values close to 1.0 (e.g., $BD_{norm}=0.975$ at the 5th {percentile} for Mfr. C), which means that their \gls{hcfirst} distributions are very similar. In contrast, $BD_{norm}$ values from different modules (purple curves) show a significantly wider distribution, especially for Mfrs. B and C (e.g., $BD_{norm}=0.66$ at the 5th {percentile} for Mfr. C).} 
{From} this {analysis,} {we conclude} that the \gls{hcfirst} 
{distribution within a subarray can be representative of other subarrays from the same DRAM module ({e.g., Mfrs.} B and C), {while} 
{the \gls{hcfirst} distribution within a subarray is often not representative of that of other subarrays for different DRAM modules.}}

\obsrefs{deeperlook:spatial:subarray_vs_bank} {and}~\ref{deeperlook:spatial:bdist} can be useful {for} improving DRAM profiling {techniques} and RowHammer {defense} mechanisms {(\secref{deeperlook:sec:implications_defense})}.

\take{{\gls{hcfirst} distribution in a subarray 1)~contains a diverse set of values and 2)~is similar to other subarrays in the same DRAM module.}}

\subsection{{{Circuit-level Justification}}}
\label{deeperlook:sec:spatial_circuit}
{We observe that RowHammer vulnerability {significantly} varies across DRAM rows, columns, {and chips, while different subarrays {in the same chip} exhibit similar vulnerability characteristics.} 

{\head{Variation across rows, columns, and chips} 
We hypothesize that {two distinct factors cause the variation in RowHammer vulnerability that we observe across rows, columns, and chips: \agy{4}{manufacturing process variation and design-induced variation}}}.

{First, \emph{manufacturing process variation} causes differences in cell size and bitline/wordline impedance values, which introduces variation in} {cell} reliability characteristics {with}in and across {DRAM} chips~{\cite{lee2015adaptivelatency, chang2016understanding, chang2017understanding, liu2012raidr, patel2017thereach, liu2013anexperimental,kim2019drange,kim2018thedram,orosa2021codic,olgun2021quactrng}}. {We hypothesize that similar imperfections in the manufacturing process (e.g., variation in cell-to-cell and cell-to-wordline spacings) cause RowHammer vulnerability to vary between cells in different DRAM chips.}

Second, {\emph{design-induced variation} causes cell access latency characteristics to vary deterministically based on a cell's physical location in the memory chip (e.g., its proximity to I/O circuitry)~\cite{kim2018solardram, lee2017designinduced}. In particular, prior work~\cite{lee2017designinduced} shows that columns closer to wordline drivers (which are typically distributed along a row) can be accessed faster. Similarly, we hypothesize that columns that are closer to repeating analog circuit elements (e.g., wordline drivers, voltage boosters) more sensitive to RowHammer disturbance than columns that are farther away from such elements.}}

{\head{Similarity across subarrays} Prior {works}~{\cite{lee2017designinduced, kim2018solardram}} demonstrate similar DRAM access latency characteristics across different subarrays. {This is because a cell's access latency is dominated by its physical distance from the peripheral structures (e.g., local sense amplifiers and wordline drivers) within the subarray{~\cite{chang2016lowcost,lee2017designinduced,lee2015adaptivelatency,chang2016understanding,kim2018solardram,lee2013tieredlatency,son2013reducing,vogelsang2010understanding}}, causing {corresponding} cells in \emph{different} subarrays to exhibit \emph{similar} access latency characteristics.}
We hypothesize that different subarrays in a DRAM chip exhibit similar RowHammer vulnerability characteristics {for {a similar} reason}.} We leave {further analysis} {and validation} of these hypotheses for future work.
\section{Implications}
\label{deeperlook:sec:implications}
The observations we make in \secref{deeperlook:sec:temperature}-\secref{deeperlook:sec:spatial} can be leveraged for both 1)~crafting more effective {RowHammer} attacks and 2)~{developing more effective and more efficient RowHammer defenses.}

\subsection{Potential Attack Improvements}
\label{deeperlook:sec:implications_attack}
Our {new observations and} characterization data can {help} improve the success probability of a RowHammer attack. We propose \param{three} attack improvements based {on} our {analyses of} temperature (\secref{deeperlook:sec:temperature}), {aggressor {row} active time (\secref{deeperlook:sec:aggressor_active_time}), and spatial variation (\secref{deeperlook:sec:spatial})}.

\head{Improvement 1} {{\obsrefs{deeperlook:temp:bounded}--\ref{deeperlook:temp:narrow}}
can be used to craft more effective RowHammer attacks}
{where} the attacker can control or monitor the DRAM temperature. {{\obsrefs{deeperlook:temp:bounded}--\ref{deeperlook:temp:narrow}}} {show} that a DRAM cell is more vulnerable to RowHammer {within} a specific temperature range.
An attacker that can monitor the DRAM temperature (e.g., a malicious employee in a data center {or} an attacker who performs a remote RowHammer attack~\cite{lipp2018nethammer, tatar2018throwhammer} on a {physically accessible} IoT {device}) can increase the chance of a {bitflip} in two ways. First, the attacker can force the sensitive data to be {stored in} the DRAM cells that are more vulnerable at the {current} {operating} temperature, using known techniques~\cite{razavi2016flip,gruss2018another}. {Second, the attacker can heat up or cool down the chip to a temperature level at which the cells {that} stor{e} sensitive data become more vulnerable to RowHammer. 
As a result, the attacker can 
{significantly reduce the hammer count, {and consequently, the attack time, necessary} to cause a {bitflip}{, {thereby} reducing the probability of being detected}.} 
{For example, without our observations, an attacker {might} choose {an aggressor row based on an \emph{uninformed}} decision {with respect to} temperature characteristics.
In such {a} case, the chosen row could require a hammer count {larger than 100K (\figref{deeperlook:fig:hcfirst_across_rows}).}
However, by leveraging our {{\obsrefs{deeperlook:temp:bounded}--\ref{deeperlook:temp:narrow}}},
an attacker can make a {more \emph{informed} decision} and {choose a row whose \gls{hcfirst} reduces by 50\%}
(\figref{deeperlook:fig:hcfirst_variation})} at the temperature level the attack {is designed to} take place.}

\head{Improvement 2} 
\obsref{deeperlook:temp:narrow} {can be used to} enable a new RowHammer attack {variant} as a temperature-dependent trigger of the main attack (which could be a RowHammer attack, or {some} other {security} attack).
{\obsref{deeperlook:temp:narrow}} demonstrates that some  DRAM  cells are  vulnerable to RowHammer in a very narrow temperature range.
{To implement a temperature-dependent trigger {using a RowHammer bitflip},} an attacker can place the victim data in a row that contains a cell that flips at the target temperature, 
{which allows the attacker to determine whether or not}
the target temperature is reached {to trigger the main attack}. 
{This could be useful for an attacker in two scenarios:
{1)~to trigger the attack only when a precise temperature is reached (e.g., triggering an attack against an IoT device in the field when the device is heated or cooled), and~2)~to identify abnormal operating conditions (e.g., triggering the attack {during peak hours} by using cells whose vulnerable temperature ranges are above the common DRAM chip temperature). For example, to detect that the temperature of a DRAM chip is precisely \SI{60}{\celsius} (above \SI{60}{\celsius}) an attacker can use the cells with a vulnerable temperature range of \SI{60}{\celsius}--\SI{60}{\celsius} ({all ranges with lower limit equal or higher than \SI{60}{\celsius}}), which are 0.3\%/0.3\%/0.3\%/0.2\% {(90.7\%/86.3\%/91.4\%/91.7\%)} of all vulnerable cells in Mfrs. A/B/C/D (\figref{deeperlook:fig:tempIntervals}).}}

\head{Improvement 3} \obsref{deeperlook:taggon:vulnerability} {shows that} keeping an aggressor row active for a longer time results in more {bitflips} and lower \gls{hcfirst} values, {which can be used to craft more powerful RowHammer attacks.} 
{{For example, an} attacker} can increase the aggressor row active time by issuing more READ commands {to the aggressor row}{, which}
can potentially 1) increase the number of {bitflips} for a given hammer count, or 2) defeat already-deployed RowHammer defenses~\mitigatingRowHammerAllCitations{} {by inducing bitflips at a smaller hammer count than the \gls{hcfirst} value used for configuring a defense mechanism{.}}
{For example, issuing $10$~to~$15$ READ commands per aggressor row activation can increase the aggressor row active time by about $5\times$, increasing \gls{ber} by {3.2$\times$--10.2$\times$}
or causing bits to flip at a hammer count {that is} 36\% smaller than the \gls{hcfirst} value {that} {may be} used {to} configure a defense mechanism {that does not consider} our Observation{~\ref{deeperlook:taggon:vulnerability}}.}

\subsection{Potential Defense Improvements}
\label{deeperlook:sec:implications_defense}
Our characterization data can potentially be used in \param{five} ways to improve RowHammer defense methods. 

\head{Improvement 1}
{\obsref{deeperlook:spatial:hcfirst_across_rows} {shows that there is a large spatial variation {in} \gls{hcfirst} across rows. A system designer} can {leverage this observation} to make existing RowHammer defense mechanisms more effective and efficient.}
{A limitation of these mechanisms {is that they are} configured for the smallest {\emph{(worst-case)}} \gls{hcfirst} across all rows in a DRAM bank {even though} {an overwhelming majority} of rows {exhibit significantly larger \gls{hcfirst} values}. This is an important limitation because, {when configured for a smaller \gls{hcfirst} value,} {the performance, {energy,} and area overheads of} {many RowHammer defense} mechanisms significantly {increase~\cite{kim2020revisiting, yaglikci2021blockhammer, park2020graphene}.} 
To {overcome this limitation,} a system designer can configure a RowHammer defense mechanism {to use} different \gls{hcfirst} values for different {DRAM rows}.}
For example, BlockHammer's~\cite{yaglikci2021blockhammer} and Graphene's~\cite{park2020graphene} area costs can reach approximately 0.6\% and 0.5\% of a high-end processor's die area~\cite{yaglikci2021blockhammer}. However, based on our \obsref{deeperlook:spatial:hcfirst_across_rows}, 95\% of DRAM rows {exhibit an \gls{hcfirst} value greater than $2\times$ the worst-case \gls{hcfirst}.}
Therefore, {both BlockHammer and Graphene
can be configured {with} {the worst-case \gls{hcfirst} for {only} 5\% of the rows and {with} $2\times$ \gls{hcfirst} for the 95\% of the rows,}
drastically reducing their area costs {down to 0.4\% and 0.1\% of the processor die area, translating to 33\% and 80\% area cost reduction, respectively.}\footnote{{{Our preliminary evaluation} estimate{s} BlockHammer's~\cite{yaglikci2021blockhammer} and Graphene's~\cite{park2020graphene} area costs for $2\times$\gls{hcfirst}, following the methodology described in BlockHammer~\cite{yaglikci2021blockhammer}.}}
Similarly, the most {area-efficient} defense mechanism PARA~\cite{kim2014flipping} incurs 28\% slowdown on average {for} benign workloads when configured for an \gls{hcfirst} of 1K~\cite{kim2020revisiting}. {T}his {large} performance overhead can be halved~\cite{kim2020revisiting} for 95\% of the rows by simply using {lower} probability thresholds for {less vulnerable} rows. {We leave the comprehensive evaluation of such improvements to future work}.}

\head{Improvement 2} \obsrefs{deeperlook:spatial:subarray_vs_bank} {and} \ref{deeperlook:spatial:bdist} on \emph{spatial {variation}} of \gls{hcfirst} {across subarrays} can be leveraged to {reduce the time required to profile a given DRAM module's RowHammer vulnerability characteristics.}
{{This is an important challenge because profiling a DRAM module's RowHammer characteristics {requires analyzing} several {environmental conditions and attack} properties (e.g., data pattern, access pattern, and temperature), {requiring {time-consuming tests that lead to long profiling times}}~\cite{kim2014flipping, kim2020revisiting, frigo2020trrespass, cojocar2020arewe, kwong2020rambleed, yang2019trapassisted, park2016experiments, park2016statistical, patel2021harp}. 
According to our \obsrefs{deeperlook:spatial:subarray_vs_bank} {and} \ref{deeperlook:spatial:bdist}, characterizing a {\emph{small subset}} of subarrays can provide approximate {yet reliable} profiling data {for an {\emph{entire}} DRAM chip}. For example, assuming that a DRAM bank contains 128~subarrays, profiling {eight randomly-chosen} subarrays reduces RowHammer characterization {time} {by {at least} an order of magnitude. This low-cost approximate profiling can be useful in two cases.} 
First, finding the \gls{hcfirst} of a DRAM row requires performing a RowHammer {test} {with} {varying} hammer counts. Profiling the \gls{hcfirst} value for a few subarrays can be used {to limit} the \gls{hcfirst} search space for {the rows in the} rest of the subarrays based on our \obsref{deeperlook:spatial:bdist}. 
Second, one can profile a few subarrays within a DRAM module and use our linear regression models {(\obsref{deeperlook:spatial:bdist})} to estimate the DRAM module's RowHammer vulnerability for} systems whose reliability and security are not {as} critical (e.g., accelerators {and} {systems} running error-resilient workloads)~\cite{luo2014characterizing, koppula2019eden, nguyen2018anapproximate, nguyen2019stdrc, tu2018rana}.}

{\head{Improvement 3} \obsrefs{deeperlook:temp:bounded} and~\ref{deeperlook:temp:narrow} show a vulnerable DRAM cell experiences bitflips at a particular temperature range. To improve a DRAM chip's reliability, {the system might incorporate a mechanism to temporarily or permanently retire DRAM rows (e.g., via software page offlining~\cite{meza2015revisiting} or hardware DRAM row remapping~\cite{yavits2020wolfram, carter1999impulse}) that are vulnerable to RowHammer within a particular operating temperature {range}. To adapt to changes in temperature, the row retirement mechanism might dynamically adjust the rows that are {retired}, potentially leveraging previously-proposed techniques (e.g., Rowclone~\cite{seshadri2013rowclone}, LISA~\cite{chang2016lowcost}, NoM~\cite{rezaei2020nomnetworkonmemory}, {FIGARO~\cite{wang2020figaro}}) to efficiently move data between these rows.}}

\head{Improvement 4} {\obsref{deeperlook:temp:ber_vs_temp}} demonstrates that overall \gls{ber} significantly increases with temperature across modules from three of the four manufacturers. {To reduce the success probability of a RowHammer attack, a system designer can improve the} cooling infrastructure for systems that use {such} DRAM modules{. Doing so} can reduce the number of RowHammer {bitflips} in a DRAM row.  
{For example,} {when} temperature {drops} from \SI{90}{\celsius} to \SI{50}{\celsius}, {\gls{ber} reduces by 25\%} {on average across DRAM modules from Mfr.~A}.  (see \figref{deeperlook:fig:ber_temp}).

\head{Improvement 5} \obsref{deeperlook:taggon:vulnerability} shows that keeping an aggressor row {active} for a longer time increases the probability of RowHammer {bitflips}.
Therefore, RowHammer defenses should take aggressor row active time into account.
{Unfortunately, monitoring {the} active time of all potential aggressor rows {throughout} an entire refresh window is not feasible for emerging lightweight on-DRAM-die RowHammer defense mechanisms~\trrCitations{}{, because such monitoring would {require} substantial storage and logic {to track} all potential aggressor rows' active times.}
{To address this issue}, the memory controller can {be modified to limit or reduce} {the} active time{s of} all rows {by changes to memory} request scheduling algorithms and{/or row buffer} policies {(e.g., via mechanisms similar to{~\cite{rixner2000memory, kim2010atlas, subramanian2016bliss, yaglikci2021blockhammer, goossens2013conservative, huan2006processor, mutlu2008parallelismaware, subramanian2014theblacklisting, mutlu2007stalltime, moscibroda2007memory, kaseridis2011minimalist})}}. {In this way}, a RowHammer defense mechanism {or the memory controller can inherently} {keep under control} an aggressor row's active time. {This is an example of a} system-DRAM cooperative scheme, similar {to} {those recommended by} prior work{~\cite{mutlu2013memory, kim2020revisiting, orosa2021codic, kim2014flipping,mutlu2017therowhammer}}.}

\head{Improvement 6} \obsrefs{deeperlook:spatial:columns} and~\ref{deeperlook:spatial:design_and_process} show that {RowHammer vulnerability exhibits significant design-induced variation across columns within a chip and manufacturing process-induced variation across chips} in a DRAM module{.} {To make error correction codes (ECC) more effective and efficient {at} correcting RowHammer bitflips,} a system designer can {1)}~design {ECC} schemes optimized for non-uniform bit error probability distribution{s across columns} {and 2)}~modify the chipkill {ECC} {mechanism}~\cite{dell1997awhite,locklear2000chipkill, jian2013adaptive} to reduce {a} system's {dependency {on}} the most vulnerable DRAM chip, {as proposed in a} {concurrent work, revisit{ing} ECC {for} RowHammer~\cite{qureshi2021rethinking}.}

\section{Summary}
\label{deeperlook:sec:conclusion}
This work provides the first study that experimentally analyzes the impact of DRAM chip temperature, aggressor row active time, and victim DRAM cell's physical location on RowHammer vulnerability, through extensive characterization of {real DRAM chips}.
We rigorously characterize 248 DDR4 and 24 DDR3 modern DRAM chips {from four major DRAM manufacturers using} {a} carefully designed methodology and metrics, {providing} 16 key observations and 6 key takeaways. {W}e highlight {three major observations:} 1)~a DRAM cell experiences RowHammer bitflips at a bounded temperature range, 2)~a DRAM row is more vulnerable to RowHammer when the {aggressor row stays active for longer}, and 3)~a small fraction of DRAM rows are significantly more vulnerable to RowHammer than the other rows within {a} DRAM module.
{We {describe and analyze} how our insights can be used to improve both RowHammer attacks and defenses.}
We hope that the {novel experimental} results and insights of our study will inspire and aid future work to develop effective {and} efficient {solutions to the RowHammer problem.}


\chapter[Understanding RowHammer under Reduced Wordline Voltage]{Understanding RowHammer under Reduced Wordline Voltage}
\label{chap:hammerdimmer}




\newcommand{\berdecravg}{\SI{15.2}{\percent}}
\newcommand{\berincravg}{\SI{-15.2}{\percent}}
\newcommand{\berdecrmax}{\SI{66.9}{\percent}}
\newcommand{\berincrmax}{\SI{11.7}{\percent}}

\newcommand{\hcfirstdecravg}{\SI{-7.4}{\percent}}
\newcommand{\hcfirstincravg}{\SI{7.4}{\percent}}
\newcommand{\hcfirstdecrmax}{\SI{9.1}{\percent}}
\newcommand{\hcfirstincrmax}{\SI{85.8}{\percent}}

\newcommand{\fracbersupportingrows}{\SI{81.2}{\percent}}
\newcommand{\fracberopposingrows}{\SI{15.4}{\percent}}
\newcommand{\fracbersmallchangerowsmfrA}{\SI{49.6}{\percent}}
\newcommand{\fracbersupportingrowsmfrA}{\SI{68.6}{\percent}}
\newcommand{\fracbersmallchangerowsmfrB}{\SI{17.0}{\percent}}
\newcommand{\fracbersupportingrowsmfrB}{\SI{75.0}{\percent}}
\newcommand{\fracbersmallchangerowsmfrC}{\SI{0.0}{\percent}}
\newcommand{\fracbersupportingrowsmfrC}{\SI{100.0}{\percent}}
\newcommand{\frachcfirstsupportingrows}{\SI{69.3}{\percent}}
\newcommand{\frachcfirstopposingrows}{\SI{14.2}{\percent}}
\newcommand{\frachcfirstsmallchangerowsmfrA}{\SI{27.1}{\percent}}
\newcommand{\frachcfirstsupportingrowsmfrA}{\SI{50.9}{\percent}}
\newcommand{\frachcfirstsmallchangerowsmfrB}{\SI{12.2}{\percent}}
\newcommand{\frachcfirstsupportingrowsmfrB}{\SI{73.5}{\percent}}
\newcommand{\frachcfirstsmallchangerowsmfrC}{\SI{7.0}{\percent}}
\newcommand{\frachcfirstsupportingrowsmfrC}{\SI{83.5}{\percent}}
\newcommand{\bersmallchange}{\SI{2}{\percent}}
\newcommand{\hcfirstsmallchange}{\SI{0}{\percent}}

\newcommand{\minnormbermfrA}{{0.43}}
\newcommand{\maxnormbermfrA}{{1.11}}
\newcommand{\minnormbermfrB}{{0.33}}
\newcommand{\maxnormbermfrB}{{1.03}}
\newcommand{\minnormbermfrC}{{0.74}}
\newcommand{\maxnormbermfrC}{{0.94}}
\newcommand{\minnormhcfirstmfrA}{{0.94}}
\newcommand{\maxnormhcfirstmfrA}{{1.52}}
\newcommand{\minnormhcfirstmfrB}{{0.92}}
\newcommand{\maxnormhcfirstmfrB}{{1.86}}
\newcommand{\minnormhcfirstmfrC}{{0.91}}
\newcommand{\maxnormhcfirstmfrC}{{1.35}}

\renewcommand{\numchips}{272}
\newcommand{\numreliablechips}{208}
\newcommand{\numunreliablechips}{64}
\newcommand{\numunreliablemodules}{5}
\newcommand{\numreliablemodules}{25}
\renewcommand{\nummodules}{30}
\newcommand{\trcdguardbandreduction}{\SI{21.9}{\percent}}
\newcommand{\trcdguardbandreduced}{\SI{34.1}{\percent}}

\newcommand{\gfrev}[1]{{#1}}
\newcommand{\agyrev}[1]{{#1}}
\newcommand{\agyrevtwo}[1]{{#1}}
\newcommand{\agycrone}[1]{#1}
\newcommand{\agycronecomment}[1]{}
\newcommand{\hluocrone}[1]{#1}
\newcommand{\hluocronecomment}[1]{}
\newcommand{\omone}[1]{#1}
\newcommand{\omonecomment}[1]{}
\newcommand{\gfo}[1]{#1}
\newcommand{\gfocomment}[1]{}
\newcommand{\atbcrcomment}[1]{}
\newcommand{\mpcronecomment}[1]{}
\newcommand{\hhone}[1]{#1}

\newcommand{\atbcrtwo}[1]{#1}
\newcommand{\agycrtwo}[1]{#1}
\newcommand{\hluocrtwo}[1]{#1}
\newcommand{\omtwo}[1]{#1}
\newcommand{\gfii}[1]{#1}
\newcommand{\gft}[1]{#1}
\newcommand{\mpt}[1]{{#1}}
\newcommand{\hhtwo}[1]{#1}

\newcommand{\agycrtwocomment}[1]{}
\newcommand{\hluocrtwocomment}[1]{}
\newcommand{\omtwocomment}[1]{}
\newcommand{\gftcomment}[1]{}
\newcommand{\atbcrtwocomment}[1]{}
\newcommand{\mpcrtwocomment}[1]{}
\newcommand\gfb[1][0]{}

\newcommand{\agycrthree}[1]{#1}
\newcommand{\agycrthreecomment}[1]{}
\newcommand{\omthree}[1]{#1}
\newcommand{\omthreecomment}[1]{}
\newcommand{\mpcrthree}[1]{#1}
\newcommand{\mpcrfour}[1]{#1}

\newcommand{\agycrfour}[1]{#1}
\newcommand{\agycrfourcomment}[1]{}
\newcommand{\omfour}[1]{#1}
\newcommand{\omfourcomment}[1]{}

\newcommand{\agycrfive}[1]{#1}
\newcommand{\agycrfivecomment}[1]{}
\newcommand{\mpcrfive}[1]{#1}
\newcommand{\mpcrfivecomment}[1]{}
\newcommand{\omfive}[1]{#1}
\newcommand{\omfivecomment}[1]{}

\newcommand{\agycrsix}[1]{#1}
\newcommand{\agycrsixcomment}[1]{}
\newcommand{\mpcrsix}[1]{#1}
\newcommand{\mpcrsixcomment}[1]{}
\newcommand{\omsix}[1]{#1}
\newcommand{\omsixcomment}[1]{}

\newcommand{\agyextone}[1]{#1}
\newcommand{\hluoextone}[1]{#1}
\newcommand{\agyextonecomment}[1]{}
\newcommand{\mpextone}[1]{#1}
\newcommand{\mpextonecomment}[1]{}
\newcommand{\omextone}[1]{#1}
\newcommand{\omextonecomment}[1]{}

\newcommand{\agyexttwo}[1]{#1}
\newcommand{\hluoexttwo}[1]{#1}
\newcommand{\agyexttwocomment}[1]{}
\newcommand{\mpexttwo}[1]{#1}
\newcommand{\mpexttwocomment}[1]{}
\newcommand{\omexttwo}[1]{#1}
\newcommand{\omexttwocomment}[1]{}

\newcommand{\agyextthree}[1]{#1}
\newcommand{\hluoextthree}[1]{#1}
\newcommand{\agyextthreecomment}[1]{}
\newcommand{\mpextthree}[1]{#1}
\newcommand{\mpextthreecomment}[1]{}
\newcommand{\omextthree}[1]{#1}
\newcommand{\omextthreecomment}[1]{}

\newcommand{\obsvref}[1]{Obsv.~\ref{#1}}
\newcommand{\obsvsref}[1]{Obsvs.~\ref{#1}}

\section{Motivation}
\label{hammerdimmer:sec:motivation}

To {enable} {effective and efficient} RowHammer {mitigation} {mechanisms}, it is {critical to develop a comprehensive understanding of how RowHammer bitflips occur~\cite{orosa2021adeeper, mutlu2017therowhammer, mutlu2019rowhammer_and}.
In this work, we observe that although \gls{vpp} is expected to affect the amount of disturbance caused by a RowHammer attack~\cite{kim2014flipping, redeker2002aninvestigation, mutlu2017therowhammer, mutlu2019rowhammer_and, walker2021ondram, yang2019trapassisted, park2016statistical, park2016experiments, orosa2021adeeper, kim2020revisiting, saroiu2022theprice, qureshi2021rethinking}, \emph{no} prior work experimentally studies its real-world impact on a DRAM chip's {RowHammer} vulnerability.\footnote{{Both {\gls{vpp}} and {\gls{vdd}} {can affect} a DRAM chip's {RowHammer vulnerability}. However, changing \gls{vdd} can negatively impact DRAM reliability in ways that are unrelated to RowHammer (e.g., I/O circuitry instabilities) because \gls{vdd} supplies power to {\emph{all}} logic elements within the DRAM chip.} In contrast, \gls{vpp} affects \emph{only} the wordline voltage, so \gls{vpp} can influence RowHammer without adverse effects {on unrelated} parts of the DRAM chip.} Therefore,} \textbf{our goal} is to understand {how \gls{vpp} affects} RowHammer vulnerability and DRAM {operation}.  

{To achieve this goal, we start with the hypothesis that \gls{vpp} can be used to reduce a DRAM chip's RowHammer vulnerability without impacting the reliability of normal DRAM operations.} Reducing a DRAM chip's RowHammer vulnerability {via} \gls{vpp} scaling has two key advantages. First, {as a circuit-level RowHammer mitigation approach, \gls{vpp} scaling} is \emph{complementary} to {existing system-level and architecture-level RowHammer mitigation mechanisms~\cite{aichinger2015ddrmemory, apple2015about, aweke2016anvil, kim2014flipping, kim2014architectural,son2017making, lee2019twice, you2019mrloc, seyedzadeh2018mitigating, vanderveen2018guardion, konoth2018zebram, park2020graphene, yaglikci2021blockhammer, kang2020cattwo, bains2015rowhammer, bains2016distributed, bains2016rowhammer, brasser2017cant, gomez2016dram_rowhammer, jedec2020jesd794c,hassan2019crow, devaux2021method, ryu2017overcoming, yang2016suppression, yang2017scanning, gautam2019rowhammering, yaglikci2021security, qureshi2021rethinking, greenfield2012throttling, marazzi2022rega, saileshwar2022randomized}.
Therefore,} \gls{vpp} scaling can be used \emph{alongside} {these mechanisms} to increase their effectiveness and/or reduce their overheads. Second, {\gls{vpp} scaling} can be implemented with a \emph{fixed hardware cost} {for a given power budget,} {irrespective of the number and types of DRAM chips used in a system.}

{We test this hypothesis through the first experimental {RowHammer} characterization study {under reduced \gls{vpp}.}
{In this study, we test} \numchips{} real DDR4 DRAM chips from 
\param{three} major DRAM manufacturers. Our study {is inspired by} state-of-the-art analytical models for RowHammer{, which suggest that the effect of RowHammer's underlying error mechanisms depends on \gls{vpp}~\cite{walker2021ondram, yang2019trapassisted, park2016statistical}. \secref{hammerdimmer:sec:vpp_with_rh} reports our findings, which} yield valuable insights into {how \gls{vpp}} impacts the circuit-level RowHammer characteristics of modern DRAM chips, both confirming our hypothesis and supporting \gls{vpp} scaling as a {promising} new dimension to{ward} robust RowHammer mitigation.}

\section{Experimental Methodology}
\label{hammerdimmer:sec:methodology}

{We {describe} our methodology for two analyses. First, we experimentally characterize the behavior of \numchips{} real DDR4 DRAM chips from three major manufacturers under reduced \gls{vpp} in terms of RowHammer vulnerability (\secref{hammerdimmer:sec:experiment_design_rowhammer}), \acrfull{trcd} (\secref{hammerdimmer:sec:experiment_design_trcd}), and data retention time (\secref{hammerdimmer:sec:experiment_design_retention}). Second, to verify our observations from real-device experiments, we investigate reduced \gls{vpp}'s effect on \emph{both} DRAM row activation and charge restoration using SPICE~\cite{corpltspice, nagel1973spice} simulations~(\secref{hammerdimmer:sec:spice_model}).}

\subsection{{Real-Device Testing Infrastructure}}
\label{hammerdimmer:sec:experimental_setup}
We conduct {real-device} characterization experiments using {an infrastructure based on} SoftMC~\cite{hassan2017softmc, safari2017softmc} \agy{3}{and DRAM Bender~\cite{olgun2023dram_bender,safari2022dram_bender}}, the state-of-the-art FPGA-based open-source {infrastructures for DRAM characterization}. 
\figref{hammerdimmer:fig:infrastructure} shows a picture of our experimental setup. {We attach {{heater pads}} to the DRAM chips that are located on both sides of a DDR4 DIMM. We use a MaxWell FT200 PID temperature controller\omone{~\cite{maxwellft20x}} connected to the {heaters pads} to maintain the DRAM chips under test at a preset temperature level with the precision of $\pm$\SI{0.1}{\celsius}.} We program a Xilinx Alveo U200 FPGA board{~\cite{xilinxu200}} with \agy{3}{DRAM Bender~\cite{olgun2023dram_bender, safari2022dram_bender}}. The FPGA board is connected to a host machine through a PCI{e} port for running our tests. We connect the DRAM module to {the} FPGA board {via} a commercial interposer board from Adexelec{~\cite{adexelecddr4sodv1}} with current measurement capability. The interposer board enforces the power to be supplied through a shunt resistor on the \gls{vpp} rail. We remove this shunt resistor to electrically disconnect the \gls{vpp} rails of the DRAM module and the FPGA board. Then, we supply power to the DRAM module's \gls{vpp} power rail from an external TTi PL068-P power supply{~\cite{ttiplplp}}{, which enables us to control \gls{vpp} \agycrone{at the precision of \SI{\pm1}{\milli\volt}.}} \agycrone{We start testing each DRAM module at the nominal \gls{vpp} of \SI{2.5}{\volt}. We gradually reduce \gls{vpp} with \SI{0.1}{\volt} steps until \gls{vppmin}.}

\begin{figure}[!ht]
    \centering
    \includegraphics[width=\linewidth]{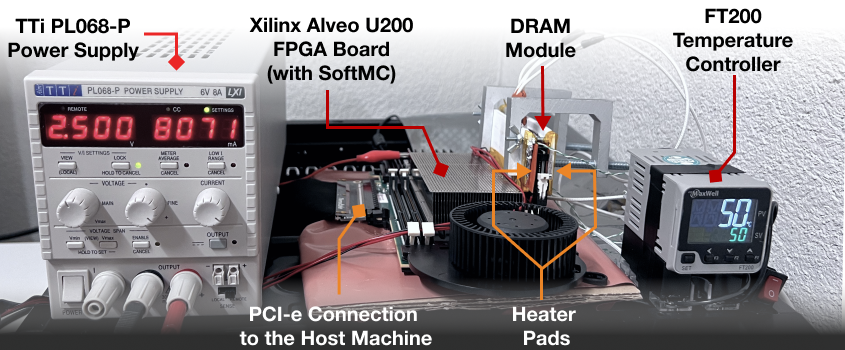}
    \caption{Our experimental setup {based on SoftMC~\cite{hassan2017softmc,safari2017softmc} and DRAM Bender~\cite{olgun2023dram_bender, safari2022dram_bender}}.}
    \label{hammerdimmer:fig:infrastructure}
\end{figure}

{To show that our observations are 
\emph{not} {specific} to a certain DRAM architecture/process but rather common across different designs and generations, we test {DDR4}} DRAM modules from all three major manufacturers {with different die revisions, purchased from the retail market}.
Table~\ref{hammerdimmer:tab:dram_chip_list} provides {the} {chip} density, die revision {(Die Rev.)}, chip organization {(Org.)}, and manufacturing date of tested DRAM modules.{\footnote{{Die Rev. and Date columns are blank if undocumented.}}} We report the manufacturing date of these modules in the form of $week-year$.
All tested modules {are listed in Table~\ref{hammerdimmer:tab:detailed_dimm_table}.}

\begin{table}[h]
    \caption{Summary of \omtwo{the tested} DDR4 DRAM chips.}
    \centering
    \footnotesize{}
    \setlength\tabcolsep{3pt} 
    \begin{tabular}{lcccccc}
        \toprule
            {{\bf Mfr.}} & \textbf{\#DIMMs} & {{\bf  \#Chips}}  & {{\bf Density}} & {{\bf Die Rev.}}& {{\bf Org.}}&
            {{\bf Date}}\\
        \midrule
        & 1 & 8 &{4Gb}&{}&{$\times$8}&48-16 \\
        {Mfr. A}& 4 &  64 &{8Gb}&{B}&{$\times$4}&11-19 \\
        \omone{(Micron)}& 3 & 24 &{4Gb}&{F}&{$\times$8}&07-21 \\
        & 2 & 16 &{4Gb}&{}&{$\times$8}&\\
        \midrule
        {}& 2 & 16 &{8Gb}&{B}&{$\times$8}&5\agycrtwo{2}-20\\
        {}& 1 & 8 &{8Gb}&{C}&{$\times$8}&19-19\\
        {Mfr. B}& 3 & 24 &{8Gb}&{D}&{$\times$8}&10-21\\
        \omone{(Samsung)}& 1 &  8 &{4Gb}&{E}&{$\times$8}&08-17\\
        & 1 &  8 &{4Gb}&{F}&{$\times$8}&02-21\\
        & 2 &  16 &{8Gb}&{}&{$\times$8}&\\
        \midrule
        {}& 2 &  16 &{16Gb}&{A}&{$\times$8}&51-20\\
        {Mfr. C}& 3 & 24 &{4Gb}&{B}&{$\times$8}&02-21\\
        \omone{(SK Hynix)}& 2 & 16 &{4Gb}&{C}&{$\times$8}&\\
        & 3 & 24 &{8Gb}&{D}&{$\times$8}&48-20\\

        \bottomrule
    \end{tabular}
    \label{hammerdimmer:tab:dram_chip_list}
\end{table}

{Table~\ref{hammerdimmer:tab:detailed_dimm_table} shows the characteristics of the {DDR4} DRAM modules we test and analyze.\footnote{All tested DRAM modules implement the DDR4 DRAM standard~\cite{jedec2020jesd794c}. We make our best effort to identify the DRAM chips used in our tests. We identify the DRAM chip density and die revision through the original manufacturer markings on the chip. For certain DIMMs we tested, the original DRAM chip markings are removed by the DIMM manufacturer. In this case, we can only identify the chip manufacturer and density by reading the information stored in the SPD. However, these DIMM manufacturers also tend to remove the die revision information in the SPD. Therefore, we \emph{cannot} identify the die revision {of five DIMMs} and {the} manufacturing date {of} six DIMMs we test, shown as `-' in the table.} For each DRAM module, we provide the 1)~DRAM chip manufacturer, 2)~{DIMM name, 3)~DIMM model,\footnote{{DIMM models CMV4GX4M1A2133C1\agyextthree{5} and F4-2400C17S-8GNT use DRAM chips from different manufacturers (i.e., Micron-SK Hynix and Samsung-SK Hynix, respectively) across different batches.}} 4)}~die density, {5)~data transfer frequency, 6)~chip organization,} {7})~die revision, specified in the module's serial presence detect (SPD) registers, {8)~manufacturing} date, specified on the module's label {in the form of $week-year$, and 9)~RowHammer vulnerability characteristics of the module}. 
{Table~\ref{hammerdimmer:tab:detailed_dimm_table} reports the RowHammer vulnerability characteristics of each DIMM under two \gls{vpp} levels: \emph{i)}~nominal \gls{vpp} (\SI{2.5}{\volt}) and \emph{ii)~}\gls{vppmin}. We quantify a DIMM's RowHammer vulnerability characteristics at a given \gls{vpp} in terms of two metrics: \emph{i)}~\gls{hcfirst} and \emph{ii)}~\gls{ber}. Based on these two metrics at nominal \gls{vpp} and \gls{vppmin}, Table~\ref{hammerdimmer:tab:detailed_dimm_table} {also} provides a \emph{recommended} \gls{vpp} level ($V_{PP_{Rec}}$) and the corresponding RowHammer characteristics in the right-most three columns.}

\newcommand{\rotcell}[1]{\rotatebox{90}{#1}}

\begin{table}[!ht]
\centering
\caption{Tested DRAM modules and their characteristics when $V_{PP}$=\SI{2.5}{\volt} (nominal) and\\$V_{PP}=V_{PPmin}$. $V_{PPmin}$ is specified for each module.}
\resizebox{\columnwidth}{!}{
\begin{tabular}{|l|c|l|ccccc||rc|c|rc|c|rc|} 
\hline
& & & & & & & & \multicolumn{2}{c|}{\textbf{$\mathbf{V_{PP}}$ = 2.5V}} & & \multicolumn{2}{c|}{$\mathbf{V_{PP} = V_{PP_{min}}}$} & & \multicolumn{2}{c|}{$\mathbf{V_{PP} = V_{PP_{Rec}}}$}\\
\rotcell{\begin{tabular}[c]{@{}c@{}}\textbf{DRAM Chip Mfr.}\end{tabular}} & \rotcell{\begin{tabular}[c]{@{}c@{}}\textbf{DIMM Name}\end{tabular}} & \textbf{DIMM Model} & \rotcell{\begin{tabular}[c]{@{}c@{}}\textbf{Die Density}\end{tabular}} & \rotcell{\begin{tabular}[c]{@{}c@{}}\textbf{Freq\agyextthree{uency} (MT/s)}\end{tabular}} & \rotcell{\textbf{Chip Org.}} & \rotcell{\begin{tabular}[c]{@{}c@{}}\textbf{Die Revision}\end{tabular}} & \rotcell{\begin{tabular}[c]{@{}c@{}}\textbf{Mfr. Date}\end{tabular}} &\rotcell{\begin{tabular}[c]{@{}l@{}}\textbf{Minimum}\\\textbf{$\mathbf{HC_{first}}$}\end{tabular}}& \textbf{$\mathbf{BER}$}& \rotcell{$\mathbf{V_{PP_{min}}}$} &\rotcell{\begin{tabular}[c]{@{}l@{}}\textbf{Minimum}\\\textbf{$\mathbf{HC_{first}}$}\end{tabular}}&\textbf{$\mathbf{BER}$}& \rotcell{\begin{tabular}[c]{@{}l@{}}\textbf{Recommended}\\\textbf{$\mathbf{V_{PP} (V_{PP_{Rec}}}$)}\end{tabular}} & \rotcell{\begin{tabular}[c]{@{}l@{}}\textbf{Minimum}\\\textbf{$\mathbf{HC_{first}}$}\end{tabular}}&\textbf{$\mathbf{BER}$}\\ 
\hline
\hline
\multirow{10}{*}{\rotcell{\begin{tabular}[c]{@{}l@{}}Mfr. A (Micron)\end{tabular}}}                                                            & A0                                                                                               & MTA18ASF2G72PZ-2G3B1QK~\cite{micronddr4mta18asf2g72pz}               & 8Gb                                                                                                & 2400                                             & x4                                                    & B                                                                                               & 11-19                                                                                            & 39.8K                                                                                         & 1.24e-03                                             & 1.4                                                                                                                                                             & 42.2K                                                                                         & 1.00e-03                                             & 1.4                                                                                                                                                 & 42.2K                                   & 1.00e-03                  \\
                                                                                                                                     & A1                                                                                               & MTA18ASF2G72PZ-2G3B1QK~\cite{micron2016ddr4mta18asf2g72pz}               & 8Gb                                                                                                & 2400                                             & x4                                                    & B                                                                                               & 11-19                                                                                            & 42.2K                                                                                         & 9.90e-04                                             & 1.4                                                                                                                                                             & 46.4K                                                                                         & 7.83e-04                                             & 1.4                                                                                                                                                 & 46.4K                                   & 7.83e-04                  \\
                                                                                                                                     & A2                                                                                               & MTA18ASF2G72PZ-2G3B1QK~\cite{micronddr4mta18asf2g72pz}               & 8Gb                                                                                                & 2400                                             & x4                                                    & B                                                                                               & 11-19                                                                                            & 41.0K                                                                                         & 1.24e-03                                             & 1.7                                                                                                                                                             & 39.8K                                                                                         & 1.35e-03                                             & 2.\agyextthree{1}                                                                                                                                                 & 4\agyextthree{2.1}K                                   & 1.\agyextthree{55e-3}                  \\
                                                                                                                                     & A3                                                                                               & CT4G4DFS8266.C8FF~\cite{crucialct4g4dfs8266}                    & 4Gb                                                                                                & 2666                                             & x8                                                    & F                                                                                               & 07-21                                                                                            & 16.7K                                                                                         & 3.33e-02                                             & 1.4                                                                                                                                                             & 16.5K                                                                                         & 3.52e-02                                             & \agyextthree{1.7}                                                                                                                                                 & \agyextthree{17.0K}                                   & 3.\agyextthree{48}e-02                  \\
                                                                                                                                     & A4                                                                                               & CT4G4DFS8266.C8FF~\cite{crucialct4g4dfs8266}                    & 4Gb                                                                                                & 2666                                             & x8                                                    & F                                                                                               & 07-21                                                                                            & 14.4K                                                                                         & 3.18e-02                                             & 1.5                                                                                                                                                             & 14.4K                                                                                         & 3.33e-02                                             & 2.5                                                                                                                                                 & 14.4K                                   & 3.18e-02                  \\
                                                                                                                                     & A5                                                                                               & CT4G4SFS8213.C8FBD1                  & 4Gb                                                                                                & 2400                                             & x8                                                    & -                                                                                               & 48-16                                                                                            & 140.7K                                                                                        & 1.39e-06                                             & 2.4                                                                                                                                                             & 145.4K                                                                                        & 3.39e-06                                             & 2.4                                                                                                                                                 & 145.4K                                  & 3.39e-06                  \\
                                                                                                                                     & A6                                                                                               & CT4G4DFS8266.C8FF~\cite{crucialct4g4dfs8266}                    & 4Gb                                                                                                & 2666                                             & x8                                                    & F                                                                                               & 07-21                                                                                            & 16.5K                                                                                         & 3.50e-02                                             & 1.5                                                                                                                                                             & 16.5K                                                                                         & 3.66e-02                                             & 2.5                                                                                                                                                 & 16.5K                                   & 3.50e-02                  \\
                                                                                                                                     & A7                                                                                               & CMV4GX4M1A2133C15~\cite{corsairskucmv4gx4m1a2133c15}                     & 4Gb                                                                                                & 2133                                             & x8                                                    & -                                                                                               & -                                                                                                & 16.5K                                                                                         & 3.42e-02                                             & 1.8                                                                                                                                                             & 16.5K                                                                                         & 3.52e-02                                             & 2.5                                                                                                                                                 & 16.5K                                   & 3.42e-02                  \\
                                                                                                                                     & A8                                                                                               & MTA18ASF2G72PZ-2G3B1QG~\cite{micronddr4mta18asf2g72pz}               & 8Gb                                                                                                & 2400                                             & x4                                                    & B                                                                                               & 11-19                                                                                            & 35.2K                                                                                         & 2.38e-03                                             & 1.4                                                                                                                                                             & 39.8K                                                                                         & 2.07e-03                                             & 1.4                                                                                                                                                 & 39.8K                                   & 2.07e-03                  \\
                                                                                                                                     & A9                                                                                               & CMV4GX4M1A2133C15~\cite{corsairskucmv4gx4m1a2133c15}                    & 4Gb                                                                                                & 2133                                             & x8                                                    & -                                                                                               & -                                                                                                & 14.3K                                                                                         & 3.33e-02                                             & 1.5                                                                                                                                                             & 14.3K                                                                                         & 3.48e-02                                             & \agyextthree{1.6}                                                                                                                                                 & 14.\agyextthree{6}K                                   & 3.\agyextthree{47}e-02                  \\ 
\hline
\multirow{10}{*}{\rotcell{\begin{tabular}[c]{@{}l@{}}Mfr. B (Samsung)\end{tabular}}}                                                           & B0                                                                                               & M378A1K43DB2-CTD~\cite{samsung2018288pin}                     & 8Gb                                                                                                & 2666                                             & x8                                                    & D                                                                                               & 10-21                                                                                            & 7.9K                                                                                          & 1.18e-01                                             & 2.0                                                                                                                                                             & 7.6K                                                                                          & 1.22e-01                                             & 2.5                                                                                                                                                 & 7.9K                                    & 1.18e-01                  \\
                                                                                                                                     & B1                                                                                               & M378A1K43DB2-CTD~\cite{samsung2018288pin}                     & 8Gb                                                                                                & 2666                                             & x8                                                    & D                                                                                               & 10-21                                                                                            & 7.3K                                                                                          & 1.26e-01                                             & 2.0                                                                                                                                                             & 7.6K                                                                                          & 1.28e-01                                             & 2.0                                                                                                                                                 & 7.6K                                    & 1.28e-01                  \\
                                                                                                                                     & B2                                                                                               & F4-2400C17S-8GNT~\cite{gskill2021f42400c17s8gnt}                     & 4Gb                                                                                                & {2400}         & x8                                                    & F                                                                                               & 02-21                                                                                            & 11.2K                                                                                         & 2.52e-02                                             & 1.6                                                                                                                                                             & 12.0K                                                                                         & 2.22e-02                                             & 1.6                                                                                                                                                 & 12.0K                                   & 2.22e-02                  \\
                                                                                                                                     & B3                                                                                               & M393A1K43BB1-CTD6Y~\cite{samsung2017288pin}                   & 8Gb                                                                                                & 2666                                             & x8                                                    & B                                                                                               & 52-20                                                                                            & 16.6K                                                                                         & 2.73e-03                                             & 1.6                                                                                                                                                             & 21.1K                                                                                         & 1.09e-03                                             & 1.6                                                                                                                                                 & 21.1K                                   & 1.09e-03                  \\
                                                                                                                                     & B4                                                                                               & M393A1K43BB1-CTD6Y~\cite{samsung2017288pin}                   & 8Gb                                                                                                & 2666                                             & x8                                                    & B                                                                                               & 52-20                                                                                            & 21.0K                                                                                         & 2.95e-03                                             & 1.8                                                                                                                                                             & 19.9K                                                                                         & 2.52e-03                                             & 2.\agyextthree{0}                                                                                                                                                 & 21.\agyextthree{1}K                                   & 2.\agyextthree{68}e-03                  \\
                                                                                                                                     & B5                                                                                               & M471A5143EB0-CPB~\cite{samsungm471a5143eb0cpb}                     & 4Gb                                                                                                & 2133                                             & x8                                                    & E                                                                                               & 08-17                                                                                            & 21.0K                                                                                         & 7.78e-03                                             & 1.8                                                                                                                                                             & 21.0K                                                                                         & 6.02e-03                                             & \agyextthree{2.0}                                                                                                                                                 & 21.\agyextthree{1}K                                   & \agyextthree{8.67}e-03               \\
                                                                                                                                     & B6                                                                                               & CMK16GX4M2B3200C16~\cite{corsaircmk16gx4m2b3200c16}                   & 8Gb                                                                                                & 3200                                             & x8                                                    & -                                                                                               & -                                                                                                & 10.3K                                                                                         & 1.14e-02                                             & 1.7                                                                                                                                                             & 10.5K                                                                                         & 9.82e-03                                             & 1.7                                                                                                                                                 & 10.5K                                   & 9.82e-03                  \\
                                                                                                                                     & B7                                                                                               & M378A1K43DB2-CTD~\cite{samsung2018288pin}                     & 8Gb                                                                                                & 2666                                             & x8                                                    & D                                                                                               & 10-21                                                                                            & 7.3K                                                                                          & 1.32e-01                                             & 2.0                                                                                                                                                             & 7.6K                                                                                          & 1.33e-01                                             & 2.0                                                                                                                                                 & 7.6K                                    & 1.33e-01                  \\
                                                                                                                                     & B8                                                                                               & CMK16GX4M2B3200C16~\cite{corsaircmk16gx4m2b3200c16}                   & 8Gb                                                                                                & 3200                                             & x8                                                    & -                                                                                               & -                                                                                                & 11.6K                                                                                         & 2.88e-02                                             & 1.7                                                                                                                                                             & 10.5K                                                                                         & 2.37e-02                                             & \agyextthree{1.8}                                                                                                                                                 & 11.\agyextthree{7}K                                   & 2.\agyextthree{5}8e-02                  \\
                                                                                                                                     & B9                                                                                               & M471A5244CB0-CRC\cite{samsung2018260pin}                     & 8Gb                                                                                                & 2133                                             & x8                                                    & C                                                                                               & 19-19                                                                                            & 11.8K                                                                                         & 2.68e-02                                             & 1.7                                                                                                                                                             & 8.8K                                                                                          & 2.39e-02                                             & \agyextthree{1.8}                                                                                                                                                 & 1\agyextthree{2.3}K                                   & 2.\agyextthree{54}e-02                  \\ 
\hline
\multirow{10}{*}{\rotcell{\begin{tabular}[c]{@{}l@{}}Mfr. C (SK Hynix)\end{tabular}}}                                                             & C0                                                                                               & F4-2400C17S-8GNT~\cite{gskill2021f42400c17s8gnt}                     & 4Gb                                                                                                & {2400}         & x8                                                    & B                                                                                               & 02-21                                                                                            & 19.3K                                                                                         & 7.29e-03                                             & 1.7                                                                                                                                                             & 23.4K                                                                                         & 6.61e-03                                             & 1.7                                                                                                                                                 & 23.4K                                   & 6.61e-03                  \\
                                                                                                                                     & C1                                                                                               & F4-2400C17S-8GNT~\cite{gskill2021f42400c17s8gnt}                     & 4Gb                                                                                                & {2400}         & x8                                                    & B                                                                                               & 02-21                                                                                            & 19.3K                                                                                         & 6.31e-03                                             & 1.7                                                                                                                                                             & 20.6K                                                                                         & 5.90e-03                                             & 1.7                                                                                                                                                 & 20.6K                                   & 5.90e-03                  \\
                                                                                                                                     & C2                                                                                               & KSM32RD8/16HDR~\cite{kingston2020ksm32rd816hdr}                       & 8Gb                                                                                                & 3200                                             & x8                                                    & D                                                                                               & 48-20                                                                                            & 9.6K                                                                                          & 2.82e-02                                             & 1.5                                                                                                                                                             & 9.2K                                                                                          & 2.34e-02                                             & 2.\agyextthree{3}                                                                                                                                                 & \agyextthree{10.0}K                                    & 2.8\agyextthree{9}e-02                  \\
                                                                                                                                     & C3                                                                                               & KSM32RD8/16HDR~\cite{kingston2020ksm32rd816hdr}                     & 8Gb                                                                                                & 3200                                             & x8                                                    & D                                                                                               & 48-20                                                                                            & 9.3K                                                                                          & 2.57e-02                                             & 1.5                                                                                                                                                             & 8.9K                                                                                          & 2.21e-02                                             & 2.\agyextthree{3}                                                                                                                                                 & 9.\agyextthree{7}K                                    & 2.\agyextthree{66}e-02                  \\
                                                                                                                                     & C4                                                                                               & HMAA4GU6AJR8N-XN~\cite{memorynethmaa4gu6ajr8nxn}                     & 16Gb                                                                                               & 3200                                             & x8                                                    & A                                                                                               & 51-20                                                                                            & 11.6K                                                                                         & 3.22e-02                                             & 1.5                                                                                                                                                             & 11.7K                                                                                         & 2.88e-02                                             & 1.5                                                                                                                                                 & 11.7K                                   & 2.88e-02                  \\
                                                                                                                                     & C5                                                                                               & HMAA4GU6AJR8N-XN~\cite{memorynethmaa4gu6ajr8nxn}                     & 16Gb                                                                                               & 3200                                             & x8                                                    & A                                                                                               & 51-20                                                                                            & 9.4K                                                                                          & 3.28e-02                                             & 1.5                                                                                                                                                             & 12.7K                                                                                         & 2.85e-02                                             & 1.5                                                                                                                                                 & 12.7K                                   & 2.85e-02                  \\
                                                                                                                                     & C6                                                                                               & CMV4GX4M1A2133C15~\cite{corsairskucmv4gx4m1a2133c15}                     & 4Gb                                                                                                & 2133                                             & x8                                                    & C                                                                                               & -                                                                                                & 14.2K                                                                                         & 3.08e-02                                             & 1.6                                                                                                                                                             & 15.5K                                                                                         & 2.25e-02                                             & 1.6                                                                                                                                                 & 15.5K                                   & 2.25e-02                  \\
                                                                                                                                     & C7                                                                                               & CMV4GX4M1A2133C15~\cite{corsairskucmv4gx4m1a2133c15}                     & 4Gb                                                                                                & 2133                                             & x8                                                    & C                                                                                               & -                                                                                                & 11.7K                                                                                         & 3.24e-02                                             & 1.6                                                                                                                                                             & 13.6K                                                                                         & 2.60e-02                                             & 1.6                                                                                                                                                 & 13.6K                                   & 2.60e-02                  \\
                                                                                                                                     & C8                                                                                               & KSM32RD8/16HDR~\cite{kingston2020ksm32rd816hdr}                       & 8Gb                                                                                                & 3200                                             & x8                                                    & D                                                                                               & 48-20                                                                                            & 11.4K                                                                                         & 2.69e-02                                             & 1.6                                                                                                                                                             & 9.5K                                                                                          & 2.57e-02                                             & 2.5                                                                                                                                                 & 11.4K                                    & 2.69e-02                  \\
                                                                                                                                     & C9                                                                                               & F4-2400C17S-8GNT~\cite{gskill2021f42400c17s8gnt}                     & 4Gb                                                                                                & {2400}         & x8                                                    & B                                                                                               & 02-21                                                                                            & 12.6K                                                                                         & 2.18e-02                                             & 1.7                                                                                                                                                             & 15.2K                                                                                         & 1.63e-02                                             & 1.7                                                                                                                                                 & 15.2K                                   & 1.63e-02                  \\
\hline
\end{tabular}
}
\label{hammerdimmer:tab:detailed_dimm_table}
\end{table}

\head{Temperature} We conduct RowHammer and \gls{trcd} tests at \SI{50}{\celsius} and retention tests at \SI{80}{\celsius} to ensure both stable and representative testing conditions.{\footnote{A recent work~\cite{orosa2021adeeper} shows a complex interaction between RowHammer and temperature{, suggesting {that one should} repeat characterization at many different temperature levels {to find the worst-case RowHammer vulnerability}. Since such characterization requires {many} months-long testing time,} we leave it to future work to study temperature, voltage, and RowHammer interaction in detail.}} 
We conduct \gls{trcd} tests at \SI{50}{\celsius} because \SI{50}{\celsius} is our infrastructure's minimum stable temperature due to cooling limitations.{\footnote{We do not repeat {the \gls{trcd}} tests at different temperature levels because prior work~\cite{chang2017understanding} shows {small} variation in \gls{trcd} with varying temperature.}} We conduct retention tests at \SI{80}{\celsius} to capture any effects of increased charge leakage~\cite{liu2013anexperimental} {at the upper bound of regular operating temperatures~\cite{jedec2020jesd794c}.\footnote{DDR4 DRAM chips are refreshed at $2\times$ the nominal refresh rate when the chip temperature reaches \SI{85}{\celsius}~\cite{jedec2020jesd794c}. Thus, we choose \SI{80}{\celsius} as a representative high temperature within the regular operating temperature range. {For a detailed analysis of the effect of temperature on data retention in DRAM, we refer the reader to~\cite{hamamoto1998onthe,liu2013anexperimental, patel2017thereach}.}}}

\head{Disabling Sources of Interference} 
{To understand fundamental device behavior in response to \gls{vpp} reduction, we} make sure that \gls{vpp} is the only {control} variable in our experiments {so that we can} {accurately} measure the effects of \gls{vpp} on 
{RowHammer, row activation latency (\gls{trcd}), and data retention time.}
To do so, we follow {four} steps{, similar to prior rigorous RowHammer~\cite{kim2020revisiting,orosa2021adeeper}, row activation latency~{\cite{lee2015adaptivelatency, chang2016understanding, chang2017understanding}}, and data retention time~\cite{liu2013anexperimental, patel2017thereach} characterization methods}. First{,} we disable DRAM refresh {to ensure no disturbance on the desired access pattern.}
Second, we ensure that during our {RowHammer and \gls{trcd}} experiments{,} {\emph{no}} bitflips occur due to {data} retention failures by conducting {each experiment} within a {time} period {of less than  \SI{30}{\milli\second}} {(i.e., much shorter than the nominal \gls{trefw} of \SI{64}{\milli\second})}. Third, we test DRAM modules 
{without} error{-}correction code (ECC) support to ensure neither on-die {ECC}~{\cite{micron2017eccbrings, nair2016xedexposing, patel2021harp, patel2020bitexact, patel2019understanding, kang2014coarchitecting, patel2021enabling}} nor rank-level ECC~\cite{cojocar2019exploiting,kim2016allinclusive} can affect {our observations} by correcting {\gls{vpp}-reduction-induced} bitflips.
{Fourth}, we \omone{disable} known on-DRAM-die RowHammer
defenses (i.e., TRR~{\cite{jedec2020jesd2095a,jedec2020jesd795,lee2014green,micron2016ddr4,hassan2021uncovering, frigo2020trrespass}}) by not issuing refresh commands throughout our tests~\cite{frigo2020trrespass,kim2020revisiting,orosa2021adeeper, hassan2021uncovering} {(as all {TRR} defenses require refresh commands to work)}. 

\head{{Data Patterns}} {We use} six commonly used data patterns~\cite{chang2016understanding,chang2017understanding,khan2014theefficacy,khan2016parbor,khan2016acase,kim2020revisiting,lee2017designinduced,mukhanov2020dstress,orosa2021adeeper, kim2014flipping, liu2013anexperimental}: row stripe (\texttt{0xFF}/\texttt{0x00}), checkerboard (\texttt{0xAA}/\texttt{0x55}), and thickchecker (\texttt{0xCC}/\texttt{0x33}). We identify {the worst-case data pattern ($WCDP$) for each row among these {six patterns} at nominal \gls{vpp} {separately} {for each of RowHammer (\secref{hammerdimmer:sec:experiment_design_rowhammer}), \acrfull{trcd} (\secref{hammerdimmer:sec:experiment_design_trcd}), and data retention time (\secref{hammerdimmer:sec:experiment_design_retention}) tests. We {use} each row{'s}
{corresponding} $WCDP$ {for a given test,} at reduced \gls{vpp} levels.}}

\subsection{\omone{RowHammer Experiments}}
\label{hammerdimmer:sec:experiment_design_rowhammer}
{We perform multiple experiments to understand how \gls{vpp} affects the RowHammer vulnerability of a DRAM chip.}

\head{Metrics}~We measure the RowHammer vulnerability of a DRAM chip {using} two {metrics}: {1)
\acrlong{hcfirst} (\acrshort{hcfirst}) and 2)~\gls{ber} {{caused} by} a double-sided RowHammer attack with a fixed hammer count of 300K per aggressor row.}{\footnote{{{We choose the 300K hammer count because 1) it is low enough to be used in a system-level RowHammer attack in a real system, and 2) it is high enough to provide us with a large number of bitflips to make meaningful observations in all DRAM modules we tested.}}}}

\head{\agycrone{WCDP}}
{We choose $WCDP$ as the data pattern {that} causes the \emph{lowest} \gls{hcfirst}. If there are multiple data patterns that {cause} the lowest \gls{hcfirst}, we choose the data pattern that causes the \emph{largest} \gls{ber} for the fixed hammer count of $300K$.}\footnote{{To investigate if $WCDP$ changes with reduced \gls{vpp},} {we repeat {$WCDP$ determination} experiments for different \gls{vpp} values for \param{16} DRAM chips.} We observe that $WCDP$ changes for \emph{only} \SI{2.4}{\percent} of tested rows,  causing less than \SI{9}{\percent} deviation in \gls{hcfirst} for \SI{90}{\percent} of the affected rows. We leave a {detailed sensitivity analysis of $WCDP$ to \gls{vpp}}
for future work.}

{\head{RowHammer Tests} Alg.~\ref{hammerdimmer:alg:test_alg} describes the core test loop of each RowHammer test that we run. The algorithm performs a \emph{double-sided} RowHammer {attack} on each row within a DRAM bank. {A double-sided RowHammer attack activates the two attacker rows that are physically adjacent to a victim row (i.e., the victim row's two immediate neighbors) in an alternating manner. 
{We define \gls{hc} as the number of times each
physically-adjacent row is activated.}
In this study, we perform double-sided attacks instead of {single-}~\cite{kim2014flipping} or many-sided attacks (e.g., as in TRRespass~\cite{frigo2020trrespass}, U-TRR~\cite{hassan2021uncovering}, and BlackSmith~\cite{jattke2022blacksmith}) because a double-sided attack is the most effective RowHammer attack when no RowHammer defense mechanism is employed{:} {it reduces \gls{hcfirst} and increases \gls{ber} compared to {both} single- {and many-}sided attack{s}}~\cite{orosa2021adeeper, kim2014flipping, kim2020revisiting, frigo2020trrespass, hassan2021uncovering, jattke2022blacksmith}. 
{Due to time limitations, 1)~we test $4K$ rows {per DRAM module} (four chunks of $1K$ rows evenly distributed across a DRAM bank) and 2)~we run each test ten times and record the smallest (largest) observed \gls{hcfirst} (\gls{ber}) to account for the worst-case.}
}}

{\head{Finding Physically Adjacent Rows} DRAM-internal address mapping schemes~\cite{cojocar2020arewe,kim2012acase} are used by DRAM manufacturers to translate {\emph{logical}} DRAM addresses (e.g., row, bank, {and} column) that are exposed over the DRAM interface (to the memory controller) to physical {DRAM} addresses {(e.g., physical location of a row)}. {Internal address mapping schemes allow 
{1)~}post-manufacturing row repair techniques to repair erroneous DRAM rows by remapping these rows to spare rows and 
{2)~}DRAM manufacturers to organize DRAM internals in a cost-optimized way, e.g., by organizing internal DRAM buffers hierarchically~\cite{khan2016parbor,vandegoor2002address}.} The mapping scheme can vary {substantially} across different DRAM {chips}~\cite{barenghi2018softwareonly,cojocar2020arewe,horiguchi1997redundancy,itoh2001vlsi,keeth2001dram_circuit,khan2016parbor,khan2017detecting,kim2014flipping,lee2017designinduced,liu2013anexperimental,patel2020bitexact,orosa2021adeeper,saroiu2022theprice,patel2022acase}. For every victim DRAM row {we test}, we identify the two {neighboring physically-adjacent} DRAM row addresses that the memory controller can use to access the {aggressor} rows in a double-sided RowHammer attack. To do so, we} reverse-engineer the physical row organization {using} techniques described in prior work{s}~\cite{kim2020revisiting, orosa2021adeeper}.

\SetAlFnt{\footnotesize}
\RestyleAlgo{ruled}
\begin{algorithm}
\caption{\omtwo{Test for} $HC_{first}$ and $BER$ \omthree{for a Given} $V_{PP}$}\label{hammerdimmer:alg:test_alg}
  \DontPrintSemicolon
  \SetKwFunction{FVPP}{set\_vpp}
  \SetKwFunction{FMain}{test\_loop}
  \SetKwFunction{FHammer}{measure\_$BER$}
  \SetKwFunction{initialize}{initialize\_row}
  \SetKwFunction{initializeaggr}{initialize\_aggressor\_rows}
  \SetKwFunction{measureber}{measure\_BER}
  \SetKwFunction{FMeasureHCfirst}{measure\_$HC_{first}$}
  \SetKwFunction{compare}{compare\_data}
  \SetKwFunction{Hammer}{hammer\_doublesided}
  \SetKwFunction{Gaggressors}{get\_aggressors}
  \SetKwFunction{FWCDP}{get\_WCDP}
  \SetKwProg{Fn}{Function}{:}{}

  \tcp{$RA_{victim}$: victim row address}
  \tcp{$WCDP$: worst-case data pattern}
  \tcp{$HC$: number \omone{of} activations per aggressor row}
  \Fn{\FHammer{$RA_{victim}$, $WCDP$, $HC$}}{
        \initialize($RA_{victim}$, $WCDP$)\;
        \omone{\initializeaggr($RA_{victim}$, \agycrtwo{bitwise\_inverse($WCDP$)})}\;
        \Hammer(\omone{$RA_{victim}$}, $HC$)\;
        $BER\omone{_{row}} =$ \compare($RA_{victim}$, $WCDP$)\;
        \KwRet $BER\omone{_{row}}$\;
  }\;
  
  \tcp{$V_{pp}$: \omone{wordline voltage for the experiment}}
  \tcp{\omone{$WCDP\_list$: the list of $WCDP$s (one $WCDP$ per row)}}
  \tcp{\omone{$row\_list$: the list of tested rows}}
  \Fn{\FMain{$V_{pp}$, \omone{$WCDP$\_list}}}{
        \FVPP($V_{pp}$)\;
        \ForEach{$RA_{victim}$ in $row\_list$}{
            \omone{$HC = 300K$ // initial hammer count to test\;
            \omone{$HC_{step} = 150K$} // how much to increment/decrement $HC$ \;
            \While{$HC_{step} > 100$}{
                $BER_{row_{max}} = 0$\;
                \For{$i\gets0$ \KwTo $num\_iterations$}{ 
                    \agycrone{$BER_{row} = ~$\measureber($RA_{victim}$, $WCDP$, $HC$)}\;
                    record\_BER($V_{pp}$, $RA_{victim}$, $WCDP$, $HC$, $BER_{row}$, $i$)\;
                    $BER_{row_{max}} =  max(BER_{row_{max}}, BER_{row})$\;
                }
                \If{$BER_{row_{max}} == 0$}{
                  \agycrtwo{$HC += HC_{step}$ // Increase HC if no bitflips occur}
                }\Else{
                  \agycrthree{$HC -= HC_{step}$ // Reduce HC if a bitflip occurs}
                }
                $HC_{step}$ = $HC_{step}/2$\;
            }
            record\_HCfirst($V_{pp}$, $RA_{victim}$, $WCDP$, $HC$)\;
        }}

  }\;
\end{algorithm}

\subsection{{Row Activation Latency Experiments}}
\label{hammerdimmer:sec:experiment_design_trcd}
{{We conduct experiments t}o find how a DRAM chip's {row activation latency (}\gls{trcd}{)} changes with {reduced} \gls{vpp}.}

\head{Metric} We {measure} \gls{trcdmin} {between a row activation and the {following} read operation {to ensure {that} there are {\emph{no}}} bitflips in the entire DRAM row}.

\head{\omone{WCDP}}
{We choose $WCDP$ as the data pattern {that} {leads to} the \emph{largest} {observed} \gls{trcdmin}.}

\head{$\mathbf{t_{RCD}}$ Tests} Alg.~\ref{hammerdimmer:alg:trcd_test_alg} describes {the core test loop of each \gls{trcd} test that we run.}
The algorithm sweeps \gls{trcd} starting from {the nominal \gls{trcd} of \SI{13.5}{\nano\second} with steps of} \SI{1.5}{\nano\second}.\footnote{{\agy{3}{DRAM Bender} can send a DRAM command every \SI{1.5}{\nano\second} due to the clock frequency limitations in the FPGA's physical {DRAM} interface.}} {We decrement (increment) \gls{trcd} by \SI{1.5}{\nano\second} until we observe at least one (no) bitflip in the entire DRAM row}
in order to {pinpoint} \gls{trcdmin}. 
To {test a DRAM row for a given \gls{trcd}, the algorithm} 1)~initializes the {row with {the row's}} $WCDP$, 2)~performs an access {using the given \gls{trcd} for each column in the row}
and 3)~checks if the access results in any {bitflips}. 
{After {testing each} column in a DRAM row, the algorithm identifies
{the row's \gls{trcdmin} as the minimum \gls{trcd} that does not cause any bitflip in the entire DRAM row.}
{{Due to time limitations, we 1)~test the same set of rows as we use in RowHammer tests {(\secref{hammerdimmer:sec:experiment_design_rowhammer})} and 2)~run each test ten times}
and record the {\emph{largest} \gls{trcdmin} for each row} across all runs.}}\footnote{{To understand whether {reliable} DRAM row activation latency changes over time, we repeat these tests for 24 DRAM chips after one week, during which the chips are tested for RowHammer vulnerability. We observe that \emph{only} {\SI{2.1}{\percent}} of tested DRAM rows experience only a small variation ($<$\SI{1.5}{\nano\second}) in \gls{trcd}. {This result is consistent with results of prior works~\cite{chang2016understanding, chang2017understanding,kim2018solardram}.}}}

\SetAlFnt{\footnotesize}
\RestyleAlgo{ruled}
\begin{algorithm}
\caption{Test {for} {Row} Activation Latency {for a Given} $V_{PP}$}\label{hammerdimmer:alg:trcd_test_alg}
  \DontPrintSemicolon
  \SetKwFunction{FVPP}{set\_vpp}
  \SetKwFunction{FMain}{test\_loop}
  \SetKwFunction{Ftrcd}{identify\_\gls{trcdmin}}
  \SetKwFunction{compare}{compare}
  \SetKwFunction{initialize}{initialize\_row}
  \SetKwFunction{initializeneighbor}{initialize\_neighboring\_rows}
  \SetKwFunction{FWCDP}{get\_WCDP}
  \SetKwProg{Fn}{Function}{:}{}
  
  \tcp{$V_{pp}$: {wordline voltage for the experiment}}
  \tcp{{$WCDP\_list$: the list of WCDPs (one WCDP per row)}}
  \tcp{{$row\_list$: the list of tested rows}}
    \Fn{\FMain{$V_{pp}$, {$WCDP\_list$, $row\_list$}}}{
    \FVPP($V_{pp}$)\;        
      \ForEach{{$RA$ in $row\_list$}}{
        \gls{trcd} = {\SI{13.5}{\nano\second}}\;
        {found\_faulty, found\_reliable = False, False\;}
        \While{{not found\_faulty or not found\_reliable}}
        {
            {is\_faulty = False\;}
            \For{{$i\gets0$ \KwTo $num\_iterations$}}{
                \ForEach{{column $C$ in row $RA$}}{
                    {\initialize($RA$, $WCDP\_list[\omone{RA}]$)}\;
                    {activate\_row($RA$, \gls{trcd}) //activate the row using \gls{trcd}}\;
                    {read\_data = read\_col($C$)}\;
                    {close\_row($RA$)}\;
                    {$BER_{col}$ = \compare(WCDP\_list[\omone{RA}], read\_data)}\;
                    {\lIf{$BER_{col}$ > 0}{is\_faulty=True}}
                }
            }
            {\lIf{is\_faulty}{
                \{\gls{trcd} += \SI{1.5}{\nano\second};
                found\_faulty = True;\}
            }
            \lElse{
                \{\gls{trcdmin} = \gls{trcd};
                \gls{trcd} -= \SI{1.5}{\nano\second};
                found\_reliable = True;\}
            }
        }
        }
        
        {record\_\gls{trcdmin}($RA$, \gls{trcdmin})}\;
    }
  }
\end{algorithm}

\subsection{{Data Retention Time Experiments}}
\label{hammerdimmer:sec:experiment_design_retention}
{We conduct {data} retention time experiments t}{o understand the effects of \gls{vpp} on DRAM cell data retention characteristics. 
{We test the same set of DRAM rows as we use in RowHammer tests {(\secref{hammerdimmer:sec:experiment_design_rowhammer})} for a set of fixed refresh windows from \SI{16}{\milli\second} to {\SI{16}{\second}} in increasing powers of two.}}

\head{Metric} We measure {\acrlong{ber} ({retention-}\gls{ber})} {due to violating a DRAM row's data retention time, using a reduced refresh rate.}

\head{{WCDP}} {We choose $WCDP$ as the data pattern which causes a bitflip at the \emph{smallest} refresh window (\gls{trefw}) among the six data patterns. If we find more than one such data {pattern}, we choose the one that {leads to} the largest \gls{ber} for \gls{trefw} of \SI{16}{\second}.}

{\head{{Data} Retention Time Tests} Alg.~\ref{hammerdimmer:alg:ref_test_alg} describes how we perform {data} retention tests to {measure} {retention-}\gls{ber} for a given \gls{vpp} and {refresh rate}.
{The algorithm} 1)~initialize{s} a DRAM row with {WCDP}, 2)~waits {as long as the given refresh window}, and {3})~read{s} and compare{s} the data in the DRAM row {to the row's initial data}{.}} 

\SetAlFnt{\footnotesize}
\RestyleAlgo{ruled}
\begin{algorithm}
\caption{Test {for} {Data} Retention Time{s} {for a Given} $V_{PP}$}\label{hammerdimmer:alg:ref_test_alg}
  \DontPrintSemicolon
  \SetKwFunction{FVPP}{set\_vpp}
  \SetKwFunction{FMain}{test\_loop}
  \SetKwFunction{Fref}{measure\_ber}
  \SetKwFunction{access}{accessDRAM}
  \SetKwFunction{initialize}{initialize\_row}
  \SetKwFunction{initializeneighbor}{initialize\_neighboring\_rows}
  \SetKwFunction{compare}{compare\_data}
  \SetKwFunction{wait}{wait}
  \SetKwFunction{FWCDP}{get\_WCDP}
  \SetKwProg{Fn}{Function}{:}{}
  
  \tcp{$V_{pp}$: {wordline voltage for the experiment}}
  \tcp{{$WCDP\_list$: the list of WCDPs (one WCDP per row)}}
  \tcp{{$row\_list$: the list of tested rows}}
  \Fn{\FMain{$V_{pp}$, {$WCDP\_list$, $row\_list$}}}{
      \FVPP($V_{pp}$)\;
      \gls{trefw} = {\SI{16}{\milli\second}}\;
      \While{{\gls{trefw} $\leq$ \SI{16}{\second}}}{
        \For{{$i\gets0$ \KwTo $num\_iterations$}}{
            \ForEach{{$RA$ in $row\_list$}}{
            {\initialize($RA$, $WCDP\_list[{RA}]$)}\;      
            {wait(\gls{trefw})}\;
            {read\_data = read\_row($RA$)}\;
            {$BER_{row}$ = \compare(WCDP\_list[{RA}], read\_data)}\;
            {record\_retention\_errors($RA$, \gls{trefw}, $BER_{row}$)}\;
            }
        }
        \gls{trefw} = \gls{trefw}$\times2$\;
      }
      
  }
\end{algorithm}

\subsection{SPICE Model}
\label{hammerdimmer:sec:spice_model}
To {provide insights into} our {real-chip-based} experimental observations {about}
{the effect of reduced \gls{vpp} on row activation latency and data retention time,}
we conduct a set of SPICE~\cite{corpltspice, nagel1973spice} simulations {to estimate the bitline and cell voltage levels} {during two relevant DRAM operations: row activation and charge restoration}. To do so, we adopt {and modify} a SPICE model used in a relevant prior work~\cite{chang2017understanding} that studies the impact of changing \gls{vdd} (but \emph{not} \gls{vpp}) on DRAM row access and refresh operations.
{Table~\ref{hammerdimmer:tab:spice-param} summarizes our SPICE model.
We use LTspice~\cite{corpltspice} with the \SI{22}{\nano\meter} PTM transistor model~\cite{sinha2012exploring, zhao2006newgeneration} {and scale} the simulation parameters
{according to the} ITRS {roadmap}~\cite{semiconductors2015itrs,vogelsang2010understanding}.\footnote{{We} do \emph{not} expect {SPICE simulation and real-world experimental results to be identical because {a} SPICE model \emph{cannot} simulate a real DRAM chip's exact behavior without}
proprietary design and manufacturing information.} {To account for manufacturing process variation, }{we perform Monte-Carlo {simulations} by randomly varying the component parameters up to \SI{5}{\percent} for each simulation run. {We} run the simulation {at \gls{vpp} levels from \SI{1.5}{\volt} to \SI{2.5}{\volt} with a step size of at \SI{0.1}{\volt}} {10K} times, {similar to prior works~\cite{hassan2019crow, luo2020clrdram}}. 
}

\begin{table}[h!]
\caption{Key parameters used in SPICE {simulations}.
}
\footnotesize
\centering
\begin{tabular}{ll}
\toprule
\head{Component} & \textbf{Parameters}  \\
\midrule
DRAM Cell          & C: 16.8 fF, R: 698 $\Omega$  \\

Bitline            & C: 100.5 fF, R: 6980 $\Omega$\\

Cell Access NMOS   & W: 55 nm, L: 85 nm    \\

Sense Amp. NMOS    & W: 1.3 um, L: 0.1 um \\

Sense Amp. PMOS    & W: 0.9 um, L: 0.1 um \\
\bottomrule
\end{tabular}
\label{hammerdimmer:tab:spice-param}
\end{table}

\subsection{{Statistical Significance of Experimental Results}}
\label{hammerdimmer:sec:statistical:sig}

To evaluate the statistical significance of our methodology, we investigate the variation in our measurements by examining the \emph{coefficient of variation (CV)}
across ten iterations. {CV is a standardized metric to measure the extent of variability in a set of measurements, in relation to the mean of the measurements. CV is calculated as the ratio of standard deviation over the mean value~\cite{everitt1998dram_circuit}. A smaller CV shows a smaller variation across measurements, indicating {higher} statistical significance.} 
The coefficient of variation is 0.08, 0.13, and 0.24 for $90^{th}$, $95^{th}$, and $99^{th}$ {percentiles} of {all of our experimental results}, respectively.

\section{RowHammer Under Reduced Wordline Voltage}
\label{hammerdimmer:sec:vpp_with_rh}

We provide the first experimental characterization of how {\acrfull{vpp}} affects the RowHammer vulnerability 
of {a} DRAM row in terms of 1)~\acrfull{ber} {(\secref{hammerdimmer:sec:vpp:ber})} and 2)~\acrfull{hcfirst} {(\secref{hammerdimmer:sec:vpp_vs_hcfirst}). To conduct this analysis, we provide experimental results {from \numchips{}} real DRAM {chips}, {using} the methodology described in \secref{hammerdimmer:sec:experimental_setup} and \secref{hammerdimmer:sec:experiment_design_rowhammer}.}

\subsection{{Effect} of Wordline Voltage on RowHammer BER}
\label{hammerdimmer:sec:vpp:ber}

\figref{hammerdimmer:fig:vpp_ber} shows the {RowHammer} \gls{ber} a DRAM row experiences at a fixed hammer count of $300K$ under different voltage levels, normalized to the row's {RowHammer} \gls{ber} at nominal \gls{vpp} (\SI{2.5}{\volt}). Each line represents a different DRAM module. The band of shade around each line marks the {\SI{90}{\percent}} confidence interval of the normalized \gls{ber} value across all tested DRAM rows. We make \obsvsref{hammerdimmer:obsv:dominant_ber} and~\ref{hammerdimmer:obsv:discrepency_ber} from \figref{hammerdimmer:fig:vpp_ber}.

\begin{figure}[!ht]
    \centering
    \includegraphics[width=0.85\linewidth]{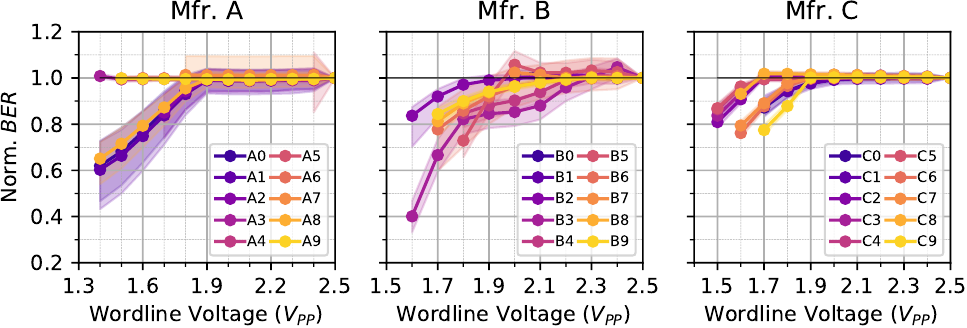}
    \caption{Normalized $BER$ values across different $V_{PP}$ levels. Each curve represents a different DRAM module.}
    \label{hammerdimmer:fig:vpp_ber}
\end{figure}

\observation{Fewer DRAM cells experience bitflips due to {RowHammer} under reduced wordline voltage. \label{hammerdimmer:obsv:dominant_ber}}

We observe that {RowHammer} \gls{ber} \emph{decreases} as \gls{vpp} reduces {in \fracbersupportingrows{} of tested rows across all tested modules}. This \gls{ber} reduction reaches up to \berdecrmax{} (B3 {at $V_{PP}=1.6V$}) with an average of \berdecravg{} {(not shown in the figure)} across all tested modules. {We} conclude that read disturbance
becomes weaker{, on average,} with reduced \gls{vpp}. 

\observation{In contrast to the dominant trend, reducing \gls{vpp} can {sometimes} increase \gls{ber}. \label{hammerdimmer:obsv:discrepency_ber}}

{We observe that \gls{ber} {increases in \fracberopposingrows{} of tested rows} with reduced \gls{vpp} {by up to {\berincrmax{}}}
{(B5 at $V_{PP} = 2.0V$)}.
We {suspect that {the} {\gls{ber} increase} {we observe} 
occurs {due to a} weakened charge restoration process rather than an {actual} increase in {read} disturbance {(due to} RowHammer{)}.} 
{\secref{hammerdimmer:sec:sideeffects_retention}} analyze{s} the impact of reduced \gls{vpp} on {the} charge restoration process.} 

\head{{Variation in \gls{ber} Reduction Across DRAM Rows}}
{We} investigate how \gls{ber} {reduction {with reduced \gls{vpp}} varies across DRAM rows. To do so, we measure \gls{ber} reduction of each DRAM row at \gls{vppmin} {(\secref{hammerdimmer:sec:experimental_setup})}.}
\figref{hammerdimmer:fig:vpp_ber_fine} shows a population density distribution of DRAM rows {(y-axis) based on their \gls{ber} at \gls{vppmin}, normalized to their \gls{ber} at {the} {nominal \gls{vpp} level} (x-axis),} for each manufacturer.
We make \obsvref{hammerdimmer:obsv:hist_ber} from \figref{hammerdimmer:fig:vpp_ber_fine}.

\begin{figure}[!h]
    \centering
    \includegraphics[width=0.85\linewidth]{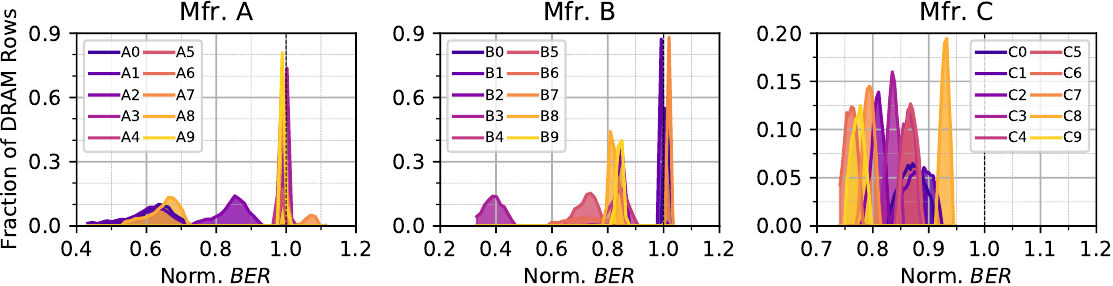}
    \caption{Population density distribution of DRAM rows based on their {normalized} $BER$ values at $V_{PPmin}$.}
    \label{hammerdimmer:fig:vpp_ber_fine}
\end{figure}

\observation{{\gls{ber} reduction with reduced \gls{vpp} varies across different DRAM rows and different manufacturers.}\label{hammerdimmer:obsv:hist_ber}}

{DRAM rows exhibit a large range of normalized \gls{ber} values {(\minnormbermfrA{}--\maxnormbermfrA{}, \minnormbermfrB{}--\maxnormbermfrB{}, and \minnormbermfrC{}--\maxnormbermfrC{} in chips from Mfrs. A, B, and C, respectively)}.}
{\gls{ber} reduction also varies across different manufacturers. For example, \gls{ber} reduces by more than \SI{5}{\percent} for \emph{all} DRAM rows {of} Mfr.~C, while \gls{ber} variation with reduced \gls{vpp} is smaller than \bersmallchange{} in \fracbersmallchangerowsmfrA{} of the rows {of} Mfr.~A.}

Based on \obsvsref{hammerdimmer:obsv:dominant_ber}--\ref{hammerdimmer:obsv:hist_ber}, we conclude that {a DRAM row's} RowHammer \gls{ber} {tends to decrease with reduced \gls{vpp}, while both the amount and the direction of change in \gls{ber} varies across different DRAM rows and manufacturers.}

\subsection{{Effect} of Wordline Voltage on the Minimum Hammer Count {Necessary} to Cause a Bitflip}
\label{hammerdimmer:sec:vpp_vs_hcfirst}

\figref{hammerdimmer:fig:vpp_hcfirst} shows the \gls{hcfirst} a DRAM row exhibits under different voltage levels, normalized to the row's \gls{hcfirst} at nominal \gls{vpp} (\SI{2.5}{\volt}). Each line represents a different DRAM module. The band of shade around each line marks the {\SI{90}{\percent}} confidence interval of the normalized \gls{hcfirst} value{s} across all tested DRAM rows {in the module}. We make \obsvsref{hammerdimmer:obsv:dominant_hcfirst} and~\ref{hammerdimmer:obsv:discrepency_hcfirst} from \figref{hammerdimmer:fig:vpp_ber}.
\begin{figure}[!ht]
    \centering
    \includegraphics[width=0.85\linewidth]{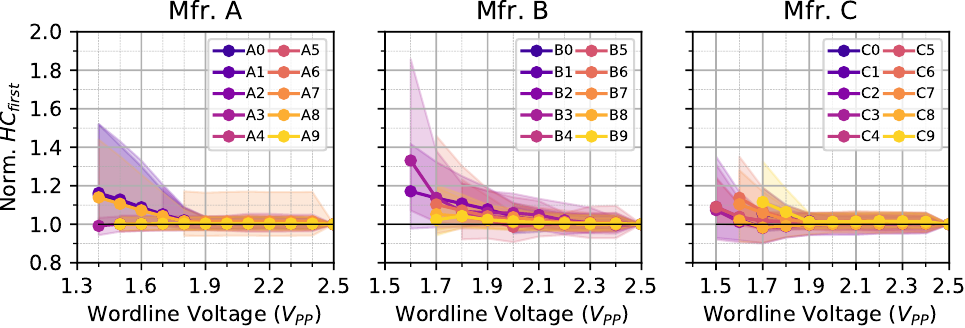}
    \caption{Normalized $HC_{first}$ values across different $V_{PP}$ levels. Each curve represents a different DRAM module.}
    \label{hammerdimmer:fig:vpp_hcfirst}
\end{figure}

\observation{DRAM cells experience RowHammer bitflips at higher hammer counts under reduced wordline voltage. \label{hammerdimmer:obsv:dominant_hcfirst}}

We observe that \gls{hcfirst} of a DRAM row increases as \gls{vpp} reduces {in} \frachcfirstsupportingrows{} of {tested rows across all tested modules.} This increase in \gls{hcfirst} reaches up to \hcfirstincrmax{} (B3 {at $V_{PP}=1.6V$}) with an average of \hcfirstincravg{} {(not shown in the figure)} across all {tested} modules. {W}e conclude that the disturbance caused by hammering a DRAM row becomes weaker with reduced \gls{vpp}. 

\observation{In contrast to the dominant trend, reducing \gls{vpp} can {sometimes} cause bitflips at lower hammer counts. \label{hammerdimmer:obsv:discrepency_hcfirst}}

We observe that \gls{hcfirst} 
reduces {in \frachcfirstopposingrows{} of tested rows with reduced \gls{vpp} by up to \hcfirstdecrmax{} (C8 at \gls{vpp}=\SI{1.6}{\volt})}. {Similar to} \obsvref{hammerdimmer:obsv:discrepency_ber}, {we suspect that this behavior is caused by the weakened charge restoration process} {(see \secref{hammerdimmer:sec:sideeffects_retention})}. 
 
\head{{Variation in \gls{hcfirst} Increase Across DRAM Rows}}
{We} investigate how \gls{hcfirst} {increase varies {with reduced \gls{vpp}} across DRAM rows. To do so, we measure \gls{hcfirst} {increase} of each DRAM row at \gls{vppmin} {(\secref{hammerdimmer:sec:experimental_setup})}.}
\figref{hammerdimmer:fig:vpp_hcfirst_fine} shows a population density distribution of DRAM rows {(y-axis) based on their \gls{hcfirst} at \gls{vppmin}, normalized to their \gls{hcfirst} at {the} {nominal \gls{vpp} level} (x-axis),} for each manufacturer.
We make \obsvref{hammerdimmer:obsv:hist_hcfirst} from \figref{hammerdimmer:fig:vpp_hcfirst_fine}.

\begin{figure}[!ht]
    \centering
    \includegraphics[width=0.85\linewidth]{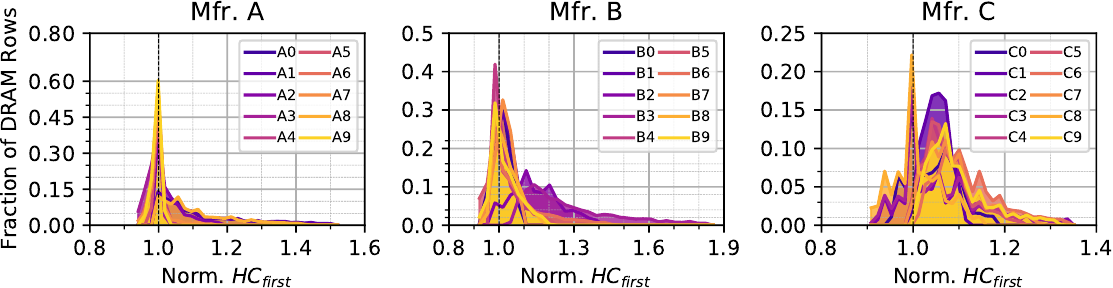}
    \caption{Population density distribution of DRAM rows based on their {normalized} $HC_{first}$ values at $V_{PPmin}$.}
    \label{hammerdimmer:fig:vpp_hcfirst_fine}
\end{figure}

\observation{{\gls{hcfirst} increase with reduced \gls{vpp} varies across different DRAM rows and different manufacturers.}\label{hammerdimmer:obsv:hist_hcfirst}}

{DRAM rows in chips from the same manufacturer exhibit a large range of normalized \gls{hcfirst} values (\minnormhcfirstmfrA{}--\maxnormhcfirstmfrA{}, \minnormhcfirstmfrB{}--\maxnormhcfirstmfrB{}, and \minnormhcfirstmfrC{}--\maxnormhcfirstmfrC{} for Mfrs. A, B, and C, respectively).}
{\gls{hcfirst} increase {also} varies across different manufacturers. For example, \gls{hcfirst} increases with reduced \gls{vpp} for
{\frachcfirstsupportingrowsmfrC{} of DRAM rows in modules from Mfr.~C, while \frachcfirstsupportingrowsmfrA{} of DRAM rows exhibit this behavior in modules from Mfr.~A.}
}

Based on \obsvsref{hammerdimmer:obsv:dominant_hcfirst}--\ref{hammerdimmer:obsv:hist_hcfirst}, we conclude that {a DRAM row's \gls{hcfirst} tends to increase with reduced \gls{vpp}, while both the amount and the direction of change in \gls{hcfirst} varies across different DRAM rows and manufacturers.}

{\head{{Summary of Findings}}
Based on our analyses on both \gls{ber} and \gls{hcfirst}, we conclude that {a DRAM chip's RowHammer vulnerability can be reduced}
by operating the chip at a \gls{vpp} level {that} is lower than the nominal \gls{vpp} value{.}}

\section{DRAM \agycrone{Reliability} Under Reduced Wordline Voltage}
\label{hammerdimmer:sec:sideeffects}

{To investigate the effect of reduced \gls{vpp} on {reliable} DRAM {operation},}
we provide the first experimental characterization of how \gls{vpp} affects the {reliability} of three \gls{vpp}-related fundamental DRAM operations: 1)~DRAM row activation {(\secref{hammerdimmer:sec:sideeffects_trcd})}, 2)~charge restoration {(\secref{hammerdimmer:sec:sideeffects_charge_restoration})}, and~3)~DRAM refresh {(\secref{hammerdimmer:sec:sideeffects_retention})}. 
To conduct these analyses{,} we provide both {1)} experimental results {from} real DRAM devices, {using} the methodology described in {\secref{hammerdimmer:sec:experimental_setup}, \secref{hammerdimmer:sec:experiment_design_trcd}, and \secref{hammerdimmer:sec:experiment_design_retention}} and 2) SPICE simulation results{,} {using} the methodology described in {\secref{hammerdimmer:sec:spice_model}}. 

\subsection{DRAM Row Activation Under Reduced Wordline Voltage}
\label{hammerdimmer:sec:sideeffects_trcd}

\head{Motivation} 
{DRAM row activation latency (\gls{trcd}) should theoretically increase with reduced \gls{vpp} (\secref{sec:background:dramvoltage}).}
{We} {investigate} how \gls{trcd} of real DRAM chips change with {reduced} \gls{vpp}.

\head{Novelty}~{We provide the first experimental analysis of} the isolated impact of \gls{vpp} on activation latency. Prior work~\cite{chang2017understanding} tests DDR3 DRAM chips under reduced {supply voltage (}\gls{vdd}{)}, which may or may not change internally{-}generated \gls{vpp} level. In contrast, we modify only wordline voltage (\gls{vpp}) without {modifying \gls{vdd} to avoid {the possibility} of negatively impacting DRAM reliability due to I/O circuitry instabilities (\secref{sec:background:dramvoltage}).}

\head{Experimental Results}~\figref{hammerdimmer:fig:vpp_trcd} demonstrates {the variation in {\gls{trcdmin} (\secref{hammerdimmer:sec:experiment_design_trcd}) on the y-axis}}
{under} {reduced} \gls{vpp} {on the x-axis}{,} {across \nummodules{} DRAM modules}. 
We annotate the nominal \gls{trcd} value (\SI{13.5}{\nano\second})~\cite{jedec2020jesd794c} with a black horizontal line. We make \obsvref{hammerdimmer:obsv:trcd_increase} from \figref{hammerdimmer:fig:vpp_trcd}.

\begin{figure}[!ht]
    \centering
    \includegraphics[width=0.85\linewidth]{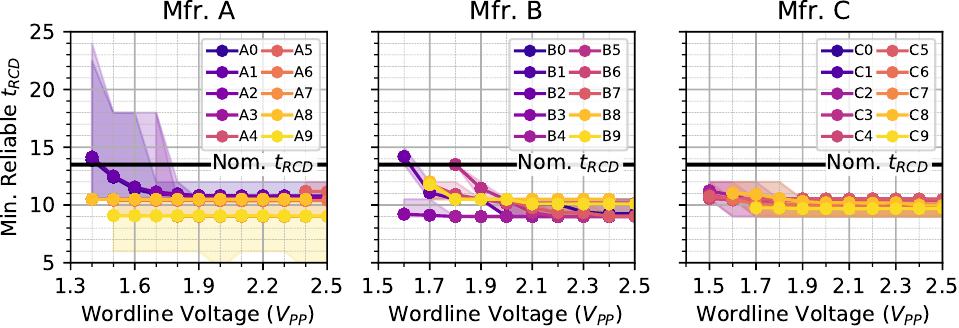}
    \caption{Minimum reliable $t_{RCD}$ values across different $V_{PP}$ levels. Each curve represents a different DRAM module.}
    \label{hammerdimmer:fig:vpp_trcd}
\end{figure}

\observation{{Reliable} row activation latency {generally} increases with reduced \gls{vpp}. {However, \numreliablechips{} {(\numreliablemodules{})} out of \numchips{} {(\nummodules{})} DRAM chips {(modules)}} complete {row} activation before the nominal activation latency.\label{hammerdimmer:obsv:trcd_increase}}

The minimum reliable activation latency {(\gls{trcdmin}) increases with reduced} \gls{vpp} across all tested modules. {\gls{trcdmin} exceeds the nominal \gls{trcd} of \SI{13.5}{\nano\second} for {\emph{only}} \numunreliablemodules{} of \nummodules{} tested modules (A0--A2, B2, and B5). Among these, modules from Mfr{.}~A and~B contain 16 and 8 chips per module. Therefore, we conclude that \numreliablechips{} of \numchips{} tested DRAM chips do \emph{not} experience bitflips when operated using nominal \gls{trcd}. We observe that since \gls{trcdmin} increases with reduced \gls{vpp}, the available \gls{trcd} guardband reduces by \trcdguardbandreduction{} with reduced \gls{vpp}{,} on average across all DRAM modules that reliably work with nominal \gls{trcd}. {{We also} observe that the three and two modules from Mfrs.~A and~B{, which exhibit \gls{trcdmin} values larger than the nominal \gls{trcd}}, 
reliably operate when {we use} a \gls{trcd} of \SI{24}{\nano\second} and \SI{15}{\nano\second}, respectively.}}

To verify our {experimental} observations and provide a deeper {insight} into the {effect of \gls{vpp} on} activation latency, we perform SPICE simulations {(}as described in \secref{hammerdimmer:sec:spice_model}{)}. \figref{hammerdimmer:fig:spice_act_waveform}a
shows a waveform of the bitline voltage during {the} row activation process. The time in the x-axis starts when {an} activation command is issued. Each color corresponds to the bitline voltage at a different \gls{vpp} level. We annotate the bitline's supply voltage (\gls{vdd}) and \gls{vthresh}. We make \obsvref{hammerdimmer:obsv:act_completes_later} from \figref{hammerdimmer:fig:spice_act_waveform}a.

\begin{figure}[!ht]
    \centering
    \includegraphics[width=0.85\linewidth]{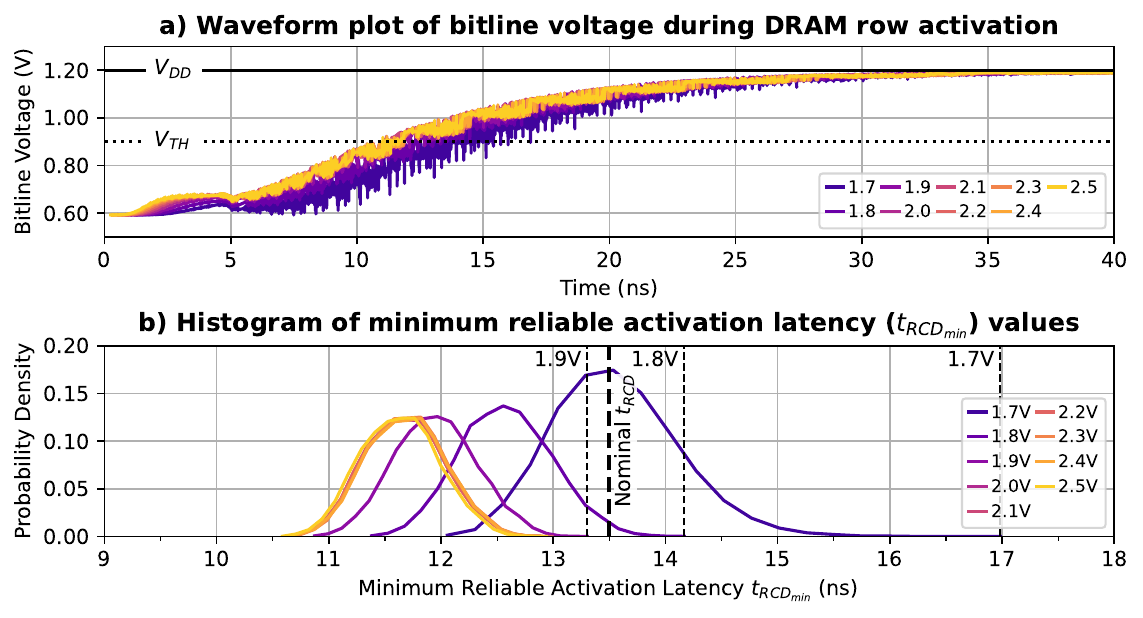}
    \caption{({a}) Waveform of the bitline voltage during row activation and ({b}) probability density distribution of {$t_{RCDmin}$} values{,} {for} different $V_{PP}$ levels.}
    \label{hammerdimmer:fig:spice_act_waveform}
\end{figure}

\observation{{Row} activation successfully completes under reduced \gls{vpp} with an increased activation latency.\label{hammerdimmer:obsv:act_completes_later}}

{{\figref{hammerdimmer:fig:spice_act_waveform}a shows} that, as \gls{vpp} decreases, the bitline voltage takes longer to {increase} to \gls{vthresh}, resulting in a slower row activation. {{For example, \gls{trcdmin} increases from} \SI{11.6}{\nano\second} to \SI{13.6}{\nano\second} {(on average across $10^4$ Monte-Carlo simulation iterations)} {when \gls{vpp} is reduced from \SI{2.5}{\volt} to} \SI{1.7}{\volt}.}
{This happens {due to} two reasons. First, a lower \gls{vpp} creates a weaker channel in the access transistor, requiring a longer time for the capacitor and bitline to share charge. Second, the charge sharing process {(0--\SI{5}{\nano\second} in \figref{hammerdimmer:fig:spice_act_waveform}a)} {leads to} a smaller {change} in bitline voltage when \gls{vpp} is reduced due to the weakened charge restoration {process} that we explain in \secref{hammerdimmer:sec:sideeffects_charge_restoration}.}}

\figref{hammerdimmer:fig:spice_act_waveform}b shows the probability density distribution of \gls{trcdmin} values {under reduced \gls{vpp}} across {a total of $10^4$} Monte-Carlo simulation {iterations} {for different \gls{vpp} levels (color-coded)}. 
Vertical lines annotate the worst-case {reliable} \gls{trcdmin} values {across all iterations of {our} Monte-Carlo simulation (\secref{hammerdimmer:sec:spice_model})} for {different} \gls{vpp} levels.
We make \obsvref{hammerdimmer:obsv:trcd_spice_agrees} from \figref{hammerdimmer:fig:spice_act_waveform}b.

\observation{{SPICE simulations agree with our activation latency-related observations based on experiments on real DRAM chips{: \gls{trcdmin} increases with reduced \gls{vpp}.}\label{hammerdimmer:obsv:trcd_spice_agrees}}}

{We} analyze the variation in 
{1})~the probability density distribution of \gls{trcdmin}, and 
{2})~the worst-case (largest) {reliable} \gls{trcdmin} value when \gls{vpp} is reduced.
\figref{hammerdimmer:fig:spice_act_waveform}b shows that the probability density distribution of {\gls{trcdmin}} both shifts to larger values and {becomes} wider {with reduced \gls{vpp}}. 
{The} {worst-case (largest) \gls{trcdmin} increases from {\SI{12.9}{\nano\second} to \SI{13.3}{\nano\second}, \SI{14.2}{\nano\second}, and \SI{16.9}{\nano\second} when \gls{vpp} is reduced from \SI{2.5}{\volt} to \SI{1.9}{\volt}}, \SI{1.8}{\volt} and \SI{1.7}{\volt}, respectively.{\footnote{{SPICE simulation results do not show reliable operation when \gls{vpp}$\leq$\SI{1.6}{\volt}, yet real DRAM chips do operate reliably as we {show} in \secref{hammerdimmer:sec:sideeffects_trcd} and \secref{hammerdimmer:sec:sideeffects_retention}.}}} For a realistic nominal value of \SI{13.5}{\nano\second}, \gls{trcd}'s guardband reduces from {\SI{4.4}{\percent} to \SI{1.5}{\percent} as \gls{vpp} reduces from \SI{2.5}{\volt} to \SI{1.9}{\volt}}. {As \secref{hammerdimmer:sec:spice_model} explains, SPICE simulation results do \emph{not} exactly match {measured} real-device characteristics {(shown in \obsvref{hammerdimmer:obsv:trcd_increase})} because a SPICE model \emph{cannot} simulate a real DRAM chip's exact behavior without proprietary design and manufacturing information.}}

{From \obsvsref{hammerdimmer:obsv:trcd_increase}--\ref{hammerdimmer:obsv:trcd_spice_agrees}, we conclude that} 
1)~the {reliable row} activation latency increases with reduced \gls{vpp}, 
2)~the increase in {reliable row} activation latency does \emph{not} immediately require increasing {the} {nominal} \gls{trcd}{, but reduces the available guardband by \SI{21.9}{\percent}}
{for 208 out of 272 tested chips}{, and 3)~observed bitflips can be {eliminated} by increasing \gls{trcd} to \SI{24}{\nano\second} and \SI{15}{\nano\second} for erroneous modules from Mfrs.~A and~B.}

\subsection{DRAM Charge Restoration Under Reduced Wordline Voltage}
\label{hammerdimmer:sec:sideeffects_charge_restoration}

\head{\js{Motivation}}
A DRAM cell's charge restoration {process} is {affected by} \gls{vpp} because, similar to {the} row activation process, a DRAM cell capacitor's charge is restored through the channel formed in the access transistor, which is controlled by the wordline. Due to access transistor's characteristics, reducing \gls{vpp} without changing \gls{vdd} reduces \gls{vgs} and forms a weaker channel. To understand the impact of \gls{vpp} reduction on {the} charge restoration process, we investigate how charge restoration of a DRAM cell varies with {reduced} \gls{vpp}.

\head{{Experimental Results}}
Since our {FPGA} infrastructure {cannot probe} a DRAM cell capacitor's voltage level, we conduct this study in {our SPICE} simulation environment (\secref{hammerdimmer:sec:spice_model}).
\figref{hammerdimmer:fig:charge_restoration_waveform}a shows the waveform plot of capacitor voltage (y-axis) over time (x-axis), following a row activation event {(at t=0)}. \figref{hammerdimmer:fig:charge_restoration_waveform}b shows the probability density distribution (y-axis) of {\gls{trasmin} to {reliably} complete {the} charge restoration process on the x-axis} under different \gls{vpp} levels.
We make \obsvsref{hammerdimmer:obsv:restoration_chargereduction} and~\ref{hammerdimmer:obsv:restoration_largetras} from \figref{hammerdimmer:fig:charge_restoration_waveform}a and~\ref{hammerdimmer:fig:charge_restoration_waveform}b. 

\begin{figure}[!ht]
    \centering
    \includegraphics[width=0.85\linewidth]{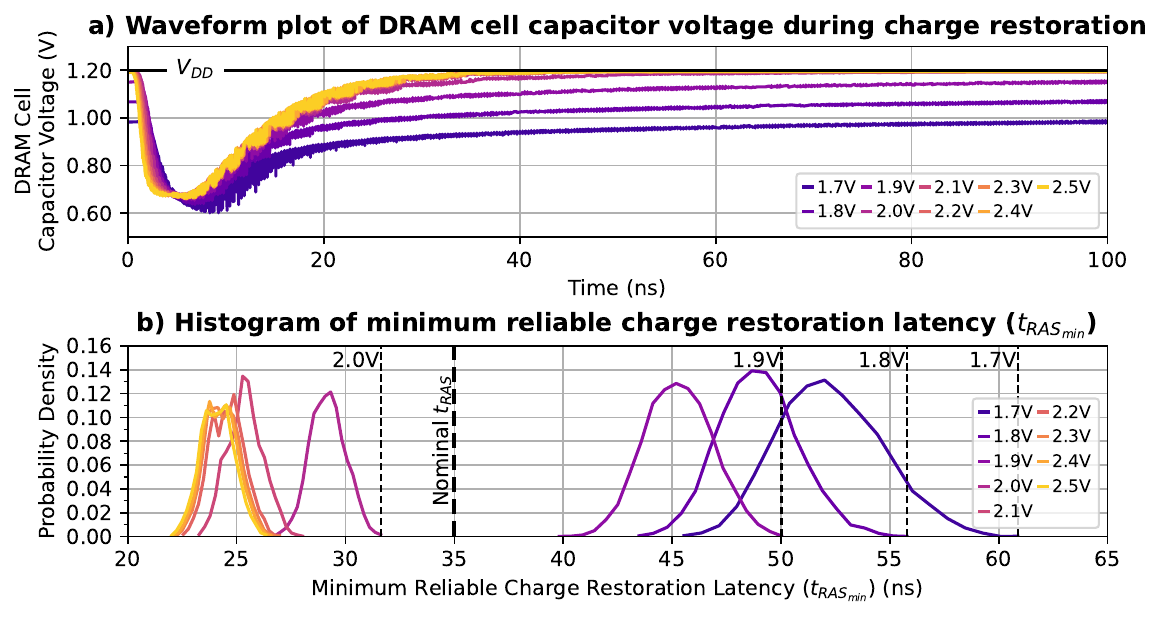}
    \caption{{(a)} Waveform of the {cell} capacitor voltage following a row activation and {(b)} probability density distribution of \agycrfour{$t_{RASmin}$} values, {for} different $V_{PP}$ levels.}
    \label{hammerdimmer:fig:charge_restoration_waveform}
\end{figure}

\observation{A DRAM cell's capacitor voltage {can} saturate at a lower voltage level when \gls{vpp} is reduced.\label{hammerdimmer:obsv:restoration_chargereduction}}

{We observe that a DRAM cell capacitor's {voltage saturates at \gls{vdd} {(\SI{1.2}{\volt})} when \gls{vpp} is \SI{2.0}{\volt} or higher.}
However, the cell capacitor's voltage {saturates at a lower voltage level}
{by} \SI{4.1}{\percent}, \SI{11.0}{\percent}, and \SI{18.1}{\percent} {when \gls{vpp} is
\SI{1.9}{\volt}, \SI{1.8}{\volt}, and \SI{1.7}{\volt}, respectively.}}
This happens because the access transistor turns off when the voltage difference between its gate and source is smaller than a threshold {level}. For example, when \gls{vpp} is set to \SI{1.7}{\volt}, the access transistor allows charge restoration until the cell voltage reaches {\SI{0.98}{\volt}}. When the cell voltage reaches this level, the voltage difference between the gate (\SI{1.7}{\volt}) and the source {(\SI{0.98}{\volt})} is not large enough to form a strong channel, {causing} the cell voltage to saturate at {\SI{0.98}{\volt}}. This reduction in voltage can potentially {1)~increase the row activation latency (\gls{trcd}) and 2)~}reduce the cell's retention time. We {1)~already account for reduced saturation voltage's effect on \gls{trcd} in \secref{hammerdimmer:sec:sideeffects_trcd} and} 2)~investigate its effect on retention time in \secref{hammerdimmer:sec:sideeffects_retention}.

\observation{The increase in a DRAM cell's charge restoration latency with {reduced} \gls{vpp} {can} increase the $t_{RAS}$ {timing parameter}, depending on the \gls{vpp} level.\label{hammerdimmer:obsv:restoration_largetras}}

{Similar to the variation in \gls{trcd} values that we discuss in \obsvref{hammerdimmer:obsv:trcd_spice_agrees}, the probability density distribution of \gls{tras} values also shifts to larger values {(i.e., \gls{tras} exceeds the nominal value when \gls{vpp} is lower than 2.0V)} and {becomes} wider {as} \gls{vpp} reduces. This happens as a {result} of reduced cell voltage, weakened channel in the access transistor, and reduced voltage level at the end of the charge sharing process, as we explain in \obsvref{hammerdimmer:obsv:trcd_spice_agrees}.}

{From \obsvsref{hammerdimmer:obsv:restoration_chargereduction} and~\ref{hammerdimmer:obsv:restoration_largetras}, we conclude that reducing \gls{vpp} {can} negatively affect the charge restoration process. {Reduced \gls{vpp}'s negative impact on charge restoration can potentially be mitigated {by} leveraging the guardbands in DRAM timing parameters~{\cite{lee2015adaptivelatency,lee2017designinduced, chang2016understanding, chang2017understanding, kim2018solardram}} and using intelligent DRAM refresh techniques, where a partially restored DRAM row can be {refreshed} more frequently, {so that the row's charge is restored before it experiences a data retention bitflip}~\cite{das2018vrldram, liu2012raidr, wang2018reducing}. We leave exploring such solutions {to} future work.}}

\subsection{DRAM Row Refresh Under Reduced Wordline Voltage}
\label{hammerdimmer:sec:sideeffects_retention}

\head{Motivation} \secref{hammerdimmer:sec:sideeffects_charge_restoration} demonstrates that the charge {re}stored in a DRAM cell {after a row activation} {can} be reduced as a result of \gls{vpp} reduction. This phenomenon is important for DRAM-based memories because reduced charge in {a} cell might reduce a DRAM cell's {data} retention time, {causing} \emph{retention bitflips} if the cell is \emph{not} refreshed more frequently. To understand the impact of \gls{vpp} reduction on real DRAM chips, we investigate the {effect of reduced \gls{vpp} on data retention related bitflips {using the methodology described in \secref{hammerdimmer:sec:experiment_design_retention}}.}

\head{Novelty} This is the first work that experimentally analyzes the isolated impact of \gls{vpp} on DRAM cell retention {times}. Prior work~\cite{chang2017understanding} tests DDR3 DRAM chips under reduced \gls{vdd}, which may or may not change {the} internally{-}generated \gls{vpp} level.

\head{{Experimental Results}}
{\figref{hammerdimmer:fig:vpp_retention} demonstrates reduced \gls{vpp}'s effect on data retention \gls{ber} {on real DRAM chips}. \figref{hammerdimmer:fig:vpp_retention}a shows how the data retention \gls{ber} (y-axis) changes with increasing refresh window (log-scaled in x-axis) for different \gls{vpp} levels (color-coded). Each curve in \figref{hammerdimmer:fig:vpp_retention}a shows the average \gls{ber} across all DRAM rows, and error bars mark {the} \SI{90}{\percent} confidence interval. {The x-axis starts from \SI{64}{\milli\second} because we do \emph{not} observe any bitflips at \gls{trefw} values {smaller} than \SI{64}{\milli\second}}. To provide deeper insight into reduced \gls{vpp}'s effect on data retention \gls{ber}, \figref{hammerdimmer:fig:vpp_retention}b demonstrates the population density distribution of data retention \gls{ber} across tested rows for {a} \gls{trefw} of \SI{4}{\second}. Dotted vertical lines mark the average \gls{ber} across rows for each \gls{vpp} level.}
We make \obsvsref{hammerdimmer:obsv:retention_vpp} and~\ref{hammerdimmer:obsv:retention_minreliable} from \figref{hammerdimmer:fig:vpp_retention}.

\begin{figure}[!ht]
    \centering
    \includegraphics[width=\linewidth]{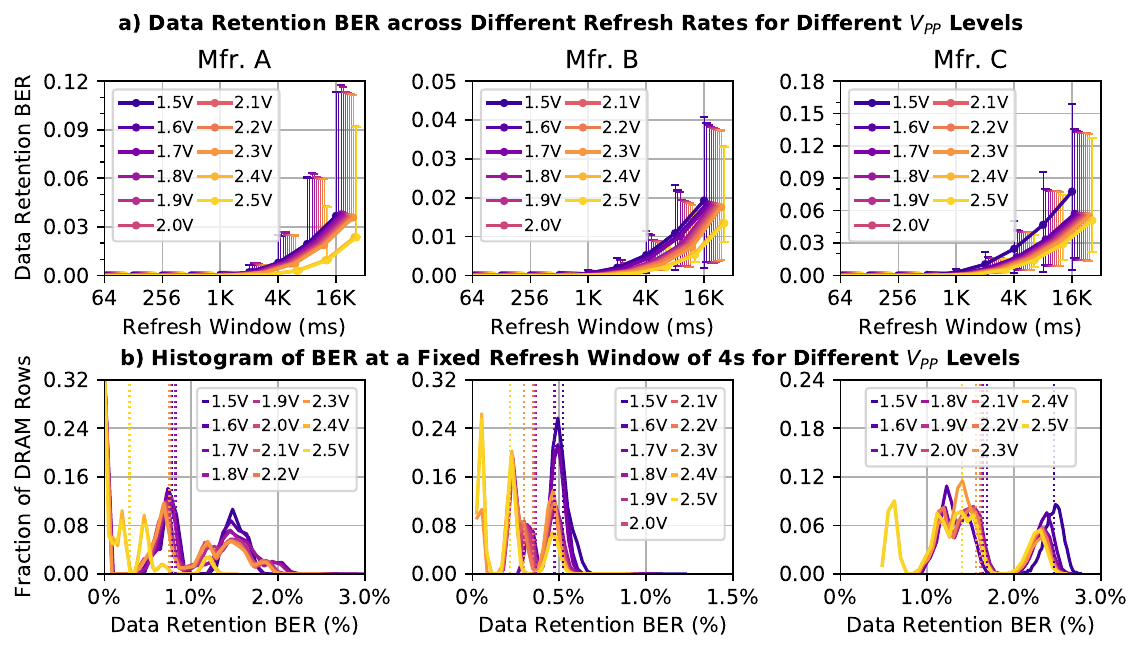}
    \caption{{Reduced $V_{PP}$'s effect on a) data retention $BER$ across different refresh rates and b) the distribution of data retention $BER$ across different DRAM rows for a fixed $t_{REFW}$ of \SI{4}{\second}.}}
    \label{hammerdimmer:fig:vpp_retention}
\end{figure}

\observation{More DRAM cells {tend to} experience {data} retention bitflips when \gls{vpp} is reduced.\label{hammerdimmer:obsv:retention_vpp}}

{\figref{hammerdimmer:fig:vpp_retention}a shows that data retention \gls{ber} curve is higher (e.g., dark-purple compared to yellow) for smaller \gls{vpp} levels (e.g., \SI{1.5}{\volt} compared to \SI{2.5}{\volt}). 
{To provide a deeper insight, \figref{hammerdimmer:fig:vpp_retention}b shows that} average data retention \gls{ber} {across all tested rows when \gls{trefw}=\SI{4}{\second}} increases from \SI{0.3}{\percent}, \SI{0.2}{\percent}, and \SI{1.4}{\percent} {for a} \gls{vpp} {of} \SI{2.5}{\volt} to \SI{0.8}{\percent}, \SI{0.5}{\percent}, and \SI{2.5}{\percent} {for a} \gls{vpp} of \SI{1.5}{\volt} for Mfrs.~A, B, and~C, respectively.}
We {hypothesize} that this happens because of the weakened charge restoration process {with reduced \gls{vpp} (\secref{hammerdimmer:sec:sideeffects_charge_restoration})}.

\observation{{Even though} DRAM cells experience retention bitflips at smaller retention times when \gls{vpp} is reduced, {23} of {30} {tested} modules {experience \emph{no} data} retention bitflips {at} the nominal refresh window (\SI{64}{\milli\second}).\label{hammerdimmer:obsv:retention_minreliable}}

{Data retention \gls{ber} is {very} low at the \gls{trefw} of \SI{64}{\milli\second} even for {a} \gls{vpp} of \SI{1.5}{\volt}. We observe that \emph{no} DRAM module from Mfr.~A exhibits a data retention bitflip at {the} \SI{64}{\ms} \gls{trefw}, and \emph{only} three and four modules from Mfrs.~B (B6, B8, and B9) and~C (C1, C3, C5, and C9) experience bitflips across all 30 DRAM modules we test.}

{We} investigate the significance of the{ {observed} data} retention bitflips {and whether it is possible to mitigate these bitflips using error correcting codes (ECC)~\cite{hamming1950error} or {other existing methods to avoid data retention bitflips (e.g., selectively refreshing a small fraction of DRAM rows at a higher refresh rate}~\cite{arahmati2014refreshing, das2018vrldram, liu2012raidr, wang2018reducing}).}
{To do so, we analyze data retention bitflips {when each tested module is operated at the module's \gls{vppmin}}
{for two \gls{trefw} values: \SI{64}{\milli\second} and \SI{128}{\milli\second}, the smallest refresh windows that yield non-zero BER for different DRAM modules.} 

{To evaluate whether data retention bitflips can be avoided using ECC, we assume a realistic data word size of 64 bits~\cite{cojocar2019exploiting, kim2016allinclusive, patel2019understanding, patel2020bitexact, patel2021harp, patel2022acase, patel2021enabling}. We make \obsvref{hammerdimmer:obsv:retention_onlyone} from this analysis.}}

\observation{{Data retention errors can be avoided using simple single error correcting codes} at the smallest \gls{trefw} that yield{s} non-zero \gls{ber}.\label{hammerdimmer:obsv:retention_onlyone}}

{We observe that \emph{no} 64-bit data word contains more than one bitflip for the smallest \gls{trefw} that yield non-zero \gls{ber}.} 
{We conclude that simple \emph{single error correction double error detection (SECDED) ECC} can correct \emph{all} erroneous data words.}

{To evaluate whether data retention bitflips can be avoided {by} selectively refreshing a small fraction of DRAM rows, we analyze the distribution of these bitflips across different DRAM rows. \figref{hammerdimmer:fig:vpp_retention_ber_hist}a (\figref{hammerdimmer:fig:vpp_retention_ber_hist}b) shows the distribution of DRAM rows {that} experience a data retention bitflip when \gls{trefw} is} \SI{64}{\milli\second} 
(\SI{128}{\milli\second})
{but \emph{not} at a smaller \gls{trefw}, based on their data retention bitflip characteristics}.
{The x-axis shows the number of 64-bit data words with one bitflip in a DRAM row.}
{The y-axis shows the fraction of DRAM rows in log-scale, exhibiting the behavior, specified in the x-axis {for different manufacturers (color-coded)}.
} 
We make \obsvref{hammerdimmer:obsv:retention_smallfraction} {from \figref{hammerdimmer:fig:vpp_retention_ber_hist}.}

\begin{figure}[!ht]
    \centering
    \includegraphics[width=0.8\linewidth]{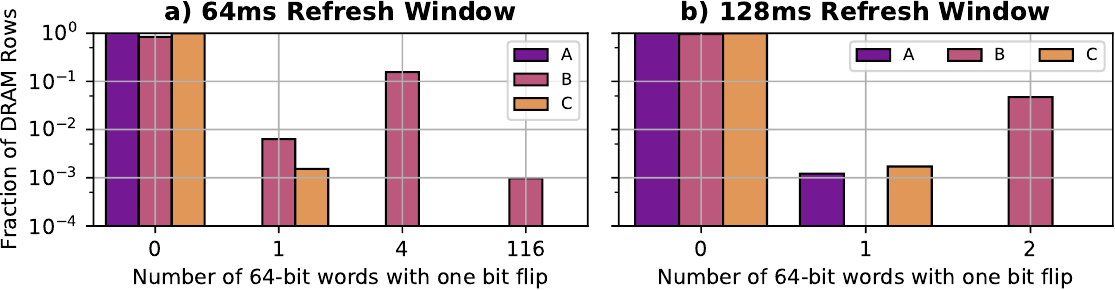}
    \caption{{Data retention bitflip characteristics of DRAM rows in DRAM modules that exhibit bitflips at (a) \SI{64}{\milli\second} and (b) \SI{128}{\milli\second} refresh windows but not at lower $t_{REFW}$ values when operated at $V_{PPmin}$. Each subplot shows the distribution of DRAM rows based on the number of erroneous {64-bit} words that the rows exhibit.}}
    \label{hammerdimmer:fig:vpp_retention_ber_hist}
\end{figure}

\observation{{Only a small fraction (\SI{16.4}{\percent} / \SI{5.0}{\percent}) of DRAM rows  contain erroneous data words at the smallest \gls{trefw} (\SI{64}{\milli\second} / \SI{128}{\milli\second}) that yields non-zero \gls{ber}.}\label{hammerdimmer:obsv:retention_smallfraction}}

{\figref{hammerdimmer:fig:vpp_retention_ber_hist}a} shows that modules from Mfr.~A do \emph{not} exhibit any bitflips when \gls{trefw} is \SI{64}{\milli\second}, while \SI{15.5}{\percent} and \SI{0.2}{\percent} of DRAM rows in modules from Mfrs.~B and~C exhibit four and one 64-{bit} words with a single bitflip, respectively{; and \SI{0.01}{\percent} of DRAM rows from Mfr.~B contain 116 data words with one bitflip.} {\figref{hammerdimmer:fig:vpp_retention_ber_hist}b} shows that \SI{0.1}{\percent}, \SI{4.7}{\percent}, and \SI{0.2}{\percent} of rows from Mfrs.~A, B, and~C contain 1, 2, and 1 erroneous {data words, respectively,}
when the refresh window is \SI{128}{\milli\second}. {We conclude that \emph{all} of these data retention bitflips can be avoided}
{by doubling the refresh rate\footnote{{We test our chips {at} fixed refresh rates in increasing powers of two (\secref{hammerdimmer:sec:experiment_design_retention}). {Therefore, our experiments do \emph{not} capture whether eliminating a bitflip is possible by increasing the}
refresh rate by less than $2\times$. We leave a finer granularity data retention time analysis to future work.}} \emph{only} for} {\SI{16.4}{\percent} / \SI{5.0}{\percent} of DRAM} rows~\cite{das2018vrldram, liu2012raidr, wang2018reducing} when \gls{trefw} is \SI{64}{\milli\second} / \SI{128}{\milli\second}.

From \obsvsref{hammerdimmer:obsv:retention_vpp}--\ref{hammerdimmer:obsv:retention_smallfraction}, we conclude that a DRAM row's {data} retention time can reduce when \gls{vpp} is reduced. However,
{1)~{most of (i.e.,} 23 out of 30{)} tested modules do \emph{not} exhibit any bitflips {at} the nominal \gls{trefw} of \SI{64}{\milli\second} and 2)~bitflips observed in seven modules can be mitigated using {existing {SECDED} ECC~\cite{hamming1950error} or selective refresh methods~\cite{das2018vrldram, liu2012raidr, wang2018reducing}}.}

\section{{Limitations of Wordline Voltage Scaling}}
\label{hammerdimmer:sec:limitations}

{We highlight \param{{four}} key limitations of \gls{vpp} scaling and our experimental characterization. 

First, in our experiments, we observe that none of the tested DRAM modules reliably operate at a \gls{vpp} lower than {a certain {voltage level, called \gls{vppmin}}}. This happens because an access transistor cannot connect the DRAM cell capacitor to the bitline when the access transistor's gate-to-source voltage difference is \emph{not} larger than the transistor's threshold voltage. Therefore, each DRAM chip has a minimum \gls{vpp} level {at} which it can reliably operate {(e.g., {lowest at} \SI{1.4}{\volt} for A0 and {highest at} {\SI{2.4}{\volt} for A5}).} 
With this limitation,} we observe \hcfirstincravg{} {/} \berdecravg{} {average} {increase /} reduction in {\gls{hcfirst} /} \gls{ber} {across all tested DRAM chips at {their respective} \gls{vppmin} {levels}. A DRAM chip's RowHammer vulnerability can potentially reduce {further} if access transistors are designed to {operate at} smaller \gls{vpp} levels.} 

Second, 
we cannot investigate the root cause of all results {we observe} since {1)~}DRAM manufacturers do \emph{not} {describe} the exact circuit design {details of} their commodity DRAM chips~\cite{patel2021enabling,saroiu2022theprice,frigo2020trrespass,patel2022acase} {{and} 2)~our infrastructure{'s} physical limitations {{prevent} us from} observing a DRAM chip's {exact internal} behavior (e.g., it is \emph{not} possible to directly measure a cell's capacitor voltage).} 

{Third, this \agy{3}{chapter} does \emph{not} thoroughly analyze the {three-way} interaction between \gls{vpp}, temperature, and RowHammer. {T}here is {already} a complex {two-way} interaction between RowHammer and temperature, requiring {studies} to test each DRAM cell at all {allowed} temperature levels~\cite{orosa2021adeeper}. Since {a three-way interaction study requires even more} characterization {that would take} several months {of} testing time, we leave it to future work to study the interaction between \gls{vpp}, temperature, and RowHammer.}

{Fourth}, we experimentally demonstrate that the RowHammer vulnerability can be mitigated by {reducing \gls{vpp} at the cost of {a} \trcdguardbandreduction{} {average} reduction in the \gls{trcd} guardband of tested DRAM chips. 
{Although reducing the guardband can hurt DRAM manufacturing yield, we leave studying \gls{vpp} reduction's effect on yield to future work because we do \emph{not} have access to DRAM manufacturers' proprietary yield statistics.}}

\section{{Key Takeaways}}
\label{hammerdimmer:sec:takeaways}

{We} summarize the key findings of our {experimental} analyses of the {\acrlong{vpp} (\gls{vpp})'s effect on the RowHammer vulnerability and reliable operation of modern DRAM chips.} 
{From our new observations, we draw two key takeaways.}

\head{Takeaway 1: Effect of \gls{vpp} on RowHammer} 
Scaling down \gls{vpp} reduces a DRAM chip's RowHammer vulnerability, such that RowHammer \gls{ber} \emph{decreases} by \berdecravg{} (up to \berdecrmax{}) and \gls{hcfirst} increases by \hcfirstincravg{} (up to \hcfirstincrmax{}) on average across all DRAM rows. Only \fracberopposingrows{} and \frachcfirstopposingrows{} of DRAM rows exhibit 
opposite \gls{ber} and \gls{hcfirst} trends, respectively (\secref{hammerdimmer:sec:vpp:ber} and \secref{hammerdimmer:sec:vpp_vs_hcfirst}).

{\head{Takeaway 2: {Effect of \gls{vpp} on} DRAM {reliability}} Reducing \gls{vpp}
{1)~reduces the guardband of row activation {latency} by \trcdguardbandreduction{} on average across tested chips}
and 2)~causes DRAM cell charge to saturate at {\SI{1}{\volt}} instead of \SI{1.2}{\volt} (\gls{vdd}) (\secref{hammerdimmer:sec:sideeffects_charge_restoration}),
leading \SI{0}{\percent}, \SI{15.5}{\percent}, and \SI{0.2}{\percent} of DRAM rows to experience {SECDED ECC-correctable} data retention {bitflips} at {the} nominal refresh window of \SI{64}{\milli\second} {in DRAM modules from Mfrs.~A, B, and~C, respectively} (\secref{hammerdimmer:sec:sideeffects_retention}).}

{{\head{{Finding} Optimal Wordline Voltage} Our {two key takeaways suggest that reducing RowHammer vulnerability of a DRAM chip {via} \gls{vpp} reduction}
can require {1)~accessing DRAM rows with a slightly larger latency, 2)~employing error correcting codes (ECC), or 3)~refreshing a small subset of rows at a higher refresh rate.}
Therefore, one can define different Pareto-optimal operating conditions for different performance and reliability requirements. For example, a security-critical system can choose a lower \gls{vpp} to reduce RowHammer vulnerability, whereas a performance-critical and error-{tolerant} system {might} prefer lower access latency over {higher RowHammer tolerance}.}} {DRAM designs and systems that are informed about the tradeoffs between \gls{vpp}, access latency, and retention time can make better{-}informed design decisions (e.g., fundamentally enable lower access latency) or employ better{-}informed memory controller policies (e.g., {using longer \gls{trcd},} {employing SECDED ECC, or} {doubling the refresh rate only for a small fraction of rows when the chip operates at} reduced \gls{vpp}).} {We believe such {designs} are important to explore in future work.} We hope that {the new insights we provide} can lead to the design of {stronger DRAM-based systems against RowHammer {along with {better-}informed DRAM{-based system} designs}.}

\section{Summary}
\label{hammerdimmer:sec:conclusion}

{{W}e present the first experimental RowHammer characterization study under reduced {\gls{vpp}}. Our results, using \numchips{} real DDR4 DRAM chips from three major manufacturers, show that {RowHammer vulnerability can be reduced by reducing} \gls{vpp}. 
Using real-device experiments and SPICE simulations, we demonstrate that although the reduced \gls{vpp} slightly worsens DRAM access latency, charge restoration process and data retention time, {most of (}\numreliablechips{} out of \numchips{}{) tested chips} reliably work under reduced \gls{vpp} {leveraging} already existing guardbands of nominal timing parameters {and employing existing ECC or selective refresh techniques}.} Our findings provide {new} insights into the increasingly {critical} RowHammer problem in modern DRAM {chips. We hope that they} lead to the design of {more robust systems against RowHammer attacks.}

\chapter[Spatial Variation-Aware Read Disturbance Defenses]{Spatial Variation-Aware Read Disturbance Defenses}
\label{chap:svard}




\newcommand{\gf}[2]{#2}
\newcommand{\yct}[2]{#2}
\newcommand{\coloredcircledletter}[2]{\tikz[baseline=(char.base)]{\node[shape=circle,inner sep=1pt,fill=#1, text=white] (char) {\footnotesize \textbf{#2}};}}
\renewcommand{\om}[2]{\ifnum#1=\value{version}\textcolor{burgundy}{#2}\else{#2}\fi}

\renewcommand{\nummodules}{\param{15}}
\renewcommand{\numchips}{\param{144}}
\renewcommand{\numworkloadmixes}{\param{120}}
\renewcommand{\wlcnt}{\numworkloadmixes{}}
\section{Motivation and Goal}
\label{svard:sec:motivation}
{Prior research experimentally demonstrates that read disturbance is {clearly a} worsening DRAM {robustness (i.e.,} reliability, security, {and safety)} {concern}~\rowHammerGetsWorseCitations{}
and many of prior {solutions}~\mitigatingRowHammerAllCitations{} will 
{incur} \emph{significant} performance, energy consumption, and hardware complexity overheads such that they become prohibitively expensive when deployed in future DRAM chips {with much larger read disturbance vulnerabilities}~\rowHammerDefenseScalingProblemsCitations{}.}

{To avoid read disturbance bitflips in future DRAM-based computing systems in an effective {and} efficient way, it is {critical} to {rigorously} gain {detailed} insights into the read disturbance phenomena 
\agy{4}{under various circumstances (e.g., the physical location of the victim row in a DRAM chip).}
{Although it might not be in the best interest of a DRAM manufacturer to make such understanding publicly available,\om{5}{\footnote{\om{5}{See~\cite{patel2022acase} for a discussion and analysis of such issues.}}} \om{4}{rigorous} research in the public domain should continue to \om{5}{enable a much more detailed and rigorous} understanding of DRAM read disturbance. This is important because a better understanding of DRAM read disturbance among the broader research community enables the development of comprehensive solutions to the problem more quickly.}
Unfortunately, despite the existing research efforts expended towards understanding read disturbance{~\understandingRowHammerAllCitations{}}, scientific literature lacks 1)~rigorous experimental observations {on the \om{4}{\emph{spatial variation \om{5}{of} read disturbance}} in \om{4}{modern} DRAM chips and 2)~a concrete methodology \om{4}{for} leveraging this variation to improve existing solutions \agy{4}{and crafting more effective attacks}.}}


{Our \emph{goal} in this \agy{3}{chapter} is {to \om{4}{close this gap\om{5}{.} We aim to empirically analyze} the spatial variation \om{4}{of} read disturbance across DRAM rows and leverage this \om{4}{analysis} to improve existing solutions}. Doing so provides us {with} a deeper understanding of {the read disturbance in DRAM chips} to enable future research {on improving the effectiveness of existing {and future solutions}. We hope \agy{4}{and expect that our} analyses will pave the way for building \agy{4}{robust (i.e., reliable, secure, and safe) systems that mitigate read disturbance at low}
performance, energy, and area overheads 
\agy{4}{while DRAM chips become more vulnerable to read disturbance over generations.}

\section{Methodology}
\label{svard:sec:methodology}

{\om{4}{We describe} our DRAM testing infrastructure and the real DDR4 DRAM chips tested.}

\subsection{DRAM Testing Infrastructure}
\label{svard:subsec:methodology_infra}

{{Fig.~\ref{deeperlook:fig:infrastructure}} shows our FPGA-based DRAM testing infrastructure for testing} real DDR4 DRAM chips. {Our infrastructure} consists of four main components: 1)~an FPGA development board (Xilinx Alveo U200~\cite{xilinxu200} for DIMMs or Bittware XUSP3S~\cite{bittware_xusp3s} for SODIMMs), programmed with DRAM Bender~\cite{olgun2023dram_bender, safari2022dram_bender} to execute our test programs, 2)~a host machine that generates the test program and collects experimental results, 3)~a thermocouple temperature sensor and a pair of heater pads pressed against the DRAM chips that {heat} up the DRAM chips to a desired temperature, and 4)~a PID temperature controller (MaxWell FT200~\cite{maxwellft20x}) that controls the heaters and keeps the temperature at the desired level with a precision of $\pm$\SI{0.5}{\celsius}.}\footnote{{To evaluate temperature stability during RowHammer tests, we perform a double-sided RowHammer test with a \agy{5}{hammer} count of 1M and traverse across all rows in round-robin fashion for 24 hours at three different temperature levels. We sample the temperature \agy{4}{of three modules (one from each manufacturer)} every 5 seconds and observe a variation within the error margin of \SI{0.2}{\celsius}, \SI{0.3}{\celsius}, and \SI{0.5}{\celsius} at \SI{35}{\celsius}, \SI{50}{\celsius}, and \SI{80}{\celsius}, respectively.}}


\head{\om{4}{Eliminating} Interference Sources}
{To observe {read disturbance induced bitflips} in circuit level,
we \om{4}{eliminate}
{perform the best effort to eliminate} 
potential sources of interference 
\om{4}{to our best ability and control,}
{by taking {four} measures,} similar \om{4}{to the} methodology used by prior works~\cite{kim2020revisiting, orosa2021adeeper, yaglikci2022understanding, hassan2021uncovering, luo2023rowpress}.
First, we disable periodic refresh during the execution of our test programs to {prevent potential on-DRAM-die {TRR} mechanisms~\cite{frigo2020trrespass, hassan2021uncovering} from refreshing victim rows so that we can} observe the DRAM chip's behavior at the circuit-level.
Second, we {strictly bound the execution time of} our test programs within {the} refresh window {of the tested DRAM chips at the tested temperature} to avoid data retention failures interfering with read disturbance failures.
{Third, we run each test ten times and record the
smallest (largest) observed \gls{hcfirst} (\gls{ber}) for each row across iterations to account for the worst-case.\footnote{{We observe a 5.7\% variation in the bit error rate across ten iterations.}}}
Fourth, we {verify} that the tested DRAM modules and chips have neither rank-level nor on-die ECC~\cite{patel2020bitexact, patel2021harp}.
{With these measures,} we directly observe and analyze all bitflips without interference.}

\subsection{Tested DDR4 DRAM Chips}
\label{svard:subsec:methodology_dramchips}
{Table~\ref{svard:tab:dram_chip_list} shows the {\numchips{} real DDR4 DRAM chips (in \nummodules{} modules)} {spanning eight different die revisions} that we test from all three major DRAM manufacturers. 
{To investigate whether our spatial variation analysis applies to different DRAM technologies, designs, and manufacturing processes,
we test various} DRAM chips with different die densities and die revisions from each DRAM chip manufacturer.}\footnote{{A DRAM chip's} technology node is {\emph{not} always} publicly available. 
We assume that two DRAM chips from the same manufacturer have the same technology node \emph{only} if they share both {1)~}the same die density and {2)~the same} die revision code. A die revision code of X indicates that there is \emph{no} public information available about the die revision (e.g., the DRAM module vendor removed the original DRAM chip manufacturer's markings, and the DRAM stepping field in the SPD is $0x00$).}

\begin{table}[h!]
  \centering
  \footnotesize
  \caption{Tested DDR4 DRAM Chips.}
    \begin{tabular}{l|lrlcc}
        &{\bf DIMM}&{\bf \# of}&{\bf Density}&{\bf Chip}&{{\bf Date}}\\
        {{\bf Mfr.}} & \textbf{ID} & {{\bf Chips}}  & {{\bf Die Rev.}}& {{\bf Org.}}& {{\bf (ww-yy)}}\\
        \hline 
        \hline 
        \multirow{3}{*}{\begin{tabular}[c]{@{}l@{}}Mfr. H \\ (SK Hynix)\end{tabular}} & H0 &  $8$ & 16Gb -- A   & x8  & 51-20 \\     
        & H1, H2, H3 &  $3\times8$ & 16Gb -- C   & x8  & 48-20   \\   
        & H4 &  $8$ &  8Gb -- D   & x8  & 48-20   \\
        \hline
                 & M0 & $4$  & 16Gb -- E   & x16 & 46-20  \\
        Mfr. M   & M1, M3 & $2\times16$ & 8Gb  -- B   & x4  & N/A   \\
        (Micron) & M2 & $16$ & 16Gb -- E   & x4  & 14-20  \\  
                 & M4 & $4$  & 16Gb -- B   & x16 & 26-21  \\
        \hline 
                  & S0, S1 & $2\times8$ & 8Gb  -- B   & x8  & 52-20  \\
        Mfr. S    & S2 &  $8$ & 8Gb  -- D   & x8  & 10-21  \\
        (Samsung) & S3 &  $8$ & 4Gb  -- F   & x8  & N/A   \\
                  & S4 & $16$ & 8Gb  -- C   & x4  & 35-21  \\
        \hline
        \hline
    \end{tabular}
    \label{svard:tab:dram_chip_list}
\end{table}

\providecommand*{\myalign}[2]{\multicolumn{1}{#1}{#2}}
\newcommand{\dimmid}[4]{\begin{tabular}[l]{@{}l@{}}#1~\cite{#2}\\#3~\cite{#4}\end{tabular}}

\agy{3}{Table~\ref{svard:tab:detailed_info} shows the characteristics of the DDR4 DRAM modules we test and analyze. We provide {the} module and chip identifiers, access frequency (Freq.), manufacturing date (Mfr. Date), {chip} density (Chip Den.), die revision {(Die Rev.)}, chip organization {(Chip Org.), and the number of rows per DRAM bank} of tested DRAM modules. We report the manufacturing date of these modules in the form of $week-year$. For each DRAM module, Table~\ref{svard:tab:detailed_info} shows the minimum (Min.), average (Avg.), and maximum (Max.) \gls{hcfirst} values across all tested rows.}
\renewcommand{\arraystretch}{1.2}
\begin{table*}[ht]
\footnotesize
\centering
\caption{Characteristics of the tested DDR4 DRAM modules.}
\label{svard:tab:detailed_info}
\resizebox{\linewidth}{!}{
\begin{tabular}{|l|l|l||c|cccc|c|ccc|}
\hline
\head{Label} & \textbf{Mfr.} & \textbf{Module Identifier} & \textbf{Freq} & \textbf{Mfr. Date} & \textbf{Chip} & \textbf{Die} & \textbf{Chip} & \textbf{\# of Rows} & \multicolumn{3}{c|}{\gls{hcfirst}}\\ 
&&\textbf{Chip Identifier}&\textbf{(MT/s)}&\textbf{ww-yy}&\textbf{Den.}&\textbf{Rev.}&\textbf{Org.}&\textbf{per Bank}&\textbf{Min.}&\textbf{Avg.}&\textbf{Max.}\\
\hline
\hline
H0 & \multirow{10}{*}{\rotcell{\begin{tabular}[c]{@{}c@{}}\textbf{SK Hynix}\end{tabular}}} & \dimmid{HMAA4GU6AJR8N-XN}{memorynethmaa4gu6ajr8nxn}{H5ANAG8NAJR-XN}{hynixh5anag8najrxn} & 3200 & 51-20 & 16Gb & A & $\times8$ & 128K & 16K & 46.2K & 96K\\
\cline{1-1}\cline{3-12}
H1 & & {\dimmid{HMAA4GU7CJR8N-XN}{memorynethmaa4gu7cjr8nxn}{H5ANAG8NCJR-XN}{hynixh5anag8ncjrxn}} & 3200 & 51-20& 16Gb & C & $\times8$ & 128K & 12K & 54.0K & 128K\\
\cline{1-1}\cline{3-12}
H2 &  & {\dimmid{HMAA4GU7CJR8N-XN}{memorynethmaa4gu7cjr8nxn}{H5ANAG8NCJR-XN}{hynixh5anag8ncjrxn}} & 3200 & 36-21 & 16Gb & C & $\times8$ & 128K & 12K & 55.4K & 128K\\
\cline{1-1}\cline{3-12}
H3 & & {\dimmid{HMAA4GU7CJR8N-XN}{memorynethmaa4gu7cjr8nxn}{H5ANAG8NCJR-XN}{hynixh5anag8ncjrxn}} & 3200 & 36-21 & 16Gb & C & $\times8$ & 128K & 12K & 57.8K & 128K\\
\cline{1-1}\cline{3-12}
H4 & & \dimmid{KSM32RD8/16HDR}{kingston2020ksm32rd816hdr}{H5AN8G8NDJR-XNC}{hynixh5an8g8ndjrxnc} & 3200 & 48-20 & 8Gb & D & $\times8$ & 64K & 16K & 38.1K & 96K \\
\hline
\hline
M0 & \multirow{10}{*}{\rotcell{\begin{tabular}[c]{@{}c@{}}\textbf{Micron}\end{tabular}}} & \dimmid{MTA4ATF1G64HZ-3G2E1}{micronmta4atf1g64hz3g2e1}{MT40A1G16KD-062E}{micronmt40a1g16kd062e} & 3200 & 46-20 & 16Gb & E & $\times16$ & 128K & 8K & 24.5K & 40K\\
\cline{1-1}\cline{3-12}
M1 &  & \dimmid{MTA18ASF2G72PZ-2G3B1QK}{micronmta18asf2g72pz2g3b1qk}{MT40A2G4WE-083E:B}{micronmt40a2g4we083eb} & 2400 & N/A & 8Gb & B & $\times4$ & 128K & 40K & 64.5K & 96K\\
\cline{1-1}\cline{3-12}
M2 & & \dimmid{MTA36ASF8G72PZ-2G9E1TI}{micronmta36asf8g72pz2g9e1ti}{MT40A4G4JC-062E:E}{micronmt40a4g4jc062ee} & 2933 &  14-20 & 16Gb & E & $\times4$ & 128K & 8K & 28.6K & 48K \\
\cline{1-1}\cline{3-12}
M3 & & \dimmid{MTA18ASF2G72PZ-2G3B1QK}{micronmta18asf2g72pz2g3b1qk}{MT40A2G4WE-083E:B}{micronmt40a2g4we083eb} & 2400 & 36-21 & 8Gb & B & $\times4$ & 128K & 56K & 90.0K & 128K\\
\cline{1-1}\cline{3-12}
M4 & & \dimmid{MTA4ATF1G64HZ-3G2B2}{micronmta4atf1g64hz3g2b2}{MT40A1G16RC-062E:B}{micronmt40a1g16rc062eb} & 3200 & 26-21 & 16Gb & B & $\times16$ & 128K & 12K & 42.2K & 96K\\
\hline 
\hline
S0 & \multirow{10}{*}{\rotcell{\begin{tabular}[c]{@{}c@{}}\textbf{Samsung}\end{tabular}}} & {\dimmid{M393A1K43BB1-CTD}{samsungm393a1k43bb1ctd}{K4A8G085WB-BCTD}{samsungk4a8g085wbbctd}} & {2666} & {52-20} & {8Gb} & {B} & {$\times8$} & 64K & 32K & 57.0K & 128K \\
\cline{1-1}\cline{3-12}
S1 & & {\dimmid{M393A1K43BB1-CTD}{samsungm393a1k43bb1ctd}{K4A8G085WB-BCTD}{samsungk4a8g085wbbctd}} & 2666 & 52-20 & 8Gb & B & $\times8$ & 64K & 24K & 59.8K & 128K\\
\cline{1-1}\cline{3-12}
S2 & & {\dimmid{M393A1K43BB1-CTD}{samsungm393a1k43bb1ctd}{K4A8G085WB-BCTD}{samsungk4a8g085wbbctd}} & 2666 & 10-21 & 8Gb & D & $\times8$ & 64K & 12K & 42.7K & 96K\\
\cline{1-1}\cline{3-12}
S3 & & \dimmid{F4-2400C17S-8GNT}{gskill2021f42400c17s8gnt}{K4A4G085WF-BCTD}{samsung2021k4a4g085wfbctd} & 2400 & 04-21 & 4Gb & F & $\times8$ & 32K & 16K & 59.2K & 128K\\
\cline{1-1}\cline{3-12}
S4 & & \dimmid{M393A2K40CB2-CTD}{samsungm393a2k40cb2ctd}{K4A8G045WC-BCTD}{samsungk4a8g045wcbctd} & 2666 & 35-21 & 8Gb & C & $\times4$ & 128K & 12K & 55.4K & 128K\\
\hline 
\end{tabular}
}

\end{table*}

{To account for in-DRAM row address mapping~\cite{kim2014flipping, smith1981laser, horiguchi1997redundancy, keeth2001dram_circuit, itoh2001vlsi, liu2013anexperimental,seshadri2015gatherscatter, khan2016parbor, khan2017detecting, lee2017designinduced, tatar2018defeating, barenghi2018softwareonly, cojocar2020arewe,  patel2020bitexact}, we reverse engineer the physical row address layout, following the prior works' methodology~\cite{kim2020revisiting, orosa2021adeeper, yaglikci2022understanding, luo2023rowpress}.}

\subsection{DRAM Testing Methodology}

\head{Metrics}
{To characterize {a DRAM module's} vulnerability to read disturbance, we examine {the change in two metrics: \agy{5}{1)~\glsfirst{hcfirst}, where we count
each pair of activations to the two neighboring rows as
one hammer (e.g., one activation each to rows N – 1 and
N +1 counts as one hammer)~\cite{kim2020revisiting}, and 2)~\glsfirst{ber}}. A higher \gls{hcfirst} \om{5}{(\gls{ber}) indicates lower (higher)} vulnerability to read disturbance.}

\head{Tests}
Alg.~\ref{svard:alg:test_alg} describes our core test loop \agy{4}{and two key functions we use: $hammer\_doublesided$ and $measure\_BER$}. 
{All our tests use \om{4}{the} double-sided hammering pattern \agy{4}{as specified in $hammer\_doublesided$ function and performed similarly by prior works}~\cite{kim2014flipping, kim2020revisiting, seaborn2015exploiting, orosa2021adeeper, luo2023rowpress}. \agy{4}{$hammer\_doublesided$} {hammers two physically adjacent (i.e., aggressor) rows to a victim row \agy{4}{($RA_{victim}\pm1$)}} in an alternating manner. {In this context, one hammer is a pair of activations} to the two aggressor rows. \agy{4}{The $HC$ parameter in Alg.~\ref{svard:alg:test_alg} defines the \agy{5}{hammer} count, i.e., the number of activations per aggressor row. The \gls{taggon} parameter in Alg.~\ref{svard:alg:test_alg} defines the time an activated aggressor row remains open.}
We perform \agy{4}{the} double-sided {hammering} {in two different ways: 1)~}with the maximum activation rate possible within DDR4 command timing {specifications}~\cite{jedec2020jesd794c,jedec2012jesd793f} as this access pattern is stated as the most effective RowHammer {access pattern} on DRAM chips when {RowHammer solutions} are disabled~\cite{kim2014flipping, kim2020revisiting, frigo2020trrespass, cojocar2020arewe, seaborn2015exploiting, orosa2021adeeper, olgun2023anexperimental}{; and 2)~with keeping aggressor rows open for longer \om{4}{than the minimum charge restoration time (\gls{taggon} > \gls{tras})} at each activation to \agy{4}{observe the effect of} RowPress, a recently demonstrated read disturbance phenomenon, \om{5}{which is different from} \agy{4}{RowHammer}~\cite{luo2023rowpress}.}}
\agy{4}{As $measure\_BER$ function (Alg.~\ref{svard:alg:test_alg}) demonstrates, we initialize 1)~two aggressor rows and one victim row with opposite data patterns to \agy{5}{exacerbate read disturbance}~\cite{kim2020revisiting, kwong2020rambleed, cojocar2019exploiting, ji2019pinpoint}, 2)~perform double-sided hammer test, and 3)~read-back the data from the victim row and compare against the victim row's initial data pattern to calculate the bit error rate (BER).}
\agy{4}{Our core test loop sweeps different \gls{taggon} values, banks, and victim row addresses. 
First, We test three different \gls{taggon} values: 1)~\SI{36}{\nano\second} as the minimum \gls{tras} value, 2)~\SI{2}{\micro\second} as a large enough time window in which a streaming access pattern can fetch the whole content in the activated aggressor row, and 3)~\SI{0.5}{\micro\second} as a more realistic time window at which a DRAM row can remain open due to high row buffer hit rate~\cite{ghose2019demystifying, luo2023rowpress}.
Second, we sweep through banks 1, 4, 10, and 15 as representative banks from each bank group~\cite{jedec2020jesd794c, olgun2023dram_bender, safari2022dram_bender}. 
Third, we test \emph{all} rows in a tested bank using 14 different \agy{5}{hammer} counts and six different data patterns.}

\SetAlFnt{\footnotesize}
\RestyleAlgo{ruled}
\begin{algorithm}
\caption{Test for profiling the spatial variation of read disturbance in DRAM}\label{svard:alg:test_alg}
  \DontPrintSemicolon
  \SetKwFunction{FVPP}{set\_vpp}
  \SetKwFunction{FMain}{test\_loop}
  \SetKwFunction{FHammer}{measure\_$BER$}
  \SetKwFunction{initialize}{initialize\_row}
  \SetKwFunction{initializeaggr}{initialize\_aggressor\_rows}
  \SetKwFunction{measureber}{measure\_BER}
  \SetKwFunction{FMeasureHCfirst}{measure\_\gls{hcfirst}}
  \SetKwFunction{compare}{compare\_data}
  \SetKwFunction{Hammer}{hammer\_doublesided}
  \SetKwFunction{Gaggressors}{get\_aggressors}
  \SetKwFunction{FWCDP}{get\_WCDP}
  \SetKwProg{Fn}{Function}{:}{}

  \tcp{$RA_{victim}$: Victim row address}
  \tcp{$WCDP$: Worst-case data pattern \om{4}{for the victim row}}
  \tcp{$HC$: \agy{5}{Hammer Count: number of activations} per aggressor row}
  \tcp{$ACT$: Row activation command to open a DRAM row}
  \tcp{$PRE$: Precharge command to close a DRAM row}
  \tcp{$WAIT$: Wait for the specified amount of time}
  \tcp{$t_{AggOn}$: Aggressor row on time}
  \tcp{$t_{RP}$: Precharge latency timing constraint}
  \Fn{\Hammer{{$RA_{victim}$}, $HC$, $t_{AggOn}$}}{
        \While{$i < HC$}{
            ACT({$RA_{victim}+1$})\; WAIT({$t_{AggOn}$})\;
            PRE({})\; WAIT({$t_{RP}$})\;
            ACT({$RA_{victim}-1$})\; WAIT({$t_{AggOn}$})\;
            PRE({})\; WAIT({$t_{RP}$})\;
            i++\;
        }
  }\;

  \Fn{\FHammer{$RA_{victim}$, $WCDP$, $HC$, $t_{AggOn}$}}{
        \initialize($RA_{victim}$, $WCDP$)\;
        {\initializeaggr($RA_{victim}$, {bitwise\_inverse($WCDP$)})}\;
        \Hammer({$RA_{victim}$}, $HC$, $t_{AggOn}$)\;
        $BER{_{row}} =$ \compare($RA_{victim}$, $WCDP$)\;
        \KwRet $BER{_{row}}$\;
  }\;

  \Fn{\FMain{}}{
    \ForEach{$t_{AggOn}$ in [36ns, 0.5us, 2us]}{
        \ForEach{$Bank$ in [1, 4, 10, 15]}{
            \ForEach{$RA_{victim}$ in $Bank$}{
                \tcp{Find the worst-case data pattern}
                \ForEach{$DP$ in [RS, RSI, CS, CSI, CB, CBI]}{
                    \FHammer{$RA_{victim}$, $DP$, 128K, $t_{AggOn}$}\;
                    WCDP = DP that causes largest BER\;
                }
    
                \tcp{Sweep the hammer count using WCDP}
                \ForEach{$HC$ in [1,2,4,8,12,16,24,32,40,48,56,64,96]K}{
                    \FHammer{$RA_{victim}$, $WCDP$, $HC$, $t_{AggOn}$}\;
                }    
            }
        }
    }
  }
\end{algorithm}

\head{Hammer Counts} We conduct our tests by using a set of \agy{5}{hammer} counts on all DRAM rows instead of finding \gls{hcfirst} precisely for each row. \agy{4}{This is because}
\gls{hcfirst} significantly varies across rows, and thus, causes a large experiment time (e.g., several weeks or even months) to find \gls{hcfirst} at high precision (e.g., \om{4}{within} $\pm 10$ hammers) for each row individually. 
Therefore, we test the DRAM chips under 14 distinct hammer counts from $1K$ to $128K$ as Alg.~\ref{svard:alg:test_alg} specifies.\footnote{{$K$ is $2^{10}$ (\emph{not} $10^3$) unless otherwise specified.}}  

\head{{Data Patterns}} {{We use} six commonly used data patterns~\cite{chang2016understanding,chang2017understanding,khan2014theefficacy,khan2016parbor,khan2016acase,kim2020revisiting,lee2017designinduced,mukhanov2020dstress,orosa2021adeeper, kim2014flipping, liu2013anexperimental}: {row stripe, checkerboard, column stripe, and the opposites of these three data patterns that are shown in \tabref{svard:tab:data_patterns} in detail}. We identify the worst-case data pattern ($WCDP$) for each row {as the data pattern that results in the largest \gls{ber} at the \agy{5}{hammer} count of $128K$}.\footnote{We find that a \agy{5}{hammer} count of 128K is both 1)~low enough to be used in a system-level attack in a real system~\cite{frigo2020trrespass}, and 2)~high enough to provide a large number of bitflips in \emph{all} DRAM modules we tested{.}}}
\agy{4}{Then, we sweep the \agy{5}{hammer} count from 1K to 96K and measure \gls{ber} for the WCDP of each row.}

\begin{table}[h]
    \centering
    \footnotesize
    \caption{Data patterns used in our tests}
    \begin{tabular}{l|cc}
     \bf{Data Pattern}   & \bf{{Aggressor Rows}} & \bf{{Victim Row}}\\
    \hline
    \hline
    {Row Stripe (RS)}         & $0xFF$ & $0x00$ \\
    {Row Stripe Inverse (RSI)} & $0x00$ & $0xFF$ \\
    \hline
    {Column Stripe (CS)}         & $0xAA$ & $0xAA$ \\
    {Column Stripe Inverse (CSI)} & $0x55$ & $0x55$ \\
    \hline
    {Checkerboard (CB)}         & $0xAA$ & $0x55$ \\
    {Checkerboard Inverse (CBI)} & $0x55$ & $0xAA$ \\
    \hline
    \hline
    
    \end{tabular}
    \label{svard:tab:data_patterns}
\end{table}


\head{Finding Physically Adjacent Rows} {DRAM-internal address mapping schemes~\cite{cojocar2020arewe, kim2012acase} are used by DRAM manufacturers to translate {\emph{logical}} DRAM addresses (e.g., row, bank, {and} column) that are exposed over the DRAM interface (to the memory controller) to physical {DRAM} addresses {(e.g., physical location of a row)}. {Internal address mapping schemes allow 
{1)~}post-manufacturing row repair techniques to repair erroneous DRAM rows by remapping \om{4}{such} rows to spare rows and 
{2)~}DRAM manufacturers organize DRAM internals in a cost-optimized way, e.g., by organizing internal DRAM buffers hierarchically~\cite{khan2016parbor,vandegoor2002address}.} The mapping scheme can substantially vary across different DRAM {chips}~\cite{barenghi2018softwareonly,cojocar2020arewe,horiguchi1997redundancy,itoh2001vlsi,keeth2001dram_circuit,khan2016parbor,khan2017detecting,kim2014flipping,lee2017designinduced,liu2013anexperimental,patel2020bitexact,orosa2021adeeper,saroiu2022theprice,patel2022acase}. For every row, we identify the two {neighboring physically-adjacent} DRAM row addresses that the memory controller can use to access the {aggressor} rows in a double-sided RowHammer attack. To do so, we reverse-engineer the physical row organization {using} techniques described in prior work{s}~\cite{kim2020revisiting, orosa2021adeeper}.}

\head{Temperature} We maintain the DRAM chip temperature at \SI{80}{\celsius}, \om{4}{which is} very close to {the maximum point of the} normal operating condition of \SI{85}{\celsius}~\cite{jedec2020jesd794c}. {We choose this temperature because prior works show that increasing temperature tends to reduce DRAM chips' overall reliability~\cite{liu2013anexperimental, orosa2021adeeper, orosa2022spyhammer, luo2023rowpress}.}{\footnote{{Prior works~\cite{orosa2021adeeper, luo2023rowpress} \om{4}{demonstrate} a complex interaction between temperature and a row's read disturbance \om{4}{(especially RowHammer)} vulnerability and suggest that each DRAM chip should be tested at all temperature levels to account for the effect of temperature. Thus, fully understanding the effects of temperature and aging requires extensive characterization studies, requiring many months-long testing time. Therefore, we leave such studies for future work.}}} {Due to time and space limitations, we leave a rigorous characterization of temperature's effect for future work, while presenting the preliminary analysis \agy{4}{where}} {we repeat double-sided RowHammer tests at \SI{50}{\celsius} on 5K randomly selected DRAM rows at nine different \agy{5}{hammer} counts. We observe that the variation in overall \gls{ber} with the effect of temperature is less than 0.5\%.}

\section{Spatial Variation in DRAM Read Disturbance}
\label{svard:sec:characterization}
{This section presents the first {rigorous} {spatial variation} analysis of read disturbance across DRAM rows. Many prior works~\cite{kim2014flipping, kim2020revisiting, park2014activeprecharge, park2016experiments, park2016statistical} analyze RowHammer vulnerability {at the} DRAM bank granularity across many DRAM modules without {providing analysis of} the variation of this vulnerability across rows.} {Recent works~\cite{orosa2021adeeper, yaglikci2022understanding, luo2023rowpress, olgun2023anexperimental} analyze the variation in RowHammer vulnerability across DRAM rows. However, these analyses are limited to a small subset of DRAM rows (4K to 9K), while a DRAM bank typically has $>16K$ DRAM rows~\cite{jedec2020jesd794c, samsungm393a2k40cb2ctd, samsungk4a8g085wbbctd, hynixh5anag8ncjrxn, hynixh5an8g8ndjrxnc, micronmta4atf1g64hz3g2e1, micronmt40a1g16kd062e, micronmta18asf2g72pz2g3b1qk, micronmta36asf8g72pz2g9e1ti, micronmta4atf1g64hz3g2b2}. Thus, \om{4}{prior} works might \emph{not} fully reflect the vulnerability profile of real DRAM chips. Fully \om{4}{characterizing and understanding} the vulnerability profile is crucial to \om{4}{avoiding} read disturbance bitflips as existing \om{4}{read disturbance} solutions must be properly configured \om{4}{based on proper characterization}~\cite{kim2014flipping, kim2014architectural, yaglikci2021blockhammer, park2020graphene, qureshi2022hydra, saxena2022aqua, saileshwar2022randomized, bostanci2024comet, olgun2024abacus}.} This section presents a more rigorous and targeted read disturbance characterization study of \numchips{} real DDR4 DRAM chips {spanning \om{5}{\param{11}} different \agy{4}{chip density and} die revisions}, following the methodology described in \secref{svard:sec:methodology}.

\subsection{Bit Error Rate Across DRAM Rows}
\label{svard:sec:char_ber}
We investigate the variation in the number of bitflips caused by read disturbance for \om{4}{a} \agy{5}{hammer} count of 128K and \gls{taggon} of \SI{36}{\nano\second}. \figref{svard:fig:ber_hist} shows the distribution of observed \gls{ber} for each DRAM row across all tested DRAM banks and modules from three main manufacturers in a box-and-whiskers plot.\footnote{\label{svard:fn:boxplot}{{The box is lower-bounded by the first quartile (i.e., the median of the first half of the ordered set of data points) and upper-bounded by the third quartile (i.e., the median of the second half of the ordered set of data points).
The \gls{iqr} is the distance between the first and third quartiles (i.e., box size).
Whiskers mark the central 1.5\gls{iqr} range, and white circles show the mean values.}}} Each of the three rows of subplots is dedicated to modules from a different manufacturer, and each subplot shows data from a different DRAM module. The x-axis shows the bank address and the y-axis shows the \glsfirst{ber}. \agy{4}{We annotate each module's name and variation across rows and banks in terms of the coefficient of variation (CV)\footnote{Coefficient of variation is the standard deviation of a distribution, normalized to the mean~\cite{faber2012statistics, biran1998thecambridge}.} at the bottom of each subplot.} We make Obsvs.~\ref{svard:obsv:ber_across_rows}-\ref{svard:obsv:ber_across_modules} from \figref{svard:fig:ber_hist}.

\begin{figure}[!ht]
    \centering
    \includegraphics[width=\linewidth]{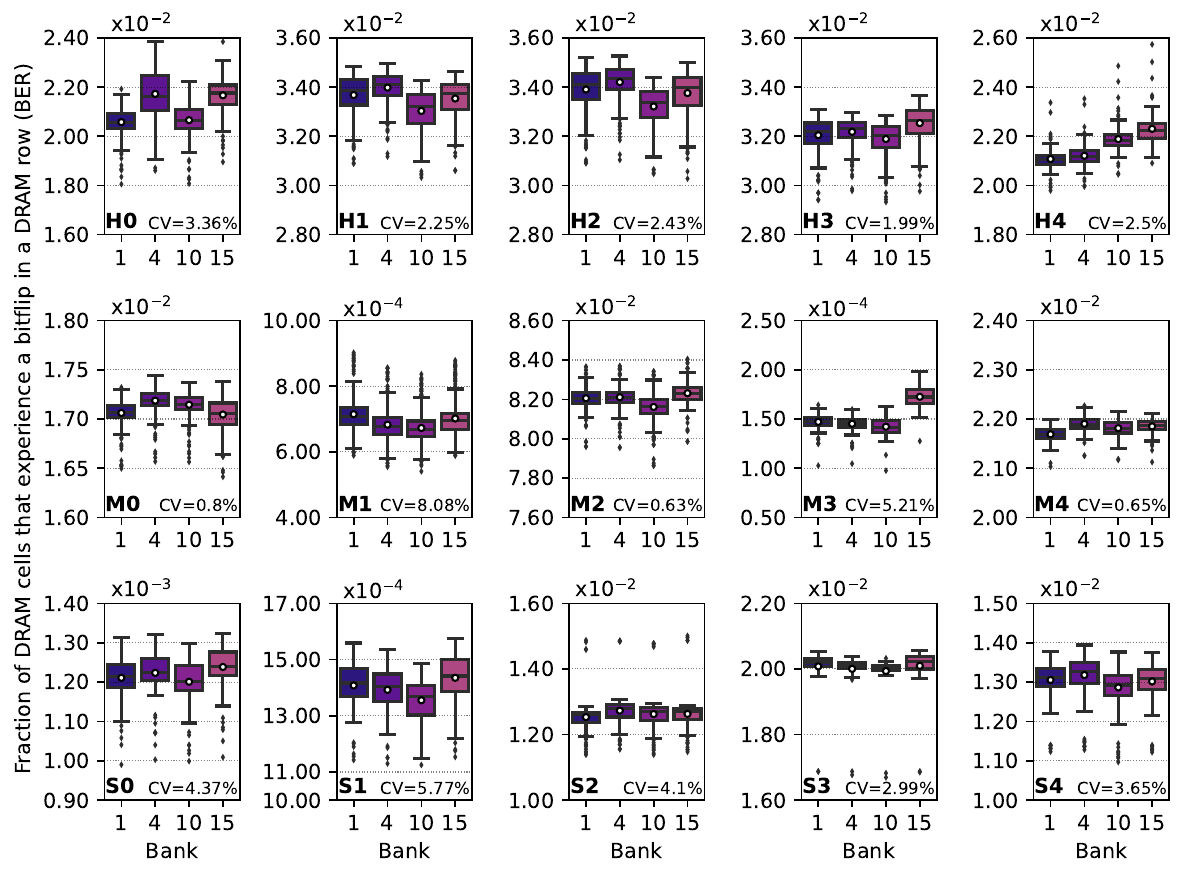}
    \caption{Distribution of $BER$ across DRAM rows and bank groups}
    \label{svard:fig:ber_hist}
\end{figure}

\observation{\gls{ber} varies across DRAM rows in a DRAM module.\label{svard:obsv:ber_across_rows}}
For example, DRAM rows in \om{4}{modules} M1 and S1 exhibit coefficient of variations (CV) of 8.08\% and 5.77\%, respectively, on average across all tested banks. 

\observation{Different banks within the same DRAM module exhibit similar \gls{ber} to each other.\label{svard:obsv:ber_across_banks}} 
As the box plots for different banks largely overlap with each other in the y-axis, we observe a smaller variation in \gls{ber} across banks compared to across rows in a bank for all tested modules except H4 and M3. 
For example, the average (minimum/maximum) \gls{ber} across all DRAM rows in four different banks of M0 are 1.71\% (1.65\%/1.73\%), 1.71\% (1.66\%/1.74\%), 1.70\% (1.64\%/1.74\%), and 1.72\% (1.66\%/1.74\%).



\observation{\gls{ber} can significantly vary across different DRAM modules from the same manufacturer.\label{svard:obsv:ber_across_modules}}
For example, modules M0, M1, and \agy{4}{M3} show \gls{ber} distributions that \agy{4}{are strictly distinct from each other}.
From Obsvs.~\ref{svard:obsv:ber_across_rows}-\ref{svard:obsv:ber_across_modules}, we draw Takeaway~\ref{svard:take:ber_across_everything}. 
\takebox{\gls{ber} significantly varies across different DRAM rows within a bank and across different modules, while different banks in a DRAM module exhibit similar \gls{ber} distributions \om{4}{to} each other.\label{svard:take:ber_across_everything}} 

To understand the spatial variation of rows with high and low \gls{ber}s, we analyze their locations within their banks. \figref{svard:fig:ber_vs_row} shows \agy{4}{how \gls{ber} varies} as the row address increases. \agy{4}{The x-axis shows a DRAM row's relative location in its bank, where 0.0 and 1.0 are the two edges of a DRAM bank. The y-axis shows the \gls{ber} the corresponding DRAM row experiences at a \agy{5}{hammer} count of 128K, normalized to the minimum \gls{ber} observed across all rows in all tested banks in a module.} Each subplot is dedicated to a different manufacturer, and each curve represents a different DRAM module. The shades around the curves show the minimum and maximum values for a given row address across different DRAM banks in a module. 
We make Obsvs.~\ref{svard:obsv:ber_row_repeating_patterns} and~\ref{svard:obsv:ber_across_chunks} from \figref{svard:fig:ber_vs_row}.

\begin{figure}[!ht]
    \centering
    \includegraphics[width=0.8\linewidth]{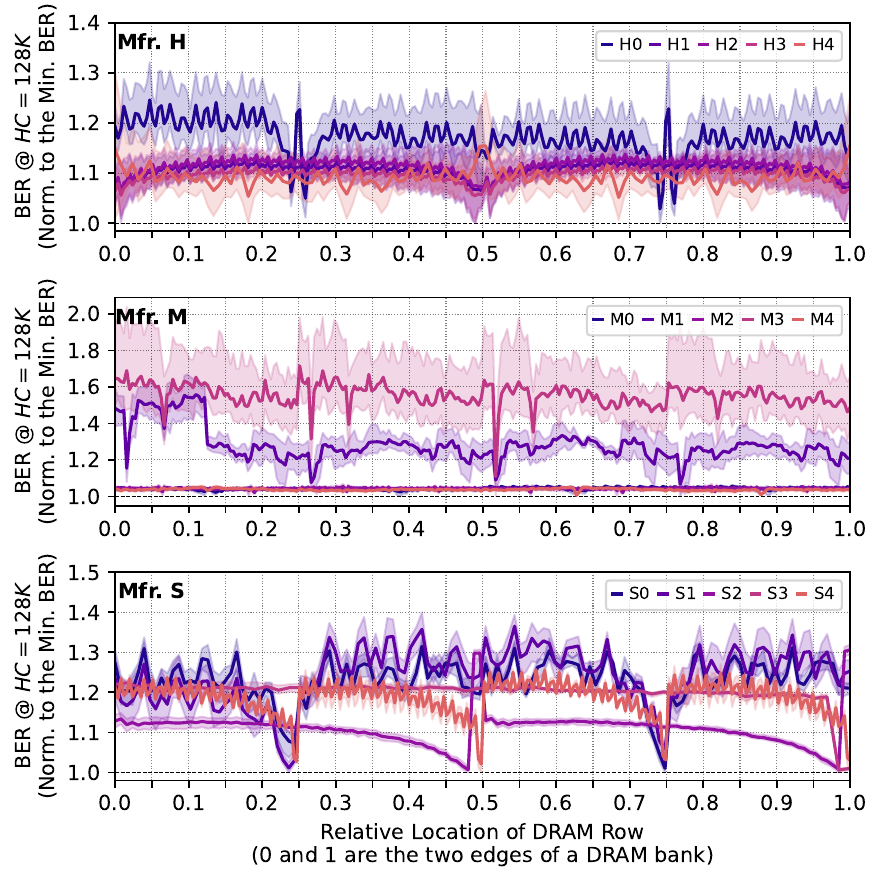}
    \caption{Distribution of $BER$ across DRAM rows}
    \label{svard:fig:ber_vs_row}
\end{figure}

\observation{\gls{ber} repeatedly increases and decreases with \agy{4}{different intervals of row distances in different DRAM modules}.\label{svard:obsv:ber_row_repeating_patterns}}
 For example, \gls{ber} curve of S4 follows a repeatedly increasing and decreasing pattern across all rows, where it shows local minimums at 0.25, 0.50, 0.75, and 1.00.
\agy{4}{We hypothesize that this regularity in \gls{ber} variation can be caused by design decisions (design-induced variation), e.g., row's distance from subarray boundaries and I/O circuitry, as discussed by prior works~\cite{lee2017designinduced, orosa2021adeeper, olgun2023anexperimental}.}

\observation{Average \gls{ber} can vary across large chunks of a DRAM bank. \label{svard:obsv:ber_across_chunks}}
For example, the average (minimum/maximum) normalized \gls{ber} \agy{4}{in the module M1} across DRAM rows between \om{4}{relative locations} 0.03 and 0.12 is 1.51 (1.31 /1.67 ) while it is 1.25 (1.00 /1.42 ) between \om{4}{relative locations} 0.20 and 1.00.
This \agy{4}{discrepancy in \gls{ber} across large chunks of rows} does \emph{not} consistently occur across all tested modules. 
Understanding the root cause of this discrepancy 
requires extensive knowledge and insights into the circuit design and manufacturing process of the particular DRAM modules exhibiting this behavior. Unfortunately, this piece of information is proprietary and \emph{not} publicly disclosed by the manufacturers. \agy{4}{We hypothesize that the root cause of this discrepancy can be the variation in the manufacturing process, leading to a part of DRAM chip being more vulnerable to read disturbance compared to other parts.}
From Obsvs.~\ref{svard:obsv:ber_row_repeating_patterns} and~\ref{svard:obsv:ber_across_chunks}, we derive Takeaway~\ref{svard:take:ber_regular}.

\takebox{\agy{4}{\gls{ber} values in a DRAM bank exhibit repeating patterns as DRAM row address increases}, and certain chunks of rows can exhibit higher \gls{ber} than the rest of the rows.\label{svard:take:ber_regular}}

\subsection{Minimum Activation Count to Induce a Bitflip}
\label{svard:sec:char_hcfirst}

We investigate the variation in \gls{hcfirst} across DRAM rows.
To do so, we repeat our tests at \param{14} different \agy{5}{hammer} counts from 1K to 128K (Algorithm~\ref{svard:alg:test_alg}). We define \agy{6}{a row's \gls{hcfirst}} as the minimum of the tested \agy{5}{hammer} counts at which the row experiences a bitflip. \figref{svard:fig:hcfirst_hist} shows the distribution of \gls{hcfirst} values across rows. Each subplot shows the distribution for a different manufacturer. The x- and y-axes show the \gls{hcfirst} values and the fraction of the DRAM rows with the specified \gls{hcfirst} value, respectively. Different colors represent different modules. The error bars mark the minimum/maximum of a given value across tested banks. Red vertical dashed line marks the minimum \gls{hcfirst} that we observe across all rows in tested modules from a manufacturer. We make Obsvs.~\ref{svard:obsv:hcfirst_rows}--\ref{svard:obsv:hcfirst_modules} \agy{4}{from \figref{svard:fig:hcfirst_hist}}.

\begin{figure}[h!]
    \centering
    \includegraphics[width=0.85\linewidth]{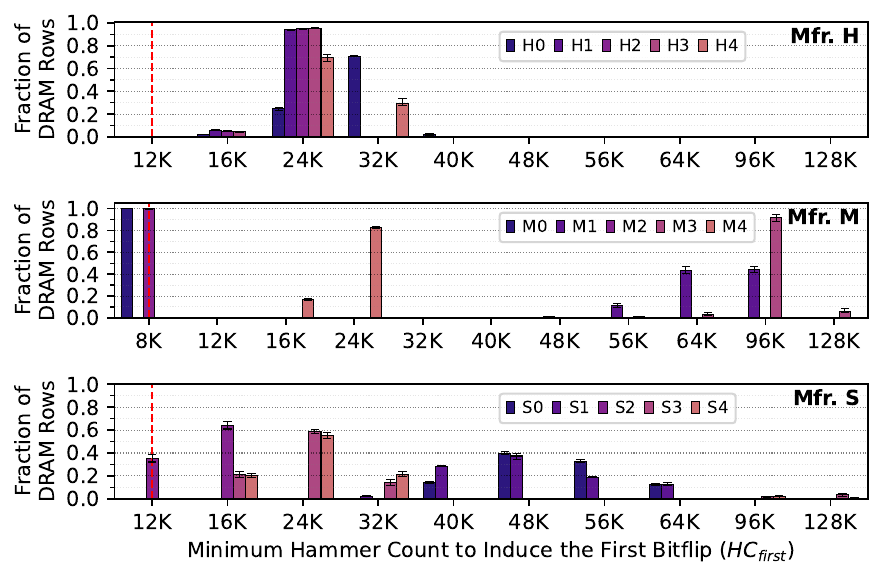}
    \caption{Distribution of $HC_{first}$ across DRAM rows}
    \label{svard:fig:hcfirst_hist}
\end{figure}

\observation{\gls{hcfirst} values significantly vary across DRAM rows but not across banks.\label{svard:obsv:hcfirst_rows}} For example, S0 and S1 contain rows that experience bitflips \agy{4}{at \agy{5}{hammer} counts of 32K and 24K, respectively, while they also have rows} that do not experience bitflips until 128K. Despite this large variation, the variation across banks is significantly low, as error bars show.


\observation{Different DRAM modules from the same manufacturer can exhibit significantly different \gls{hcfirst} distributions.\label{svard:obsv:hcfirst_modules}} For example, rows from M0 and M4 exhibit \gls{hcfirst} values from 8K to 40K and 12K to 96K, respectively. 

\takebox{\gls{hcfirst} varies \om{4}{significantly} across different DRAM rows within a DRAM bank and across different DRAM modules, while different banks in a DRAM module exhibit similar \gls{hcfirst} distributions with each other.\label{svard:take:hcfirst_across_everything}} 

To understand the spatial variation in \gls{hcfirst} across rows, we investigate \agy{4}{how a row's \gls{hcfirst} changes with the row's location within the DRAM bank. \figref{svard:fig:hcfirst_vs_row} shows the row's relative location on the x-axis and its \gls{hcfirst}, normalized to the minimum \gls{hcfirst} observed in the corresponding module.}
Each subplot corresponds to a different manufacturer, and different modules are color-coded. We make Obsvs.~\ref{svard:obsv:hcfirst_everything_everywhere} and \ref{svard:obsv:hcfirst_row_repeating_patterns} from \figref{svard:fig:hcfirst_vs_row}.
    \begin{figure}[h!]
        \centering
        \includegraphics[width=0.85\linewidth]{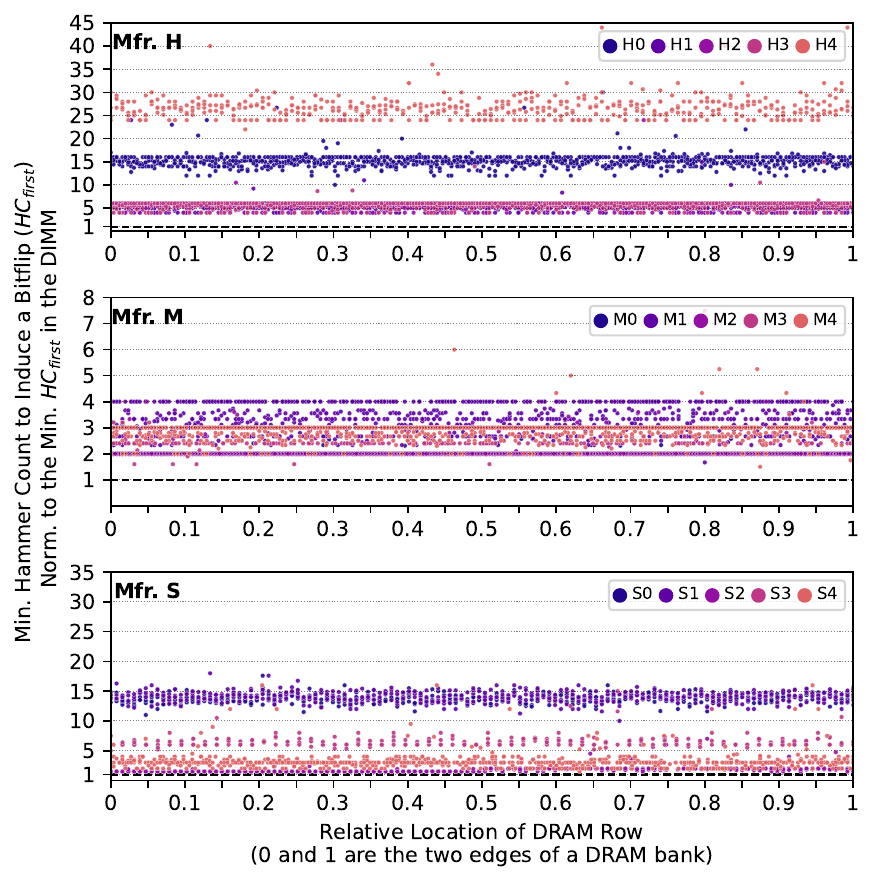}
        \caption{Distribution of $HC_{first}$ across DRAM rows}
        \label{svard:fig:hcfirst_vs_row}
    \end{figure}

\observation{\agy{4}{\gls{hcfirst} values vary significantly across rows.}\label{svard:obsv:hcfirst_everything_everywhere}}
For example, the \agy{4}{module H0 exhibits \gls{hcfirst} values that are between $8\times$ and $20\times$ the minimum \gls{hcfirst} observed in the bank between relative row addresses 0.02 and 0.03.}

\observation{Variation in \gls{hcfirst} does not exhibit a \agy{4}{regular} trend as the row address increases.\label{svard:obsv:hcfirst_row_repeating_patterns}} \agy{4}{For example, the data points of modules H4 \agy{5}{concentrate at the y-axis values of $24\times$ and $32\times$ across \emph{all} rows in a bank with \emph{no} regular transition pattern across them}.} \om{4}{This observation \om{5}{is discrepant with} Obsv.~\ref{svard:obsv:ber_row_repeating_patterns} we have for \gls{ber}.} \agy{4}{\om{5}{The} discrepancy across Obsvs.~\ref{svard:obsv:ber_row_repeating_patterns} and~\ref{svard:obsv:hcfirst_row_repeating_patterns} shows that although read disturbance vulnerability varies regularly across rows in terms of the fraction of DRAM cells experiencing bitflips, \agy{5}{the \gls{hcfirst} values across the weakest DRAM cells do \emph{not} exhibit such a regular variation pattern}.} 
From Obsvs.~\ref{svard:obsv:hcfirst_everything_everywhere} and \ref{svard:obsv:hcfirst_row_repeating_patterns}, we derive Takeaway~\ref{svard:take:hcfirst_across_rows}.

\takebox{\gls{hcfirst} varies \agy{4}{significantly and irregularly across rows and banks in a DRAM module.}\label{svard:take:hcfirst_across_rows}}

\subsection{Effect of RowPress}
\label{svard:sec:char_access_pattern}
We analyze the effect of the recently discovered read disturbance phenomenon, RowPress~\cite{luo2023rowpress}, on the \gls{hcfirst} distribution. To do so, we repeat our tests with \gls{taggon} configurations of \SI{0.5}{\micro\second} and \SI{2}{\micro\second} instead of \SI{36}{\nano\second}.\footnote{We choose these \gls{taggon} values because they are large enough to show the effects of RowPress and realistic, such that an adversarial access pattern can easily force these \gls{taggon} values by accessing different cachelines in a DRAM row. We do \emph{not} sweep all possible \gls{taggon} values due to the limitations in experiment time.}
\agy{4}{\figref{svard:fig:hcfirst_tAggOn} shows 
a box-and-whiskers plot\footref{svard:fn:boxplot} of the \gls{hcfirst} distribution across all rows in all tested modules under the three different \gls{taggon} values we test. The x-axis shows the \gls{taggon} values, and the y-axis shows the \gls{hcfirst} values. Different subplots show modules from different manufacturers.
We make Obsvs.~\ref{svard:obsv:hcfirst_taggon_decreasing} and~\ref{svard:obsv:hcfirst_taggon_heterogeneity} from \figref{svard:fig:hcfirst_tAggOn}.}

\begin{figure}[h!]
    \centering
    \includegraphics[width=0.8\linewidth]{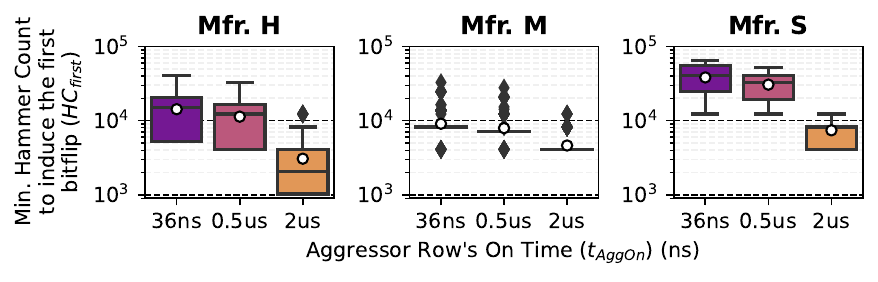}
    \caption{\om{4}{Effect of $t_{AggOn}$ on $HC_{first}$}}
    \label{svard:fig:hcfirst_tAggOn}
\end{figure}

\observation{\gls{hcfirst} decreases with increasing \gls{taggon} for the vast majority of DRAM rows.\label{svard:obsv:hcfirst_taggon_decreasing}} 
\agy{4}{We observe that both mean values and the box (\gls{iqr}) boundaries decrease on the y-axis when \gls{taggon} increases on the x-axis.}

\observation{\gls{hcfirst} values vary significantly across DRAM rows even when \gls{taggon} is \SI{2}{\micro\second}.\label{svard:obsv:hcfirst_taggon_heterogeneity}}
\agy{4}{For example, \gls{hcfirst} distribution across rows in module H2 exhibits the coefficient of variation (CV) values of 25.0\%, 23.0\%, and 30.4\% for \gls{taggon} values of \SI{36}{\nano\second}, \SI{0.5}{\micro\second}, and \SI{2}{\micro\second}.}

From Obsvs.~\ref{svard:obsv:hcfirst_taggon_decreasing} and~\ref{svard:obsv:hcfirst_taggon_heterogeneity}, we draw Takeaway~\ref{svard:take:taggon_hcfirst}.

\takebox{\agy{4}{\gls{hcfirst} values reduce as \gls{taggon} increases and vary significantly across rows for large \gls{taggon} values (e.g., \SI{2}{\micro\second}).}\label{svard:take:taggon_hcfirst}}

\subsection{Spatial Features}
\label{svard:sec:char_feature_selection}
This section investigates \agy{7}{the predictability of a DRAM row's vulnerability to read disturbance by the row's spatial features.}
To do so, we consider a set of features that might affect a DRAM row's reliable operation based on the findings of prior works~\cite{lee2015adaptivelatency, chang2017understanding, chang2016understanding, lee2017designinduced, kim2018solardram, orosa2021adeeper}: 1)~bank address, 2)~row address, 3)~subarray address, 4)~row's distance to the sense amplifiers, i.e., subarray boundaries.
To perform this analysis, subarray boundary identification is critical. Unfortunately, this information is \emph{not} publicly available. To address this problem, we reverse-engineer the subarray boundaries. 

\subsubsection{Subarray Reverse Engineering}
\label{svard:sec:char_subarray_reverse_engineering}
We leverage two key insights.

\head{Key Insight 1}
First, a \om{4}{row located at a subarray boundary} can be disturbed by hammering or pressing its neighboring rows \emph{only} on one side of the row instead of both sides. \om{4}{Exploiting} this observation, we cluster the DRAM rows based on row address and the number of rows that single-sided hammering or pressing a given row affects.
We do \agy{4}{so} using \om{4}{the} k-means clustering algorithm\cite{hartigan1979algorithm}. Because the number of subarrays is initially unknown, we sweep the parameter k and choose the best k value based on the clustering's silhouette score~\cite{rousseeuw1987silhouettes}. 
As a representative example, \figref{svard:fig:silhouette_score} shows the silhouette score of the classification of DRAM rows into subarrays using k-means. We sweep the parameter k on the x-axis and show the silhouette score on the y-axis. \agy{4}{Different curves represent different modules from Mfr. S.}

\begin{figure}[h!]
    \centering
    \includegraphics[width=0.8\linewidth]{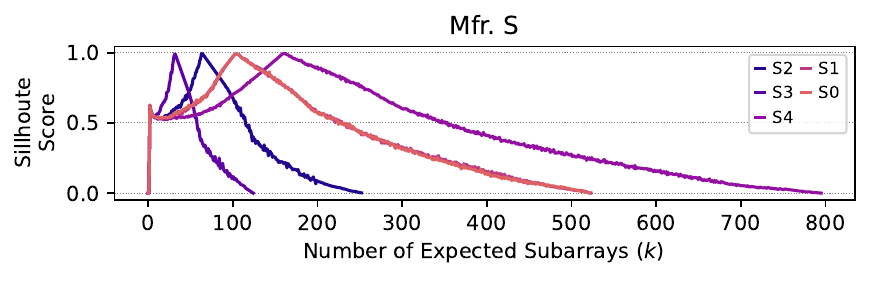}
    \caption{Silhouette score of classification of DRAM rows into subarrays using k-means}
    \label{svard:fig:silhouette_score}
\end{figure}

From \figref{svard:fig:silhouette_score}, we observe that the silhouette score reaches a global maximum and decreases monotonically as we sweep the k-parameter. Based on this observation, we hypothesize that the k-value at the global maximum is the number of subarray boundaries in a DRAM bank, and each cluster for this k-value is a subarray containing the rows in the cluster.

\head{Key Insight 2} Since DRAM rows share a local bitline within a subarray, it is possible to copy one row's (i.e., source row) data to another row (i.e., destination row) within the same subarray (i.e., also known as \om{4}{the intra-subarray} RowClone operation~\cite{seshadri2013rowclone}). Prior works~\cite{yaglikci2022hira, gao2019computedram, gao2022fracdram, olgun2022pidram, yuksel2023pulsar, yuksel2024functionallycomplete} already show that it is possible to perform RowClone in off-the-shelf DRAM chips by violating timing constraints \om{4}{such that two rows are activated in quick succession}.
We conduct RowClone tests following the prior work's methodology~\cite{gao2019computedram}. \agy{4}{If the source row's content is successfully copied to the destination row with \emph{no} bitflips, both the source and destination rows have to be in the same subarray. However, the opposite case (an unsuccessful RowClone operation) does \emph{not} necessarily mean that the two rows are in different subarrays. This is because intra-subarray RowClone is \emph{not} officially supported \om{5}{in existing DRAM chips}, and thus \emph{not} guaranteed to work reliably across all rows in a subarray.}

\agy{4}{We first identify the candidates of subarray boundaries using \om{5}{Key Insight 1} based on the single-sided RowHammer tests. Second, we test these subarray boundaries using \om{5}{Key Insight 2} based on the intra-subarray RowClone tests such that a successful intra-subarray RowClone operation invalidates a candidate subarray boundary since the source and the destination rows have to be in the same subarray, and thus there \emph{cannot} be a subarray boundary in between those two rows.}
Our analysis identifies differently sized subarrays (from 330 to 1027 rows per subarray) and different numbers of subarrays (from 32 to 206 subarrays per bank) across the tested chips. Unfortunately, we do \emph{not} have the ground truth design to verify our results.

\subsubsection{\agy{7}{Predictability} Analysis}
\label{svard:sec:spatial_features_correlation_analysis}
We \agy{4}{analyze} the predictability \agy{4}{between a DRAM row's} spatial features and \agy{4}{the row's \gls{hcfirst} value.}
\agy{4}{As spatial features,} we take each bit in the binary representation of \agy{4}{a DRAM row's} four properties: 1)~bank address, 2)~row address, 3)~subarray address, and 4)~row's distance to the sense amplifiers. 
\agy{4}{We use each of the spatial features for each DRAM row to predict the row's \gls{hcfirst} among 14 tested \agy{5}{hammer} counts}. 
We compare the prediction and real experiment results to create the confusion matrix~\cite{aalst2010process} and calculate the F1 score~\cite{aalst2010process} for each feature.

\agy{4}{\figref{svard:fig:hcfirst_correlations} shows the fraction of spatial features that \emph{strongly correlate} with the row's \gls{hcfirst}. \agy{5}{We consider a spatial feature's correlation with \gls{hcfirst} \om{6}{to be} stronger if predicting \gls{hcfirst} based on the spatial feature results in a larger F1 score.}
The x-axis sweeps the F1 score threshold from 0 to 1, and the y-axis shows the fraction of spatial features that correlate with \gls{hcfirst} with a larger F1 score than the corresponding F1 score threshold. Each subplot shows DRAM modules from a different manufacturer, and each curve represents a different module.} 

\begin{figure}[h!]
    \centering
    \includegraphics[width=\linewidth]{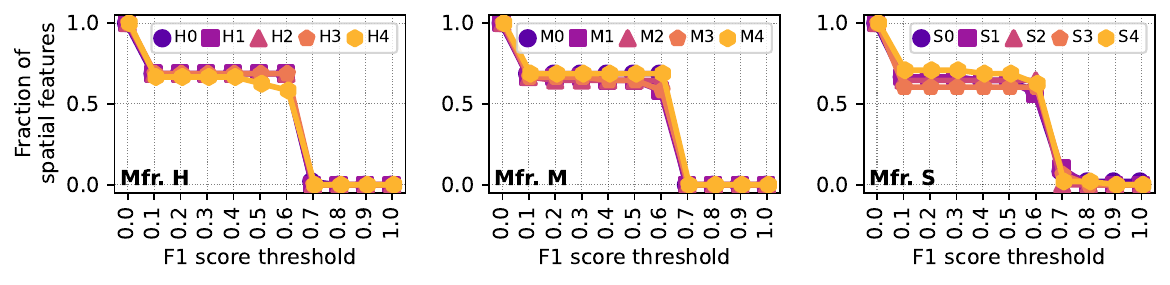}
    \caption{Fraction of spatial features vs F1 score threshold}
    \label{svard:fig:hcfirst_correlations}
\end{figure}

\agy{4}{We make three observations from \figref{svard:fig:hcfirst_correlations}.
First, the fraction of spatial features drastically drops when F1 score threshold is increased from 0.6 to 0.7 for \emph{all} modules. 
Second, \emph{no} spatial feature strongly correlates with \gls{hcfirst} when F1 score threshold is chosen as 0.8. 
Third, \emph{only} four modules (S0, S1, S3, and S4) out of \param{15} tested modules have spatial features correlating with \gls{hcfirst} with an F1 score above 0.7 \agy{5}{(not shown in the figure)}.}\footnote{\agy{5}{We empirically choose the threshold of 0.7 to filter out spatial features exhibiting a weak correlation with \gls{hcfirst}
\agy{5}{and provide few stronger features.}}}
{\tabref{svard:tab:correlated_features} shows \agy{5}{the set of spatial features that result in an F1 score above 0.7.} 
Ba, Ro, Sa, and Dist. columns show such spatial features from the bank address, row address, subarray address, and the row's distance to its local sense amplifiers, respectively.} The F1 score column shows the average F1 score for the module across all specified features.


\begin{table}[h!]
  \centering
  \footnotesize
  \caption{Spatial features that \agy{7}{predict} $HC_{first}$ \agy{7}{with} an F1 score > 0.7}
    \begin{tabular}{l|llllr}
        {{\bf Module}} & \textbf{Ba} & \textbf{Ro} & {{\bf Sa}}  & {{\bf Dist.}} & {{\bf F1 Score}}\\
        \hline 
        S0 & & Bits 7 and 8 &  Bit 0 & Bit 7 & 0.77\\     
        S1 & & Bits 7, 8, 10, and 12 &  Bit 0 & & 0.71\\     
        S3 & & Bit 10 &  Bits 1 and 2 & & 0.75\\     
        S4 & & &  Bit 0 & & 0.76\\     
        \hline
    \end{tabular}
    \label{svard:tab:correlated_features}
\end{table}

\agy{4}{We make two observations from \tabref{svard:tab:correlated_features}. First,}  
the average F1 score among these features does \emph{not} exceed 0.77 for any tested module. 
\agy{4}{Second, such spatial features mostly come from row and subarray address bits, while \emph{no} bank bit results in an F1 score larger than 0.7. From these two observations, we draw Takeaway~\ref{svard:take:correlations}.} 

\takebox{\agy{4}{Spatial features of DRAM rows \agy{5}{correlate well with their \gls{hcfirst} values in four out of 15 tested DRAM modules.}}\label{svard:take:correlations}}





\subsection{{Repeatability and The Effect of Aging}}
\label{svard:sec:repeatability_and_aging}
A DRAM row's read disturbance vulnerability can change over time. A rigorous aging study requires extensively characterizing many DRAM chips many times over a large time span. Due to time and space limitations, we leave such studies for future work while presenting a preliminary analysis \agy{4}{on module H3} as our best effort. We repeat our experiments on one of the tested modules after 68 days of keeping the module under double-sided RowHammer tests at \SI{80}{\celsius}. 
\agy{4}{\figref{svard:fig:aging} demonstrates the effect of aging on \gls{hcfirst} in a scatter plot. The x-axis and the y-axis show \gls{hcfirst} values before and after aging, respectively. The size of each marker and annotated text near each data point represent the population of DRAM rows at the given before- and after-aging \gls{hcfirst} values, normalized to the total population of rows at the \gls{hcfirst} before aging, i.e., the population at each x-tick sums up to 1.0. The straight black line marks $y=x$ points, where \gls{hcfirst} does \emph{not} change after aging.}  
We make \obssref{svard:obsv:nonzero_aging} and \ref{svard:obsv:weaker_ages_more} from \figref{svard:fig:aging}. 

\begin{figure}[h!]
    \centering
    \includegraphics[width=0.7\linewidth]{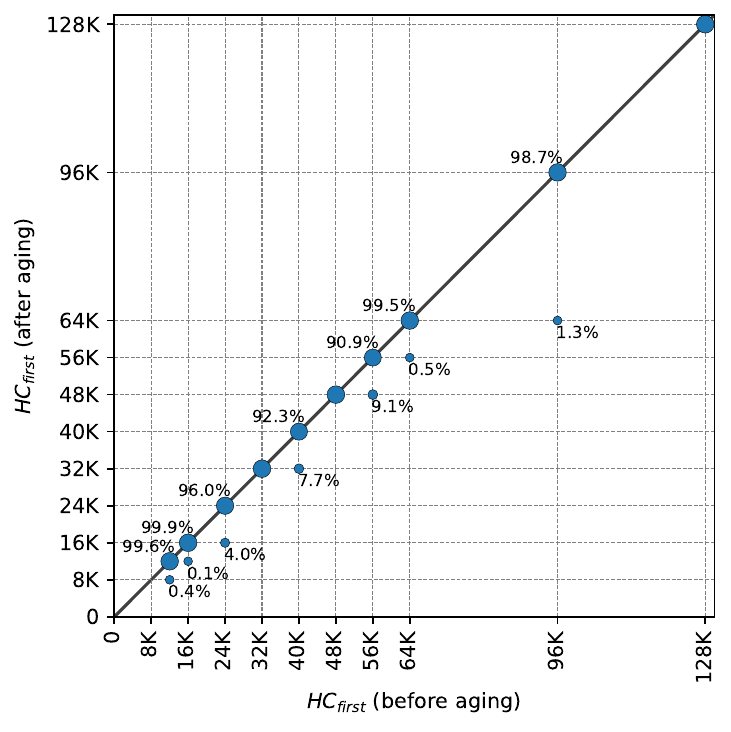}
    \caption{Effect of aging (68 days \agy{4}{using} double-sided RowHammer \agy{4}{test at} \SI{80}{\celsius}) on $HC_{first}$}
    \label{svard:fig:aging}
\end{figure}

\observation{\agy{4}{A non-zero fraction of DRAM rows exhibit lower \gls{hcfirst} values after aging.}\label{svard:obsv:nonzero_aging}} \agy{4}{For example, 0.4\% of DRAM rows with an \gls{hcfirst} of 12K before aging experience bitflips at a \agy{5}{hammer} count of 8K after aging. Therefore, configuring a read disturbance solution for a threshold of 12K is \emph{not} safe for those 0.4\% of the rows, and thus, the \gls{hcfirst} values need to be updated \om{4}{online} in the field.}
We believe this \om{5}{result} makes a \om{5}{strong case} for periodic \om{5}{online} testing of DRAM chips, \om{5}{as also proposed by prior works}~\cite{lee2017designinduced, khan2016parbor, khan2017detecting, liu2013anexperimental, khan2014theefficacy, qureshi2015avatar, patel2017thereach, patel2021harp}.

\observation{Rows with the smallest \gls{hcfirst} values (weakest rows) get affected by aging, unlike the rows with highest \gls{hcfirst} values (strongest rows).\label{svard:obsv:weaker_ages_more}}
For example, \gls{hcfirst} values vary with aging on the left-hand-side of the figure while the rows that show an \gls{hcfirst} value of 128K exhibit no change in their \gls{hcfirst} value. \agy{4}{This indicates that the worst-case \gls{hcfirst} (the lowest \gls{hcfirst} across \om{5}{\emph{all}} rows) changes with aging, and thus aging can jeopardize the security guarantees of existing solutions that are configured based on static \om{5}{identification of the worst-case \gls{hcfirst}}.}
Thus, measuring \gls{hcfirst} under the effect of aging is a challenge for existing solutions \om{4}{and we call \om{6}{future} research on this topic}.
\agy{4}{Based on \obssref{svard:obsv:nonzero_aging} and \ref{svard:obsv:weaker_ages_more}, we draw Takeaway~\ref{svard:take:aging}.} 

\takebox{Determining \gls{hcfirst} values statically \om{4}{is} \emph{not} completely safe, and finding the worst-case \gls{hcfirst} is an open research problem and challenge for existing solutions due to the variation in minimum \gls{hcfirst} values as a result of aging.\label{svard:take:aging}}


\section{{Svärd: Spatial Variation Aware Read Disturbance Defenses}}
\label{svard:sec:adaptation}

We propose a new mechanism Svärd.
The goal of Svärd is to reduce the performance overheads induced by existing read disturbance solutions. Svärd achieves this goal by leveraging the variation in read disturbance vulnerability {across DRAM rows} (\secref{svard:sec:characterization}) to \om{4}{dynamically} tune the aggressiveness of existing read disturbance solutions.

\subsection{High-Level Overview}
\label{svard:sec:adaptation_overview}
{\figref{svard:fig:svard_overview}\agycomment{4}{ToDo: Revise the figure} shows Svärd's high-level overview for its memory controller \agy{6}{(MC)}-based implementation. Svärd can similarly be implemented within the DRAM chip (\secref{svard:sec:adaptation_implementation}).}

\begin{figure}[h!]
    \centering
    \includegraphics[width=0.9\linewidth]{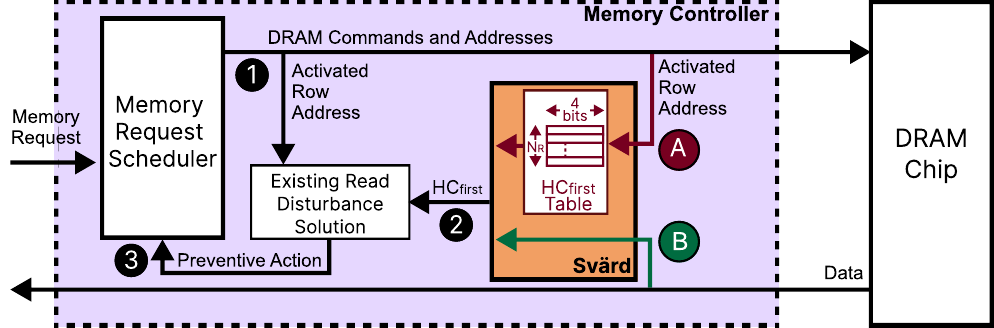}
    \caption{\om{6}{Overview of Svärd MC}-based implementation}
    \label{svard:fig:svard_overview}
\end{figure}

When a DRAM row is activated \circled{1}, {both \om{4}{an existing} read disturbance solution and Svärd are provided with the activated row address. The existing solution computes \om{4}{a} value (e.g., a random number~\cite{kim2014flipping} or estimated activation count~\cite{qureshi2022hydra, yaglikci2021blockhammer, saxena2022aqua}) to compare against a threshold to decide whether to take a preventive action. Meanwhile \circled{2},} Svärd provides the read disturbance solution with an \gls{hcfirst} value based on the activated row's vulnerability level. Then \circled{3}, the read disturbance solution uses this \gls{hcfirst} value to decide \om{4}{whether or not} to perform a preventive action.
{By providing the \gls{hcfirst} value based on the row's characteristics, Svärd tunes the existing solution's aggressiveness dynamically.} Therefore, the read disturbance solution acts \om{4}{either more or less} aggressively when \om{4}{a row with high or low} vulnerability \om{4}{is} accessed. \om{4}{The} read disturbance solution does \emph{not} perform \om{4}{a} preventive action (e.g., refresh victim rows, throttle accesses to the aggressor row, or relocate the aggressor row's content to a far place from the victim row) if the accessed rows do \emph{not} need the preventive action to avoid bitflips. {Svärd maintains a few \om{5}{bits} (e.g., 4 bits) to specify the \gls{hcfirst} classification of each DRAM row. To do so, Svärd can store and obtain the necessary \om{5}{classification} metadata in various ways, including but not limited to: \coloredcircledletter{burgundy}{A}~implementing an \gls{hcfirst} table within the memory controller \agy{6}{that stores as many entries as the number of rows ($N_{R}$)} and \coloredcircledletter{cadmiumgreen}{B}~fetching the classification data along with the first read from the metadata bits stored in DRAM.} \agy{4}{\om{5}{Svärd's} classification \om{5}{meta}data storage can be optimized by using Bloom filters, similar to prior work~\cite{liu2012raidr, seshadri2012theevictedaddress}.}

\subsection{Implementation Options}
\label{svard:sec:adaptation_implementation}
Svärd can be implemented where the existing read disturbance solution is implemented. \om{5}{We} explain two implementations of Svärd that support \om{4}{read disturbance solutions implemented in} 1)~\om{4}{the} memory controller and 2)~in DRAM \om{4}{chips}. However, there is a large design space for Svärd's implementation options that we foreshadow and leave for future research.

\head{Memory Controller} Many prior works~\mcBasedRowHammerMitigations{} propose implementing read disturbance solutions in the memory controller where they can observe and enhance all memory requests. To support these solutions, Svärd maintains a data structure that stores the read disturbance profile \om{4}{of DRAM rows} in the memory controller. Svärd observes the row activation ($ACT$) commands that the memory request scheduler issues and uses the activated row address to query the read disturbance profile. In parallel, the read disturbance solution also executes its algorithm (e.g., \om{4}{generates} a random number or \om{4}{increments} the corresponding activation counters). 
Svärd provides the \agy{4}{read disturbance} solution with \om{4}{a more precise} threshold corresponding to the activated row's vulnerability level. \agy{4}{The read disturbance solution uses this more precise threshold to decide whether or not to perform a preventive action.} Svärd's implementation can follow one of many common practices of storing metadata in computing systems. Svärd can store \om{5}{its} metadata within 1)~a \om{5}{hardware data structure (e.g., a table or Bloom filters)} in the memory controller, 2)~the integrity check bits in the DRAM array~\cite{jedec2020jesd795, patel2019understanding, patel2020bitexact, patel2021harp, patel2022acase, qureshi2021rethinking, fakhrzadehgan2022safeguard}, or 3)~a dedicated memory space in the DRAM array with an optional caching mechanism in the memory controller, \agy{5}{similar to prior works~\cite{qureshi2022hydra, hong2018attache, meza2012enabling}}.

\head{DRAM \om{4}{Chip}} \agy{4}{\om{5}{Various} prior works propose to mitigate read disturbance within the DRAM chip}~\inDRAMRowHammerMitigations{}. 
\agy{4}{These read disturbance solutions observe memory access patterns and perform preventive actions within the DRAM chip, \om{5}{transparently} to the rest of the system. To support these read disturbance solutions, Svärd can be implemented within the DRAM chip. Similar to \om{6}{the} memory controller-based implementation, when a DRAM row is activated, Svärd provides the read disturbance solution with a more precise threshold corresponding to the activated row's vulnerability level.}
Svärd can store the necessary metadata within the DRAM array or the activation counters and access when the row is accessed. 

\subsection{Security}
\label{svard:sec:adaptation_security}
{Svärd does \emph{not} \om{4}{affect} the security guarantees of existing read disturbance solutions. 
\agy{4}{Existing read disturbance solutions provide their security guarantees for each DRAM row, conservatively assuming that all DRAM rows are as vulnerable to read disturbance as the most vulnerable (weakest) row.}
As a result, they overprotect the rows that are stronger than the weakest row. With Svärd, these solutions still provide the same security \om{5}{guarantees} for the weakest DRAM rows \om{4}{while at the same time avoiding the overprotection of} the rows that are stronger than the weakest rows without compromising their security guarantees. This is because Svärd tunes the aggressiveness of \om{5}{existing} solutions based on the vulnerability level of \om{4}{each row}.}

\subsection{Hardware Complexity of \om{6}{Storing the} Read Disturbance Vulnerability Profile}

\agy{4}{The hardware complexity of storing the read disturbance vulnerability profile depends on the size of the profile's metadata. The size of this metadata can be reduced by grouping DRAM rows using their spatial features if there is a strong correlation between the spatial features of a DRAM row and the row's \gls{hcfirst}. \secref{svard:sec:spatial_features_correlation_analysis} shows that such correlation exists only for a subset of the tested modules. To make Svärd widely applicable to all DRAM modules, including the ones that do not exhibit such a strong correlation, we evaluate the hardware complexity of storing the read disturbance vulnerability profile for the worst-case, where we cluster rows into several vulnerability bins, and store a bin id separately for each DRAM row.} We evaluate two different implementations of this metadata storage: 1) a table in the memory controller and 2) dedicated bits within the data integrity metadata~\cite{jedec2020jesd795, patel2019understanding, patel2020bitexact, patel2021harp, patel2022acase, qureshi2021rethinking, fakhrzadehgan2022safeguard} in DRAM. In both cases, we store a vulnerability bin identifier per DRAM row. As the number of bins is smaller than 16, we represent a bin with a 4-bit identifier. 
For the area overhead analysis, we assume a DRAM bank size of 64K DRAM rows and a DRAM row size of 8KB. 

First, for the table implementation, we use CACTI~\cite{balasubramonian2017cacti} and estimate an area cost of \SI{0.056}{\milli\meter\squared} per DRAM bank. When configured for a dual rank system with 16 banks at each rank, the table implementation consumes an overall area overhead of \SI{0.86}{\percent} of the chip area of a high-end Intel Xeon processor with four memory channels~\cite{wikichipcascade}. This table's access latency is \SI{0.47}{\nano\second}, which can be overlapped with the latency of a row activation, e.g., \SI{\approx14}{\nano\second}~\cite{samsung2017288pin}.  

Second, storing this metadata as part of the data integrity bits within DRAM requires dedicating four additional bits for an 8KB-large DRAM row. Therefore, it increases the DRAM array size by \SI{0.006}{\percent} with a conservative estimate where each row's width is extended to store four more bits. In this implementation, because the metadata is implemented as part of the data integrity bits, Svärd reads a data word and the metadata in parallel. Therefore, it does \emph{not} increase the memory access latency. \agy{5}{Since the metadata is stored in the DRAM array, the existing read disturbance solution needs to prevent read disturbance bitflips in the metadata. To do so, the read disturbance solution performs its preventive actions (e.g., refreshing a potential victim row) also on the DRAM cells that store the metadata.}

\section{Performance Evaluation}
\label{svard:sec:evaluation}

\subsection{Methodology}
\label{svard:sec:evaluation_methodology}

\head{Simulation Environment} {We evaluate Svärd's performance impact via {cycle-level simulations} using Ramulator~\cite{safariramulator, Kim2016Ramulator, safari2023ramulator2, luo2023ramulator}.
Table~\ref{svard:tab:configs} shows the simulated system configuration.}

\begin{table}[h]
\centering
\caption{Simulated system configuration}
\label{svard:tab:configs}
\resizebox{0.6\linewidth}{!}{
\footnotesize
\begin{tabular}{ll}
\hline
\head{Processor}                                                   & \begin{tabular}[c]{@{}l@{}} 1 or 8 cores, 3.2GHz clock frequency,\\ 4-wide issue, 128-entry instruction window\end{tabular}  \\ \hline
\head{DRAM}                                                        & \begin{tabular}[c]{@{}l@{}}DDR4, 1 channel, 2 rank/channel, 4 bank groups,\\ 4 banks/bank group, 128K rows/bank\end{tabular}  \\ \hline
\begin{tabular}[c]{@{}l@{}}\textbf{Memory Ctrl.}\end{tabular} & \begin{tabular}[c]{@{}l@{}}64-entry read and write requests queues,\\Scheduling policy: FR-FCFS~\cite{rixner2000memory,zuravleff1997controller} \\with a column cap of 16~\cite{mutlu2007stalltime},\\Address mapping: MOP\cite{kaseridis2011minimalist}\end{tabular}   \\ \hline
\head{Last-Level Cache}& \begin{tabular}[c]{@{}l@{}} 2 MiB per core \end{tabular}  \\ \hline
\end{tabular}
}
\end{table}

{In our evaluations, we assume a realistic system with eight cores connected to a memory rank with four bank groups, each containing four banks (16~banks in total). The {memory controller} employs {the} FR-FCFS~\cite{rixner2000memory, rixner2004memory, zuravleff1997controller} scheduling algorithm with {the} open-{row} policy.}

\head{Comparison Points}
{Our baseline does \emph{not} implement any read disturbance solution and accesses memory in \om{4}{accordance with} DDR4 specifications~\cite{jedec2020jesd794c}. We evaluate Svärd with {five} state-of-the-art solutions: \agy{6}{AQUA~\cite{saxena2022aqua}, BlockHammer~\cite{yaglikci2021blockhammer}, Hydra~\cite{qureshi2022hydra}, PARA~\cite{kim2014flipping}, and RRS~\cite{saileshwar2022randomized}.}


\head{Workloads}
{We execute {\wlcnt{} 8-core multiprogrammed workload mixes}, randomly chosen from {{five} benchmark suites: {SPEC CPU2006~\cite{stspec}}, SPEC CPU2017~\cite{st2017spec}, {TPC~\cite{transaction}, MediaBench~\cite{fritts2009mediabench}, and YCSB~\cite{cooper2010benchmarking}}}.
We simulate these
workloads until each core executes 200M instructions with a warmup period of 100M, similar to prior work~\cite{kim2020revisiting, yaglikci2021blockhammer, yaglikci2022hira}.}
 
\head{Metrics}
{{We evaluate Svärd's impact on \emph{system throughput} (in terms of weighted speedup~\cite{snavely2000symbiotic, eyerman2008systemlevel, michaud2012demystifying}), \emph{job turnaround time} (in terms of harmonic speedup~\cite{luo2001balancing,eyerman2008systemlevel}), and \emph{fairness} {(in terms of maximum slowdown~\cite{kim2010thread, kim2010atlas,subramanian2014theblacklisting,subramanian2016bliss, subramanian2013mise, mutlu2007stalltime, subramanian2015theapplication, ebrahimi2010fairness, ebrahimi2011prefetchaware, das2009applicationaware, das2013applicationtocore, yaglikci2021blockhammer}})}.}

\head{Read Disturbance Vulnerability Profile}
To evaluate Svärd, we use each manufacturer's read disturbance vulnerability profile based on \om{4}{our} real chip characterization results \om{4}{(\secref{svard:sec:characterization})}. Our evaluation spans seven different \agy{4}{worst-case} \gls{hcfirst} values \agy{4}{from 4K down to 64 to evaluate system performance for} modern and future DRAM chips \agy{4}{as technology node scaling over generations exacerbates read disturbance vulnerability. To apply our read disturbance vulnerability profile to future DRAM chips, we scale down all observed \gls{hcfirst} values}
such that the minimum \agy{4}{(worst-case) \gls{hcfirst} value in the read disturbance vulnerability profile becomes} equal to the \gls{hcfirst} value used \agy{4}{for evaluating the read disturbance solution \emph{without} Svärd.} 

\subsection{Performance Analysis}
\label{svard:sec:evaluation_results}
\figref{svard:fig:mitigation_perf} shows how system performance varies when \yct{4}{five} state-of-the-art read disturbance solutions are employed with and without Svärd. Each column of subplots evaluates a different read disturbance solution. We sweep \gls{hcfirst} on the x-axis from 4K down to 64. We show the three performance metrics (weighted speedup, harmonic speedup, and max. slowdown) normalized to the baseline where \agy{4}{\emph{no} read disturbance solution} is implemented. We annotate each configuration of Svärd in the form of Svärd-[DRAM Mfr], \agy{4}{where we choose a representative module from each manufacturer}. Each marker and shade show the average and the minimum-maximum span of performance measurements across \numworkloadmixes{} workload mixes. We make \obssref{svard:obsv:allgood}-\ref{svard:obsv:MfrSgood} from \figref{svard:fig:mitigation_perf}.   

\begin{figure*}[h!]
    \centering
    \includegraphics[width=\linewidth]{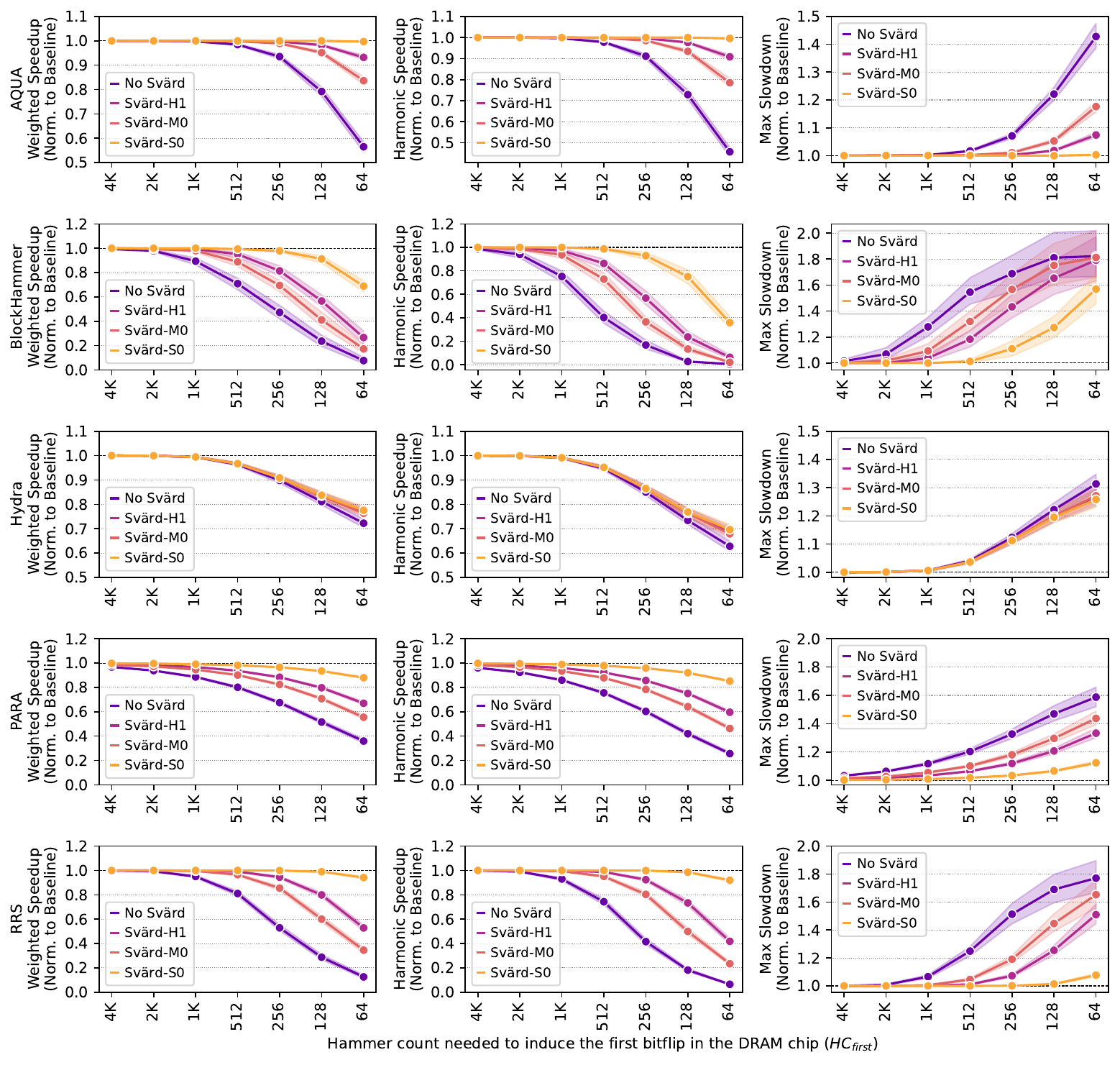}
    \caption{\agy{4}{Performance} overheads of \agy{4}{AQUA~\cite{saxena2022aqua},} {BlockHammer~\cite{yaglikci2021blockhammer},} Hydra~\cite{qureshi2022hydra}, PARA~\cite{kim2014flipping}, and RRS~\cite{saileshwar2022randomized} with and without Svärd}
    \label{svard:fig:mitigation_perf}
\end{figure*}

\observation{Svärd consistently improves system performance in terms of all three metrics when used with any of the \yct{4}{five} tested solutions for all \gls{hcfirst} values below 1K.\label{svard:obsv:allgood}} Svärd configurations result in clearly \om{4}{higher} values for weighted and harmonic speedups and lower values for maximum slowdown, \agy{4}{compared to the respective solution \emph{without} Svärd}. 
{For \agy{4}{an \gls{hcfirst} value} of 128 (64), Svärd significantly increases system performance over \yct{4}{AQUA~\cite{saxena2022aqua},} BlockHammer~\cite{yaglikci2021blockhammer}, Hydra~\cite{qureshi2022hydra}, PARA~\cite{kim2014flipping}, and RRS~\cite{saileshwar2022randomized} by
$1.23\times$ ($1.63\times$),
$2.65\times$ ($4.88\times$),
$1.03\times$ ($1.07\times$),
$1.57\times$ ($1.95\times$),
and
$2.76\times$ ($4.80\times$),
respectively, on average across \wlcnt{} evaluated workloads and three read disturbance profiles of modules S0, M0, and H1.}
Svärd's performance benefits are relatively smaller for Hydra~\cite{qureshi2022hydra} compared to other \agy{4}{evaluated read disturbance solutions}.This is because Hydra's performance overheads are \emph{not} dominated by the preventive refresh operations but the off-chip counter transfer between Hydra's two key components: the counter cache table within the memory controller and the per-row counter table within the DRAM chip. Svärd reduces Hydra's preventive refresh operations, but \emph{not} the off-chip counter transfers. We leave \om{5}{for future work} the Hydra-specific optimizations that Svärd might \agy{4}{enable}.


\observation{Svärd performs the best for the \gls{hcfirst} distribution profile \agy{4}{of S0 among the three evaluated modules.}\label{svard:obsv:MfrSgood}} 
For an \gls{hcfirst} of 64, Svärd reduces \agy{4}{AQUA's~\cite{saxena2022aqua},}
BlockHammer's~\cite{yaglikci2021blockhammer},
Hydra's~\cite{qureshi2022hydra},
PARA's~\cite{kim2014flipping}, and
RRS's~\cite{saileshwar2022randomized} performance overheads of
43.51\%,
92.29\%,
27.75\%,
64.08\%,
87.40\%,
down to
0.32\% / 16.36\% / 6.81\%,
31.15\% / 82.68\% / 73.32\%,
22.44\% / 23.66\% / 22.84\%,
12.05\% / 44.48\% / 33.02\%,
5.83\% / 65.44\% / 47.17\%,
for \agy{4}{modules} S0 / M0 / H1, respectively.
\agy{6}{From \obssref{svard:obsv:allgood} and~\ref{svard:obsv:MfrSgood}, we draw Takeaway~\ref{svard:take:main_perf}.}




\takebox{Svärd effectively reduces the performance degradation that read disturbance solutions inflict on a system.\label{svard:take:main_perf}}

\head{{Adversarial Access Patterns}}
We investigate Svärd's performance benefits under two adversarial access patterns that exacerbate the performance overheads of Hydra~\cite{qureshi2022hydra} and RRS~\cite{saileshwar2022randomized}. Hydra's adversarial access pattern maximizes the evictions in the counter cache and causes an additional DRAM row activation for each row activation in the steady state. RRS's adversarial access pattern keeps hammering a DRAM row to maximize the number of row swap operations. \figref{svard:fig:adversarial} shows the slowdown caused by Hydra (\figref{svard:fig:adversarial}a) and RRS (\figref{svard:fig:adversarial}b) when these \agy{4}{read disturbance solutions are} used with different Svärd configurations for \agy{4}{an \gls{hcfirst}} of 64. The x-axis shows Svärd's configurations, and the y-axis shows the measured slowdown (higher is worse), normalized to the slowdown of the \agy{4}{evaluated read disturbance solution \emph{without} Svärd (i.e., No Svärd).}
\agy{5}{We make \obssref{svard:obsv:adv_one} and~\ref{svard:obsv:adv_two}} from \figref{svard:fig:adversarial}.

\begin{figure}[h!]
    \centering
    \includegraphics[width=0.9\linewidth]{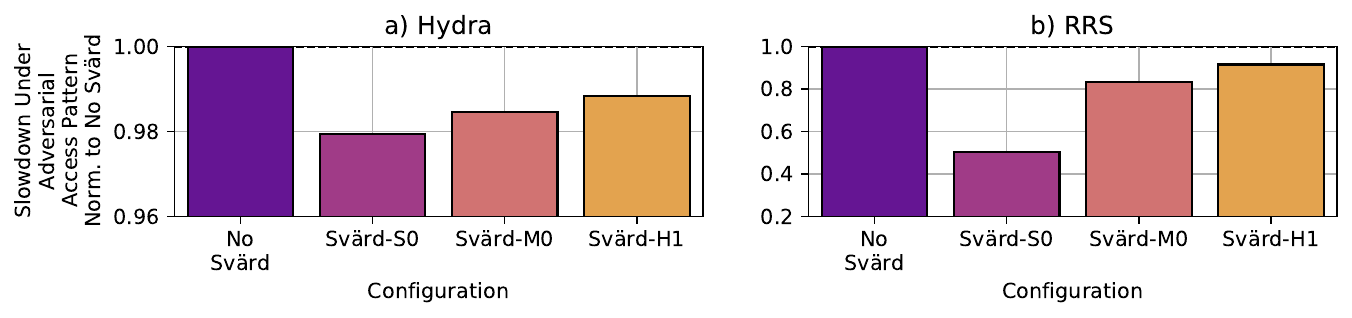}
    \caption{Effect of adversarial access patterns on Svärd's performance \agy{4}{when used with a) Hydra~\cite{qureshi2022hydra} and b) RRS~\cite{saileshwar2022randomized}}}
    \label{svard:fig:adversarial}
\end{figure}

\observation{Svärd reduces both Hydra's and RRS's performance overheads under adversarial access patterns as all bars except \emph{No Svärd} are below 1.0 for both Hydra and RRS.\label{svard:obsv:adv_one}}
For example, Hydra's and RRS's performance overheads with no Svärd are 73.1\% and 95.6\%, respectively (not shown in the Figure), while Svärd reduces these overheads down to 71.6\% and 48.2\%, respectively, with Mfr. S's profile. 

\observation{Similar to \obsref{svard:obsv:MfrSgood}, \om{4}{Svärd provides the best performance overhead reduction with module S0's profile}
among the three tested profiles.\label{svard:obsv:adv_two}}
The bars for \agy{4}{Svärd-S0} exhibit the lowest slowdown in both \figref{svard:fig:adversarial}a and b. 

From these two \obssref{svard:obsv:adv_one} and~\ref{svard:obsv:adv_two}, we draw Takeaway~\ref{svard:take:adv}.
\takebox{Svärd mitigates the performance overheads of Hydra and RRS under adversarial access patterns.\label{svard:take:adv}}

\section{Summary}
\label{svard:sec:conclusion}
 
{This paper tackles the shortcomings of existing RowHammer solutions}
{by leveraging the spatial variation of read disturbance vulnerability across different memory locations within a memory module.
To do so, we
1)~present the first rigorous real DRAM chip characterization study of the spatial variation of read disturbance and
2)~propose Svärd, a new mechanism that dynamically adapts the aggressiveness of existing solutions to the read disturbance vulnerability of potential victim rows. 
Our experimental characterization on \numchips{} real DDR4 DRAM chips, {spanning \param{11} different \agy{4}{density and} die revisions}, demonstrates a large variation in \agy{4}{read disturbance} vulnerability across different memory locations within a module. 
By learning and leveraging this large spatial variation in read disturbance vulnerability across DRAM rows, Svärd reduces the performance overheads of state-of-the-art \agy{4}{DRAM read disturbance} solutions, leading to large performance benefits. We hope and expect that the understanding we develop via our experimental characterization and \om{4}{the resulting} Svärd \om{4}{technique} will inspire DRAM vendors and system designers to efficiently and scalably \om{4}{enable} robust (i.e., reliable, secure, and safe) operation \om{4}{as} DRAM technology node scaling exacerbates read disturbance.}

\chapter[Hidden Row Activation to Reduce Time Spent for Refresh]{Hidden Row Activation to Reduce Time Spent for Refresh}
\label{chap:hira}




\newcommand{\agycrbtodo}[1]{}
\newcommand{\agycrb}[1]{#1}
\newcommand{\atacrb}[1]{#1}
\newcommand{\omcrb}[1]{#1}
\newcommand{\agycrbcomment}[1]{}
\newcommand{\atacrbcomment}[1]{}
\newcommand{\omcrbcomment}[1]{}

\newcommand{\omcri}[1]{#1}
\newcommand{\omcricomment}[1]{}

\newcommand{\agycrutodo}[1]{}
\newcommand{\agycru}[1]{#1}
\newcommand{\atacru}[1]{#1}
\newcommand{\omcru}[1]{#1}
\newcommand{\agycrucomment}[1]{}
\newcommand{\atacrucomment}[1]{}
\newcommand{\omcrucomment}[1]{}

\newcommand{\agyarbtodo}[1]{}
\newcommand{\agyarb}[1]{#1}
\newcommand{\ataarb}[1]{#1}
\newcommand{\omarb}[1]{#1}
\newcommand{\agyarbcomment}[1]{}
\newcommand{\ataarbcomment}[1]{}
\newcommand{\omarbcomment}[1]{}

\newcommand{\cqlabel}[1]{}
\newcommand{\iqlabel}[1]{}
\newcommand{\shom}[1]{#1}
\newcommand{\sho}[1]{#1}
\newcommand{\micrev}[1]{#1}
\newcommand{\dmpa}[1]{#1}

\newcommand{\memacc}[0]{{memory} access}
\newcommand{\memaccs}[0]{\memacc{}es}
\newcommand{\dura}[0]{\gls{hira}}
\newcommand{\duras}[0]{\gls{hirasched}}
\newcommand{\doduo}[0]{\dura{}}
\newcommand{\doduodram}[0]{\duras{}}
\newcommand{\hira}[0]{\dura{}}
\newcommand{\hiraop}[0]{\dura{} operation}
\newcommand{\hirasched}[0]{\duras{}}
\newacronym{hirasched}{HiRA{-}MC}{the {HiRA Memory Controller}}
\newcommand{\rowrefred}[0]{\param{\SI{3}{\nano\second}}}
\newcommand{\systemperf}[0]{\param{\SI{12.5}{\percent}}}
\newcommand{\periodicreduction}[0]{\SI{35.4}{\percent}}
\newcommand{\reactivereduction}[0]{\SI{11.4}{\percent}}
\newcommand{\rhperf}[0]{\param{\SI{78}{\percent}}}
\newcommand{\chipcnt}[0]{\param{56}}
\newcommand{\nchips}[0]{\chipcnt{}}
\renewcommand{\wlcnt}[0]{\param{{125}}}
\newcommand{\ncores}[0]{\param{8}}
\newcommand{\revengtime}[0]{\param{?s}}
\newcommand{\halfvdd}[0]{$V_{DD}/2$}
\newcommand{\covref}[0]{\param{\SI{32}{\percent}}}
\newcommand{\covacc}[0]{\param{\SI{32}{\percent}}}
\newcommand{\tablesizerefptr}[0]{\param{N}}

\newcommand{\equ}[2]{
\vspace{-0.75em}
\begin{equation}
    #1
    \label{#2}
    \vspace{0.25em}
\end{equation}
}

\renewcommand{\hcfirst}[0]{N_{RH}}
\newcommand{\hcdeadline}[0]{N_{RefSlack}}
\newacronym{hcdeadline}{$N_{RefSlack}:t_{RefSlack}/t_{RC}$}{the maximum {number of activations} that an {attacker} can {perform within a \gls{trrslack}}}
\providecommand{\nummodules}{25}
\providecommand{\numchips}{256}
\providecommand{\trc}[0]{t_{RC}}
\providecommand{\tfaw}[0]{t_{FAW}}
\providecommand{\trefw}[0]{t_{REFW}}
\providecommand{\trefi}[0]{t_{REFI}}
\newcommand{\trfc}[0]{t_{RFC}}
\newcommand{\pth}[0]{p_{th}}
\newcommand{\pf}[0]{p_{failure}}
\newcommand{\prh}[0]{p_{RH}}

\section{HiRA: Hidden Row Activation}
\label{hira:sec:doduo}

\head{{Overview}}
{We develop the \gls{hira} operation} for concurrently activating two DRAM rows within a {DRAM} bank. 
\gls{hira} overlaps the latency of refreshing a DRAM row with the latency of refreshing or 
{activating} 
another DRAM row in the same DRAM bank.
\figref{hira:fig:cmdseq} demonstrates how {a} \gls{hira} {operation} is performed by issuing a carefully{-}engineered sequence of $ACT~RowA$, \gls{pre}{, and $ACT~RowB$} commands with two customized timing parameters: $t_1$ ({$ACT~RowA$} to $PRE$ latency) and $t_2$ ($PRE$ to {$ACT~RowB$} latency). {A \hiraop{}'s first \gls{act} refreshes $RowA$ and the second \gls{act} refreshes $RowB$ {and} opens it for column accesses. Since \gls{act} and \gls{pre} commands} {are already implemented in off-the-shelf DRAM chips{,}}  \hira{} \emph{does not} require modifications to the DRAM chip circuitry.

{At a high level, a \hiraop{} 1)~activates $RowA$, 2)~precharges the bank \emph{without} disconnecting $RowA$ from its local row buffer, and 3)~activates $RowB$. In doing so, it allows the memory controller to 1)~perform two refresh operations on $RowA$ and $RowB$ with a latency significantly smaller than two times the \gls{trc} (i.e., refresh-refresh parallelization) and~2)~{activate} $RowB$ for column accesses (i.e., only $RowB$'s local row buffer gets connected to the bank I/O after performing a \hiraop{}) {{concurrently} with refreshing $RowA$} (i.e., refresh-access parallelization).}

\begin{figure*}[!ht]
    \centering
        \includegraphics[width=0.9\linewidth]{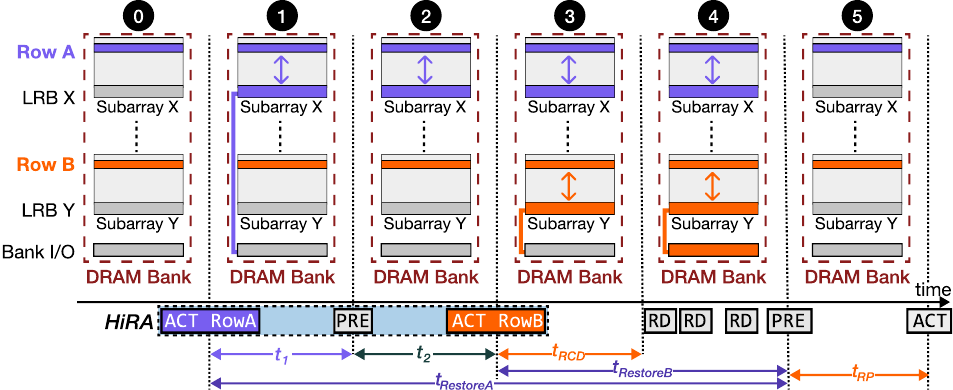}
    \caption{Performing a HiRA operation and its effects on a DRAM bank. Command timings are not {to scale}. {LRB: Local Row Buffer}}
    \label{hira:fig:cmdseq}
\end{figure*}

\head{\gls{hira} Operation Walk-Through} \figref{hira:fig:cmdseq} demonstrates how a \doduo{} operation is performed and how it affects the state of a DRAM bank.
Initially (\circled{0}) the DRAM bank {is in} precharged {state and thus there is no active row}. \doduo{} begins {by} issuing an {\gls{act}} command targeting $RowA$, which connects $RowA$'s cells to \emph{local row buffer X}~(\circled{1}). Then, a precharge command is issued to disconnect \emph{local row buffer {(LRB)} X} {from the} \emph{{bank I/O}}~(\circled{2}). This precharge operation is interrupted by issuing a new row activation, targeting $RowB$ {in a completely separate subarray Y}~(\circled{3}), to avoid breaking the connection between the \emph{local row buffer X} and $RowA$. {Therefore, the sense amplifiers in the local row buffer X continue charge restoration of $RowA$. Thus, \gls{hira} overlaps the latency of refreshing $RowA$ with the latency of activating $RowB$.} It is important that the subarray that contains $RowB$ (\emph{subarray~Y}) is physically isolated from the subarray that contains $RowA$ (\emph{subarray~X}), such that subarrays X and Y do not share any bitline or sense amplifier and thus activating $RowB$ does not {affect} the voltage levels on subarray~X's bitlines~(\circled{3}). The {\hiraop{}} completes when the second row activation is issued, {after which both $RowA$ and $RowB$ are connected to their local row buffers without corrupting each other's data~(\circled{3})}. Following a \doduo{} operation, $RowB$'s content can be read by issuing {\gls{rd}} commands once \gls{trcd} is satisfied~(\circled{4}). To close both $RowA$ and $RowB$, issuing one precharge command is enough~(\circled{5}).\footnote{{Our experiments verify that {issuing one precharge command is enough} {to {reliably} close \emph{both} rows} in {all} \chipcnt{} real DRAM chips {we test}. We hypothesize that issuing a \gls{pre} command disables all wordlines and precharges all bitlines in a DRAM bank because {the precharge command} is \emph{not} provided with a row address~\cite{jedec2008jesd79f, jedec2021jesd235d, jedec2020jesd794c, jedec2017jesd2094b, jedec2020jesd2095a, jedec2020jesd795, micron2014sdram}.}}

\head{Charge Restoration after \gls{hira}}
\figref{hira:fig:cmdseq} highlights the charge restoration time that $RowA$ and $RowB$ experience as $t_{RestoreA}$ and $t_{RestoreB}$, respectively.
To ensure charge restoration {happens correctly} for $RowA$ and $RowB$, both $t_{RestoreA}$ and $t_{RestoreB}$ should be larger than or equal to the {existing} \gls{tras} {timing parameter}~\cite{jedec2008jesd79f, jedec2021jesd235d, jedec2020jesd794c, jedec2017jesd2094b, jedec2020jesd2095a, jedec2020jesd795, micron2014sdram}. Because we do \emph{not} modify the timing constraints of the second \gls{pre} command (\circled{5}), {existing DRAM} timing restrictions already ensure that $t_{RestoreB}$ is larger than or equal to the nominal \gls{tras}. Since $t_{RestoreA}$ is already larger than $t_{RestoreB}$ {(because $RowA$ is activated earlier than $RowB$)}, we conclude that {charge restoration happens correctly for} both rows.

\head{\doduo{}'s Novelty} \doduo{}'s command sequence ($ACT$-$PRE$-$ACT$) is similar to the command sequences used in {multiple} prior works~\cite{gao2019computedram, olgun2021quactrng, olgun2022pidram}. {These prior works use the $ACT$-$PRE$-$ACT$ command sequence to activate two rows in the \emph{same} subarray for various purposes {(which we explain below)}. In contrast, \doduo{}'s purpose is to activate two rows in \emph{different} subarrays such that we can
{refresh a DRAM row {concurrently} with refreshing or activating another row in the same bank.}}

First, ComputeDRAM~\cite{gao2019computedram} {and PiDRAM~\cite{olgun2022pidram}} perform {an} $ACT$-$PRE$-$ACT$ command sequence to enable bulk data copy across DRAM rows {in the same subarray} (also known as RowClone~\cite{seshadri2013rowclone}) in {off-the-shelf} DRAM chips. 
Second, QUAC-TRNG~\cite{olgun2021quactrng} uses $ACT$-$PRE$-$ACT$ command sequence for performing an operation called \emph{{quadruple row activation}}, which concurrently activates four rows whose addresses vary only in the least significant two bits. 
{1)~}{The} RowClone~\cite{seshadri2013rowclone} implementation{s} of {both} {ComputeDRAM~\cite{gao2019computedram} and PiDRAM~\cite{olgun2022pidram}} and {2)~}QUAC-TRNG's~\cite{olgun2021quactrng} quadruple row activation require using two rows within the {\emph{same}} subarray, so that the bitlines and local sense amplifiers are used for sharing the electrical charge across activated DRAM rows. Therefore, these works do \emph{not} {activate} DRAM rows {in \emph{different} subarrays.} 
In contrast, \doduo{} {exclusively} targets two {rows in different subarrays}, so that it enables the memory controller to perform two key operations that were \emph{not} {known to be} possible before {on off-the-shelf DRAM chips}: 1)~concurrently refreshing two rows, and 2)~refreshing one row while {activating another row in a different subarray.}

\head{HiRA's Main Benefit} \gls{hira} largely overlaps {a DRAM row's charge restoration latency ($t_{RestoreA}$} in \figref{hira:fig:cmdseq}) with the {latency of another row's activation and charge restoration (\gls{trcd} and $t_{RestoreB}$ in \figref{hira:fig:cmdseq}, respectively)}. {Doing so allows \gls{hira} to reduce the latency {of two operations. First, \gls{hira} reduces the latency of} a memory {access} request that is scheduled immediately after a refresh operation. {With \gls{hira}, such {a} request experiences a latency of $t_1+t_2$ (\circled{1} and \circled{2} in \figref{hira:fig:cmdseq}), which can be as small as \SI{6}{\nano\second} (\secref{hira:sec:experiments_coverage}), instead of the nominal row cycle {time} of} \SI{46.25}{\nano\second} (\gls{trc}~\cite{jedec2020jesd794c, micron2014sdram}). {Second, \gls{hira} reduces the overall latency of} refreshing two DRAM rows in the same bank. {With \gls{hira}, such {an} operation takes \emph{only} \SI{38}{\nano\second} ({{{\SI{6}{\nano\second} for {the} \hiraop{} {to complete} (\secref{hira:sec:experiments_coverage}) and \SI{32}{\nano\second} to ensure that $t_{RestoreB}$ is large enough to complete charge restoration~\cite{jedec2020jesd794c, micron2014sdram}}}) instead of}} {the nominal latency of} \SI{78.25}{\nano\second}.{\footnote{{To refresh two rows using nominal timing parameters, {a conventional} memory controller 1)~activates the first row and waits until charge restoration is complete ($t_{RAS}=32ns$), 2)~precharges the bank and waits until all bitlines are ready for the next row activation ($t_{RP}=14.25ns$), and~3)~activates the second row and waits until charge restoration is complete ($t_{RAS}=32ns$)~\cite{jedec2020jesd794c, micron2014sdram}.\label{hira:fn:reftworows}}}}}

\head{\gls{hira} Operating Conditions}
{A} \gls{hira} operation works {reliably} if {four} conditions are satisfied.
{First,} $t_1$ should be large enough {so that the sense amplifiers are enabled before the {precharge} command is issued {($PRE$ in \figref{hira:fig:cmdseq})}.} 
{Second,}~$t_2$ should be small enough {so that the second activate command ($ACT~RowB$ in \figref{hira:fig:cmdseq}) interrupts the precharge operation \emph{before} $RowA$'s wordline is disabled, allowing charge restoration on $RowA$ to complete correctly.} 
{Third,}~$t_2$ should be large enough to disconnect the local row {buffer X from {the} bank I/O logic if \gls{hira} is performed for refresh-access parallelization{, so that {future} column accesses are performed {only on $RowB$ (LRB~Y).}} This constraint does not apply to refresh-refresh parallelization because the bank I/O logic is not used during refresh.
Fourth,}~$RowA$ and $RowB$ should be located in two different subarrays that are physically isolated from each other, such that they do not share any sense amplifier or bitline.  

\section{HiRA in {Off-the-Shelf} DRAM Chips}
\label{hira:sec:characterization}

{In this section, we demonstrate that \gls{hira} works {reliably} on \chipcnt{} real DDR4 DRAM chips.
{Table~\ref{hira:tab:dram_chip_list} provides {the} {chip} density, die revision {(Die Rev.)}, chip organization {(Org.)}, and manufacturing date of tested DRAM modules {where DRAM chips are manufactured by SK Hynix}.{\footnote{{We observe that \gls{hira} reliably works only in DRAM chips from SK Hynix (similar to QUAC-TRNG~\cite{olgun2021quactrng}) out of {40, 40, and 56} DRAM chips that we test from three major {DRAM} manufacturers: Micron, Samsung, and SK Hynix{, respectively}. {\secref{hira:sec:limitations} discusses why we do \emph{not} observe a successful HiRA operation in DRAM chips manufactured by Micron and Samsung.}
{A~is} F4-2400C17S-8GNT from GSKill~\cite{gskill2021f42400c17s8gnt}, {B is} KSM32RD8/16HDR from Kingston~\cite{datasheetksm32rd8}, and {C is} HMAA4GU6AJR8N-XN from SK Hynix~\cite{memorynethmaa4gu6ajr8nxn}{.}
}}} We report the manufacturing date of these modules in the form of $week-year$.}

\begin{table}[h]
    \caption{Summary of the tested DDR4 DRAM chips {and {key} experimental results}}
    \centering
    \footnotesize{}
    \setlength\tabcolsep{3pt} 
    \begin{tabular}{l|l|cccc|cc}
        
        \toprule
            {\textbf{Model}} & 
            \begin{tabular}[l]{@{}l@{}}\textbf{DIMM} \textbf{Mfr.}\end{tabular}& \begin{tabular}[c]{@{}c@{}}\textbf{Chip}\\\textbf{\textbf{Capacity}}\end{tabular}& \begin{tabular}[c]{@{}c@{}}\textbf{Die}\\\textbf{Rev.}\end{tabular}& \begin{tabular}[c]{@{}c@{}}\textbf{Chip}\\\textbf{Org.}\end{tabular}& \begin{tabular}[c]{@{}c@{}}\textbf{Mfr.}\\\textbf{Date}\end{tabular}& \begin{tabular}[c]{@{}c@{}}\textbf{HiRA}\\\textbf{Cov.$^*$}\end{tabular}& \begin{tabular}[c]{@{}c@{}}\textbf{Norm.}\\$\boldsymbol{N_{RH}^{**}}$\end{tabular}\\
        \midrule
        A0 & \multirow{2}{*}{GSKill~\cite{gskill2021f42400c17s8gnt}} & \multirow{2}{*}{4Gb}&\multirow{2}{*}{B}&\multirow{2}{*}{$\times$8}&\multirow{2}{*}{42--20} &\SI{25.0}{\percent}&1.90\\
        A1 & & & & & &\SI{26.6}{\percent}&1.94\\
        \midrule
        B0 & \multirow{2}{*}{Kingston~\cite{datasheetksm32rd8}} & \multirow{2}{*}{8Gb}&\multirow{2}{*}{D}&\multirow{2}{*}{$\times$8}&\multirow{2}{*}{48--20} &\SI{32.6}{\percent}&1.89\\
        B1 & & & & & &\SI{31.6}{\percent}&1.91\\
        \midrule
        C0 & \multirow{3}{*}{SK Hynix~\cite{memorynethmaa4gu6ajr8nxn}} & \multirow{3}{*}{4Gb}&\multirow{3}{*}{F}&\multirow{3}{*}{$\times$8}&\multirow{3}{*}{51--20} &{\SI{35.3}{\percent}}&{1.89}\\
        C1 & & & & & &\SI{38.4}{\percent}&1.88\\
        C2 & & & & & &\SI{36.1}{\percent}&1.96\\
        \bottomrule
    \end{tabular}
    \begin{flushleft}
    $^*$ {HiRA Cov. stands for HiRA coverage results, presented in \secref{hira:sec:experiments_coverage}.}\\
    $^{**}$ {Norm. $N_{RH}$ is {the} normalized RowHammer threshold, shown in \secref{hira:sec:experiments_chargerestoration}.}\\
    \agyarb{Table~\ref{hira:tab:detailed_info} in Appendix A shows the minimum and the maximum values for both HiRA Cov. and Norm. $N_{RH}$ across all tested rows per DRAM module.}
    \end{flushleft}
    \label{hira:tab:dram_chip_list}
\end{table}

We conduct experiments {in three steps (\secref{hira:sec:experiments_coverage}-\secref{hira:sec:variation_across_banks})} to evaluate the {feasibility,} reliability,  {benefits} and limitations of HiRA on real DRAM chips.}

\subsection{Testing Infrastructure}
We conduct experiments {on \param{\chipcnt{}} real DRAM chips\footnote{\label{hira:fn:testedrows}Due to time limitations, we conduct our tests on the 1)~first 2K, 2)~last 2K, and 3)~middle 2K rows of {Bank 0} in each DRAM chip, similar to~\cite{kim2014flipping, orosa2021adeeper, yaglikci2022understanding}.}} using \agy{3}{DRAM Bender~\cite{olgun2023dram_bender, safari2022dram_bender}, which is based on} a modified version of SoftMC~\cite{hassan2017softmc, safari2017softmc} {and supports DDR4 modules}. 
\figref{deeperlook:fig:infrastructure} shows a picture of our experimental setup.
We {{use the}} Xilinx Alveo {U200 FPGA} board~{\cite{xilinxu200},} {programmed} {with} \agy{3}{DRAM Bender} {to} precisely issue {DRAM} command{s}.{\footnote{{\agy{3}{DRAM Bender} works with a minimum clock cycle of \SI{3}{\nano\second} on Alveo U200~\cite{xilinxu200} and thus issues a DRAM command every \SI{1.5}{\nano\second} in {the} double data rate domain.}}} {The host machine generates the sequence of DRAM commands that we issue to the DRAM {module}.} To avoid {fluctuations} in ambient temperature, we {place the DRAM module clamped with a pair of heaters on both sides. The heaters are controlled by a MaxWell FT200~\cite{maxwellft20x}} temperature controller that {keeps} DRAM chips at $\pm$\param{\SI{0.1}{\celsius}} neighborhood of the {target} temperature.


\head{Data Patterns} {Our tests} use four data patterns {that are used by prior works~\cite{khan2014theefficacy,liu2013anexperimental,patel2017thereach, kim2020revisiting, chang2016understanding, chang2017understanding, lee2017designinduced,khan2016parbor,khan2016acase,khan2017detecting,lee2015adaptivelatency}}: 1)~all ones {($0xFF$)}, 2)~all zeros ($0x00$), 3)~alternating ones and zeros, i.e., checkerboard ($0xAA$), and 4)~the inverse checkerboard ($0x55$).

\head{Disabling Sources of Interference}
To directly observe whether HiRA reliably works {at the} circuit-level,
we {disable} all {known} sources of interference {(i.e., we prevent other DRAM error mechanisms (e.g., retention errors~\cite{liu2013anexperimental,khan2014theefficacy,meza2015revisiting,patel2017thereach,qureshi2015avatar}) or error correction from interfering with a \hiraop{}'s results)} {in three steps, similar to prior works~\cite{orosa2021adeeper, kim2020revisiting, yaglikci2022understanding}.} 
First, we disable all DRAM self-regulation events {(e.g., DRAM Refresh{)}} {and error mitigation mechanisms (e.g., error correction codes and RowHammer defense mechanisms)}~\cite{jedec2020jesd794c,hassan2017softmc,xilinxultrascale}) except {calibration related events (e.g., ZQ calibration, {which is required}} for signal integrity~\cite{jedec2020jesd794c,hassan2017softmc}). {Second, we {conduct each} test within a relatively short period of time {(\SI{10}{\milli\second})} such that we do \emph{not} observe retention errors}. 
{Third, we conduct each test for ten iterations to reduce noise in our measurements.}

\subsection{HiRA's Coverage}
\label{hira:sec:experiments_coverage}
{\gls{hira} works if the two rows that \gls{hira} opens do \emph{not} corrupt each other's data. Therefore, it is important to carefully choose two DRAM rows for \gls{hira} such that the rows are electrically isolated from each other, i.e., do \emph{not} share a bitline or sense amplifier. {The \emph{goal} of our first experiment is to find all combinations of DRAM row pairs that \gls{hira} can {concurrently} activate.} 
{To this end, we define HiRA's \emph{coverage} for a given row ($RowA$) {in a given bank {($BankX$)}} as the {fraction} of other DRAM rows {within $BankX$} which \gls{hira} can {reliably} activate {concurrently with} $RowA$.}
Algorithm~\ref{hira:algorithm:hira} shows the experiment} {to find HiRA's \emph{coverage {for $RowA$.}}
{{To test a pair of DRAM rows $RowA$ and $RowB$ within $BankX$, first, we initialize the two rows using inverse data patterns (lines 7{--}8). Second, we perform \gls{hira} (lines 11--13) and close both rows (line 16). Third, we check whether there is a bitflip in either of the rows (lines 19{--}20). Fourth, if} performing \gls{hira} does \emph{not} cause bitflips in {either} of the rows for any tested data pattern, we increment a counter called $row\_count$ {(line 25). Fifth, we calculate} \gls{hira}'s coverage for $RowA$ {as the fraction of $RowB$s that \gls{hira} can concurrently activate with $RowA$ (line 26).}}
}

\SetAlFnt{\footnotesize}
\RestyleAlgo{ruled}
\begin{algorithm}
\caption{Testing HiRA's Coverage {for {a given} RowA}}\label{hira:algorithm:hira}
  \DontPrintSemicolon
  \SetKwProg{Fn}{Function}{:}{}
  
  \For{RowA in Tested Rows {in BankX}}{
    {row\_count = 0}\;
    \For{RowB in Tested Rows {in BankX}}{
        success = True\;
        \For{datapattern in [$0xFF$, $0x00$, $0xAA$, $0x55$]}
        {
        \textbf{\emph{\# Initialize {the two rows with} inverse data patterns}}\;
        initialize($RowA$, datapattern)\;
        initialize($RowB$, !datapattern)\;
        \;
        \textbf{\emph{\# Perform HiRA}}\;
        act({$BankX$,}$RowA$, wait=$t_{1}$)\;
                pre({$BankX$,} wait=$t_{2}$)\;
        act({$BankX$,}$RowB$, wait=$t_{RAS}$)\; 
                \;
        \textbf{\emph{\# Clos{e} both rows}}\;
        pre({$BankX$,} wait=$t_{RP}$)\;
        \;
        \textbf{\emph{\# Read back the two rows and check for bitflips}}\;
        $RowA$\_pass = compare\_data(datapattern, $RowA$)\; 
        $RowB$\_pass = compare\_data(!datapattern, $RowB$)\;
        \;
        \textbf{\emph{\# Fail if there is at least one bitflip}}\;
        \lIf{!($RowA$\_pass AND $RowB$\_pass)}{success = false}
        }
        \lIf{success == true}{{row\_count}++}
    }
    {HiRA\_coverage[$RowA$] = row\_count / NumberOfTestedRows}\;
  }
\end{algorithm}

{\figref{hira:fig:coverage} shows
the distribution of {\gls{hira} coverage across} {tested DRAM rows{\footref{hira:fn:testedrows}}}
in a box and whiskers plot\footnote{\label{hira:fn:boxplot}{A box-and-whiskers plot emphasizes the important metrics of a dataset's distribution. The box is lower-bounded by the first quartile (i.e., the median of the first half of the ordered set of data points) and upper-bounded by the third quartile (i.e., the median of the second half of the ordered set of data points).
The \gls{iqr} is the distance between the first and third quartiles (i.e., box size).
Whiskers show the minimum and maximum values.}}  {for different combinations of $t_1$ {(x-axis)} and $t_2$ {({colored boxes})} timing parameters}.
{The y-axis shows \gls{hira} coverage across tested rows.}
}

\begin{figure}[!t]
    \centering
    \includegraphics[width=0.7\linewidth]{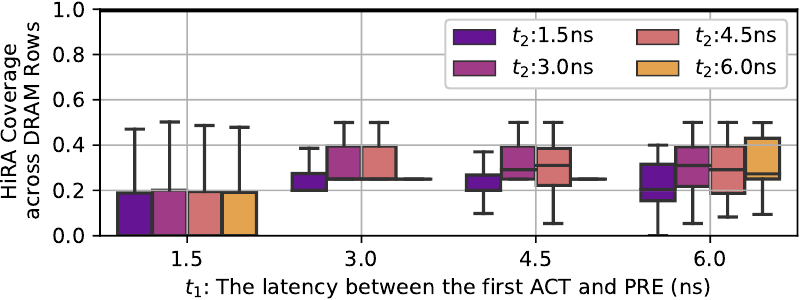}
    \caption{HiRA's coverage {across tested DRAM rows} for different $t_1$ (x-axis) and $t_2$ ({colored boxes}) timing parameter {combinations}}
    \label{hira:fig:coverage}
\end{figure}

We make three observations from \figref{hira:fig:coverage}.
{First, if $t_1$ is \SI{3}{\nano\second} or \SI{4.5}{\nano\second}, a {given} DRAM row's refresh operation can {always} be performed concurrently with at least another DRAM row's {refresh or activation} (i.e., there are no DRAM rows with a \gls{hira} coverage of $0$) {for all tested $t_2$ values.}}
{Second,
{\gls{hira} reliably parallelizes a tested row's {refresh operation} {with refresh or {activation of} {any of the} \covref{hira:}}}
of the other rows\footnote{{The minimum \gls{hira} coverage we observe across all tested rows is \SI{25}{\percent} when $t_1$ is \SI{3}{\nano\second} and $t_2$ is either \SI{3}{\nano\second} or \SI{4.5}{\nano\second}.}} when $t_1$ is \SI{3}{\nano\second} and $t_2$ is either \SI{3}{\nano\second} or \SI{4.5}{\nano\second}.
{Third, we observe that {\gls{hira} coverage} can be $0$ if $t_1$ is chosen too small (e.g., \SI{1.5}{\nano\second}) or too large (e.g., \SI{6}{\nano\second}), {meaning that at least one tested DRAM row's refresh \emph{cannot} be concurrently performed with refresh{ing or activating} another tested DRAM row.} We hypothesize that}
this happens because {1)~\SI{1.5}{\nano\second} is \emph{not} long enough to enable sense amplifiers and 2)~\SI{6}{\nano\second} is \emph{not} short enough for $t_1$ to interrupt row activation before the local row buffer is connected to the bank I/O or \SI{1.5}{\nano\second} is too short for $t_2$ to disconnect $RowA$'s local row buffer from the bank I/O.} Both design-{induced variation}~\cite{lee2017designinduced} and manufacturing process-induced variation~\cite{chang2016understanding, lee2017designinduced} in row activation latency can cause this behavior.
From these {three} observations, we conclude that it is possible to refresh {a given DRAM row concurrently} with {refreshing or activating} \SI{32}{\percent} of the other DRAM rows on average when both $t_1$ and $t_2$ are \SI{3}{\nano\second}.} 
{When \gls{hira} is used with the configuration of $t_1 = t_2 = 3ns$, the latency of refreshing two rows is \emph{only} \SI{38}{\nano\second} ($t_1 + t_2 + t_{RAS}$), while {refreshing two rows with standard DRAM commands takes} \SI{78.25}{\nano\second} for 1)~restoring the charge of the first row ($t_{RAS}$), 2)~precharging bitlines to prepare for second activation ($t_{RP}$), and 3)~restoring the charge of the second row ($t_{RAS}$).{\footref{hira:fn:reftworows}}}
{Therefore, \gls{hira} reduces the latency of refreshing two rows by \SI{51.4}{\percent}.}

\subsection{{Verifying} HiRA's {Second Row Activation}}
\label{hira:sec:experiments_chargerestoration}
Observing \emph{no bitflips} for a pair of rows tested using Algorithm~\ref{hira:algorithm:hira} indicates either of {the} two cases: 1)~\gls{hira} successfully works or 2)~\gls{hira} activates only {the first} row because {the} DRAM chip simply ignores the second \gls{act} command. {The \emph{goal} of our second experiment is to} test whether the DRAM chip ignores or performs \gls{hira}'s second row activation command.
{To this end,} we hammer the two adjacent rows\footnote{{DRAM manufacturers {use {DRAM-internal} mapping} {schemes} to internally translate memory-controller-visible row addresses to physical row {addresses~\cite{kim2014flipping, smith1981laser, horiguchi1997redundancy, keeth2001dram_circuit, itoh2001vlsi, liu2013anexperimental,seshadri2015gatherscatter, khan2016parbor, khan2017detecting, lee2017designinduced, tatar2018defeating, barenghi2018softwareonly, cojocar2020arewe,  patel2020bitexact, yaglikci2021blockhammer, orosa2021adeeper}{,} which can vary across different DRAM modules}. We reconstruct this mapping using single-sided RowHammer, similar to prior works~\cite{kim2014flipping, kim2020revisiting, orosa2021adeeper, yaglikci2022understanding}, so {that} we can hammer aggressor rows that are physically adjacent to a victim row.}} (i.e., aggressor rows) of a given victim row to induce RowHammer bitflips in the victim row (i.e., double-sided RowHammer~\cite{kim2014flipping, kim2020revisiting}). {During the test}, we try refreshing the victim row by using {\gls{hira}'s} \emph{second} row activation command. We hypothesize that {if \gls{hira}'s second row activation is \emph{not} ignored {(i.e., if \gls{hira} works)}, then}
{the minimum number of aggressor row activations required to induce the first RowHammer bitflip, i.e., RowHammer threshold {(\secref{sec:background_rowhammer})}, increases compared to the RowHammer threshold measured \emph{without} using \gls{hira}.}
{W}e \emph{measure} RowHammer threshold {of a given victim row} via binary-search (similar to prior works~\cite{kim2020revisiting, orosa2021adeeper, yaglikci2022understanding}).
{Algorithm~\ref{hira:algorithm:rowhammer} shows how we perform a RowHammer {test} {for a given victim row} at a given hammer count ($HC$) with and without a \gls{hira} operation.}

\SetAlFnt{\footnotesize}
\RestyleAlgo{ruled}
\begin{algorithm}[!ht]
\caption{{{Verifying} HiRA's Second {Row Activation}}}\label{hira:algorithm:rowhammer}
  \DontPrintSemicolon
  \SetKwProg{Fn}{Function}{:}{}
  
  \For{with\_HiRA in [False, True]}{
    \textbf{\emph{{\# Step 1: Initialize DRAM rows}}}\;
    \textbf{\emph{\# Initialize the victim row with the specified data pattern}}\;
    initialize(victim\_row, datapattern)\;
    \textbf{\emph{\# Initialize a dummy row for \gls{hira}{'s first ACT}}}\;
    initialize(\gls{hira}\_dummy\_row, !datapattern)\;
    \textbf{\emph{\# Initialize the two aggressor rows with {inverse} data pattern}}\;
    initialize(aggr\_row\_a, !datapattern)\;
    initialize(aggr\_row\_b, !datapattern)\;
        \;
    \textbf{\emph{\# Step {2}: Hammer each aggressor row $HC/2$ times}}\;
    \For{for act\_cnt = 0 to $HC/2$}{
        act({BankX,} aggr\_row\_a, wait=$t_{RAS}$)\; 
        pre({BankX,} wait=$t_{RP}$)\;
        act({BankX,} aggr\_row\_b, wait=$t_{RAS}$)\;
        pre({BankX,} wait=$t_{RP}$)\;
    }
    \textbf{\emph{\# Step {3}: Perform HiRA or wait}}\;
    \If{with\_HiRA}{
        act({BankX,} \gls{hira}\_dummy\_row, wait=$t_{1}$)\;   
        pre({BankX,} wait=$t_{2}$)\;
        act({BankX,} victim\_row, wait=$t_{RAS}$)\; 
        pre({BankX,} wait=$t_{RP}$)\;
    }\Else{ \textbf{\emph{{\# Without HiRA}}}\;
        wait($t_{1}$+$t_{2}$+$t_{RAS}$+$t_{RP}$);\;
    }
    \textbf{\emph{\# Step {4}: Hammer each aggressor row $HC/2$ times}}\;
    \For{for act\_cnt = 0 to $HC/2$}{
        act({BankX,} aggr\_row\_a, wait=$t_{RAS}$)\; 
        pre({BankX,} wait=$t_{RP}$)\;
        act({BankX,} aggr\_row\_b, wait=$t_{RAS}$)\;
        pre({BankX,} wait=$t_{RP}$)\;
    }
    {\textbf{\emph{\# Step {5}: Check for bitflips on the victim row}}\;}
    {bitflips} = {check\_bitflips(datapattern, victim\_row)}\;
  }
\end{algorithm}

{We conduct the RowHammer test in {{five}} steps.}
{First, we initialize four DRAM rows in a given DRAM bank ({BankX}): the given victim row, a dummy row which \gls{hira} can concurrently refresh with the given victim row, and the two aggressor rows. We initialize the victim row with the specified data pattern and the other three rows with the inverse data pattern (lines {3--9}).}
{Second, we hammer {each aggressor row $HC/2$} times {({lines 12--16})}. Third, we either perform {{(lines 19--23)}} a \gls{hira} operation {({\emph{with}} \gls{hira})} or {wait {(line {26})} exactly the same amount of time as performing \gls{hira} would take {({\emph{without}} \gls{hira})}}. Fourth, we hammer {both aggressor} rows $HC/2$ times {(lines {30--33})}.} {If \gls{hira}'s second row activation is \emph{not} ignored,} {then the victim row would be refreshed, and thus} we {would} observe a significant increase in measured RowHammer threshold values {in the test {\emph{with} HiRA}, compared to the {test \emph{without} HiRA}}. {Fifth, we read the victim row to check if the RowHammer test causes any bitflip (line 36).}

\figref{hira:fig:norm_hcfirst} {shows} how {a DRAM row's} RowHammer threshold varies when the row is refreshed using \gls{hira}. \figref{hira:fig:norm_hcfirst}{a and~\ref{hira:fig:norm_hcfirst}b show the histogram of absolute and normalized RowHammer threshold values, respectively.} 
We report the normalized values relative to {tests without} \gls{hira}. 

\begin{figure}[!ht]
    \vspace{1em}
    \centering
    \includegraphics[width=0.8\linewidth]{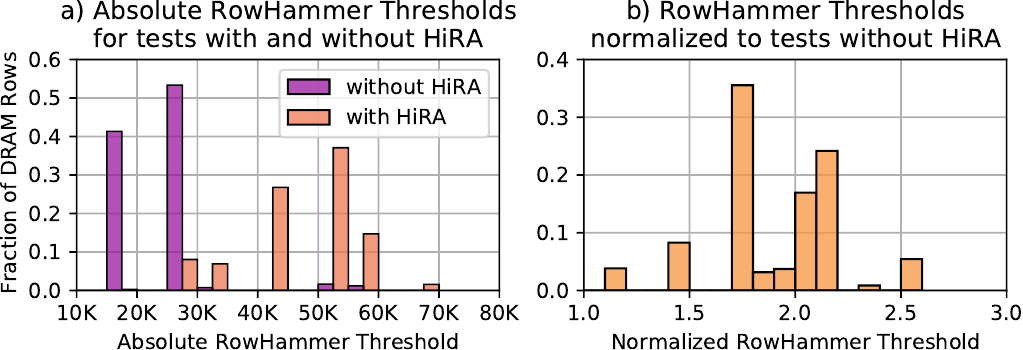}
    \caption{Variation in RowHammer threshold {due to} HiRA{'s second {row} activation}}
    \label{hira:fig:norm_hcfirst}
\end{figure}

{We make two observations. First, \figref{hira:fig:norm_hcfirst}a shows that RowHammer threshold is $27.2K$ / $51.0K$ on average across tested rows when tested \emph{without} / \emph{with} \gls{hira}.
Second, \figref{hira:fig:norm_hcfirst}b shows that} RowHammer threshold increases {by $1.9\times$ on average across tested DRAM rows and} by more than 1.7$\times$ for the vast majority {(\SI{88.1}{\percent})} of {tested} rows. {Based on these {two observations},} we conclude that {\gls{hira} works in 56 tested DRAM chips (\tabref{hira:tab:dram_chip_list}) such that} \gls{hira}{'s second row activation, targeting the victim row, is \emph{not} ignored, and thus the victim row is successfully activated} {concurrently} with {the dummy row}.

\subsection{Variation Across DRAM Banks}
\label{hira:sec:variation_across_banks}
To {investigate} the variation in \gls{hira}'s {coverage and verify \gls{hira}'s second row activation} across DRAM banks, {we repeat {the} tests {that we explain in \secref{hira:sec:experiments_coverage} and \secref{hira:sec:experiments_chargerestoration}} for {\emph{all}} 16 banks {of}} three DRAM modules{: A0, B0, and C0 {(Table~\ref{hira:tab:dram_chip_list})}.}

\subsubsection{HiRA's Coverage}
We observe that the pairs {of rows that {\gls{hira}} can {concurrently refresh and activate}} are \emph{identical} across {all 16} DRAM banks in all three modules.
{Based on this observation, we {hypothesize} that \gls{hira}'s coverage {largely} depends on the DRAM circuit design, which should be {a design-induced phenomenon} across all DRAM banks.}

\subsubsection{{Verifying} HiRA's {Second Row Activation}}
{To verify that HiRA's second row activation works across all 16 DRAM banks, we repeat the tests shown in \algref{hira:algorithm:rowhammer}.
\figref{hira:fig:bank_variation} shows how a DRAM row's RowHammer threshold varies when the victim row is activated using \gls{hira}'s second activation during a RowHammer attack (similar to \figref{hira:fig:norm_hcfirst}b). The x-axis and different box colors show the module's name and DRAM bank, respectively. The y-axis shows the measured RowHammer threshold in the tests \emph{with \gls{hira}}, normalized to the tests \emph{without \gls{hira}}.}
{E}ach box in \figref{hira:fig:bank_variation} shows the distribution's \gls{iqr} and whiskers show the minimum and maximum values.{\footref{hira:fn:boxplot}}

\begin{figure}[!ht]
    \centering
    \includegraphics[width=0.8\linewidth]{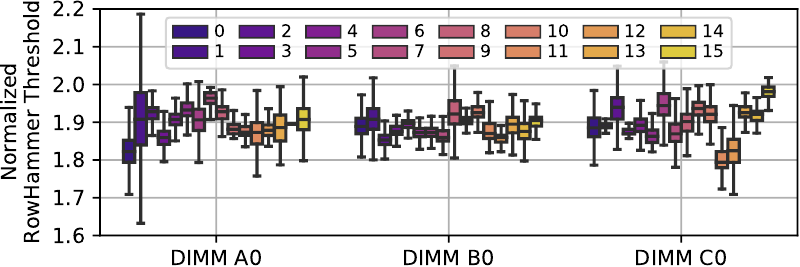}
    \caption{Variation in {normalized RowHammer threshold across banks {in three modules} {due to HiRA's second row activation}}}
    \label{hira:fig:bank_variation}
\end{figure}

{We make three observations from \figref{hira:fig:bank_variation}. First, the {normalized RowHammer threshold values are larger than \SI{1.56}{\times} {across \emph{all} banks in \emph{all} three DRAM modules}. Second, RowHammer threshold increases by \SI{1.89}{\times}, average{d} across all banks in all three modules{,} when the victim row is refreshed using \gls{hira}. {Third,} the average RowHammer threshold increase in a DRAM bank varies between \SI{1.80}{\times} and \SI{1.97}{\times} {across \emph{all} banks in \emph{all} three modules}.
{Therefore, we conclude that \gls{hira}'s second row activation is \emph{not} ignored in \emph{any} bank.}}}

\subsection{Summary of Experimental Results}
\label{hira:sec:appendix_chips}
\providecommand*{\myalign}[2]{\multicolumn{1}{#1}{#2}}

\agyarb{Table~\ref{hira:tab:detailed_info} shows the characteristics of the DDR4 DRAM modules we test and analyze. {We provide {the} access frequency (Freq.), manufacturing date (Date Code), {chip} capacity (Chip Cap.), die revision {(Die Rev.)}, and chip organization {(Chip Org.)} of tested DRAM modules. We report the manufacturing date of these modules in the form of $week-year$. For each DRAM module, Table~\ref{hira:tab:detailed_info} shows two \gls{hira} characteristics in terms of minimum (Min.), average (Avg.) and maximum (Max.) values across all tested rows: 
1)~HiRA Coverage: the fraction of DRAM rows
within a bank which HiRA can reliably activate concurrently
with refreshing a given row (\secref{hira:sec:experiments_coverage})
and 2)~Norm. $N_{RH}$: the increase in the RowHammer threshold when \gls{hira}'s second row activation is used for refreshing the victim row (\secref{hira:sec:experiments_chargerestoration}).}}
\setlength{\tabcolsep}{5.5pt}
\begin{table*}[ht]
\footnotesize
\centering
\caption{Characteristics of the tested DDR4 DRAM modules.}
\label{hira:tab:detailed_info}
\resizebox{\columnwidth}{!}{
\begin{tabular}{l|l|l||ccccc|ccc|ccc}
\head{Module} & \textbf{Module} & \textbf{Module Identifier} & \textbf{Freq} & \textbf{Date} & \textbf{Chip} & \textbf{Die} & \textbf{Chip} & \multicolumn{3}{c|}{\textbf{HiRA Coverage}} & \multicolumn{3}{c}{\textbf{Norm. $N_{RH}$}} \\ 
\head{Label}&\textbf{Vendor}&\textbf{Chip Identifier}&\textbf{(MT/s)}&\textbf{Code}&\textbf{Cap.}&\textbf{Rev.}&\textbf{Org.}&\textbf{Min.}&\textbf{Avg.}&\textbf{Max.}&\textbf{Min.}&\textbf{Avg.}&\textbf{Max.}\\
\hline
\hline
A0 & \multirow{2}{*}{G.SKILL} & \multirow{2}{*}{\begin{tabular}[l]{@{}l@{}}DWCW (Partial Marking)$^*$\\F4-2400C17S-8GNT~\cite{gskill2021f42400c17s8gnt}\end{tabular}} & \multirow{2}{*}{2400} & \multirow{2}{*}{42-20} & \multirow{2}{*}{4Gb} & \multirow{2}{*}{B} & \multirow{2}{*}{x8} & \SI{24.8}{\percent} & \SI{25.0}{\percent} & \SI{25.5}{\percent} & 1.75&1.90&2.52\\
A1 & & & & & & & & \SI{24.9}{\percent} & \SI{26.6}{\percent} & \SI{28.3}{\percent} & 1.72 & 1.94 & 2.55 \\ \hline
B0 & \multirow{2}{*}{Kingston} & \multirow{2}{*}{\begin{tabular}[l]{@{}l@{}}H5AN8G8NDJR-XNC\\KSM32RD8/16HDR~\cite{datasheetksm32rd8}\end{tabular}} & \multirow{2}{*}{2400} & \multirow{2}{*}{48-20} & \multirow{2}{*}{4Gb} & \multirow{2}{*}{D} & \multirow{2}{*}{x8} & \SI{25.1}{\percent} & \SI{32.6}{\percent} & \SI{36.8}{\percent} & 1.71&1.89&2.34\\
B1 & & & & & & & & \SI{25.0}{\percent} & \SI{31.6}{\percent} & \SI{34.9}{\percent} & 1.74&1.91&2.51\\ \hline
C0 & \multirow{3}{*}{SK Hynix} & \multirow{3}{*}{\begin{tabular}[l]{@{}l@{}}H5ANAG8NAJR-XN\\HMAA4GU6AJR8N-XN~\cite{memorynethmaa4gu6ajr8nxn}\end{tabular}} & \multirow{3}{*}{2400} & \multirow{3}{*}{51-20} & \multirow{3}{*}{4Gb} & \multirow{3}{*}{F} & \multirow{3}{*}{x8} & \SI{25.3}{\percent} & \SI{35.3}{\percent} & \SI{39.5}{\percent} & 1.47 & 1.89 & 2.23 \\ 
C1 & & & & & & & & \SI{29.2}{\percent} & \SI{38.4}{\percent} & \SI{49.9}{\percent} & 1.09 & 1.88 & 2.27\\ 
C2 & & & & & & & & \SI{26.5}{\percent} & \SI{36.1}{\percent} & \SI{42.3}{\percent} & 1.49 & 1.96 & 2.58\\ \hline
\hline
\end{tabular}
}
\begin{flushleft}
$^*$ The chip identifier is partially removed on these modules. We infer the chip manufacturer and die revision based on the remaining part of the chip identifier.

\end{flushleft}
\end{table*}
\section{HiRA-MC: HiRA Memory Controller}
\label{hira:sec:implementation}

{\Gls{hirasched} {aims} to improve overall system performance. {To do so, \gls{hirasched} queues each periodic and preventive refresh request with a deadline and takes one of three possible actions in decreasing priority order: 1)~{concurrently perform a} refresh operation with a \memacc{} {(refresh-access parallelization)} before the refresh operation's deadline; 2)~{concurrently perform a} refresh operation with another refresh operation {(refresh-refresh parallelization)} if no \memacc{} can be parallelized until the refresh operation's deadline; or 3)~perform a refresh operation {by} its deadline if the refresh {operation \emph{cannot}} be {concurrently performed} with {a} \memacc{} {or} another refresh.} \gls{hirasched} intelligently schedules refresh operations from within the memory controller \emph{without} {requiring any} modification to off-the-shelf DRAM chips.}

\figref{hira:fig:mechanism} shows \gls{hirasched}'s components and {their} interaction with the memory request scheduler. {First, we give an overview of \gls{hirasched} where we introduce its components.} Then, we explain how \gls{hirasched}'s {components} interact with the memory request scheduler in {performing} four {key operations}.
\begin{figure}[t]
    \centering
        \includegraphics[width=0.8\linewidth]{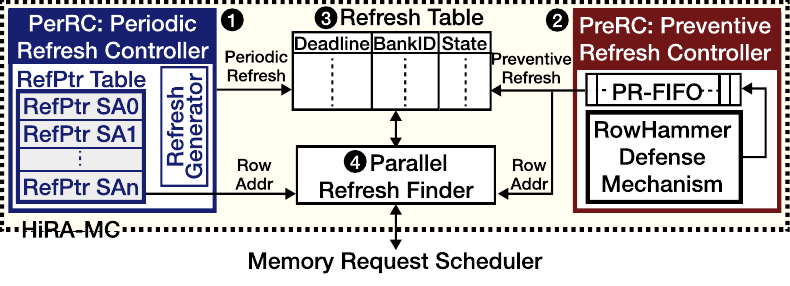}
    \caption{HiRA-MC's components}
    \label{hira:fig:mechanism}
\end{figure}

{\head{\gls{hirasched} Overview}} {\gls{hirasched} consists of four main components: \emph{Periodic Refresh Controller ({PeriodicRC})}, \emph{Preventive Refresh Controller ({PreventiveRC})}, \emph{Refresh Table}, and \emph{{Concurrent} Refresh Finder}.} 
{\circled{1}~\emph{{PeriodicRC}} generates {a \emph{periodic}} refresh request {for each DRAM row} to maintain data integrity {in {the} presence of} DRAM cell charge leakage. {To leverage \gls{hira}'s {subarray-level} parallelism, {{PeriodicRC}} maintains a table called \emph{RefPtr table}. RefPtr table has an entry per subarray, which contains a pointer to the next row to be refreshed within the corresponding subarray.} 
\circled{2}~\emph{{PreventiveRC}} employs a refresh-based RowHammer defense mechanism {(e.g., PARA~\cite{kim2014flipping})} to generate {a} \emph{preventive} refresh request {for a {victim} DRAM row.}
{{There {might not be \emph{any} memory access} requests {that can be} parallelize{d} with a periodic or preventive refresh} {when} {the} refresh request is generated {(i.e., there {might not be \emph{any}} load or store memory requests waiting to be served by the memory controller)}. To address this issue,} both {PeriodicRC} and {PreventiveRC} {assign {each refresh} request a \emph{deadline} that specifies the {timestamp until which} the refresh request \emph{must} be {performed.
The deadline is determined using a configuration parameter called \gls{trrslack}.}
{\circled{3}~\emph{The Refresh Table} stores generated periodic and preventive refresh requests along with their deadline, {target} bank id, and {refresh type} (invalid, periodic, or preventive).}}}
{\circled{4}~\emph{The {Concurrent} Refresh Finder} identifies the refresh requests that can be parallelized with \memacc{} requests among the refresh requests stored in the Refresh Table.} 
To serve a refresh {request} {concurrently} with a \memacc{} request, {the {Concurrent} Refresh Finder} observes the \memacc{} requests that the memory request scheduler\footnote{{The \emph{memory request scheduler} is the component {that is} responsible for scheduling {DRAM requests, using a scheduling algorithm (e.g., FR-FCFS~\cite{rixner2000memory, rixner2004memory, zuravleff1997controller} or PAR-BS~\cite{mutlu2008parallelismaware})}, {and issuing} DRAM commands to serve {those} requests{.} 
}} issues. If there is a refresh request {that} can be parallelized with {a} {memory} request, the {Concurrent} Refresh Finder replaces the {memory} request's row activation command with a \gls{hira} operation, such that \gls{hira}'s first ACT targets the {row {that needs to be refreshed} and \gls{hira}'s second ACT targets the row {that needs to be accessed}}. 
If \gls{hirasched} \emph{cannot} perform a pending refresh request {concurrently} with a \memacc{} until the refresh request's deadline, {the {Concurrent} Refresh Finder} searches for another refresh request {within the Refresh Table} to parallelize {the refresh request with}. If possible, \gls{hirasched} performs a \gls{hira} operation to {concurrently} refresh two rows. {If {the} refresh {request} \emph{cannot} be parallelized with an access or another refresh}, \gls{hirasched} activates the row that needs to be refreshed and precharges {the bank} using nominal timing parameters.  

\subsection{{HiRA-MC: Key Operations}}

\subsubsection{Generating Periodic Refresh Requests} 
{The Periodic Refresh Controller} periodically {{generates refresh requests}}. {{{PeriodicRC}} faces two main challenges in scheduling \gls{hira} operations due to two fundamental differences between \gls{hira} and \gls{ref} operations. 
First, a \gls{ref} command refreshes several rows in a DRAM bank as a batch~\cite{jedec2020jesd794c,jedec2020jesd795,liu2012raidr}{. In contrast,} using the \gls{hira} operation, the memory controller needs to issue an $ACT$ command for each {refreshed} DRAM row. Therefore, using \gls{hira} increases {DRAM} bus utilization compared to using \gls{ref} commands. For example, to refresh 64K rows in a bank of a DDR4 DRAM chip in \SI{64}{\milli\second}, the memory controller issues 8K \gls{ref} commands (once every \SI{7.8}{\micro\second}~\cite{jedec2020jesd794c}), {indicating} that each \gls{ref} command refreshes eight rows {in one bank}. To ensure the same refresh rate {as the baseline}, {{PeriodicRC}} schedules 64K \gls{hira} operations (once every \SI{975}{\nano\second}).} 
{Second, issuing a \gls{ref} command triggers refresh operations in \emph{all} banks in a rank {(assuming all-bank refresh, as in DDR4~\cite{jedec2020jesd794c})}. {In contrast}, \gls{hira} operations are performed separately for each DRAM bank because they use already defined $ACT$ and $PRE$ commands {at} row- and bank-level, respectively.} 
Therefore, {frequently issued} \gls{hira} command sequence{s} can occupy the command bus more than \gls{ref} commands. {{{For example, \gls{hirasched} needs} to perform 128 \gls{hira} operations {to refresh the same number of rows as one \gls{ref} command does {in current systems},} assuming {that a single \gls{ref} command refreshes eight rows from each of the 16 banks in a rank as in DDR4~\cite{jedec2020jesd794c}.}}} 
To avoid overwhelming the command bus with \gls{hira} {operations}, {{PeriodicRC}} spreads the command bus pressure of \gls{hira} command sequences over time {by generating} {\gls{ref} {requests}} for each bank {with the same period, starting at different time offsets{. For example, assuming that
1)~each bank receives a \gls{ref} request once in every \SI{975}{\nano\second} and 2)~there are 16 banks,}
{{PeriodicRC}} {generates a refresh request every \SI{60.9}{\nano\second} (\SI{975}{\nano\second}/16 banks) targeting a different bank}.}
{{PeriodicRC}} inserts {the generated} {\gls{ref} request} into the Refresh Table {with {the request's}} {1)~\emph{deadline}, which is a timestamp pointing {to} the time that is \gls{trrslack} later than the request's generation time,}
{2)~\emph{BankID}, which is the target bank of the refresh request, and 3)~{\emph{refresh type}}, which is set to \emph{Periodic} to indicate that the refresh request will perform a \emph{periodic} refresh operation.}

\subsubsection{Generating {Preventive} Refresh Requests {for {RowHammer}}}
\gls{hirasched} is not a RowHammer defense mechanism by itself, but {it provides parallelism support for} all memory controller-based {preventive} refresh mechanisms, which mitigate {the} RowHammer {effect} on victim rows by {timely} refreshing the victim rows~\rhdefrefresh{}.
{\gls{hirasched} overlaps} the latency of a {preventive} refresh operation with {another periodic/preventive refresh} or a \memacc{}. 
{To {do so, {{PreventiveRC}} generates preventive refreshes with a large enough} \gls{trrslack} without compromising the security guarantees {of RowHammer defense mechanisms}. To achieve this,}
{{PreventiveRC} assumes the worst case{,} where an attack fully exploit{s} \gls{trrslack} to maximize {the} {hammer} count such that the attack performs \gls{trrslack}/\gls{trc} additional activations during \gls{trrslack} (after the preventive refresh is generated and before it is performed). To account for {such case},}
{{{PreventiveRC}} {employs state-of-the-art RowHammer defense mechanisms\cite{park2020graphene, kang2020cattwo, lee2019twice, seyedzadeh2018mitigating, qureshi2022hydra, kim2022mithril,kim2014flipping, son2017making, you2019mrloc} {with a slightly increased aggressiveness in performing preventive refreshes}.
When implemented in {PreventiveRC}, the \emph{counter-based} RowHammer defense mechanisms~\cite{park2020graphene, kang2020cattwo, lee2019twice, seyedzadeh2018mitigating, qureshi2022hydra, kim2022mithril} should be configured such that the mechanism triggers a preventive refresh at a {hammer} count that is \gls{trrslack}/\gls{trc} activations smaller than the mechanism's original {hammer} count threshold {(typically, the {RowHammer threshold} of the DRAM module)}. Therefore, even if the attack performs the maximum number of activations after the preventive refresh is generated, the total {hammer} count does not exceed the RowHammer defense mechanism's original threshold before the preventive refresh is performed.
When implemented in {PreventiveRC}, the \emph{probabilistic} RowHammer defense mechanisms~\cite{kim2014flipping, son2017making, you2019mrloc} should be configured with an increased probability threshold to maintain the same security guarantees as the original {mechanism} in the presence of \gls{trrslack}. \secref{hira:sec:security_analysis} explains how to increase the probability threshold to account for \gls{trrslack}.}}

When the employed RowHammer defense mechanism generates a {preventive} refresh {request}, {{PreventiveRC}} 1)~enqueues the refresh operation in a first-in-first-out queue, called \emph{PR-FIFO} {(\circled{2} in \figref{hira:fig:mechanism})}, 2)~creates an entry in the Refresh Table {(\circled{3} in \figref{hira:fig:mechanism})} with the preventive refresh's deadline and bank id, and 3)~{sets the request's {refresh type} to \emph{Preventive} to indicate that the refresh request will perform a \emph{preventive} refresh operation}.

\subsubsection{Finding {Concurrent} Refresh Operations}
\label{hira:sec:finding-concurrent-refresh}
{\figref{hira:fig:mechanism_flowdiagram} shows how \gls{hirasched}'s {Concurrent} Refresh Finder interacts with the memory request scheduler {in two different cases: 1)~when the memory request scheduler issues a \gls{pre} command to prepare the bank for activating a $RowA$ (\circled{1} in \figref{hira:fig:mechanism_flowdiagram}) and 2)~when an internal timer periodically initiates a process that performs refreshes by their deadline 
(\circled{4} in \figref{hira:fig:mechanism_flowdiagram}).}}

\begin{figure}[ht]
    \centering
    \includegraphics[width=0.8\linewidth]{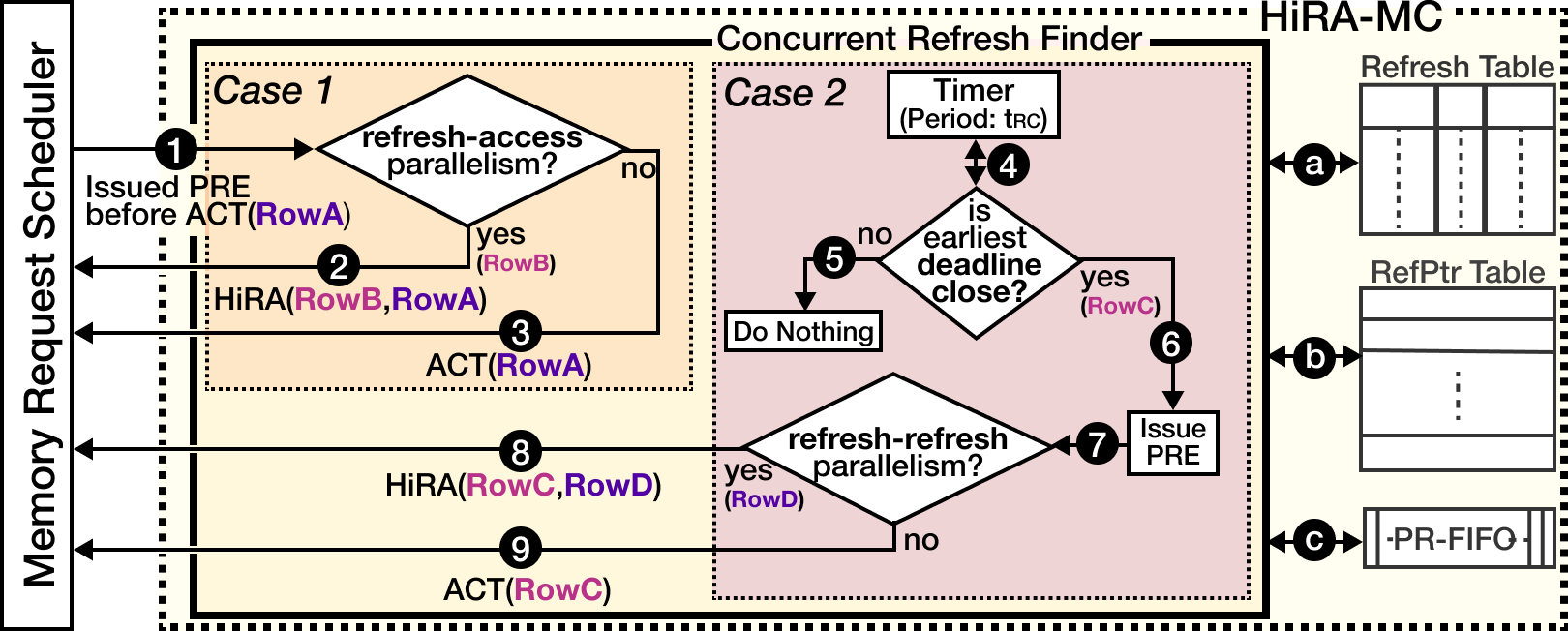}
    \caption{The Concurrent Refresh Finder's interaction with the memory request scheduler}
    \label{hira:fig:mechanism_flowdiagram}
\end{figure}

{\head{Case~1}}
To find an opportunity {to} {concurrently refresh a DRAM row} with a \memacc{} {(refresh-access parallelism)}, {the {Concurrent} Refresh Finder} observes the commands that the memory request scheduler issues.
{When the memory request scheduler issues a \gls{pre} command to pre{charge} the DRAM bank {before} activating a DRAM row ($RowA$)~\circled{1}, the Concurrent Refresh Finder} {searches} the Refresh Table ({by iterating over the Refresh Table entries} in {the} order {of increasing deadlines) {to find a refresh operation} {that targets} the precharged bank}~(\circled{a}).

{{If the Refresh Table entry with the {earliest} deadline is a periodic refresh,}}
{{the Concurrent Refresh Finder} {accesses} the RefPtr Table~(\circled{b}) {to find} {a} subarray} 
{1)~{where} refreshing a DRAM row} can be parallelized with the activation {of $RowA$} {{and 2)~that has the smallest number of DRAM rows} refreshed within the current refresh window.}
By doing so, \gls{hirasched} aims {to} advance the refresh pointers of all subarrays in a balanced manner {while leveraging the subarray-level parallelism.}

{If the Refresh Table entry with the {earliest} deadline is a preventive refresh,}
{the Concurrent Refresh Finder} checks whether the request at the head of {P}R-FIFO can be {refreshed concurrently} with {activating $RowA$}~(\circled{c}).

If \gls{hirasched} finds a periodic or preventive refresh {{that targets} $RowB$} {and can be} {concurrently perform{ed}} with an {activation to $RowA$~(\circled{2})}, the memory request scheduler issues {a} \gls{hira} {operation} such that the first {and the second} \gls{act} {commands} target {$RowB$ and {$RowA$}, respectively{, so that $RowB$ is refreshed concurrently with activating $RowA$.}} 

If there is \emph{no} opportunity {{to}} {concurrently perform a periodic or preventive refresh with the activation of $RowA$ (\circled{3}),}
the memory request scheduler {issues} a regular \gls{act} {command} targeting $RowA$. {In this case, 1)~the {DRAM row} activation is performed \emph{without} leveraging \gls{hira}'s parallelism} 
{because refresh-access parallelism is not possible and 2)~the} {memory access request is prioritized over the queued refresh requests}
{to improve system performance when queued refresh requests can be delayed until their deadline.}

{\head{Case 2}}
{To guarantee that each periodic and preventive refresh is performed {timely (i.e., by its deadline)}, the Concurrent Refresh Finder periodically checks if there is a refresh operation that is close to its deadline (i.e., whose deadline is {earlier} than \gls{trc})~(\circled{4}). If there is {\emph{no} queued periodic or preventive refresh whose deadline is close~(\circled{5}), the Concurrent Refresh Finder does \emph{not} take any action{. In doing so, \gls{hirasched}} 1)~does \emph{not} interfere with the memory access requests and 
2)~{opportunistically leaves refresh requests in the Refresh Table such that {the refresh requests} can be concurrently performed with a memory access by their deadlines.}}

If there is a refresh {request {(targeting a $RowC$)}} that needs to be performed {within the next \gls{trc} time window~(\circled{6}),} the Concurrent Refresh Finder precharges {the target} bank {of the refresh operation} if {the bank} is open and~(\circled{7})~tries leveraging refresh-refresh parallelism {by searching for a queued refresh operation that can be concurrently performed with refreshing $RowC$.}
{If {there is a refresh request {(}targeting a $RowD${),} which can be concurrently performed with refreshing $RowC$ (refresh-refresh parallelism), \gls{hirasched} forces the memory request scheduler to issue a \gls{hira} operation such that the two activations of \gls{hira} target $RowC$ and $RowD$, {respectively}~(\circled{8}).}
{If such refresh-refresh parallelism is not possible~(\circled{9}),} the memory request scheduler issues a regular \gls{act} command {targeting $RowC$} to perform the refresh operation {because 1)~refresh-refresh parallelism is \emph{not} possible and 2)~delaying $RowC$'s refresh {would violate its deadline and could} have caused bitflips}.}}

\subsubsection{Maintaining the Parallelism Information}
{T}o know {if a} DRAM row can be {concurrently} activated with {another} DRAM row, the memory controller needs to {determine   
whether the two rows are located in subarrays {that} do \emph{not} share {bitlines} or sense {amplifiers}.}
The memory controller can {learn
which subarrays do not share {bitlines or a sense amplifiers} with another subarray {(i.e., determine the \emph{boundaries of a subarray})}} {in two ways. First, the memory controller can} perform a one-time reverse engineering (e.g., by testing for HiRA's coverage as we do in \secref{hira:sec:experiments_coverage}). {Second}, DRAM manufacturers can {expose this information} to the memory controller by {using} mode status registers (MSRs)~\cite{jedec2020jesd794c} in the DRAM chip.
{Once the memory controller obtains the subarray boundaries, it} maintains them
in a table {called \emph{Subarray {P}airs {T}able (SPT)}} {implemented as an on-chip {SRAM} storage}. {{SPT has an entry for each subarray ($S_A$). {{This entry}} contains a list of subarrays which do \emph{not} share bitlines or sense amplifiers with $S_A$. Therefore, \gls{hira} operation can be performed targeting a DRAM row in $S_A$ and another DRAM row in any of the listed subarrays.}}

\subsection{Power Constraints}
Each refresh and row activation in a HiRA operation is counted as a row activation with respect to {the} \gls{tfaw} constraint {in DDRx DRAM chips}. We respect \gls{tfaw} in our performance evaluation, such that within a given \gls{tfaw}, at most four activations are performed in a DRAM rank {(as required by {DRAM} datasheets {(e.g.,~\cite{jedec2020jesd794c,jedec2020jesd795,jedec2017jesd2094b,jedec2020jesd2095a})})}, {thereby ensuring} that the row activations are performed within the power budget of a DRAM rank.

\subsection{Compatibility with Off-the-{Shelf} DRAM Chips}
We experimentally demonstrate that \gls{hira} works on {all} 56 {real} DRAM chips {that we test} {(\secref{hira:sec:characterization})};
and \gls{hirasched} does \emph{not} require any modifications to these {real} DRAM chips to {enable} refresh-refresh and refresh-access parallelization.

\subsection{Compatibility with {Different Computing} System{s}}
{We discuss \gls{hirasched}'s compatibility with three {types} of computing systems: 1)~FPGA-based systems {(e.g., PiDRAM \cite{olgun2022pidram})}, 2)~contemporary processors, and 3)~systems with programmable memory controllers~{\cite{hussain2014pmss, bojnordi2012pardis}}. First, {\gls{hirasched} can be easily integrated into all existing FPGA-based systems that use DRAM to store data~{\cite{olgun2022pidram, xilinx2021adaptable, xilinx2021fpgas}} {{by} implementing \gls{hirasched} in RTL}.}}
{Second, contemporary processors require modifications to their memory controller logic to implement \gls{hirasched}. Implementing \gls{hirasched} is a design-time decision that requires balancing manufacturing cost with periodic and {preventive} refresh overhead reduction benefits. 
{We show that \gls{hirasched} significantly improves system performance (\secref{hira:sec:doduoref} and~\secref{hira:sec:doduopara}) at low chip area cost (\secref{hira:sec:hardware_complexity}) and thus {can be relatively easily integrated into contemporary processors}.}
Third, systems that employ programmable controllers~\cite{hussain2014pmss,bojnordi2012pardis} can be {relatively} easily modified to implement \gls{hirasched} {by} programming {the \gls{hira} operation and implementing \gls{hirasched}'s components using the ISA of programmable memory controllers~\cite{hussain2014pmss,bojnordi2012pardis}.}}

\section{Hardware Complexity}
\label{hira:sec:hardware_complexity}

{We evaluate the hardware complexity of implementing \gls{hirasched} in a processor,} {using CACTI 7.0~\cite{balasubramonian2017cacti}} to model {\gls{hirasched}'s components (}Refresh Table{, RefPtr Table, and} {P}R-FIFO). We use {CACTI's} \SI{22}{\nano\meter} {technology node to model SRAM array{s} for each component's on-chip data storage.} {Table~\ref{hira:tab:scheduler-area} shows the area cost {and the access latency} of each component.}

\begin{table}[h] 
    \caption{{The} area {cost (per DRAM rank)} {and access latency of HiRA-MC's components}}
    \label{hira:tab:scheduler-area}
    \centering
    \footnotesize
    \begin{tabular}{l r r r}
        \textbf{\gls{hirasched} Component} & \textbf{Area ($mm^2$)} & \textbf{Area (\SI{}{\percent}{\textsuperscript{*}})} & \textbf{Access Latency}\\
    \hline
    Refresh Table & {$0.00031$} & <\SI{0.0001}{\percent} & \SI{0.07}{\nano\second}\\
    RefPtr Table & $0.00683$ & \SI{0.0017}{\percent} & \SI{0.12}{\nano\second}\\
    PR-FIFO & $0.00029$ & <\SI{0.0001}{\percent} & \SI{0.07}{\nano\second}\\
    {Subarray Pairs Table (SPT)} & {$0.00180$} & \SI{0.0005}{\percent} & \SI{0.09}{\nano\second}\\
    \hline
    {Overall} & {$0.00923$} & {\SI{0.0023}{\percent}} & {\textsuperscript{**}}{\SI{6.31}{\nano\second}}\\

    \end{tabular}
    \footnotesize
    \\
    \textsuperscript{*}{Normalized to the die area of} a 22nm Intel processor~\cite{wikichipcore}.
    \\
    \textsuperscript{**}Calculated as the {overall latency of serially accessing 1)~the SPT, 2)~the Refresh Table, {and 3)~the RefPtr Table for {68} times}} {(\secref{hira:sec:hirasched-access-latency})}.
\end{table}

\head{Refresh Table} {In this analysis, we assume a \gls{trrslack} of 4\gls{trc} {because 1)~increasing \gls{trrslack} increases the hardware complexity of \gls{hirasched} (by increasing the number of entries in the Refresh Table and the PR-FIFO) and 2)~a \gls{trrslack} of 4\gls{trc} {already} provides {as large} performance benefit {as a \gls{trrslack} of 8\gls{trc}} (\secref{hira:sec:doduoref} and~\secref{hira:sec:doduopara}).}
Within a time window of {{4}}\gls{trc}, {\gls{hirasched} can generate at most} {{4}} {periodic {refresh requests per \emph{rank}} and {4} preventive} refresh requests per \emph{{b}ank} {(64 preventive refresh requests per \emph{rank})}. Therefore, a Refresh Table with {{68}} entries {per rank}  is enough to store all generated refresh requests.}
Each entry {consists} of 1) 1{0} bits to store the deadline,\footnote{
{A {10}-bit number can represent the number of clock cycles within a \gls{trrslack} of 4\gls{trc} (\SI{185}{\nano\second}~\cite{jedec2020jesd794c}), assuming a memory controller clock frequency of \SI{3}{\giga\hertz}.}}
2) 4 bits to store the bank id, and 3) 2 bits to store the refresh {type} (Periodic, Preventive, or Invalid). {Our analysis shows that Refresh Table consumes \emph{only} \SI{0.00031}{\milli\meter\squared} chip area per rank and can be accessed in}
\SI{0.07}{\nano\second}.

{\head{RefPtr Table} We model a 2048-entry RefPtr Table (128 entries per bank and 16 banks {per} rank). We assume that there can be as many as 1024 rows in a DRAM subarray. Thus, each RefPtr Table entry contains 10 bits to point to a row in a subarray.} {Based on our analysis,} {RefPtr Table{'s size is \SI{0.00683}{\milli\meter\squared} chip area per rank and {it} can be accessed in} \SI{0.12}{\nano\second}}.

\head{PR-FIFO}
We model a {4}-entry PR-FIFO per DRAM bank, {assuming the worst case, where the RowHammer defense mechanism generates a preventive refresh for {every} performed row activation}.
{PR-FIFO's chip area cost is \SI{0.00029}{\milli\meter\squared} per rank and access {latency is} \SI{0.07}{\nano\second}.}

\head{Subarray Pairs Table (SPT)} 
{For 128 subarrays per bank, our} analysis shows that this table can be accessed in \SI{0.09}{\nano\second} and consumes {only \SI{0.0018}{\milli\metre\squared}} {chip area} {per DRAM rank.}

\subsection{HiRA-MC's Overall Area Overhead and Access Latency}
\label{hira:sec:hirasched-overall}
\label{hira:sec:hirasched-access-latency}

{\gls{hirasched} takes only {$0.00923mm^2$} chip area {per DRAM rank}. This area corresponds to \SI{0.0023}{\percent} of the chip area of a \SI{22}{\nano\meter} Intel processor~\cite{wikichipcore}.}

In the worst-case, \gls{hirasched} traverses the Refresh Table to search for refresh-access parallelization opportunities. {During traversal, within a \gls{trp} time window,} \gls{hirasched} accesses the Refresh Table and {the} SPT 6{8}~times {to iterate over all Refresh Table entries. {{Iterating over} {Refresh Table and SPT {entries}} in a pipelined manner results in an {overall} latency of} {\SI{6.19}{\nano\second}}. {If \gls{hirasched} finds a periodic refresh request, it accesses} RefPtr Table once to {get the address of the row that needs to be refreshed}
{(see \secref{hira:sec:finding-concurrent-refresh})}}{, which takes \SI{0.12}{\nano\second}}. {If \gls{hirasched} finds a preventive refresh, it accesses the head of the PR-FIFO, which takes \SI{0.07}{\nano\second}.}
{Therefore}, the overall access latency of \gls{hirasched} is \SI{6.31}{\nano\second}, {which is significantly smaller than the nominal \gls{trp} of \SI{14.5}{\nano\second}.}
{W}e conclude that {\gls{hirasched} completes all search operations {with} a significantly smaller latency than {the latency of} a precharge operation,} 
and thus it does \emph{not} {cause} additional latency for memory {access} requests.

\section{Evaluation Methodology}
\label{hira:sec:usecases}

We evaluate \gls{hirasched} {via} two case studies {focusing on high-density DRAM chips}: 1)~refreshing very high capacity DRAM chips and 2)~protecting DRAM chips with {high} RowHammer vulnerability. We demonstrate for each {study} that \gls{hirasched} {significantly} improves system performance by {leveraging \gls{hira}{'s} {ability to} {concurrently refresh a row {while} refreshing or activating another row.}}

\head{{Simulation Environment}}
To evaluate the performance impact of \gls{hirasched} under each use-case, we conduct {cycle-level simulations}, using Ramulator~\cite{safariramulator, Kim2016Ramulator}.
Our baseline leverages the regular {rank-level} \gls{ref} commands, periodically issued at every \gls{trefi} {with a latency of \gls{trfc} in respect to} DDR4 specifications~\cite{jedec2020jesd794c}.
Table~\ref{hira:tab:system_configuration} {shows the simulated system configuration.} 
In our evaluations, we assume a realistic system with \ncores{} cores, connected to a memory rank with four bank groups each of which contains four banks (16~banks in total). 
{We execute \wlcnt{} 8-core multiprogrammed workloads, randomly chosen from SPEC CPU2006~\cite{stspec} benchmarks. We simulate these
workloads until each core executes 200M instructions with a warmup period of 100M instructions, similar to prior work~\cite{kim2020revisiting, yaglikci2021blockhammer}.}
The {memory controller} employs {the} FR-FCFS~\cite{rixner2000memory, rixner2004memory, zuravleff1997controller} scheduling algorithm with {the} open-{row} policy. 
{We assume that a refresh to a DRAM row can be served concurrently with a refresh or an access to \covref{hira:} of the rows within the same DRAM bank,}
based on our experimental results {(\secref{hira:sec:experiments_coverage})}. We {measure system performance in terms of} weighted speedup~\cite{eyerman2008systemlevel, snavely2000symbiotic}.

 \newcolumntype{C}[1]{>{\let\newline\\\arraybackslash\hspace{2pt}}m{#1}}
 \begin{table}[ht]
 \vspace{2mm}
 \caption{{Simulated} system configuration}
 \footnotesize
\centering
 \begin{tabular}{l|l}
 \hline
 \multirow{1}{*}{\textbf{Processor}} & 3.2GHz, 8~core, 4-wide issue, {128-entry} {instr.} window\\ \hline
 \textbf{Last-Level Cache} & {64-byte} cache line, 8-way {set-associative, 8MB} \\ \hline
 \multirow{3}{*}{\textbf{Memory Controller}} & 64-entry each read and write request queues\\&Scheduling policy: FR-FCFS~\cite{rixner2000memory, rixner2004memory, zuravleff1997controller}\\& Address mapping: MOP~\cite{kaseridis2011minimalist} \\ \hline
 \multirow{2}{*}{\textbf{Main Memory}} & DDR4-2400~\cite{jedec2020jesd794c}, 1 channel{$^*$}, 1 rank{$^*$}, 4 bank groups\\&4 banks/bank group {(16 banks per rank)}, {64K} rows/bank\\ \hline
 \multirow{2}{*}{\textbf{Timing Parameters}} & $t_1=t_2=3ns$, $t_{RC}=46.25ns$, $t_{FAW}=16ns$\\& $ {t_{RefSlack}} \in \{0, 2t_{RC}, 4t_{RC}, 8t_{RC}\}$ \\ \hline\hline
 \end{tabular}
 \begin{flushleft}
 \vspace{-0.5em}
 {$^*$\secref{hira:sec:doduoref} and~\secref{hira:sec:doduopara} assume a 1-channel 1-rank system. \secref{hira:sec:sensitivity} presents a sensitivity analysis for 1, 2, 4, and 8 channels / ranks.}
 \end{flushleft}
 \label{hira:tab:system_configuration}
 \end{table}
 
{\head{Adapting Baseline Refresh for High-Capacity DRAM Chips}}
Across different generations of DRAM protocols~\cite{jedec2020jesd2095a, jedec2020jesd795, jedec2020jesd794c, jedec2017jesd2094b, jedec2021jesd235d} the minimum and maximum \glsfirst{trefi} does not significantly change, while the standards allow the manufacturer to define the necessary \glsfirst{trfc} value based on the time required to complete a refresh operation. 
{As DRAM capacity increases, more DRAM rows need to be refreshed when a \gls{ref} command is issued, {which increases} \gls{trfc}~\cite{nguyen2018nonblocking}.}
{To estimate \gls{trfc} for a given density, we use the state-of-the-art regression model~\cite{nguyen2018nonblocking} for projecting \gls{trfc} with increased {chip capacity ($C_{chip}$), {as shown in Expression~\ref{hira:exp:trfc}}:}}
\vspace{-1em}

\begin{equation}
\trfc{} = 110 \times C_{chip}^{0.6}
\label{hira:exp:trfc}
\end{equation}

\newif\iflmh
\lmhfalse

\section{{Periodic Refresh Results}}
\label{hira:sec:doduoref}

{We evaluate \gls{hira}'s performance when \gls{hira} is used for performing periodic refreshes in high{-}capacity DRAM chips instead of {using conventional} \gls{ref} commands {used in current systems}.}
We sweep DRAM chip {capacity} and {quantify} the performance overhead of periodic refresh operations {{on} {a} baseline {system that performs rank-level \gls{ref} operations} and four configurations of \gls{hira} {with different deadlines}}. 
\figref{hira:fig:multi_core_refresh} demonstrates system performance (y-axis) for different DRAM chip capacities from \SI{2}{Gb} to \SI{128}{Gb} (x-axis).

We denote \gls{hira}'s different configurations as \gls{hira}-N, where N {specifies} {the} {\gls{trrslack} configuration} 
in terms of the number of row activations that can be performed {within a \gls{trrslack}, i.e., \gls{trrslack} of \gls{hira}-N is {$N\times{}\trc{}$}.}

For example, \gls{hira}-2 {schedules} each refresh request with a {\gls{trrslack}} of {$2\times{}\trc{}$}{, whereas \gls{hira}-0 schedules refresh requests with a {\gls{trrslack}} of 0 (i.e., the refresh operation must be performed {immediately after it is generated by \gls{hirasched}})}.
{\figref{hira:fig:multi_core_refresh}a shows system performance in terms of weighted speedup, normalized to an ideal system that we call \emph{No Refresh}, where the system does \emph{not} need to perform any periodic refreshes.}

\begin{figure}[!ht]
    \centering
    \includegraphics[width=0.7\linewidth]{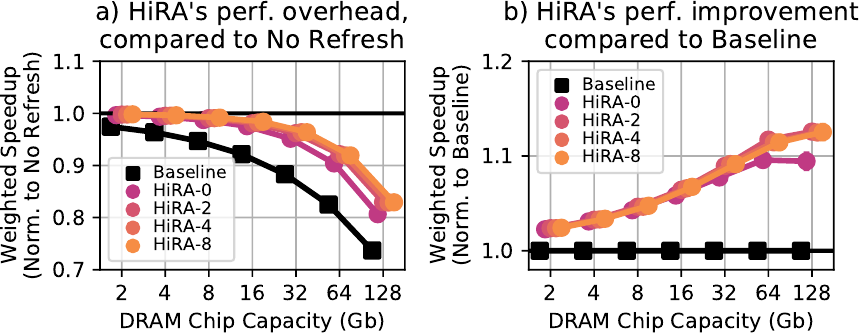}
    \caption{{HiRA's impact on system performance for 8-core multiprogrammed workloads with increasing {DRAM} chip capacity, compared to a)~an ideal system called \emph{No Refresh} that performs \emph{no} periodic refreshes and~b)~{the} Baseline system that uses conventional REF commands to perform periodic refreshes}}
    \label{hira:fig:multi_core_refresh}
\end{figure}

We make {two} observations {from \figref{hira:fig:multi_core_refresh}a.}
First, {using \gls{ref} commands to perform} periodic refresh operations {({as done in} baseline)} significantly degrades system performance as DRAM chip {capacity} increases. {For example,} periodic refresh operations cause \param{\SI{26.3}{\percent}} system performance degradation for refreshing \param{128Gb} DRAM chips on average across all {evaluated} workloads.
Second, \gls{hira} (\gls{hira}-2) significantly reduces the performance degradation caused by periodic refresh operations by \periodicreduction{} (from {\SI{26.3}{\percent}} down to \param{\SI{17.0}{\percent}}), on average across all {evaluated} workloads {for a DRAM chip capacity of 128Gb}. 

{\figref{hira:fig:multi_core_refresh}b shows system performance in terms of weighted speedup, normalized to the \emph{baseline}. We make three observations from \figref{hira:fig:multi_core_refresh}b.}
{First, \gls{hira} significantly improves system performance. For example, \gls{hira}-2 provides \SI{12.6}{\percent} performance improvement over the baseline on average across all evaluated workloads for a DRAM chip capacity of 128Gb.}
{{Second}, \gls{hira}'s performance benefits increase with \gls{trrslack} {up to a certain {value} of \gls{trrslack}}. For example, for a DRAM chip capacity of 128Gb, \gls{hira}-0 and \gls{hira}-2 provide \SI{9.4}{\percent} and \SI{12.6}{\percent} performance improvement over {the} baseline, respectively.} This is because as \gls{trrslack} increases, \gls{hirasched} can find more opportunities to perform each queued refresh operation concurrently with refreshing or accessing another DRAM row. We observe that increasing \gls{trrslack} from {$2\times{}\trc{}$} to $8\times{}\trc{}$ does \emph{not} significantly improve system performance (i.e., curves for \gls{hira}-2, \gls{hira}-4, \gls{hira}-8 overlap with each other) on average across all {evaluated} workloads. This is because a \gls{trrslack} of {$2\times{}\trc{}$} is large enough to {perform periodic refreshes concurrently with memory accesses or other refreshes}.
{{Third},} \gls{hira}'s performance improvement increases with DRAM chip capacity. For example, \gls{hira}-2's performance improvement increases from \SI{2.4}{\percent} for 2Gb chips to \SI{12.6}{\percent} for 128Gb chips on average across all {evaluated} workloads.

{Based on these observations, we conclude that \gls{hira} significantly improves system performance by reducing the performance overhead of periodic refresh operations{, and HiRA's benefits increase with DRAM chip capacity.}}

\section{RowHammer {Preventive Refresh Results}}
\label{hira:sec:doduopara}

Modern DRAM chips, including the ones that are marketed as RowHammer{-safe~\rhsafe{}}, are shown to be even more vulnerable to RowHammer {(at the circuit level)} than their predecessors~\rhworse{}. Therefore, {it is critical for a RowHammer defense mechanism to efficiently scale} with worsening RowHammer vulnerability. Among {many} RowHammer defense mechanisms {(e.g.,~\rhdef{})}, we {find} \emph{Probabilistic Row Activation (PARA)}~\cite{kim2014flipping} as the most lightweight and {hardware-}scalable RowHammer defense {due to} two reasons.
First, PARA{'s hardware cost does \emph{not} increase when {it is} scaled {to work on chips that have} higher RowHammer vulnerability{. This is} because PARA} {is a \emph{stateless} mechanism {that} refreshes a} potential victim row with a low probability, {defined as} \gls{pth}, when a DRAM row is activated{,} {with \emph{no} need for maintaining {any} metadata}.
Second, PARA{,
as a memory controller-based mechanism which is implemented solely in the processor chip,} easily adapts to the RowHammer {vulnerability} of a {given} DRAM chip by programming \gls{pth} after {the processor is} {deployed}. {Unlike PARA,} other defenses {(e.g.,~\rhdef{})} are {usually} configured for a particular RowHammer vulnerability {level} at {the} processor chip's design time and {they} cannot be easily reconfigured for {a {new} DRAM chip's} {RowHammer} vulnerability. {This is because these mechanisms require implementing as many {hardware} counters as needed to accurately identify a RowHammer attack for a given RowHammer threshold, and thus they {likely} need more counters to reliably work for smaller RowHammer thresholds; unfortunately, the number of {hardware} counters cannot be {easily} {increased} after deployment}. 

{When scaled {to work on a DRAM chip {that has a} higher} RowHammer vulnerability, PARA refreshes a victim row with a {higher} probability, {thereby} inducing a larger {system} performance overhead~{\cite{kim2020revisiting,yaglikci2021blockhammer}}. \gls{hirasched} {reduces PARA's} {performance overhead}, by leveraging the parallelism \gls{hira} provides. \secref{hira:sec:security_analysis} explains how we configure \gls{pth} when PARA is used with \gls{hira}. Then, \secref{hira:sec:uctwo_eval} evaluates {\gls{hira}'s performance when it is used for performing PARA's preventive refreshes.}}

\subsection{Security Analysis} 
\label{hira:sec:security_analysis}

{\subsubsection{Threat Model}
We assume a comprehensive {RowHammer} threat model,\cqlabel{hira:CQ4} similar to {that assumed by} state-of-the-art works~\cite{kim2014flipping, park2020graphene, yaglikci2021blockhammer}, in which the attacker can 1)~fully utilize memory bandwidth, 2)~precisely time each memory request, and 3)~comprehensively and accurately know details of the memory controller and the DRAM chip. We do not consider any hardware or software component to be trusted or safe except {we assume that the DRAM commands issued by the memory controller are performed within the DRAM chip as intended.}}

\subsubsection{Revisiting PARA's configuration methodology}
\label{hira:sec:revisitingpara}
Kim et al.~\cite{kim2014flipping} {configure} PARA's probability threshold (\gls{pth}), assuming {that} the attacker hammers an aggressor row {\emph{only enough times, but no more}}.
{With} more than an order of magnitude decrease in {the} {RowHammer threshold} in the last decade\cite{kim2020revisiting, frigo2020trrespass, hassan2021uncovering, orosa2021adeeper, yaglikci2022understanding}, an attacker can {complete a RowHammer attack 144 times {within a refresh window of \SI{64}{\milli\second}},\footnote{{\Gls{hcfirst} for modern DRAM chips is as low as 9600~\cite{kim2020revisiting}. Assuming a \gls{trc} of \SI{46.25}{\nano\second}, performing 9600 row activations can be completed \emph{only} in \SI{444}{\micro\second}, which is 1/144.14 of a nominal refresh window of \SI{64}{\milli\second}. {As such, an attacker can perform 9600 activations for 144 times within a \SI{64}{\milli\second} refresh window.}}}}
{requiring {a} revisit {of PARA's configuration methodology}.\footnote{A concurrent work also revisits PARA's configuration methodology~\cite{saroiu2022howto}.}}

{The rest of this section explains how we {calculate}} \gls{pth} for a given RowHammer threshold.
\figref{hira:fig:para_fsm} shows the probabilistic state machine that we use to 
calculate the {hammer count, which we define as the number of aggressor row activations} that may affect a victim row.Initially the {hammer count} is zero (state~0). When an {aggressor row} {is} activated, PARA {triggers a \emph{preventive} refresh with a probability of $\pth{}$. Since there are two rows that are adjacent to the activated row, PARA} refreshes {the} victim row with a probability of $\pth{}/2$, {in which case the {hammer} count is reset.} {Therefore,} the {hammer} count {{is incremented} with a probability of $1-\pth{}/2$ upon an {aggressor row} activation}. An attack is considered to be \emph{successful} if {its hammer count} reaches the RowHammer threshold {($N_{RH}$)} within a \glsfirst{trefw}. 

\begin{figure}[!ht]
    \centering
    \includegraphics[width=0.8\linewidth]{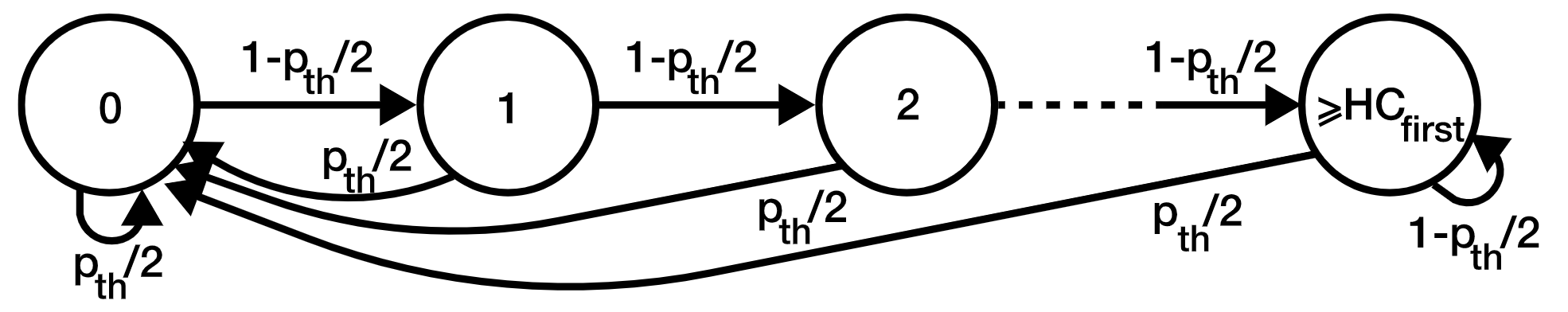}
    \caption{Probabilistic state machine of {hammer} count in a PARA-protected system}    \label{hira:fig:para_fsm}
\end{figure}

{Because the time delay between two row activations targeting the same {bank} \emph{cannot} be smaller than \gls{trc}, an} attacker can perform {a maximum of} $\trefw{}/\trc{}$ state transitions within a \gls{trefw}. 
To account for all possible access patterns, we model a \emph{successful RowHammer {access pattern}} as a set of failed attempts{, where} the victim row is refreshed before the {hammer} count reaches {the} RowHammer threshold, followed by a successful attempt, {where} the victim row is {\emph{not}} refreshed until the {hammer} count reaches {the} RowHammer threshold. {To {calculate} \gls{pth}{,} we follow a {five-}step approach. First, we calculate the probability of a failed attempt ($p_f$) and a successful attempt. Second, we calculate {the probability of observing a number ($N_f$) of consecutive failed attempts.}Third, we calculate
{\emph{the {overall} RowHammer success probability ($p_{RH}$)} as the overall probability of {\emph{all} possible successful RowHammer access patterns.}}
Fourth, we extend the probability calculation to account for \gls{trrslack}. Fifth, we calculate \gls{pth} for a given failure probability target.}

\head{{Step {1}: Failed and successful attempts}}
\expref{hira:exp:failedattempt} {shows} $p_{f}(HC)$: the probability of a {\emph{failed} attempt with a given hammer count ($HC$). The attempt contains 1)~$HC$ consecutive aggressor row activations that do \emph{not} trigger a preventive} refresh, $(1-\pth{}/2)^{HC}$, {where $HC$ is smaller than the RowHammer threshold ($N_{RH}$)}, {and 2)~}an {aggressor row} activation {that triggers} a {preventive} refresh $(\pth{}/2)$.

\equ{p_{f}(HC) = (1-\pth{}/2)^{HC}\times\pth{}/2,~where~1\leq HC<N_{RH}}{hira:exp:failedattempt}

{Similarly, we calculate the probability of a \emph{successful} attempt which has $N_{RH}$ consecutive aggressor row activations that do \emph{not} trigger a preventive refresh {as} $(1-p_{th}/2)^{N_{RH}}$.}

\head{{Step {2}: {The probability of} ${\bm{N_f}}$ {consecutive} failed attempts}}
{{Since a failed attempt may have a hammer count value in the range $[1, N_{RH})$, we account} for all possible hammer count {values} that {a failed attempt might have}.}
{\expref{hira:exp:nconsecutivefailedattempts} shows the probability of {a given number of (}$N_f${)} consecutive failed attempts.}
\equ{\prod_{i=1}^{N_{f}} p_f(HC_i)=\prod_{i=1}^{N_{f}}((1-\pth{}/2)^{HC_i}\times\pth{}/2)~, where~1\leq HC_{i} < \hcfirst{}}{hira:exp:nconsecutivefailedattempts}

\head{{Step 3: {Overall} RowHammer success probability}}
{To find the overall RowHammer success probability, we 1)~calculate the probability of the \emph{successful RowHammer access pattern {($p_{success}(N_{f})$),} which consists of $N_f$ consecutive failed attempts and one successful attempt} for each possible value that $N_f$ can take and 2)~sum the probability of all possible successful RowHammer access patterns: $\sum{p_{success}(N_f)}$.}
To do so, {we} multiply \expref{hira:exp:nconsecutivefailedattempts} with the probability of a successful attempt: $(1-p_{th}/2)^{N_{RH}}$.
{\expref{hira:exp:successfulattack}} shows {how we calculate $p_{success}(N_{f})$.} We derive \expref{hira:exp:successfulattack} by evaluating the product on both terms in \expref{hira:exp:nconsecutivefailedattempts}: $p_{th}/2$ and $(1-p_{th}/2)^{HC_i}$.

\equ{p_{success}(N_{f}) = (1-{\pth{}}/{2})^{\sum_{i=1}^{N_{f}} {HC_{i}}} \times ({\pth{}}/{2})^{N_{f}} \times (1-{\pth{}}/{2})^{\hcfirst{}}}{hira:exp:successfulattack}

To account for the worst-case, we maximize 
$p_{success}(N_{f})$ {by choosing} the worst possible value for each $HC_{i}$ {value}. \textbf{Intuitively}, {the number of activations in a failed attempt should be minimized. Since} a failed attempt {has to perform at least one activation,
we conclude that all failed attempts} fail after only one row activation in the worst case. 
\head{Mathematically,} {our goal is} to maximize $p_{success}(N_{f})$. {{Since} $\pth{}$ is a value between zero and one,} we minimize the term $\sum_{i=1}^{N_{f}} {HC_{i}}$ to maximize $p_{success}(N_{f})$. Thus, we derive \expref{hira:exp:prhn} {by choosing $HC_i=1$ to achieve} {the maximum (worst-case) $p_{success}(N_{f})$.}

\equ{
p_{success}(N_{f}) = (1-{\pth{}}/{2})^{N_{f}+\hcfirst{}} \times ({\pth{}}/{2})^{N_{f}}}{hira:exp:prhn}

{{\expref{hira:exp:prh}} shows {{the overall} RowHammer success probability} ($p_{RH}$), as the sum of all possible $p_{success}(N_f)$ values.}
{$N_{f}$ can be as small as 0 if {the} RowHammer attack does \emph{not} include any failed attempt and as large as} the maximum number of failed attempts that can fit in a refresh window {(\gls{trefw}) {together with a successful attempt}}.
Since $HC_i=1$ in the worst case, each failed attempt {costs only two row} activations ($2\times\trc{}$){:} {one aggressor row activation and one activation for preventively refreshing the victim row}. Thus, the execution time of $N_{f}$ {failed} attempts, {followed by one successful attempt is} $(2N_{f}+\hcfirst{})\times\trc{}$. {Therefore,} the maximum {value $N_{f}$ can take within a refresh window (\gls{trefw})} is ${N_{f}}_{max} = ((\trefw{}/\trc{})-\hcfirst{})/2$.

\equ{p_{RH} = \sum_{N_{f}=0}^{{N_{f}}_{max}} p_{success}(N_{f}),~~~{N_{f}}_{max} = (\trefw{}/\trc{}-\hcfirst{})/2}{hira:exp:prh}

Using {\expsref{hira:exp:prhn}} and~\ref{hira:exp:prh}, we compute the {{overall} RowHammer success probability} for a given {PARA} probability threshold ($p_{th}$).

\head{{Step 4: Accounting for \gls{trrslack}}}
The original PARA proposal~\cite{kim2014flipping} performs a {preventive} refresh immediately after the activated row is closed. However, \gls{hirasched} allows a {preventive} refresh to be queued {for a time window as long as \gls{trrslack}}. {Since the aggressor rows can be activated while a preventive refresh request is queued, we update {~\expsref{hira:exp:prhn}} and~\ref{hira:exp:prh}, assuming the worst case, where an aggressor row is activated as many times as possible {within \gls{trrslack} ({i.e.,} the maximum amount of time} the preventive refresh is queued).}
To do so, we {update} \expref{hira:exp:prh}: {we reduce} {the RowHammer threshold ($N_{RH}$)} by {\gls{hcdeadline}}. Thus, we calculate ${N_{f}}_{max}$ and $\prh{}$ as shown in \expsref{hira:exp:nmax} and~\ref{hira:exp:prhdl}, respectively. 

\begin{equation}
{N_{f}}_{max} = ((\trefw{}/\trc{})-\hcfirst{}-{\hcdeadline{}})/2
\label{hira:exp:nmax}
\end{equation}

\equ{p_{RH} = \sum_{N_{f}=0}^{{N_{f}}_{max}} (1-{\pth{}}/{2})^{{N_{f}}+\hcfirst{}-{\hcdeadline{}}} \times ({\pth{}}/{2})^{N_{f}}}{hira:exp:prhdl}

\head{{Step 5: Finding $\bf{p_{th}}$}}
{We iteratively evaluate \expsref{hira:exp:nmax} and \ref{hira:exp:prhdl} to find} \gls{pth} for a {target} {{overall} RowHammer success probability of {$10^{-15}$}, as a typical consumer memory reliability target {(see, e.g.,~\cite{cai2012error, cai2017error, jedec2012jep122g, luo2016enabling, patel2017thereach, yaglikci2021blockhammer, kim2020revisiting, micheloni2015apparatus})}}.

\subsubsection{{Results}}

{We refer to {the} original PARA work~\cite{kim2014flipping} as PARA-Legacy. PARA-Legacy calculates {{the overall} RowHammer success probability} as ${p_{RH_{Legacy}}=(1-p_{th}/2)^{N_{RH}}}$ with an optimistic assumption that the attacker hammers an aggressor row \emph{only enough times, but no more}. To mathematically compare {{the overall} RowHammer success probability} that we calculate ($p_{RH}$) with $p_{RH_{Legacy}}$, we reorganize \expref{hira:exp:prhdl}, which already includes ${p_{RH_{Legacy}}=(1-p_{th}/2)^{N_{RH}}}$, and derive \expref{hira:exp:prhdiff}.}

\equ{
p_{RH} =k \times p_{RH_{Legacy}},~where~k = (1-\frac{\pth{}}{2})^{-{\hcdeadline}}\sum_{N_{f}=0}^{{N_{f}}_{max}}(\frac{\pth{}}{2}(1-\frac{\pth{}}{2}))^{{N_{f}}}
}{hira:exp:prhdiff}

{
\expref{hira:exp:prhdiff} shows that $p_{RH}$ {is a multiple of} $p_{RH_{Legacy}}$ by a factor of $k$, {where $k$ depends on a given system's properties (e.g., $N_{f_{max}}$)} and PARA's configuration (e.g., \gls{pth}).
{To understand the difference between $p_{RH}$ and $p_{RH_{Legacy}}$, we evaluate \expref{hira:exp:prhdiff} for different RowHammer threshold values.\footnote{{We calculate for $N_{RefSlack}=0$, $t_{REFW}=64ms$, and $t_{RC}=46.25ns$.}}}
For {old DRAM chips (manufactured in} 2010-2013)~\cite{kim2014flipping}, $k$ is 1.0005 {(for $N_{RH}=50K$ and $p_{th}=0.001$~\cite{kim2014flipping})}, causing \emph{only} \SI{0.05}{\percent} variation in PARA's reliability target. {However,for future DRAM chips with $N_{RH}$ values of 1024 and 64 {($p_{th}$ values of 0.4730 and 0.8341)}, $k$ becomes 1.0331 and 1.3212, respectively. Therefore, the difference between {the two probabilities, $p_{RH}$ and $p_{RH_{Legacy}}$,} significantly increases as RowHammer worsens.}}

\figref{hira:fig:para_thresholds}{a} show{s} how {decreasing RowHammer threshold, i.e., worsening RowHammer vulnerability {(x-axis)}, change{s}} {PARA's {probability} threshold (\gls{pth}) {(y-axis)}}. {The dashed curve shows} PARA-Legacy's probability threshold (calculated using ${p_{RH_{Legacy}}=(1-p_{th}/2)^{N_{RH}}}$), {whereas the other curves show $p_{th}$ for different \gls{trrslack}} {values which we} {calculate} using {\expref{hira:exp:prhdl}}.

\begin{figure}[!ht]
    \centering
    \includegraphics[width=0.8\linewidth]{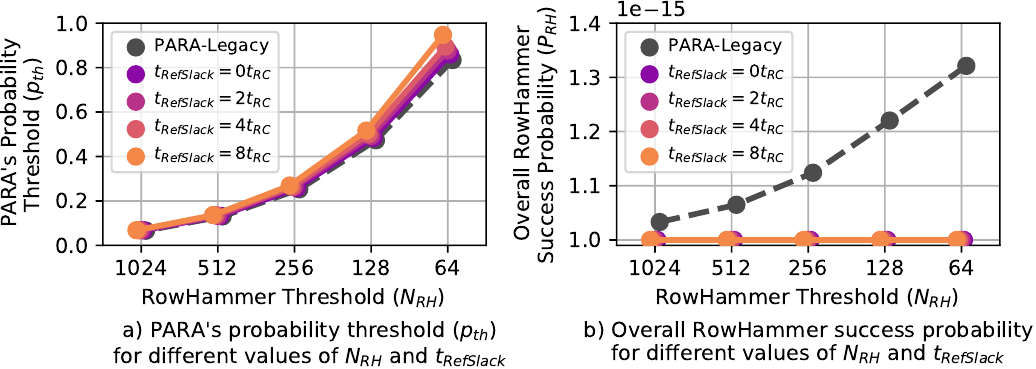}
    \caption{{PARA configurations for different RowHammer thresholds ($N_{RH}$) and $t_{RefSlack}$ values: a)~PARA's probability threshold ($p_{th}$}) and b)~{{overall} RowHammer success probability} ($p_{RH}$)}
    \label{hira:fig:para_thresholds}
\end{figure}

{We make two observations from \figref{hira:fig:para_thresholds}a.} First, to maintain {a $10^{-15}$} RowHammer {success} {probability}, $p_{th}$ significantly increases for smaller RowHammer thresholds.
{For example,} $p_{th}$ {increases} from 0.068 to 0.860 {(\gls{trrslack}=0)} when the RowHammer threshold reduces from 1024 to 64. This is because as the RowHammer threshold reduces, fewer activations {are} enough for an attack to induce bitflips. {Thus,} PARA needs to perform preventive refreshes more aggressively.
{Second,}
$p_{th}$ {increases with} \gls{trrslack}, e.g., when {the} RowHammer threshold is 128, $p_{th}$ should be 0.48, 0.49, 0.50, and 0.52 for {\gls{trrslack} values of 0, 2\gls{trc}, 4\gls{trc}, and 8\gls{trc}, respectively.} 
This is because a larger {\gls{trrslack} {allows reaching a higher hammer count, requiring PARA to perform preventive refreshes more aggressively.}} 

\figref{hira:fig:para_thresholds}{b} show{s} how {decreasing RowHammer threshold ($N_{RH}$) change{s}} {the {{overall} RowHammer success probability} ($p_{RH}$).} {We calculate \emph{all} $p_{RH}$ values by evaluating \expref{hira:exp:prhdl} using the $p_{th}$ values in \figref{hira:fig:para_thresholds}a.} {The dashed curve shows} PARA-Legacy's $p_{RH}$, {whereas the other curves show PARA's $p_{RH}$ for different \gls{trrslack} configurations.}
{We make two observations from \figref{hira:fig:para_thresholds}b.}
{{First}, configuring $p_{th}$ as described in PARA-Legacy~\cite{kim2014flipping} {1)~}results in a larger {{overall} RowHammer success probability} than {the consumer memory reliability target ($10^{-15}$)}, and {2)~}the difference between $p_{RH}$ and the $10^{-15}$ increases as the RowHammer vulnerability increases (i.e., the RowHammer threshold reduces). For example, the $p_{th}$ values {that PARA-Legacy calculates}
{targeting a $10^{-15}$ {overall} RowHammer success probability} {for RowHammer thresholds of 1024 and 64,} result in {{overall} RowHammer success probability {values} of} {$1.03\times{}10^{-15}$} and {$1.32\times{}10^{-15}$}, {respectively.}
This happens because PARA-Legacy assumes that the attacker performs \emph{{only}} as many {aggressor row} activations as the RowHammer threshold ($N_{RH}$) within a refresh window, even though {increasingly} more {aggressor} row activations can {be performed} in a refresh window {as $N_{RH}$ reduces}.}
{Second}, {the} {$p_{th}$ values} {that we calculate using \expref{hira:exp:prhdl}} significantly reduce the {{overall} RowHammer success probability} {compared to PARA-Legacy} {(and maintains a $p_{RH}$ of $10^{-15}$ across all RowHammer thresholds)} {because \expref{hira:exp:prhdl}, {to calculate $p_{RH}$,} takes into account all aggressor row activations that can be performed in a refresh window.}

{{W}e conclude that as RowHammer threshold decreases, PARA-Legacy{'s $p_{th}$ values} result in a significantly larger {{overall} RowHammer success probability} than the {consumer memory reliability target ($10^{-15}$)}, while $p_{th}$ {values} calculated using \expref{hira:exp:prhdl} maintain the {{{overall} RowHammer success probability} at $10^{-15}$}. }

\subsection{Performance of PARA with HiRA}
\label{hira:sec:uctwo_eval}
{We evaluate \gls{hira}'s performance {benefits} when it is used {to} perform PARA's preventive refreshes. {We} us{e} the evaluation methodology described in \secref{hira:sec:usecases}. {We calculate \gls{pth} values \expref{hira:exp:prhdl}.}
{{W}e evaluate PARA's {impact on} performance when it is used \emph{with} and \emph{without} \gls{hira} for different RowHammer thresholds}.}

\figref{hira:fig:para_perf_multicore}{a} shows {the performance of 1)~a system that implements PARA without \gls{hira}, labeled as PARA, and 2)~a system that implements PARA with {four different \gls{trrslack} configurations of} \gls{hira}, labeled {using the HiRA-N notation (\gls{trrslack} of HiRA-N is $N\times{}\trc{}$)} as HiRA-{0, HiRA-2, HiRA-4, and HiRA-8}, {normalized to {the} baseline that does \emph{not} perform any preventive refresh operations (i.e., does \emph{not} implement PARA).}}

\begin{figure}[!ht]
    \centering
    \includegraphics[width=0.8\linewidth]{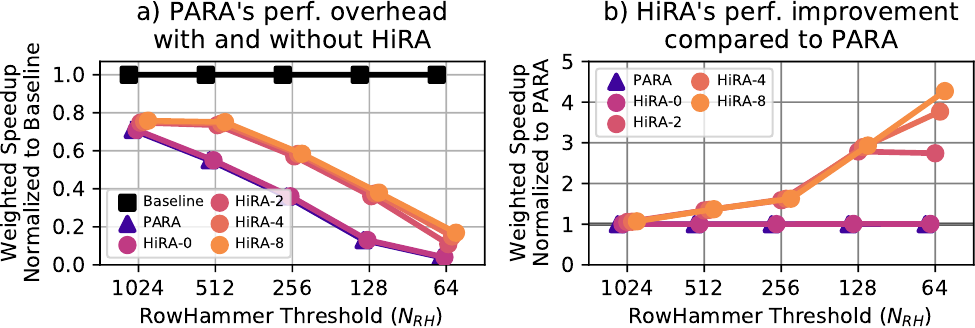}
    \caption{HiRA's {impact on} system performance {for 8-core multiprogrammed workloads with increasing RowHammer vulnerability (i.e., {decreasing} $N_{RH}$) compared to a) Baseline system with \emph{no} RowHammer defense and b) a system that implements PARA.}}
    \label{hira:fig:para_perf_multicore}
\end{figure}

{From \figref{hira:fig:para_perf_multicore}a, we observe that}
PARA induces {\SI{29.0}{\percent}} slowdown on system performance on average across all {evaluated} workloads when it is configured for a RowHammer threshold of 1024. \gls{hira}-2 reduces PARA's performance overhead down to {\SI{25.2}{\percent}}, which results in a performance improvement of {\SI{5.4}{\percent}} compared to PARA.
Similarly,  when configured for a RowHammer threshold of 64, \gls{hira}-4 {increases system performance by $3.73\times$ compared to PARA {as it} reduces PARA's performance overhead by \reactivereduction{} (from} \SI{96.0}{\percent} down to \SI{85.1}{\percent}{)}. 
{This happens because HiRA reduces the latency of preventive refreshes by concurrently performing them with refreshing or accessing other rows in the same bank.}

{\figref{hira:fig:para_perf_multicore}b shows the performance of the system that implements PARA with \gls{hira} (labeled using the HiRA-N notation), normalized to the performance of the system that implements PARA without \gls{hira} (labeled as PARA).}
{We make two observations from \figref{hira:fig:para_perf_multicore}b.}
{First, \gls{hira}'s performance improvement increases with higher RowHammer vulnerability, i.e., smaller RowHammer threshold.}
{For example, when compared to PARA (\figref{hira:fig:para_perf_multicore}b), \gls{hira}-2 provides a speedup of $2.75\times$ on average across all {evaluated} workloads when RowHammer threshold is 64, which is significantly larger than \gls{hira}-2's performance improvement of \SI{5.4}{\percent} when RowHammer threshold is 1024.}
{This is because PARA generates preventive refreshes more aggressively as RowHammer threshold reduces (\secref{hira:sec:revisitingpara}), which increases PARA's memory bandwidth utilization and provides HiRA with a larger number of preventive refreshes to parallelize with other accesses and refreshes.}
{Second, configuring \gls{hira} with a larger \gls{trrslack} improves system performance. For example, when {the} RowHammer threshold is \SI{64}{\milli\second}, \gls{hira}-0, \gls{hira}-2, \gls{hira}-4, and \gls{hira}-8 improve system performance by \SI{0.6}{\percent}, $2.75\times$, $3.73\times$, and $4.23\times$, respectively, on average across all {evaluated} workloads, compared to PARA without \gls{hira} (\figref{hira:fig:para_perf_multicore}b). This happens because \gls{hirasched} can find a parallelization opportunity for a queued preventive refresh with a larger probability when there is a larger \gls{trrslack}.}

{Based on {our} observations, we conclude that \gls{hira} significantly reduces PARA's system performance degradation.}

\section{\sho{Sensitivity Studies}}
\label{hira:sec:sensitivity}
\sho{We analyze how HiRA's performance changes {with} 1)~number of channels and 2)~number of ranks per channel. To {evaluate} high-end system configurations, we sweep the number of channels and ranks from one to eight, {inspired by} commodity systems~\cite{intel3rdgen, amdepyc, micron2020tn4040, jedec2020jesd794c, micron2014sdram, micron2014tn4003}.}

\subsection{HiRA with Periodic Refresh}
\sho{\figref{hira:fig:ref_channel} shows how increasing \shom{the} number of channels {(x-axis)} affects HiRA's performance for two configurations (HiRA-2 and HiRA-4), compared to the baseline, where rows are periodically refreshed using rank-level $REF$ command. {The y-axis shows system performance in terms of {average} weighted speedup across 125 evaluated workloads,} normalized to the baseline's performance at \shom{the} 1-channel 1-rank configuration.}
{Three subplots show the results for 2Gb (left), 8Gb (middle), and 32Gb (right) {DRAM} chip capacity.}

\begin{figure}[!ht]
    \centering
    \includegraphics[width=0.8\linewidth]{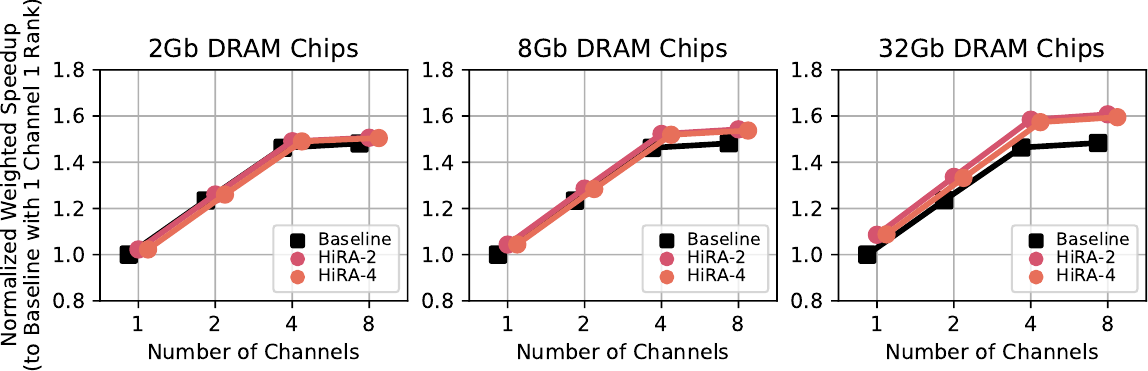}
    \caption{\sho{Effect of \shom{channel count} on system performance for Baseline and HiRA}}
    \label{hira:fig:ref_channel}
\end{figure}

\sho{We make three observations.
First, both {the} baseline and HiRA {provide} \shom{higher} performance with \shom{more} channels. For example, HiRA and {the} baseline exhibit speedups of $1.60\times$ and $1.48\times$, \shom{respectively,} when the number of channels \shom{increases} from one to eight for a DRAM chip {capacity} of 32Gb. This is because memory-level parallelism {increases with more channels}. Performance overhead\shom{s} of rank-level refresh and HiRA do \emph{not} increase with \shom{more} channels because different channels do not share command, address, or data bus{es}, thereby allowing {different channels} to be \shom{accessed} simultaneously.  
Second, {at smaller channel counts,} the effect of channel count {on performance} is greater. For example, the slope\shom{s} of the line plots are steeper in-between one and four channels than in-between four and eight channels. This happens because {the evaluated} workloads do not exhibit \shom{sufficient} memory-level parallelism to fully leverage the available parallelism \shom{with} {more than} four channels. 
Third, both HiRA-2 and HiRA-4 configurations exhibit {significant} speedup {over} {the} baseline for \emph{all} channel counts. For example, \gls{hira}-2 improves the performance of a system using 32Gb DRAM chips with eight channels by \SI{8.1}{\percent} compared to {the baseline with} 8-channels.} 
{We conclude that \gls{hira} provides significant performance benefits for high-capacity DRAM chips {even with a large number of channels}.}

{{\figref{hira:fig:ref_rank}} shows how increasing \shom{the} number of {ranks} (x-axis) affects HiRA's performance benefits. The y-axis shows system performance using the same metric as \figref{hira:fig:ref_channel} uses. {Three subplots show the results for 2Gb (left), 8Gb (middle), and 32Gb (right) {DRAM} chip capacity.}}

\begin{figure}[!ht]
    \centering
    \includegraphics[width=0.8\linewidth]{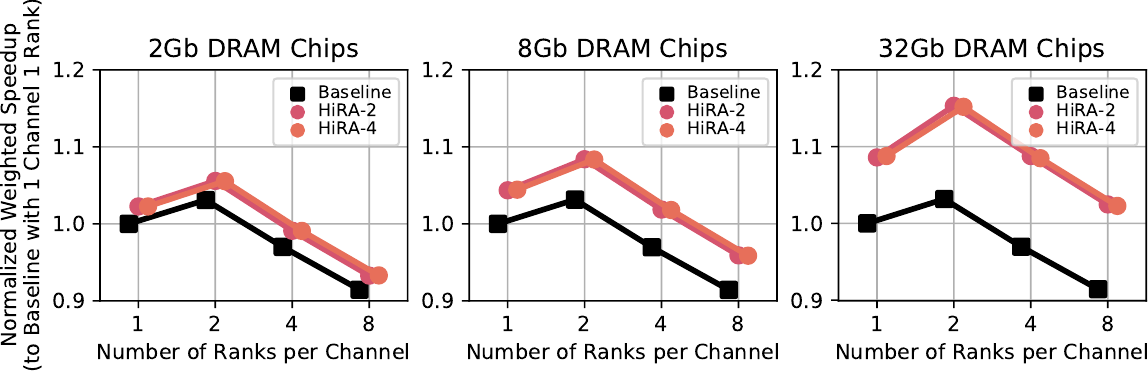}
    \caption{\sho{Effect of {rank count} on system performance for Baseline and HiRA}}
    \label{hira:fig:ref_rank}
\end{figure}

\sho{We make three observations.
First, increasing the number of ranks from one to two increases system performance {(e.g., by \SI{3}{\percent} {and} \SI{15.3}{\percent} for {the} baseline {and} \gls{hira}-2, {respectively,} for a chip capacity of 32Gb)}. This is because the {evaluated} workloads leverage the {higher} rank-level parallelismSecond, \shom{unlike {with} channels,} further {increasing} the number of ranks beyond two \emph{slows down} the system for both \shom{the} baseline and HiRA {by \SI{11.7}{\percent} and \SI{11.1}{\percent}, respectively, on average {as} number of ranks increases from 2 to 8}. This happens because multiple ranks share a single command bus and {together} occupy the command bus for refresh operations, {making the command bus a bottleneck}. Third, HiRA {provides} higher performance than \shom{the} baseline for \emph{all} {evaluated} rank configurations.}
{For example, \gls{hira}-2 provides \SI{12.1}{\percent} performance improvement over {the} baseline {even} for an 8-rank system with 32Gb DRAM chips.}
{We conclude that \gls{hira} provides significant performance benefits for high-capacity DRAM chips {compared to the baseline even with a large number of ranks}.}

\subsection{HiRA with Preventive Refresh}

\sho{\figref{hira:fig:para_channel} shows how increasing \shom{the} {channel count} {(x-axis)} affects 
{PARA's impact on system performance when used without {\gls{hira}} (labeled as PARA) and with HiRA (labeled as HiRA-N, where N represents the \gls{trrslack} configuration as in \secref{hira:sec:doduoref} and~\secref{hira:sec:uctwo_eval}).}
{The y-axis reports system performance in terms of average weighted speedup across 125 workloads,} normalized to the baseline 1-channel 1-rank system with \emph{no} RowHammer defense mechanism.
{Three subplots show the results for RowHammer thresholds ($N_{RH}$) of 1024 (left), 256 (middle), and 64 (right).}}

\begin{figure}[!ht]
    \centering
    \includegraphics[width=0.8\linewidth]{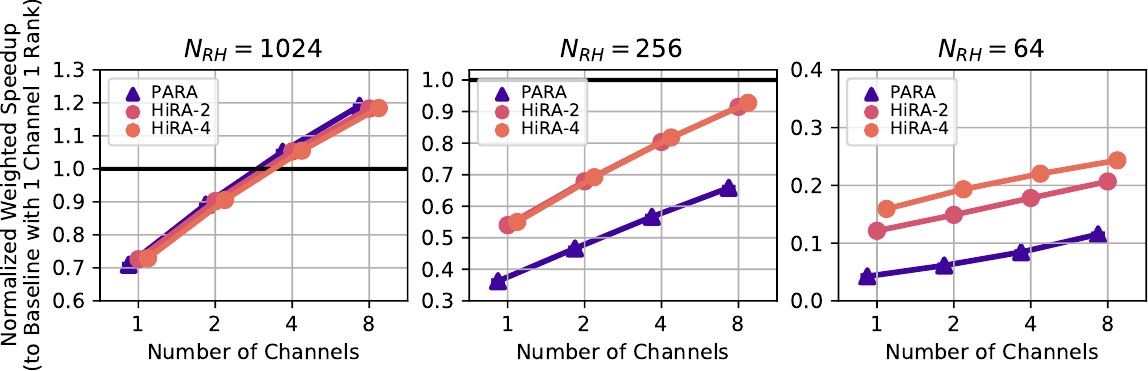}
        \caption{\sho{Effect of \shom{channel count} on system performance {with} PARA and HiRA}}
    \label{hira:fig:para_channel}
\end{figure}

\sho{We make {three} observations.
First, {system} performance increases with \shom{channel count} when \shom{PARA} is used {(with and without HiRA)}. For example, increasing \shom{the} number of channels from one to eight {improves the performance of the system with PARA {and \gls{hira}-2} by \SI{67.6}{\percent}}
{and by \SI{63}{\percent}, respectively,}
when the RowHammer threshold is 1024. This is because memory accesses are distributed across a larger number of banks \shom{given more channels}, thereby reducing the congestion in banks, and thus the \shom{number of} row buffer conflicts. As a result, {the evaluated} workloads perform fewer row activations, and PARA generates fewer {preventive} refreshes.
Second, {at smaller RowHammer thresholds, \gls{hira} significantly improves system performance even with a large number of channels.} 
{For example,}
PARA \shom{{causes}}
{\SI{88.5}{\percent}} performance reduction on an eight-channel system when the RowHammer threshold is 64. HiRA{-2 and HiRA-4 reduce} this performance overhead {to {\SI{79.3}{\percent} and \SI{75.7}{\percent}}, respectively,} by {performing preventive refreshes concurrently with refreshing or activating other rows in the same bank.}
{Third, \gls{hira} improves system performance compared to PARA for all {evaluated} channel counts. This happens because \gls{hira} reduces the performance overhead of PARA's preventive refreshes for each memory channel regardless of the system's channel count.
}
{We conclude that \gls{hira} provides significant performance benefits even with {a} {large number of channels.}}
}

{\figref{hira:fig:para_rank} shows how increasing \shom{the} {rank count} {(x-axis)} affects 
{PARA's impact on system performance when used without {\gls{hira}} and with HiRA.}
{The y-axis shows system performance {using the same metric as \figref{hira:fig:para_channel} uses}.
{Three subplots show the results for RowHammer thresholds ($N_{RH}$) of 1024 (left), 256 (middle), and 64 (right).}}}

\begin{figure}[!ht]
    \centering
    \includegraphics[width=0.8\linewidth]{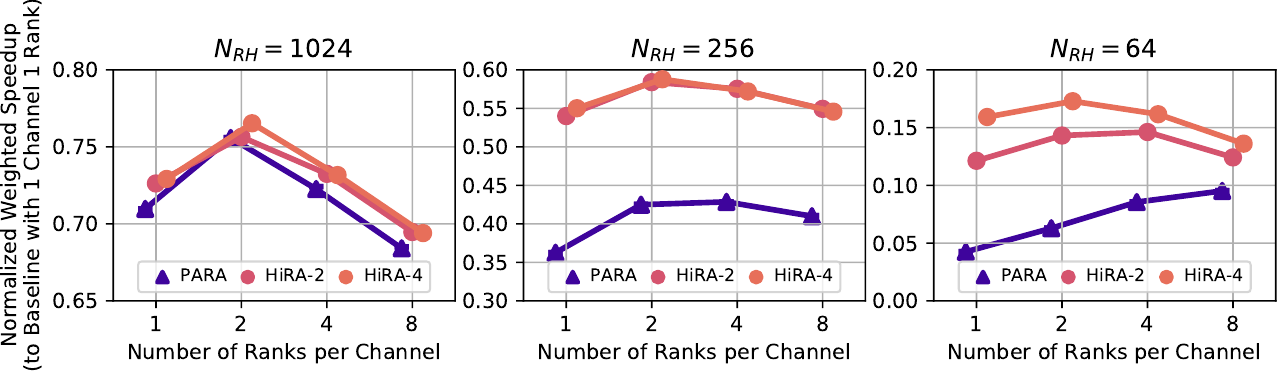}
    \caption{\sho{Effect of \shom{{rank} count} on system performance {with} PARA and HiRA}}
    \label{hira:fig:para_rank}
\end{figure}

We make three observations.
First, similar to \figref{hira:fig:ref_rank}, increasing the number of ranks from one to two increases system performance for all three mechanisms {(e.g., by \SI{6.5}{\percent} and \SI{4.9}{\percent} for PARA and HiRA-4 when $N_{RH}=1024$)} across all shown RowHammer thresholds due to {the higher} rank-level parallelism.
Second, further {increasing} the number of ranks beyond two ranks reduces HiRA's benefits over PARA. {{This happens because} increasing the {rank count}
beyond two {increases the command bus bandwidth usage of periodic refresh requests.}{Third, despite the performance reduction at high rank counts, \gls{hira} significantly improves system performance compared to PARA. For example, \gls{hira}-2 (\gls{hira}-4) {improves system performance by} \SI{30.5}{\percent} (\SI{42.9}{\percent}) compared to PARA on an 8-rank system with a RowHammer threshold of 64.}
{Based on these observations, we conclude that \gls{hira} provides significant performance benefits over PARA even when a large number of ranks {share the command bus.}}}

\section{Major Results}
\label{hira:sec:takeaways}
{We summarize} the {major} observations from {four main} evaluations in this \agy{3}{chapter}. 
First, {by using \gls{hira}, it is possible to reliably} refresh a DRAM row {concurrently with}
refreshing or {activating} another DRAM row within the same bank {in} off-the-shelf DRAM chips {(\secref{hira:sec:characterization})}. {\secref{hira:sec:characterization}} experimentally demonstrates {on \chipcnt{} real DRAM chips} that 
{\gls{hira} {can} reliably parallelize a DRAM row's {refresh operation} {with refresh or {activation of} {any of the}}} \covref{hira:} of the rows within the same bank.
Second, \gls{hirasched} reduces the {overall latency of {refreshing} two DRAM rows within the same bank} by \param{\SI{51.4}{\percent}} (\secref{hira:sec:characterization}).
{Third, {\gls{hira} {significantly improves system performance by} reducing} the performance {degradation} {caused by} periodic refreshes across \emph{all} system configurations we evaluate (\secref{hira:sec:doduoref}). 
{Fourth}, \gls{hirasched} {significantly} {improves system} performance {by reducing the performance overhead of PARA's preventive refreshes across \emph{all} system configurations we evaluate (\secref{hira:sec:doduopara})}.
{Our major results show that \gls{hira}}
can {effectively and robustly} improve {system performance} by reducing the time spent for \emph{both} periodic refreshes and {preventive} refreshes without compromising system reliability or security. We hope that our findings inspire DRAM manufacturers and {standards bodies} to {explicitly and} {properly} support \gls{hira} in future DRAM chips.}
\section{Limitations}
\label{hira:sec:limitations}
\micrev{We identify HiRA's limitations under {{three}} categories.}

\micrev{First, we experimentally demonstrate that HiRA is supported by real DDR4 DRAM chips. However, we cannot verify the {\emph{exact}} {operation} of HiRA {(i.e., how HiRA is enabled)} in those chips for two reasons: 1) \emph{no} public documentation discloses or verifies HiRA in real DRAM chips and 2) we do \emph{not} have access to DRAM manufacturers' proprietary circuit designs.}

\micrev{Second, all DRAM chips that exhibit successful HiRA operation are manufactured by {SK Hynix ({the second largest} DRAM manufacturer {that has \SI{27.4}{\percent} of {the} DRAM market share~\cite{statista2022dram_manufacturers}})}. We also conducted experiments using {40} DRAM chips from {{each of the} two} other manufacturers {(Samsung and Micron)} for which we observed \emph{no} successful HiRA operation. {We hypothesize that the DRAM chips from these other manufacturers \emph{ignore} the \gls{pre} or the second \gls{act} command of HiRA's command sequence when \gls{tras} and \gls{trp} timing parameters are greatly violated ({e.g.,} the DRAM chip acts as if it did not receive the \gls{pre} or the second \gls{act} commands).} Therefore, HiRA is currently limited to DRAM chips that {can successfully perform \gls{hira}} operation{s}. {We believe that other DRAM chips are fundamentally capable of \gls{hira} since \gls{hira} is consistent with fundamental operational principles of modern DRAM.} We hope that this work inspires future DRAM designs {that} explicitly support HiRA, given that HiRA 1) provides significant performance benefits and 2) is already {possible} in real DRAM chips, {even though DRAM chips are not even designed to support it}.}
\micrev{Third, performing periodic refresh using HiRA results in higher memory command bus utilization compared to using {conventional} {REF} commands. HiRA issues a row activation (ACT) {command} and a precharge (PRE) command to refresh a {\emph{{single}}} DRAM row, while \emph{multiple} DRAM rows are refreshed when a {single} REF command is issued. Even though HiRA overlaps the latency of row activation and precharge operations with the latency of other refresh or access operations, it still uses the command bus bandwidth to transmit {\gls{act} and \gls{pre}} commands to DRAM chips. {As the} number of ranks and banks per DRAM channel {increases}, HiRA's command bus utilization can cause memory access requests to experience larger delays compared to using {REF commands} \sho{({as we evaluate in} \secref{hira:sec:sensitivity})}. {However, \gls{hira} still provides \SI{12.1}{\percent} system performance benefit over a baseline memory controller that uses {REF commands} even in an 8-rank system {with high} command bus utilization.}}

{We {conclude} that none of these limitations fundamentally prevent a system designer from {using existing} DRAM chips {that can reliably perform HiRA operations} and {thus}, {benefit from} HiRA's {refresh-refresh and refresh-access parallelization}.} 

\section{Summary}
\label{hira:sec:conclusion}

{We introduce} \gls{hira}, {a new {DRAM} operation} {that can} {reliably parallelize a DRAM row's refresh operation with refresh or activation of another row within the same bank.}
\gls{hira} achieves this by {activating} two electrically-isolated rows {in quick succession,} allowing them to be refreshed{/{activated}} without disturbing each other.
{{W}e show that \gls{hira} 1)~works reliably in 56 real off-the-shelf DRAM chips}{, using already{-}available (i.e., standard) \gls{act} and \gls{pre} DRAM commands{,} by violating timing constraints and 2)~reduces} the {overall} latency {of refreshing two rows} by \SI{51.4}{\percent}.
To leverage {the parallelism} \gls{hira} {provides,}
we {design} \glsfirst{hirasched}.  
\gls{hirasched} modifies the memory request scheduler to perform \gls{hira} operations when a periodic or {RowHammer-preventive} refresh can be {performed} concurrently with another {refresh {or row activation}} {to} the same bank. {{Our system-level evaluations} show that} {\gls{hirasched}} {increases system performance by \SI{12.6}{\percent} and $3.73\times$ {{as it reduces}}}
the performance {degradation {due to}} periodic and {preventive} refreshes{,}
respectively.
{We conclude that \gls{hira} {{1)~}already works in off-the-shelf DRAM chips and can be used to} significantly reduce the {performance degradation caused by} both periodic and preventive refreshes {and 2)~provides higher performance benefits in higher{-}capacity DRAM chips.}}
{We hope that our findings will inspire DRAM {manufacturers and} {standards bodies} to {explicitly} {and properly} support \gls{hira} in future DRAM {chips and standards}.}

\chapter[BlockHammer]{BlockHammer}
\label{chap:blockhammer}




\newif\ifgraphene
\graphenetrue

\newcommand{\mpm}[1]{#1}

\renewcommand{\rowhammer}{{RowHammer}}
\newcommand{\rowhamemr}{\rowhammer{}}
\renewcommand{\blockhammer}{{BlockHammer}}
\newcommand{\rowblocker}{{RowBlocker}}
\newcommand{\procthrottler}{{{Attack}Throttler}}
\newcommand{\hammerthrottler}{{{Attack}Throttler}}
\newcommand{\rbbl}{{RowBlocker-BL}}
\newcommand{\rbhb}{{RowBlocker-HB}}
\newcommand{\prohit}{{PRoHIT}}
\newcommand{\para}{{PARA}}
\newcommand{\cbt}{{CBT}}
\newcommand{\twice}{{TWiCe}}
\newcommand{\graphene}{{Graphene}}
\newcommand{\mrloc}{{MRLoc}}
\newcommand{\baseline}{{baseline}}

\newcommand{\desirable}{desirable}
\newcommand{\benign}{benign}

\renewcommand{\cmark}{\ding{51}}%
\newcommand{\cmarki}{\ding{51}*}%
\renewcommand{\xmark}{\ding{55}}%
\newcommand{\xmarki}{\ding{55}*}%

\newcommand{\bhadj}{\desirable{}}
\newcommand{\abhadj}{a~\desirable{}}
\newcommand{\devagno}{DRAM-agnostic}
\newcommand{\nonmalic}{\benign{}}
\newcommand{\actstate}{{\emph{active}}}
\newcommand{\passtate}{{\emph{passive}}}

\newcommand{\act}{$ACT$\xspace}
\newcommand{\pre}{$PRE$\xspace}
\newcommand{\dramrefcmd}{$REF$\xspace}

\renewcommand{\trc}{$t_{RC}$\xspace}
\newcommand{\trrd}{$t_{RRD}$}
\renewcommand{\tfaw}{$t_{FAW}$}
\renewcommand{\trefw}{$t_{REFW}$\xspace}
\renewcommand{\trefi}{$t_{REFI}$\xspace}
\newcommand{\tw}{$t_{W}$\xspace}
\renewcommand{\tras}{$t_{RAS}$\xspace}
\newcommand{\tai}{$t_{AI}$\xspace}
\newcommand{\tailong}{the minimum \rowhammer{}-safe activation interval\xspace}
\newcommand{\mac}{$MAC$\xspace}
\newcommand{\macw}{$MAC_{W}$\xspace}
\newcommand{\macrefw}{$MAC_{REFW}$\xspace}
\newcommand{\maclong}{the maximum activation count that can be issued without refreshing the adjacent rows\xspace}
\newcommand{\nth}{\nbl{}}
\newcommand{\nbl}{$N_{BL}$\xspace}
\newcommand{\nthlong}{the threshold number of activations for labeling a row as a potentially aggressor row\xspace}
\newcommand{\nep}{$\nepn{}$}
\newcommand{\nbf}{$N_{BF}$\xspace}
\newcommand{\nbflong}{size of \blockhammer's observation window\xspace}
\newcommand{\nrh}{$N_{RH}$}
\newcommand{\trr}{$TRR$\xspace}
\newcommand{\trrlong}{Target Row Refresh\xspace}
\newcommand{\rltl}{RLTL\xspace}
\newcommand{\rltllong}{row-level temporal locality\xspace}
\newcommand{\rburst}{$\rburstn{}$}
\newcommand{\tdelay}{$\tdelayn{}$}
\newcommand{\tdelaymin}{$\tdelayminn{}$}
\newcommand{\rhli}{$RHLI$}

\newcommand{\nepn}{N_{ep}}
\newcommand{\nbfn}{N_{BF}\xspace}
\newcommand{\trefwn}{t_{REFW}\xspace}
\newcommand{\twn}{t_{W}\xspace}
\newcommand{\nthn}{\nbln{}}
\newcommand{\nbln}{N_{BL}\xspace}
\newcommand{\nrhn}{N_{RH}\xspace}
\newcommand{\trcn}{t_{RC}\xspace}
\newcommand{\trrdn}{t_{RRD}}
\newcommand{\tfawn}{t_{FAW}}
\newcommand{\trasn}{t_{RAS}\xspace}
\newcommand{\macn}{MAC\xspace}
\newcommand{\macwn}{MAC_W\xspace}
\newcommand{\macnewn}{MAC_{new}}
\newcommand{\macnew}{$\macnewn{}$}
\newcommand{\thammern}{T_{hammer}\xspace}
\newcommand{\thammer}{$\thammern$\xspace}
\newcommand{\tremainingn}{T_{remaining}\xspace}
\newcommand{\tremaining}{$\tremainingn$\xspace}
\newcommand{\tbbfn}{t_{BF}}
\newcommand{\tbbf}{$\tbbfn{}$}
\newcommand{\tbfn}{t_{CBF}}
\newcommand{\tbf}{$\tbfn{}$}
\newcommand{\macbfn}{MAC_{BF}\xspace}
\newcommand{\macbf}{$\macbfn$\xspace}
\newcommand{\cbfan}{CBF_{A}}
\newcommand{\cbfa}{$\cbfan$}
\newcommand{\cbfbn}{CBF_{B}}
\newcommand{\cbfb}{$\cbfbn$}
\newcommand{\tain}{t_{AI}\xspace}
\newcommand{\macrefwn}{MAC_{REFW}\xspace}
\newcommand{\nmaxn}{N_{max}\xspace}
\newcommand{\nmax}{$\nmaxn$\xspace}
\newcommand{\taiavgn}{t_{AI{avg}}\xspace}
\newcommand{\taiavg}{$\taiavgn$\xspace}
\newcommand{\naggrn}{N_{aggr}\xspace}
\newcommand{\naggr}{$\naggrn$\xspace}
\newcommand{\nhbfn}{\frac{1}{2}\nbfn{}}
\newcommand{\nhbf}{$\nhbfn{}$}
\newcommand{\rburstn}{r_{Burst}}
\newcommand{\tdelayn}{t_{Delay}}
\newcommand{\tdelayminn}{t_{Delay_{min}}}
\newcommand{\tepochn}{t_{ep}}
\newcommand{\tepoch}{$\tepochn{}$}
\newcommand{\tcbfn}{\tbfn{}}
\newcommand{\tcbf}{\tbf{}}
\newcommand{\nrhbf}{$\nrhbfn{}$}
\newcommand{\nrhbfn}{\nrhn{}_{BF}}
\newcommand{\nrhtuned}{$\nrhtunedn{}$}
\newcommand{\nrhtunedn}{{N_{RH}}^*}

\newcommand{\noutn}{N_{o}}
\newcommand{\naggn}{N_{ep}}
\newcommand{\nepmaxin}{N_{ep_{max}}}
\newcommand{\nepmaxi}{$\nepmaxn{}$}
\newcommand{\nepmax}{$\nepmaxin{}$}
\newcommand{\npren}{N_{ep-1}}
\newcommand{\nthpn}{{N_{BL}}^{*}}
\newcommand{\nblpn}{{N_{BL}}^{*}}
\newcommand{\nout}{$\noutn{}$}
\newcommand{\nagg}{$\naggn{}$}
\newcommand{\npre}{$\npren{}$}
\newcommand{\nthp}{$\nthpn{}$}
\newcommand{\nblp}{$\nblpn{}$}
\newcommand{\ten}{T_{epoch}}
\newcommand{\te}{$\ten{}$}

\newcommand{\agysc}[2]{#1}
\newcommand{\agyscsc}[2]{#1}
\newcommand{\agyscscnotodo}[2]{{{#1}}}

\newcommand{\agypr}[2]{#1}
\newcommand{\agyprpr}[2]{#1}
\newcommand{\agyprprnotodo}[2]{#1}
\newcommand{\agycinline}[1]{}
\newcommand{\agyc}[1]{}
\renewcommand{\atbc}[1]{}
\newcommand{\agycr}[1]{}
\newcommand{\agytdinline}[1]{}
\newcommand{\mpc}[1]{}
\newcommand{\jkc}[1]{}
\newcommand{\sgc}[1]{}
\newcommand{\omc}[1]{}
\newcommand{\fw}[1]{}

\newcommand{\agyprm}[2]{{\small \textcolor{red!50}{[removed text here]}}{\todo[size=\small,color=red!40]{\textbf{\textcolor{black}{#2}}}{}}}

\newcommand{\subsecref}[1]{Section~\ref{#1}}
\newcommand{\subsubsecref}[1]{Section~\ref{#1}}
\newcommand{\equref}[1]{Equation~\ref{#1}}
\newcommand{\exttwo}[1]{{#1}}

\renewcommand{\om}[1]{#1}
\section{BlockHammer}
\label{blockhammer:sec:blockhammer}
\blockhammer{} is designed to (1) {scale efficiently as DRAM chips become increasingly vulnerable to \rowhammer{}} and (2) be compatible with commodity DRAM chips. 
\blockhammer{} consists of two components.
The first component, \rowblocker{} {(\secref{blockhammer:sec:rowblocker})}, prevents any possibility of a {\rowhammer{}} bitflip
by making it impossible to access a DRAM row at a high enough rate to induce \rowhammer{} bitflips. \rowblocker{} achieves this by efficiently tracking row activation rates using Bloom filters and throttling the row activations that target rows with high activation rates.
We implement \rowblocker{} entirely within the memory controller, ensuring \rowhammer{}-safe operation without {any proprietary information about or modifications to the DRAM chip}. Therefore, \rowblocker{} is compatible with all commodity DRAM chips.
The second component, \hammerthrottler{} (\secref{blockhammer:sec:hammerthrottler}), alleviates the performance degradation a \rowhammer{} attack can impose upon benign applications by selectively reducing the memory bandwidth usage of \emph{only} threads that \hammerthrottler{} identifies {as likely} \rowhammer{} attack{s} 
{(i.e., \emph{attacker threads}).
{By doing so, \hammerthrottler{} provides a larger memory bandwidth to benign applications compared to a baseline system {that does not} throttle {attacker threads}. {As DRAM chips become more vulnerable to \rowhammer{},} \hammerthrottler{} throttles {attacker threads} more aggressively, freeing even more memory bandwidth for benign applications to use.}
{\blockhammer{} (\rowblocker{} + \hammerthrottler{}) achieves both of {its {design} goals}.}} 

\subsection{\rowblocker{}}
\label{blockhammer:sec:mech_overview}
\label{blockhammer:sec:rowblocker}
\rowblocker{}'s goal is to proactively throttle row activations {in an efficient manner} to avoid any possibility of a \rowhammer{} attack. \rowblocker{} achieves this by overcoming two challenges regarding performance and area overheads.

{First, achieving low performance {overhead} is a key challenge for a throttling mechanism because many benign applications tend to repeatedly activate a DRAM row that they have recently activated~\cite{k2015memory, kim2012acase, hassan2016chargecache, hassan2019crow}. This can potentially {cause} a throttling mechanism to mistakenly throttle benign applications, thereby degrading system performance}.
{To ensure {throttling} \emph{only} applications that might cause \rowhammer{} bitflips,} \rowblocker{} throttles the row activations targeting \emph{only} rows whose activation rates are above a {given} threshold. To this end, \rowblocker{} implements two components as shown in \figref{blockhammer:fig:overview}:
(1)~a {per-bank} blacklisting mechanism, \rbbl{}, which blacklists {all} rows with an activation rate {greater} than a predefined threshold called {the} \emph{blacklisting threshold} (\nbl{}); and
(2)~a {per-rank} activation history buffer, \rbhb{}, which tracks the most recently activated rows. \rowblocker{} enforces a time delay between two consecutive activations targeting a row \emph{only if} the row is \emph{blacklisted}.
By doing so, \rowblocker{} {is less likely to throttle} a benign application's row activations.

Second, achieving low area overhead is a key challenge for a throttling mechanism because throttling requires tracking all row activations {throughout} an entire refresh window \emph{without} losing information of any row activation. \rowblocker{} implements its blacklisting mechanism, \rbbl{}, by using area-efficient \emph{{counting Bloom filters}}~\cite{bloom1970spacetime, fan2000summary} to track row activation rates. \rbbl{} maintains two counting Bloom filters in a {time-interleaved} manner to track row activation rates for large time windows without missing any row that should be blacklisted. We explain how counting Bloom filters work and how \rbbl{} employs them in \secref{blockhammer:sec:mech_detect}.

\begin{figure}[ht!]
    \centering
    \includegraphics[width=0.75\linewidth]{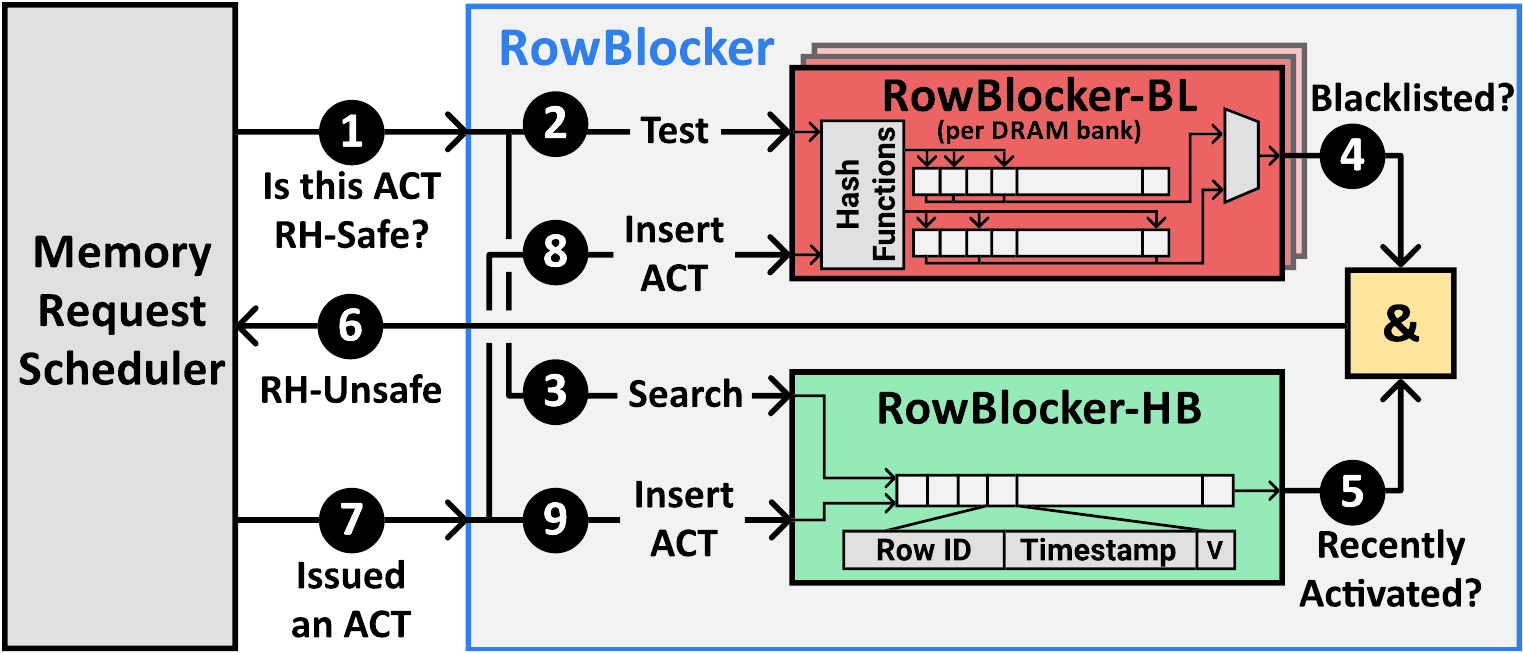}
    \caption{High-level overview of \rowblocker{} {(per DRAM rank)}. {An ACT is accompanied by its row address.}}
    \label{blockhammer:fig:overview}
    
\end{figure}
 

\nind
\head{High-Level Overview of \rowblocker{}} 
\rowblocker{} modifies the memory request scheduler to temporarily block {(i.e., delay)} an activation that targets a \emph{blacklisted} and {\emph{recently-activated}} row {until {the activation} {can be safely performed}}. {{By blocking such {row activations}}, \rowblocker{} ensures that} no row can be activated at a high enough rate to induce \rowhammer{} bitflips. 
When the memory request scheduler attempts to schedule a row activation command {to a bank}, it queries \rowblocker{} (\circled{1}) to check if the row activation is \rowhammer{}-safe. This simultaneously triggers two lookup operations. First, \rowblocker{} checks the \rbbl{} to see if the row to be activated is blacklisted (\circled{2}). 
A row is blacklisted if {its activation rate exceeds}
a given threshold. We discuss how \rbbl{} estimates the activation rate of a row in \secref{blockhammer:sec:mech_cbf}.
Second, \rowblocker{} checks \rbhb{} to see if the row has been recently activated (\circled{3}).
{If a row is both blacklisted (\circled{4}) \emph{and} recently activated (\circled{5}), \rowblocker{} responds to the memory request scheduler with {a} \emph{{\rowhammer{}}-unsafe} signal (\circled{6}), consequently blocking the row activation. 
Blocking such a row activation is essential because allowing further activations to a blacklisted and {recently-activated} row could increase the {row's} overall activation rate and thus result in \rowhammer{} bitflips.}
The memory request scheduler does \emph{not} issue a row activation 
if \rowblocker{} returns \emph{unsafe}. {However, it} keeps issuing the \emph{\rowhammer{}-safe} requests. {This scheduling {decision}}
effectively prioritizes {\rowhammer{}-safe} memory accesses {over unsafe ones}. 
{An unsafe row activation becomes safe again as soon as a certain amount of time (\tdelay{}) passes after its latest activation, {effectively limiting the row's average} {activation rate to a \rowhammer{}-safe {value}.}
After \tdelay{} is satisfied, \rbhb{} no longer reports that the row {has been} recently activated (\circled{5}), thereby allowing the memory request scheduler to issue the row activation (\circled{6}).}
{When the memory request scheduler issues a row activation~(\circled{7}), it simultaneously updates both \rbbl{}~(\circled{8}) and \rbhb{}~(\circled{9}). 
We explain how \rbbl{} and \rbhb{} work in \secref{blockhammer:sec:mech_detect} and~\ref{blockhammer:sec:mech_prevent}, respectively.}

\subsubsection{{\rbbl{} Mechanism}}
\label{blockhammer:sec:mech_detect}
\label{blockhammer:sec:mech_cbf}
\label{blockhammer:sec:mech_switch}
\rbbl{} uses two counting Bloom filters (CBF) in a time-interleaved fashion to decide whether a row should be blacklisted. Each CBF takes turns to make the blacklisting decision.
A row is blacklisted when its activation rate exceeds a configurable threshold, \agypr{which we call the \emph{blacklisting threshold} {(\nbl{})}}{IQA5}. When a CBF blacklists a row, {any further activations targeting the row} are throttled
until the end of the CBF's turn.
In this subsection, we {describe} how a CBF works, how we use two CBFs to avoid stale blacklists, and how {the} two CBFs never fail to blacklist an aggressor row.

\nind
\head{{Bloom Filter}}
A Bloom filter~\cite{bloom1970spacetime} is a space-efficient probabilistic data structure {that is} used for testing {whether} a set contains a particular element. 
A Bloom filter consists of a set of hash functions and a {bit array} {on which it performs} three operations: {\emph{clear}, \emph{insert}, and \emph{test}}.
Clearing a Bloom filter zeroes its {bit array}.
To insert/test an element,
each hash function evaluates an index 
into the {bit array} for the element, {using an {identifier} for the element}.
Inserting {an element} sets the bits that the hash functions point to. Testing for {an element} checks whether all these bits are set.
Since a hash function can yield the same {set of} {indices} for different elements {(i.e., aliasing)}, testing a Bloom filter can return true for an element that was never inserted (i.e., false positive).
However, the \emph{test} {operation} never returns false for an inserted element (i.e., no false negatives).
A Bloom filter eventually saturates (i.e., {always returns true when tested for any element})
{if elements {are continually} inserted,}
{which requires periodically clearing the filter and losing all inserted elements.}

\nind{}\textbf{Unified Bloom Filter (UBF).} {UBF~\cite{li2012compression} is} a Bloom filter variant that allows a system to continuously track a set of elements that are inserted into a Bloom filter within the most recent time window of a fixed length (i.e., a \emph{rolling time window}). {Using a conventional Bloom filter} to track a rolling time window could result in data loss whenever {the Bloom filter is} cleared, as the clearing eliminates the elements that still fall within the rolling time window.
{Instead,} {UBF} {continuously {tracks} insertions in a rolling time window by {maintaining}} \emph{two} Bloom filters {and using them in} a {time-interleaved} manner. {UBF {inserts every element into} both filters,} 
while {the filters} take turns in responding {to} {\emph{test}} queries across consecutive limited time windows {(i.e., {\emph{epochs}})}. {UBF clears the filter which responds to {\emph{test}} queries at the end of an epoch and redirects the {\emph{test}} queries to the other filter for the next epoch. Therefore, each filter is cleared every {other epoch (i.e., the filter's lifetime is two epochs).}}
By doing so, UBF ensures no false negatives for the elements that are inserted in {a rolling time window of up to two epochs}.

\nind
\head{{Counting Bloom Filter (CBF)}}
{To track}
\emph{the number of times} an element {is} inserted 
into the filter, {another Bloom filter variant, called}
\emph{counting} Bloom filters {(CBF)}~\cite{fan2000summary}, 
{replaces} the {bit array} with a {\emph{counter} array}.
{Inserting an element in a CBF} \emph{increments} all of its corresponding counters. Testing an element returns the \emph{minimum} value {among} all {of {the element's}} {corresponding} counters, which represents an \emph{upper bound} on the number of times {an element} was inserted into the filter.
Due to aliasing, the test result can be \emph{larger} than the true insertion count, but it \emph{cannot} be {smaller} than that because counters are \emph{never decremented} (i.e., false positives are possible, but false negatives are not).

\nind
\head{{Combining} {UBF and} CBF for Blacklisting} 
{To estimate row activation rates with low area cost, \rbbl{} combines {the ideas of} UBF and CBF to form our \emph{dual} counting Bloom filter (D-CBF). D-CBF maintains \emph{two} CBFs in the time-interleaved manner of UBF.}
{On every row activation, \rbbl{} inserts the {activated row's} address into {both} CBFs. \rbbl{} considers a row {to be} \emph{blacklisted} when {the row's} activation {count} exceeds {the blacklisting threshold} (\nbl{}) {in a rolling time window}.} 
{\figref{blockhammer:fig:bloomfilter_timing} illustrates how \rbbl{} uses a D-CBF over time.} 
{\rbbl{} {designates one of the CBFs as \emph{active} and the other as \emph{passive}}.}
{At {any given time,} only the \emph{active} CBF responds to {\emph{test}} queries.  When a {\emph{clear}} signal is received, D-CBF (1)~clears only the active filter (e.g., \cbfa{} at \circled{3}) and (2)~swaps the active and passive filters (e.g., \cbfa{} becomes passive and \cbfb{} becomes active at \circled{3}). 
\rbbl{} blacklists a row if {the row's} activation count in the active CBF exceeds
the blacklisting threshold (\nbl{}).
}


\nind
\head{{D-CBF Operation} Walk-Through}
{{We walk through {D-CBF operation in} \figref{blockhammer:fig:bloomfilter_timing}} 
from the perspective of a DRAM row.}
{The counters that correspond to {the row} in} both filters (\cbfa{} and \cbfb{}) are initially {zero} (\circled{1}).
\cbfa{} is the \emph{active} filter, while \cbfb{} is the \emph{passive} {filter}.
As the row's activation count accumulates and reaches \nbl{} (\circled{2}), both \cbfa{} and \cbfb{} decide to blacklist the row. \rowblocker{} applies the active filter's decision (\cbfa{}) and blacklists the row.
As the counter values do not decrease, the row remains blacklisted until the end of {Epoch~1}. Therefore, a minimum delay is enforced between consecutive activations {of} this row between \circled{2} and \circled{3}. At the end of {Epoch~1} (\circled{3}), \cbfa{} is cleared, and \cbfb{} becomes the active filter. Note that \cbfb{} immediately blacklists the {row,
as the counter values corresponding to the row in \cbfb{} are still larger than \nbl{}}. Meanwhile, assuming that the row continues to be activated, the counters in \cbfa{} {again} reach \nbl{} (\circled{4}). {At} the end of {Epoch~2} (\circled{5}), \cbfa{} becomes the active filter again and immediately blacklists the row. {By following this scheme, D-CBF blacklists the row as long as the row's activation count exceeds \nbl{} in an epoch.}
{Assuming that} the row's activation count does not exceed \nbl{} within Epoch~3, starting from \circled{6}, the row is {no longer blacklisted}. 
Time-interleaving across {the} two CBFs ensures that \blockhammer{} maintains a \emph{fresh} blacklist {that} never incorrectly excludes a DRAM row that needs to be blacklisted.
\secref{blockhammer:sec:mech_security} provides a generalized analytical proof of \blockhammer{}'s security guarantees that comprehensively studies all possible
{row activation patterns across all epochs}.

\begin{figure}[ht!]
    \centering
    \includegraphics[width=0.8\linewidth]{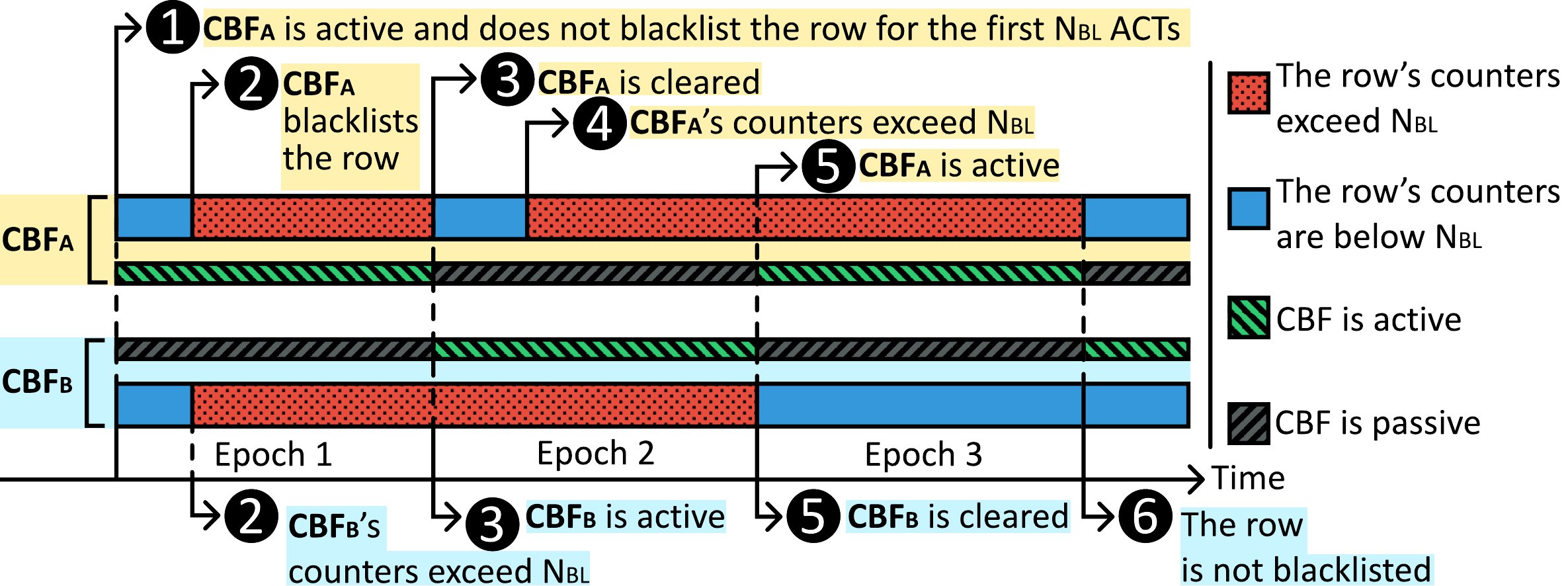}
    \caption{{{D-CBF} operation from a DRAM row's perspective}.}
    
    \label{blockhammer:fig:bloomfilter_timing}
\end{figure}

To prevent any specific row from being repeatedly blacklisted due to its CBF counters aliasing with those of an aggressor row (i.e., due to a false positive), \rbbl{} alters the hash functions that each CBF uses whenever the CBF is cleared.
{To achieve this, \rbbl{} replaces the hash function's seed with a {new} random{ly-generated} value, as we explain {next}}. 
Consequently, an aggressor row aliases with a different set of rows after every {\emph{clear}} operation. 

\pagebreak
\nind
\head{Implementing Counting Bloom Filters}
{To periodically send a \emph{clear} signal to D-CBF,} \rbbl{} implements a clock register that stores the timestamp of the {latest \emph{clear} operation}.
{In our implementation, each CBF} contains 1024 elements of 12-bit saturating counters to count up to the blacklisting threshold~\nbl{}. We employ four area- and latency-efficient H3-class hash functions that consist of {simple} static bit-shift and {mask} 
operations~\cite{carter1979universal}. {We hardwire the static shift operation, so it does not require any logic gates. The mask} operation performs a {bitwise} exclusive-OR on the shifted element (i.e., row address) and a seed. To alter the hash function {when a CBF is cleared},
{\rowblocker{} simply replaces the hash function's seed value with a randomly-generated value}. 

\subsubsection{{\rbhb{} Mechanism}}
\label{blockhammer:sec:mech_prevent}
\label{blockhammer:sec:mech_rahb}
{\rbhb{}'s goal is to ensure {that} a blacklisted row {cannot be} activated often enough to cause a bitflip.}
{To ensure this,} {\rbhb{} delays a subsequent activation to a blacklisted row until the row's last activation becomes older than a certain amount of time that we call \tdelay{}.} 
{To do so, \rbhb{} maintains a first-in-first-out history buffer that stores a record of all row activations in the last \tdelay{} time window.}
{When \rowblocker{} queries \rbhb{} with a row address (\circled{3} in Fig.~\ref{blockhammer:fig:overview}), \rbhb{} searches the row address in the history buffer and {sets the} {\emph{``Recently Activated?''}} signal {to true} if the row address is found.}

\nind
\head{Implementing \rbhb{}}
{{We implement} a per-DRAM-{rank} history buffer as a circular queue using a head and a tail pointer. Each entry of this buffer stores (1)~a row {{ID} (which is unique in the rank)}, (2)~a timestamp of when the entry was inserted into the buffer, and (3)~a valid bit.}
{The head and the tail pointers address the oldest and the youngest entries in the history buffer, {respectively}.} {When the memory request scheduler issues a row activation (\circled{7} in \figref{blockhammer:fig:overview}), \rbhb{} inserts a new entry with the activated row address, the current timestamp, and a valid bit {set to} {logic `1'} into the history buffer and updates the tail pointer.}
{\rbhb{} checks the timestamp of the oldest entry, {indicated by the head pointer}, every cycle. When the oldest entry becomes as old as \tdelay{}, \rbhb{} invalidates {the entry} by resetting its valid bit to {logic `0'} and updates the head pointer.}
{To test whether a row is recently activated (\circled{3} in \figref{blockhammer:fig:overview}), \rbhb{} looks up the tested row address in each \emph{valid} entry (i.e., {an} entry with a valid bit set to one) in parallel. {To search the history buffer {with} low latency, we keep row addresses in a content-addressable memory array.} 
Any matching \emph{valid} entry means that} the row has been activated {within the last \smash{\tdelay{}} time window, so}
the {new} activation should not be issued if the row is blacklisted by \rbbl{}.
We {size} the {history buffer} {to be} large enough to contain the {worst-case} number of row
activations that need to be {tested}.
{The number of activations that can be performed in a DRAM rank is bounded by {the} timing parameter \tfaw{}~\dramStandardCitations{}, which defines a rolling time window that can contain {at most} four row activations. Therefore, within a \smash{\tdelay{}} time window{,} there can be at most \smash{$\lceil4\times\tdelayn{}/\tfawn{}\rceil$} row} activations.

\nind
\head{{Determining How Long to Delay an Unsafe Activation}} 
To avoid \rowhammer{} {bitflips}, a row's activation {count should not exceed the \rowhammer{} threshold (\nrh{})} 
{within a refresh window (\trefw{}).}
\rowblocker{} satisfies this upper bound activation rate within each CBF's {lifetime (\tbf{}), which is the time window between two \emph{clear} operations applied to a CBF (e.g., Epochs 1 and 2 for \cbfb{} and Epochs 2 and 3 for \cbfa{} in \figref{blockhammer:fig:bloomfilter_timing}).}
{To ensure an upper bound activation rate of \nrh{}/\trefw{} at all times, \rowblocker{} does not allow a row to be activated more than $(\tbfn{}/\trefwn{})\times\nrhn{}$ times within a \tbf{} time window.} 
In the worst-case {access pattern within a CBF's lifetime,} a row is activated \nbl{} times at the very beginning of the \tbf{}
time window {as rapidly as possible}, taking a total time of $\nbln{} \times \trcn{}$. In this case, \rowblocker{} evenly distributes the activations that it can allow (i.e., $(\tbfn{}/\trefwn{})\times\nrhn{}-\nbln{}$) {throughout} the rest of the window (i.e., $\tbfn{} - (\nbln{} \times \trcn{})$).
Thus, we define \tdelay{} as shown in Equation~\ref{blockhammer:equ:tdelay}.
\begin{equation}
  \tdelayn{} = \frac{\tbfn{} - (\nbln{} \times \trcn{})}{(\tbfn{}/\trefwn{})\times\nrhn{} - \nbln{}}
  \label{blockhammer:equ:tdelay}
\end{equation}
\subsubsection{Configuration}
{\rowblocker{} has} three {tunable} configuration parameters {that collectively define \rowblocker{}'s false positive {rate} and area characteristics}: (1)~{the CBF} size{:} the number of counters in a CBF; (2)~\tbf{}: the CBF lifetime; and (3)~\nbl{}: the blacklisting threshold. 
{Configuring the CBF size directly impacts the CBF's area and false positive rate (i.e., the fraction of mistakenly blacklisted row activations) because the CBF size determines {both} the CBF's physical storage requirements and the likelihood of unique row addresses aliasing to the same counters. Configuring \nbl{} and \tbf{} determines the penalty of each false positive and the area cost of \rbhb{}'s history buffer, because \nbl{} and \tbf{} jointly determine the delay between activations required for \rowhammer{}-safe operation (via Equation~\ref{blockhammer:equ:tdelay}) and the maximum number of rows that \rowblocker{} must track within each epoch.}

To determine suitable values for each of the three parameters, we follow a three-step methodology that minimizes the cost of false positives for a given area budget. First, we empirically choose the CBF size based on false positive rates observed in our experiments (\secref{blockhammer:sec:methodology} discusses our experimental configuration). 
{We choose a CBF size of 1K counters {because} we observe that reducing the CBF size below 1K {significantly} increases the false positive rate due to aliasing.}






{Second, we {configure} \nbl{} {based on} three goals:\fw{NBL vs Area Overhead} (1)~\nbl{} should be smaller than the \rowhammer{} threshold to prevent \rowhammer{} bitflips; (2)~\nbl{} should be significantly {larger} than the per-row activation counts that benign applications exhibit {in order to ensure that \rowblocker{} does not blacklist benign applications' row activations, even when accounting for false positives due to Bloom filter aliasing;}
and (3)~\nbl{} should be as low as possible to {minimize} \smash{\tdelay{}} ({i.e., the time delay penalty for all activations to blacklisted rows, including those due to false positives) per \equref{blockhammer:equ:tdelay}.}}
To {balance these three goals,} we analyze the memory access patterns of 125 eight-core multiprogrammed {workloads}, each of which consists of eight {randomly-chosen} benign {threads}.
We simulate these {workloads} {using {cycle-level} simulation~\cite{Kim2016Ramulator, safariramulator}} for 200M instructions with a warmup {period} of 100M instructions on a \SI{3.2}{\giga\hertz} system with 16MB of last-level cache. We measure per-row activation rates by counting the activations that each row experiences within {a} \SI{64}{\milli\second} time window {(i.e., one refresh window)} starting {from} the row's first activation. We observe that benign {threads} reach up to 78, 109, and 314 activations per row in a \SI{64}{\milli\second} time window for {the} 95th, 99th, and 100th percentile of the set of DRAM rows that are accessed at least once. 
{Based on these observations}, we set \nbl{} to 8K for a \rowhammer{} threshold of 32K, providing {(1)~\rowhammer{}-safe operation, (2)~an} ample margin for row activations from benign threads {to achieve a low false positive rate (less than 0.01\%, as shown in \secref{blockhammer:sec:evaluation_tech_scaling}), and (3)~a} reasonable {worst-case} \tdelay{} penalty {of {\SI{7.7}{\micro\second}}} for activations to blacklisted rows.
 

Third, we use Equation~\ref{blockhammer:equ:tdelay} to choose a value for \tbf{} such that the resulting {\smash{\tdelay{}}} does not excessively penalize a mistakenly blacklisted row (i.e., a false positive). 
{Increasing \tbf{} both (1)~decreases \smash{\tdelay{}} (via \equref{blockhammer:equ:tdelay}) and (2)~extends the length of time for which a row is blacklisted. Therefore, we set \tbf{} equal to \trefw{}, which achieves as low a {\tdelay{}} as possible without blacklisting a row past the point at which its victim rows have already been refreshed.}
{We present} the final values we choose for all \blockhammer{} parameters {in conjunction with the DRAM timing parameters we use} 
in \tabref{blockhammer:tab:tuned_params} after explaining how \blockhammer{} addresses {many-sided \rowhammer{} attacks in \secref{blockhammer:sec:many_sided_attacks}}. 

\nind
\head{Tuning for Different DRAM Standards} The values in Table~\ref{blockhammer:tab:tuned_params} depend on {three} timing constraints defined by the memory standard: (1)~the minimum delay between activations to the same bank (\trc{}), (2)~the refresh window (\trefw{}), and (3)~the four-activation window (\tfaw{}). The delay enforced by \blockhammer{} (\smash{\tdelay{}}) scales linearly with \trefw{}, while it is marginally affected by \trc{} (\equref{blockhammer:equ:tdelay}). \trefw{} remains constant at \SI{64}{\milli\second} across DDRx standards from  DDR~\cite{jedec2008jesd79f} to DDR4~\cite{jedec2020jesd794c}, while \trc{} has marginally reduced from \SI{55}{\nano\second} to \SI{46.25}{\nano\second}~\dramStandardCitations{}. Therefore, \smash{\tdelay{}} increases only marginally across several DDR generations. In LPDDR4, \trefw{} is halved, which allows a reduction in {\smash{\tdelay{}}, and thus the} latency penalty of a blacklisted row. \tfaw{} affects {only} the size of the history buffer, {and its value varies} between \SIrange{30}{45}{\nano\second} across modern DRAM standards~\dramStandardCitations{}.

\subsection{AttackThrottler}
\label{blockhammer:sec:mech_perfattacks}
\label{blockhammer:sec:hammerthrottler}

{\hammerthrottler{}'s goal is to mitigate} the system-wide performance degradation that a \rowhammer{} attack could {inflict upon benign applications. 
\hammerthrottler{} achieves this by using memory access patterns to {(1)}~identify and {(2)}~throttle threads that potentially {induce} a \rowhammer{} attack.}
{First, to identify potential {\rowhammer{} attack} threads, \hammerthrottler{} exploits the fact that a {\rowhammer{} attack} thread inherently attempts to issue more activations to a blacklisted row than a benign application would.}
{Thus}, \hammerthrottler{} tracks {the exact number of} times each thread {performs a row activation to a {blacklisted row in each bank}.}
{Second, \hammerthrottler{} applies a quota to the total number of in-flight memory requests allowed for {\emph{any}} thread that is {identified} to be a {potential} attacker (i.e., that frequently activates blacklisted rows). {Because} {such} a thread activates blacklisted rows more often, \hammerthrottler{} reduces the thread's quota, reducing its memory bandwidth utilization. Doing so frees up}
{memory resources for concurrently-running benign applications that are \emph{not} repeatedly activating (i.e., hammering) blacklisted rows.}

\subsubsection{{Identifying Ongoing \rowhammer{} Attacks}}

\hammerthrottler{} {identifies {threads} that {exhibit} memory access patterns {similar to a \rowhammer{} attack} by monitoring a new metric called the \emph{\rowhammer{} likelihood index} (\rhli{}), which {quantifies} {the similarity between a given thread's memory access pattern and a real \rowhammer{} attack}.}
{\hammerthrottler{} calculates \rhli{} for each {<}thread, DRAM bank{>} pair. {\rhli{} is defined} as the number of {blacklisted row activations} the thread performs {to} the DRAM bank, normalized to the maximum
number of times a blacklisted row can be activated in a \blockhammer{}-protected system. As we describe in \secref{blockhammer:sec:rowblocker}, a row's activation count {during one CBF} lifetime is bounded by the \rowhammer{} threshold, scaled to a CBF's lifetime (i.e., $\nrhn{} \times (\tbfn{}/\trefwn{})$). Therefore, a blacklisted row that {has already been} activated \nbl{} times cannot be activated more than {$\nrhn{}\times(\tbfn{}/\trefwn{})-\nbln{}$} times. Thus, {\hammerthrottler{}} calculates \rhli{} {as shown in} Equation~\ref{blockhammer:equ:rhli}, {during a CBF's lifetime}.}
\begin{equation}
  RHLI = \frac{Blacklisted~Row~Activation~Count}{\nrhn{}\times(\tbfn{}/\trefwn{})-\nbln{}}
  \label{blockhammer:equ:rhli}
\end{equation}
{The} {\rhli{} {of a {<}thread, bank{>} pair is 0 when a thread certainly does \emph{not} perform {a} \rowhammer{} attack on the bank. As a {<}thread, bank{>} pair's \rhli{} {reaches} 1, the thread {is} more likely to induce \rowhammer{} bitflips in the bank. 


\rhli{} never exceeds 1 in a \blockhammer{}-protected system because 
\hammerthrottler{} completely blocks a thread's memory accesses {to} a bank (i.e., applies a quota of zero to them) when the {<thread, bank>} pair's \rhli{} reaches 1, {as we describe in \subsubsecref{blockhammer:sec:ht_throttling}}.}} 
{\rhli{} can be used independently from \blockhammer{} as a metric quantifying a thread's potential to be a \rowhammer{} attack, as we discuss in \secref{blockhammer:sec:rhli_system_level}}.

To demonstrate example \rhli{} values, we conduct {cycle-level} simulations on a set of 125~{multiprogrammed workloads}, each of which consists of one RowHammer attack thread and seven benign threads {randomly-selected from the set of workloads we describe in \secref{blockhammer:sec:methodology}.} 
We measure {the \rhli{} values} of benign {threads and \rowhammer{} attacks} {for \blockhammer{}'s two modes: (1)~\emph{observe-only} and (2)~\emph{full-functional}.
In \emph{observe-only} mode, \blockhammer{} computes \rhli{} but does not interfere with memory requests. In this mode, only \rowblocker{}'s blacklisting logic (\rbbl{}) and \hammerthrottler{}'s counters are functional, allowing \blockhammer{} to blacklist row addresses and measure \rhli{} per thread {without blocking any row activations}. In \emph{full-functional} mode, \blockhammer{} operates normally, i.e., {it} detects {the threads performing \rowhammer{} attacks}, throttles their requests, and ensures {that} no row's activation rate exceeds {the} \rowhammer{} threshold.} 
{We set the blacklisting threshold to {512}~activations} in a \SI{16}{\milli\second} time window. 
\newif\ifrhliplot
\rhliplotfalse
\ifrhliplot
Figure~\ref{blockhammer:fig:rli} reports measured \rhli{} values that \sg{\rowhammer{} attacks exhibit, with error bars showing the maximum and minimum observed values.}
\agy{We do not show benign applications in Figure~\ref{blockhammer:fig:rli} because they exhibit zero \rhli{} as the rows they access are never blacklisted.}
\begin{figure}[h!]
    \centering
    \includegraphics[width=\columnwidth]{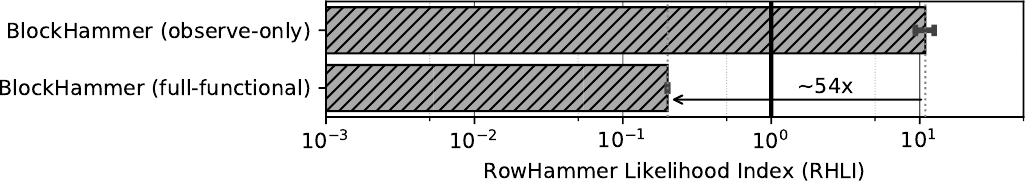}
    \caption{RowHammer Likelihood Index (RHLI) of benign applications and real RowHammer attacks when \blockhammer{} is {in} \emph{observe-only} and {\emph{full-functional}} modes.}
    \label{blockhammer:fig:rli}
\end{figure}

{We make two observations from Figure~\ref{blockhammer:fig:rli}. First, when \blockhammer{} is in \emph{observe-only} mode, 
while \rowhammer{} attacks reach an \rhli{} value of 10.9 on average. This observation shows that} \rhli{} can be used to identify a \rowhammer{} attack thread. Second, {in {\emph{full-functional}} mode,} \blockhammer{} significantly reduces \rhli{} values of \rowhammer{} attacks (by {54x}) by throttling their memory accesses, {without affecting benign applications' \rhli{} values, which stay at zero}.
\else 
{We make two observations from these experiments. First, benign applications exhibit zero \rhli{} because their row activation counts {never exceed} the blacklisting threshold. {On the other hand,} \rowhammer{} attacks reach an average (maximum, minimum) \rhli{} value of 10.9 (15.5, 6.9) in observe-only mode, showing that {an \rhli{} greater than 1 reliably {distinguishes} a} \rowhammer{} attack thread. Second, when in full-functional mode, \blockhammer{} {reduces an attack's \rhli{} by 54x on average, effectively reducing the \rhli{} of all \rowhammer{} attacks to below 1.}} {\blockhammer{} does not affect} {benign applications' \rhli{} values, which stay at zero}.
\fi 

\hammerthrottler{} calculates \rhli{} separately for each {<thread, bank>} pair. {To do so,}
{\hammerthrottler{} maintains two counters per {<thread, bank>} pair, using the same time-interleaving mechanism as the dual counting Bloom filters (D-CBFs) in \rowblocker{} (see \secref{blockhammer:sec:mech_detect}).
At any given time, one of the counters is designated as the active counter, while the other is designated as the passive counter. Both counters are incremented when {the thread activates a} blacklisted row {in the bank}. {Only the} active counter is used to calculate \rhli{} {at any point in time}.
{When} \rowblocker{} clears its active filter {for a given bank}, \hammerthrottler{} clears {each thread's active counter {corresponding to} the bank and swaps the active and passive counters.}

We implement \hammerthrottler{}'s counters as saturating counters}
because \rhli{} never exceeds 1 in a \blockhammer{}-{protected} system. Therefore, a{n} \hammerthrottler{} counter saturates at the \rowhammer{} threshold normalized to a CBF's lifetime, which we calculate as $\nrhn{}\times(\tbfn{}/\trefwn{})$.
{For the configuration we provide in Table~\ref{blockhammer:tab:tuned_params}}, \hammerthrottler{}'s counters require {only} four bytes of additional storage in the memory controller for each {<thread, bank>} pair 
{(e.g., 512~bytes in total for an eight-thread system with a 16-bank DRAM rank).}

\subsubsection{{Throttling {\rowhammer{}} Attack {Threads}}}
\label{blockhammer:sec:ht_throttling} 
\hammerthrottler{} throttles any thread  with {a non-zero} \rhli{}.
{To do so, \hammerthrottler{} limits the in-flight request count of each {<thread, bank>} pair by applying a quota inversely proportional to the {<thread, bank>} pair's \rhli{}.}
{Whenever a thread reaches its quota, the thread is {\emph{not}} allowed to make} {a} new memory request {to the shared caches} {or directly to the main memory} until {one of} its in-flight requests {is} completed.
If the thread continues to activate blacklisted rows {in a bank, its \rhli{} increases and consequently} its quota decreases{.}
This slows down the {\rowhammer{} attack thread} while freeing up additional memory bandwidth for {concurrently-running} benign threads that experience no throttling due to their {zero} \rhli{}. In this way, \blockhammer{} mitigates the performance overhead that a \rowhammer{} attack could inflict upon benign applications.

\subsubsection{{Exposing RHLI to the System Software}}
\label{blockhammer:sec:rhli_system_level}
{Although \blockhammer{} operates independently from the system software, e.g., the operating system (OS), \blockhammer{} can optionally expose its {per-DRAM-bank, per-thread \rhli{} values} to the OS. The OS can then use {this information} to mitigate
an ongoing 
\rowhammer{} attack at the software level. For example, the OS might kill or deschedule an attacking thread to prevent it from negatively impacting the system's performance and energy.} We leave {the study of OS-level mechanisms} using \rhli{} for future work.

\section{Many-Sided \rowhammer{} Attacks}
\label{blockhammer:sec:many_sided_attacks}
Hammering an aggressor row can disturb physically nearby rows even if they are not {immediately} adjacent~\cite{kim2014flipping, kim2020revisiting}, allowing {\emph{many-sided}} attacks that hammer {\emph{multiple}} DRAM rows to induce \rowhammer{} bitflips as a result of their cumulative disturbance~\cite{frigo2020trrespass}. Kim et al.~\cite{kim2014flipping} report that an aggressor row's impact decreases based on its physical distance to the victim row (e.g., by an order of magnitude per row) and disappears after a certain distance (e.g., 6 rows~\cite{kim2014flipping, frigo2020trrespass, kim2020revisiting}).

To address many-sided \rowhammer{} attacks, we conservatively add up the effect of each row to reduce \blockhammer{}'s \rowhammer{} threshold (\nrh{}), such that the cumulative effect of concurrently hammering each row \nrhtuned{} times becomes equivalent to hammering only an immediately-adjacent row \nrh{} times.
{We} calculate \nrhtuned{} {using} three parameters:
(1)~\nrh{}: the \rowhammer{} threshold for hammering a single row; (2)~{blast radius ($r_{blast}$)}: the
{maximum physical distance (in terms of rows) from the aggressor row at which RowHammer bitflips can be observed;}
and (3)~{blast impact factor {($c_{k}$)}}: {the} ratio {between the} activation counts required to {induce a bitflip in a victim row {by hammering {($i$)~an} immediately-adjacent row and {($ii$)~a} row} at {a} distance {of} $k$ rows away{.}} {W}e calculate the disturbance {that} hammering a row $N$ times {causes for} a victim row {that is physically located} $k$ {rows away as:} $N \times c_{k}$.
{\equref{blockhammer:equ:tai_naggr} shows how we calculate \nrhtuned{} in terms of \nrh{}, {$c_{k}$}, and $r_{blast}$. 
{We set \nrhtuned{} such that, even when all rows within the blast radius of a victim row (i.e., $r_{blast}$ rows on both sides of the victim row) are hammered for \nrhtuned{} times, their cumulative disturbance (i.e., $2\times(\nrhtunedn{}\times c_{1}+\nrhtunedn{}\times c_{2}+...+\nrhtunedn{}\times c_{r_{blast}})$) on the victim row will not exceed the disturbance of hammering an immediately-adjacent row \smash{\nrh{}} times.}}
\begin{equation}
\footnotesize
\nrhtunedn{} = \frac{\nrhn{}}{2\sum_{1}^{r_{blast}}{c_{k}}}, \quad \mathrm{where}
\begin{cases}
      c_{k} = 1, & \text{if $k = 1$}\\
      0 < c_{k} < 1, & \text{if $r_{blast} \geq k > 1$ }\\
      c_{k} = 0, & \text{if $k > r_{blast}$}
\end{cases}
\label{blockhammer:equ:tai_naggr}
\vspace{-0.5\baselineskip}
\end{equation}%
{$r_{blast}=6$ and {\smash{$c_{k}={0.5}^{k-1}$}} are the {worst-case} values observed {in modern DRAM chips} based on experimental results presented in prior characterization studies~\cite{kim2014flipping, kim2020revisiting}, {which} characterize {more than 1500} real DRAM chips from different vendors, standards, and generations from {2010} to 2020. To support {a} DRAM chip {with {these} worst-case characteristics},}
we find that $\nrhtunedn{}$ should equal $0.2539\times\nrhn{}$ using \equref{blockhammer:equ:tai_naggr}.
Similarly, to {configure \blockhammer{} for} double-sided attacks {{(which is} the attack model that state-of-the-art \rowhammer{} mitigation mechanisms address}~\cite{kim2014flipping, son2017making, you2019mrloc, seyedzadeh2018mitigating, lee2019twice, park2020graphene}{)}, we calculate \nrhtuned{} as half of \nrh{} 
(i.e., {$r_{blast} = c_{k} = 1$}).
{Table~\ref{blockhammer:tab:tuned_params}
presents \blockhammer{}'s configuration
for timing specifications of a commodity DDR4 DRAM chip~\cite{jedec2020jesd794c} and
a realistic \rowhammer{} threshold of 32K~\cite{kim2020revisiting}, tuned to address double-sided attacks.}

\begin{table}[ht]
\footnotesize
\renewcommand{\arraystretch}{1.05}
\setlength{\tabcolsep}{3pt}
\centering
\caption{{Example \blockhammer{} parameter values based on DDR4 specifications~\cite{jedec2017jesd794b, jedec2020jesd794c} and \rowhammer{} vulnerability~\cite{kim2020revisiting}.}}
 \label{blockhammer:tab:tuned_params}
\begin{tabular}{l|llllll}
\head{Component} & \multicolumn{5}{l}{\textbf{Parameters}} \\ \hline
\multirow{2}{*}{DRAM Features} & \nrh{} &: {32K} & Banks &: 16 & \trc{}  &: \SI{46.25}{\nano\second}  \\ 
                               & {\nrhtuned{}} &: {16K} & \trefw{} &: \SI{64}{\milli\second}  & \tfaw{}  &: \SI{35}{\nano\second} \\ 
\hline
\multirow{3}*{\setlength\tabcolsep{0pt}\begin{tabular}{l}{\rbbl{}}\end{tabular}} & \nbl{}  &: 8K  & \tbf{} &: {\SI{64}{\milli\second}\quad} & \tdelay{}\footnote{This is the {theoretical} maximum delay that a row activation can {experience}. 
Benign workloads {actually} experience {smaller delays of} up to \SI{1.7}{\micro\second}, \SI{3.9}{\micro\second}, {and \SI{7.6}{\micro\second} for P50, P90, and P100} of the row activations (see Section~\ref{blockhammer:sec:evaluation_false_positives}).} &: {\SI{7.7}{\micro\second}} \\ 
 & \multicolumn{2}{l}{CBF size}   & \multicolumn{4}{l}{: {1K} counters per CBF} {(per-bank)} \\
& \multicolumn{2}{l}{CBF Hashing} & \multicolumn{4}{l}{: 4 H3-class functions~\cite{carter1979universal} per CBF} \\
\hline
{\rbhb{}} & \multicolumn{2}{l}{Hist. buffer size} & \multicolumn{4}{l}{: {887} entries per {rank (16 banks)}} \\ 
\hline
\hammerthrottler{} & \multicolumn{6}{l}{2 counters per <thread, bank> pair}\\
\hline
 \end{tabular}
 
\end{table}

\section{Security Analysis}
\label{blockhammer:sec:mech_security}
{W}e use the \emph{proof by contradiction} method to prove that no \rowhammer{} attack can defeat \blockhammer{} {(i.e.,} activate a DRAM row more than \nrh{} times in a refresh window{)}. To do so, we begin with the assumption that there exists an access pattern that can exceed \nrh{} by defeating \blockhammer{}. Then, we mathematically represent all possible distributions of row activations and define the constraints for activating a {row more} than \nrh{} times in a refresh window. Finally, we show that it is impossible to satisfy these constraints, and thus, no {such} access pattern that can defeat \blockhammer{} {exists}. Due to space constraints, we briefly summarize all steps of the proof.

\nind
\head{Threat Model} We assume a {comprehensive} threat model in which the attacker can (1)~fully utilize memory bandwidth, (2)~precisely time each memory request, and (3)~comprehensively {and accurately} know details of the memory controller, \blockhammer{}, and DRAM implementation. 
{In addressing this threat model, 
{we do}
not consider any hardware or software component {to be} \emph{trusted} or \emph{safe} except {for} the memory controller, {the} DRAM chip, and the physical interface between those two.}

\nind
\head{Crafting an Attack} 
We model a generalized {memory} access pattern that a \rowhammer{} attack can exhibit from the perspective of an aggressor row. We represent {an attack's}
{row activation pattern in} 
a series of epochs, each of which is bounded by \rowblocker{}'s D-CBF {\emph{clear}} commands {to either CBF (i.e., half of the CBF lifetime or $\tbfn{}/2$), }%
as shown in \figref{blockhammer:fig:bloomfilter_timing}. 
According to the {time-interleaving mechanism} {(explained in Section~\ref{blockhammer:sec:mech_detect})}, the active CBF blacklists a row based on the row's total activation count in the current and previous epochs {to} limit the number of activations to the row. To demonstrate {that} \rowblocker{} effectively limits the number of activations to a row, {and therefore prevents} {all possible \rowhammer{} attacks,} 
we model all possible activation patterns targeting a DRAM row at the granularity of a single epoch. From the perspective of a CBF, each epoch can be classified based on the number of activations that the aggressor can receive in the previous (\npre{}) and current (\nagg{}) epochs. We identify five possible epoch types (i.e., $T_{0}-T_{4}$), which we list in Table~\ref{blockhammer:tab:security_accpatterns}. {The table} {shows} (1)~the {range of row} activation counts in the previous epoch (\npre{}), (2)~the {range of row} activation counts in the current epoch (\nagg{}), {and (3)~the maximum {possible} row activation count in the current epoch ($\naggn{}_{max}$).} 

\begin{table}[h]
\centering
\caption{{Five possible epoch types that span all possible memory access patterns, {defined by the number of row activations the aggressor row can receive in the previous epoch (\npre{}) and in the current epoch (\nagg{}). $\naggn{}_{max}$ shows the maximum value of \nagg{}.}}}
 \label{blockhammer:tab:security_accpatterns}
 \begin{tabular}{c|ccl}
  Epoch Type    & \npre{}        &           \nagg{}                  & $\naggn{}_{max}$ \\ 
  \hline
  $T_{0}$ &                &        $\naggn{} < \nblpn{}$       & $\nblpn{} - 1$      \\ 
  $T_{1}$ & $  < \nthn{} $ & $\nblpn{} \leq \naggn{} < \nthn{}$ & $\nthn{} - 1$    \\ 
  $T_{2}$ &                &      $\naggn{} \geq \nthn{}$       & $\tepochn{}/\tdelayn{} - (1-\trcn{}/\tdelayn{})\nblpn{}$    \\  \hline
  $T_{3}$ & \multirow{2}{*}{$\geq \nbln{}$} &      $ \naggn{} < \nbln{} $        & $\nbln{} - 1$    \\ 
  $T_{4}$ &                &      $\naggn{} \geq \nthn{}$       & $\tepochn{}/\tdelayn{} $    \\ \hline
 \end{tabular}
 
\end{table}

{{The epoch type indicates the recent activation rate of the aggressor row, and \rowblocker{} uses this information to determine whether {or not} to blacklist} the aggressor row {in the current and next epochs}. A $T_0$ epoch indicates that the row was activated {fewer} than $\nbln{}$ times in the previous epoch (i.e., $\npren{} < \nbln{}$) and {fewer} than $\nbln{}-\npren{}$ times (denoted as \nblp{} for simplicity) in the current epoch. Since the row was activated fewer times than the {blacklisting} threshold, the row {is} not blacklisted in the current epoch. Compared to $T_0$, a $T_1$ epoch indicates that the row was activated greater than \nblp{} times but fewer than \nbl{} times in the current epoch. Since the activation count exceeds the threshold \nblp{} but not \nbl{}, the row is blacklisted in the current epoch. {When a $T_1$ type epoch finishes, the row starts the next epoch as \emph{not blacklisted} because the row's activation count is {lower} than \nbl{}.} Compared to $T_1$, a $T_2$ epoch indicates that the row's activation count in the current epoch exceeds \nbl{}. As the activation count exceeds the {blacklisting threshold} \nbl{}, the row is blacklisted in the current and next epochs.}

{A $T_3$ epoch indicates that the row's activation count in the previous epoch exceeded \nbl{} and the row is activated fewer times than \nbl{} times in the current epoch. In this case, the row is blacklisted in the current epoch, but no longer blacklisted in the {beginning of the} next epoch. Compared to $T_3$, a $T_4$ epoch indicates that the row is activated more than \nbl{} times in the current epoch. The row is blacklisted in both current and next epochs, as its activation rate is too high and {could} lead to a successful \rowhammer{} attack {if not blacklisted}.}

{We calculate the upper bound for the total activation count an attacker can reach during the current epoch (shown under \nepmax{} in  Table~\ref{blockhammer:tab:security_accpatterns}).} In {the} $T_{0}$, $T_1$, or $T_3$ {epochs}, by definition, a row's activation count cannot exceed $\nblpn{}-1$, $\nbln{}-1$, and $\nbln{}-1$, respectively.
{{In {a $T_{4}$ epoch}, the row is} already blacklisted from the beginning ($N_{0} \geq \nbln{}$). Therefore, {the row can be activated at most once in every \tdelay{} time window, resulting in an upper bound activation count of $\tepochn{}/\tdelayn{}$.} 
{In {a $T_{2}$ epoch}, a row can be activated} \nblp{} times at a time interval as small as \trc{}, which takes $t_{1}=\nblpn{}\times\trcn{}$ time. Then, the row is blacklisted and further activations are performed with a minimum interval of \tdelay{}, which takes $t_{2}=(\nepmaxin{}-\nblpn{})\times\tdelayn{}$ time. Since all {of} these activations need to fit into the epoch's time window, we solve the equation $\tepochn{}=t_{1}+t_{2}$ for \nep{}, and derive \nepmax{} for an epoch of type $T_{2}$ as shown in Table~\ref{blockhammer:tab:security_accpatterns}.} 

\nind
\head{Constraints of a Successful \rowhammer{} Attack}
{\lineskip=0pt\lineskiplimit=-\maxdimen{%
We mathematically represent a hypothetically successful \rowhammer{} attack as a permutation of many epochs. We denote the number of instances for an epoch type $i$ as $n_i$ and the maximum activation count the epoch $i$ can reach as $\nepmaxin(i)$.
{To be successful, the \rowhammer{} attack must satisfy} three constraints, {which} we present in Table~\ref{blockhammer:tab:security_constraints}.} 
(1)~{The attacker should activate a}n aggressor row more than \nrh{} times within a refresh window (\trefw{}). 
(2)~Each epoch type can be preceded only by a {subset} of epoch types.\footnote{{Since we define epoch types based on activation counts in both the previous and current epochs, we note that consecutive epochs are dependent and therefore limited: an epoch of type $T_{0}$, $T_1$, or $T_2$ can be preceded only by an epoch of type $T_{0}$, $T_1$, or $T_3$, while an epoch of type $T_3$ or $T_4$ can be preceded only by an epoch of type $T_2$ or $T_4$}.}
Therefore, 
an epoch type $T_x$ {cannot occur more} times {than} the total number of {instances of all epoch types that can {precede} epoch type $T_x$.}
(3)~An epoch cannot occur for a negative number of times. 
}%

\begin{table}[h]
 \centering
\caption{{Necessary} constraints of a successful attack.}
 \label{blockhammer:tab:security_constraints}
 \begin{tabular}{rll}
 \hline
{(1)} & $\nrhn{} \leq \sum{(n_{i}\times\nepmaxin{})}$, & $\trefwn{} \geq \tepochn{}\times\sum{n_{i}}$\\
  {(2)} & $n_{0,1,2} \leq n_{0} + n_{1} + n_{3};$ & $n_{3,4}   \leq n_{2} + n_{4};$\\
  {(3)} & $\forall n_{i} \geq 0$ & \\
  \hline
 \end{tabular}
 
\end{table}


We use an analytical solver~\cite{wolframresearchwolframalpha} to {identify} a set of $n_{i}$ values that meets all constraints in Table \ref{blockhammer:tab:security_constraints} for the {\blockhammer{}} configuration we {provide} in Table~\ref{blockhammer:tab:tuned_params}. 
{We find that there exist{s} no combination of $n_{i}$ values that satisfy these constraints. {Therefore, we conclude} that no access pattern exists that}
can activate an aggressor row more than \nrh{} times within a refresh window {in a \blockhammer{}-protected system.} 
\begin{table*}[ht]
    \centering
    \caption{{{Per-rank area, {access energy,} and {static} power} of \blockhammer{} {vs. state-of-the-art \rowhammer{} mitigation mechanisms.}}}
    \label{blockhammer:tab:area_cost_analysis}
    \footnotesize
    \renewcommand{\arraystretch}{1} 
    \resizebox{\linewidth}{!}{
    \begin{tabular}{l|rrrrrr|rrrrrr}
    & \multicolumn{6}{c}{\textbf{\nrh{}=32K*}} & \multicolumn{6}{c}{\textbf{\nrh{}=1K}} \\
    & \textit{} &  & & & Access & Static & \textit{} &  & & & Access & Static\\
    Mitigation  & \textit{SRAM} & \textit{CAM} & \multicolumn{2}{c}{\textit{Area}} & Energy & Power & \textit{SRAM} & \textit{CAM} & \multicolumn{2}{c}{\textit{Area}} & Energy & Power\\
    Mechanism & KB & KB & mm$\mathbf{^2}$ & \% CPU & (\SI{}{\pico\joule}) & (\SI{}{\milli\watt}) & KB & KB & mm$\mathbf{^2}$ & \% CPU & (\SI{}{\pico\joule}\om{)} & (\SI{}{\milli\watt})\\ \hline
        \textbf{\blockhammer{}} & \textbf{51.48} & \textbf{1.73} & \textbf{$\mathbf{0.14}$} & \textbf{0.06} & \textbf{20.30} & \textbf{22.27} & \textbf{441.33} & \textbf{55.58} & \textbf{$\mathbf{1.57}$} & \textbf{0.64} & \textbf{99.64} & \textbf{220.99} \\
        \quad D-CBF & 48.00 & - & $0.11$ & 0.04 & 18.11 & 19.81 & 384.00 & - & $0.74$ & 0.30 & 86.29 & 158.46\\
        \quad Hash functions & - & - & $<0.01$ & $<0.01$ & - & - & - & - & $<0.01$ & $<0.01$ & - & -\\
        \quad History buffer & 1.73 & 1.73 & $0.03$ & 0.01 & 1.83 & 2.05 & 55.58 & 55.58 & $0.83$ & 0.34 & 12.99 & 62.12 \\
        \quad \hammerthrottler{} & 1.75 & - & $<0.01$ & $<0.01$ & 0.36 & 0.41 & 1.75 & - & $<0.01$ & $<0.01$ & 0.36 & 0.41\\
        \textbf{PARA~\cite{kim2014flipping} }   & \textbf{-} & \textbf{-} & \textbf{$\mathbf{<0.01}$} & \textbf{-} & \textbf{-} & \textbf{-} & \textbf{-} & \textbf{-} & \textbf{$\mathbf{<0.01}$} & \textbf{-} & \textbf{-} & \textbf{-} \\
        \textbf{ProHIT~\cite{son2017making}* } & \textbf{-} & \textbf{0.22} & \textbf{$\mathbf{<0.01}$} & \textbf{<0.01} & \textbf{3.67} & \textbf{0.14} & \textbf{$\times$} & \textbf{$\times$} & \textbf{$\times$} & \textbf{$\times$} & \textbf{$\times$} & \textbf{$\times$}\\
        \textbf{MrLoc~\cite{you2019mrloc}* }  & \textbf{-} & \textbf{0.47} & \textbf{$\mathbf{<0.01}$} & \textbf{<0.01} & \textbf{4.44} & \textbf{0.21} & \textbf{$\times$} & \textbf{$\times$} & \textbf{$\times$} & \textbf{$\times$} & \textbf{$\times$} & \textbf{$\times$}\\
        \textbf{CBT~\cite{seyedzadeh2018mitigating} } & \textbf{16.00} & \textbf{8.50} & \textbf{$\mathbf{0.20}$} & \textbf{0.08} & \textbf{9.13} & \textbf{35.55} & \textbf{512.00} & \textbf{272.00} & \textbf{$\mathbf{3.95}$} & \textbf{1.60} & \textbf{127.93} & \textbf{535.50} \\
        \textbf{TWiCE~\cite{lee2019twice} } & \textbf{23.10} & \textbf{14.02} & \textbf{$\mathbf{0.15}$} & \textbf{0.06} & \textbf{7.99} & \textbf{21.28} & \textbf{738.32} & \textbf{448.27} & \textbf{$\mathbf{5.17}$} & \textbf{2.10} & \textbf{124.79} & \textbf{631.98}\\
        \textbf{Graphene~\cite{park2020graphene} } & \textbf{-} & \textbf{5.22} & \textbf{$\mathbf{0.04}$} & \textbf{0.02} & \textbf{40.67} & \textbf{3.11} & \textbf{-} & \textbf{166.03} & \textbf{$\mathbf{1.14}$} & \textbf{0.46} & \textbf{917.55} & \textbf{93.96} \\
        \hline
        \multicolumn{13}{l}{{*~\prohit{}~\cite{son2017making} and \mrloc{}~\cite{you2019mrloc} do {\emph{not}} provide a concrete discussion on how to adjust their empirically-determined parameters for}}\\
        \multicolumn{13}{l}{{different \nrh{} {values}. Therefore, we 1)~report their values for a fixed design point that each paper provides for \nrh{}=2K and 2)~mark}}\\
        \multicolumn{13}{l}{values we cannot estimate using an $\times$.}\\
    \end{tabular}
    }
    
\end{table*}
\section{Hardware Complexity Analysis}
\label{blockhammer:sec:hardware_complexity}

{We evaluate \blockhammer{}'s {(1)}~chip area, {static power, and access energy} {consumption} using CACTI~\cite{muralimanohar2009cacti} and {(2)~circuit latency using} Synopsys DC~\cite{synopsyssynopsys}. We demonstrate that \blockhammer{}'s physical costs are competitive with state-of-the-art \rowhammer{} mitigation mechanisms.}

\subsection{Area, Static Power, and Access Energy}
\label{blockhammer:sec:blockhammer_areaoverhead}

{\lineskiplimit=-\maxdimen%
Table~\ref{blockhammer:tab:area_cost_analysis} shows an area, {static power, and access energy} cost analysis {of} \blockhammer{}  
alongside six state-of-the-art \rowhammer{} mitigation
mechanisms~\cite{kim2014flipping,son2017making,you2019mrloc,seyedzadeh2018mitigating, lee2019twice, park2020graphene}, one of which is concurrent work {with} \blockhammer{} (Graphene~\cite{park2020graphene}).
{We perform this analysis {at} two \rowhammer{} thresholds {(\nrh{})}: 32K and 1K.\footnote{We {configure each mechanism} as we describe in Section~\ref{blockhammer:sec:methodology}.}}}

\nind
\head{{Main Components of \blockhammer{}}} {\blockhammer{} combines two mechanisms: \rowblocker{} and \hammerthrottler{}.} \rowblocker{}, {as shown in Figure~\ref{blockhammer:fig:overview},} 
{consists of two components} (1)~{\rbbl{}, which implements} {a dual counting Bloom filter {for each DRAM bank}}, and (2)~{\rbhb{}, which implements} a row activation history buffer {for each DRAM rank}. 
When configured {to handle} a \rowhammer{} threshold {(\nrh{})} of 32K, {as shown in Table~\ref{blockhammer:tab:tuned_params},} each counting Bloom filter has {1024} {13-bit} counters, stored in an SRAM array. These {counters} are indexed by four H3-class hash functions~\cite{carter1979universal}{, which introduce negligible area overhead (discussed in Section~\ref{blockhammer:sec:mech_detect})}. 
\rbhb{}'s history buffer holds {887~entries} per DRAM rank. {E}ach {entry} contains 32~bits for a row ID, a timestamp, and a valid bit.
{\hammerthrottler{}} uses {two counters per thread per DRAM bank}
to measure the \rhli{} of each <thread, bank> pair.
We estimate {\blockhammer{}'s} overall area overhead as \smash{\SI{0.14}{\milli\meter\squared}} per DRAM rank, for a 16-bank DDR4 memory. 
{For {a high-end} 28-core {Intel Xeon processor} system with four memory channels and single-rank {DDR4} DIMMs, \blockhammer{} consumes approximately \smash{\SI{0.55}{\milli\meter\squared}}, which translates to {only} 0.06\% of {the}
{CPU} die area~\cite{wikichipcascade}.}
{When configured for an \nrh{} of 1K, {we reduce} \blockhammer{}{'s} blacklisting threshold (\nbl{}) from 8K to 512, reducing the CBF counter width from {13~bits to 9~bits}. To avoid false positives {at} the reduced blacklisting threshold, we increase {the} CBF size to 8K. With this modification, \blockhammer{}'s D-CBF consumes \smash{\SI{0.74}{\milli\meter\squared}}. Reducing \nrh{} mandates larger time delays between subsequent row activations targeting a blacklisted row, thereby increasing the history buffer's size from 887 to 27.8K~entries, which translates to \smash{\SI{0.83}{\milli\meter\squared}} chip area. Therefore, \blockhammer{}'s total area overhead {at an \nrh{} of 1K} is \smash{\SI{1.57}{\milli\meter\squared}} or 0.64\% of the {CPU} die area~\cite{wikichipcascade}.}

\nind 
{\textbf{Area Comparison.} 
{\graphene{}, \twice{}, and \cbt{} need to store 5.22KB, 37.12KB, and 24.50KB}} of metadata in the memory controller per DRAM {rank, for the same 16-bank DDR4 memory}, which translates to {similarly {low area overheads of}} 0.02\%, 0.06\%, {and} 0.08\% {of the CPU die area}, respectively. {Graphene's area overhead per byte of metadata is larger than other mechanisms because \graphene{} is fully implemented with CAM logic, as shown in Table~\ref{blockhammer:tab:area_cost_analysis}.} {\para{}, \prohit{}, and \mrloc{} are extremely area efficient compared to other mechanisms because they {are probabilistic mechanisms~\cite{kim2014flipping, son2017making, you2019mrloc}}, and thus do not need to store {kilobytes} of metadata to track row activation rates.}

{We repeat our area overhead analysis for future DRAM chips by scaling the \rowhammer{} threshold down to 1K. While \blockhammer{} {consumes \smash{\SI{1.57}{\milli\meter\squared}} of chip area {to {prevent {bitflips}} at this lower threshold},}
\twice{}'s {and \cbt{}'s} area overhead {increases} to 3.3x and 2.5x of \blockhammer{}'s.
{We} conclude that \blockhammer{} scales better than {both \cbt{} and} \twice{} in terms of area overhead.}
{\graphene{}'s} area overhead does not scale as efficiently as \blockhammer{} with decreasing \rowhammer{} threshold, and becomes comparable to \blockhammer{} when configured for a \rowhammer{} threshold of 1K. 

\nind 
{\textbf{Static Power and Access Energy Comparison.}
When configured for {an} \nrh{} {of} 32K, {\blockhammer{} consumes \SI{20.30}{\pico\joule} {per access}, which is half of \graphene{}'s access {energy; and} \SI{22.27}{\milli\watt} of static power, which is 
{63\% of \cbt{}'s.}}
\blockhammer{}'s {static} power consumption scales more efficiently {than {that of} \cbt{} and \twice{}} {as {\nrh{}} decreases {to 1K}}, {whereas} 
\cbt{} and \twice{} {consume 2.42x and 2.86x} {the} static power {of \blockhammer{}, respectively.} 
Similarly, \graphene{}'s access energy and static power drastically increase by 22.56x and 30.2x, respectively, when {\nrh{}} scales down to 1K. As a result, \graphene{} {consumes $9.21\times$ of \blockhammer{}'s {access energy}.}
}
\subsection{Latency Analysis} 
\label{blockhammer:sec:blockhammer_latencyoverhead}
We implement \blockhammer{} in Verilog HDL and synthesize {our design using} Synopsys DC~\cite{synopsyssynopsys} {with} a \SI{65}{\nano\meter} process technology to evaluate {the} latency impact on memory accesses. According to our RTL model\om{,} {which we open source~\cite{safari2021blockhammer}}, \blockhammer{} responds {to} {an} \emph{{``{Is this ACT} \rowhammer{}-safe?''}} query (\circled{1} in \figref{blockhammer:fig:overview}) in only \SI{0.97}{\nano\second}. {This} latency {can be hidden because it is} {one-to-}two orders of magnitude smaller than the row access latency (e.g., \SIrange{45}{50}{\nano\second}) that DRAM standards (e.g., DDRx, LPDDRx, GDDRx) enforce~\dramStandardCitations{}. 
\section{Experimental Methodology}
\label{blockhammer:sec:methodology}
\label{blockhammer:sec:evaluation_methodology_simulation}
We evaluate \blockhammer{}'s {effect} on a typical DDR4-based memory subsystem's performance and energy consumption {as compared to six prior} \rowhammer{} mitigation mechanisms~\cite{kim2014flipping,son2017making,you2019mrloc,seyedzadeh2018mitigating,lee2019twice,park2020graphene}.
We use Ramulator~\cite{Kim2016Ramulator,safariramulator}
for performance evaluation and DRAMPower~\cite{chdrampower_opensource} to estimate {DRAM} energy consumption. \exttwo{We open-source our infrastructure, which implements both BlockHammer and six state-of-the-art RowHammer mitigation mechanisms~\cite{safari2021blockhammer}}.
\tabref{blockhammer:tab:system_configuration} shows our system configuration.

 \newcolumntype{C}[1]{>{\let\newline\\\arraybackslash\hspace{0pt}}m{#1}}
 \begin{table}[h]
 \footnotesize
\centering
 \caption{{Simulated} system configuration.}
 \label{blockhammer:tab:system_configuration}
 \begin{tabular}{l|C{5.8cm}}
 \hline
 \textbf{Processor} & {\SI{3.2}{\giga\hertz}, \{1,8\}~core, 4-wide issue, {128-entry} instr. window}\\ \hline
 \textbf{Last-Level Cache} & {64-byte} cache line, 8-way {set-associative, \SI{16}{\mega\byte}} \\ \hline
 \textbf{Memory Controller} & {64-entry each read and write request queues; Scheduling policy: FR-FCFS~\cite{rixner2000memory, rixner2004memory, zuravleff1997controller}; Address mapping: MOP~\cite{kaseridis2011minimalist}} \\ \hline
 \textbf{Main Memory} & DDR4, 1 channel, 1 rank, 4 bank groups, 4 banks/bank group, {64K} rows/bank\\ \hline
 \end{tabular}
 \end{table}

\nind
\head{Attack Model}
{W}e compare BlockHammer under the same {\rowhammer{}} attack model ({i.e.,} double-sided attacks \cite{kim2014flipping}) as prior works use~\cite{lee2019twice, seyedzadeh2018mitigating, park2020graphene, kim2014flipping, son2017making, you2019mrloc}. {To do so, we halve the \rowhammer{} threshold that \blockhammer{} uses to account for the cumulative disturbance effect of both aggressor rows (i.e., }
$N_{RH}*=N_{RH}/2$).
In Sections \ref{blockhammer:sec:evaluation_single_core} and \ref{blockhammer:sec:evaluation_multi_core}, we set $N_{RH}*=16K$ (i.e., $N_{RH}=32K$), which is the minimum \rowhammer{} threshold that \twice{}~\cite{lee2019twice} supports~\cite{kim2020revisiting}.
{In Section~\ref{blockhammer:sec:evaluation_tech_scaling}, we} conduct an $N_{RH}$ scaling study for double-sided attacks, {across a} range of $32K > N_{RH} > 1K$, {using parameters provided in Table~\ref{blockhammer:tab:allconfigs}.} 

\nind
\head{{Comparison Points}}
We compare \blockhammer{} to 
a baseline system with no \rowhammer{} {mitigation and to}
{six} {state-of-the-art \rowhammer{} mitigation} mechanisms {that provide \rowhammer{}-safe operation:} {three {are probabilistic mechanisms~\cite{kim2014flipping, son2017making, you2019mrloc} and another three are deterministic mechanisms~\cite{seyedzadeh2018mitigating, lee2019twice, park2020graphene}.}}
(1)~\para{}~\cite{kim2014flipping} mitigates \rowhammer{} by injecting an adjacent row activation with a low probability whenever the memory controller {closes a row following an activation}.
{We 
tune} \para{}'s probability threshold for a given \rowhammer{} threshold to meet a desired failure probability (we use \smash{$10^{-15}$} as a typical consumer memory reliability target~\cite{cai2012error, cai2017error, jedec2012jep122g, luo2016enabling, patel2017thereach}) in a refresh window (\SI{64}{\milli\second}).
(2)~\prohit{}~\cite{son2017making} implements a history table of recent row activations to extend \para{} by {reducing the probability} threshold for more frequently activated rows.
We configure \prohit{} using the default probabilities and parameters provided in \cite{son2017making}.
(3)~\mrloc{}~\cite{you2019mrloc} {extends \para{} by keeping a record of recently-refreshed potential} victim rows in a queue and dynamically {adjusts} the probability threshold, which it uses to decide whether {or} not to refresh the victim row, 
based on {the row's temporal} locality information. We implement \mrloc{} by using the empirically-determined parameters {provided in \cite{you2019mrloc}.}
(4) \cbt{}~\cite{seyedzadeh2017counterbased} proposes a tree of counters to count the activations for non-uniformly-sized disjoint memory regions, {each} of which is halved in size (i.e., moved to the next level of the tree) every time its activation count reaches a predefined threshold. After being halved a predefined number of times (i.e., after becoming a leaf node in the tree), all rows in the memory region {are} refreshed.
We implement \cbt{} with a six-level tree that contains 125~counters, {and exponentially increase the threshold values across tree levels} 
from 1K to the \rowhammer{} threshold (\nrh{}), {as described in \cite{seyedzadeh2018mitigating}}.
(5)~\twice{} uses a table of counters to track the activation count of every row. Aiming {for} an area-efficient implementation, \twice{} periodically prunes the activation records of the rows whose activation counts {{cannot reach a {high enough} value to} cause bitflips}. 
{We implement and configure \twice{} for a \rowhammer{} threshold of 32K using the methodology described in the original paper~\cite{lee2019twice}. Unfortunately, \twice{} faces scalability challenges due to time consuming pruning operations, as described in \cite{kim2020revisiting}. To scale \twice{} for smaller \rowhammer{} thresholds, we follow the same methodology as Kim et al.{~\cite{kim2020revisiting}}.}
(6)~\graphene{}~\cite{park2020graphene} {adopts Misra-Gries, a frequent-element detection algorithm~\cite{misra1982finding},  to detect the} most frequently activated rows in a given time window. Graphene maintains a set of counters where it keeps the address and activation count of frequently activated rows. Whenever a row's counter reaches a multiple of a predefined threshold value, \graphene{} refreshes its adjacent rows. We configure \graphene{} by evaluating the equations provided in the original work~\cite{park2020graphene} for a given RowHammer threshold.


\head{BlockHammer's Configurations}
{Table \ref{blockhammer:tab:allconfigs} shows BlockHammer's configuration parameters used for each \rowhammer{} threshold (\nrh{}) in Sections~\ref{blockhammer:sec:blockhammer_areaoverhead} and~\ref{blockhammer:sec:evaluation_tech_scaling}.}

\begin{table}[h!]
    \footnotesize
    \centering
    \caption{\blockhammer{}'s configuration parameters used for different $\mathbf{\nrhn{}}$ values.}
    \label{blockhammer:tab:allconfigs}
    \begin{tabular}{r||rrrr}
        \emph{$\mathbf{\nrhn{}}$} & {\nrhtuned{}} & {CBF Size} & {\nbl{}} & {\tbf{}} \\ \hline\hline
        \textbf{32K} & 16K & 1K & 8K & \SI{64}{\milli\second} \\
        \textbf{16K} &  8K & 1K & 4K & \SI{64}{\milli\second} \\
        \textbf{ 8K} &  4K & 1K & 2K & \SI{64}{\milli\second} \\
        \textbf{ 4K} &  2K & 2K & 1K & \SI{64}{\milli\second} \\
        \textbf{ 2K} &  1K & 4K & 512& \SI{64}{\milli\second} \\
        \textbf{ 1K} & 512 & 8K & 256& \SI{64}{\milli\second} \\
         \hline
    \end{tabular}
    
\end{table}

\head{Workloads}
We evaluate \blockhammer{} and state-of-the-art \rowhammer{} mitigation mechanisms with {280} {(30~single-core and 250~multiprogrammed)} workloads. 
{W}e use {22} memory-intensive benign applications {from the} SPEC CPU2006 benchmark {suite}~\cite{stspec}, {four disk I/O {applications}
from {the} YCSB benchmark {suite}~\cite{cooper2010benchmarking}, two network I/O {applications} from a commercial network chip~\cite{semiconductorsqoriq}, and two synthetic {microbenchmarks}
that mimic non-temporal data copy.}
{We categorize these benign applications based on their row buffer conflicts per kilo instruction ($RBCPKI$) into three categories: {\emph{L}} ($RBCPKI<1$), {\emph{M}} ($1<RBCPKI<5$), and {\emph{H}} ($RBCPKI>5$). \emph{RBCPKI} is {an} indicator of row activation rate, {which is the key workload property that triggers} \rowhammer{} mitigation mechanisms.}
{There are 12, 9, and 9 applications in the {\emph{L}, \emph{M}, and \emph{H}} categories, respectively, {as listed in Table~\ref{blockhammer:tab:workload_list}}}.
{To mimic a double-sided \rowhammer{} attack, we use a synthetic trace that activates two rows {in each bank} as fast as possible by alternating between them at every row activation (i.e., $R_{A}$, $R_{B}$, $R_{A}$, $R_{B}$, ...).}

{Table~\ref{blockhammer:tab:workload_list} {lists the} 30 benign applications we use for cycle-level simulations. We report last-level cache misses ($MPKI$) and row buffer conflicts ($RBCPKI$) per kilo instructions for each application. Non-temporal data copy, YCSB Disk I/O, and network accelerator applications do not have an $MPKI$ value because they directly access main memory.} 
\begin{table}[h!]
    \footnotesize 
    \centering
    \caption{Benign applications used in cycle-level simulations.}
    \label{blockhammer:tab:workload_list}
    \begin{tabular}{l|l|l||rr}
\head{Category}                & \textbf{Benchmark Suite}            & \textbf{Application}    &   \textbf{MPKI} & \textbf{RBCPKI} \\ \hline \hline
\multirow{12}{*}{L}   & \multirow{10}{*}{SPEC2006} & 444.namd       &    0.1 &   0.0  \\
                        &                            & 481.wrf        &    0.1 &   0.0  \\
                        &                            & 435.gromacs    &    0.2 &   0.0  \\
                        &                            & 456.hmmer      &    0.1 &   0.0  \\
                        &                            & 464.h264ref    &    0.1 &   0.0  \\
                        &                            & 447.dealII     &    0.1 &   0.0  \\
                        &                            & 403.gcc        &    0.2 &   0.1  \\
                        &                            & 401.bzip2      &    0.3 &   0.1  \\
                        &                            & 445.gobmk      &    0.4 &   0.1  \\
                        &                            & 458.sjeng      &    0.3 &   0.2  \\ \cline{2-2}
                        & Non-Temp. Data Copy        & movnti.rowmaj  &    -   &   0.2  \\ \cline{2-2}
                        & \multirow{4}{*}{YCSB Disk I/O}      & ycsb.A         &    -   &   0.4  \\ \cline{1-1}\cline{3-5}
\multirow{9}{*}{M} &                            & ycsb.F         &    -   &   1.0  \\
                        &                            & ycsb.C         &    -   &   1.0  \\
                        &                            & ycsb.B         &    -   &   1.1  \\ \cline{2-2}
                        & \multirow{12}{*}{SPEC2006} & 471.omnetpp    &    1.3 &   1.2  \\
                        &                            & 483.xalancbmk  &    8.5 &   2.4  \\
                        &                            & 482.sphinx3    &    9.6 &   3.7  \\
                        &                            & 436.cactusADM  &   16.5 &   3.7  \\
                        &                            & 437.leslie3d   &    9.9 &   4.6  \\
                        &                            & 473.astar      &    5.6 &   4.8  \\ \cline{1-1}\cline{3-5}
\multirow{9}{*}{H}   &                            & 450.soplex     &   10.2 &   7.1  \\
                        &                            & 462.libquantum &   26.9 &   7.7  \\
                        &                            & 433.milc       &   13.6 &  10.9  \\
                        &                            & 459.GemsFDTD   &   20.6 &  15.3  \\
                        &                            & 470.lbm        &   36.5 &  24.7  \\
                        &                            & 429.mcf        &  201.7 &  62.3  \\ \cline{2-2}
                        & Non-Temp. Data Copy        & movnti.colmaj  &  -     &  30.9  \\ \cline{2-2}
                        & \multirow{2}{*}{Network accelerator}   & freescale1     &  -     & 336.8  \\
                        &                            & freescale2     &  -     & 370.4  \\ \hline 

    \end{tabular}
    
\end{table}

We randomly combine these {single-core} {workloads} to create {two types of multiprogrammed workload{s}}: (1)~{125~{workloads}} with \emph{{no \rowhammer{} attack}}, {each} including eight \nonmalic{} {threads}; and (2)~{{125~{workloads}} with a \emph{\rowhammer{} attack present}}, {each} including one \rowhammer{} {attack and} seven \nonmalic{} {threads}. {We simulate
each {{multiprogrammed} workload} until each benign thread executes at least 200~million
instructions. For all configurations, we warm up the
caches by fast-forwarding 100~million instructions,} {as done in prior work~\cite{kim2020revisiting}}.

\nind
\head{Performance {and {DRAM} Energy} Metrics}
{We evaluate \blockhammer{}'s impact on \emph{system throughput} (in terms of weighted speedup~\cite{snavely2000symbiotic, eyerman2008systemlevel, michaud2012demystifying}), \emph{job turnaround time} (in terms of harmonic speedup~\cite{luo2001balancing,eyerman2008systemlevel}), and \emph{fairness} {(in terms of maximum slowdown~\cite{kim2010thread, kim2010atlas,subramanian2014theblacklisting,subramanian2016bliss, subramanian2013mise, mutlu2007stalltime, subramanian2015theapplication, ebrahimi2010fairness, ebrahimi2011prefetchaware, das2009applicationaware, das2013applicationtocore}})}.
{Because the performance of a \rowhammer{} attack should not be accounted for in the {performance evaluation},
we calculate all three metrics only for benign {applications}.}
{To evaluate DRAM energy consumption, we compare the total energy consumption that DRAMPower provides in Joules. DRAM energy consumption includes both benign and \rowhammer{} attack requests.}
{Each data point shows the average value across
all workloads, with minimum and maximum values {depicted} {using}
error bars.}
\section{Performance and Energy Evaluation}
\label{blockhammer:sec:overhead_analysis}
\label{blockhammer:sec:evaluation}

{{W}e evaluate the performance and energy overheads of \blockhammer{} and {six} state-of-the-art \rowhammer{} mitigation mechanisms.}
{First, we evaluate all mechanisms with single-core {applications} and show that \blockhammer{} exhibits {no} performance 
and energy overheads, compared to a baseline system without any \rowhammer{} mitigation.
Second, we evaluate \blockhammer{} with multiprogrammed {workloads} and show that, by throttling an attack's requests, \blockhammer{} significantly improves the performance of benign applications {by 45.4\% on average (with a maximum of 61.9\%),}
compared to 
{both {the baseline system} and a system with the prior best-performing state-of-the-art \rowhammer{} mitigation mechanism.}
Third, we compare \blockhammer{} {with} state-of-the-art \rowhammer{} mitigation mechanisms {when applied to} future DRAM chips {that} are projected to be more vulnerable to \rowhammer{}. {We} show that \blockhammer{} {is competitive with state-of-the-art mechanisms at \rowhammer{} thresholds as low as 1K when there is no attack in the system, and provides significantly higher performance and lower DRAM energy consumption than state-of-the-art mechanisms when a \rowhammer{} attack is present}. 
{Fourth}, we {analyze} \blockhammer{}'s {internal mechanisms.}}

\subsection{Single-Core {Applications}}
\label{blockhammer:sec:evaluation_single_core}
{\figref{blockhammer:fig:single_core_performance_overhead_wo_hammer} presents the execution time and energy of benign applications (grouped into three categories based on their {\emph{RBCPKI}; see Section~\ref{blockhammer:sec:methodology}}) when executed on a single-core system that uses BlockHammer versus six state-of-the-art mitigation {mechanisms}, normalized to a baseline system that does not employ any RowHammer mitigation mechanism.}

\begin{figure}[h]
    \centering
    \includegraphics[width=0.8\linewidth]{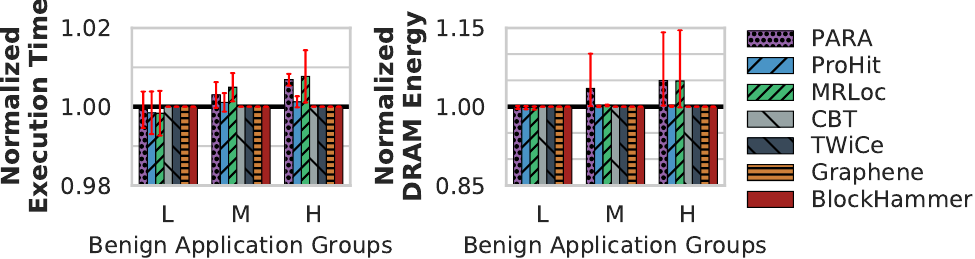}
    \caption{Execution time and DRAM energy consumption for {benign} single-core {applications}, normalized to baseline.}
    \label{blockhammer:fig:single_core_performance_overhead_wo_hammer}
\end{figure}

{We observe} that \blockhammer{} {introduces no} performance {and {DRAM} energy} overheads on benign applications {compared to the baseline configuration}.
This is because benign applications' per-row activation rates {never} exceed \blockhammer{}'s blacklisting threshold (\nbl{}).
{In {contrast}, 
{\para{}/\mrloc{}} exhibit {0.7\%/0.8\%} performance and {4.9\%/4.9\%} energy overheads for high \emph{RBCPKI} {applications,} on average.} {\cbt{}, \twice{}, and \graphene{} do not perform any victim row refreshes in these {applications} because none of {the applications} activate a row at a high enough rate to trigger victim row refreshes.}
{We} conclude that \blockhammer{} {does not incur performance or DRAM energy overheads for single-core benign applications.} 

\subsection{Multiprogrammed Workload{s}}
\label{blockhammer:sec:evaluation_multi_core}
{\figref{blockhammer:fig:8_cores_mech_compare} presents the performance and DRAM energy impact of \blockhammer{} and six state-of-the-art mechanisms\footnote{{We label Graphene as ``Graph'' and \blockhammer{} as ``BH'' for brevity.}}
on an eight-core system, {normalized to {the} baseline{.} 
We show results for two types of workload{s}:
(1)~\emph{No \rowhammer{} Attack}, where all eight applications in the {workload} are benign; and
(2)~\emph{\rowhammer{} Attack Present}, where one of the eight applications in the {workload} is a malicious thread performing a \rowhammer{} attack, running alongside seven benign applications.
We make four observations from the figure.}}

\begin{figure}[h]
    \centering
    \includegraphics[width=0.8\linewidth]{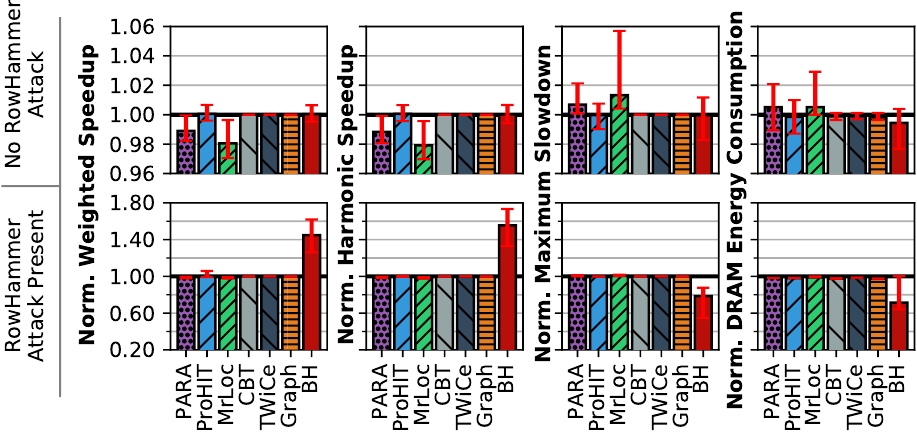}
    \caption{{Performance and {DRAM} energy {consumption} for multiprogrammed {workloads}},  
    normalized to {baseline}.}
    \label{blockhammer:fig:8_cores_mech_compare}
    \vspace{0.25\baselineskip}
\end{figure}



\nind
{\textbf{No \rowhammer{} Attack.}}
First, \blockhammer{} {has {a very small} performance overhead} 
for multiprogrammed {workloads} {when there is no \rowhammer{} attack present. {\blockhammer{} incurs} less than 0.5\%, 0.6\%, and 1.2\%
overhead in terms of weighted speedup, harmonic speedup, and maximum slowdown, respectively, compared to the baseline system with no \rowhammer{} mitigation.} In comparison,
{\prohit{}, \cbt{}, \twice{}, and \graphene{} do not perform enough refresh operations to have an impact on system performance, while \para{} and \mrloc{} incur 
1.2\% and 2.0\% 
performance (i.e., weighted speedup) overheads on average, 
respectively}.
{Second,}
\blockhammer{} \emph{reduces} {average} DRAM energy consumption by 
0.6\%,
while for the worst {workload} we observe, it increases energy consumption by up to 
0.4\%.
This is {because \blockhammer{} (1)~increases the standby energy consumption by delaying requests and (2)~reduces} the energy consumed for row activation and precharge operations {by batching delayed requests and servicing them when their target row is activated}. 
In comparison\om{,} {{\prohit{}}, \cbt{}, \twice{}, and \graphene{} \emph{increase} {average} DRAM energy consumption by less than 0.1\%}, while {\para{} and \mrloc{} \emph{increase} average DRAM energy consumption {by 0.5\%, as a result of}
the unnecessary row refreshes that these mitigation mechanisms must perform.}


\nind
{\textbf{\rowhammer{} Attack Present.}}
{{{Third}, unlike any other \rowhammer{} mitigation mechanism,}} \blockhammer{} {\emph{reduces}} the performance degradation 
{inflicted on benign applications when one of the applications in the {workload} is a \rowhammer{} attack.}
{By throttling the attack, \blockhammer{} significantly improves the performance of benign {applications},}
{{with a} 45.0\% (up to 61.9\%) and 56.2\% (up to 73.4\%) increase in weighted and harmonic speedups and 22.7\% (up to 45.4\%) decrease in maximum slowdown on average, respectively.} 
{In contrast, \para{}, \prohit{}, and \mrloc{} incur 1.3\%, 0.1\% and {1.7\% performance overheads, on average}, respectively, while {the average performance overheads of} \cbt{}, \twice{}, and \graphene{} are all less than 0.1\%. 
{Fourth}, \blockhammer{} {\emph{reduces}} DRAM energy consumption by {28.9\%} on average (up to 33.8\%). In contrast, {all} other state-of-the-art mechanisms {\emph{increase}} DRAM energy consumption ({by} up to 0.4\%).}
 {\blockhammer{}} significantly improves performance and DRAM energy because it increases the row buffer locality that benign {applications} experience {by throttling the attacker
 ({the row buffer hit rate increases by 177\% on average, and 23\%} of row buffer conflicts {are converted} {to row buffer misses}).}

We conclude that \blockhammer{} {(1)~introduces} {very {low}} performance and {DRAM} energy {overheads} 
when there is
no \rowhammer{} attack and {(2)~significantly} improves 
{benign application performance and DRAM energy consumption} 
{when a \rowhammer{} attack is present.} 
\subsection{{Effect of {Worsening} \rowhammer{} {Vulnerability}}}
\label{blockhammer:sec:evaluation_tech_scaling}

{\lineskiplimit=-\maxdimen%
{We analyze {how} \blockhammer{}'s {impact on} performance {and DRAM energy consumption} 
{scales as DRAM chips become increasingly  vulnerable to \rowhammer{} (i.e., as the \rowhammer{} threshold, {\nrh{},} decreases)}.}
{We compare \blockhammer{} with three state-of-the-art \rowhammer{} mitigation mechanisms, {which are shown to be the most viable mechanisms when the \rowhammer{} threshold decreases~\cite{kim2020revisiting, park2020graphene}}: {\para{}~\cite{kim2014flipping}, \twice{}~\cite{lee2019twice},\footnote{{As described in \secref{blockhammer:sec:methodology}, \twice{} faces latency issues, preventing it from scaling when $\nrhn{}<32K$~\cite{kim2020revisiting}. Our scalability analysis assumes a \twice{} variation that solves this issue, the same as \twice{}-Ideal in \cite{kim2020revisiting}}.} and \graphene{}~\cite{park2020graphene}.}}
{We {analyze} the scalability of these mechanisms down to \nrh{}$=1024$, which is approximately an order of magnitude smaller than the minimum observed \nrh{} reported in {current} literature {(i.e., 9600)}~\cite{kim2020revisiting}}. 
%
\figref{blockhammer:fig:multi_core_performance_tech_scaling} shows the performance {and energy overheads of each mechanism for {our} multiprogrammed {workloads}  as \nrh{} decreases, normalized to the baseline system with no \rowhammer{} mitigation}. {We make two observations from {Figure~\ref{blockhammer:fig:multi_core_performance_tech_scaling}}.}}

\begin{figure}[ht]
    \centering
    \includegraphics[width=0.8\columnwidth]{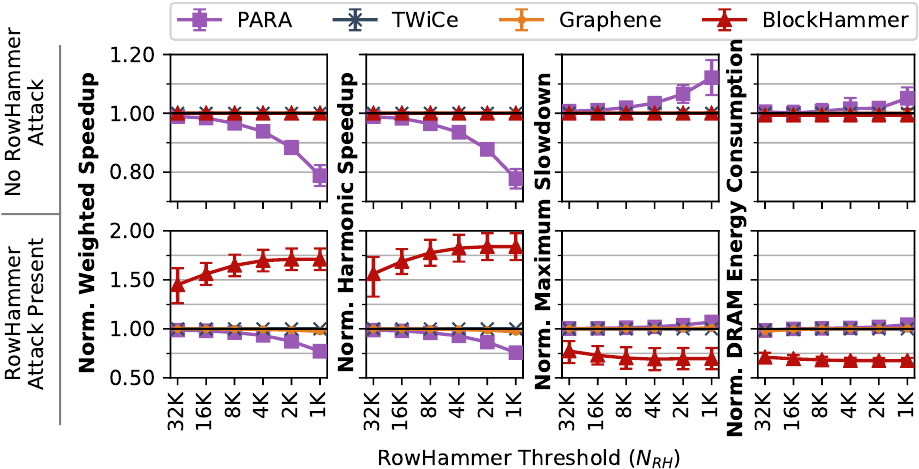}
    \vspace{-5mm}
    \caption{Performance {and DRAM Energy} for different \nrh{} values, normalized to baseline (\nrh{} \emph{decreases} along the x-axis).}
    \label{blockhammer:fig:multi_core_performance_tech_scaling}
    \vspace{2mm}
\end{figure}

\nind
{\textbf{No \rowhammer{} Attack.}}
{First, {{\blockhammer{}'s} performance and {DRAM} energy {consumption}} {are} better 
than \para{} {and competitive with other mechanisms} as \nrh{} decreases. When \nrh{}=1024, 
{the average performance and {DRAM} energy overheads of \blockhammer{}, \graphene{}, and \twice{} are 
less than 0.6\% because 
they do not act aggressively enough to cause significant performance or energy overheads.
On the other hand, \para{} performs reactive refreshes more aggressively with increasing \rowhammer{} vulnerability, which {leads to a} performance overhead {of} 
21.2\% and 22.3\% (weighted and harmonic speedup) and an energy overhead of 5.1\% on average.
}

\nind 
{\textbf{\rowhammer{} Attack Present.}}
Second, \blockhammer{}'s {performance and {DRAM} energy benefits increase}
as \nrh{} decreases. At \nrh{}=1024, \blockhammer{} {more aggressively throttles a \rowhammer{} attack and mitigates {the} performance degradation {of} benign {applications}. As a result, compared to the baseline, \blockhammer{} improves {average} {performance} by 71.0\% and 83.9\% {(weighted} and harmonic {speedups)} while reducing the maximum slowdown and DRAM energy consumption by 30.4\% and 32.4\%, respectively. {In contrast,} the additional refresh operations that \graphene{} and \twice{} perform {cause} 2.9\% and 0.9\% {average} performance {degradation} {and 0.4\% and 0.2\% DRAM energy increase} for benign application{s,} respectively.} {\blockhammer{} is the only \rowhammer{} mitigation mechanism that improves performance and energy when a \rowhammer{} attack {is} present in the system.}

{We conclude that (1)~\blockhammer{}'s performance and energy overheads {remain} negligible at reduced \rowhammer{} thresholds {as low as \nrh{}=1K {when there is no \rowhammer{} attack},} and (2)~\blockhammer{} scalably provides {much} higher performance and lower energy {consumption} {than all} state-of-the-art mechanisms when
{a \rowhammer{} attack is present.}
}
}

\subsection{{Analysis of \blockhammer{} Internal Mechanisms}}
\label{blockhammer:sec:evaluation_false_positives}
{\blockhammer{}'s impact on performance and {DRAM} energy depends on (1)~the false positive rate of {the} blacklisting mechanism
and {(2)~}the false positive penalty {resulting from} {delaying row activations}.}\fw{w/ and w/o \hammerthrottler{} added to the gdoc}  {We {calculate} (1)~the} false positive rate {as} the
{number of}
row {activations that are {mistakenly delayed} by \blockhammer{}'s Bloom filters
{(i.e., activations to rows that would not have been blacklisted if the filters had no aliasing)}
{as a fraction of} all activations, 
and (2)~{the} false positive penalty as the additional time delay a mistakenly-delayed row activation suffers from}.
{We find that for a configuration where} \nrh{}=32K, \blockhammer{}'s false positive rate is
{{0.010\%, and {it} increases to only} 0.012\% when \nrh{} is scaled down to 1K.}
{Therefore, \blockhammer{} successfully avoids delaying more than 99.98\% of benign row activations.}
Even though we set \smash{\tdelay{}} to {\SI{7.7}{\micro\second}}, we observe \SI{1.7}{\micro\second}, {\SI{3.9}{\micro\second}, {and \SI{7.6}{\micro\second}} of {delay} for the 50th, 90th, {and 100th} percentile of} mistakenly-delayed activations (which {are} only 0.012\% of all activations). 
{Note that the worst-case latency we observe is at least two orders of magnitude smaller than}
typical quality-of-service targets, which are {on} the order of milliseconds~\cite{kasture2016tailbench}.
{Therefore, we} believe that \blockhammer{} is unlikely to {introduce} quality-of-service violations with {its low} worst-case latency ({on the order of \si{\micro\second}}) and very 
low {false positive} rate (0.012\%).

\section{Comparison of Mitigation Mechanisms}
\label{blockhammer:sec:qualitative_analysis}
We qualitatively compare \blockhammer{} and a number of published \rowhammer{} mitigation mechanisms, which we classify into four high-level approaches:
($i$)~\emph{increased refresh rate},
($ii$)~\emph{physical isolation},
($iii$)~\emph{reactive refresh},
and ($iv$)~\emph{proactive throttling}. 
We evaluate \rowhammer{} mitigation mechanisms across four dimensions: \emph{comprehensive protection}, \emph{compatibility with commodity DRAM chips}, \emph{scaling with \rowhammer{} vulnerability}, and \emph{deterministic protection}.
Table~\ref{blockhammer:tab:current_mechanisms} summarizes {our comprehensive} qualitative evaluation.

\newcommand{\astfootnote}[1]{
\let\oldthefootnote=\thefootnote
\setcounter{footnote}{0}
\renewcommand{\thefootnote}{\fnsymbol{footnote}}
\footnote{#1}
\let\thefootnote=\oldthefootnote
}

\begin{table}[htb]
\centering
\footnotesize
\renewcommand{\arraystretch}{1}
\setlength\tabcolsep{5pt}
\caption{{Comparison of RowHammer mitigation mechanisms.}}
\label{blockhammer:tab:current_mechanisms}
\begin{tabular}{l|l||c|c|c|c}
 \multicolumn{2}{c||}{}
    & \multirow{8}{*}{\thead{Comprehensive Protection}}
    & \multirow{8}{*}{\thead{Compatible w/ Commodity\\DRAM Chips}}
    & \multirow{8}{*}{\thead{Scaling with RowHammer Vulnerability}}
    & \multirow{8}{*}{\thead{Deterministic Protection}} \\
\multicolumn{2}{c||}{} & & & & \\
\multicolumn{2}{c||}{} & & & & \\
\multicolumn{2}{c||}{} & & & & \\
\multicolumn{2}{c||}{} & & & & \\
\multicolumn{2}{c||}{} & & & & \\
\multicolumn{2}{c||}{} & & & & \\
\head{Approach} & \textbf{Mechanism} &  & &  &   \\
\hline{} 
\multirow{2}*{\setlength\tabcolsep{0pt}\begin{tabular}{l}Increased Refresh Rate\end{tabular}}&Kim et al.~\cite{kim2014flipping} & \cmark  & \cmark  & \xmark & \cmark  \\
& Apple Update~\cite{apple2015about} & \cmark  & \cmark  & \xmark & \cmark  \\
\hline
\multirow{3}*{\setlength\tabcolsep{0pt}\begin{tabular}{l}Physical Isolation\end{tabular}} & CATT~\cite{brasser2017cant}      & \xmark & \xmark & {\xmark} & \cmark  \\ %
 & GuardION~\cite{vanderveen2018guardion}    & \xmark & \xmark & {\xmark} & \cmark \\
  & ZebRAM~\cite{konoth2018zebram}   & \xmark & \xmark & {\xmark} & \cmark \\ %
          \hline
          & ANVIL~\cite{aweke2016anvil}     & \xmark & \xmark & {\xmark} & \cmark  \\ %
          & \para{}~\cite{kim2014flipping}  & \cmark & \xmark & {\xmark} & \xmark  \\ %
\multirow{3}*{\setlength\tabcolsep{0pt}\begin{tabular}{l}Reactive Refresh\end{tabular}}  & \prohit{}~\cite{son2017making}  & \cmark & \xmark & \xmark & \xmark  \\ %
   & \mrloc{}~\cite{you2019mrloc}    & \cmark & \xmark & \xmark & \xmark \\ %
          & \cbt{}~\cite{seyedzadeh2018mitigating} & \cmark & \xmark & {\xmark} & \cmark   \\ %
          & \twice{}~\cite{lee2019twice}    & \cmark & \xmark & {\xmark} & \cmark  \\ %
          & \graphene{}~\cite{park2020graphene} & \cmark & \xmark & {\cmark} & \cmark  \\
          \hline
\multirow{3}*{\setlength\tabcolsep{0pt}\begin{tabular}{l}Proactive Throttling\end{tabular}} & Naive Thrott.~\cite{mutlu2018rowhammer} & \cmark & \cmark & \xmark & \cmark  \\ %
& Thrott. Supp.~\cite{greenfield2012throttling}& \cmark & \xmark & \xmark & \cmark  \\ \cline{2-6}
          & \textbf{BlockHammer}                                     & \cmark & \cmark & {\cmark} & \cmark  \\ \hline
\end{tabular}
\end{table}

\nind
\head{1. Comprehensive Protection} A \rowhammer{} mitigation mechanism should comprehensively prevent {\emph{all} potential} \rowhammer{} bitflips regardless of the methods that an attacker may use to {hammer a DRAM row.} 
Unfortunately, {four} key \rowhammer{} mitigation mechanisms~\cite{konoth2018zebram, vanderveen2018guardion, brasser2017cant, aweke2016anvil}
are effective only against a limited threat model and 
have already been defeated by recent attacks~\cite{qiao2016anew, gruss2016rowhammerjs_a, gruss2018another, cojocar2019exploiting, zhang2019telehammer, kwong2020rambleed}
{because they (1)~trust system components (e.g., hypervisor) that can be used to perform a \rowhammer{} attack~\cite{konoth2018zebram, vanderveen2018guardion}; (2)~disregard practical methods (e.g., flipping opcode bits within the attacker's memory space~\cite{brasser2017cant}) that can be used to gain root privileges; or (3)~detect \rowhammer{} attacks by relying on hardware performance counters (e.g., LLC miss rate~\cite{aweke2016anvil}), which
can be oblivious to several attack models~\cite{vanderveen2016drammer_deterministic, qiao2016anew, gruss2018another, tatar2018throwhammer}.}
In contrast, \blockhammer{} comprehensively prevents \rowhammer{} bitflips by monitoring all memory accesses from within the memory controller, even if the entire software stack is compromised and the attacker possesses knowledge about all hardware/software implementation details (e.g., the DRAM chip's \rowhammer{} vulnerability characteristics, \blockhammer{}'s configuration parameters).

\head{2. Compatibility with Commodity DRAM Chips} {Especially} given that recent works~\cite{cojocar2020arewe, frigo2020trrespass, kim2020revisiting} experimentally observe \rowhammer{} bitflips on {cutting-edge}
commodity DRAM chips, including ones that are marketed as \rowhammer{}-{free}~\cite{frigo2020trrespass, cojocar2020arewe, kim2020revisiting}, it is {important} for a \rowhammer{} mitigation mechanism to be compatible with {\emph{all}} commodity
DRAM chips, current and future. To achieve this, a \rowhammer{} mitigation mechanism should \emph{not} (1)~rely on any proprietary information that DRAM vendors do not share, and (2)~require any modifications to DRAM chip design.
Unfortunately, both physical isolation and reactive refresh mechanisms need to be fully aware of the internal physical layout of DRAM rows or require modifications to DRAM chip design either (1)~to ensure that isolated memory regions are not physically close to each other~\cite{konoth2018zebram, brasser2017cant, vanderveen2018guardion} or (2)~to identify victim rows that need to be refreshed~\cite{greenfield2012throttling, kim2014architectural, bains2015rowhammer, bains2016rowhammer, bains2016distributed, aweke2016anvil, kim2014flipping, son2017making, you2019mrloc, seyedzadeh2017counterbased, seyedzadeh2018mitigating, kang2020cattwo, lee2019twice, park2020graphene}.
In contrast, designing \blockhammer{} requires knowledge of only six readily-available DRAM parameters:
(1)~\trefw{}: the refresh window, 
(2)~\trc{}: the ACT-to-ACT latency, 
(3)~\tfaw{}: the four-activation window, 
(4)~\nrh{}: the \rowhammer{} threshold,
(5)~the blast radius, and 
(6)~the blast impact factor. 
Among these parameters, \trefw{}, \trc{}, and \tfaw{} are publicly available in datasheets~\dramStandardCitations{}. 
\nrh{}, the blast radius, and the blast impact factor
can be obtained from prior characterization works~\cite{kim2014flipping, kim2020revisiting, frigo2020trrespass}.
Therefore, \blockhammer{} is compatible with all commodity DRAM chips because it does not need any proprietary information about or any modifications to commodity DRAM chips. 

\nind
\head{3. Scaling with Increasing RowHammer Vulnerability}
Since main memory is
a {growing}
system performance and
energy bottleneck~\cite{wulf1995hitting, sites1996itsthe, wilkes2001thememory, mutlu2020amodern, mutlu2021amodern, mutlu2013memory, mutlu2014research, kanev2015profiling, wang2014bigdatabench, boroum2018google, oliveira2021anew, gomezluna2021benchmarkingarxiv},
a \rowhammer{} mitigation mechanism {should exhibit acceptable performance and energy overheads at low area cost when configured for more vulnerable DRAM chips.}



\emph{Increasing the refresh rate}~\cite{kim2014flipping, apple2015about} is {\emph{already} a prohibitively expensive} solution {for modern DRAM chips with a} \rowhammer{} threshold of 32K. {This is} because the latency of refreshing rows at a high enough rate to prevent bitflips overwhelms DRAM's availability, increasing its average performance overhead to 78\%, as shown in \cite{kim2020revisiting}.


\emph{~Physical isolation}~\cite{konoth2018zebram, vanderveen2018guardion, brasser2017cant} {requires reserving as many rows as {twice} the \emph{blast radius}} {(up to 12 in modern DRAM chips~\cite{kim2020revisiting})} to isolate sensitive data from a potential attacker's memory space. This is expensive for most modern systems where memory capacity is {critical.}
{As} the blast radius {has increased} by 33\%
from 2014~\cite{kim2014flipping} to 2020~\cite{kim2020revisiting}, physical isolation mechanisms can require reserving even more rows when configured for future DRAM chips, {further} reducing the total amount of secure memory available to the system. 
%

{\emph{Reactive refresh} mechanisms~\cite{greenfield2012throttling, kim2014architectural, bains2015rowhammer, bains2016rowhammer, bains2016distributed, aweke2016anvil, kim2014flipping, son2017making, you2019mrloc, seyedzadeh2017counterbased, seyedzadeh2018mitigating, kang2020cattwo, lee2019twice, park2020graphene} generally incur increasing performance, energy, and/or area overheads at lower \rowhammer{} thresholds {when configured for more vulnerable DRAM chips.} 
{ANVIL} samples hardware performance counters on the order of \SI{}{\milli\second} for a \rowhammer{} threshold (\nrh{}) of 110K~\cite{aweke2016anvil}. However, a \rowhammer{} attack can successfully induce bitflips in less than \SI{50}{\micro\second} when \nrh{} is reduced to 1K, which significantly increases {ANVIL}'s sampling rate, and thus, its performance {and energy} overheads.
\prohit{} and \mrloc{}~\cite{son2017making, you2019mrloc} do not provide a concrete discussion on how to adjust their empirically-determined parameters, so we cannot demonstrate how their overheads scale as DRAM chips become more vulnerable to RowHammer.
\twice{}~\cite{lee2019twice} faces design challenges to protect DRAM chips when reducing \nrh{} below $32\mathrm{K}$}, as described in Section~\ref{blockhammer:sec:methodology}. Assuming that \twice{} overcomes {its design} challenges (as {also assumed} by prior work~\cite{kim2020revisiting}){,} we scale \twice{} down to \nrh{}$=1\mathrm{K}$ along with three other state-of-the-art mechanisms~\cite{seyedzadeh2018mitigating, kim2014flipping, park2020graphene}. Table~\ref{blockhammer:tab:area_cost_analysis} shows that the {CPU die area}, access energy, and static power consumption of \twice{}~\cite{lee2019twice}/\cbt{}~\cite{seyedzadeh2018mitigating} drastically increase by {35x/20x,} 15.6x/14.0x, and 29.7x/15.1x, respectively, when \nrh{} is reduced from 32K to 1K.
In contrast, \blockhammer{} {consumes} only {{30\%/40\%,} 79.8\%/77.8\%, 35\%/41.3\% of \twice{}/\cbt{}'s} {CPU die area,} access energy, and static power, respectively, when configured for \nrh{}$=1\mathrm{K}$.
Section~\ref{blockhammer:sec:evaluation_tech_scaling} shows that \para{}'s average performance and DRAM energy overheads reach 21.2\% and 22.3\%, respectively, when configured for \nrh{}$=1\mathrm{K}$.
{We observe that \graphene{} and \blockhammer{} are the two most scalable mechanisms with worsening \rowhammer{} vulnerability. When configured for \nrh=1K, \blockhammer{} (1)~consumes only 11\% of \graphene{}'s access energy (see Table~\ref{blockhammer:tab:area_cost_analysis}) and}
(2)~improves benign applications' performance by 71.0\% and reduces DRAM energy consumption by 32.4\% on average, while \graphene{} incurs 2.9\% {performance and 0.4\% DRAM energy} overheads, as shown in Section~\ref{blockhammer:sec:evaluation_tech_scaling}.
{Na{\"i}ve} {\emph{proactive throttling}~\cite{greenfield2012throttling, kim2014flipping, mutlu2018rowhammer}} either (1)~blocks all activations targeting a row until the end of the refresh window {once the row's activation count reaches the} \rowhammer{} threshold, or (2)~statically extends each row's activation interval so that no {row's activation count can ever exceed the} \rowhammer{} threshold. The first method has a high area overhead because it requires implementing a counter for each DRAM row~\cite{kim2014flipping, mutlu2018rowhammer}, while
the second method prohibitively increases \trc{}~\dramStandardCitations{} 
(e.g., 42.2x/1350.4x for a DRAM chip with \nrh{}=32K/1K)~\cite{kim2014flipping, mutlu2018rowhammer}.
{\blockhammer{} is the first efficient and scalable proactive throttling-based \rowhammer{} prevention technique.}

\nind
\head{4. Deterministic Prevention} {To effectively prevent all \rowhammer{} bitflips, a \rowhammer{} mitigation mechanism should be deterministic, meaning that it should ensure \rowhammer{}-safe operation at all times}
{because} it is important to guarantee zero chance of a security failure for a critical system
whose failure or malfunction may result in severe consequences (e.g., related to loss of lives, environmental damage, or economic loss)~\cite{aven2009identification}.
PARA~\cite{kim2014flipping}, ProHIT~\cite{son2017making}, and MRLoc~\cite{you2019mrloc} are probabilistic by design, and therefore cannot reduce the probability of a successful \rowhammer{} attack to zero like \cbt{}~\cite{seyedzadeh2018mitigating}, \twice{}~\cite{lee2019twice}, and \graphene{}~\cite{park2020graphene} {potentially} can.
\blockhammer{} has the capability to provide zero probability for a successful \rowhammer{} attack by guaranteeing that no row can be activated at a unsafe rate.

\section{Summary}
\label{blockhammer:sec:conclusion}

We introduce \blockhammer{}, a new \rowhammer{} detection and prevention mechanism that 
uses area-efficient Bloom filters to track and proactively throttle memory accesses that can potentially induce \rowhammer{} bitflips. \blockhammer{} operates entirely from within the memory controller, comprehensively protecting a system from all \rowhammer{} bitflips at low area, energy, and performance cost. Compared to existing \rowhammer{} mitigation mechanisms, \blockhammer{} is the first one that (1)~prevents \rowhammer{} bitflips efficiently and scalably without knowledge of or {modification} to DRAM internals, (2)~provides all four desired characteristics of a \rowhammer{} mitigation mechanism (as we describe in Section~\ref{blockhammer:sec:qualitative_analysis}), and (3)~improves the performance and energy consumption of a system that is under attack. We believe that \blockhammer{} provides a new direction in \rowhammer{} prevention and hope that it enables researchers and engineers to develop low-cost \rowhammer{}-free systems going forward. \exttwo{To further aid future research and development, we make BlockHammer's source code freely and openly available~\cite{safari2021blockhammer}.}

\ifcameraready
    \setcounter{version}{99}
\else
    \setcounter{version}{9}
\fi

\chapter{Conclusions and Future Directions}
\label{chap:conc}

In summary, the goal of this dissertation is
\agy{5}{to
1)~build a detailed understanding of DRAM read disturbance, and
2)~mitigate DRAM read disturbance efficiently and scalably without requiring proprietary knowledge of DRAM chip internals by leveraging our detailed understanding.}
\agy{5}{To achieve this goal, we conduct a set of research projects based on the thesis statement that}
\emph{``We can mitigate DRAM read disturbance efficiently and scalably by 1)~building a detailed understanding of DRAM read disturbance, 2)~leveraging insights into modern DRAM chips and memory controllers, and 3)~devising novel solutions that do not require proprietary knowledge of DRAM chip internals.''} To this end, we combine experimental studies, statistical analyses, and architecture-level mechanisms.  

First, we build a detailed understanding of RowHammer vulnerability. To this end, we present the first rigorous experimental characterization study on the sensitivities of DRAM read disturbance to temperature \agy{5}{(\secref{deeperlook:sec:temperature})}, memory access patterns \agy{5}{(\secref{deeperlook:sec:temporal})}, victim DRAM cell's physical location \agy{5}{(\secref{deeperlook:sec:spatial} and \secref{svard:sec:characterization})}, and wordline voltage \agy{5}{(\secref{hammerdimmer:sec:vpp_with_rh})}. 
We find that a DRAM read disturbance bitflip is more likely to occur 1)~in a bounded temperature range specific to each DRAM cell, 2)~if the aggressor row remains
active for a longer time when activated, 3)~in certain physical regions of the DRAM module, and
4)~when the aggressor row's wordline is asserted with a higher voltage. 
We describe and analyze the implications of our findings on future DRAM read disturbance attacks and defenses. We hope that the {novel experimental} results and insights of our study will inspire and aid future work to develop effective {and} efficient {solutions to the DRAM read disturbance problem.

Second, we enable efficient and scalable DRAM read disturbance mitigation by leveraging two insights into DRAM chips and memory controllers: the spatial variation in read disturbance vulnerability across DRAM rows \agy{5}{(\secref{svard:sec:adaptation})} and subarray-level parallelism in off-the-shelf DRAM chips \agy{5}{(\secref{hira:sec:doduo})}. 
To leverage the spatial variation in DRAM read disturbance across DRAM rows, we propose Svärd, a new mechanism that dynamically adapts the aggressiveness of existing DRAM read disturbance solutions based on our experimental observations. 
By learning and leveraging spatial variation in read disturbance vulnerability across DRAM rows, Svärd reduces the performance overheads of state-of-the-art {DRAM read disturbance} solutions, leading to large system performance benefits.  
To leverage subarray-level parallelism in off-the-shelf DRAM chips, {we introduce} \gls{hira}, {a new {DRAM} operation} {that can} {reliably parallelize a DRAM row's refresh operation with the refresh or the activation of another row within the same bank.}
\gls{hira} achieves this by {activating} two electrically-isolated rows {in quick succession,} allowing them to be refreshed{/{activated}} without disturbing each other.
{{W}e show that \gls{hira} 1)~works reliably in real off-the-shelf DRAM chips}{, using already{-}available (i.e., standard) \gls{act} and \gls{pre} DRAM commands{,} by violating timing constraints and 2)~significantly reduces} the time spent for refresh operations.
To leverage {the parallelism} \gls{hira} {provides,}
we {design} \gls{hirasched}.  
\gls{hirasched} modifies the memory request scheduler to perform \gls{hira} operations when a periodic or {RowHammer-preventive} refresh can be {performed} concurrently with another {refresh {or row activation}} {to} the same bank. {{Our system-level evaluations} show that} {\gls{hirasched}} {significantly increases system performance by {{reducing}}}
the performance {degradation {due to}} periodic and {preventive} refreshes.
{We show that \gls{hira} significantly reduces the {performance degradation caused by} both periodic and preventive refreshes, compatible with off-the-shelf DRAM chips, and 2)~provides higher performance benefits in higher{-}capacity DRAM chips.}

Third, we show that it is possible to prevent DRAM read disturbance bitflips efficiently and scalably with \emph{no} proprietary knowledge of or modifications to DRAM chips \agy{5}{(\secref{blockhammer:sec:blockhammer})}. To this end, we propose BlockHammer, a new DRAM read disturbance solution that proactively throttles memory accesses that can potentially induce read disturbance bitflips. \blockhammer{} operates entirely from within the memory controller, comprehensively protecting a system from all \rowhammer{} bitflips at low area, energy, and performance overhead. Compared to existing DRAM read disturbance solutions, \blockhammer{} is the first one that 1)~prevents read disturbance bitflips efficiently and scalably without the knowledge of or {modifications} to DRAM internals, 2)~provides all four desired characteristics of a DRAM read disturbance solution, and 3)~improves the performance and energy consumption of a system that is under attack. We believe that \blockhammer{} provides a new direction in DRAM read disturbance prevention and hope that it enables researchers and engineers to develop systems immune to DRAM read disturbance going forward. {To further aid future research and development, we make BlockHammer's source code freely and openly available~\cite{safari2021blockhammer}.}

\section{Future Research Directions}

Although this dissertation focuses on understanding various aspects of DRAM read disturbance and proposes several new ideas to solve DRAM read disturbance more efficiently and scalably, we believe that this work is applicable in a more general sense and opens up new research directions. This section
reviews promising directions for future work.

\subsection{Further Understanding DRAM Read Disturbance}
\label{future:sec:characterization}

Even though there are various detailed characterization studies performed to understand various properties of DRAM read disturbance~\understandingRowHammerAllCitations{}, there are still unknown aspects of DRAM read disturbance properties/sensitivities and the manifestations of
such properties in cutting-edge and future DRAM chips. 
It is critical to fundamentally understand the various properties of DRAM read disturbance under different operating conditions and memory access patterns to develop
fully-secure and efficient solutions.
We highlight two outstanding research questions.

\subsubsection{\agy{6}{Practically and Accurately Measuring DRAM Read Disturbance Vulnerability}}
\label{future:sec:worstcase}
\agy{6}{A DRAM read disturbance bitflip occurs when the cumulative effect of RowHammer and RowPress is large enough to induce a bitflip.}
\agy{6}{Existing solutions estimate this cumulative effect in terms of row activation count, and assume that a bitflip can occur when the row activation count exceeds a threshold value called read disturbance threshold.\footnote{\agy{6}{To account for RowPress, existing solutions either reduce their read disturbance threshold conservatively~\cite{luo2023rowpress} or keep incrementing the row activation count as the row remains open for longer time~\cite{luo2023rowpress, qureshi2024impress}.}} Unfortunately, identifying the read disturbance threshold for a DRAM cell is \emph{not} straightforward.} 
Our experimental characterization studies show that DRAM read disturbance significantly varies \agy{6}{with 1)~temperature (\secref{deeperlook:sec:temperature})), 2)~memory access patterns (\secref{deeperlook:sec:temporal}), 3)~physical location of a DRAM cell (\secref{deeperlook:sec:spatial}) and \secref{svard:sec:characterization}, and 4)~data patterns~\cite{kim2014flipping, kim2020revisiting, hassan2021uncovering, orosa2021adeeper}. Unfortunately, the best practice for testing DRAM read disturbance is to test each DRAM cell} for every possible temperature, memory access pattern, and data pattern to find the \agy{6}{read disturbance threshold}, which is \agy{6}{prohibitively expensive} \agy{6}{because such a testing procedure} drastically increases time-to-market\agy{6}{.}
\agy{6}{Therefore, to ensure robustness in modern memory systems,}
more research is needed to quickly and practically \agy{6}{measure read disturbance vulnerability}. 

\subsubsection{The Effect of Aging on DRAM Read Disturbance}
\label{future:sec:aging}
We present preliminary results showing that there are DRAM rows that exhibit lower \gls{hcfirst} values after \agy{6}{rigorously testing for 68 days} \agy{5}{(\secref{svard:sec:repeatability_and_aging})}. \agy{6}{Our results motivate future research to understand how DRAM read disturbance vulnerability varies with aging. Unfortunately,} \emph{no} work rigorously characterizes and explains how DRAM read disturbance changes with aging. Therefore, we call for further research on understanding the effects of DRAM aging on DRAM read disturbance.  

\subsubsection{\agy{6}{Temperature's Effect on Spatial Variation of Read Disturbance across DRAM Cells}}
\agy{6}{We investigate the effect of temperature on DRAM read disturbance (\secref{deeperlook:sec:temperature}) and spatial variation in read disturbance vulnerability across DRAM rows (\secref{deeperlook:sec:spatial} and \secref{svard:sec:characterization}) independently, i.e., our analysis sweeps either temperature or a DRAM cell's physical location. Our temperature study shows that the vulnerable temperature range significantly varies across DRAM cells, and thus it can affect the spatial variation of DRAM read disturbance. Therefore, our spatial variation analysis can be improved by extending the analysis with a temperature sweep.}

\subsubsection{\agy{6}{DRAM Read Disturbance under Reduced Supply and Wordline Voltage}}
\agy{6}{Our voltage scaling analysis (\secref{hammerdimmer:sec:vpp_with_rh}) shows that when DRAM chip is provided with a lower wordline voltage, fewer read disturbance bitflips occur and they occur at higher hammer counts. Wordline voltage ($V_{PP}$) drives the wordline, and thus the gate of the access transistor in a DRAM cell, while the rest of the circuitry (e.g., bitlines, sense amplifiers, and precharge circuitry), including the drain and the source of the access transistors is driven by the supply voltage ($V_{DD}$). Therefore, reducing wordline voltage \emph{without} changing the supply voltage reduces the access transistor's gate-to-source voltage ($V_{GS}$), which is $V_{GS} = V_{PP} - V_{DD}$. This reduction in $V_{GS}$ voltage leads to a smaller charge stored in the DRAM cell capacitor, and thus increases the row activation latency and reduces the data retention time. Therefore, to assess the potential of voltage scaling to reduce read disturbance vulnerability, it is important to investigate DRAM read disturbance, row activation latency, and data retention time while simultaneously reducing $V_{PP}$ and $V_{DD}$ \emph{without} reducing $V_{GS}$. Unfortunately, \emph{no} prior work investigates this direction, and thus we encourage future research to do so.}

\subsection{Mitigating DRAM Read Disturbance at Low Cost}
\label{future:sec:low_cost}
Even though our mechanisms (Svärd, HiRA, and BlockHammer, \agy{6}{respectively presented in \secref{svard:sec:adaptation}, \secref{hira:sec:implementation}, and \secref{blockhammer:sec:blockhammer}}) are significantly less expensive than the state-of-the-art, they do \emph{not} completely alleviate the performance, energy, and hardware overheads of DRAM read disturbance mitigation. Further research is needed to solve DRAM read disturbance at \agy{6}{even lower} cost from \agy{6}{four} aspects: 1)~\agy{6}{accurately tracking aggressor row activations at low hardware overhead}, \agy{6}{2)~updating row activation counters at low latency,} 3)~performing preventive actions at low performance overhead, \agy{6}{and 4)~profiling DRAM read disturbance at low performance overhead}. 

\subsubsection{Low Cost \agy{6}{Aggressor Row Tracking}} 
To efficiently and scalably prevent DRAM read disturbance bitflips, it is important to accurately detect the memory accesses that might cause bitflips so that preventive actions are performed \emph{only} when necessary. Unfortunately, the hardware cost of such accurate access pattern detection increases with memory scaling for two reasons. First, as DRAM chips become more vulnerable to read disturbance, fewer memory accesses can induce bitflips, and thus, many more rows can concurrently be subject to bitflips until they are periodically refreshed or accessed. Therefore, tracking mechanisms must keep track of accesses targeting an increasingly large number of DRAM rows as DRAM read disturbance is exacerbated over generations, leading to increased metadata storage for this tracking information. Second, as the demand for memory capacity and bandwidth increases, memory systems contain increasingly more banks. Unfortunately, the hardware complexity of existing detection mechanisms \agy{5}{significantly} increases with the number of banks in the memory system. \agy{6}{Two of our recent works (\secref{appendix:otherworks}) address this issue.
\agy{6}{First, CoMeT~\cite{bostanci2024comet} adopts Count-Min-Sketch algorithm to track aggressor rows with high accuracy at low cost. CoMeT significantly reduces the hardware complexity of state-of-the-art aggressor row trackers.}
Second, ABACuS~\cite{olgun2024abacus} provides} a leap in reducing the rampant rise of hardware complexity, transitioning the counter cost scaling from a linear to a logarithmic relationship with increasing bank count.

We invite future research aiming for \agy{6}{aggressor row tracking} cost that \agy{6}{scales} sub-logarithmically with the bank count. In conclusion}, it is an open research problem to achieve accurate detection at low cost. 

\subsubsection{\agy{6}{Performing Preventive Actions at Low Performance Overhead}}
Several preventive actions have been proposed to avoid DRAM read disturbance bitflips, including 1)~refreshing potential victim rows~\refreshBasedRowHammerDefenseCitations{} and 2)~copying or moving aggressor rows to different physical locations~\migrationBasedRowHammerDefenseCitations{}.
These operations occupy the corresponding DRAM bank, and thus, the corresponding bank \emph{cannot} service memory requests. 
\agy{7}{The performance impact of such preventive actions can be reduced in various ways, e.g., by performing them in the idle time of a memory region, reducing the latency of such operations by leveraging the safety margins embedded in DRAM timing parameters, and overlapping their latency with the latency of other memory accesses. Doing so often requires implementing on-DRAM-die maintenance mechanisms, which may require modifications to the DRAM communication protocols (e.g., DDR4~\cite{jedec2012jesd794}, DDR5~\cite{jedec2020jesd795}). Dependency to such modifications significantly increases their time-to-market~\cite{hassan2024acase, hassan2024selfmanaging, patel2022acase}. \agy{8}{For example, moving from DDR4~\cite{jedec2012jesd794} to DDR5~\cite{jedec2020jesd795} took 8 years and it took about 10 years
to introduce per row activation counting (PRAC) mechanism~\cite{jedec2024jesd795c} for on-DRAM-die RowHammer mitigation after the first demonstration of RowHammer as a widespread reliability and security problem~\cite{kim2014flipping}.} To address this issue, one of our works, Self-Managing DRAM (SMD)~\cite{hassan2022acase, hassan2024acase, safari2022selfmanaging} significantly improves the flexibility of DRAM communication protocols. SMD leverages the key insight that we can enable easy-adoption of new on-DRAM-die maintenance mechanisms at the cost of adding a simple not-acknowledgement (NACK) signal for row activations (ACT-NACK). \agy{8}{A DRAM chip that requires to perform maintenance internally and autonomously (e.g., RowHammer prevention, refresh, and scrubbing), issues the ACT-NACK signal when it receives an activate command from the memory controller, meaning that it is busy and cannot respond.} SMD showcases its flexibility by \agy{8}{implementing} three \agy{8}{major} maintenance operations \agy{8}{autonomously in a DRAM chip}: \agy{9}{1)~periodic refresh~\cite{saino2000impact, liu2012raidr, liu2013anexperimental} or heterogeneous refresh~\cite{liu2012raidr}}, 2)~RowHammer-preventive refresh~\refreshBasedRowHammerDefenseCitations{}, and 3)~memory scrubbing~\cite{gong2018duoexposing, jacob2010memory, meza2015revisiting, mukherjee2004cache, rooney2019micron, saleh1990reliability, schroeder2009dram_errors, siddiqua2017lifetime}. In line with SMD's findings, the April 2024 update to the DDR5 protocol~\cite{jedec2024jesd795c} introduces the addition of an alert back-off signal, similar to SMD's ACT-NACK signal roughly two years after SMD's first \agy{8}{public release}~\cite{hassan2022acase}. We hope and expect future research to introduce even better flexibility in DRAM communication protocols for wider adoption of efficient on-DRAM-die maintenance mechanisms.}

\textbf{\agy{7}{Leveraging subarray-level parallelism to perform preventive actions at low performance overhead.}}
As our \agy{6}{observation in \secref{hira:sec:doduo} that leads to HiRA}~\cite{yaglikci2022hira} exemplifies, the latency of preventive actions can be overlapped with the latency of memory accesses by leveraging the subarray-level parallelism in DRAM chips. 
\agy{6}{SMD~\cite{hassan2024acase, hassan2024selfmanaging}, explores the benefits of leveraging subarray-level parallelism for performing periodic and RowHammer-preventive refreshes, similar to prior works on reducing performance overheads of periodic refresh operations (e.g.,~\cite{kim2012acase, chang2014improving}), and significantly improves system performance. Despite its obvious performance benefits~\salprefs{}, leveraging subarray-level parallelism faces two key challenges to be adopted in modern DRAM chips. 
First, density-optimized open-bitline DRAM array architecture shares sense amplifiers across two adjacent subarrays, and thus electrically isolating adjacent subarrays from each other requires a detailed circuit design of a low cost implementation. \agy{7}{Unfortunately, current academic studies are limited to the DRAM technology libraries in the public domain, which might \emph{not} reflect the design challenges of modern DRAM chips~\cite{marazzi2024hifidram}.} Second, DRAM row activation is a power-hungry operation, and simultaneously activating multiple rows might require revisiting the power-delivery circuitry in DRAM chips. Our recent study, SiMRA~\cite{yuksel2024simra}, \agy{7}{presents a preliminary analysis on the power consumption of} simultaneously activating multiple rows within the same subarray, \agy{7}{and shows} \emph{only} \agy{7}{a marginal increase in the} power consumption of DRAM chips \agy{7}{with increasing number of simultaneously activated rows}. \agy{7}{Although our preliminary results are promising, the literature needs a deeper understanding of power consumption breakdown in modern off-the-shelf DRAM chips, which requires 1)~\agy{8}{rigorous modeling based on a better understanding of} DRAM circuit design \agy{8}{parameters and characteristics} and 2)~rigorous characterization of modern off-the-shelf DRAM chips.}
Therefore, we encourage \agy{8}{the} DRAM industry \agy{8}{and the DRAM-based system industry} to support such research by providing access to DRAM design and manufacturing technology libraries, \agy{8}{better access to infrastructures that enable testing of a wide variety of DRAM chips,} \agy{7}{and expect future research to \agy{8}{continue to} rigorously characterize modern off-the-shelf DRAM chips.}}

\subsubsection{\agy{6}{Latency Reduction for Row Activation Counter Updates}}
\agy{6}{To help mitigate read disturbance, the latest DDR5 specifications (as of April 2024) introduced a new RowHammer mitigation framework, called Per Row Activation Counting ($PRAC$)~\cite{jedec2024jesd795c}.
$PRAC$ enables the DRAM chip to accurately track row activations by allocating an activation counter per DRAM row. Our recent paper~\cite{canpolat2024understanding} analyzes $PRAC$ and shows that $PRAC$ reduces system performance by 10\% even when there is \emph{no} RowHammer attack present because $PRAC$ increases critical DRAM access latency parameters due to the additional time required to increment activation counters. We call for future research on DRAM array and periphery architecture to reduce the latency of incrementing row activation counters.}

\subsubsection{\agy{6}{Online Read Disturbance Profiling }}
\agy{6}{DRAM read disturbance profile 1)~is costly to generate and 2)~can change with aging (\secref{future:sec:aging}). Therefore, it is important to develop accurate and fast profiling procedures that can be periodically repeated in the field. This profiling should be performed with minimal impact to DRAM's availability and system performance with \emph{no} effect on data integrity. We invite future research to explore such methodologies \agy{7}{(similar to prior research in DRAM retention time profiling~\cite{patel2017thereach})}.}

\subsection{Fairness, Quality of Service, and Denial of Service Challenges as DRAM Read Disturbance Worsens}
\label{future:sec:fairness_and_qos}

Actions \agy{6}{to prevent read disturbance bitflips} (e.g., refreshing victim rows) occupy channels, ranks, and banks and thus reduce the available memory bandwidth significantly. Consequently, DRAM read disturbance mitigation can inflict performance degradation on concurrently running benign applications and thus worsen fairness and quality of service. To make matters worse, a malicious user can cause this performance degradation on demand by triggering existing mitigation mechanisms to perform preventive actions. As a result, DRAM read disturbance mitigation mechanisms can be exploited to mount memory performance and even denial of service attacks~\agy{6}{\cite{moscibroda2007memory}} on the memory system, \agy{5}{as our recent works demonstrate~\cite{canpolat2024understanding, canpolat2024breakhammer}}. These fairness, quality of service, and denial of service challenges \agy{6}{are already primary concerns for a wide variety of systems from mobile devices to servers and} become increasingly \agy{5}{more important in scaled-out memory systems, including} disaggregated memory systems~\agy{6}{\cite{lim2008understanding,lim2009disaggregated, calciu2021rethinking}}. To overcome this challenge, future research is needed on 1)~\agy{6}{accurate} detection of processes that exploit DRAM read disturbance mechanisms and 2)~countermeasures targeting such processes. 
\agy{6}{Our recent work, BreakHammer~\cite{canpolat2024breakhammer} addresses this issue by scoring hardware threads based on how frequently they trigger preventive actions (e.g., refresh victim rows) and throttling their memory accesses accordingly. BreakHammer exposes scores of hardware threads to the system software similar to hardware performance counters as an optional integration. We encourage future research to extend BreakHammer to system software to perform throttling at the granularity of processes, address spaces, and users.} 

\section{Concluding Remarks}

In this dissertation, we investigate various aspects of DRAM read disturbance and propose mechanisms to efficiently and scalably prevent DRAM read disturbance bitflips without the knowledge of or modifications to proprietary DRAM chip internals as DRAM chips become more and more vulnerable to read disturbance over generations.   
We build a detailed understanding of the DRAM read disturbance's sensitivities to
temperature (\secref{deeperlook:sec:temperature}),
memory access patterns (\secref{deeperlook:sec:temporal}),
victim DRAM cell's physical location {(\secref{deeperlook:sec:spatial} and \secref{svard:sec:characterization})}, and wordline voltage {(\secref{hammerdimmer:sec:vpp_with_rh})}.
We propose three new mechanisms to mitigate DRAM read disturbance efficiently and scalably: 
1)~Svärd~\agy{5}{(\secref{svard:sec:adaptation})} leverages the spatial variation of read disturbance vulnerability across DRAM rows;
2)~HiRA~\agy{5}{(\secref{hira:sec:doduo})} leverages the subarray-level parallelism to parallelize refresh operations with memory accesses or other refresh operations in off-the-shelf DRAM chips; and
3)~BlockHammer~\agy{5}{(\secref{blockhammer:sec:blockhammer})} selectively throttle unsafe memory accesses to prevent read disturbance bitflips without the knowledge of or modifications to proprietary DRAM internals.   

\agy{5}{We hope and expect that the detailed understanding we develop via our rigorous experimental characterization \agy{6}{of real DRAM chips} and the mechanisms we propose to efficiently and scalably mitigate DRAM read disturbance with \emph{no} proprietary knowledge of DRAM chip internals will inspire DRAM manufacturers and system designers to {enable} robust (i.e., reliable, secure, and safe) memory systems {as} DRAM technology node scaling exacerbates read disturbance.}
\agy{6}{We also hope that future directions we provide will uncover novel findings to solve the DRAM read disturbance problem more efficiently.}

\appendix
\chapter{Other Works of the Author}
\label{appendix:otherworks}
\agy{5}{Besides the works presented in this dissertation, I had the opportunity to contribute on several different areas during my doctoral studies in collaboration with researchers from ETH Zürich, Carnegie Mellon University, University of Illinois Urbana-Champaign, Galicia Supercomputing Center, University of Toronto, Barcelona Supercomputing Center, TOBB University of Economics and Technology, Intel, Huawei, NVIDIA, University of Cyprus, and University of Connecticut}. In this chapter, I acknowledge these works in five categories. 

\textbf{DRAM characterization projects.}
In collaboration with Kevin Chang, we build a concrete understanding of reduced voltage operation in modern DRAM devices~\cite{chang2017understanding}. 
In collaboration with Saugata Ghose, we investigate the power consumption of DRAM chips~\cite{ghose2018what}.
In collaboration with Jeremie S. Kim, we investigate how DRAM read disturbance vulnerability change across generations~\cite{kim2020revisiting}.
In collaboration with Ataberk Olgun, we build 1) DRAM Bender~\cite{olgun2023dram_bender}, an extensible and versatile FPGA-based infrastructure to easily test state-of-the-art DRAM chips; 2)~QUAC-TRNG~\cite{olgun2021quactrng}, a high-throughput random number generation mechanism using quadruple row activation in commodity DRAM chips; and 3)~a detailed understanding of DRAM read disturbance in high-bandwidth memory chips~\cite{olgun2023anexperimental, olgun2024read}. 
In collaboration with Haocong Luo, we investigate the memory access pattern dependency of DRAM read disturbance based on our findings in~\cite{orosa2021adeeper} and discover a new DRAM read disturbance phenomenon that we call RowPress~\cite{luo2023rowpress}. \agy{5}{Our analyses on RowPress later on yielded a more advanced RowHammer--RowPress hybrid access pattern that can induce bitflips in significantly sooner than RowPress-only access patterns.
In collaboration with Ismail Emir Yuksel, we investigate the multiple row activation capabilities of real off-the-shelf DRAM chips in two works which 1)~enables to simultaneously open up to 32 DRAM rows in a subarray and significantly improves the success rate of bulk bitwise majority operations~\cite{yuksel2023pulsar, yuksel2024simra}; and 2)~enables the NOT-operation \agy{6}{(as well as NAND and NOR operations)} and thus functional completeness in real off-the-shelf DRAM chips by leveraging the inverter circuitry in DRAM sense amplifiers~\cite{yuksel2024functionallycomplete}.}
\agy{6}{Under my co-supervision, Yahya Can Tu\u{g}rul investigates the effect of reducing charge restoration time on DRAM read disturbance and data retention time~\cite{tugrul2024understanding}. We show that 1)~refreshing a victim row with partial charge restoration at a reduced refresh latency results in either \emph{no} change or a small reduction in the minimum hammer count to induce the first read disturbance bitflip. Based on our experimental findings, we propose a new mechanism, PaCRAM, that reduces the refresh latency at the cost of reducing the hammer count threshold accordingly \emph{without} compromising security. PaCRAM significantly improves system performance by reducing time spent on refresh operations.}

\textbf{Architecture solutions for DRAM read disturbance.}
In collaboration with Ataberk Olgun, we design ABACuS~\cite{olgun2024abacus}, a new DRAM read disturbance mitigation mechanism that scalably tracks row activation counts with the number of banks by sharing row activation counters across banks.   
In collaboration with F. Nisa Bostanci, we design CoMeT~\cite{bostanci2024comet}, a new DRAM read disturbance detection and mitigation mechanism using the count-min-sketch algorithm. 
In addition, I \agy{7}{have \agy{8}{had} the pleasure of co-supervising} two of O{\u{g}}uzhan Canpolat's projects.
The first project~\cite{canpolat2024understanding} is the first to analyze the security benefits and performance, energy, and cost overheads of a new framework, Per-Row Activation Counting (PRAC), introduced in the JEDEC DDR5 specifications' April 2024 update~\cite{jedec2024jesd795c}. Our analysis identifies PRAC's outstanding problems and foreshadows future research directions.
The second project~\cite{canpolat2024breakhammer} demonstrates that it is possible to exploit several existing RowHammer defenses to mount memory performance attacks, and proposes a new mechanism called BreakHammer, which identifies and throttles the hardware threads that \agy{6}{would otherwise} reduce memory performance by exploiting existing RowHammer defenses. 

\textbf{New DRAM architectures.}
In collaboration with Hasan Hassan, we design 1)~CROW~\cite{hassan2019crow}, a low-cost substrate for improving DRAM performance, energy efficiency, and reliability, and \agy{5}{2)~Self-Managing DRAM~\cite{hassan2022acase, hassan2024acase}, a DRAM architecture that eases the adoption of new in-DRAM maintenance mechanisms with \emph{no} \agy{6}{modifications} to the DRAM communication protocol except the addition of a single ACT-NACK signal.}
\agy{5}{As a team work with Jeremie S. Kim, Fabrice Devaux, and Onur Mutlu, we investigate the security limitations and overheads of the first disclosed DRAM industry solution to DRAM read disturbance, Silver Bullet~\cite{devaux2021method}. Our study~\cite{yaglikci2021security} mathematically demonstrates that Silver Bullet can securely prevent RowHammer attacks and concludes that Silver Bullet is a promising RowHammer prevention mechanism that can be configured to operate securely against RowHammer attacks at various efficiency-area tradeoff points, supporting relatively small hammer count values (e.g., 1000) and Silver Bullet table sizes \agy{6}{(e.g., 1.06 KB)}.}
In collaboration with Haocong Luo, we design CLR-DRAM~\cite{luo2020clrdram}, a low-cost DRAM architecture that enables dynamic capacity-latency tradeoff. 
\agy{5}{In collaboration with Geraldo Francisco de Oliveira, we design MIMDRAM~\cite{deoliveira2024mimdram}, which enables multiple-instruction-multiple-data (MIMD) computation using DRAM cells' analog computation capability.}

\textbf{Energy-efficiency oriented architecture solutions.}
In collaboration with Skanda Koppula, we design EDEN~\cite{koppula2019eden}, energy-efficient and high-performance neural network inference using approximate DRAM. 
In collaboration with Jawad Haj-Yahya, we design 1)~SysScale~\cite{hajyahya2020sysscale} that exploits multi-domain dynamic voltage and frequency scaling for energy-efficient mobile processors and 2)~DarkGates~\cite{hajyahya2022darkgates}, a hybrid power gating architecture that mitigates the performance limitation of dark silicon in high-performance processors. 

\textbf{Other security-oriented works.}
In collaboration with Jawad Haj-Yahya, we design IChannels~\cite{hajyahya2021ichannels} that exploits current management mechanisms to create covert channels in modern processors. 
In collaboration with F. Nisa Bostanci, we design DR-STRaNGe~\cite{bostanci2022drstrange} an end-to-end system design for DRAM-based true random number generators.
In collaboration with Jeremie S. Kim, we analyze the security of the Silver Bullet technique for RowHammer prevention~\cite{yaglikci2021security}.
In collaboration with Onur Mutlu and Ataberk Olgun, we present a survey on fundamentally understanding and solving RowHammer~\cite{mutlu2023fundamentally}.

\chapter{Complete List of the Author's Contributions}
\label{appendix:complete_list}
This section lists the author's contributions to the literature in reverse chronological order under three categories:
1)~major contributions that the author led (\secref{appendix:sec:first_author_contributions}), 
2)~co-supervised contributions by the author (\secref{appendix:sec:co_supervised_contributions}), 
and~3)~other contributions (\secref{appendix:sec:other_contributions}).

\section{Major Contributions \agy{7}{Led by the Author}}
\label{appendix:sec:first_author_contributions}

\begin{enumerate}
    \item A. Giray Ya\u{g}l{\i}k\c{c}{\i}, Yahya Can Tugrul, Geraldo Francisco de Oliveira Junior, Ismail Yuksel, Ataberk Olgun, Haocong Luo, and Onur Mutlu, \emph{``Spatial Variation-Aware Read Disturbance Defenses: Experimental Analysis of Real DRAM Chips and Implications on Future Solutions,''} in HPCA, 2024.
    
    \item A. Giray Ya\u{g}l{\i}k\c{c}{\i}, Ataberk Olgun, Minesh Patel, Haocong Luo, Hasan Hassan, Lois Orosa, Oguz Ergin, and Onur Mutlu, \emph{``HiRA: Hidden Row Activation for Reducing Refresh Latency of Off-the-Shelf DRAM Chips,''} in MICRO, 2022. 
    
    \item A. Giray Ya\u{g}l{\i}k\c{c}{\i}, Haocong Luo, Geraldo F. de Oliveira, Ataberk Olgun, Minesh Patel, Jisung Park, Hasan Hassan, Jeremie S. Kim, Lois Orosa, and Onur Mutlu, \emph{``Understanding RowHammer Under Reduced Wordline Voltage: An Experimental Study Using Real DRAM Devices,''} in DSN, 2022.
    
    \item Lois Orosa*, A. Giray Ya\u{g}l{\i}k\c{c}{\i}*, Haocong Luo, Ataberk Olgun, Jisung Park, Hasan Hassan, Minesh Patel, Jeremie S. Kim, and Onur Mutlu, \emph{``A Deeper Look into RowHammer's Sensitivities: Experimental Analysis of Real DRAM Chips and Implications on Future Attacks and Defenses,''} in MICRO, 2021.
    
    \item A. Giray Ya\u{g}l{\i}k\c{c}{\i}, Minesh H. Patel, Jeremie S. Kim, Lois Orosa, Roknoddin Azizibarzoki, Hasan Hassan, Ataberk Olgun, Jisung Park, Konstantinos Kanellopoulos, Taha Shahroodi, Saugata Ghose, and Onur Mutlu, \emph{``BlockHammer: Preventing RowHammer at Low Cost by Blacklisting Rapidly-Accessed DRAM Rows,''} in HPCA, 2021.
    Implementation available based on \href{https://github.com/CMU-SAFARI/Ramulator}{Ramulator}: \href{https://github.com/CMU-SAFARI/BlockHammer}{https://github.com/CMU-SAFARI/BlockHammer} 
    and as a plugin of \href{https://github.com/CMU-SAFARI/ramulator2}{Ramulator2}: \href{https://github.com/CMU-SAFARI/ramulator2/tree/main/src/dram_controller/impl/plugin/blockhammer}{https://github.com/CMU-SAFARI/ramulator2/tree/main/src/dram\_controller/impl/plugin/blockhammer}.
    
    \item A. Giray Ya\u{g}l{\i}k\c{c}{\i}, Jeremie S. Kim, Fabrice Devaux, and Onur Mutlu, \emph{``Security Analysis of the Silver Bullet Technique for RowHammer Prevention,''} arXiv, 2021.
\end{enumerate}

\section{Co-supervised Contributions}
\label{appendix:sec:co_supervised_contributions}

\begin{enumerate}
    \item O\u{g}uzhan Canpolat, A. Giray Ya\u{g}l{\i}k\c{c}{\i}, Ataberk Olgun, Ismail Emir Yüksel, Yahya Can Tu\u{g}rul, Konstantinos Kanellopoulos, O\u{g}uz Ergin, Onur Mutlu, \emph{``BreakHammer: Enabling Scalable and Low Overhead RowHammer Mitigations via Throttling Preventive Action Triggering Threads,''} to appear in MICRO. (Preprint on arXiv:2404.13477 [cs.CR]), 2024. 

    \item O\u{g}uzhan Canpolat, A. Giray Ya\u{g}l{\i}k\c{c}{\i}, Geraldo F. Oliveira, Ataberk Olgun, O\u{g}uz Ergin Onur Mutlu, \emph{``Understanding the Security Benefits and Overheads of Emerging Industry Solutions to DRAM Read Disturbance,''} in DRAMSec, 2024. 
    Implementation available: \href{https://github.com/CMU-SAFARI/ramulator2}{https://github.com/CMU-SAFARI/ramulator2}

    \item Yahya Can Tu\u{g}rul, A. Giray Ya\u{g}l{\i}k\c{c}{\i}, Ismail Emir Yüksel, Ataberk Olgun, O\u{g}uzhan Canpolat, F. Nisa Bostanc\i, Mohammad Sadrosadati, O\u{g}uz Ergin, and Onur Mutlu, \emph{``Understanding RowHammer Under Reduced Refresh Latency: Experimental Analysis of Real DRAM Chips and Implications on Future Solutions,''} [Under Submission] (Preprint on arXiv), 2024.
\end{enumerate}

\section{Other Contributions}
\label{appendix:sec:other_contributions}

\begin{enumerate}
  \item Hasan Hassan, Ataberk Olgun, A. Giray Ya\u{g}l{\i}k\c{c}{\i}, Haocong Luo, Onur Mutlu, \emph{``Self-Managing DRAM: A Low-Cost Framework for Enabling Autonomous and Efficient in-DRAM Operations,''} to appear in MICRO, 2024. (Preprint: arXiv:2207.13358 [cs.AR])
  Artifact available: \href{https://github.com/CMU-SAFARI/SelfManagingDRAM}{https://github.com/CMU-SAFARI/SelfManagingDRAM}
  
  \item Ataberk Olgun, Yahya Can Tu\u{g}rul, Nisa Bostanci, Ismail Emir Yuksel, Haocong Luo, Steve Rhyner, A. Giray Ya\u{g}l{\i}k\c{c}{\i}, Geraldo F. Oliveira, and Onur Mutlu, \emph{``ABACuS: All-Bank Activation Counters for Scalable and Low Overhead RowHammer Mitigation,''} USENIX Security, 2024.
  Artifact available: \href{https://github.com/CMU-SAFARI/ABACuS}{https://github.com/CMU-SAFARI/ABACuS}
  
  \item Lois Orosa, Ulrich Ruhrmair, A Giray Yaglikci, Haocong Luo, Ataberk Olgun, Patrick Jattke, Minesh Patel, Jeremie S. Kim, Kaveh Razavi, and Onur Mutlu, \emph{``SpyHammer: Understanding and Exploiting RowHammer Under Fine-Grained Temperature Variations,''} IEEE Access, June 2024.
  
  \item Ataberk Olgun, Majd Osseiran, A. Giray Ya\u{g}l{\i}k\c{c}{\i}, Yahya Can Tuğrul, Haocong Luo, Steve Rhyner, Behzad Salami, Juan Gómez Luna, Onur Mutlu, \emph{``Read Disturbance in High Bandwidth Memory: A Detailed Experimental Study on HBM2 DRAM Chips,''} in DSN, 2024.
  Artifact available: \href{https://github.com/CMU-SAFARI/HBM-Read-Disturbance}{https://github.com/CMU-SAFARI/HBM-Read-Disturbance}
  
  \item Haocong Luo, Ismail Emir Yüksel, Ataberk Olgun, A Giray Ya\u{g}l{\i}k\c{c}{\i}, Mohammad Sadrosadati, Onur Mutlu, \emph{``An Experimental Characterization of Combined RowHammer and RowPress Read Disturbance in Modern DRAM Chips,''} in DSN (Disrupt), 2024.
  
  \item Ismail Emir Yuksel, Yahya Can Tuğrul, Nisa Bostanci, Geraldo Francisco de Oliveira Junior, A. Giray Ya\u{g}l{\i}k\c{c}{\i}, Ataberk Olgun, Melina Soysal, Haocong Luo, Juan Gómez Luna, Mohammad Sadrosadati, Onur Mutlu, \emph{``Simultaneous Many-Row Activation in Off-the-Shelf DRAM Chips: Experimental Characterization and Analysis,''} in DSN, 2024.
  Artifact available: \href{https://github.com/CMU-SAFARI/SiMRA-DRAM}{https://github.com/CMU-SAFARI/SiMRA-DRAM}
  
  \item Geraldo F. Oliveira, Ataberk Olgun, A. Giray Ya\u{g}l{\i}k\c{c}{\i}, F. Nisa Bostanc{\i}, Juan Gómez-Luna, Saugata Ghose, Onur Mutlu, \emph{``MIMDRAM: An End-to-End Processing-Using-DRAM System for High-Throughput, Energy-Efficient and Programmer-Transparent Multiple-Instruction Multiple-Data Computing,''} in HPCA, 2024.
  Artifact available: \href{https://github.com/CMU-SAFARI/MIMDRAM}{https://github.com/CMU-SAFARI/MIMDRAM}
  
  \item Ismail E. Yüksel, Yahya C. Tu\u{g}rul, Ataberk Olgun, F. Nisa Bostanc{\i}, A. Giray Ya\u{g}l{\i}k\c{c}{\i}, Geraldo F. Oliveira, Haocong Luo, Juan Gomez-Luna, Mohammad Sadrosadati, and Onur Mutlu, \emph{``Functionally-Complete Boolean Logic in Real DRAM Chips: Experimental Characterization and Analysis,''} in HPCA, 2024.
  Artifact available: \href{https://github.com/CMU-SAFARI/FCDRAM}{https://github.com/CMU-SAFARI/FCDRAM}
  
  \item F. Nisa Bostanc{\i}, Ismail E. Yüksel, Ataberk Olgun, Konstantinos Kanellopoulos, Yahya Can Tu\u{g}rul, A. Giray Ya\u{g}l{\i}k\c{c}{\i}, Mohammad Sadrosadati, and Onur Mutlu, \emph{``CoMeT: Count-Min-Sketch-based Row Tracking to Mitigate RowHammer at Low Cost,''} in HPCA, 2024.
  Artifact available: \href{https://github.com/CMU-SAFARI/CoMeT}{https://github.com/CMU-SAFARI/CoMeT}

  \item Konstantinos Kanellopoulos, F. Nisa Bostanci, Ataberk Olgun, A. Giray Ya\u{g}l{\i}k\c{c}{\i}, Ismail Emir Yüksel, Nika Mansouri Ghiasi, Zülal Bingöl, Mohammad Sadrosadati, and Onur Mutlu. \emph{``Amplifying Main Memory-Based Timing Covert and Side Channels using Processing-in-Memory Operations,''} arXiv:2404.11284 [cs.CR], 2024.
  
  \item Haocong Luo, Ataberk Olgun, A. Giray Ya\u{g}l{\i}k\c{c}{\i}, Yahya Can Tuğrul, Steve Rhyner, Meryem Banu Cavlak, Joël Lindegger, Mohammad Sadrosadati, and Onur Mutlu. \emph{``RowPress: Amplifying Read Disturbance in Modern DRAM Chips,''} in ISCA, 2023.
  Artifact available: \href{https://github.com/CMU-SAFARI/RowPress}{https://github.com/CMU-SAFARI/RowPress}
  
  \item Ataberk Olgun, Majd Osseiran, A. Giray Ya\u{g}l{\i}k\c{c}{\i}, Yahya Can Tu\u{g}rul, Haocong Luo, Steve Rhyner, Behzad Salami, Juan Gomez Luna, and Onur Mutlu, \emph{``An Experimental Analysis of RowHammer in HBM2 DRAM Chips,''} in DSN (Disrupt), 2023.
  Artifact available: \href{https://github.com/CMU-SAFARI/HBM-Read-Disturbance}{https://github.com/CMU-SAFARI/HBM-Read-Disturbance}
  
  \item Onur Mutlu, Ataberk Olgun, and A. Giray Ya\u{g}l{\i}k\c{c}{\i}, \emph{``Fundamentally Understanding and Solving RowHammer,''} Invited Special Session Paper at ASP-DAC, 2023.
  
  \item Ataberk Olgun, Hasan Hassan, A. Giray Ya\u{g}l{\i}k\c{c}{\i}, Yahya Can Tu\u{g}rul, Lois Orosa, Haocong Luo, Minesh Patel, O\u{g}uz Ergin, Onur Mutlu, \emph{``DRAM Bender: An Extensible and Versatile FPGA-based Infrastructure to Easily Test State-of-the-art DRAM Chips,''} IEEE TCAD, 2023.
  Implementation available: \href{https://github.com/CMU-SAFARI/DRAM-Bender}{https://github.com/CMU-SAFARI/DRAM-Bender}

  \item Haocong Luo, Yahya Tu\u{g}rul Can, F. Nisa Bostanc{\i}, Ataberk Olgun, A. Giray Ya\u{g}l{\i}k\c{c}{\i}, Onur Mutlu \emph{``Ramulator 2.0: A Modern, Modular, and Extensible DRAM Simulator,''}, IEEE CAL, 2023.
  Available: \href{https://github.com/CMU-SAFARI/ramulator2}{https://github.com/CMU-SAFARI/ramulator2}https://github.com/CMU-SAFARI/ramulator2.

  \item F. Nisa Bostanci, Ataberk Olgun, Lois Orosa, A. Giray Ya\u{g}l{\i}k\c{c}{\i}, Jeremie S. Kim, Hasan Hassan, O\u{g}uz Ergin, and Onur Mutlu, \emph{``DR-STRaNGe: End-to-End System Design for DRAM-based True Random Number Generators,''} HPCA, 2022.

  \item Jawad Haj Yahya, Jeremie S. Kim, A. Giray Ya\u{g}l{\i}k\c{c}{\i}, Jisung Park, Efraim Rotem, Yanos Sazeides, and Onur Mutlu, \emph{``DarkGates: A Hybrid Power-Gating Architecture to Mitigate the Performance Impact of Dark-Silicon in High Performance Processors,''} HPCA, 2022.

  \item Minesh Patel, Taha Shahroodi, Aditya Manglik, A. Giray Ya\u{g}l{\i}k\c{c}{\i}, Ataberk Olgun, Haocong Luo, and Onur Mutlu. \emph{``A Case for Transparent Reliability in DRAM Systems,''} arXiv:2204.10378 [cs.AR], 2022.

  \item İsmail Emir Yüksel, Ataberk Olgun, Behzad Salami, F. Nisa Bostanc\i, Yahya Can Tu\u{g}rul, A. Giray Ya\u{g}l{\i}k\c{c}{\i}, Nika Mansouri Ghiasi, Onur Mutlu, and O\u{g}uz Ergin. \emph{``TuRaN: True Random Number Generation Using Supply Voltage Underscaling in SRAMs,''} arXiv:2211.10894 [cs.AR], 2022.

  \item Jawad Haj-Yahya, Jeremie S. Kim, A. Giray Ya\u{g}l{\i}k\c{c}{\i}, Ivan Puddu, Lois Orosa, Juan Gomez Luna, Mohammed Alser, and Onur Mutlu, \emph{``IChannels: Exploiting Current Management Mechanisms to Create Covert Channels in Modern Processors,''} ISCA, 2021.

  \item Ataberk Olgun, Minesh Patel, A. Giray Ya\u{g}l{\i}k\c{c}{\i}, Haocong Luo, Jeremie S. Kim, F. Nisa Bostanci, Nandita Vijaykumar, O\u{g}uz Ergin, and Onur Mutlu, \emph{``QUAC-TRNG: High-Throughput True Random Number Generation Using Quadruple Row Activation in Commodity DRAM Chips,''} ISCA, 2021.
  Artifact available: \href{https://github.com/CMU-SAFARI/QUAC-TRNG}{https://github.com/CMU-SAFARI/QUAC-TRNG}

  \item Jeremie S. Kim, Minesh Patel, A. Giray Ya\u{g}l{\i}k\c{c}{\i}, Hasan Hassan, Roknoddin Azizi, Lois Orosa, and Onur Mutlu, \emph{``Revisiting RowHammer: An Experimental Analysis of Modern Devices and Mitigation Techniques,''} in ISCA, 2020.

  \item Jawad Haj-Yahya, Mohammad Alser, Jeremie Kim, A. Giray Ya\u{g}l{\i}k\c{c}{\i}, Nandita Vijaykumar, Efraim Rotem, and Onur Mutlu, \emph{``SysScale: Exploiting Multi-domain Dynamic Voltage and Frequency Scaling for Energy Efficient Mobile Processors,''} in ISCA, 2020.  
  
  \item Haocong Luo, Taha Shahroodi, Hasan Hassan, Minesh Patel, A. Giray Ya\u{g}l{\i}k\c{c}{\i}, Lois Orosa, Jisung Park, and Onur Mutlu, \emph{``CLR-DRAM: A Low-Cost DRAM Architecture Enabling Dynamic Capacity-Latency Trade-Off,''} in ISCA, 2020.

  \item Skanda Koppula, Lois Orosa, A. Giray Ya\u{g}l{\i}k\c{c}{\i}, Roknoddin Azizi, Taha Shahroodi, Konstantinos Kanellopoulos, and Onur Mutlu, \emph{``EDEN: Energy-Efficient, High-Performance Neural Network Inference Using Approximate DRAM,''} in MICRO, 2019.

  \item Hasan Hassan, Minesh Patel, Jeremie S. Kim, A. Giray Ya\u{g}l{\i}k\c{c}{\i}, Nandita Vijaykumar, Nika Mansouri Ghiasi, Saugata Ghose, Onur Mutlu, \emph{``CROW: A Low-Cost Substrate for Improving DRAM Performance, Energy Efficiency, and Reliability,''} in ISCA, 2019.
  
  \item Saugata Ghose, A. Giray Ya\u{g}l{\i}k\c{c}{\i}, Raghav Gupta, Donghyuk Lee, K. Kudrolli, William X. Liu, Hasan Hassan, Kevin K. Chang, Niladrish Chatterjee, Aditya Agrawal, Michael O'Connor, and Onur Mutlu, \emph{``What Your DRAM Power Models Are Not Telling You: Lessons from a Detailed Experimental Study,''} in SIGMETRICS, 2018.
  Simulator available: \href{https://github.com/CMU-SAFARI/VAMPIRE}{https://github.com/CMU-SAFARI/VAMPIRE}
  
  \item Kevin Chang, A. Giray Ya\u{g}l{\i}k\c{c}{\i}, Saugata Ghose, Aditya Agrawal, Niladrish Chatterjee, Abishek Kashyap, Donghyuk Lee, Michael O’Connor, Hasan Hassan, and Onur Mutlu. \emph{``Understanding Reduced-Voltage Operation in Modern DRAM Devices: Experimental Characterization, Analysis, and Mechanisms,''} in SIGMETRICS, 2017.
  Artifact available: \href{https://github.com/CMU-SAFARI/DRAM-Voltage-Study}{https://github.com/CMU-SAFARI/DRAM-Voltage-Study}
\end{enumerate}

\chapter{Built Infrastructures}

\agy{6}{This chapter showcases the infrastructures we have built throughout my journey.}

\section{DDR3 and DDR4 Power Measurement Infrastructure}
\agy{6}{We build a DDR3 SODIMM power consumption measurement infrastructure~\cite{ghose2018what}. \figref{fig:riser-board} shows the DDR3 SO-DIMM riser board with current sensing capability. We use a commercial riser board, JET-5467A~\cite{mfactorsjet5467a}, which uses a shunt resistor to measure the current.} 

\begin{figure}[ht]
    \centering
    \begin{subfigure}[b]{0.495\linewidth}
        \centering
        \rotatebox{180}{\includegraphics[width=\linewidth]{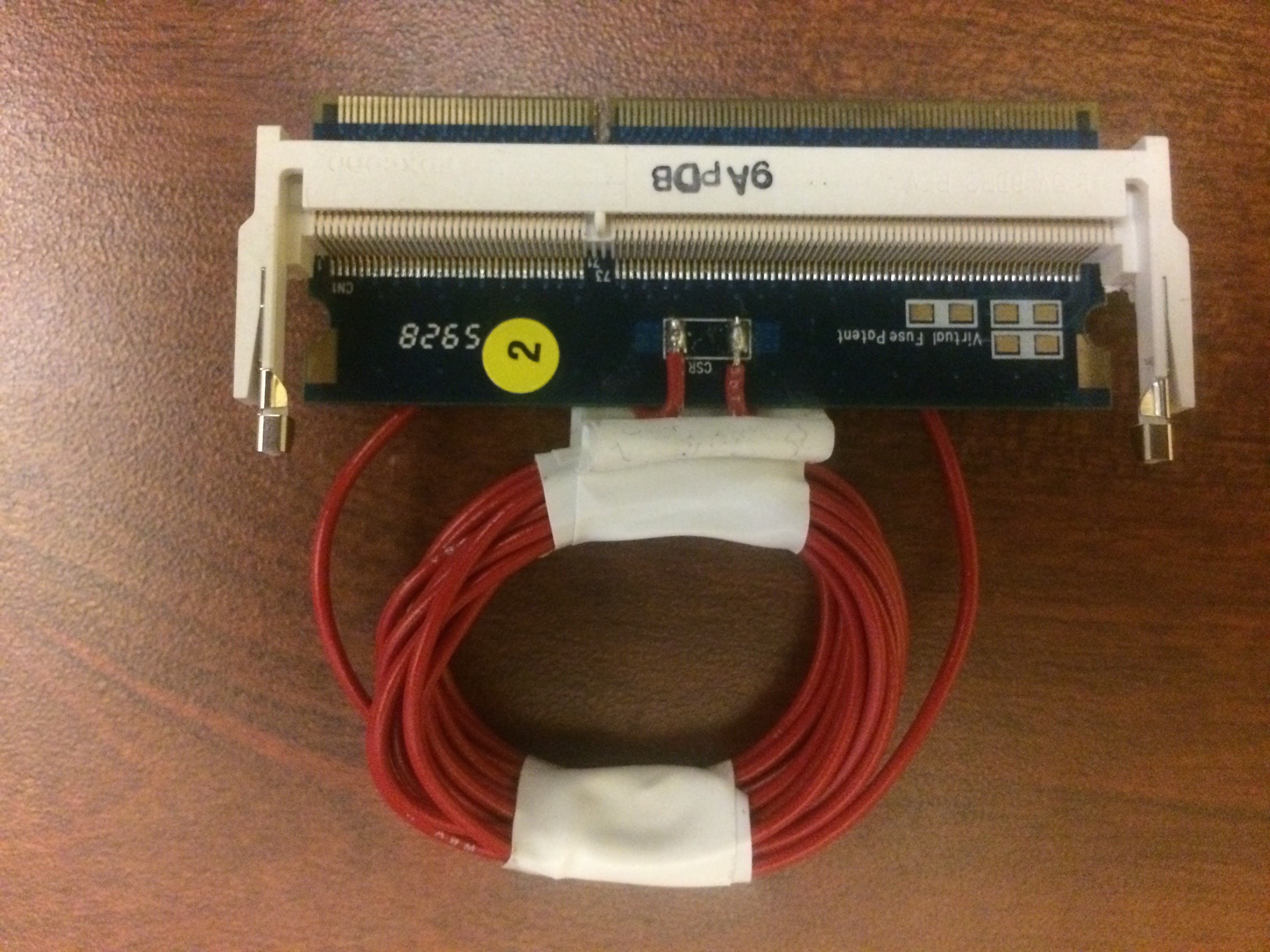}}
        \caption{DDR3 SO-DIMM riser board}
    \end{subfigure}
    \hfill
    \begin{subfigure}[b]{0.495\linewidth}
        \centering
        \rotatebox{180}{\includegraphics[width=\linewidth]{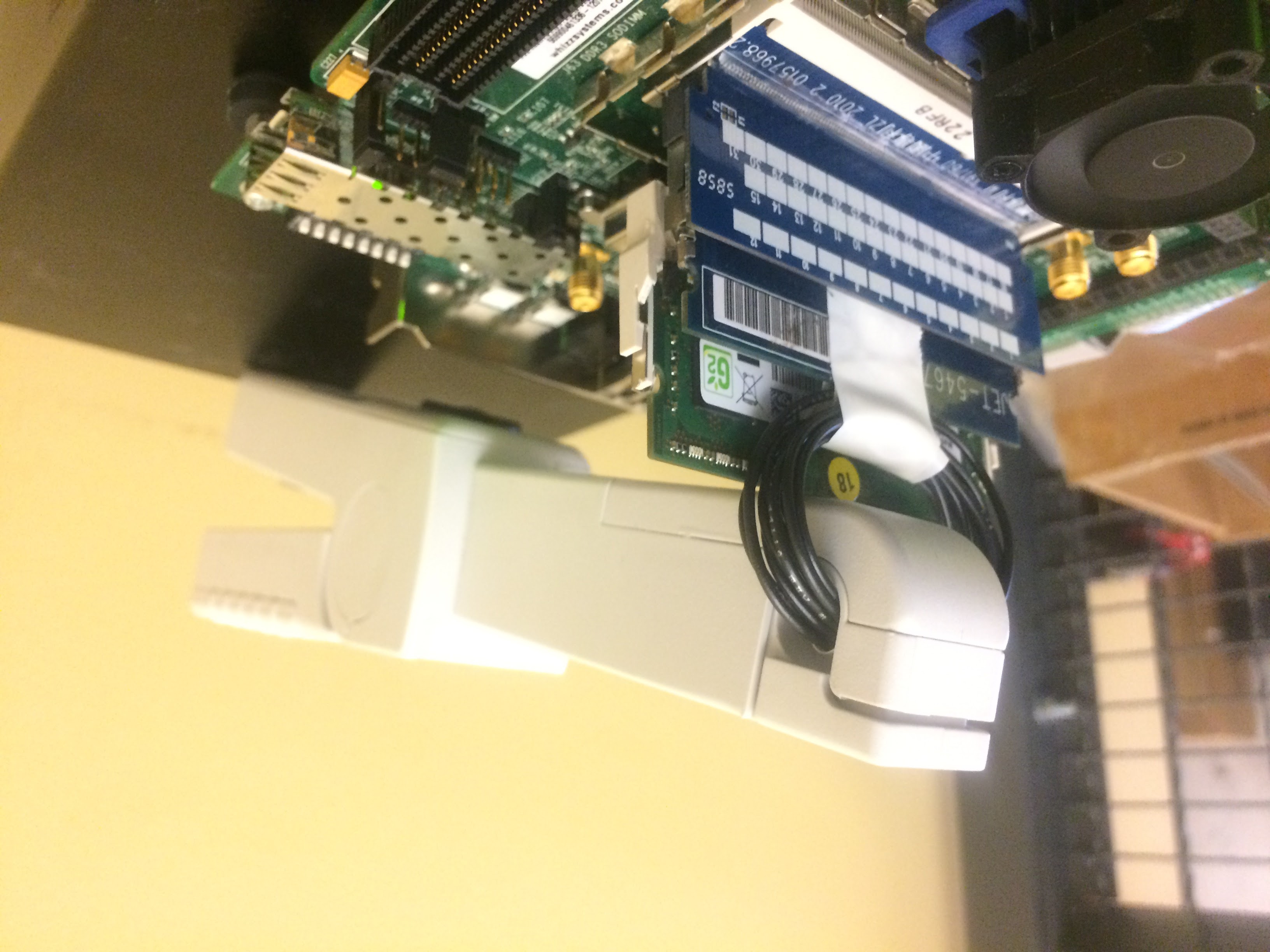}}
        \caption{Current measurement probe}
    \end{subfigure}
    \caption{DDR3 DRAM power measurement infrastructure}
    \label{fig:riser-board}
\end{figure}

\agy{6}{To improve measurement accuracy, we remove the shunt resistor provided on the extender and add in a 5-coil wire (\figref{fig:riser-board}a).\footnote{\agy{6}{The coiled red wire in \figref{fig:riser-board}a is a 20-coil wire, but we use a 5-coil version of it in our measurements.}} 
\figref{fig:riser-board}b shows how we insert the coil into a Keysight 34134A high-precision DC current probe~\cite{technologies34134a}, which is coupled to a Keysight 34461A high-precision multimeter~\cite{technologies34461a}.}
\agy{7}{This infrastructure enabled the experimental results in~\cite{ghose2018what, yuksel2024simra}.}\footnote{To support DDR4 DRAM chips, we use DDR4 riser boards with current sensing capability (e.g.,~\cite{adexelecddr4sodv1}). As the coil wire approach is unreliable with DDR4 DRAM chips, we directly measure the voltage on the shunt resistors.}

\section{DDR4 DRAM Testing Infrastructure}
\label{infrastructures:sec:ddr4drambender}
\agy{6}{\figref{fig:ddr4-setup} shows our FPGA-based DRAM testing infrastructure for testing real DDR4 DRAM chips. We port our DRAM Bender~\cite{hassan2017softmc, safari2017softmc, safari2022dram_bender, olgun2023dram_bender} design to two different FPGA boards.
We use Bittware XUSP3S~\cite{bittware_xusp3s} for SODIMMs (\figref{fig:ddr4-setup}a) and Xilinx Alveo U200~\cite{xilinxu200} for DIMMs (\figref{fig:ddr4-setup}b).}

\begin{figure}[ht]
    \centering
    \begin{subfigure}[b]{0.495\linewidth}
        \centering
        \rotatebox{0}{\includegraphics[width=\linewidth]{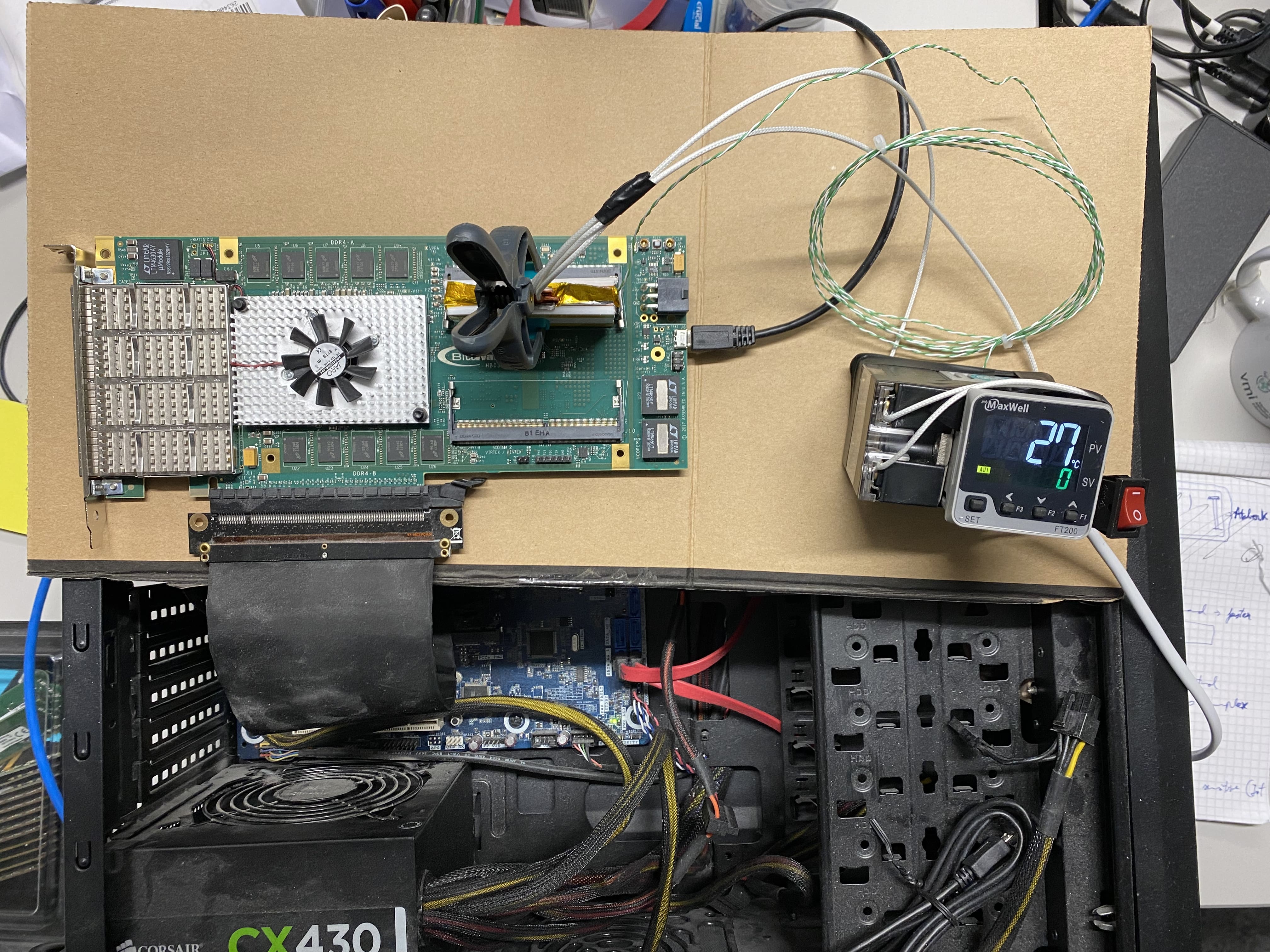}}
        \caption{Bittware XUSP3S FPGA~\cite{bittware_xusp3s}}
    \end{subfigure}
    \hfill
    \begin{subfigure}[b]{0.495\linewidth}
        \centering
        \rotatebox{0}{\includegraphics[width=\linewidth]{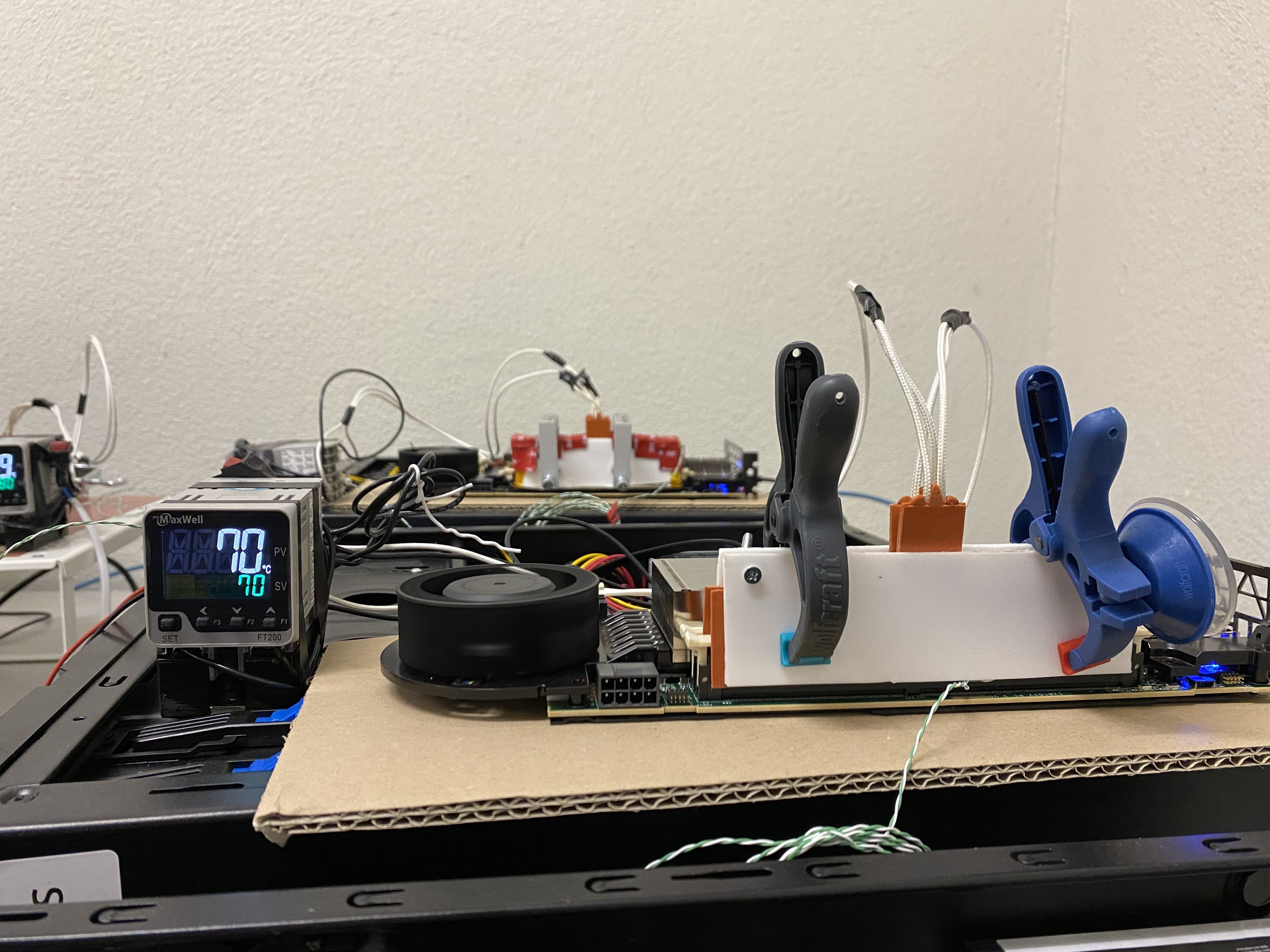}}
        \caption{Xilinx Alveo U200 FPGA~\cite{xilinxu200}}
    \end{subfigure}
    \caption{DDR4 DRAM testing infrastructure}
    \label{fig:ddr4-setup}
\end{figure}

We use a host machine connected to our FPGA boards through a PCIe port~\cite{xilinx2011virtex6} (\figref{deeperlook:fig:infrastructure}c) to
1)~run our tests, 2)~monitor and adjust the temperature of DRAM chips in cooperation with the temperature controller, and 3)~process the experimental data.
As described in \secref{deeperlook:sec:testing_infrastructure}, this infrastructure provides us with precise control over DDR4 DRAM command timings at the granularity of \SI{1.5}{\nano\second} and temperature at the granularity of \SI{0.5}{\celsius}.
\agy{7}{This infrastructure enabled the experimental results provided in \secref{deeperlook:sec:temperature}--\secref{deeperlook:sec:spatial}, \secref{hammerdimmer:sec:vpp_with_rh}--\secref{hammerdimmer:sec:sideeffects}, \secref{svard:sec:characterization}, \secref{hira:sec:characterization} of this thesis and also in~\cite{frigo2020trrespass, hassan2021uncovering, kim2020revisiting, luo2023rowpress, orosa2022spyhammer, olgun2021quactrng, yuksel2023pulsar, luo2024experimental, yuksel2024simra, yuksel2024functionallycomplete, olgun2023anexperimental, olgun2024read, tugrul2024understanding}.}

\section{DDR4 Voltage Scaling Infrastructure}
\agy{9}{As \secref{hammerdimmer:sec:methodology} explains, we built a voltage scaling infrastructure that allows us to change the wordline voltage of DDR4 DRAM chips. \figref{fig:vpp-hacked-riser} shows our modifications on a DDR4 DIMM riser board with current sensing capability to drive the $V_{PP}$ power rail using an external power supply.}

\begin{figure}
    \centering
    \includegraphics[width=0.8\linewidth]{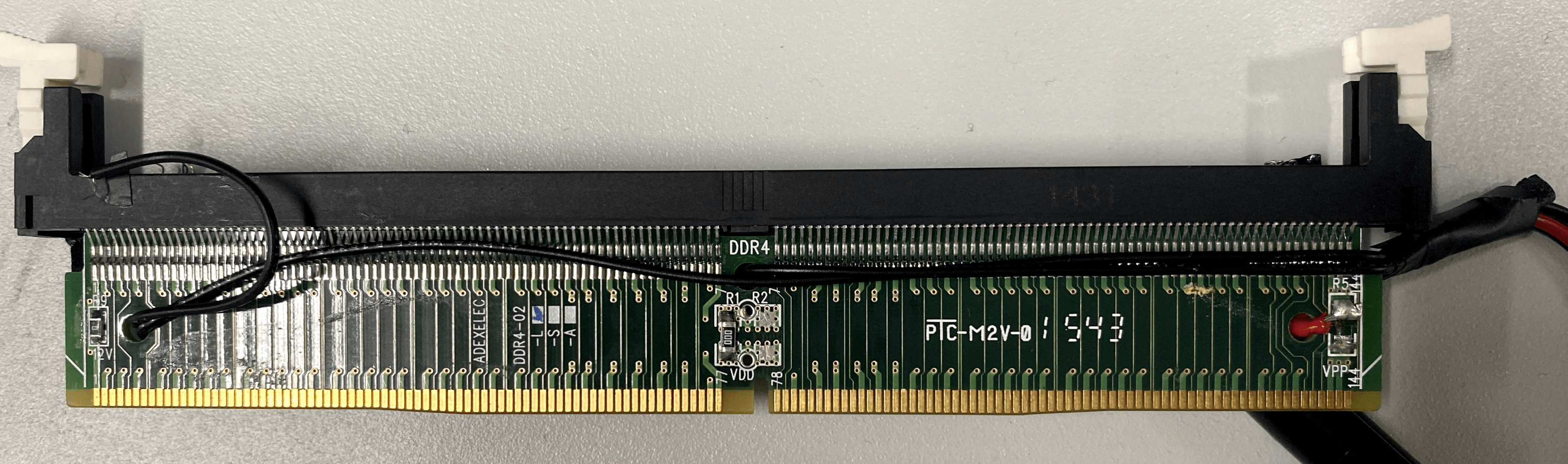}
    \caption{DDR4 DIMM riser board with current sensing capability that we modify to drive $V_{PP}$ power rail with an external power supply}
    \label{fig:vpp-hacked-riser}
\end{figure}

\agy{9}{This riser board originally has a shunt resistor soldered as a surface mount device on the $V_{PP}$ power rail. We removed this shunt resistor and exposed its two pads. The bottom and the top pads visible on the right-hand side of \figref{fig:vpp-hacked-riser} are connected to the FPGA- and the DIMM-sides of the riser board, respectively. We solder a wire (the red wire in the picture) to the top pad of the $V_{PP}$ rail, driven by the external power supply. We also connect the grounds of the power supply and the riser board via a wire to ensure they share a common ground. We follow a similar methodology for DDR4 SODIMM riser boards as well. This infrastructure enabled the experimental results provided in \secref{hammerdimmer:sec:vpp_with_rh}--\secref{hammerdimmer:sec:sideeffects} of this thesis and also in~\cite{yuksel2024simra}.}

\section{HBM2 DRAM Testing Infrastructure}
\agy{9}{\figref{fig:hbm2-setup} shows our FPGA-based DRAM testing infrastructure for testing real HBM2 DRAM chips. \figref{fig:hbm2-setup}a and \figref{fig:hbm2-setup}b shows the same setup from top and front, respectively.}

\begin{figure}[ht]
    \centering
    \begin{subfigure}[b]{0.495\linewidth}
        \centering
        \rotatebox{0}{\includegraphics[width=\linewidth]{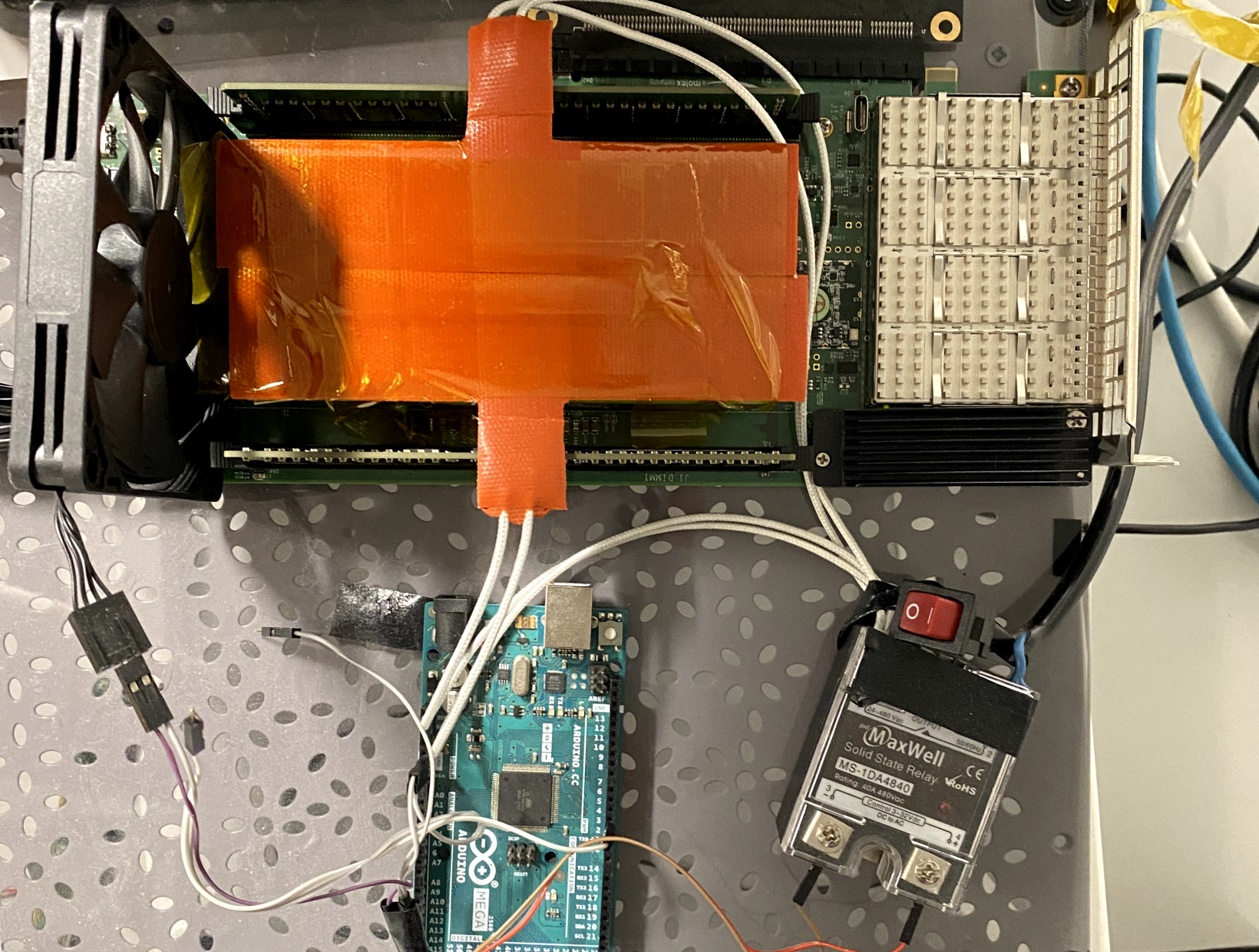}}
        \caption{Top view}
    \end{subfigure}
    \hfill
    \begin{subfigure}[b]{0.495\linewidth}
        \centering
        \rotatebox{0}{\includegraphics[width=\linewidth]{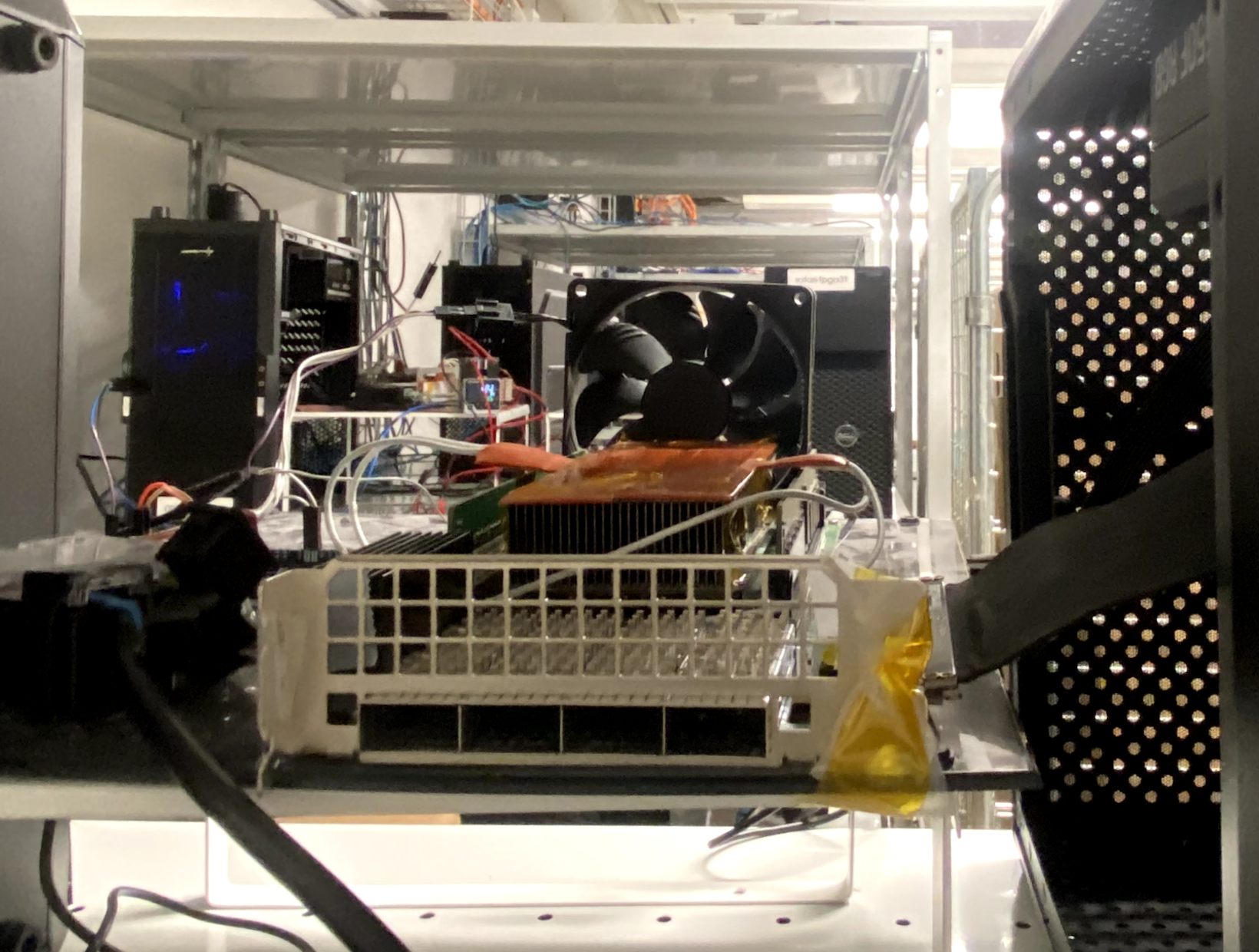}}
        \caption{Front view}
    \end{subfigure}
    \caption{HBM2 DRAM testing infrastructure}
    \label{fig:hbm2-setup}
\end{figure}

\pagebreak
\agy{9}{We port our DRAM Bender design~\cite{hassan2017softmc, safari2017softmc, safari2022dram_bender, olgun2023dram_bender} to Xilinx Alveo U50 FPGA board~\cite{xilinxu50}.
Similar to our DDR4 DRAM Bender setup (\secref{infrastructures:sec:ddr4drambender}),
this infrastructure provides precise control over HBM2 DRAM command timings at the granularity of \SI{1.67}{\nano\second}. 
\agy{8}{We run our tests using a host machine connected to our FPGA boards through a PCIe port~\cite{xilinx2011virtex6}. The host machine monitors the DRAM chip's temperature by reading a DRAM-internal temperature sensor. We feed the measured temperature to an Arduino Mega microcontroller board~\cite{arduinomega}, which controls a couple of heater pads and a fan. The heater pads are fixed on the heat sink to heat the chip, and the fan is located perpendicular to the heat sink to provide cooling with airflow through the heat sink. The microcontroller implements a PID control loop to stabilize the temperature by turning on/off the heater pads and adjusting the fan speed via standard pulse-width-modulated signals. Our measurements show that we control the chip's temperature with a precision of $\pm$\SI{2}{\celsius}. \agy{7}{This infrastructure enabled the experimental results provided in~\cite{olgun2023anexperimental, olgun2024read}.}}}

\balance
\begin{singlespace}
\bibliographystyle{unsrt}
\bibliography{main}

\begin{thebibliography}{100}

\bibitem{xilinxu200}
{Xilinx}.
\newblock {Xilinx Alveo U200 FPGA Board}.
\newblock
  \url{https://www.xilinx.com/products/boards-and-kits/alveo/u200.html}.

\bibitem{hassan2017softmc}
Hasan Hassan, Nandita Vijaykumar, Samira Khan, Saugata Ghose, Kevin Chang,
  Gennady Pekhimenko, Donghyuk Lee, Oguz Ergin, and Onur Mutlu.
\newblock {SoftMC: A Flexible and Practical Open-Source Infrastructure for
  Enabling Experimental DRAM Studies}.
\newblock In {\em HPCA}, 2017.

\bibitem{safari2017softmc}
{SAFARI Research Group}.
\newblock {SoftMC --- GitHub Repository}.
\newblock \url{https://github.com/CMU-SAFARI/softmc}, 2017.

\bibitem{olgun2023dram_bender}
Ataberk Olgun, Hasan Hassan, A~Giray Ya{\u{g}}l{\i}k{\c{c}}{\i}, Yahya~Can
  Tu{\u{g}}rul, Lois Orosa, Haocong Luo, Minesh Patel, O{\u{g}}uz Ergin, and
  Onur Mutlu.
\newblock {DRAM Bender: An Extensible and Versatile FPGA-Based Infrastructure
  to Easily Test State-of-the-Art DRAM Chips}.
\newblock {\em IEEE TCAD}, 2023.

\bibitem{safari2022dram_bender}
{SAFARI Research Group}.
\newblock {DRAM Bender --- GitHub Repository}.
\newblock \url{https://github.com/CMU-SAFARI/DRAM-Bender}, 2022.

\bibitem{saxena2022aqua}
Anish Saxena, Gururaj Saileshwar, Prashant~J. Nair, and Moinuddin Qureshi.
\newblock {AQUA: Scalable Rowhammer Mitigation by Quarantining Aggressor Rows
  at Runtime}.
\newblock In {\em MICRO}, 2022.

\bibitem{yaglikci2021blockhammer}
A~Giray Ya{\u{g}}l{\i}k{\c{c}}{\i}, Minesh Patel, Jeremie~S. Kim, Roknoddin
  Azizibarzoki, Ataberk Olgun, Lois Orosa, Hasan Hassan, Jisung Park,
  Konstantinos Kanellopoullos, Taha Shahroodi, Saugata Ghose, and Onur Mutlu.
\newblock {BlockHammer: Preventing RowHammer at Low Cost by Blacklisting
  Rapidly-Accessed DRAM Rows}.
\newblock In {\em HPCA}, 2021.

\bibitem{qureshi2022hydra}
Moinuddin Qureshi, Aditya Rohan, Gururaj Saileshwar, and Prashant~J Nair.
\newblock {Hydra: Enabling Low-Overhead Mitigation of Row-Hammer at Ultra-Low
  Thresholds via Hybrid Tracking}.
\newblock In {\em ISCA}, 2022.

\bibitem{kim2014flipping}
Y.~{Kim}, R.~{Daly}, J.~{Kim}, C.~{Fallin}, J.~H. {Lee}, D.~{Lee},
  C.~{Wilkerson}, K.~{Lai}, and O.~{Mutlu}.
\newblock {Flipping Bits in Memory Without Accessing Them: An Experimental
  Study of DRAM Disturbance Errors}.
\newblock In {\em ISCA}, 2014.

\bibitem{saileshwar2022randomized}
Gururaj Saileshwar, Bolin Wang, Moinuddin Qureshi, and Prashant~J Nair.
\newblock {Randomized Row-Swap: Mitigating Row Hammer by Breaking Spatial
  Correlation Between Aggressor and Victim Rows}.
\newblock In {\em ASPLOS}, 2022.

\bibitem{jedec2017jesd794b}
{JEDEC}.
\newblock {JESD79-4B: DDR4 SDRAM Standard}, 2017.

\bibitem{jedec2020jesd794c}
{JEDEC}.
\newblock {\em {JESD79-4C: DDR4 SDRAM Standard}}, 2020.

\bibitem{kim2020revisiting}
Jeremie~S. Kim, Minesh Patel, Abdullah~Giray Ya\u{g}l{\i}k\c{c}{\i}, Hasan
  Hassan, Roknoddin Azizi, Lois Orosa, and Onur Mutlu.
\newblock {Revisiting RowHammer: An Experimental Analysis of Modern Devices and
  Mitigation Techniques}.
\newblock In {\em ISCA}, 2020.

\bibitem{dennard1968fieldeffect}
Robert~H. Dennard.
\newblock {Field-Effect Transistor Memory}.
\newblock US Patent 3387286A, 1968.

\bibitem{dennard1974design}
Robert~H Dennard, Fritz~H Gaensslen, Hwa-Nien Yu, V~Leo Rideout, Ernest
  Bassous, and Andre~R LeBlanc.
\newblock {Design of Ion-Implanted MOSFET's with Very Small Physical
  Dimensions}.
\newblock {\em JSSC}, 1974.

\bibitem{markoff2019ibms}
John Markoff.
\newblock {IBM's Robert H. Dennard and the Chip That Changed the World}.
\newblock
  \url{https://www.ibm.com/blogs/think/2019/11/ibms-robert-h-dennard-and-the-chip-that-changed-the-world/},
  2019.

\bibitem{electronics2018memory}
{Memory Lane}.
\newblock {\em {Nature Electronics}}, 2018.

\bibitem{hennessy2011computer}
John~L Hennessy and David~A Patterson.
\newblock {\em {Computer Architecture: A Quantitative Approach}}.
\newblock Elsevier, 2011.

\bibitem{meena2014overview}
Jagan~Singh Meena, Simon~Min Sze, Umesh Chand, and Tseung-Yuen Tseng.
\newblock {Overview Of Emerging Nonvolatile Memory Technologies}.
\newblock {\em Nanoscale Research Letters}, 2014.

\bibitem{mutlu2013memory}
Onur Mutlu.
\newblock {Memory Scaling: A Systems Architecture Perspective}.
\newblock In {\em {MemCon}}, 2013.

\bibitem{mutlu2015themain}
Onur Mutlu, Justin Meza, and Lavanya Subramanian.
\newblock {The Main Memory System: Challenges and Opportunities}.
\newblock {\em Communications of the KIISE}, 2015.

\bibitem{chang2017thesis}
Kevin~K. Chang.
\newblock {\em {Understanding and Improving Latency of DRAM-Based Memory
  Systems}}.
\newblock PhD thesis, Carnegie Mellon University, 2017.

\bibitem{kang2014coarchitecting}
Uksong Kang, Hak-Soo Yu, Churoo Park, Hongzhong Zheng, John Halbert, Kuljit
  Bains, S~Jang, and Joo~Sun Choi.
\newblock {Co-Architecting Controllers and DRAM to Enhance DRAM Process
  Scaling}.
\newblock In {\em {The Memory Forum}}, 2014.

\bibitem{mandelman2002challenges}
J~A Mandelman, R.~H. Dennard, G.~B. Bronner, J.~K. DeBrosse, R.~Divakaruni,
  Y~Li, and C.~J. Radens.
\newblock {Challenges and Future Directions for the Scaling of Dynamic
  Random-Access Memory (DRAM)}.
\newblock {\em IBM JRD}, 2002.

\bibitem{childers2015achieving}
Bruce~R Childers, Jun Yang, and Youtao Zhang.
\newblock {Achieving Yield, Density and Performance Effective DRAM at Extreme
  Technology Sizes}.
\newblock In {\em MEMSYS}, 2015.

\bibitem{mutlu2014research}
Onur Mutlu and Lavanya Subramanian.
\newblock {Research Problems and Opportunities in Memory Systems}.
\newblock {\em SUPERFRI}, 2014.

\bibitem{mutlu2017therowhammer}
Onur Mutlu.
\newblock {The RowHammer Problem and Other Issues We May Face as Memory Becomes
  Denser}.
\newblock In {\em DATE}, 2017.

\bibitem{mutlu2019rowhammer_a}
Onur Mutlu and Jeremie Kim.
\newblock {RowHammer: A Retrospective}.
\newblock {\em IEEE TCAD Special Issue on Top Picks in Hardware and Embedded
  Security}, 2019.

\bibitem{mutlu2019rowhammer_and}
Onur Mutlu.
\newblock {RowHammer and Beyond}.
\newblock In {\em COSADE}, 2019.

\bibitem{mutlu2023fundamentally}
Onur Mutlu, Ataberk Olgun, and A.~Giray Ya{\u{g}}l{\i}k{\c{c}}{\i}.
\newblock {Fundamentally Understanding and Solving RowHammer}.
\newblock In {\em ASP-DAC}, 2023.

\bibitem{aga2017when}
Misiker~Tadesse Aga, Zelalem~Birhanu Aweke, and Todd Austin.
\newblock {When Good Protections Go Bad: Exploiting Anti-DoS Measures to
  Accelerate Rowhammer Attacks}.
\newblock In {\em HOST}, 2017.

\bibitem{agarwal2018rowhammer_for}
S.~Agarwal, H.~Dixit, D.~Datta, M.~Tran, D.~Houssameddine, D.~Shum, and
  F.~Benistant.
\newblock {Rowhammer for Spin Torque based Memory: Problem or Not?}
\newblock In {\em INTERMAG}, 2018.

\bibitem{barenghi2018softwareonly}
Alessandro Barenghi, Luca Breveglieri, Niccol{\`o} Izzo, and Gerardo Pelosi.
\newblock {Software-Only Reverse Engineering of Physical DRAM Mappings for
  Rowhammer Attacks}.
\newblock In {\em IVSW}, 2018.

\bibitem{bhattacharya2016curious}
Sarani Bhattacharya and Debdeep Mukhopadhyay.
\newblock {Curious Case of RowHammer: Flipping Secret Exponent Bits Using
  Timing Analysis}.
\newblock In {\em CHES}, 2016.

\bibitem{bhattacharya2018advanced}
Sarani Bhattacharya and Debdeep Mukhopadhyay.
\newblock {Advanced Fault Attacks in Software: Exploiting the Rowhammer Bug}.
\newblock In {\em Fault Tolerant Architectures for Cryptography and Hardware
  Security}. 2018.

\bibitem{bosman2016dedup}
Erik Bosman, Kaveh Razavi, Herbert Bos, and Cristiano Giuffrida.
\newblock {Dedup Est Machina: Memory Deduplication as an Advanced Exploitation
  Vector}.
\newblock In {\em IEEE S\&P}, 2016.

\bibitem{brasser2017cant}
Ferdinand Brasser, Lucas Davi, David Gens, Christopher Liebchen, and Ahmad-Reza
  Sadeghi.
\newblock {Can't Touch This: Software-Only Mitigation Against Rowhammer Attacks
  Targeting Kernel Memory}.
\newblock In {\em USENIX Security}, 2017.

\bibitem{burleson2016invited}
Wayne Burleson, Onur Mutlu, and Mohit Tiwari.
\newblock {Invited: Who is the Major Threat to Tomorrow's Security? You, the
  Hardware Designer}.
\newblock In {\em DAC}, 2016.

\bibitem{carre2018openssl}
Sebastien Carre, Matthieu Desjardins, Adrien Facon, and Sylvain Guilley.
\newblock {OpenSSL Bellcore's Protection Helps Fault Attack}.
\newblock In {\em DSD}, 2018.

\bibitem{cohen2022hammerscope}
Yaakov Cohen, Kevin~Sam Tharayil, Arie Haenel, Daniel Genkin, Angelos~D
  Keromytis, Yossi Oren, and Yuval Yarom.
\newblock {HammerScope: Observing DRAM Power Consumption Using Rowhammer}.
\newblock In {\em CCS}, 2022.

\bibitem{cojocar2019exploiting}
Lucian Cojocar, Kaveh Razavi, Cristiano Giuffrida, and Herbert Bos.
\newblock {Exploiting Correcting Codes: On The Effectiveness Of ECC Memory
  Against Rowhammer Attacks}.
\newblock In {\em IEEE S\&P}, 2019.

\bibitem{cojocar2020arewe}
Lucian Cojocar, Jeremie Kim, Minesh Patel, Lillian Tsai, Stefan Saroiu, Alec
  Wolman, and Onur Mutlu.
\newblock {Are We Susceptible to Rowhammer? An End-to-End Methodology for Cloud
  Providers}.
\newblock In {\em IEEE S\&P}, 2020.

\bibitem{deridder2021smash}
Finn de~Ridder, Pietro Frigo, Emanuele Vannacci, Herbert Bos, Cristiano
  Giuffrida, and Kaveh Razavi.
\newblock {SMASH: Synchronized Many-Sided Rowhammer Attacks from JavaScript}.
\newblock In {\em {USENIX Security}}, 2021.

\bibitem{fahrjr2022when}
Michael Fahr~Jr, Hunter Kippen, Andrew Kwong, Thinh Dang, Jacob Lichtinger,
  Dana Dachman-Soled, Daniel Genkin, Alexander Nelson, Ray Perlner, Arkady
  Yerukhimovich, and Daniel Apon.
\newblock {When Frodo Flips: End-to-End Key Recovery on FrodoKEM via
  Rowhammer}.
\newblock {\em CCS}, 2022.

\bibitem{fournaris2017exploiting}
Apostolos~P Fournaris, Lidia Pocero~Fraile, and Odysseas Koufopavlou.
\newblock {Exploiting Hardware Vulnerabilities to Attack Embedded System
  Devices: A Survey of Potent Microarchitectural Attacks}.
\newblock {\em Electronics}, 2017.

\bibitem{frigo2018grand}
Pietro Frigo, Cristiano Giuffrida, Herbert Bos, and Kaveh Razavi.
\newblock {Grand Pwning Unit: Accelerating Microarchitectural Attacks with the
  GPU}.
\newblock In {\em {IEEE S\&P}}, 2018.

\bibitem{frigo2020trrespass}
Pietro Frigo, Emanuele Vannacci, Hasan Hassan, Victor {van der Veen}, Onur
  Mutlu, Cristiano Giuffrida, Herbert Bos, and Kaveh Razavi.
\newblock {TRRespass: Exploiting the Many Sides of Target Row Refresh}.
\newblock In {\em {IEEE S\&P}}, 2020.

\bibitem{genssler2022onthe}
Paul~R. Genssler, Victor~M. van Santen, Jörg Henkel, and Hussam Amrouch.
\newblock {On the Reliability of FeFET On-Chip Memory}.
\newblock {\em TC}, 2022.

\bibitem{gruss2015rowhammerjs_a}
Daniel Gruss, Cl{\'e}mentine Maurice, and Stefan Mangard.
\newblock {Rowhammer.js: A Remote Software-Induced Fault Attack in JavaScript}.
\newblock {\em CoRR}, 2015.

\bibitem{gruss2018another}
Daniel Gruss, Moritz Lipp, Michael Schwarz, Daniel Genkin, Jonas Juffinger,
  Sioli O'Connell, Wolfgang Schoechl, and Yuval Yarom.
\newblock {Another Flip in the Wall of Rowhammer Defenses}.
\newblock In {\em IEEE S\&P}, 2018.

\bibitem{hassan2021uncovering}
Hasan Hassan, Yahya~Can Tugrul, Jeremie~S. Kim, Victor van~der Veen, Kaveh
  Razavi, and Onur Mutlu.
\newblock {Uncovering In-DRAM RowHammer Protection Mechanisms: A New
  Methodology, Custom RowHammer Patterns, and Implications}.
\newblock In {\em MICRO}, 2021.

\bibitem{hong2019terminal}
Sanghyun Hong, Pietro Frigo, Yi\u{g}itcan Kaya, Cristiano Giuffrida, and Tudor
  Dumitras.
\newblock {Terminal Brain Damage: Exposing the Graceless Degradation in Deep
  Neural Networks Under Hardware Fault Attacks}.
\newblock In {\em USENIX Security}, 2019.

\bibitem{jang2017sgxbomb}
Yeongjin Jang, Jaehyuk Lee, Sangho Lee, and Taesoo Kim.
\newblock {SGX-Bomb: Locking Down the Processor via Rowhammer Attack}.
\newblock In {\em SysTEX}, 2017.

\bibitem{jattke2022blacksmith}
Patrick Jattke, Victor van~der Veen, Pietro Frigo, Stijn Gunter, and Kaveh
  Razavi.
\newblock {Blacksmith: Scalable Rowhammering in the Frequency Domain}.
\newblock In {\em {IEEE S\&P}}, 2022.

\bibitem{ji2019pinpoint}
Sangwoo Ji, Youngjoo Ko, Saeyoung Oh, and Jong Kim.
\newblock {Pinpoint Rowhammer: Suppressing Unwanted Bit Flips on Rowhammer
  Attacks}.
\newblock In {\em ASIACCS}, 2019.

\bibitem{khan2018analysis}
Mohammad Nasim~Imtiaz Khan and Swaroop Ghosh.
\newblock {Analysis of Row Hammer Attack on STTRAM}.
\newblock In {\em ICCD}, 2018.

\bibitem{kogler2022halfdouble}
Andreas Kogler, Jonas Juffinger, Salman Qazi, Yoongu Kim, Moritz Lipp, Nicolas
  Boichat, Eric Shiu, Mattias Nissler, and Daniel Gruss.
\newblock {Half-Double: Hammering From the Next Row Over}.
\newblock In {\em USENIX Security}, 2022.

\bibitem{kwong2020rambleed}
Andrew Kwong, Daniel Genkin, Daniel Gruss, and Yuval Yarom.
\newblock {RAMBleed: Reading Bits in Memory Without Accessing Them}.
\newblock In {\em IEEE S\&P}, 2020.

\bibitem{li2014write}
Haitong Li, Hong-Yu Chen, Zhe Chen, Bing Chen, Rui Liu, Gang Qiu, Peng Huang,
  Feifei Zhang, Zizhen Jiang, Bin Gao, Lifeng Liu, Xiaoyan Liu, Shimeng Yu,
  H.-S.~Philip Wong, and Jinfeng Kang.
\newblock {Write Disturb Analyses on Half-Selected Cells of Cross-Point RRAM
  Arrays}.
\newblock In {\em IRPS}, 2014.

\bibitem{lipp2018nethammer}
Moritz Lipp, Misiker~Tadesse Aga, Michael Schwarz, Daniel Gruss, Cl{\'e}mentine
  Maurice, Lukas Raab, and Lukas Lamster.
\newblock {Nethammer: Inducing Rowhammer Faults Through Network Requests}.
\newblock In {\em EuroS\&PW}, 2020.

\bibitem{liu2022generating}
Liang Liu, Yanan Guo, Yueqiang Cheng, Youtao Zhang, and Jun Yang.
\newblock {Generating Robust DNN with Resistance to Bit-Flip Based Adversarial
  Weight Attack}.
\newblock {\em IEEE TC}, 2022.

\bibitem{ni2018write}
Kai Ni, Xueqing Li, Jeffrey~A. Smith, Matthew Jerry, and Suman Datta.
\newblock {Write Disturb in Ferroelectric FETs and Its Implication for 1T-FeFET
  AND Memory Arrays}.
\newblock {\em IEEE EDL}, 2018.

\bibitem{orosa2021adeeper}
Lois Orosa, A~Giray Ya{\u{g}}l{\i}k{{c}}{\i}, Haocong Luo, Ataberk Olgun,
  Jisung Park, Hasan Hassan, Minesh Patel, Jeremie~S. Kim, and Onur Mutlu.
\newblock {A Deeper Look into RowHammer's Sensitivities: Experimental Analysis
  of Real DRAM Chips and Implications on Future Attacks and Defenses}.
\newblock In {\em MICRO}, 2021.

\bibitem{orosa2022spyhammer}
Lois Orosa, Ulrich R{\"u}hrmair, A~Giray Ya{\u{g}}l{\i}k{\c{c}}{\i}, Haocong
  Luo, Ataberk Olgun, Patrick Jattke, Minesh Patel, Jeremie Kim, Kaveh Razavi,
  and Onur Mutlu.
\newblock {SpyHammer: Using RowHammer to Remotely Spy on Temperature}.
\newblock {\em IEEE Access}, 2024.

\bibitem{park2016experiments}
Kyungbae Park, Chulseung Lim, Donghyuk Yun, and Sanghyeon Baeg.
\newblock {Experiments and Root Cause Analysis for Active-Precharge Hammering
  Fault in DDR3 SDRAM under 3$\times$ nm Technology}.
\newblock {\em Microelectronics Reliability}, 2016.

\bibitem{lim2017active}
Chulseung Lim, Kyungbae Park, and Sanghyeon Baeg.
\newblock {Active Precharge Hammering to Monitor Displacement Damage Using
  High-Energy Protons in 3x-nm SDRAM}.
\newblock {\em TNS}, 2017.

\bibitem{park2016statistical}
Kyungbae Park, Donghyuk Yun, and Sanghyeon Baeg.
\newblock {Statistical Distributions of Row-Hammering Induced Failures in DDR3
  Components}.
\newblock {\em Microelectronics Reliability}, 2016.

\bibitem{pessl2016drama_exploiting}
Peter Pessl, Daniel Gruss, Cl{\'e}mentine Maurice, Michael Schwarz, and Stefan
  Mangard.
\newblock {DRAMA: Exploiting DRAM Addressing for Cross-CPU Attacks}.
\newblock In {\em USENIX Security}, 2016.

\bibitem{poddebniak2018attacking}
Damian Poddebniak, Juraj Somorovsky, Sebastian Schinzel, Manfred Lochter, and
  Paul R{\"o}sler.
\newblock {Attacking Deterministic Signature Schemes Using Fault Attacks}.
\newblock In {\em EuroS\&P}, 2018.

\bibitem{qiao2016anew}
Rui Qiao and Mark Seaborn.
\newblock {A New Approach for RowHammer Attacks}.
\newblock In {\em HOST}, 2016.

\bibitem{rakin2022deepsteal}
Adnan~Siraj Rakin, Md~Hafizul~Islam Chowdhuryy, Fan Yao, and Deliang Fan.
\newblock {DeepSteal: Advanced Model Extractions Leveraging Efficient Weight
  Stealing in Memories}.
\newblock In {\em IEEE S\&P}, 2022.

\bibitem{razavi2016flip}
Kaveh Razavi, Ben Gras, Erik Bosman, Bart Preneel, Cristiano Giuffrida, and
  Herbert Bos.
\newblock {Flip Feng Shui: Hammering a Needle in the Software Stack}.
\newblock In {\em USENIX Security}, 2016.

\bibitem{ryu2017overcoming}
Seong-Wan Ryu, Kyungkyu Min, Jungho Shin, Heimi Kwon, Donghoon Nam, Taekyung
  Oh, Tae-Su Jang, Minsoo Yoo, Yongtaik Kim, and Sungjoo Hong.
\newblock {Overcoming the Reliability Limitation in the Ultimately Scaled DRAM
  Using Silicon Migration Technique by Hydrogen Annealing}.
\newblock In {\em IEDM}, 2017.

\bibitem{safari2014rowhammer}
{SAFARI Research Group}.
\newblock {RowHammer --- GitHub Repository}.
\newblock \url{https://github.com/CMU-SAFARI/rowhammer}, 2014.

\bibitem{seaborn2015exploiting}
Mark Seaborn and Thomas Dullien.
\newblock {Exploiting the DRAM Rowhammer Bug to Gain Kernel Privileges}.
\newblock
  \url{http://googleprojectzero.blogspot.com.tr/2015/03/exploiting-dram-rowhammer-bug-to-gain.html},
  2015.

\bibitem{tatar2018defeating}
Andrei Tatar, Cristiano Giuffrida, Herbert Bos, and Kaveh Razavi.
\newblock {Defeating Software Mitigations Against Rowhammer: A Surgical
  Precision Hammer}.
\newblock In {\em RAID}, 2018.

\bibitem{tatar2018throwhammer}
Andrei Tatar, Radhesh~Krishnan Konoth, Elias Athanasopoulos, Cristiano
  Giuffrida, Herbert Bos, and Kaveh Razavi.
\newblock {Throwhammer: Rowhammer Attacks Over the Network and Defenses}.
\newblock In {\em {USENIX} {ATC}}, 2018.

\bibitem{tobah2022spechammer}
Youssef Tobah, Andrew Kwong, Ingab Kang, Daniel Genkin, and Kang~G. Shin.
\newblock {SpecHammer: Combining Spectre and Rowhammer for New Speculative
  Attacks}.
\newblock In {\em IEEE S\&P}, 2022.

\bibitem{tol2022toward}
M~Caner Tol, Saad Islam, Berk Sunar, and Ziming Zhang.
\newblock {Toward Realistic Backdoor Injection Attacks on DNNs Using
  RowHammer}.
\newblock {arXiv:2110.07683 [cs.LG]}, 2022.

\bibitem{tol2023dont}
M~Caner Tol, Saad Islam, Andrew~J Adiletta, Berk Sunar, and Ziming Zhang.
\newblock {Don't Knock! Rowhammer at the Backdoor of DNN Models}.
\newblock In {\em DSN}, 2023.

\bibitem{vanderveen2016drammer_deterministic}
Victor {van der Veen}, Yanick Fratantonio, Martina Lindorfer, Daniel Gruss,
  Clementine Maurice, Giovanni Vigna, Herbert Bos, Kaveh Razavi, and Cristiano
  Giuffrida.
\newblock {Drammer: Deterministic Rowhammer Attacks on Mobile Platforms}.
\newblock In {\em CCS}, 2016.

\bibitem{vanderveen2018guardion}
Victor van~der Veen, Martina Lindorfer, Yanick Fratantonio,
  Harikrishnan~Padmanabha Pillai, Giovanni Vigna, Christopher Kruegel, Herbert
  Bos, and Kaveh Razavi.
\newblock {GuardION: Practical Mitigation of DMA-Based Rowhammer Attacks on
  ARM}.
\newblock In {\em DIMVA}, 2018.

\bibitem{walker2021ondram}
Andrew~J. Walker, Sungkwon Lee, and Dafna Beery.
\newblock {On DRAM RowHammer and the Physics on Insecurity}.
\newblock {\em IEEE TED}, 2021.

\bibitem{weissman2020jackhammer}
Zane Weissman, Thore Tiemann, Daniel Moghimi, Evan Custodio, Thomas Eisenbarth,
  and Berk Sunar.
\newblock {JackHammer: Efficient Rowhammer on Heterogeneous FPGA--CPU
  Platforms}.
\newblock arXiv:1912.11523 [cs.CR], 2020.

\bibitem{xiao2016onebit}
Yuan Xiao, Xiaokuan Zhang, Yinqian Zhang, and Radu Teodorescu.
\newblock {One Bit Flips, One Cloud Flops: Cross-VM Row Hammer Attacks and
  Privilege Escalation}.
\newblock In {\em USENIX Security}, 2016.

\bibitem{yaglikci2022understanding}
A.~Giray Ya{\u{g}}l{\i}k{\c{c}}{\i}, Haocong Luo, Geraldo~F Oliveira, Ataberk
  Olgun, Minesh Patel, Jisung Park, Hasan Hassan, Jeremie~S Kim, Lois Orosa,
  and Onur Mutlu.
\newblock {Understanding RowHammer Under Reduced Wordline Voltage: An
  Experimental Study Using Real DRAM Devices}.
\newblock In {\em DSN}, 2022.

\bibitem{yang2019trapassisted}
Thomas Yang and Xi-Wei Lin.
\newblock {Trap-Assisted DRAM Row Hammer Effect}.
\newblock {\em EDL}, 2019.

\bibitem{yao2020deephammer}
Fan Yao, Adnan~Siraj Rakin, and Deliang Fan.
\newblock {Deephammer: Depleting the Intelligence of Deep Neural Networks
  Through Targeted Chain of Bit Flips}.
\newblock In {\em USENIX Security}, 2020.

\bibitem{yun2018study}
Donghyuk Yun, Myungsang Park, Chulseung Lim, and Sanghyeon Baeg.
\newblock {Study of TID Effects on One Row Hammering Using Gamma in DDR4
  SDRAMs}.
\newblock In {\em IRPS}, 2018.

\bibitem{zhang2018triggering}
Zhenkai Zhang, Zihao Zhan, Daniel Balasubramanian, Xenofon Koutsoukos, and
  Gabor Karsai.
\newblock {Triggering Rowhammer Hardware Faults on ARM: A Revisit}.
\newblock In {\em ASHES}, 2018.

\bibitem{zhang2020pthammer}
Zhi Zhang, Yueqiang Cheng, Dongxi Liu, Surya Nepal, Zhi Wang, and Yuval Yarom.
\newblock {PTHammer: Cross-User-Kernel-Boundary Rowhammer Through Implicit
  Accesses}.
\newblock In {\em MICRO}, 2020.

\bibitem{zhang2022implicit}
Zhi Zhang, Wei He, Yueqiang Cheng, Wenhao Wang, Yansong Gao, Dongxi Liu, Kang
  Li, Surya Nepal, Anmin Fu, and Yi~Zou.
\newblock {Implicit Hammer: Cross-Privilege-Boundary Rowhammer through Implicit
  Accesses}.
\newblock {\em IEEE TDSC}, 2022.

\bibitem{zheng2023trojvit}
Mengxin Zheng, Qian Lou, and Lei Jiang.
\newblock {TrojViT: Trojan Insertion in Vision Transformers}.
\newblock In {\em CVPR}, 2023.

\bibitem{aydin2022cyber}
Hakan Aydin and Ahmet Sertba{\c{s}}.
\newblock {Cyber Security in Industrial Control Systems (ICS): A Survey of
  RowHammer Vulnerability}.
\newblock {\em Applied Computer Science}, 2022.

\bibitem{mus2022jolt}
Koksal Mus, Yark{\i}n Dor{\"o}z, M~Caner Tol, Kristi Rahman, and Berk Sunar.
\newblock {Jolt: Recovering TLS Signing Keys via Rowhammer Faults}.
\newblock {\em Cryptology ePrint Archive}, 2022.

\bibitem{wang2022research}
Jianxin Wang, Hongke Xu, Chaoen Xiao, Lei Zhang, and Yuzheng Zheng.
\newblock {Research and Implementation of Rowhammer Attack Method Based on
  Domestic NeoKylin Operating System}.
\newblock In {\em ICFTIC}, 2022.

\bibitem{lefforge2023reverse}
Sam Lefforge.
\newblock {Reverse Engineering Post-Quantum Cryptography Schemes to Find
  Rowhammer Exploits}.
\newblock Master's thesis, University of Arkansas, 2023.

\bibitem{fahr2022theeffects}
Michael~Jacob Fahr.
\newblock {The Effects of Side-Channel Attacks on Post-Quantum Cryptography:
  Influencing FrodoKEM Key Generation Using the Rowhammer Exploit}.
\newblock Master's thesis, University of Arkansas, Fayetteville, 2022.

\bibitem{kaur2022workinprogress}
Anandpreet Kaur, Pravin Srivastav, and Bibhas Ghoshal.
\newblock {Work-in-Progress: DRAM-MaUT: DRAM Address Mapping Unveiling Tool for
  ARM Devices}.
\newblock In {\em CASES}, 2022.

\bibitem{cai2022onthe}
Kunbei Cai, Zhenkai Zhang, and Fan Yao.
\newblock {On the Feasibility of Training-Time Trojan Attacks through
  Hardware-Based Faults in Memory}.
\newblock In {\em HOST}, 2022.

\bibitem{li2022cyberradar}
Dawei Li, Di~Liu, Yangkun Ren, Ziyi Wang, Yu~Sun, Zhenyu Guan, Qianhong Wu, and
  Jianwei Liu.
\newblock {CyberRadar: A PUF-Based Detecting and Mapping Framework for Physical
  Devices}.
\newblock arXiv:2201.07597 [cs.CR], 2022.

\bibitem{li2023fphammer}
Dawei Li, Di~Liu, Yangkun Ren, Ziyi Wang, Yu~Sun, Zhenyu Guan, Qianhong Wu, and
  Jianwei Liu.
\newblock {FPHammer: A Device Identification Framework based on DRAM
  Fingerprinting}.
\newblock In {\em TrustCom}, 2023.

\bibitem{roohi2022efficient}
Arman Roohi and Shaahin Angizi.
\newblock {Efficient Targeted Bit-Flip Attack Against the Local Binary Pattern
  Network}.
\newblock In {\em HOST}, 2022.

\bibitem{staudigl2022neurohammer}
Felix Staudigl, Hazem Al~Indari, Daniel Sch{\"o}n, Dominik Sisejkovic, Farhad
  Merchant, Jan~Moritz Joseph, Vikas Rana, Stephan Menzel, and Rainer Leupers.
\newblock {NeuroHammer: Inducing Bit-Flips in Memristive Crossbar Memories}.
\newblock In {\em DATE}, 2022.

\bibitem{yang2022sociallyaware}
Li-Hsing Yang, Shin-Shan Huang, Tsai-Ling Cheng, Yi-Ching Kuo, and Jian-Jhih
  Kuo.
\newblock {Socially-Aware Collaborative Defense System against Bit-Flip Attack
  in Social Internet of Things and Its Online Assignment Optimization}.
\newblock In {\em ICCCN}, 2022.

\bibitem{islam2022signature}
Saad Islam, Koksal Mus, Richa Singh, Patrick Schaumont, and Berk Sunar.
\newblock {Signature Correction Attack on Dilithium Signature Scheme}.
\newblock In {\em Euro S\&P}, 2022.

\bibitem{tomita2022extracting}
Chihiro Tomita, Makoto Takita, Kazuhide Fukushima, Yuto Nakano, Yoshiaki
  Shiraishi, and Masakatu Morii.
\newblock {Extracting the Secrets of OpenSSL with RAMBleed}.
\newblock {\em Sensors}, 2022.

\bibitem{france2022modeling}
Lo{\"\i}c France, Florent Bruguier, Maria Mushtaq, David Novo, and Pascal
  Benoit.
\newblock {Modeling Rowhammer in the gem5 Simulator}.
\newblock In {\em CHES}, 2022.

\bibitem{kurmus2017from}
Anil Kurmus, Nikolas Ioannou, Matthias Neugschwandtner, Nikolaos Papandreou,
  and Thomas Parnell.
\newblock {From Random Block Corruption to Privilege Escalation: A Filesystem
  Attack Vector for RowHammer-like Attacks}.
\newblock In {\em USENIX WOOT}, 2017.

\bibitem{silberschatz2018operating}
Abraham Silberschatz, Peter~B. Galvin, and Greg Gagne.
\newblock {\em {Operating System Concepts}}.
\newblock Wiley, 2018.

\bibitem{luo2023rowpress}
Haocong Luo, Ataberk Olgun, Abdullah~Giray Ya\u{g}l{\i}k\c{c}{\i}, Yahya~Can
  Tu\u{g}rul, Steve Rhyner, Meryem~Banu Cavlak, Joël Lindegger, Mohammad
  Sadrosadati, and Onur Mutlu.
\newblock {RowPress: Amplifying Read Disturbance in Modern DRAM Chips}.
\newblock In {\em ISCA}, 2023.

\bibitem{mutlu2018rowhammer}
Onur Mutlu.
\newblock {RowHammer}.
\newblock
  \url{https://people.inf.ethz.ch/omutlu/pub/onur-Rowhammer-TopPicksinHardwareEmbeddedSecurity-November-8-2018.pdf},
  2018.

\bibitem{park2020graphene}
Yeonhong Park, Woosuk Kwon, Eojin Lee, Tae~Jun Ham, Jung~Ho Ahn, and Jae~W Lee.
\newblock {Graphene: Strong yet Lightweight Row Hammer Protection}.
\newblock In {\em MICRO}, 2020.

\bibitem{lenovo2015rowhammer}
{Lenovo}.
\newblock {Row Hammer Privilege Escalation}.
\newblock \url{https://support.lenovo.com/us/en/product_security/row_hammer},
  2015.

\bibitem{enterprise2015hpmoonshot}
{Hewlett-Packard Enterprise}.
\newblock {HP Moonshot Component Pack Version 2015.05.0}.
\newblock
  \url{http://h17007.www1.hp.com/us/en/enterprise/servers/products/moonshot/component-pack/index.aspx},
  2015.

\bibitem{lee2019twice}
Eojin Lee, Ingab Kang, Sukhan Lee, G.~Edward Suh, and Jung~Ho Ahn.
\newblock {TWiCe: Preventing Row-Hammering by Exploiting Time Window Counters}.
\newblock In {\em ISCA}, 2019.

\bibitem{seyedzadeh2017counterbased}
Seyed~Mohammad Seyedzadeh, Alex~K. Jones, and Rami Melhem.
\newblock {Counter-Based Tree Structure for Row Hammering Mitigation in DRAM}.
\newblock {\em IEEE CAL}, 2017.

\bibitem{seyedzadeh2017mitigating}
Seyed~Mohammad Seyedzadeh, Donald Kline~Jr, Alex~K Jones, and Rami Melhem.
\newblock {Mitigating Bitline Crosstalk Noise in DRAM Memories}.
\newblock In {\em MEMSYS}, 2017.

\bibitem{seyedzadeh2018mitigating}
S.~M. {Seyedzadeh}, A.~K. {Jones}, and R.~{Melhem}.
\newblock {Mitigating Wordline Crosstalk Using Adaptive Trees of Counters}.
\newblock In {\em ISCA}, 2018.

\bibitem{vig2018rapid}
Saru Vig, Sarani Bhattacharya, Debdeep Mukhopadhyay, and Siew-Kei Lam.
\newblock {Rapid Detection of Rowhammer Attacks Using Dynamic Skewed Hash
  Tree}.
\newblock In {\em HASP}, 2018.

\bibitem{irazoqui2016mascat}
Gorka Irazoqui, Thomas Eisenbarth, and Berk Sunar.
\newblock {MASCAT: Stopping Microarchitectural Attacks Before Execution}.
\newblock {\em IACR Cryptology}, 2016.

\bibitem{kang2020cattwo}
Ingab Kang, Eojin Lee, and Jung~Ho Ahn.
\newblock {CAT-TWO: Counter-Based Adaptive Tree, Time Window Optimized for DRAM
  Row-Hammer Prevention}.
\newblock {\em {IEEE} Access}, 2020.

\bibitem{kim2022mithril}
Michael~Jaemin Kim, Jaehyun Park, Yeonhong Park, Wanju Doh, Namhoon Kim,
  Tae~Jun Ham, Jae~W Lee, and Jung~Ho Ahn.
\newblock {Mithril: Cooperative Row Hammer Protection on Commodity DRAM
  Leveraging Managed Refresh}.
\newblock In {\em HPCA}, 2022.

\bibitem{kim2014architectural}
Dae-Hyun Kim, Prashant~J Nair, and Moinuddin~K Qureshi.
\newblock {Architectural Support for Mitigating Row Hammering in DRAM
  Memories}.
\newblock {\em IEEE CAL}, 2014.

\bibitem{bains2015method}
Kuljit~S Bains, John~B Halbert, Suneeta Sah, and Zvika Greenfield.
\newblock {Method, Apparatus and System for Providing a Memory Refresh}.
\newblock US Patent: 9,030,903, 2015.

\bibitem{bains2016distributed}
Kuljit~S Bains and John~B Halbert.
\newblock {Distributed Row Hammer Tracking}.
\newblock US Patent: 9,299,400, 2016.

\bibitem{bains2016rowhammer}
Kuljit~S Bains and John~B Halbert.
\newblock {Row Hammer Monitoring Based on Stored Row Hammer Threshold Value}.
\newblock US Patent: 10,083,737, 2016.

\bibitem{aweke2016anvil}
Zelalem~Birhanu Aweke, Salessawi~Ferede Yitbarek, Rui Qiao, Reetuparna Das,
  Matthew Hicks, Yossi Oren, and Todd Austin.
\newblock {ANVIL: Software-Based Protection Against Next-Generation Rowhammer
  Attacks}.
\newblock In {\em ASPLOS}, 2016.

\bibitem{apple2015about}
{Apple Inc.}
\newblock {About the Security Content of Mac EFI Security Update 2015-001}.
\newblock \url{https://support.apple.com/en-us/HT204934}, 2015.

\bibitem{son2017making}
Mungyu Son, Hyunsun Park, Junwhan Ahn, and Sungjoo Yoo.
\newblock {Making DRAM Stronger Against Row Hammering}.
\newblock In {\em DAC}, 2017.

\bibitem{you2019mrloc}
Jung~Min You and Joon-Sung Yang.
\newblock {MRLoc: Mitigating Row-Hammering Based on Memory Locality}.
\newblock In {\em DAC}, 2019.

\bibitem{yaglikci2021security}
A.~Giray Ya{\u{g}}l{\i}k{\c{c}}{\i}, Jeremie~S. Kim, Fabrice Devaux, and Onur
  Mutlu.
\newblock {Security Analysis of the Silver Bullet Technique for RowHammer
  Prevention}.
\newblock arXiv:2106.07084 [cs.CR], 2021.

\bibitem{loughlin2021stop}
Kevin Loughlin, Stefan Saroiu, Alec Wolman, and Baris Kasikci.
\newblock {Stop! Hammer Time: Rethinking Our Approach to Rowhammer
  Mitigations}.
\newblock In {\em HotOS}, 2021.

\bibitem{devaux2021method}
Fabrice Devaux and Renaud Ayrignac.
\newblock {Method and Circuit for Protecting a DRAM Memory Device from the Row
  Hammer Effect}.
\newblock US Patent: 10,885,966, 2021.

\bibitem{wang2021discreetpara}
Yicheng Wang, Yang Liu, Peiyun Wu, and Zhao Zhang.
\newblock {Discreet-PARA: Rowhammer Defense with Low Cost and High Efficiency}.
\newblock In {\em ICCD}, 2021.

\bibitem{marazzi2023protrr}
Michele Marazzi, Flavien Solt, Patrick Jattke, Kubo Takashi, and Kaveh Razavi.
\newblock {ProTRR: Principled yet Optimal In-DRAM Target Row Refresh}.
\newblock In {\em {IEEE S\&P}}, 2023.

\bibitem{zhang2022softtrr}
Zhi Zhang, Yueqiang Cheng, Minghua Wang, Wei He, Wenhao Wang, Surya Nepal,
  Yansong Gao, Kang Li, Zhe Wang, and Chenggang Wu.
\newblock {SoftTRR: Protect Page Tables against Rowhammer Attacks Using
  Software-Only Target Row Refresh}.
\newblock In {\em USENIX ATC}, 2022.

\bibitem{joardar2022learning}
Biresh~Kumar Joardar, Tyler~K Bletsch, and Krishnendu Chakrabarty.
\newblock {Learning to Mitigate RowHammer Attacks}.
\newblock In {\em DATE}, 2022.

\bibitem{yaglikci2022hira}
A~Giray Ya{\u{g}}lik{c}i, Ataberk Olgun, Minesh Patel, Haocong Luo, Hasan
  Hassan, Lois Orosa, O{\u{g}}uz Ergin, and Onur Mutlu.
\newblock {HiRA: Hidden Row Activation for Reducing Refresh Latency of
  Off-the-Shelf DRAM Chips}.
\newblock In {\em MICRO}, 2022.

\bibitem{saroiu2022howto}
Stefan Saroiu and Alec Wolman.
\newblock {How to Configure Row-Sampling-Based Rowhammer Defenses}.
\newblock {\em DRAMSec}, 2022.

\bibitem{bostanci2024comet}
F.~Nisa Bostancı, Ismail~Emir Yüksel, Ataberk Olgun, Konstantinos
  Kanellopoulos, Yahya~Can Tugrul, A.~Giray Yaglıkçı, Mohammad Sadrosadati,
  and Onur Mutlu.
\newblock {CoMeT: Count-Min-Sketch-Based Row Tracking to Mitigate RowHammer at
  Low Cost}.
\newblock In {\em HPCA}, 2024.

\bibitem{olgun2024abacus}
Ataberk Olgun, Yahya~Can Tugrul, F.~Nisa Bostancı, Ismail~Emir Yüksel,
  Haocong Luo, Steve Rhyner, A.~Giray Yaglıkçı, Geraldo~F. Oliveira, and
  Onur Mutlu.
\newblock {ABACuS: All-Bank Activation Counters for Scalable and Low Overhead
  RowHammer Mitigation}.
\newblock In {\em USENIX Security}, 2024.

\bibitem{joardar2022machine}
Biresh~Kumar Joardar, Tyler~K. Bletsch, and Krishnendu Chakrabarty.
\newblock {Machine Learning-Based Rowhammer Mitigation}.
\newblock {\em TCAD}, 2022.

\bibitem{vanderveen2024dynamic}
Victor van~der Veen, Pankaj Deshmukh, Behnam Dashtipour, and David Hartley.
\newblock {Dynamic Rowhammer Management}.
\newblock US Patent App. 17/940,430, 2024.

\bibitem{vanderveen2024dram}
Victor van~der Veen, Pankaj Deshmukh, Behnam Dashtipour, David Hartley, and
  Mosaddiq Saifuddin.
\newblock {Dynamic Random Access Memory (DRAM) Row Hammering Mitigation}.
\newblock US Patent App. 17/890,022, 2024.

\bibitem{jedec2008ddr3}
JEDEC.
\newblock {\em {DDR3 SDRAM Specification}}, 2008.

\bibitem{jedec2008jesd79f}
{{JEDEC}}.
\newblock {\em {JESD79F: Double Data Rate (DDR) SDRAM Standard}}, 2008.

\bibitem{jedec2012ddr4}
JEDEC.
\newblock {\em {DDR4 SDRAM Specification}}, 2012.

\bibitem{jedec2010jesd218}
JEDEC.
\newblock {\em {JESD218: Solid-State Drive (SSD) Requirements and Endurance
  Test Method}}, 2010.

\bibitem{jedec2010jesd219}
JEDEC.
\newblock {\em {JESD219: Solid-State Drive (SSD) Endurance Workloads}}, 2010.

\bibitem{jedec2013standard}
{JEDEC}.
\newblock {\em {Standard No. 21C. DDR3 SDRAM Unbuffered DIMM Design
  Specification}}, 2013.

\bibitem{jedec2014lowpower}
JEDEC.
\newblock {Low Power Double Data Rate 4 (LPDDR4) SDRAM Specification}.
\newblock {\em JEDEC Standard JESD209--4B}, 2014.

\bibitem{jedec2020jesd795}
{JEDEC}.
\newblock {\em {JESD79-5: DDR5 SDRAM Standard}}, 2020.

\bibitem{jedec2021jesd250c}
{JEDEC}.
\newblock {\em {JESD250C: Graphics Double Data Rate 6 (GDDR6) Standard}},
  {2021}.

\bibitem{jedec2016jesd232a}
{JEDEC}.
\newblock {\em {JESD232A: Graphics Double Data Rate (GDDR5X) Standard}},
  {2016}.

\bibitem{jedec2022jesd238}
JEDEC.
\newblock {High Bandwidth Memory DRAM (HBM3)}.
\newblock {\em JEDEC Standard JESD238}, 2022.

\bibitem{jedec2021jesd235d}
{JEDEC}.
\newblock {\em {JESD235D: High Bandwidth Memory DRAM (HBM1, HBM2)}}, 2021.

\bibitem{micron2014tn4003}
{Micron}.
\newblock {TN-40-03: DDR4 Networking Design Guide}, 2014.

\bibitem{micron2014sdram}
{Micron}.
\newblock {\em {SDRAM, 4Gb: x4, x8, x16 DDR4 SDRAM Features}}, 2014.

\bibitem{smith1981laser}
R.T. Smith, J.D. Chlipala, J.F.M. Bindels, R.G. Nelson, F.H. Fischer, and T.F.
  Mantz.
\newblock {Laser Programmable Redundancy and Yield Improvement in a 64K DRAM}.
\newblock {\em JSSC}, 1981.

\bibitem{horiguchi1997redundancy}
Masashi Horiguchi.
\newblock {Redundancy Techniques for High-Density DRAMs}.
\newblock In {\em ISIS}, 1997.

\bibitem{keeth2001dram_circuit}
B.~Keeth and R.J. Baker.
\newblock {\em {DRAM Circuit Design: A Tutorial}}.
\newblock IEEE Press, 2001.

\bibitem{keeth2007dram_circuit}
Brent Keeth, R~Jacob Baker, Brian Johnson, and Feng Lin.
\newblock {\em {DRAM Circuit Design: Fundamental and High-Speed Topics}}.
\newblock IEEE Press, 2007.

\bibitem{itoh2001vlsi}
Kiyoo Itoh.
\newblock {\em {VLSI Memory Chip Design}}.
\newblock Springer, 2001.

\bibitem{liu2013anexperimental}
Jamie Liu, Ben Jaiyen, Yoongu Kim, Chris Wilkerson, Onur Mutlu, J~Liu,
  B~Jaiyen, Y~Kim, C~Wilkerson, and O~Mutlu.
\newblock {An Experimental Study of Data Retention Behavior in Modern DRAM
  Devices}.
\newblock In {\em ISCA}, 2013.

\bibitem{seshadri2015gatherscatter}
Vivek Seshadri, Thomas Mullins, Amirali Boroumand, Onur Mutlu, Phillip~B
  Gibbons, Michael~A Kozuch, and Todd~C Mowry.
\newblock {Gather-Scatter DRAM: In-DRAM Address Translation to Improve the
  Spatial Locality of Non-Unit Strided Accesses}.
\newblock In {\em MICRO}, 2015.

\bibitem{khan2016parbor}
Samira Khan, Donghyuk Lee, and Onur Mutlu.
\newblock {PARBOR: An Efficient System-Level Technique to Detect Data-Dependent
  Failures in DRAM}.
\newblock In {\em DSN}, 2016.

\bibitem{khan2017detecting}
Samira Khan, Chris Wilkerson, Zhe Wang, Alaa~R Alameldeen, Donghyuk Lee, and
  Onur Mutlu.
\newblock {Detecting and Mitigating Data-Dependent DRAM Failures by Exploiting
  Current Memory Content}.
\newblock In {\em MICRO}, 2017.

\bibitem{lee2017designinduced}
Donghyuk Lee, Samira Khan, Lavanya Subramanian, Saugata Ghose, Rachata
  Ausavarungnirun, Gennady Pekhimenko, Vivek Seshadri, and Onur Mutlu.
\newblock {Design-Induced Latency Variation in Modern DRAM Chips:
  Characterization, Analysis, and Latency Reduction Mechanisms}.
\newblock In {\em SIGMETRICS}, 2017.

\bibitem{patel2020bitexact}
Minesh Patel, Jeremie Kim, Taha Shahroodi, Hasan Hassan, and Onur Mutlu.
\newblock {Bit-Exact ECC Recovery (BEER): Determining DRAM On-Die ECC Functions
  by Exploiting DRAM Data Retention Characteristics}.
\newblock In {\em {MICRO}}, 2020.

\bibitem{patel2024rethinking}
Minesh Patel, Taha Shahroodi, Aditya Manglik, Abdullah~Giray
  Ya{\u{g}}l{\i}k{\c{c}}{\i}, Ataberk Olgun, Haocong Luo, and Onur Mutlu.
\newblock {Rethinking the Producer-Consumer Relationship in Modern DRAM-Based
  Systems}.
\newblock {\em IEEE Access}, 2024.

\bibitem{patel2024rethinkingarxiv}
Minesh Patel, Taha Shahroodi, Aditya Manglik, Abdullah~Giray
  Ya{\u{g}}l{\i}k{\c{c}}{\i}, Ataberk Olgun, Haocong Luo, and Onur Mutlu.
\newblock {Rethinking the Producer-Consumer Relationship in Modern DRAM-Based
  Systems}.
\newblock arXiv:2401.16279 [cs.AR], 2024.

\bibitem{yaglikci2024spatial}
A.~Giray Ya{\u{g}}l{\i}k{\c{c}}{\i}, Yahya~Can Tu{\u{g}}rul, Geraldo~F
  Oliveira, Ismail~Emir Yüksel, Ataberk Olgun, Haocong Luo, and Onur Mutlu.
\newblock {Spatial Variation-Aware Read Disturbance Defenses: Experimental
  Analysis of Real DRAM Chips and Implications on Future Solutions}.
\newblock In {\em HPCA}, 2024.

\bibitem{kim2012acase}
Yoongu Kim, Vivek Seshadri, Donghyuk Lee, Jamie Liu, and Onur Mutlu.
\newblock {A Case for Exploiting Subarray-Level Parallelism (SALP) in DRAM}.
\newblock In {\em ISCA}, 2012.

\bibitem{chang2014improving}
Kevin~K Chang, Donghyuk Lee, Zeshan Chishti, Alaa~R Alameldeen, Chris
  Wilkerson, Yoongu Kim, and Onur Mutlu.
\newblock {Improving DRAM Performance by Parallelizing Refreshes with
  Accesses}.
\newblock In {\em HPCA}, 2014.

\bibitem{wang2020figaro}
Yaohua Wang, Lois Orosa, Xiangjun Peng, Yang Guo, Saugata Ghose, Minesh Patel,
  Jeremie~S Kim, Juan~G{\'o}mez Luna, Mohammad Sadrosadati, Nika~Mansouri
  Ghiasi, and Onur Mutlu.
\newblock {FIGARO: Improving System Performance via Fine-Grained In-DRAM Data
  Relocation and Caching}.
\newblock In {\em MICRO}, 2020.

\bibitem{zhang2014cream}
Tao Zhang, Matt Poremba, Cong Xu, Guangyu Sun, and Yuan Xie.
\newblock {CREAM: A Concurrent-Refresh-Aware DRAM Memory Architecture}.
\newblock In {\em HPCA}, 2014.

\bibitem{seshadri2013rowclone}
Vivek Seshadri, Yoongu Kim, Chris Fallin, Donghyuk Lee, Rachata
  Ausavarungnirun, Gennady Pekhimenko, Yixin Luo, Onur Mutlu, Phillip~B
  Gibbons, Michael~A Kozuch, and Todd Mowry.
\newblock {RowClone: Fast and Energy-Efficient In-DRAM Bulk Data Copy and
  Initialization}.
\newblock In {\em MICRO}, 2013.

\bibitem{seshadri2018rowclone}
Vivek Seshadri, Yoongu Kim, Chris Fallin, Donghyuk Lee, Rachata
  Ausavarungnirun, Gennady Pekhimenko, Yixin Luo, Onur Mutlu, Phillip~B.
  Gibbons, Michael~A. Kozuch, and Todd~C. Mowry.
\newblock {RowClone: Accelerating Data Movement and Initialization Using DRAM}.
\newblock arXiv:1805.03502 [cs.AR], 2018.

\bibitem{safari2021blockhammer}
{SAFARI Research Group}.
\newblock {BlockHammer --- GitHub Repository}.
\newblock \url{https://github.com/CMU-SAFARI/blockhammer}, 2021.

\bibitem{luo2023ramulator}
Haocong Luo, Yahya~Can Tugrul, F.~Nisa Bostancı, Ataberk Olgun, A.~Giray
  Yaglıkçı, and Onur Mutlu.
\newblock {Ramulator 2.0: A Modern, Modular, and Extensible DRAM Simulator}.
\newblock {\em IEEE CAL}, 2023.

\bibitem{safari2023ramulator2}
{SAFARI}.
\newblock {Ramulator 2.0}.
\newblock \url{https://github.com/CMU-SAFARI/ramulator2}, 2023.

\bibitem{olgun2024read}
Ataberk Olgun, Majd Osseiran, Abdullah~Giray Ya{\u{g}}l{\i}k{\c{c}}{\i},
  Yahya~Can Tugrul, Haocong Luo, Steve Rhyner, Behzad Salami, Juan Gomez~Luna,
  and Onur Mutlu.
\newblock {Read Disturbance in High Bandwidth Memory: A Detailed Experimental
  Study on HBM2 DRAM Chips}.
\newblock In {\em DSN}, 2024.

\bibitem{luo2024experimental}
Haocong Luo, Ismail~Emir Y{\"u}ksel, Ataberk Olgun, A~Giray
  Ya{\u{g}}l{\i}k{\c{c}}{\i}, Mohammad Sadrosadati, and Onur Mutlu.
\newblock {An Experimental Characterization of Combined RowHammer and RowPress
  Read Disturbance in Modern DRAM Chips}.
\newblock In {\em DSN (Disrupt)}, 2024.

\bibitem{olgun2023anexperimental}
Ataberk Olgun, Majd Osseiran, Abdullah~Giray Ya{\u{g}}l{\i}k{\c{c}}{\i},
  Yahya~Can Tugrul, Haocong Luo, Steve Rhyner, Behzad Salami, Juan Gomez~Luna,
  and Onur Mutlu.
\newblock {An Experimental Analysis of RowHammer in HBM2 DRAM Chips}.
\newblock In {\em DSN (Disrupt)}, 2023.

\bibitem{yang2017scanning}
Chia-Ming Yang, Chen-Kang Wei, Hsiu-Pin Chen, Jian-Shing Luo, Yu~Jing Chang,
  Tieh-Chiang Wu, and Chao-Sung Lai.
\newblock {Scanning Spreading Resistance Microscopy for Doping Profile in
  Saddle-Fin Devices}.
\newblock {\em IEEE Transactions on Nanotechnology}, 2017.

\bibitem{lim2018study}
Chulseung Lim, Kyungbae Park, Geunyong Bak, Donghyuk Yun, Myungsang Park,
  Sanghyeon Baeg, Shi-Jie Wen, and Richard Wong.
\newblock {Study of Proton Radiation Effect to Row Hammer Fault in DDR4
  SDRAMs}.
\newblock {\em Microelectronics Reliability}, 2018.

\bibitem{zhou2023doublesided}
Longda Zhou, Jie Li, Zheng Qiao, Pengpeng Ren, Zixuan Sun, Jianping Wang,
  Blacksmith Wu, Zhigang Ji, Runsheng Wang, Kanyu Cao, and Ru~Huang.
\newblock {Double-Sided Row Hammer Effect in Sub-20 nm DRAM: Physical
  Mechanism, Key Features and Mitigation}.
\newblock In {\em IRPS}, 2023.

\bibitem{zhou2024unveiling}
Longda Zhou, Sheng Ye, Runsheng Wang, and Zhigang Ji.
\newblock {Unveiling RowPress in Sub-20 nm DRAM Through Comparative Analysis
  With Row Hammer: From Leakage Mechanisms to Key Features}.
\newblock {\em IEEE TED}, 2024.

\bibitem{zhou2024understanding}
Longda Zhou, Jie Li, Pengpeng Ren, Sheng Ye, Da~Wang, Zheng Qiao, and Zhigang
  Ji.
\newblock {Understanding the Physical Mechanism of RowPress at the Device-Level
  in Sub-20 nm DRAM}.
\newblock In {\em IRPS}, 2024.

\bibitem{li2024understanding}
Jie Li, Longda Zhou, Sheng Ye, Zheng Qiao, and Zhigang Ji.
\newblock {Understanding the Competitive Interaction in Leakage Mechanisms for
  Effective Row Hammer Mitigation in Sub-20nm DRAM}.
\newblock {\em IEEE EDL}, 2024.

\bibitem{lang2023blaster}
Zhenrong Lang, Patrick Jattke, Michele Marazzi, and Kaveh Razavi.
\newblock {Blaster: Characterizing the Blast Radius of Rowhammer}.
\newblock In {\em DRAMSec}. ETH Zurich, 2023.

\bibitem{greenfield2012throttling}
Zvika Greenfield and Tomer Levy.
\newblock {Throttling Support for Row-Hammer Counters}.
\newblock US Patent 9251885B2, 2012.

\bibitem{wi2023shadow}
Minbok Wi, Jaehyun Park, Seoyoung Ko, Michael~Jaemin Kim, Nam~Sung Kim, Eojin
  Lee, and Jung~Ho Ahn.
\newblock {SHADOW: Preventing Row Hammer in DRAM with Intra-Subarray Row
  Shuffling}.
\newblock In {\em HPCA}, 2023.

\bibitem{zhou2023dnndefender}
Ranyang Zhou, Sabbir Ahmed, Adnan~Siraj Rakin, and Shaahin Angizi.
\newblock {DNN-Defender: An In-DRAM Deep Neural Network Defense Mechanism for
  Adversarial Weight Attack}.
\newblock arXiv:2305.08034 [cs.CR], 2023.

\bibitem{woo2023scalable}
Jeonghyun Woo, Gururaj Saileshwar, and Prashant~J. Nair.
\newblock {Scalable and Secure Row-Swap: Efficient and Safe Row Hammer
  Mitigation in Memory Systems}.
\newblock In {\em HPCA}, 2023.

\bibitem{bock2019riprh}
Carsten Bock, Ferdinand Brasser, David Gens, Christopher Liebchen, and
  Ahamd-Reza Sadeghi.
\newblock {RIP-RH: Preventing Rowhammer-Based Inter-Process Attacks}.
\newblock In {\em Asia-CCS}, 2019.

\bibitem{konoth2018zebram}
Radhesh~Krishnan Konoth, Marco Oliverio, Andrei Tatar, Dennis Andriesse,
  Herbert Bos, Cristiano Giuffrida, and Kaveh Razavi.
\newblock {ZebRAM: Comprehensive and Compatible Software Protection Against
  Rowhammer Attacks}.
\newblock In {\em OSDI}, 2018.

\bibitem{hassan2019crow}
H.~{Hassan}, M.~{Patel}, J.~S. {Kim}, A.~G. {Ya\u{g}l{\i}k\c{c}{\i}},
  N.~{Vijaykumar}, N.~{Mansouri Ghiasi}, S.~{Ghose}, and O.~{Mutlu}.
\newblock {CROW: A Low-Cost Substrate for Improving DRAM Performance, Energy
  Efficiency, and Reliability}.
\newblock In {\em ISCA}, 2019.

\bibitem{qazi2021halfdouble}
Salman Qazi, Yoongu Kim, Nicolas Boichat, Eric Shiu, and Mattias Nissler.
\newblock {Half-Double: Next-Row-Over Assisted Rowhammer}.
\newblock
  \url{https://github.com/google/hammer-kit/blob/main/20210525_half_double.pdf},
  2021.

\bibitem{qazi2021introducing}
Salman Qazi, Yoongu Kim, Nicolas Boichat, Eric Shiu, and Mattias Nissler.
\newblock {Introducing Half-Double New Hammering Technique for DRAM RowHammer
  Bug}.
\newblock
  \url{https://security.googleblog.com/2021/05/introducing-half-double-new-hammering.html},
  2021.

\bibitem{jedec2016jesd212c}
{JEDEC}.
\newblock {\em {Graphics Double Data Rate (GDDR5) SGRAM Standard}}, 2016.

\bibitem{micron2016ddr4}
{Micron}.
\newblock {DDR4 SDRAM Datasheet}.
\newblock In {\em Micron}, 2016.

\bibitem{jedec2012jesd793d}
{JEDEC}.
\newblock {JESD79-3D: DDR3 SDRAM Standard}, 2012.

\bibitem{jedec2012lowpower}
JEDEC.
\newblock {\em {Low Power Double Data Rate 3 (LPDDR3)}}, 2012.

\bibitem{jedec2017jesd2094b}
{JEDEC}.
\newblock {\em {JESD209-4B: Low Power Double Data Rate 4 (LPDDR4) Standard}},
  2017.

\bibitem{jedec2020jesd2095a}
{JEDEC}.
\newblock {\em {JESD209-5A: LPDDR5 SDRAM Standard}}, 2020.

\bibitem{frank2001device}
D.J. Frank, R.H. Dennard, E.~Nowak, P.M. Solomon, Y.~Taur, and Hon-Sum~Philip
  Wong.
\newblock {Device Scaling Limits of Si MOSFETs and Their Application
  Dependencies}.
\newblock {\em Proc.\ of the IEEE}, 2001.

\bibitem{lee2011simultaneous}
Dong-Su Lee, Young-Hyun Jun, and Bai-Sun Kong.
\newblock {Simultaneous Reverse Body and Negative Word-Line Biasing Control
  Scheme for Leakage Reduction of DRAM}.
\newblock {\em IEEE JSSC}, 2011.

\bibitem{redeker2002aninvestigation}
Michael Redeker, Bruce~F Cockburn, and Duncan~G Elliott.
\newblock {An Investigation into Crosstalk Noise in DRAM Structures}.
\newblock In {\em MTDT}, 2002.

\bibitem{yang2016suppression}
Chia Yang, Chen~Kang Wei, Yu~Jing Chang, Tieh~Chiang Wu, Hsiu~Pin Chen, and
  Chao~Sung Lai.
\newblock {Suppression of RowHammer Effect by Doping Profile Modification in
  Saddle-Fin Array Devices for Sub-30-nm DRAM Technology}.
\newblock {\em TDMR}, 2016.

\bibitem{gautam2019rowhammering}
S.~K. Gautam, S.~K. Manhas, Arvind Kumar, Mahendra Pakala, and Yiehm Ellie.
\newblock {Row Hammering Mitigation Using Metal Nanowire in Saddle Fin DRAM}.
\newblock {\em IEEE TED}, 2019.

\bibitem{jiang2021quantifying}
Yichen Jiang, Huifeng Zhu, Dean Sullivan, Xiaolong Guo, Xuan Zhang, and Yier
  Jin.
\newblock {Quantifying RowHammer Vulnerability for DRAM Security}.
\newblock In {\em DAC}, 2021.

\bibitem{park2014activeprecharge}
Kyungbae Park, Sanghyeon Baeg, ShiJie Wen, and Richard Wong.
\newblock {Active-Precharge Hammering on a Row-Induced Failure in DDR3 SDRAMs
  Under 3x nm Technology}.
\newblock In {\em IIRW}, 2014.

\bibitem{jung2014optimized}
Matthias Jung, Christian Weis, Norbert Wehn, Mohammadsadegh Sadri, and Luca
  Benini.
\newblock {Optimized Active and Power-Down Mode Refresh Control in 3D-DRAMs}.
\newblock In {\em VLSI-SoC}, 2014.

\bibitem{hamamoto1995well}
T~Hamamoto, S~Sugiura, and S~Sawada.
\newblock {Well Concentration: A Novel Scaling Limitation Factor Derived From
  DRAM Retention Time and Its Modeling}.
\newblock In {\em IEDM}, 1995.

\bibitem{hamamoto1998onthe}
T.~Hamamoto, S.~Sugiura, and S.~Sawada.
\newblock {On the Retention Time Distribution of Dynamic Random Access Memory
  (DRAM)}.
\newblock {\em IEEE TED}, 1998.

\bibitem{yaney1987ametastable}
David~S Yaney, Chih-Yuan Lu, Ross~A Kohler, Michael~J Kelly, and James~T
  Nelson.
\newblock {A Meta-Stable Leakage Phenomenon in DRAM Charge Storage - Variable
  Hold Time}.
\newblock In {\em IEDM}, 1987.

\bibitem{shirley2014copula}
C~Glenn Shirley and W~Robert Daasch.
\newblock {Copula Models of Correlation: A DRAM Case Study}.
\newblock In {\em TC}, 2014.

\bibitem{weis2015retention}
Christian Weis, Matthias Jung, Peter Ehses, Cristiano Santos, Pascal Vivet,
  Sven Goossens, Martijn Koedam, and Norbert Wehn.
\newblock {Retention Time Measurements and Modelling of Bit Error Rates of Wide
  I/O DRAM in MPSoCs}.
\newblock In {\em DATE}, 2015.

\bibitem{jung2015omitting}
Matthias Jung, \'{E}der Zulian, Deepak~M. Mathew, Matthias Herrmann, Christian
  Brugger, Christian Weis, and Norbert Wehn.
\newblock {Omitting Refresh: A Case Study for Commodity and Wide I/O DRAMs}.
\newblock In {\em MEMSYS}, 2015.

\bibitem{weis2015thermal}
Christian Weis, Matthias Jung, Omar Naji, Cristiano Santos, Pascal Vivet, and
  Andreas Hansson.
\newblock {Thermal Aspects and High-Level Explorations of 3D Stacked DRAMs}.
\newblock In {\em ISVLSI}, 2015.

\bibitem{baek2014refresh}
Seungjae Baek, Sangyeun Cho, and Rami Melhem.
\newblock {Refresh Now and Then}.
\newblock {\em IEEE TC}, 2014.

\bibitem{khan2014theefficacy}
Samira Khan, Donghyuk Lee, Yoongu Kim, Alaa~R Alameldeen, Chris Wilkerson, and
  Onur Mutlu.
\newblock {The Efficacy of Error Mitigation Techniques for DRAM Retention
  Failures: A Comparative Experimental Study}.
\newblock In {\em SIGMETRICS}, 2014.

\bibitem{venkatesan2006retentionaware}
Ravi~K Venkatesan, Stephen Herr, and Eric Rotenberg.
\newblock {Retention-Aware Placement in DRAM (RAPID): Software Methods for
  Quasi-Non-Volatile DRAM}.
\newblock In {\em HPCA}, 2006.

\bibitem{weis2017dramspec_a}
Christian Weis, Abdul Mutaal, Omar Naji, Matthias Jung, Andreas Hansson, and
  Norbert Wehn.
\newblock {DRAMSpec: A High-Level DRAM Timing, Power and Area Exploration
  Tool}.
\newblock {\em IJPP}, 2017.

\bibitem{khan2016acase}
Samira Khan, Chris Wilkerson, Donghyuk Lee, Alaa~R Alameldeen, and Onur Mutlu.
\newblock {A Case for Memory Content-Based Detection and Mitigation of
  Data-Dependent Failures in DRAM}.
\newblock {\em IEEE {CAL}}, 2016.

\bibitem{qureshi2015avatar}
M.K. Qureshi, Dae-Hyun Kim, S.~Khan, P.J. Nair, and O.~Mutlu.
\newblock {AVATAR: A Variable-Retention-Time (VRT) Aware Refresh for DRAM
  Systems}.
\newblock In {\em DSN}, 2015.

\bibitem{sutar2016dpuf}
Soubhagya Sutar, Arnab Raha, and Vijay Raghunathan.
\newblock {D-PUF: An Intrinsically Reconfigurable DRAM PUF for Device
  Authentication in Embedded Systems}.
\newblock In {\em CASES}, 2016.

\bibitem{kim2009anew}
Kinam Kim and Jooyoung Lee.
\newblock {A New Investigation of Data Retention Time in Truly Nanoscaled
  DRAMs}.
\newblock In {\em EDL}, 2009.

\bibitem{kong2008analysis}
Wei Kong, Paul~C Parries, G~Wang, and Subramanian~S Iyer.
\newblock {Analysis of Retention Time Distribution of Embedded DRAM-A New
  Method to Characterize Across-Chip Threshold Voltage Variation}.
\newblock In {\em ITC}, 2008.

\bibitem{lieneweg1998assessment}
Udo Lieneweg, D~Nguyen, and B~Blaes.
\newblock {Assessment of DRAM Reliability from Retention Time Measurements}.
\newblock {\em Flight Readiness Technol. Assessment NASA EEE Parts Prog.},
  1998.

\bibitem{patel2017thereach}
Minesh Patel, Jeremie~S Kim, and Onur Mutlu.
\newblock {The Reach Profiler (REAPER): Enabling the Mitigation of DRAM
  Retention Failures via Profiling at Aggressive Conditions}.
\newblock {\em ISCA}, 2017.

\bibitem{lee2016reducing}
Donghyuk Lee, Samira~Manabi Khan, Lavanya Subramanian, Rachata Ausavarungnirun,
  Gennady Pekhimenko, Vivek Seshadri, Saugata Ghose, and Onur Mutlu.
\newblock {Reducing DRAM Latency by Exploiting Design-Induced Latency Variation
  in Modern DRAM Chips}.
\newblock arXiv:1610.09604 [cs.AR], 2016.

\bibitem{chandrasekar2014exploiting}
Karthik Chandrasekar, Sven Goossens, Christian Weis, Martijn Koedam, Benny
  Akesson, Norbert Wehn, and Kees Goossens.
\newblock {Exploiting Expendable Process-Margins in DRAMs for Run-Time
  Performance Optimization}.
\newblock In {\em DATE}, 2014.

\bibitem{chang2016understanding}
Kevin~K. Chang, Abhijith Kashyap, Hasan Hassan, Saugata Ghose, Kevin Hsieh,
  Donghyuk Lee, Tianshi Li, Gennady Pekhimenko, Samira Khan, and Onur Mutlu.
\newblock {Understanding Latency Variation in Modern DRAM Chips: Experimental
  Characterization, Analysis, and Optimization}.
\newblock In {\em SIGMETRICS}, 2016.

\bibitem{lee2015adaptivelatency}
Donghyuk Lee, Yoongu Kim, Gennady Pekhimenko, Samira Khan, Vivek Seshadri,
  Kevin Chang, and Onur Mutlu.
\newblock {Adaptive-Latency DRAM: Optimizing DRAM Timing for the Common-Case}.
\newblock In {\em HPCA}, 2015.

\bibitem{kim2018thedram}
Jeremie~S. Kim, Minesh Patel, Hasan Hassan, and Onur Mutlu.
\newblock {The DRAM Latency PUF: Quickly Evaluating Physical Unclonable
  Functions by Exploiting the Latency-Reliability Tradeoff in Modern Commodity
  DRAM Devices}.
\newblock In {\em HPCA}, 2018.

\bibitem{kim2018solardram}
Jeremie~S Kim, Minesh Patel, Hasan Hassan, and Onur Mutlu.
\newblock {Solar-DRAM: Reducing DRAM Access Latency by Exploiting the Variation
  in Local Bitlines}.
\newblock In {\em ICCD}, 2018.

\bibitem{kim2019drange}
Jeremie~S. Kim, Minesh Patel, Hasan Hassan, Lois Orosa, and Onur Mutlu.
\newblock {D-RaNGe: Using Commodity DRAM Devices to Generate True Random
  Numbers with Low Latency and High Throughput}.
\newblock In {\em HPCA}, 2019.

\bibitem{talukder2019exploiting}
B.~M.~S. {Bahar Talukder}, J.~{Kerns}, B.~{Ray}, T.~{Morris}, and M.~T.
  {Rahman}.
\newblock {Exploiting DRAM Latency Variations for Generating True Random
  Numbers}.
\newblock In {\em ICCE}, 2019.

\bibitem{talukder2018ldpuf}
Bashir M. Sabquat~Bahar Talukder, Biswajit Ray, Mark~Mohammad Tehranipoor,
  Domenic Forte, and Md.~Tauhidur Rahman.
\newblock {LDPUF: Exploiting DRAM Latency Variations to Generate Robust Device
  Signatures}.
\newblock arXiv:1808.02584 [cs.CR], 2018.

\bibitem{talukder2019prelatpuf}
BMS~Bahar Talukder, Biswajit Ray, Domenic Forte, and Md~Tauhidur Rahman.
\newblock {PreLatPUF: Exploiting DRAM Latency Variations for Generating Robust
  Device Signatures}.
\newblock {\em IEEE Access}, 7, 2019.

\bibitem{mukhanov2020dstress}
Lev Mukhanov, Dimitrios~S Nikolopoulos, and Georgios Karakonstantis.
\newblock {DStress: Automatic Synthesis of DRAM Reliability Stress Viruses
  Using Genetic Algorithms}.
\newblock In {\em MICRO}, 2020.

\bibitem{jung2016reverse}
Matthias Jung, Carl~C Rheinl{\"a}nder, Christian Weis, and Norbert Wehn.
\newblock {Reverse Engineering of DRAMs: Row Hammer with Crosshair}.
\newblock In {\em MEMSYS}, 2016.

\bibitem{patel2019understanding}
Minesh Patel, Jeremie~S. Kim, Hasan Hassan, and Onur Mutlu.
\newblock {Understanding and Modeling On-Die Error Correction in Modern DRAM:
  An Experimental Study Using Real Devices}.
\newblock In {\em DSN}, 2019.

\bibitem{patel2021enabling}
Minesh Patel.
\newblock {\em {Enabling Effective Error Mitigation in Memory Chips That Use
  On-Die Error-Correcting Codes}}.
\newblock PhD thesis, ETH Zürich, 2021.

\bibitem{patel2021harp}
Minesh Patel, Geraldo~F. Oliveira, and Onur Mutlu.
\newblock {HARP: Practically and Effectively Identifying Uncorrectable Errors
  in Main Memory Chips That Use On-Die ECC}.
\newblock In {\em MICRO}, 2021.

\bibitem{ghose2018what}
Saugata Ghose, A.~Giray Ya{\u{g}}l{\i}k{\c{c}}{\i}, Raghav Gupta, Donghyuk Lee,
  Kais Kudrolli, William Liu, Hasan Hassan, Kevin Chang, Niladrish Chatterjee,
  Aditya Agrawal, Mike O'Connor, and Onur Mutlu.
\newblock {What Your DRAM Power Models Are Not Telling You: Lessons from a
  Detailed Experimental Study}.
\newblock In {\em SIGMETRICS}, 2018.

\bibitem{david2011memory}
Howard David, Chris Fallin, Eugene Gorbatov, Ulf~R Hanebutte, and Onur Mutlu.
\newblock {Memory Power Management via Dynamic Voltage/Frequency Scaling}.
\newblock In {\em ICAC}, 2011.

\bibitem{deng2011memscale}
Qingyuan Deng, David Meisner, Luiz Ramos, Thomas~F Wenisch, and Ricardo
  Bianchini.
\newblock {MemScale: Active Low-Power Modes for Main Memory}.
\newblock In {\em ASPLOS}, 2011.

\bibitem{chang2017understanding}
Kevin~K Chang, A~Giray Ya{\u{g}}l{\i}k{\c{c}}{\i}, Saugata Ghose, Aditya
  Agrawal, Niladrish Chatterjee, Abhijith Kashyap, Donghyuk Lee, Mike O'Connor,
  Hasan Hassan, and Onur Mutlu.
\newblock {Understanding Reduced-Voltage Operation in Modern DRAM Devices:
  Experimental Characterization, Analysis, and Mechanisms}.
\newblock In {\em SIGMETRICS}, 2017.

\bibitem{gao2019computedram}
Fei Gao, Georgios Tziantzioulis, and David Wentzlaff.
\newblock {ComputeDRAM: In-Memory Compute Using Off-the-Shelf DRAMs}.
\newblock In {\em MICRO}, 2019.

\bibitem{olgun2021quactrng}
Ataberk Olgun, Minesh Patel, A~Giray Ya{\u{g}}l{\i}k{\c{c}}{\i}, Haocong Luo,
  Jeremie~S Kim, Nisa Bostanc{\i}, Nandita Vijaykumar, O{\u{g}}uz Ergin, and
  Onur Mutlu.
\newblock {QUAC-TRNG: High-Throughput True Random Number Generation Using
  Quadruple Row Activation in Commodity DRAM Chips}.
\newblock In {\em {ISCA}}, 2021.

\bibitem{olgun2022pidram}
Ataberk Olgun, Juan~G{\'o}mez Luna, Konstantinos Kanellopoulos, Behzad Salami,
  Hasan Hassan, Oguz Ergin, and Onur Mutlu.
\newblock {PiDRAM: A Holistic End-to-End FPGA-Based Framework for
  Processing-in-DRAM}.
\newblock {\em ACM TACO}, 2022.

\bibitem{gao2022fracdram}
Fei Gao, Georgios Tziantzioulis, and David Wentzlaff.
\newblock {FracDRAM: Fractional Values in Off-the-Shelf DRAM}.
\newblock In {\em MICRO}, 2022.

\bibitem{yuksel2023pulsar}
Ismail~Emir Yuksel, Yahya~Can Tugrul, F.~Nisa Bostanci, A.~Giray
  Ya{\u{g}}l{\i}k{\c{c}}{\i}, Ataberk Olgun, Geraldo~F. Oliveira, Melina
  Soysal, Haocong Luo, Juan~Gomez Luna, Mohammad Sadrosadati, and Onur Mutlu.
\newblock {PULSAR: Simultaneous Many-Row Activation for Reliable and
  High-Performance Computing in Off-the-Shelf DRAM Chips}.
\newblock arXiv:2312.02880 [cs.AR], 2023.

\bibitem{yuksel2024functionallycomplete}
Ismail~Emir Yuksel, Yahya~Can Tugrul, Ataberk Olgun, F.~Nisa Bostanci, A.~Giray
  Ya{\u{g}}l{\i}k{\c{c}}{\i}, Geraldo~F. Oliveira, Haocong Luo, Juan~Gomez
  Luna, Mohammad Sadrosadati, and Onur Mutlu.
\newblock {Functionally-Complete Boolean Logic in Real DRAM Chips: Experimental
  Characterization and Analysis}.
\newblock In {\em {HPCA}}, 2024.

\bibitem{nam2024dramscope}
Hwayong Nam, Seungmin Baek, Minbok Wi, Michael~Jaemin Kim, Jaehyun Park, Chihun
  Song, Nam~Sung Kim, and Jung~Ho Ahn.
\newblock {DRAMScope: Uncovering DRAM Microarchitecture and Characteristics by
  Issuing Memory Commands}.
\newblock In {\em ISCA}, 2024.

\bibitem{marazzi2024hifidram}
Michele Marazzi, Tristan Sachsenweger, Flavien Solt, Peng Zeng, Kubo Takashi,
  Maksym Yarema, and Kaveh Razavi.
\newblock {HiFi-DRAM: Enabling High-Fidelity DRAM Research by Uncovering Sense
  Amplifiers with IC Imaging}.
\newblock In {\em ISCA}, 2024.

\bibitem{schroeder2009dram_errors}
Bianca Schroeder, Eduardo Pinheiro, and Wolf-Dietrich Weber.
\newblock {DRAM Errors in the Wild: A Large-Scale Field Study}.
\newblock In {\em SIGMETRICS}, 2009.

\bibitem{hwang2012cosmic}
Andy~A Hwang, Ioan~A Stefanovici, and Bianca Schroeder.
\newblock {Cosmic Rays Don't Strike Twice: Understanding the Nature of DRAM
  Errors and the Implications for System Design}.
\newblock In {\em ASPLOS}, 2012.

\bibitem{sridharan2012astudy}
Vilas Sridharan and Dean Liberty.
\newblock {A Study of DRAM Failures in the Field}.
\newblock In {\em {SC}}, 2012.

\bibitem{sridharan2015memory}
Vilas Sridharan, Nathan DeBardeleben, Sean Blanchard, Kurt~B Ferreira, Jon
  Stearley, John Shalf, and Sudhanva Gurumurthi.
\newblock {Memory Errors in Modern Systems: The Good, the Bad, and the Ugly}.
\newblock In {\em ASPLOS}, 2015.

\bibitem{sridharan2013feng}
Vilas Sridharan, Jon Stearley, Nathan DeBardeleben, Sean Blanchard, and
  Sudhanva Gurumurthi.
\newblock {Feng Shui of Supercomputer Memory: Positional Effects in DRAM and
  SRAM Faults}.
\newblock In {\em SC}, 2013.

\bibitem{meza2015revisiting}
Justin Meza, Qiang Wu, Sanjeev Kumar, and Onur Mutlu.
\newblock {Revisiting Memory Errors in Large-Scale Production Data Centers:
  Analysis and Modeling of New Trends from the Field}.
\newblock In {\em {DSN}}, 2015.

\bibitem{bautistagomez2016unprotected}
Leonardo Bautista-Gomez, Ferad Zyulkyarov, Osman Unsal, and Simon
  McIntosh-Smith.
\newblock {Unprotected Computing: A Large-Scale Study Of DRAM Raw Error Rate on
  a Supercomputer}.
\newblock In {\em SC}, 2016.

\bibitem{siddiqua2013analysis}
Taniya Siddiqua, Athanasios~E. Papathanasiou, Arijit Biswas, Sudhanva
  Gurumurthi, Intel Corp, and Teradata Aster.
\newblock {Analysis and Modeling of Memory Errors From Large-Scale Field Data
  Collection}.
\newblock In {\em SELSE}, 2013.

\bibitem{meza2018large}
Justin~J Meza.
\newblock {\em {Large Scale Studies of Memory, Storage, and Network Failures in
  a Modern Data Center}}.
\newblock PhD thesis, Carnegie Mellon University, 2018.

\bibitem{zhang2021quantifying}
Da~Zhang, Gagandeep Panwar, Jagadish~B Kotra, Nathan DeBardeleben, Sean
  Blanchard, and Xun Jian.
\newblock {Quantifying Server Memory Frequency Margin and Using It to Improve
  Performance in HPC Systems}.
\newblock In {\em ISCA}, 2021.

\bibitem{maiz2003characterization}
Jose Maiz, Scott Hareland, Kevin Zhang, and Patrick Armstrong.
\newblock {Characterization of Multi-Bit Soft Error Events in Advanced SRAMs}.
\newblock In {\em IEDM}, 2003.

\bibitem{autran2009altitude}
Jean-Luc Autran, P~Roche, S~Sauze, G~Gasiot, Daniela Munteanu, P~Loaiza,
  M~Zampaolo, and J~Borel.
\newblock {Altitude and Underground Real-Time SER Characterization of CMOS 65
  nm SRAM}.
\newblock {\em IEEE Trans. Nucl. Sci.}, 2009.

\bibitem{radaelli2005investigation}
Daniele Radaelli, Helmut Puchner, Skip Wong, and Sabbas Daniel.
\newblock {Investigation of Multi-Bit Upsets in a 150 nm Technology SRAM
  Device}.
\newblock {\em IEEE Trans. Nucl. Sci.}, 2005.

\bibitem{cai2011fpga}
Yu~Cai, Erich~F Haratsch, Mark McCartney, and Ken Mai.
\newblock {FPGA-Based Solid-State Drive Prototyping Platform}.
\newblock In {\em FCCM}. IEEE, 2011.

\bibitem{cai2015read}
Yu~Cai, Yixin Luo, Saugata Ghose, and Onur Mutlu.
\newblock {Read Disturb Errors in MLC NAND Flash Memory: Characterization,
  Mitigation, and Recovery}.
\newblock In {\em DSN}, 2015.

\bibitem{luo2015warm}
Yixin Luo, Yu~Cai, Saugata Ghose, Jongmoo Choi, and Onur Mutlu.
\newblock {WARM: Improving NAND Flash Memory Lifetime with Write-Hotness Aware
  Retention Management}.
\newblock {\em {MSST}}, 2015.

\bibitem{cai2015data}
Yu~Cai, Yixin Luo, Erich~F Haratsch, Ken Mai, and Onur Mutlu.
\newblock {Data Retention in MLC NAND Flash Memory: Characterization,
  Optimization, and Recovery}.
\newblock In {\em HPCA}, 2015.

\bibitem{cai2014neighborcell}
Yu~Cai, Gulay Yalcin, Onur Mutlu, Erich~F Haratsch, Osman Unsal, Adrian
  Cristal, and Ken Mai.
\newblock {Neighbor-Cell Assisted Error Correction for MLC NAND Flash
  Memories}.
\newblock In {\em SIGMETRICS}, 2014.

\bibitem{cai2013program}
Yu~Cai, Onur Mutlu, Erich~F Haratsch, and Ken Mai.
\newblock {Program Interference in MLC NAND Flash Memory: Characterization,
  Modeling, and Mitigation}.
\newblock In {\em ICCD}, 2013.

\bibitem{cai2013error}
Yu~Cai, Gulay Yalcin, Onur Mutlu, Erich~F Haratsch, Adrian Cristal, Osman~S
  Unsal, and Ken Mai.
\newblock {Error Analysis and Retention-Aware Error Management for NAND Flash
  Memory}.
\newblock In {\em ITJ}, 2013.

\bibitem{cai2013threshold}
Yu~Cai, Erich~F Haratsch, Onur Mutlu, and Ken Mai.
\newblock {Threshold Voltage Distribution in MLC NAND Flash Memory:
  Characterization, Analysis, and Modeling}.
\newblock In {\em DATE}, 2013.

\bibitem{cai2012flash}
Yu~Cai, Gulay Yalcin, Onur Mutlu, Erich~F Haratsch, Adrian Cristal, Osman~S
  Unsal, and Ken Mai.
\newblock {Flash Correct-And-Refresh: Retention-Aware Error Management for
  Increased Flash Memory Lifetime}.
\newblock In {\em ICCD}, 2012.

\bibitem{cai2012error}
Yu~Cai, Erich~F Haratsch, Onur Mutlu, and Ken Mai.
\newblock {Error Patterns in MLC NAND Flash Memory: Measurement,
  Characterization, and Analysis}.
\newblock In {\em DATE}, 2012.

\bibitem{meza2015alargescale}
Justin Meza, Qiang Wu, Sanjev Kumar, and Onur Mutlu.
\newblock {A Large-Scale Study of Flash Memory Errors in the Field}.
\newblock In {\em SIGMETRICS}, 2015.

\bibitem{schroeder2016flash}
Bianca Schroeder, Raghav Lagisetty, and Arif Merchant.
\newblock {Flash Reliability in Production: The Expected and the Unexpected}.
\newblock In {\em USENIX FAST}, 2016.

\bibitem{luo2016enabling}
Yixin Luo, Saugata Ghose, Yu~Cai, Erich~F Haratsch, and Onur Mutlu.
\newblock {Enabling Accurate and Practical Online Flash Channel Modeling for
  Modern MLC NAND Flash Memory}.
\newblock In {\em JSAC}, 2016.

\bibitem{narayanan2016ssdfailures}
Iyswarya Narayanan, Di~Wang, Myeongjae Jeon, Bikash Sharma, Laura Caulfield,
  Anand Sivasubramaniam, Ben Cutler, Jie Liu, Badriddine~M. Khessib, and Vaid
  Kushagra.
\newblock {SSD Failures in Datacenters: What, When and Why?}
\newblock In {\em SIGMETRICS}, 2016.

\bibitem{fukami2017improving}
Aya Fukami, Saugata Ghose, Yixin Luo, Yu~Cai, and Onur Mutlu.
\newblock {Improving the Reliability of Chip-Off Forensic Analysis of NAND
  Flash Memory Devices}.
\newblock In {\em Digital Investigation}, 2017.

\bibitem{cai2017error}
Yu~Cai, Saugata Ghose, Erich~F Haratsch, Yixin Luo, and Onur Mutlu.
\newblock {Error Characterization, Mitigation, and Recovery in
  Flash-Memory-Based Solid-State Drives}.
\newblock {\em Proc.\ of the IEEE}, 2017.

\bibitem{cai2017vulnerabilities}
Yu~Cai, Saugata Ghose, Yixin Luo, Ken Mai, Onur Mutlu, and Erich~F. Haratsch.
\newblock {Vulnerabilities in MLC NAND Flash Memory Programming: Experimental
  Analysis, Exploits, and Mitigation Techniques}.
\newblock In {\em HPCA}, 2017.

\bibitem{luo2018heatwatch}
Yixin Luo, Saugata Ghose, Yu~Cai, Erich~F Haratsch, and Onur Mutlu.
\newblock {HeatWatch: Improving 3D NAND Flash Memory Device Reliability by
  Exploiting Self-Recovery and Temperature Awareness}.
\newblock In {\em HPCA}, 2018.

\bibitem{luo2018improving}
Yixin Luo, Saugata Ghose, Yu~Cai, Erich~F Haratsch, and Onur Mutlu.
\newblock {Improving 3D NAND Flash Memory Lifetime by Tolerating Early
  Retention Loss and Process Variation}.
\newblock {\em POMACS}, 2018.

\bibitem{kim2020evanesco}
Myungsuk Kim, Jisung Park, Genhee Cho, Yoona Kim, Lois Orosa, Onur Mutlu, and
  Jihong Kim.
\newblock {Evanesco: Architectural Support for Efficient Data Sanitization in
  Modern Flash-Based Storage Systems}.
\newblock In {\em ASPLOS}, 2020.

\bibitem{park2021reducing}
Jisung Park, Myungsuk Kim, Myoungjun Chun, Lois Orosa, Jihong Kim, and Onur
  Mutlu.
\newblock {Reducing Solid-State Drive Read Latency by Optimizing Read-Retry}.
\newblock In {\em ASPLOS}, 2021.

\bibitem{luo2018thesis}
Yixin Luo.
\newblock {\em {Architectural Techniques for Improving NAND Flash Memory
  Reliability}}.
\newblock PhD thesis, Carnegie Mellon University, 2018.

\bibitem{cai2012nand}
Yu~Cai.
\newblock {\em {NAND Flash Memory: Characterization, Analysis, Modelling, and
  Mechanisms}}.
\newblock PhD thesis, Carnegie Mellon University, 2012.

\bibitem{bairavasundaram2008ananalysis}
Lakshmi~N Bairavasundaram, Andrea~C Arpaci-Dusseau, Remzi~H Arpaci-Dusseau,
  Garth~R Goodson, and Bianca Schroeder.
\newblock {An Analysis of Data Corruption in The Storage Stack}.
\newblock {\em TOS}, 2008.

\bibitem{bairavasundaram2007ananalysis}
Lakshmi~N Bairavasundaram, Garth~R Goodson, Shankar Pasupathy, and Jiri
  Schindler.
\newblock {An Analysis Of Latent Sector Errors in Disk Drives}.
\newblock In {\em SIGMETRICS}, 2007.

\bibitem{pinheiro2007failure}
Eduardo Pinheiro, Wolf-Dietrich Weber, and Luiz~Andr{\'e} Barroso.
\newblock {Failure Trends in a Large Disk Drive Population}.
\newblock In {\em FAST}, 2007.

\bibitem{schroeder2007understandingdisk}
Bianca Schroeder and Garth~A Gibson.
\newblock {Understanding Disk Failure Rates: What Does an MTTF of 1,000,000
  Hours Mean to You?}
\newblock {\em ACM TOS}, 2007.

\bibitem{schroeder2007understanding}
Bianca Schroeder and Garth~A Gibson.
\newblock {Understanding Failures in Petascale Computers}.
\newblock In {\em Journal of Physics: Conference Series}, 2007.

\bibitem{pirovano2004reliability}
Agostino Pirovano, Andrea Redaelli, Fabio Pellizzer, Federica Ottogalli, Marina
  Tosi, Daniele Ielmini, Andrea~L Lacaita, and Roberto Bez.
\newblock {Reliability Study of Phase-Change Nonvolatile Memories}.
\newblock {\em TDMR}, 2004.

\bibitem{zhang2012memory}
Zhe Zhang, Weijun Xiao, Nohhyun Park, and David~J Lilja.
\newblock {Memory Module-Level Testing and Error Behaviors for Phase Change
  Memory}.
\newblock In {\em ICCD}, 2012.

\bibitem{bains2015rowhammer}
Kuljit Bains, John Halbert, Christopher Mozak, Theodore Schoenborn, and Zvika
  Greenfield.
\newblock {Row Hammer Refresh Command}.
\newblock US Patents: 9,117,544 10,210,925, 2015.

\bibitem{aichinger2015ddrmemory}
Barbara Aichinger.
\newblock {DDR Memory Errors Caused by Row Hammer}.
\newblock In {\em HPEC}, 2015.

\bibitem{lee2021cryoguard}
Gyu-Hyeon Lee, Seongmin Na, Ilkwon Byun, Dongmoon Min, and Jangwoo Kim.
\newblock {CryoGuard: A Near Refresh-Free Robust DRAM Design for Cryogenic
  Computing}.
\newblock In {\em ISCA}, 2021.

\bibitem{juffinger2023csirowhammercryptographic}
Jonas Juffinger, Lukas Lamster, Andreas Kogler, Maria Eichlseder, Moritz Lipp,
  and Daniel Gruss.
\newblock {CSI: Rowhammer--Cryptographic Security and Integrity against
  Rowhammer}.
\newblock In {\em IEEE S\&P}, 2023.

\bibitem{enomoto2022efficient}
Shuhei Enomoto, Hiroki Kuzuno, and Hiroshi Yamada.
\newblock {Efficient Protection Mechanism for CPU Cache Flush Instruction Based
  Attacks}.
\newblock {\em IEICE Transactions on Information and Systems}, 2022.

\bibitem{manzhosov2022revisiting}
Evgeny Manzhosov, Adam Hastings, Meghna Pancholi, Ryan Piersma, Mohamed
  Tarek~Ibn Ziad, and Simha Sethumadhavan.
\newblock {Revisiting Residue Codes for Modern Memories}.
\newblock In {\em MICRO}, 2022.

\bibitem{ajorpaz2022evax}
Samira~Mirbagher Ajorpaz, Daniel Moghimi, Jeffrey~Neal Collins, Gilles Pokam,
  Nael Abu-Ghazaleh, and Dean Tullsen.
\newblock {EVAX: Towards a Practical, Pro-Active \& Adaptive Architecture for
  High Performance \& Security}.
\newblock In {\em MICRO}, 2022.

\bibitem{hassan2024acase}
Hasan Hassan, Ataberk Olgun, A~Giray Ya{\u{g}}l{\i}k{\c{c}}{\i}, Haocong Luo,
  and Onur Mutlu.
\newblock {A Case for Self-Managing DRAM Chips: Improving Performance,
  Efficiency, Reliability, and Security via Autonomous In-DRAM Maintenance
  Operations}.
\newblock In {\em MICRO (Preprint on arXiv:2207.13358 [cs.AR])}, 2024.

\bibitem{zhang2020leveraging}
Zhenkai Zhang, Zihao Zhan, Daniel Balasubramanian, Bo~Li, Peter Volgyesi, and
  Xenofon Koutsoukos.
\newblock {Leveraging EM Side-Channel Information to Detect Rowhammer Attacks}.
\newblock In {\em IEEE S\&P}, 2020.

\bibitem{han2021surround}
Jin-Woo Han, Jungsik Kim, Dafna Beery, K.~Deniz Bozdag, Peter Cuevas, Amitay
  Levi, Irwin Tain, Khai Tran, Andrew~J. Walker, Senthil~Vadakupudhu Palayam,
  Antonio Arreghini, Arnaud Furnémont, and M.~Meyyappan.
\newblock {Surround Gate Transistor With Epitaxially Grown Si Pillar and
  Simulation Study on Soft Error and Rowhammer Tolerance for DRAM}.
\newblock {\em IEEE TED}, 2021.

\bibitem{fakhrzadehgan2022safeguard}
Ali Fakhrzadehgan, Yale~N. Patt, Prashant~J. Nair, and Moinuddin~K. Qureshi.
\newblock {SafeGuard: Reducing the Security Risk from Row-Hammer via Low-Cost
  Integrity Protection}.
\newblock In {\em HPCA}, 2022.

\bibitem{saroiu2022theprice}
Stefan Saroiu, Alec Wolman, and Lucian Cojocar.
\newblock {The Price of Secrecy: How Hiding Internal DRAM Topologies Hurts
  Rowhammer Defenses}.
\newblock In {\em IRPS}, 2022.

\bibitem{loughlin2022moesiprime}
Kevin Loughlin, Stefan Saroiu, Alec Wolman, Yatin~A. Manerkar, and Baris
  Kasikci.
\newblock {MOESI-Prime: Preventing Coherence-Induced Hammering in Commodity
  Workloads}.
\newblock In {\em ISCA}, 2022.

\bibitem{zhou2022ltpim}
Ranyang Zhou, Sepehr Tabrizchi, Arman Roohi, and Shaahin Angizi.
\newblock {LT-PIM: An LUT-Based Processing-in-DRAM Architecture with RowHammer
  Self-Tracking}.
\newblock {\em IEEE CAL}, 2022.

\bibitem{didio2023copyonflip}
Andrea Di~Dio, Koen Koning, Herbert Bos, and Cristiano Giuffrida.
\newblock {Copy-on-Flip: Hardening ECC Memory Against Rowhammer Attacks}.
\newblock In {\em NDSS}, 2023.

\bibitem{sharma2022areview}
Sonia Sharma, Debdeep Sanyal, Arpit Mukhopadhyay, and Ramij~Hasan Shaik.
\newblock {A Review on Study of Defects of DRAM-RowHammer and Its Mitigation}.
\newblock {\em Journal For Basic Sciences}, 2022.

\bibitem{park2022rowhammer_reduction}
Jin~Hyo Park, Su~Yeon Kim, Dong~Young Kim, Geon Kim, Je~Won Park, Sunyong Yoo,
  Young-Woo Lee, and Myoung~Jin Lee.
\newblock {Row Hammer Reduction Using a Buried Insulator in a Buried Channel
  Array Transistor}.
\newblock {\em IEEE TED}, 2022.

\bibitem{kim2023a11v}
Woongrae Kim, Chulmoon Jung, Seongnyuh Yoo, Duckhwa Hong, Jeongjin Hwang,
  Jungmin Yoon, Ohyong Jung, Joonwoo Choi, Sanga Hyun, Mankeun Kang, Sangho
  Lee, Dohong Kim, Sanghyun Ku, Donhyun Choi, Nogeun Joo, Sangwoo Yoon, Junseok
  Noh, Byeongyong Go, Cheolhoe Kim, Sunil Hwang, Mihyun Hwang, Min Seol-Yi,
  Hyungmin Kim, Sanghyuk Heo, Yeonsu Jang, Kyoungchul Jang, Shinho Chu, Yoonna
  Oh, Kwidong Kim, Junghyun Kim, Soohwan Kim, Jeongtae Hwang, Sangil Park,
  Junphyo Lee, Inchul Jeong, Joohwan Cho, and Jonghwan Kim.
\newblock {A 1.1 V 16Gb DDR5 DRAM with Probabilistic-Aggressor Tracking,
  Refresh-Management Functionality, Per-Row Hammer Tracking, a Multi-Step
  Precharge, and Core-Bias Modulation for Security and Reliability
  Enhancement}.
\newblock In {\em ISSCC}, 2023.

\bibitem{guderamarao2023defending}
C~Gude~Ramarao, K~Tejesh Kumar, G~Ujjinappa, and B~Vasu~Deva Naidu.
\newblock {Defending SoCs with FPGAs from Rowhammer Attacks}.
\newblock {\em Material Science}, 2023.

\bibitem{guha2022criticality}
Krishnendu Guha and Amlan Chakrabarti.
\newblock {Criticality Based Reliability from Rowhammer Attacks in
  Multi-User-Multi-FPGA Platform}.
\newblock In {\em VLSID}, 2022.

\bibitem{france2022reducing}
Lo{\"\i}c France, Florent Bruguier, David Novo, Maria Mushtaq, and Pascal
  Benoit.
\newblock {Reducing the Silicon Area Overhead of Counter-Based Rowhammer
  Mitigations}.
\newblock In {\em 18th CryptArchi Workshop}, 2022.

\bibitem{arikan2022processor}
Kerem Ar{\i}kan, Alessandro Palumbo, Luca Cassano, Pedro Reviriego, Salvatore
  Pontarelli, Giuseppe Bianchi, O{\u{g}}uz Ergin, and Marco Ottavi.
\newblock {Processor Security: Detecting Microarchitectural Attacks via
  Count-Min Sketches}.
\newblock {\em VLSI}, 2022.

\bibitem{saxena2023ptguard}
Anish Saxena, Gururaj Saileshwar, Jonas Juffinger, Andreas Kogler, Daniel
  Gruss, and Moinuddin Qureshi.
\newblock {PT-Guard: Integrity-Protected Page Tables to Defend Against
  Breakthrough Rowhammer Attacks}.
\newblock In {\em DSN}, 2023.

\bibitem{bennett2021panopticon}
Tanj Bennett, Stefan Saroiu, Alec Wolman, and Lucian Cojocar.
\newblock {Panopticon: A Complete In-DRAM Rowhammer Mitigation}.
\newblock In {\em DRAMSec}, 2021.

\bibitem{gomez2016dram_rowhammer}
Hector Gomez, Andres Amaya, and Elkim Roa.
\newblock {DRAM Row-hammer Attack Reduction Using Dummy Cells}.
\newblock In {\em NORCAS}, 2016.

\bibitem{woo2023rampart}
Steven~C Woo, Wendy Elsasser, Mike Hamburg, Eric Linstadt, Michael~R Miller,
  Taeksang Song, and James Tringali.
\newblock {RAMPART: RowHammer Mitigation and Repair for Server Memory Systems}.
\newblock In {\em MEMSYS}, 2023.

\bibitem{gautam2018improvement}
Satendra~Kumar Gautam, Arvind Kumar, and Sanjeev~Kumar Manhas.
\newblock {Improvement of Row Hammering Using Metal Nanoparticles in DRAM—A
  Simulation Study}.
\newblock {\em IEEE EDL}, 2018.

\bibitem{jedec2024jesd795c}
{JEDEC}.
\newblock {\em {{JESD79-5c: DDR5 SDRAM Standard}}}, 2024.

\bibitem{canpolat2024understanding}
O{\u{g}}uzhan Canpolat, A.~Giray Ya{\u{g}}l{\i}k{\c{c}}{\i}, Geraldo~F
  Oliveira, Ataberk Olgun, O{\u{g}}uz Ergin, and Onur Mutlu.
\newblock {Understanding the Security Benefits and Overheads of Emerging
  Industry Solutions to DRAM Read Disturbance}.
\newblock In {\em DRAMSec}, 2024.

\bibitem{hassan2024selfmanaging}
Hasan Hassan, Ataberk Olgun, A~Giray Ya{\u{g}}l{\i}k{\c{c}}{\i}, Haocong Luo,
  and Onur Mutlu.
\newblock {A Case for Self-Managing DRAM Chips: Improving Performance,
  Efficiency, Reliability, and Security via Autonomous In-DRAM Maintenance
  Operations}.
\newblock In {\em MICRO}, 2024.

\bibitem{qureshi2024moat}
Moinuddin Qureshi and Salman Qazi.
\newblock {MOAT: Securely Mitigating Rowhammer with Per-Row Activation
  Counters}.
\newblock arXiv:2407.09995 [cs.CR], 2024.

\bibitem{safari2024ramulator2}
{SAFARI}.
\newblock {Ramulator 2.0}.
\newblock \url{https://github.com/CMU-SAFARI/ramulator2}, 2024.

\bibitem{jaleel2024pride}
Aamer Jaleel, Gururaj Saileshwar, Stephen~W Keckler, and Moinuddin Qureshi.
\newblock {PrIDE: Achieving Secure Rowhammer Mitigation with Low-Cost In-DRAM
  Trackers}.
\newblock In {\em ISCA}, 2024.

\bibitem{qureshi2024mint}
Moinuddin Qureshi, Salman Qazi, and Aamer Jaleel.
\newblock {MINT: Securely Mitigating Rowhammer with a Minimalist In-DRAM
  Tracker}.
\newblock In {\em MICRO}, 2024.

\bibitem{qureshi2024impress}
Moinuddin Qureshi, Anish Saxena, and Aamer Jaleel.
\newblock {ImPress: Securing DRAM Against Data-Disturbance Errors via Implicit
  Row-Press Mitigation}.
\newblock In {\em MICRO}, 2024.

\bibitem{rixner2000memory}
S.~Rixner, W.~J. Dally, U.~J. Kapasi, P.~Mattson, and J.~D. Owens.
\newblock {Memory Access Scheduling}.
\newblock In {\em ISCA}, 2000.

\bibitem{rixner2004memory}
Scott Rixner.
\newblock {Memory Controller Optimizations for Web Servers}.
\newblock In {\em MICRO}, 2004.

\bibitem{moscibroda2007memory}
Thomas Moscibroda and Onur Mutlu.
\newblock {Memory Performance Attacks: Denial of Memory Service in Multi-Core
  Systems}.
\newblock In {\em USENIX Security}, 2007.

\bibitem{mutlu2007stalltime}
Onur Mutlu and Thomas Moscibroda.
\newblock {Stall-Time Fair Memory Access Scheduling for Chip Multiprocessors}.
\newblock In {\em MICRO}, 2007.

\bibitem{mutlu2008parallelismaware}
Onur Mutlu and Thomas Moscibroda.
\newblock {Parallelism-Aware Batch Scheduling: Enhancing Both Performance and
  Fairness of Shared DRAM Systems}.
\newblock In {\em ISCA}, 2008.

\bibitem{lee2008prefetchaware}
Chang~Joo Lee, Onur Mutlu, Veynu Narasiman, and Yale~N Patt.
\newblock {Prefetch-Aware DRAM Controllers}.
\newblock In {\em MICRO}, 2008.

\bibitem{kim2010atlas}
Yoongu Kim, Dongsu Han, Onur Mutlu, and Mor Harchol-Balter.
\newblock {ATLAS: A Scalable and High-Performance Scheduling Algorithm for
  Multiple Memory Controllers}.
\newblock In {\em HPCA}, 2010.

\bibitem{kim2010thread}
Yoongu Kim, Michael Papamichael, Onur Mutlu, and Mor Harchol-Balter.
\newblock {Thread Cluster Memory Scheduling: Exploiting Differences in Memory
  Access Behavior}.
\newblock In {\em MICRO}, 2010.

\bibitem{subramanian2014theblacklisting}
Lavanya Subramanian, Donghyuk Lee, Vivek Seshadri, Harsha Rastogi, and Onur
  Mutlu.
\newblock {The Blacklisting Memory Scheduler: Achieving High Performance and
  Fairness at Low Cost}.
\newblock In {\em ICCD}, 2014.

\bibitem{subramanian2016bliss}
Lavanya Subramanian, Donghyuk Lee, Vivek Seshadri, Harsha Rastogi, and Onur
  Mutlu.
\newblock {BLISS: Balancing Performance, Fairness and Complexity in Memory
  Access Scheduling}.
\newblock {\em {TPDS}}, 2016.

\bibitem{ausavarungnirun2012staged}
Rachata Ausavarungnirun, Kevin Kai-Wei Chang, Lavanya Subramanian, Gabriel~H
  Loh, and Onur Mutlu.
\newblock {Staged Memory Scheduling: Achieving High Performance and Scalability
  in Heterogeneous Systems}.
\newblock In {\em ISCA}, 2012.

\bibitem{ebrahimi2011parallel}
Eiman Ebrahimi, Rustam Miftakhutdinov, Chris Fallin, Chang~Joo Lee, Jos{\'e}~A
  Joao, Onur Mutlu, and Yale~N Patt.
\newblock {Parallel Application Memory Scheduling}.
\newblock In {\em MICRO}, 2011.

\bibitem{ebrahimi2010fairness}
Eiman Ebrahimi, Chang~Joo Lee, Onur Mutlu, and Yale~N Patt.
\newblock {Fairness via Source Throttling: A Configurable and High-Performance
  Fairness Substrate for Multi-Core Memory Systems}.
\newblock In {\em ASPLOS}, 2010.

\bibitem{ebrahimi2011prefetchaware}
Eiman Ebrahimi, Chang~Joo Lee, Onur Mutlu, and Yale~N. Patt.
\newblock {Prefetch-Aware Shared Resource Management for Multi-Core Systems}.
\newblock In {\em ISCA}, 2011.

\bibitem{nychis2012onchip}
George Nychis, Chris Fallin, Thomas Moscibroda, Onur Mutlu, and Srinivasan
  Seshan.
\newblock {On-Chip Networks from a Networking Perspective: Congestion and
  Scalability in Many-Core Interconnects}.
\newblock In {\em SIGCOMM}, 2012.

\bibitem{nychis2010next}
George Nychis, Chris Fallin, Thomas Moscibroda, and Onur Mutlu.
\newblock {Next Generation On-Chip Networks: What Kind of Congestion Control Do
  We Need?}
\newblock In {\em HOTNETS}, 2010.

\bibitem{chang2012hatheterogeneous}
Kevin~KaiWei Chang, Rachata Ausavarungnirun, Chris Fallin, and Onur Mutlu.
\newblock {HAT: Heterogeneous Adaptive Throttling for On-Chip Networks}.
\newblock In {\em SBAC-PAD}, 2012.

\bibitem{usui2016dash}
Hiroyuki Usui, Lavanya Subramanian, Kevin Kai-Wei Chang, and Onur Mutlu.
\newblock {DASH: Deadline-Aware High-Performance Memory Scheduler for
  Heterogeneous Systems with Hardware Accelerators}.
\newblock {\em TACO}, 2016.

\bibitem{lugo2022survey}
Tamara Lugo, Santiago Lozano, Javier Fern{\'a}ndez, and Jesus Carretero.
\newblock {A Survey of Techniques for Reducing Interference in Real-Time
  Applications on Multicore Platforms}.
\newblock {\em IEEE Access}, 10, 2022.

\bibitem{kim2016bounding}
Hyoseung Kim, Dionisio De~Niz, Bj\"{o}rn Andersson, Mark Klein, Onur Mutlu, and
  Ragunathan Rajkumar.
\newblock {Bounding and Reducing Memory Interference Delay in COTS-Based
  Multi-Core Systems}.
\newblock {\em {RTS}}, 2016.

\bibitem{hassan2015aframework}
Mohamed Hassan, Hiren Patel, and Rodolfo Pellizzoni.
\newblock {A Framework for Scheduling DRAM Memory Accesses for Multi-Core
  Mixed-Time Critical Systems}.
\newblock In {\em RTAS}, 2015.

\bibitem{zhou2016mitts}
Yanqi Zhou and David Wentzlaff.
\newblock {MITTS: Memory Inter-Arrival Time Traffic Shaping}.
\newblock In {\em ISCA}, 2016.

\bibitem{farshchi2020bru}
Farzad Farshchi, Qijing Huang, and Heechul Yun.
\newblock {BRU: Bandwidth Regulation Unit for Real-Time Multicore Processors}.
\newblock In {\em RTAS}, 2020.

\bibitem{sun2015response}
Youcheng Sun and Giuseppe Lipari.
\newblock {Response Time Analysis with Limited Carry-In for Global Earliest
  Deadline First Scheduling}.
\newblock In {\em RTSS}, 2015.

\bibitem{liu2012raidr}
Jamie Liu, Ben Jaiyen, Richard Veras, and Onur Mutlu.
\newblock {RAIDR: Retention-Aware Intelligent DRAM Refresh}.
\newblock In {\em ISCA}, 2012.

\bibitem{lin2012secret}
Chung-Hsiang Lin, De-Yu Shen, Yi-Jung Chen, Chia-Lin Yang, and Michael Wang.
\newblock {SECRET: Selective Error Correction for Refresh Energy Reduction in
  DRAMs}.
\newblock In {\em ICCD}, 2012.

\bibitem{nair2013archshield}
Prashant~J Nair, Dae-Hyun Kim, and Moinuddin~K Qureshi.
\newblock {ArchShield: Architectural Framework for Assisting DRAM Scaling by
  Tolerating High Error Rates}.
\newblock In {\em ISCA}, 2013.

\bibitem{baek2013refresh}
Seungjae Baek, Sangyeun Cho, and Rami Melhem.
\newblock {Refresh Now and Then}.
\newblock {\em IEEE TC}, 2013.

\bibitem{isen2009eskimoenergy}
Ciji Isen and Lizy John.
\newblock {ESKIMO-Energy Savings Using Semantic Knowledge of Inconsequential
  Memory Occupancy for DRAM Subsystem}.
\newblock In {\em MICRO}, 2009.

\bibitem{liu2011flikker}
Song Liu, Karthik Pattabiraman, Thomas Moscibroda, and Benjamin~G. Zorn.
\newblock {Flikker: Saving DRAM Refresh-Power through Critical Data
  Partitioning}.
\newblock In {\em ASPLOS}, 2011.

\bibitem{jafri2021refresh}
Syed M. A.~H. Jafri, Hasan Hassan, Ahmed Hemani, and Onur Mutlu.
\newblock {Refresh Triggered Computation: Improving the Energy Efficiency of
  Convolutional Neural Network Accelerators}.
\newblock {\em TACO}, 2021.

\bibitem{katayama1999faulttolerant}
Yasunao Katayama, Eric~J Stuckey, Sumio Morioka, and Zhao Wu.
\newblock {Fault-Tolerant Refresh Power Reduction of DRAMs for
  Quasi-Nonvolatile Data Retention}.
\newblock In {\em DFT}, 1999.

\bibitem{wilkerson2010reducing}
Chris Wilkerson, Alaa~R. Alameldeen, Zeshan Chishti, Wei Wu, Dinesh Somasekhar,
  and Shih-lien Lu.
\newblock {Reducing Cache Power with Low-Cost, Multi-Bit Error-Correcting
  Codes}.
\newblock In {\em ISCA}, 2010.

\bibitem{ghosh2007smart}
Mrinmoy Ghosh and Hsien-Hsin~S Lee.
\newblock {Smart Refresh: An Enhanced Memory Controller Design for Reducing
  Energy in Conventional and 3D Die-Stacked DRAMs}.
\newblock In {\em MICRO}, 2007.

\bibitem{song2000method}
S.~P. Song.
\newblock {Method and System for Selective DRAM Refresh to Reduce Power
  Consumption}, 2000.

\bibitem{emma2008rethinking}
Philip~G Emma, William~R Reohr, and Mesut Meterelliyoz.
\newblock {Rethinking Refresh: Increasing Availability and Reducing Power in
  DRAM for Cache Applications}.
\newblock {\em IEEE Micro}, 2008.

\bibitem{mukhanov2019workloadaware}
Lev Mukhanov, Konstantinos Tovletoglou, Hans Vandierendonck, Dimitrios~S
  Nikolopoulos, and Georgios Karakonstantis.
\newblock {Workload-Aware DRAM Error Prediction Using Machine Learning}.
\newblock In {\em IISWC}, 2019.

\bibitem{hong2018eareccaided}
Jeongkyu Hong, Hyeonggyu Kim, and Soontae Kim.
\newblock {EAR: ECC-Aided Refresh Reduction through 2-D Zero Compression}.
\newblock In {\em PACT}, 2018.

\bibitem{kraft2018improving}
Kira Kraft, Chirag Sudarshan, Deepak~M Mathew, Christian Weis, Norbert Wehn,
  and Matthias Jung.
\newblock {Improving the Error Behavior of DRAM by Exploiting Its Z-Channel
  Property}.
\newblock In {\em DATE}, 2018.

\bibitem{park2011power}
Hyunsun Park, Sungjoo Yoo, and Sunggu Lee.
\newblock {Power Management of Hybrid DRAM/PRAM-Based Main Memory}.
\newblock In {\em DAC}, 2011.

\bibitem{wang2018content}
Shibo Wang, Mahdi~Nazm Bojnordi, Xiaochen Guo, and Engin Ipek.
\newblock {Content Aware Refresh: Exploiting the Asymmetry of DRAM Retention
  Errors to Reduce the Refresh Frequency of Less Vulnerable Data}.
\newblock {\em IEEE TC}, 2018.

\bibitem{kim2020chargeaware}
Seikwon Kim, Wonsang Kwak, Changdae Kim, Daehyeon Baek, and Jaehyuk Huh.
\newblock {Charge-Aware DRAM Refresh Reduction with Value Transformation}.
\newblock In {\em HPCA}, 2020.

\bibitem{mrozek2010analysis}
Ireneusz Mrozek.
\newblock {Analysis of Multibackground Memory Testing Techniques}.
\newblock {\em IJAMCS}, 2010.

\bibitem{mrozek2019multirun}
Ireneusz Mrozek.
\newblock {\em {Multi-Run Memory Tests for Pattern Sensitive Faults}}.
\newblock Springer, 2019.

\bibitem{luo2020clrdram}
Haocong Luo, Taha Shahroodi, Hasan Hassan, Minesh Patel, Abdullah~Giray
  Ya{\u{g}}l{\i}k{\c{c}}{\i}, Lois Orosa, Jisung Park, and Onur Mutlu.
\newblock {CLR-DRAM: A Low-Cost DRAM Architecture Enabling Dynamic
  Capacity-Latency Trade-Off}.
\newblock In {\em ISCA}, 2020.

\bibitem{kim2000dynamic}
Joohee Kim and Marios~C. Papaefthymiou.
\newblock {Dynamic Memory Design for Low Data-Retention Power}.
\newblock In {\em PATMOS}, 2000.

\bibitem{kim2003blockbased}
Joohee Kim and M.C. Papaefthymiou.
\newblock {Block-Based Multiperiod Dynamic Memory Design for Low Data-Retention
  Power}.
\newblock {\em TVLSI}, 2003.

\bibitem{yanagisawa1988semiconductor}
K.~Yanagisawa.
\newblock {Semiconductor Memory}.
\newblock {US Patent 4,736,344}, 1988.

\bibitem{ohsawa1998optimizing}
Taku Ohsawa, Koji Kai, and Kazuaki Murakami.
\newblock {Optimizing the DRAM Refresh Count for Merged DRAM/Logic LSIs}.
\newblock In {\em ISLPED}, 1998.

\bibitem{nair2014refresh}
Prashant~J Nair, Chia-Chen Chou, and Moinuddin~K Qureshi.
\newblock {Refresh Pausing in DRAM Memory Systems}.
\newblock {\em TACO}, 2014.

\bibitem{orosa2021codic}
Lois Orosa, Yaohua Wang, Mohammad Sadrosadati, Jeremie~S. Kim, Minesh Patel,
  Ivan Puddu, Haocong Luo, Kaveh Razavi, Juan Gómez-Luna, Hasan Hassan, Nika
  Mansouri-Ghiasi, Saugata Ghose, and Onur Mutlu.
\newblock {CODIC: A Low-Cost Substrate for Enabling Custom In-DRAM
  Functionalities and Optimizations}.
\newblock In {\em ISCA}, 2021.

\bibitem{choi2020reducing}
Haerang Choi, Dosun Hong, Jaesung Lee, and Sungjoo Yoo.
\newblock {Reducing DRAM Refresh Power Consumption by Runtime Profiling of
  Retention Time and Dual-Row Activation}.
\newblock {\em MICPRO}, 2020.

\bibitem{mukundan2013understanding}
Janani Mukundan, Hillery Hunter, Kyu-hyoun Kim, Jeffrey Stuecheli, and
  Jos{\'e}~F Mart{\'\i}nez.
\newblock {Understanding and Mitigating Refresh Overheads in High-Density DDR4
  DRAM Systems}.
\newblock In {\em ISCA}, 2013.

\bibitem{stuecheli2010elastic}
Jeffrey Stuecheli, Dimitris Kaseridis, Hillery C.Hunter, and Lizy~K. John.
\newblock {Elastic Refresh: Techniques to Mitigate Refresh Penalties in High
  Density Memory}.
\newblock In {\em MICRO}, 2010.

\bibitem{pan2019thecolored}
Xing Pan and Frank Mueller.
\newblock {The Colored Refresh Server for DRAM}.
\newblock In {\em ISORC}, 2019.

\bibitem{pan2019hiding}
Xing Pan and Frank Mueller.
\newblock {Hiding DRAM Refresh Overhead in Real-Time Cyclic Executives}.
\newblock In {\em RTSS}, 2019.

\bibitem{kotra2017hardwaresoftware}
Jagadish~B Kotra, Narges Shahidi, Zeshan~A Chishti, and Mahmut~T Kandemir.
\newblock {Hardware-Software Co-Design to Mitigate DRAM Refresh Overheads: A
  Case for Refresh-Aware Process Scheduling}.
\newblock {\em ASPLOS}, 2017.

\bibitem{hassan2016chargecache}
Hasan Hassan, Gennady Pekhimenko, Nandita Vijaykumar, Vivek Seshadri, Donghyuk
  Lee, Oguz Ergin, and Onur Mutlu.
\newblock {ChargeCache: Reducing DRAM Latency by Exploiting Row Access
  Locality}.
\newblock In {\em HPCA}, 2016.

\bibitem{das2018vrldram}
Anup Das, Hasan Hassan, and Onur Mutlu.
\newblock {VRL-DRAM: Improving DRAM Performance via Variable Refresh Latency}.
\newblock In {\em DAC}, 2018.

\bibitem{wang2018reducing}
Yaohua Wang, Arash Tavakkol, Lois Orosa, Saugata Ghose, Nika~Mansouri Ghiasi,
  Minesh Patel, Jeremie~S Kim, Hasan Hassan, Mohammad Sadrosadati, and Onur
  Mutlu.
\newblock {Reducing DRAM Latency via Charge-Level-Aware Look-Ahead Partial
  Restoration}.
\newblock In {\em {MICRO}}, 2018.

\bibitem{zhang2016restore}
Xianwei Zhang, Youtao Zhang, Bruce~R. Childers, and Jun Yang.
\newblock {Restore Truncation for Performance Improvement in Future DRAM
  Systems}.
\newblock In {\em HPCA}, 2016.

\bibitem{shin2014nuat}
Wongyu Shin, Jeongmin Yang, Jungwhan Choi, and Lee-Sup Kim.
\newblock {NUAT: A Non-Uniform Access Time Memory Controller}.
\newblock In {\em HPCA}, 2014.

\bibitem{lee2013tieredlatency}
Donghyuk Lee, Yoongu Kim, Vivek Seshadri, Jamie Liu, Lavanya Subramanian, and
  Onur Mutlu.
\newblock {Tiered-Latency DRAM: A Low Latency and Low Cost DRAM Architecture}.
\newblock In {\em {HPCA}}, 2013.

\bibitem{shin2015dramlatency_optimization}
Wongyu Shin, Jungwhan Choi, Jaemin Jang, Jinwoong Suh, Youngsuk Moon, Yongkee
  Kwon, and Lee-Sup Kim.
\newblock {DRAM-Latency Optimization Inspired by Relationship between
  Row-Access Time and Refresh Timing}.
\newblock {\em IEEE TC}, 2015.

\bibitem{mathew2017using}
Deepak~M. Mathew, \'{E}der~F. Zulian, Matthias Jung, Kira Kraft, Christian
  Weis, Bruce Jacob, and Norbert Wehn.
\newblock {Using Run-Time Reverse-Engineering to Optimize DRAM Refresh}.
\newblock In {\em MEMSYS}, 2017.

\bibitem{mathew2020using}
Deepak~M. Mathew, Matthias Jung, Christian Weis, and Norbert Wehn.
\newblock {Using Run-Time Reverse Engineering to Optimize DRAM Refresh}.
\newblock US Patent 10,622,054B2, 2017.

\bibitem{chen2005modeling}
Qikai Chen et~al.
\newblock {Modeling and Testing of SRAM for New Failure Mechanisms Due to
  Process Variations in Nanoscale CMOS}.
\newblock In {\em VTS}, 2005.

\bibitem{guo2009largescale}
Zheng Guo et~al.
\newblock {Large-Scale SRAM Variability Characterization in 45 nm CMOS}.
\newblock {\em JSSC}, 2009.

\bibitem{kim2011variationaware}
Daeyeon Kim et~al.
\newblock {Variation-Aware Static and Dynamic Writability Analysis for
  Voltage-Scaled Bit-Interleaved 8-T SRAMs}.
\newblock In {\em ISLPED}, 2011.

\bibitem{lee2008drain}
Yung-Huei Lee, Neal Mielke, William McMahon, Yin-Lung~R. Lu, Qingru Meng, and
  Linda Jiang.
\newblock {Drain Read Disturb Assessment of NOR Flash Memory}.
\newblock In {\em VLSI-TSA}, 2008.

\bibitem{cooke2007theinconvenient}
Jim Cooke.
\newblock {The Inconvenient Truths of NAND Flash Memory}.
\newblock In {\em Flash Memory Summit}, 2007.

\bibitem{grupp2009characterizing}
Laura~M. Grupp, Adrian~M. Caulfield, Joel Coburn, Steven Swanson, Eitan
  Yaakobi, Paul~H. Siegel, and Jack~K. Wolf.
\newblock {Characterizing Flash Memory: Anomalies, Observations, and
  Applications}.
\newblock In {\em MICRO}, 2009.

\bibitem{ha2013areaddisturb}
Keonsoo Ha, Jaeyong Jeong, and Jihong Kim.
\newblock {A Read-Disturb Management Technique for High-Density NAND Flash
  Memory}.
\newblock In {\em {APSys}}, 2013.

\bibitem{mielke2008biterror}
Neal Mielke, Todd Marquart, Ning Wu, Jeff Kessenich, Hanmant Belgal, Eric
  Schares, Falgun Trivedi, Evan Goodness, and Leland~R. Nevill.
\newblock {Bit Error Rate in NAND Flash Memories}.
\newblock In {\em IEEE International Reliability Physics Symposium}, 2008.

\bibitem{sugahara2014memory}
Takahiko Sugahara and Tetsuo Furuichi.
\newblock {Memory Controller for Suppressing Read Disturb When Data is
  Repeatedly Read Out}.
\newblock US Patent 8,725,952, 2014.

\bibitem{cai2017errors}
Yu~Cai, Saugata Ghose, Erich~F Haratsch, Yixin Luo, and Onur Mutlu.
\newblock {Errors in Flash-Memory-Based Solid-State Drives: Analysis,
  Mitigation, and Recovery}.
\newblock {\em arXiv:1711.11427 [cs.AR]}, 2017.

\bibitem{watanabe2018system}
Hikaru Watanabe, Yoshiaki Deguchi, Atsuro Kobayashi, Chihiro Matsui, and Ken
  Takeuchi.
\newblock {System-Level Read Disturb Suppression Techniques of TLC NAND Flash
  Memories for Read-Hot/Cold Data Mixed Applications}.
\newblock {\em Solid-State Electronics}, 2018.

\bibitem{cai2018errors}
Yu~Cai, Saugata Ghose, Erich~F Haratsch, Yixin Luo, and Onur Mutlu.
\newblock {Errors in Flash-Memory-Based Solid-State Drives: Analysis,
  Mitigation, and Recovery}.
\newblock {\em Inside Solid State Drives}, 2018.

\bibitem{mutlu2018guest}
Onur Mutlu, Saugata Ghose, and Rachata Ausavarungnirun.
\newblock {Guest Editor Introduction: Recent Advances in DRAM and Flash Memory
  Architectures}.
\newblock {\em IPSI TIR}, 14, 2018.

\bibitem{jiang2003crosstrack}
Wen Jiang et~al.
\newblock {Cross-Track Noise Profile Measurement for Adjacent-Track
  Interference Study and Write-Current Optimization in Perpendicular
  Recording}.
\newblock {\em Journal of Applied Physics}, 2003.

\bibitem{tang2008understanding}
Yuhui Tang et~al.
\newblock {Understanding Adjacent Track Erasure in Discrete Track Media}.
\newblock {\em Transactions on Magnetics}, 2008.

\bibitem{wood2009thefeasibility}
R~Wood et~al.
\newblock {The Feasibility of Magnetic Recording at 10 Terabits Per Square Inch
  on Conventional Media}.
\newblock {\em Transactions on Magnetics}, 2009.

\bibitem{lee2009architecting}
Benjamin~C Lee, Engin Ipek, Onur Mutlu, and Doug Burger.
\newblock {Architecting Phase Change Memory as a Scalable DRAM Alternative}.
\newblock In {\em ISCA}, 2009.

\bibitem{zhou2009adurable}
Ping Zhou, Bo~Zhao, Jun Yang, and Youtao Zhang.
\newblock {A Durable and Energy Efficient Main Memory Using Phase Change Memory
  Technology}.
\newblock In {\em ISCA}, 2009.

\bibitem{qureshi2009scalable}
Moinuddin~K Qureshi, Vijayalakshmi Srinivasan, and Jude~A Rivers.
\newblock {Scalable High Performance Main Memory System Using Phase-Change
  Memory Technology}.
\newblock In {\em ISCA}, 2009.

\bibitem{qureshi2009enhancing}
Moinuddin~K. Qureshi, John Karidis, Michele Franceschini, Vijayalakshmi
  Srinivasan, Luis Lastras, and Bulent Abali.
\newblock {Enhancing Lifetime and Security of Phase Change Memories via
  Start-Gap Wear Leveling}.
\newblock In {\em MICRO}, 2009.

\bibitem{wong2010phase}
H-S~Philip Wong, Simone Raoux, SangBum Kim, Jiale Liang, John~P Reifenberg,
  Bipin Rajendran, Mehdi Asheghi, and Kenneth~E Goodson.
\newblock {Phase Change Memory}.
\newblock {\em Proc. IEEE}, 2010.

\bibitem{raoux2008phasechange}
S.~Raoux, G.~W. Burr, M.~J. Breitwisch, C.~T. Rettner, Y.-C. Chen, R.~M.
  Shelby, M.~Salinga, D.~Krebs, S.-H. Chen, H.-L. Lung, and C.~H. Lam.
\newblock {Phase-Change Random Access Memory: A Scalable Technology}.
\newblock {\em IBM JRD}, 2008.

\bibitem{lee2010phase}
Benjamin~C Lee, Engin Ipek, Onur Mutlu, and Doug Burger.
\newblock {Phase Change Memory Architecture and the Quest for Scalability}.
\newblock In {\em CACM}, 2010.

\bibitem{lee2010phasechange}
Benjamin~C Lee, Ping Zhou, Jun Yang, Youtao Zhang, Bo~Zhao, Engin Ipek, Onur
  Mutlu, and Doug Burger.
\newblock {Phase-Change Technology and the Future of Main Memory}.
\newblock {\em IEEE Micro}, 2010.

\bibitem{yoon2012rowbuffer}
HanBin Yoon, Justin Meza, Rachata Ausavarungnirun, Rachael~A Harding, and Onur
  Mutlu.
\newblock {Row Buffer Locality Aware Caching Policies for Hybrid Memories}.
\newblock In {\em ICCD}, 2012.

\bibitem{yoon2014efficient}
Hanbin Yoon, Justin Meza, Naveen Muralimanohar, Norman~P. Jouppi, and Onur
  Mutlu.
\newblock {Efficient Data Mapping and Buffering Techniques for Multi-Level Cell
  Phase-Change Memories}.
\newblock {\em TACO}, 2014.

\bibitem{chen2010advances}
E.~Chen, D.~Apalkov, Z.~Diao, A.~Driskill-Smith, D.~Druist, D.~Lottis,
  V.~Nikitin, X.~Tang, S.~Watts, S.~Wang, S.~A. Wolf, A.~W. Ghosh, J.~W. Lu,
  S.~J. Poon, M.~Stan, W.~H. Butler, S.~Gupta, C.~K.~A. Mewes, Tim Mewes, and
  P.~B Visscher.
\newblock {Advances and Future Prospects of Spin-Transfer Torque Random Access
  Memory}.
\newblock {\em IEEE Transactions on Magnetics}, 2010.

\bibitem{kultursay2013evaluating}
Emre K{\"{u}}lt{\"{u}}rsay, Mahmut Kandemir, Anand Sivasubramaniam, and Onur
  Mutlu.
\newblock {Evaluating STT-RAM as an Energy-Efficient Main Memory Alternative}.
\newblock In {\em ISPASS}, 2013.

\bibitem{wong2012metaloxide}
H-S~Philip Wong, Heng-Yuan Lee, Shimeng Yu, Yu-Sheng Chen, Yi~Wu, Pang-Shiu
  Chen, Byoungil Lee, Frederick~T Chen, and Ming-Jinn Tsai.
\newblock {Metal--Oxide RRAM}.
\newblock {\em Proc. IEEE}, 2012.

\bibitem{staudigl2024itsgetting}
Felix Staudigl, Hazem Al~Indari, Daniel Sch{\"o}n, Hsin-Yu Chen, Dominik
  Sisejkovic, Jan~Moritz Joseph, Vikas Rana, Stephan Menzel, Amelie Hagelauer,
  and Rainer Leupers.
\newblock {It's Getting Hot in Here: Hardware Security Implications of Thermal
  Crosstalk on ReRAMs}.
\newblock {\em IEEE Transactions on Reliability}, 2024.

\bibitem{kumar2023fault}
Ankit Kumar, Robin Degraeve, Arthur Beckers, Andrea Fantini, Ingrid
  Verbauwhede, Dimitri Linten, and Gouri~S Kar.
\newblock {Fault Attack Investigation on TaOx Resistive-RAM for Cyber Secure
  Application}.
\newblock {\em IEEE Transactions on Electron Devices}, 2023.

\bibitem{meza2013acase}
Justin Meza, Yixin Luo, Samira Khan, Jishen Zhao, Yuan Xie, and Onur Mutlu.
\newblock {A Case for Efficient Hardware/Software Cooperative Management of
  Storage and Memory}.
\newblock In {\em WEED}, 2013.

\bibitem{he2023whistleblower}
Wei He, Zhi Zhang, Yueqiang Cheng, Wenhao Wang, Wei Song, Yansong Gao, Qifei
  Zhang, Kang Li, Dongxi Liu, and Surya Nepal.
\newblock {WhistleBlower: A System-Level Empirical Study on RowHammer}.
\newblock {\em IEEE TC}, 2023.

\bibitem{baeg2022estimation}
Sanghyeon Baeg, Donghyuk Yun, Myungsun Chun, and Shi-Jie Wen.
\newblock {Estimation of the Trap Energy Characteristics of Row Hammer-Affected
  Cells in Gamma-Irradiated DDR4 DRAM}.
\newblock {\em IEEE Transactions on Nuclear Science}, 2022.

\bibitem{xilinx2012ml605}
{Xilinx}.
\newblock {\em {ML605 Hardware User Guide}}, 2012.

\bibitem{xilinx2011virtex6}
{Xilinx}.
\newblock {\em {Virtex-6 FPGA Integrated Block for PCI Express}}, 2011.

\bibitem{jedec1995eiajesd511}
{JEDEC}.
\newblock {\em {JESD51-1: Integrated Circuits Thermal Measurement Method -
  Electrical Test Method (Single Semiconductor Device)}}, 1995.

\bibitem{maxwellft20x}
{Maxwell}.
\newblock {FT20X User Manual}.
\newblock \url{https://www.maxwell-fa.com/upload/files/base/8/m/311.pdf}.

\bibitem{instruments2014rs485}
{Texas Instruments}.
\newblock {RS-485}.
\newblock
  \url{https://web.archive.org/web/20180517101401/http://www.ti.com/lit/sg/slyt484a/slyt484a.pdf},
  2014.

\bibitem{micron2002tn0008}
{Micron Technology}.
\newblock {TN-00-08: Thermal Applications}, 2002.

\bibitem{xilinxultrascale}
{Xilinx}.
\newblock {UltraScale Architecture-Based FPGAs Memory IP v1.4}.
\newblock
  \url{https://www.xilinx.com/support/documentation/ip_documentation/ultrascale_memory_ip/v1_4/pg150-ultrascale-memory-ip.pdf}.

\bibitem{hamming1950error}
Richard~W Hamming.
\newblock {Error Detecting and Error Correcting Codes}.
\newblock {\em The Bell System Technical Journal}, 1950.

\bibitem{bose1960onaclass}
Raj~Chandra Bose and Dwijendra~K Ray-Chaudhuri.
\newblock {On a Class of Error Correcting Binary Group Codes}.
\newblock {\em {Information and Control}}, 1960.

\bibitem{hocquenghem1959codes}
Alexis Hocquenghem.
\newblock {Codes Correcteurs d'Erreurs}.
\newblock {\em Chiffers}, 1959.

\bibitem{reed1960polynomial}
Irving~S Reed and Gustave Solomon.
\newblock {Polynomial Codes over Certain Finite Fields}.
\newblock {\em SIAM}, 1960.

\bibitem{kim2016allinclusive}
Jungrae Kim, Michael Sullivan, Sangkug Lym, and Mattan Erez.
\newblock {All-Inclusive ECC: Thorough End-to-End Protection for Reliable
  Computer memory}.
\newblock In {\em ISCA}, 2016.

\bibitem{micron2017eccbrings}
{Micron Technology Inc}.
\newblock {ECC Brings Reliability and Power Efficiency to Mobile Devices}.
\newblock Technical report, {Micron Technology Inc}, 2017.

\bibitem{nair2016xedexposing}
Prashant~J Nair, Vilas Sridharan, and Moinuddin~K Qureshi.
\newblock {XED: Exposing On-Die Error Detection Information for Strong Memory
  Reliability}.
\newblock In {\em ISCA}, 2016.

\bibitem{lee2014green}
J.~Lee.
\newblock {Green Memory Solution}.
\newblock {Samsung Electronics, Investor’s Forum}, 2014.

\bibitem{micron2016ddr4mta18asf2g72pz}
{Micron}.
\newblock {DDR4 SDRAM RDIMM MTA18ASF2G72PZ – 16GB}, 2016.

\bibitem{samsung2021k4a4g085wfbctd}
{Samsung}.
\newblock {K4A4G085WF-BCTD Specification}.
\newblock
  \url{https://www.samsung.com/semiconductor/dram/ddr4/K4A4G085WF-BCTD/}, 2021.

\bibitem{gskill2021f42400c17s8gnt}
{G.SKILL}.
\newblock {F4-2400C17S-8GNT Specification}.
\newblock
  \url{https://www.gskill.com/specification/165/186/1535961538/F4-2400C17S-8GNT-Specification},
  2021.

\bibitem{kingston2021kvr24n17s88}
{Kingston}.
\newblock {KVR24N17S8/8 Specification}.
\newblock \url{https://www.kingston.com/datasheets/KVR24N17S8_8.pdf}, 2021.

\bibitem{crucial2017mt41k512m8da107p}
{Crucial}.
\newblock {MT41K512M8DA-107:P}.
\newblock
  \url{https://media-www.micron.com/-/media/client/global/documents/products/data-sheet/dram/ddr3/4gb_ddr3l.pdf?rev=8d4b345161424b60bbe4886434cbccf4},
  2017.

\bibitem{samsung2013m471b5173qh0}
{Samsung}.
\newblock {M471B5173QH0 Specification}.
\newblock
  \url{https://www.samsung.com/semiconductor/global.semi/file/resource/2017/11/135V_DDR3_4Gb_Qdie_UnbufferedSODIMM_Rev121.pdf},
  2013.

\bibitem{hynix2013ddr3lHMT451S6BFR8A}
{SK Hynix}.
\newblock {DDR3L SDRAM Unbuffered SODIMMsBased on 4Gb B-die}.
\newblock
  \url{https://www.samsung.com/semiconductor/global.semi/file/resource/2017/11/135V_DDR3_4Gb_Qdie_UnbufferedSODIMM_Rev121.pdf},
  2013.

\bibitem{Tukey1977Exploratory}
John Tukey.
\newblock {\em {Exploratory Data Analysis}}.
\newblock Pearson, 1977.

\bibitem{hofmann2017lettervalue}
Heike Hofmann, Hadley Wickham, and Karen Kafadar.
\newblock {Letter-Value Plots: Boxplots for Large Data}.
\newblock {\em Journal of Computational and Graphical Statistics}, 2017.

\bibitem{biran1998thecambridge}
Everitt Biran.
\newblock {The Cambridge Dictionary of Statistics}.
\newblock {\em {Cambridge University Press}}, 1998.

\bibitem{son2013reducing}
Young~Hoon Son, O~Seongil, Yuhwan Ro, Jae~W Lee, and Jung~Ho Ahn.
\newblock {Reducing Memory Access Latency with Asymmetric DRAM Bank
  Organizations}.
\newblock In {\em ISCA}, 2013.

\bibitem{vogelsang2010understanding}
Thomas Vogelsang.
\newblock {Understanding the Energy Consumption of Dynamic Random Access
  Memories}.
\newblock In {\em MICRO}, 2010.

\bibitem{chang2016lowcost}
Kevin~K Chang, Prashant~J Nair, Donghyuk Lee, Saugata Ghose, Moinuddin~K
  Qureshi, and Onur Mutlu.
\newblock {Low-Cost Inter-Linked Subarrays (LISA): Enabling Fast Inter-Subarray
  Data Movement in DRAM}.
\newblock In {\em HPCA}, 2016.

\bibitem{wright1921correlation}
Sewall Wright.
\newblock {Correlation and Causation}.
\newblock {\em Agricultural Research}, 1921.

\bibitem{bhattacharyya1943onameasure}
Anil Bhattacharyya.
\newblock {On a Measure of Divergence between Two Statistical Populations
  Defined by Their Probability Distributions}.
\newblock {\em Bull. Calcutta Math. Soc.}, 1943.

\bibitem{naseredini2022alarm}
Amir Naseredini, Martin Berger, Matteo Sammartino, and Shale Xiong.
\newblock {ALARM: Active LeArning of Rowhammer Mitigations}.
\newblock \url{https://users.sussex.ac.uk/~mfb21/rh-draft.pdf}, 2022.

\bibitem{hong2023dsac}
Seungki Hong, Dongha Kim, Jaehyung Lee, Reum Oh, Changsik Yoo, Sangjoon Hwang,
  and Jooyoung Lee.
\newblock {DSAC: Low-Cost Rowhammer Mitigation Using In-DRAM Stochastic and
  Approximate Counting Algorithm}.
\newblock arXiv:2302.03591 [cs.CR], 2023.

\bibitem{marazzi2022rega}
Michele Marazzi, Patrick Jattke, Flavien Solt, and Kaveh Razavi.
\newblock {REGA: Scalable Rowhammer Mitigation with Refresh-Generating
  Activations}.
\newblock In {\em {IEEE S\&P}}, 2022.

\bibitem{verma2023defense}
Akash Verma, Victor van~der Veen, Joona Kannisto, and Marcel Selhorst.
\newblock {Defense Against Row Hammer Attacks}.
\newblock US Patent App. 17/842,606, 2023.

\bibitem{luo2014characterizing}
Yixin Luo, Sriram Govindan, Bikash Sharma, Mark Santaniello, Justin Meza, Aman
  Kansal, Jie Liu, Badriddine Khessib, Kushagra Vaid, and Onur Mutlu.
\newblock {Characterizing Application Memory Error Vulnerability to Optimize
  Datacenter Cost via Heterogeneous-Reliability Memory}.
\newblock In {\em DSN}, 2014.

\bibitem{koppula2019eden}
Skanda Koppula, Lois Orosa, A~Giray Ya{\u{g}}l{\i}k{\c{c}}{\i}, Roknoddin
  Azizi, Taha Shahroodi, Konstantinos Kanellopoulos, and Onur Mutlu.
\newblock {EDEN: Enabling Energy-Efficient, High-Performance Deep Neural
  Network Inference Using Approximate DRAM}.
\newblock In {\em MICRO}, 2019.

\bibitem{nguyen2018anapproximate}
Duy~Thanh Nguyen, Hyun Kim, Hyuk-Jae Lee, and Ik-Joon Chang.
\newblock {An Approximate Memory Architecture for a Reduction of Refresh Power
  Consumption in Deep Learning Applications}.
\newblock In {\em ISCAS}, 2018.

\bibitem{nguyen2019stdrc}
Duy-Thanh Nguyen, Nhut-Minh Ho, and Ik-Joon Chang.
\newblock {St-DRC: Stretchable DRAM Refresh Controller with No Parity-Overhead
  Error Correction Scheme for Energy-Efficient DNNs}.
\newblock In {\em DAC}, 2019.

\bibitem{tu2018rana}
Fengbin Tu, Weiwei Wu, Shouyi Yin, Leibo Liu, and Shaojun Wei.
\newblock {RANA: Towards Efficient Neural Acceleration with Refresh-Optimized
  Embedded DRAM}.
\newblock In {\em ISCA}, 2018.

\bibitem{yavits2020wolfram}
Leonid Yavits, Lois Orosa, Suyash Mahar, Jo{\~a}o~Dinis Ferreira, Mattan Erez,
  Ran Ginosar, and Onur Mutlu.
\newblock {WoLFRaM: Enhancing Wear-Leveling and Fault Tolerance in Resistive
  Memories Using Programmable Address Decoders}.
\newblock In {\em ICCD}, 2020.

\bibitem{carter1999impulse}
John Carter, Wilson Hsieh, Leigh Stoller, Mark Swanson, Lixin Zhang, Erik
  Brunvand, Al~Davis, Chen-Chi Kuo, Ravindra Kuramkote, Michael Parker, Lambert
  Schaelicke, and Terry Tateyama.
\newblock {Impulse: Building a Smarter Memory Controller}.
\newblock In {\em HPCA}, 1999.

\bibitem{rezaei2020nomnetworkonmemory}
Seyyed Hossein~SeyyedAghaei Rezaei, Mehdi Modarressi, Rachata Ausavarungnirun,
  Mohammad Sadrosadati, Onur Mutlu, and Masoud Daneshtalab.
\newblock {NoM: Network-on-Memory for Inter-Bank Data Transfer in Highly-Banked
  Memories}.
\newblock {\em IEEE CAL}, 2020.

\bibitem{goossens2013conservative}
Sven Goossens, Benny Akesson, and Kees Goossens.
\newblock {Conservative Open-Page Policy for Mixed Time-Criticality Memory
  Controllers}.
\newblock In {\em DATE}, 2013.

\bibitem{huan2006processor}
Dandan Huan, Zusong Li, Weiwu Hu, and Zhiyong Liu.
\newblock {Processor Directed Dynamic Page Policy}.
\newblock In {\em ACSAC}, 2006.

\bibitem{kaseridis2011minimalist}
Dimitris Kaseridis, Jeffrey Stuecheli, and Lizy~Kurian John.
\newblock {Minimalist Open-Page: A DRAM Page-Mode Scheduling Policy for the
  Many-Core Era}.
\newblock In {\em MICRO}, 2011.

\bibitem{dell1997awhite}
Timothy~J Dell.
\newblock {A White Paper on the Benefits of Chipkill-Correct ECC for PC Server
  Main Memory}.
\newblock {\em IBM Microelectronics Division}, 1997.

\bibitem{locklear2000chipkill}
David Locklear.
\newblock {Chipkill Correct Memory Architecture}.
\newblock {\em {Dell Enterprise Systems Group, Technology Brief}}, 2000.

\bibitem{jian2013adaptive}
Xun Jian and Rakesh Kumar.
\newblock {Adaptive Reliability Chipkill Correct (ARCC)}.
\newblock In {\em HPCA}, 2013.

\bibitem{qureshi2021rethinking}
Moinuddin Qureshi.
\newblock {Rethinking ECC in the Era of Row-Hammer}.
\newblock {\em {DRAMSec}}, 2021.

\bibitem{corpltspice}
Linear~Technology Corp.
\newblock {LTspice IV}.
\newblock \url{http://www.linear.com/LTspice}.

\bibitem{nagel1973spice}
Laurence~W. Nagel and D.O. Pederson.
\newblock {SPICE (Simulation Program with Integrated Circuit Emphasis)}.
\newblock Technical Report No. UCB/ERL M382, UC Berkeley, 1973.

\bibitem{adexelecddr4sodv1}
Adexelec.
\newblock {DDR4-SOD-V1 260-pin 1.2V, DDR4 SODIMM Vertical Extender with CSR
  Option}.
\newblock \url{http://www.adexelec.com/ddr4-sod-v1}.

\bibitem{ttiplplp}
TTi.
\newblock {PL \& PL-P Series DC Power Supplies Data Sheet - Issue 5}.
\newblock
  \url{https://resources.aimtti.com/datasheets/AIM-PL+PL-P_series_DC_power_supplies_data_sheet-Iss5.pdf}.

\bibitem{micronddr4mta18asf2g72pz}
{Micron}.
\newblock {DDR4 SDRAM RDIMM MTA18ASF2G72PZ – 16GB}.
\newblock
  \url{https://www.micro-semiconductor.com/datasheet/7c-MTA18ASF2G72PZ-2G9E1.pdf}.

\bibitem{crucialct4g4dfs8266}
{Crucial}.
\newblock {CT4G4DFS8266}.
\newblock \url{https://www.crucial.com/memory/eol_ddr4/ct4g4dfs8266}.

\bibitem{corsairskucmv4gx4m1a2133c15}
{CORSAIR}.
\newblock {SKU CMV4GX4M1A2133C15 Specification}.
\newblock \url{https://tinyurl.com/CMV4GX4M1A2133C15}.

\bibitem{samsung2018288pin}
{Samsung}.
\newblock {288pin Unbuffered DIMM based on 8Gb D-die, Rev 1.1}.
\newblock
  \url{https://semiconductor.samsung.com/resources/data-sheet/DDR4_8Gb_D_die_Unbuffered_DIMM_Rev1.1_Jun.18.pdf},
  2018.

\bibitem{samsung2017288pin}
{Samsung}.
\newblock {288pin Registered DIMM based on 8Gb B-die, Rev 1.91}.
\newblock
  \url{https://semiconductor.samsung.com/resources/data-sheet/20170731_DDR4_8Gb_B_die_Registered_DIMM_Rev1.91_May.17.pdf},
  2017.

\bibitem{samsungm471a5143eb0cpb}
Samsung.
\newblock {M471A5143EB0-CPB Specifications}.
\newblock
  \url{https://semiconductor.samsung.com/dram/module/sodimm/m471a5143eb0-cpb/}.

\bibitem{corsaircmk16gx4m2b3200c16}
CORSAIR.
\newblock {CMK16GX4M2B3200C16}.
\newblock
  \url{https://www.corsair.com/eu/en/Categories/Products/Memory/VENGEANCE-LPX/p/CMK16GX4M2B3200C16}.

\bibitem{samsung2018260pin}
Samsung.
\newblock {260pin Unbuffered SODIMM based on 8Gb C-die}.
\newblock
  \url{https://semiconductor.samsung.com/resources/data-sheet/DDR4_8Gb_C_die_Unbuffered_SODIMM_Rev1.5_Apr.18.pdf},
  2018.

\bibitem{kingston2020ksm32rd816hdr}
Kingston.
\newblock {KSM32RD8/16HDR Specifications}.
\newblock \url{https://www.kingston.com/dataSheets/KSM32RD8_16HDR.pdf}, 2020.

\bibitem{memorynethmaa4gu6ajr8nxn}
Memory.NET.
\newblock {HMAA4GU6AJR8N-XN Specifications}.
\newblock
  \url{https://memory.net/product/hmaa4gu6ajr8n-xn-sk-hynix-1x-32gb-ddr4-3200-udimm-pc4-25600u-dual-rank-x8-module/}.

\bibitem{vandegoor2002address}
Ad~J Van De~Goor and Ivo Schanstra.
\newblock {Address and Data Scrambling: Causes and Impact on Memory Tests}.
\newblock In {\em DELTA}, 2002.

\bibitem{patel2022acase}
Minesh Patel, Taha Shahroodi, Aditya Manglik, A.~Giray
  Ya{\u{g}}l{\i}k{\c{c}}{\i}, Ataberk Olgun, Haocong Luo, and Onur Mutlu.
\newblock {A Case for Transparent Reliability in DRAM Systems}.
\newblock arXiv:2204.10378 [cs.AR], 2022.

\bibitem{sinha2012exploring}
Saurabh Sinha, Greg Yeric, Vikas Chandra, Brian Cline, and Yu~Cao.
\newblock {Exploring Sub-20nm FinFET Design with Predictive Technology Models}.
\newblock In {\em DAC}, 2012.

\bibitem{zhao2006newgeneration}
Wei Zhao and Yu~Cao.
\newblock {New Generation of Predictive Technology Model for Sub-45 nm Early
  Design Exploration}.
\newblock {\em IEEE TED}, 2006.

\bibitem{semiconductors2015itrs}
{International Technology Roadmap for Semiconductors}.
\newblock {ITRS Reports}.
\newblock \url{http://www.itrs2.net/itrs-reports.html}, 2015.

\bibitem{everitt1998dram_circuit}
Brian Everitt.
\newblock {\em {DRAM Circuit Design: Fundamental and High-Speed Topics}}.
\newblock Cambridge University Press, 1998.

\bibitem{arahmati2014refreshing}
Amir Rahmati, Matthew Hicks, Daniel Holcomb, and Kevin Fu.
\newblock {Refreshing Thoughts on DRAM: Power Saving vs. Data Integrity}.
\newblock In {\em WACAS}, 2014.

\bibitem{bittware_xusp3s}
{Xilinx}.
\newblock {Bittware XUSP3S FPGA Board}.
\newblock \url{https://www.bittware.com/fpga/xus-p3s/}.

\bibitem{hynixh5anag8najrxn}
SK~Hynix.
\newblock {H5ANAG8NAJR-XN Specifications}.
\newblock \url{https://www.memory-distributor.com/h5anag8najr-xnc.html}.

\bibitem{memorynethmaa4gu7cjr8nxn}
Memory.NET.
\newblock {HMAA4GU7CJR8N-XN Specifications}.
\newblock
  \url{https://memory.net/product/hmaa4gu7cjr8n-xn-sk-hynix-1x-32gb-ddr4-3200-ecc-udimm-pc4-25600e-dual-rank-x8-module/}.

\bibitem{hynixh5anag8ncjrxn}
SK~Hynix.
\newblock {H5ANAG8NCJR-XN Specifications}.
\newblock \url{https://www.memory-distributor.com/h5anag8ncjr-xnc.html}.

\bibitem{hynixh5an8g8ndjrxnc}
SK~Hynix.
\newblock {H5AN8G8NDJR-XNC Specifications}.
\newblock \url{https://www.memory-distributor.com/h5an8g8ndjr-xnc.html}.

\bibitem{micronmta4atf1g64hz3g2e1}
Micron.
\newblock {MTA4ATF1G64HZ-3G2E1 Specifications}.
\newblock
  \url{https://media-www.micron.com/-/media/client/global/documents/products/data-sheet/modules/sodimm/ddr4/atf4c1gx64hz.pdf?rev=f436f917f8d74c08bf32a576a15b5e66}.

\bibitem{micronmt40a1g16kd062e}
Micron.
\newblock {MT40A1G16KD-062E Specifications}.
\newblock
  \url{https://www.micron.com/-/media/client/global/documents/products/data-sheet/dram/ddr4/16gb_ddr4_sdram.pdf}.

\bibitem{micronmta18asf2g72pz2g3b1qk}
Micron.
\newblock {MTA18ASF2G72PZ-2G3B1QK Specifications}.
\newblock
  \url{https://media-www.micron.com/-/media/client/global/documents/products/data-sheet/modules/rdimm/ddr4/asf18c2gx72pz.pdf?rev=6ef1f46c2b2e4e95824c859fd05c01b5}.

\bibitem{micronmt40a2g4we083eb}
Micron.
\newblock {MT40A2G4WE-083E:B Specifications}.
\newblock
  \url{https://www.micron.com/products/dram/ddr4-sdram/part-catalog/mt40a2g4we-083e}.

\bibitem{micronmta36asf8g72pz2g9e1ti}
Micron.
\newblock {MTA36ASF8G72PZ-2G9E1TI Specifications}.
\newblock
  \url{https://www.micron.com/products/dram-modules/rdimm/part-catalog/mta36asf8g72pz-2g9/mta36asf8g72pz-2g9e1}.

\bibitem{micronmt40a4g4jc062ee}
Micron.
\newblock {MT40A4G4JC-062E:E Specifications}.
\newblock
  \url{https://eu.mouser.com/datasheet/2/671/mict_s_a0010972464_1-2291055.pdf}.

\bibitem{micronmta4atf1g64hz3g2b2}
Micron.
\newblock {MTA4ATF1G64HZ-3G2B2 Specifications}.
\newblock
  \url{https://media-www.micron.com/-/media/client/global/documents/products/data-sheet/modules/sodimm/ddr4/atf4c1gx64hz.pdf?rev=f436f917f8d74c08bf32a576a15b5e66}.

\bibitem{micronmt40a1g16rc062eb}
Micron.
\newblock {MT40A1G16RC-062E:B Specifications}.
\newblock
  \url{https://www.micron.com/products/dram/ddr4-sdram/part-catalog/mt40a1g16rc-062e}.

\bibitem{samsungm393a1k43bb1ctd}
Samsung.
\newblock {M393A1K43BB1-CTD Specifications}.
\newblock
  \url{https://semiconductor.samsung.com/dram/module/rdimm/m393a1k43bb1-ctd/}.

\bibitem{samsungk4a8g085wbbctd}
Samsung.
\newblock {K4A8G085WB-BCTD Specifications}.
\newblock
  \url{https://semiconductor.samsung.com/dram/ddr/ddr4/k4a8g085wb-bctd/}.

\bibitem{samsungm393a2k40cb2ctd}
Samsung.
\newblock {M393A2K40CB2-CTD Specifications}.
\newblock
  \url{https://semiconductor.samsung.com/dram/module/rdimm/m393a2k40cb2-ctd/}.

\bibitem{samsungk4a8g045wcbctd}
Samsung.
\newblock {K4A8G045WC-BCTD Specifications}.
\newblock
  \url{https://semiconductor.samsung.com/emea/dram/ddr/ddr4/k4a8g045wc-bctd/}.

\bibitem{jedec2012jesd793f}
{JEDEC}.
\newblock {\em {JESD79-3F: DDR3 SDRAM Specification}}, 2012.

\bibitem{ghose2019demystifying}
Saugata Ghose, Tianshi Li, Nastaran Hajinazar, Damla~Senol Cali, and Onur
  Mutlu.
\newblock {Demystifying Complex Workload--DRAM Interactions: An Experimental
  Study}.
\newblock In {\em {SIGMETRICS}}, 2019.

\bibitem{faber2012statistics}
Michael~Havbro Faber.
\newblock {\em {Statistics and Probability Theory}}.
\newblock Springer, 2012.

\bibitem{hartigan1979algorithm}
John~A Hartigan and Manchek~A Wong.
\newblock {Algorithm AS 136: A K-Means Clustering Algorithm}.
\newblock {\em J. R. Stat. Soc. C-Appl.}, 1979.

\bibitem{rousseeuw1987silhouettes}
Peter~J Rousseeuw.
\newblock {Silhouettes: A Graphical Aid to the Interpretation and Validation of
  Cluster Analysis}.
\newblock {\em J. Comput. Appl. Math.}, 1987.

\bibitem{aalst2010process}
Wil~M.P. van~der Aalst.
\newblock {\em {Process Mining Discovery, Conformance and Enhancement of
  Business Processes}}.
\newblock Springer, 2010.

\bibitem{seshadri2012theevictedaddress}
Vivek Seshadri, Onur Mutlu, Michael~A. Kozuch, and Todd~C. Mowry.
\newblock {The Evicted-Address Filter: A Unified Mechanism to Address Both
  Cache Pollution and Thrashing}.
\newblock In {\em PACT}, 2012.

\bibitem{hong2018attache}
Seokin Hong, Prashant~J. Nair, Bulent Abali, Alper Buyuktosunoglu, Kyu-Hyoun
  Kim, and Michael~B. Healy.
\newblock {Attaché: Towards Ideal Memory Compression by Mitigating Metadata
  Bandwidth Overheads}.
\newblock In {\em MICRO}, 2018.

\bibitem{meza2012enabling}
Justin Meza, Jichuan Chang, HanBin Yoon, Onur Mutlu, and Parthasarathy
  Ranganathan.
\newblock {Enabling Efficient and Scalable Hybrid Memories Using
  Fine-Granularity DRAM Cache Management}.
\newblock {\em IEEE {CAL}}, 2012.

\bibitem{balasubramonian2017cacti}
Rajeev Balasubramonian, Andrew~B. Kahng, Naveen Muralimanohar, Ali Shafiee, and
  Vaishnav Srinivas.
\newblock {CACTI 7: New Tools for Interconnect Exploration in Innovative
  Off-Chip Memories}.
\newblock {\em ACM TACO}, 2017.

\bibitem{wikichipcascade}
WikiChip.
\newblock {Cascade Lake SP - Intel}.
\newblock \url{https://en.wikichip.org/wiki/intel/cores/cascade\_lake\_sp}.

\bibitem{safariramulator}
{SAFARI Research Group}.
\newblock {Ramulator --- GitHub Repository}.
\newblock \url{https://github.com/CMU-SAFARI/ramulator}.

\bibitem{Kim2016Ramulator}
Yoongu Kim, Weikun Yang, and Onur Mutlu.
\newblock {Ramulator: A Fast and Extensible DRAM Simulator}.
\newblock {\em IEEE CAL}, 2016.

\bibitem{zuravleff1997controller}
William~K Zuravleff and Timothy Robinson.
\newblock {Controller for a Synchronous DRAM That Maximizes Throughput by
  Allowing Memory Requests and Commands to Be Issued Out of Order}, 1997.
\newblock US Patent 5,630,096.

\bibitem{stspec}
{Standard Performance Evaluation Corp.}
\newblock {SPEC CPU 2006}.
\newblock \url{http://www.spec.org/cpu2006/}.

\bibitem{st2017spec}
{Standard Performance Evaluation Corp.}
\newblock {SPEC CPU 2017}.
\newblock \url{http://www.spec.org/cpu2017}, 2017.

\bibitem{transaction}
Transaction Processing~Performance Council.
\newblock {\em {TPC-C, TPC-H}}.

\bibitem{fritts2009mediabench}
Jason~E. Fritts, Frederick~W. Steiling, Joseph~A. Tucek, and Wayne Wolf.
\newblock {MediaBench II Video: Expediting the next Generation of Video Systems
  Research}.
\newblock {\em MICPRO}, 2009.

\bibitem{cooper2010benchmarking}
Brian Cooper, Adam Silberstein, Erwin Tam, Raghu Ramakrishnan, and Russell
  Sears.
\newblock {Benchmarking Cloud Serving Systems with YCSB}.
\newblock In {\em SoCC}, 2010.

\bibitem{snavely2000symbiotic}
Allan Snavely and Dean~M Tullsen.
\newblock {Symbiotic Job Scheduling for A Simultaneous Multithreaded
  Processor}.
\newblock In {\em ASPLOS}, 2000.

\bibitem{eyerman2008systemlevel}
Stijn Eyerman and Lieven Eeckhout.
\newblock {System-Level Performance Metrics for Multiprogram Workloads}.
\newblock {\em IEEE Micro}, 2008.

\bibitem{michaud2012demystifying}
Pierre Michaud.
\newblock {Demystifying Multicore Throughput Metrics}.
\newblock {\em {IEEE CAL}}, 2012.

\bibitem{luo2001balancing}
Kun Luo, Jayanth Gummaraju, and Manoj Franklin.
\newblock {Balancing Thoughput and Fairness in SMT Processors}.
\newblock In {\em ISPASS}, 2001.

\bibitem{subramanian2013mise}
Lavanya Subramanian, Vivek Seshadri, Yoongu Kim, Ben Jaiyen, and Onur Mutlu.
\newblock {MISE: Providing Performance Predictability and Improving Fairness in
  Shared Main Memory Systems}.
\newblock In {\em HPCA}, 2013.

\bibitem{subramanian2015theapplication}
Lavanya Subramanian, Vivek Seshadri, Arnab Ghosh, Samira Khan, and Onur Mutlu.
\newblock {The Application Slowdown Model: Quantifying and Controlling the
  Impact of Inter-Application Interference at Shared Caches and Main Memory}.
\newblock In {\em MICRO}, 2015.

\bibitem{das2009applicationaware}
Reetuparna Das, Onur Mutlu, Thomas Moscibroda, and Chita~R Das.
\newblock {Application-Aware Prioritization Mechanisms for On-Chip Networks}.
\newblock In {\em MICRO}, 2009.

\bibitem{das2013applicationtocore}
Reetuparna Das, Rachata Ausavarungnirun, Onur Mutlu, Akhilesh Kumar, and Mani
  Azimi.
\newblock {Application-to-Core Mapping Policies to Reduce Memory System
  Interference in Multi-Core Systems}.
\newblock In {\em HPCA}, 2013.

\bibitem{datasheetksm32rd8}
Kingston.
\newblock {KSM32RD8/16HDR Specifications}.
\newblock \url{https://www.kingston.com/dataSheets/KSM32RD8_16HDR.pdf}, 2020.

\bibitem{hussain2014pmss}
Tassadaq Hussain, Amna Haider, and Eduard Ayguadé.
\newblock {PMSS: A Programmable Memory System and Scheduler for Complex Memory
  Patterns}.
\newblock {\em JPDC}, 2014.

\bibitem{bojnordi2012pardis}
Mahdi~Nazm Bojnordi and Engin Ipek.
\newblock {PARDIS: A Programmable Memory Controller for the DDRx Interfacing
  Standards}.
\newblock In {\em ISCA}, 2012.

\bibitem{xilinx2021adaptable}
{Xilinx}.
\newblock {Adaptable Accelerator Cards for Data Center Workloads}, 2021.

\bibitem{xilinx2021fpgas}
{Xilinx {Inc.}}
\newblock {FPGAs and 3D ICs}, 2021.

\bibitem{wikichipcore}
WikiChip.
\newblock {Core i7-5960X Extreme Edition - Intel}.

\bibitem{nguyen2018nonblocking}
Kate Nguyen, Kehan Lyu, Xianze Meng, Vilas Sridharan, and Xun Jian.
\newblock {Nonblocking Memory Refresh}.
\newblock In {\em ISCA}, 2018.

\bibitem{jedec2012jep122g}
{JEDEC}.
\newblock {\em {JEP122G: Failure Mechanisms and Models for Semiconductor
  Devices}}, 2012.

\bibitem{micheloni2015apparatus}
R.~Micheloni, P.Z. Onufryk, A.~Marelli, C.I.W. Norrie, and I.~Jaser.
\newblock {Apparatus and Method Based on LDPC Codes for Adjusting a Correctable
  Raw Bit Error Rate Limit in a Memory System}.
\newblock {US Patent 9,092,353}, 2015.

\bibitem{intel3rdgen}
{Intel Inc.}
\newblock {3rd Gen Intel Xeon Scalable Processors}.
\newblock
  \url{https://www.intel.com/content/dam/www/public/us/en/documents/a1171486-icelake-productbrief-updates-r1v2.pdf}.

\bibitem{amdepyc}
{AMD Inc.}
\newblock {AMD EPYC™ 7003 Series Processors}.
\newblock
  \url{https://www.amd.com/system/files/documents/amd-epyc-7003-series-datasheet.pdf}.

\bibitem{micron2020tn4040}
{Micron}.
\newblock {TN-40-40: DDR4 Point-to-Point Design Guide}.
\newblock
  \url{https://media-www.micron.com/-/media/client/global/documents/products/technical-note/dram/tn4040_ddr4_point_to_point_design_guide.pdf?rev=d58bc222192d411aae066b2577a12677},
  2020.

\bibitem{statista2022dram_manufacturers}
Statista.
\newblock {DRAM Manufacturers Revenue Share Worldwide From 2011 to 2022, by
  Quarter}.
\newblock
  \url{https://www.statista.com/statistics/271726/global-market-share-held-by-dram-chip-vendors-since-2010/},
  2022.

\bibitem{k2015memory}
Mahmut Kandemir, Hui Zhao, Xulong Tang, and Mustafa Karakoy.
\newblock {Memory Row Reuse Distance and Its Role in Optimizing Application
  Performance}.
\newblock In {\em SIGMETRICS}, 2015.

\bibitem{bloom1970spacetime}
Burton~H Bloom.
\newblock {Space/Time Trade-Offs in Hash Coding with Allowable Errors}.
\newblock {\em CACM}, 1970.

\bibitem{fan2000summary}
Li~Fan, Pei Cao, Jussara Almeida, and Andrei~Z. Broder.
\newblock {Summary Cache: A Scalable Wide-Area Web Cache Sharing Protocol}.
\newblock {\em Transactions on Networking}, 2000.

\bibitem{li2012compression}
Zhaogeng Li, Jun Bi, Sen Wang, and Xiaoke Jiang.
\newblock {Compression of Pending Interest Table with Bloom Filter in Content
  Centric Network}.
\newblock In {\em CFI}, 2012.

\bibitem{carter1979universal}
J~Carter and M~Wegman.
\newblock {Universal Classes of Hash Functions}.
\newblock {\em JCSS}, 1979.

\bibitem{wolframresearchwolframalpha}
{Wolfram Research, Inc.}
\newblock {WolframAlpha}.
\newblock \url{http://www.wolframalpha.com/}.

\bibitem{muralimanohar2009cacti}
Naveen Muralimanohar, Rajeev Balasubramonian, and Norman~P. Jouppi.
\newblock {CACTI 6.0: A Tool to Model Large Caches}.
\newblock 2009.

\bibitem{synopsyssynopsys}
{Synopsys, Inc.}
\newblock {Synopsys Design Compiler}.
\newblock
  \url{https://www.synopsys.com/support/training/rtl-syndesign-compiler-rtl-synthesis.html}.

\bibitem{chdrampower_opensource}
Karthik Chandrasekar, Christian Weis, Yonghui Li, Sven Goossens, Matthias Jung,
  Omar Naji, Benny Akesson, Norbert Wehn, and Kees Goossens.
\newblock {DRAMPower: Open-Source DRAM Power \& Energy Estimation Tool}.
\newblock \url{http://www.drampower.info/}.

\bibitem{misra1982finding}
Jayadev Misra and David Gries.
\newblock {Finding Repeated Elements}.
\newblock {\em {Science of Computer Programming}}, 1982.

\bibitem{semiconductorsqoriq}
{NXP Semiconductors}.
\newblock {QorIQ Processing Platforms: 64-Bit Multicore SoCs}.
\newblock
  \url{https://www.nxp.com/products/processors-and-microcontrollers/applications-processors/qoriq-platforms:QORIQ_HOME}.

\bibitem{kasture2016tailbench}
Harshad Kasture and Daniel Sanchez.
\newblock {TailBench: A Benchmark Suite and Evaluation Methodology for
  Latency-Critical Applications}.
\newblock In {\em IISWC}, 2016.

\bibitem{gruss2016rowhammerjs_a}
Daniel Gruss, Cl{\'e}mentine Maurice, and Stefan Mangard.
\newblock {Rowhammer.js: A Remote Software-Induced Fault Attack in Javascript}.
\newblock In {\em DIMVA}, 2016.

\bibitem{zhang2019telehammer}
Zhi Zhang, Yueqiang Cheng, Dongxi Liu, Surya Nepal, and Zhi Wang.
\newblock {TeleHammer: A Stealthy Cross-Boundary Rowhammer Technique}.
\newblock arXiv:1912.03076 [cs.CR], 2019.

\bibitem{wulf1995hitting}
Wm.~A. Wulf and Sally~A. McKee.
\newblock {Hitting the Memory Wall: Implications of the Obvious}.
\newblock {\em SIGARCH Computer Architecture News}, 23, 1995.

\bibitem{sites1996itsthe}
Richard Sites.
\newblock {It's the Memory, Stupid}.
\newblock {\em Microprocessor Report}, 1996.

\bibitem{wilkes2001thememory}
Maurice~V. Wilkes.
\newblock {The Memory Gap and the Future of High Performance Memories}.
\newblock {\em SIGARCH Computer Architecture News}, 29, 2001.

\bibitem{mutlu2020amodern}
Onur Mutlu, Saugata Ghose, Juan Gomez-Luna, and Rachata Ausavarungnirun.
\newblock {A Modern Primer on Processing in Memory}.
\newblock In {\em {arXiv:2012.03112 [cs.AR]}}. 2020.

\bibitem{mutlu2021amodern}
Onur Mutlu, Saugata Ghose, Juan G{\'o}mez-Luna, and Rachata Ausavarungnirun.
\newblock {A Modern Primer on Processing in Memory}.
\newblock In {\em Emerging Computing: From Devices to Systems --- Looking
  Beyond Moore and Von Neumann}. 2021.

\bibitem{kanev2015profiling}
Svilen Kanev, Juan~Pablo Darago, Kim Hazelwood, Parthasarathy Ranganathan, Tipp
  Moseley, Gu-Yeon Wei, and David Brooks.
\newblock {Profiling a Warehouse-Scale Computer}.
\newblock In {\em ISCA}, 2015.

\bibitem{wang2014bigdatabench}
Lei Wang, Jianfeng Zhan, Chunjie Luo, Yuqing Zhu, Qiang Yang, Yongqiang He,
  Wanling Gao, Zhen Jia, Yingjie Shi, Shujie Zhang, Chen Zheng, Gang Lu, Kent
  Zhan, Xiaona Li, and Bizhu Qiu.
\newblock {BigDataBench: A Big Data Benchmark Suite from Internet Services}.
\newblock In {\em HPCA}, 2014.

\bibitem{boroum2018google}
Amirali Boroumand, Saugata Ghose, Youngsok Kim, Rachata Ausavarungnirun, Eric
  Shiu, Rahul Thakur, Daehyun Kim, Aki Kuusela, Allan Knies, Parthasarathy
  Ranganathan, and Onur Mutlu.
\newblock {Google Workloads for Consumer Devices: Mitigating Data Movement
  Bottlenecks}.
\newblock In {\em ASPLOS}, 2018.

\bibitem{oliveira2021anew}
Geraldo~F. Oliveira, Juan Gómez-Luna, Lois Orosa, Saugata Ghose, Nandita
  Vijaykumar, Ivan Fernandez, Mohammad Sadrosadati, and Onur Mutlu.
\newblock {A New Methodology and Open-Source Benchmark Suite for Evaluating
  Data Movement Bottlenecks: A Near-Data Processing Case Study}.
\newblock In {\em SIGMETRICS}, 2021.

\bibitem{gomezluna2021benchmarkingarxiv}
Juan G{\'o}mez-Luna, Izzat~El Hajj, Ivan Fernandez, Christina Giannoula,
  Geraldo~F Oliveira, and Onur Mutlu.
\newblock {Benchmarking a New Paradigm: An Experimental Analysis of a Real
  Processing-in-Memory Architecture}.
\newblock arXiv:2105.03814 [cs.AR], 2021.

\bibitem{aven2009identification}
Terje Aven.
\newblock {Identification of Safety and Security Critical Systems and
  Activities}.
\newblock {\em Reliability Engineering \& System Safety}, 2009.

\bibitem{jedec2012jesd794}
JEDEC.
\newblock {JESD79-4: DDR4 SDRAM Standard}.
\newblock {\em Joint Electron Device Engineering Council}, 2012.

\bibitem{hassan2022acase}
Hasan Hassan, Ataberk Olgun, A~Giray Ya{\u{g}}l{\i}k{\c{c}}{\i}, Haocong Luo,
  and Onur Mutlu.
\newblock {A Case for Self-Managing DRAM Chips: Improving Performance,
  Efficiency, Reliability, and Security via Autonomous In-DRAM Maintenance
  Operations}.
\newblock {\em arXiv:2207.13358v1 [cs.AR]}, 2022.

\bibitem{safari2022selfmanaging}
{SAFARI Research Group}.
\newblock {Self-Managing DRAM (SMD) Source Code}.
\newblock \url{https://github.com/CMU- SAFARI/SelfManagingDRAM}, 2022.

\bibitem{saino2000impact}
K.~Saino, S.~Horiba, S.~Uchiyama, Y.~Takaishi, M.~Takenaka, T.~Uchida,
  Y.~Takada, K.~Koyama, H.~Miyake, and C.~Hu.
\newblock {Impact of Gate-Induced Drain Leakage Current on the Tail
  Distribution of DRAM Data Retention Time}.
\newblock In {\em IEDM}, 2000.

\bibitem{gong2018duoexposing}
Seong-Lyong Gong, Jungrae Kim, Sangkug Lym, Michael Sullivan, Howard David, and
  Mattan Erez.
\newblock {DUO: Exposing On-Chip Redundancy to Rank-Level ECC for High
  Reliability}.
\newblock In {\em HPCA}, 2018.

\bibitem{jacob2010memory}
Bruce Jacob, Spencer Ng, and David Wang.
\newblock {\em {Memory Systems: Cache, DRAM, Disk}}.
\newblock 2010.

\bibitem{mukherjee2004cache}
Shubhendu~S Mukherjee, Joel Emer, Tryggve Fossum, and Steven~K Reinhardt.
\newblock {Cache Scrubbing in Microprocessors: Myth or Necessity?}
\newblock In {\em SDC}, 2004.

\bibitem{rooney2019micron}
Randall Rooney and Neal Koyle.
\newblock {Micron DDR5 SDRAM: New Features}.
\newblock Technical report, 2019.

\bibitem{saleh1990reliability}
Abdallah~M Saleh, Juan~J Serrano, and Janak~H Patel.
\newblock {Reliability of Scrubbing Recovery-Techniques for Memory Systems}.
\newblock {\em TR}, 1990.

\bibitem{siddiqua2017lifetime}
Taniya Siddiqua, Vilas Sridharan, Steven~E Raasch, Nathan DeBardeleben, Kurt~B
  Ferreira, Scott Levy, Elisabeth Baseman, and Qiang Guan.
\newblock {Lifetime Memory Reliability Data from the Field}.
\newblock In {\em DFT}, 2017.

\bibitem{yuksel2024simra}
Ismail~Emir Yuksel, Yahya~Can Tu\u{g}rul, F~Nisa~Bostanci, Geraldo~F Oliveira,
  A~Giray~Ya{\u{g}}l{\i}k{\c{c}}{\i}, Ataberk Olgun, Melina Soysal, Haocong
  Luo, Juan G{\'o}mez-Luna, Mohammad Sadrosadati, and Onur Mutlu.
\newblock {Simultaneous Many-Row Activation in Off-the-Shelf DRAM Chips:
  Experimental Characterization and Analysis}.
\newblock In {\em {DSN}}, 2024.

\bibitem{canpolat2024breakhammer}
O{\u{g}}uzhan Canpolat, A.~Giray Ya{\u{g}}l{\i}k{\c{c}}{\i}, Ataberk Olgun,
  Ismail~Emir Yüksel, Yahya~Can Tu{\u{g}}rul, Konstantinos Kanellopoulos,
  O{\u{g}}uz Ergin, and Onur Mutlu.
\newblock {BreakHammer: Enabling Scalable and Low Overhead RowHammer
  Mitigations via Throttling Preventive Action Triggering Threads}.
\newblock In {\em MICRO}, 2024.

\bibitem{lim2008understanding}
Kevin Lim, Parthasarathy Ranganathan, Jichuan Chang, Chandrakant Patel, Trevor
  Mudge, and Steve Reinhardt.
\newblock {Understanding and Designing New Server Architectures for Emerging
  Warehouse-Computing Environments}.
\newblock In {\em ISCA}, 2008.

\bibitem{lim2009disaggregated}
Kevin Lim, Jichuan Chang, Trevor Mudge, Parthasarathy Ranganathan, Steven~K
  Reinhardt, and Thomas~F Wenisch.
\newblock {Disaggregated Memory for Expansion and Sharing in Blade Servers}.
\newblock In {\em ISCA}, 2009.

\bibitem{calciu2021rethinking}
Irina Calciu, M~Talha Imran, Ivan Puddu, Sanidhya Kashyap, Hasan~Al Maruf, Onur
  Mutlu, and Aasheesh Kolli.
\newblock {Rethinking Software Runtimes for Disaggregated Memory}.
\newblock In {\em ASPLOS}, 2021.

\bibitem{tugrul2024understanding}
Yahya~Can Tu\u{g}rul, Giray Ya\u{g}likci, Ismail~Emir Yuksel, Ataberk Olgun,
  O\u{g}uzhan Canpolat, Nisa Bostanci, Mohammad Sadrosadati, O\u{g}uz Ergin,
  and Onur Mutlu.
\newblock {Understanding RowHammer Under Reduced Refresh Latency: Experimental
  Analysis of Real DRAM Chips and Implications on Future Solutions}.
\newblock {\em arXiv}, 2024.

\bibitem{deoliveira2024mimdram}
Geraldo~F Oliveira, Ataberk Olgun, Abdullah~Giray Ya{\u{g}}l{\i}k{\c{c}}{\i},
  F~Nisa Bostanc{\i}, Juan G{\'o}mez-Luna, Saugata Ghose, and Onur Mutlu.
\newblock {MIMDRAM: An End-to-End Processing-Using-DRAM System for
  High-Throughput, Energy-Efficient and Programmer-Transparent
  Multiple-Instruction Multiple-Data Computing}.
\newblock In {\em HPCA}, 2024.

\bibitem{hajyahya2020sysscale}
Jawad Haj-Yahya, Mohammed Alser, Jeremie Kim, A.~Giray Yaglıkçı, Nandita
  Vijaykumar, Efraim Rotem, and Onur Mutlu.
\newblock {SysScale: Exploiting Multi-Domain Dynamic Voltage and Frequency
  Scaling for Energy Efficient Mobile Processors}.
\newblock In {\em ISCA}, 2020.

\bibitem{hajyahya2022darkgates}
Jawad~Haj Yahya, Jeremie~S Kim, A~Giray Ya{\u{g}}l{\i}k{\c{c}}{\i}, Jisung
  Park, Efraim Rotem, Yanos Sazeides, and Onur Mutlu.
\newblock {DarkGates: A Hybrid Power-Gating Architecture to Mitigate the
  Performance Impact of Dark-Silicon in High-Performance Processors}.
\newblock In {\em HPCA}, 2022.

\bibitem{hajyahya2021ichannels}
Jawad Haj-Yahya, Lois Orosa, Jeremie~S Kim, Juan~G{\'o}mez Luna, A~Giray
  Ya{\u{g}}l{\i}k{\c{c}}{\i}, Mohammed Alser, Ivan Puddu, and Onur Mutlu.
\newblock {IChannels: Exploiting Current Management Mechanisms to Create Covert
  Channels in Modern Processors}.
\newblock In {\em ISCA}, 2021.

\bibitem{bostanci2022drstrange}
F.~Nisa Bostanc{\i}, Ataberk Olgun, Lois Orosa, A.~Giray
  Ya{\u{g}}l{\i}k{\c{c}}{\i}, Jeremie~S. Kim, Hasan Hassan, O{\u{g}}uz Ergin,
  and Onur Mutlu.
\newblock {DR-STRaNGe: End-to-End System Design for DRAM-Based True Random
  Number Generators}.
\newblock In {\em HPCA}, 2022.

\bibitem{mfactorsjet5467a}
MFactors.
\newblock {JET-5467A Product Page}.
\newblock
  \url{http://www.mfactors.com/jet-5467a-ddr3-sodimm-extender-with-currentsensing/}.

\bibitem{technologies34134a}
{Keysight Technologies}.
\newblock {34134A AC/DC DMM Current Probe: User's Guide}.
\newblock \url{https://literature.cdn.keysight.com/
  litweb/pdf/34134-90001.pdf}, 2009.

\bibitem{technologies34461a}
{Keysight Truevolt Series Digital Multimeters: Operating and Service Guide}.
\newblock {34461A 6.5 Digit Multimeter, Truevolt DMM}.
\newblock \url{https://literature.cdn.keysight.com/litweb/pdf/34460-90901.pdf},
  2017.

\bibitem{xilinxu50}
{Xilinx}.
\newblock {Xilinx Alveo U50 FPGA Board}.
\newblock \url{https://www.xilinx.com/products/boards-and-kits/alveo/u50.html}.

\bibitem{arduinomega}
Arduino.
\newblock {Arduino MEGA Documentation}.
\newblock \url{https://docs.arduino.cc/hardware/mega-2560/}, 2009.

\end{thebibliography}
\end{singlespace}

\bookmarksetup{startatroot}
\end{document}